\documentclass[a4paper,twoside]{ociamthesis}

\usepackage[style=alphabetic, sorting=nyt, backend=biber, doi=false, isbn=false]{biblatex}
\newcommand*{\bibtitle}{References}

\numberwithin{equation}{section}

\DeclareSymbolFontAlphabet{\mathbbvar}{bbold}
\DeclareSymbolFontAlphabet{\mathbb}{AMSb}
\newcommand{\M}{\mathcal{M}}
\newcommand{\BU}{\mathbf{B}U(1)_\mathrm{conn}}
\newcommand{\Coo}{\mathcal{C}^{\infty}}
\newcommand{\Aut}{\mathrm{Aut}}
\newcommand{\Diff}{\mathrm{Diff}}
\newcommand{\SDiff}{\mathrm{SDiff}}
\newcommand{\ISO}{\mathrm{ISO}}
\newcommand{\SO}{\mathrm{SO}}
\newcommand{\GL}{\mathrm{GL}}

\newcommand{\B}{\mathbf{B}}

\newcommand{\II}{\mathrm{II}}
\newcommand{\IIA}{\mathrm{IIA}}
\newcommand{\IIB}{\mathrm{IIB}}

\usepackage{slashed}
\newcommand{\Lim}[1]{\raisebox{0.5ex}{\scalebox{0.8}{$\displaystyle \lim_{#1}\;$}}}
\newcommand{\di}{\mathrm{d}}

\newcommand{\bosonic}[1]{\overset{\rightsquigarrow}{#1}}

\newcommand{\tr}{\mathrm{tr}}
\newcommand{\String}{\mathrm{String}}
\newcommand{\Grpd}{\infty\mathbf{Grpd}}
\newcommand{\Struc}{\mathrm{Struc}}
\newcommand{\Cartan}{\mathrm{Cartan}}
\newcommand{\Func}{\mathbf{Func}}
\newcommand{\SugraIIA}{\mathrm{Sugra}_{\IIA}}
\newcommand{\SugraIIB}{\mathrm{Sugra}_{\IIB}}

\newcommand{\tenhie}{\mathscr{T\!\!H}}
\newcommand{\ind}[1]{\mathcal{#1}}
\newcommand{\bigsp}[1]{\underline{#1}}
\newcommand{\frgt}{\mathrm{frgt}}

\DeclareRobustCommand\longtwoheadrightarrow
    {\relbar\joinrel\twoheadrightarrow}
\DeclareRobustCommand\longhookrightarrow
     {\lhook\joinrel\longrightarrow}
\newcommand{\xtwoheadrightarrow}[2][]{%
  \mathrel{\ooalign{$\xrightarrow[#1\mkern4mu]{#2\mkern4mu}$\cr%
  \hidewidth$\rightarrow\mkern4mu$}}
}
\theoremstyle{definition}
\newtheorem{theorem}{Lemma}[section]
\theoremstyle{definition}
\newtheorem{definition}[theorem]{Definition}
\theoremstyle{definition}
\newtheorem{post}[theorem]{Postulate}
\theoremstyle{definition}
\newtheorem{remark}[theorem]{Remark}
\theoremstyle{definition}
\newtheorem{example}[theorem]{Example}
\theoremstyle{definition}
\newtheorem{digression}[theorem]{Digression}
\tikzcdset{%
    triple line/.code={\tikzset{%
        double equal sign distance, 
        double=\pgfkeysvalueof{/tikz/commutative diagrams/background color}}},
    quadruple line/.code={\tikzset{%
        double equal sign distance, 
        double=\pgfkeysvalueof{/tikz/commutative diagrams/background color}}},
    Rrightarrow/.code={\tikzcdset{triple line}\pgfsetarrows{tikzcd implies cap-tikzcd implies}},
    RRightarrow/.code={\tikzcdset{quadruple line}\pgfsetarrows{tikzcd implies cap-tikzcd implies}}
}    
\newcommand*{\tarrow}[2][]{\arrow[Rrightarrow, #1]{#2}\arrow[dash, shorten >= 0.5pt, #1]{#2}}

\makeatletter
\providecommand{\leftsquigarrow}{%
  \mathrel{\mathpalette\reflect@squig\relax}%
}
\newcommand{\reflect@squig}[2]{%
  \reflectbox{$\m@th#1\rightsquigarrow$}%
}
\makeatother
\usepackage{hyperref}
\usetikzlibrary{decorations.markings}
\usepackage{pdflscape}
\usepackage{newunicodechar}
\newunicodechar{よ}{\text{\usefont{U}{min}{m}{n}\symbol{'210}}}
\DeclareFontFamily{U}{min}{}
\DeclareFontShape{U}{min}{m}{n}{<-> udmj30}{}
\usepackage{CJKutf8}
\newcommand{\chinese}[1]{\begin{CJK}{UTF8}{bsmi}#1\end{CJK}}
\usepackage{bookmark}

\title{Extended Field Theories \\ as higher Kaluza-Klein theories}
\author{Luigi Alfonsi}

\degree{Degree of Doctor of Philosophy}
\degreedate{April 19, 2021}

\begin{document}

\setlength{\textbaselineskip}{15pt plus 0.5pt minus 0.5pt}

\setlength{\frontmatterbaselineskip}{15pt plus1pt minus1pt}

\setlength{\baselineskip}{\textbaselineskip}


\setcounter{secnumdepth}{2}
\setcounter{tocdepth}{2}


\begin{romanpages}

\maketitle

\begin{dedication}
{\small
To my \textit{piccola}\\
\vspace{1cm}
A mamma e pap\`{a}\\
\vspace{1cm}
A nonna e nonno}
\end{dedication}

\begin{statementoriginality}
 	\vspace{-0.6cm}

I, Luigi Alfonsi, confirm that the research included within this thesis is my own work or that where it has been carried out in collaboration with, or supported by others, that this is duly acknowledged below and my contribution indicated. Previously published material is also acknowledged below. \vspace{0.1cm} 

\noindent I attest that I have exercised reasonable care to ensure that the work is original, and does not to the best of my knowledge break any UK law, infringe any third party’s copyright or other Intellectual Property Right, or contain any confidential material. \vspace{0.1cm} 

\noindent I accept that the College has the right to use plagiarism detection software to check the electronic version of the thesis.\vspace{0.1cm} 

\noindent I confirm that this thesis has not been previously submitted for the award of a degree by this or any other university.\vspace{0.1cm} 

\noindent The copyright of this thesis rests with the author and no quotation from it or information derived from it may be published without the prior written consent of the author. \vspace{0.2cm}

\begin{flushright}

\includegraphics[scale=0.5]{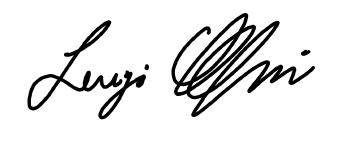}

April 19, 2021 
\end{flushright}

\vspace{0.1cm}

\subsection*{Details of collaborations and publications}
The work presented in this thesis is based on research carried out both independently and in collaboration with my supervisor Professor David S. Berman. In particular, it is based on the material published in the following papers: 
\begin{itemize}
    \item[\cite{Alf19}] {\sc Luigi Alfonsi}, \textit{Global Double Field Theory is Higher Kaluza‐Klein Theory}, \\Fortsch. Phys. 68 (2020) 3-4, 2000010, {\tt arXiv:1912.07089 [hep-th]}
    \item[\cite{Alf20}] {\sc Luigi Alfonsi}, \textit{The puzzle of global Double Field Theory: open problems and the case for a Higher Kaluza-Klein perspective}, Fortsch. Phys. 69 (2021) 7, 2000102,  {\tt arXiv:2007.04969 [hep-th]}
    \item[\cite{Alfonsi:2021bot}] {\sc Luigi Alfonsi and David S. Berman}, \textit{Double Field Theory and Geometric Quantisation},  JHEP 06 (2021) 059, {\tt  arXiv:2101.12155 [hep-th]}
    \item[\cite{Alfonsi:2021uwh}] {\sc Luigi Alfonsi}, \textit{Towards an extended/higher correspondence -- Generalised geometry, bundle gerbes and global Double Field Theory}, {\tt arXiv:2102.10970 [hep-th]}
\end{itemize}
Additionally, chapter \ref{ch:7} is mostly based on unpublished material. Finally, the paper \cite{Alfonsi:2020lub} was published in JHEP 07 (2020) in collaboration with Chris D. White and Sam Wikeley, but the material therein has not been included in this thesis.

\end{statementoriginality}

\pdfbookmark[chapter]{Abstract}{abstract}
\begin{abstract}
	

Extended Field Theories (ExFTs) include Double Field Theory (DFT) and Exceptional Field Theory, which are respectively the T- and U-duality covariant formulations of the supergravity limit of String Theory and M-theory. Extended Field Theories do not live on spacetime, but on an extended spacetime, locally modelled on the space underlying the fundamental representation of the duality group. Despite its importance in M-theory, however, the global understanding of Extended Field Theories is still an open problem.\vspace{0.1cm}

\noindent In this thesis we propose a global geometric formulation of Extended Field Theory. Recall that ordinary Kaluza‐Klein theory unifies a metric with a gauge field on a principal bundle. We propose a generalisation of the Kaluza‐Klein principle which unifies a metric and a higher gauge field on a principal infinity-bundle. This is achieved by introducing an atlas for the principal infinity-bundle, whose local charts can be naturally identified with the ones of Extended Field Theory. Thus, DFT is interpreted as a higher Kaluza-Klein theory set on the total space of a bundle gerbe underlying Kalb‐Ramond field.\vspace{0.1cm}

\noindent As first application, we define the higher Kaluza‐Klein monopole by naturally generalising the ordinary Gross‐Perry monopole. Then we show that this monopole is exactly the NS5‐brane of String Theory.\vspace{0.1cm}

\noindent Secondly, we show that our higher geometric formulation gives automatically rise to global abelian T‐duality and global Poisson-Lie T-duality. In particular, we globally recover the abelian T-fold and we define the notion of Poisson-Lie T-fold.\vspace{0.1cm}

\noindent Crucially, we will investigate the global geometric formulation of tensor hierarchies and gauged supergravity. In particular, we will provide a global formulation of generalised Scherk-Schwarz reductions and we will discuss the global non-geometric properties of tensor hierarchies. \vspace{0.1cm}

\noindent Finally, we explore the T-duality covariant geometric quantisation of DFT by transgressing its underlying bundle gerbe to a U(1)-bundle on the loop space of its base manifold.

\end{abstract}

\begin{acknowledgements}

I would like, first, to express my gratitude to my supervisor Professor David S. Berman, without whose support and encouragement the work presented here could not exist. 
During my time as a Ph.D. student, I constantly benefited from his lively guidance, vast knowledge of the field and extraordinary intuition.

\noindent I also had the pleasure to collaborate with Chris D. White and Sam Wikeley, whom I would like to thank especially.

\noindent I would like to thank Christian S\"{a}mann, Franco Pezzella, Emanuel Malek, Urs Schreiber, Hisham Sati, Francesco Genovese, Severin Bunk, Richard Szabo, David Svoboda, Daniel Thompson, Mark Bugden, Meer Ashwinkumar, Jeong-Hyuck Park, Vincenzo Marotta, Lennart Schmidt and Marco Zambon.

\noindent I would like to thank the organisers Vicente Cortés, Liana David and Carlos Shahbazi of the workshop \href{https://www.math.uni-hamburg.de/projekte/gg2020/}{\textit{Generalized Geometry and Applications 2020}} at University of Hamburg.
I would also like to thank the organisers of the \href{https://sites.google.com/view/egseminars}{\textit{Exceptional Geometry Seminar Series}}.

\noindent Finally, I want to thank my fellow Ph.D. students at Queen Mary University of London and all the academics at \href{https://www.qmul.ac.uk/spa/strings}{Centre for Research in String Theory} for making it fun to work here.
\end{acknowledgements}

\dominitoc 
\flushbottom

\pdfbookmark[chapter]{Contents}{toc}
\tableofcontents

\listoffigures
	\mtcaddchapter

\listoftables
	\mtcaddchapter


\end{romanpages}

\flushbottom

\begin{savequote}[8cm]
The supreme task of the physicist is the discovery of the most general elementary laws from which the world-picture can be deduced logically. \\
But there is no logical way to the discovery of these elemental laws. \\
There is only the way of intuition, which is helped by a feeling for the order lying behind the appearance, and this Einfühlung [literally: `feeling one's way in'] is developed by experience.
  \qauthor{--- Albert Einstein, Preface to \textit{Where is Science Going?}}
\end{savequote}
  
\chapter{\label{ch:1-intro}Introduction} 

\minitoc

\section{String Theory and duality}

\noindent String Theory is our most promising theory of quantum gravity and, even more interestingly, of unification of the fundamental forces. 
The core idea of String Theory is to replace the worldlines of the various species of particles from particle physics by the worldsheets of spinning strings. Such a worldsheet is an Euclidean Riemann surface, equipped with a $2$d superconformal field theory. The spectrum of excitations of the spinning string reproduces the properties of the ordinary particles. \vspace{0.2cm}

\begin{figure}[ht!]\begin{center}
\includegraphics[scale=0.25]{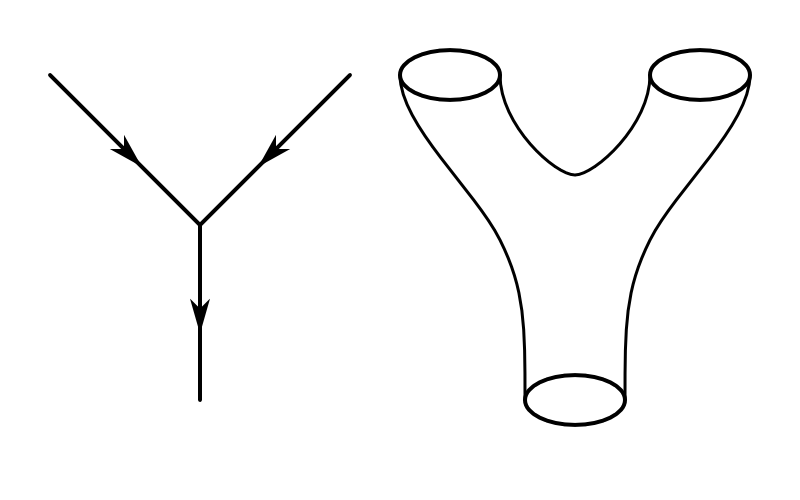}
\caption[Worldsheets in String Theory]{Worldlines of particles are replaced by a worldsheets in String Theory.}
\end{center}\end{figure}

\noindent One of the most characteristic and fascinating features of String Theory is the natural appearance of a particular kind of duality, \textit{T-duality}, which follows from the extended nature of the string, in contrast to the infinitesimal size of a particle. This additional, hidden symmetry of the theory has the potential to revolutionise our perspective on physics. T-duality, in fact, implies that different spacetimes, populated by seemingly different fields, are actually different faces of the same fundamental physics. We have barely scratched the surface, but what the existence of T-duality suggests is that our familiar notion of spacetime must irredeemably break down close to the string scale. 
The immediate consequence is that the geometry on which our current understanding of spacetime is based needs to be revolutionised to take into account the quantum effects at the string scale. \vspace{0.2cm}


\subsection{T-duality}

In this subsection we will concretely introduce T-duality on the circle, with a perspective inspired by \cite{Deligne:1999qp}.\vspace{0.2cm}

\noindent Let us consider a $\sigma$-model on a $2$-dimensional worldsheet $\Sigma$ equipped with a worldsheet metric $h$, which allows to define the Hodge operator $\star$. Let the target space be a circle $S^1$. We define over this surface a scalar field 
\begin{equation}
    \phi:\,\Sigma\;\longrightarrow\; S^1\,=\,\mathbb{R}/2\pi\mathbb{Z}
\end{equation}
that represents the position of a string on the circle. The action of this $\sigma$-model will be
\begin{equation}
S[\phi] \;=\; \frac{R^2}{4\pi}\int_\Sigma\di\phi\wedge\star\di\phi
\end{equation}
that determines the classical equations of motion $\di\star\di\phi=0$. Now we want to define a more general theory that can be reduced to this one. Let us consider a trivial $S^1$-bundle on $\Sigma$ equipped with a connection $A$ and define a covariant derivative $D_A\phi:=\di\phi+A$. For the new theory we start with the action
\begin{equation}\label{eq:newthone}
S'[\phi,A] \;=\; \frac{R^2}{4\pi}\int_\Sigma D_A\phi\wedge\star D_A\phi,
\end{equation}
but we note that there is not a mechanism that imposes $A=0$ to recover the original action. Since the bundle is trivial we can globally define a curvature as $F_A=\di A$ and add another term to the action \eqref{eq:newthone}:
\begin{equation}
\mathbb{S}[\phi,A,\widetilde{\phi}] \;=\; \frac{R^2}{4\pi}\int_\Sigma D_A\phi\wedge\star D_A\phi + \frac{1}{2\pi}\int_\Sigma\widetilde{\phi} F,
\end{equation}
where $\widetilde{\phi}$ is a new auxiliary field $\widetilde{\phi}:\Sigma\rightarrow S^1$. The action $\mathbb{S}[\phi,A,\widetilde{\phi}]$ is equivalent to $S[\phi]$. This can by easily shown by integrating out the extra fields $\widetilde{\phi}$, which plays the role of a Lagrange multiplier. We would immediately obtain $F_A=0$, so that we can choose a gauge where $A=0$ and recover the original $\sigma$-model.
However, we can make a different choice: we can integrate out the gauge fields first and then choose a gauge so that the original field vanishes, i.e. $\phi=0$. 
By following this prescription, we end up with the action functional
\begin{equation}
    \widetilde{S}[\widetilde{\phi}] \;=\; \frac{1}{4\pi}\!\left(\frac{\alpha'}{R}\right)^{\!\!2}\!\int_\Sigma\di\widetilde{\phi}\wedge\star\di\widetilde{\phi}
\end{equation}
Thus, we end up with a $\sigma$-model of the same form but with a new radius $\widetilde{R}:= \alpha'/R$. 

\noindent Schematically, we have the following relations:
\begin{equation}
    \begin{tikzcd}[row sep={12ex,between origins}, column sep={11ex,between origins}]
     & \mathbb{S}[\phi,A,\widetilde{\phi}] \arrow[dl, bend right, "\substack{\text{integrate out }\widetilde{\phi}\\\text{gauge fix }A=0}"']\arrow[dr, bend left, "\substack{\text{integrate out }A\\\text{gauge fix }\phi=0}"] & \\
     S[\phi] \arrow[rr, dotted, "\text{T-duality}"'] & & \widetilde{S}[\widetilde{\phi}].
    \end{tikzcd}
\end{equation}

\noindent A quick inspection of the mass spectrum of our original string, together with the Level Matching Condition, reveals that it is of the form
\begin{equation}\label{eq:spectrum}
\begin{aligned}
    M^2 \;&=\; \left(\frac{n}{R}\right)^{\!\!2} + \left(\frac{wR}{\alpha'}\right)^{\!\!2} + \frac{2}{\alpha'}\big(N+\widetilde{N}-2\big), \\
    N-\widetilde{N} \;&=\; nw,
\end{aligned}
\end{equation}
where $n,w\in\mathbb{Z}$ are respectively the quantum number of momentum and the winding number of the $\sigma$-model on the circle, and $N,\widetilde{N}\in\mathbb{Z}$ are the left-moving and right-moving oscillators. Notice that the spectrum \eqref{eq:spectrum} is invariant under the symmetry group $\mathbb{Z}_2$ of transformations given as follows:
\begin{equation}
    \begin{aligned}
    R \;&\longmapsto\; \frac{\alpha'}{R}, \\
    n \;&\longmapsto\; w, \\[0.1cm]
    w \;&\longmapsto\; n.
    \end{aligned}
\end{equation}
Crucially, this transformation maps the data of a string $S[\phi]$ to the data of its T-dual $\widetilde{S}[\widetilde{\phi}]$.

\subsection{Buscher's rules}

Let our target space $M$ now be a general smooth manifold equipped with a metric $g$ and a Kalb-Ramond field, i.e. a local $2$-form $B$.
The Polyakov action which includes the background fields $(g,B)$ is the following functional:
\begin{equation}\label{eq:introaction}
S[X,g,B] \;=\; \frac{1}{4\pi}\int_\Sigma g_{\mu\nu}\di X^\mu\wedge\star\di X^\nu + \frac{1}{4\pi}\int_\Sigma X^\ast B,
\end{equation}
where $X:\Sigma \longhookrightarrow M$ is an embedding of the worldsheet $\Sigma$ in the target space $M$ and $X^\ast$ is its pullback.\vspace{0.2cm}

\noindent The second term of the action \eqref{eq:introaction} is called Wess-Zumino action term, i.e.
\begin{equation}
S_{\mathrm{WZ}}[B] \;=\; \frac{1}{4\pi}\int_\Sigma X^\ast B  \;=\; \frac{1}{4\pi}\int_{X(\Sigma)} B.
\end{equation}
The Kalb-Ramond field $B$ will be defined up to a gauge transformation $B'=B+\di\lambda$ for any $1$-form $\lambda$, thus local $2$-form $B$ cannot generally be a globally-defined $2$-form field. However, the $3$-form $H:=\di B=\di B'$, known as the Kalb-Ramond field flux, is well-defined everywhere. Thus we can use Stoke's lemma and rewrite
\begin{equation}
S_{\mathrm{WZ}}[B]\;=\; \frac{1}{4\pi}\int_{V}\di X^\ast B \;=\; \frac{1}{4\pi}\int_V X^\ast H,
\end{equation}
where $V$ is a manifold such that its boundary is the worldsheet, i.e. $\partial V=\Sigma$. Since there is not a unique manifold equipped with this property, this introduces an ambiguity. Let us consider two manifolds $V,V'$ with opposite orientations such that $\partial V=\partial V'=\Sigma$. As observed by \cite{Hul06x}, we can define the functional
\begin{equation}
\Delta[H] \;:=\; \frac{1}{4\pi}\bigg(\int_VX^\ast H-\int_{V'}X^\ast H\bigg) \;=\; \frac{1}{4\pi}\int_{V-V'}X^\ast H \;=\; \frac{1}{4\pi}\int_{X(V-V')}H.
\end{equation}
This ambiguity in the definition of $V$ cannot influence the classical equations of motion, but we can investigate its effect on the quantum theory. In fact, if we consider the partition function $\mathcal{Z}=\int\mathcal{D}X\exp i(S_{\mathrm{WZ}}[B]+\Delta[H])$, we must have that $\Delta[H]\in 2\pi\mathbb{Z}$ is an integer, because there cannot be dependence on the particular choice of $3$-cycle. In other words we obtain a generalisation of the Dirac quantisation condition
\begin{equation}
\frac{1}{4\pi^2}\int_{c_3}H\;\in\;\mathbb{Z}
\end{equation}
for any choice of $3$-cycle $c_3\in H_3(M,\mathbb{Z})$ on the target manifold $M$.\vspace{0.2cm}

\noindent  Let us now consider a target space of the form $M:=M_0\times S^1$, where $M_0$ is a smooth manifold, and let us call $X^\circ:\Sigma\longrightarrow S^1$ the coordinate on the circle.
Just like in the case of the previous subsection, we can introduce the connection $A$ of trivial $S^1$-bundle and gauge our $\sigma$-model action \eqref{eq:introaction}. Thus we obtain
\begin{equation*}
\mathbb{S}[X,A,\widetilde{X},g,B] \,=\, \frac{1}{4\pi}\int_\Sigma \bigg( g_{\mu\nu} D_AX^\mu\wedge\star D_A X^\nu + B_{\mu\nu} D_AX^\mu\wedge D_A X^\nu \bigg) + \frac{1}{2\pi}\int_\Sigma\widetilde{X}_\circ F,
\end{equation*}
where $\widetilde{X}_\circ$ is the Lagrange multiplier.
As usual, this action is equivalent to the action \eqref{eq:introaction}. Similarly to the previous subsection, let us integrate out the gauge field $A$ and then gauge fix $X^\circ=0$. Buscher \cite{Bus1, Bus2} observed that we end up with the action functional
\begin{equation}
    \widetilde{S}[\widetilde{X},\widetilde{g},\widetilde{B}] \;=\; \frac{1}{4\pi}\int_\Sigma \widetilde{g}^{\mu\nu}\di \widetilde{X}_\mu\wedge\star\di \widetilde{X}_\nu + \frac{1}{4\pi}\int_\Sigma \widetilde{X}^\ast \widetilde{B},
\end{equation}
which is of the same form of \eqref{eq:introaction}, but the new target space $\widetilde{M}=M_0\times \widetilde{S}^1$ has a different metric and Kalb-Ramond field. These new background fields will be given by so-called Buscher's rules \cite{Bus1, Bus2}, i.e.
\begin{equation}\label{eq:introbusch}
    \begin{aligned}
    \widetilde{g}_{\circ\circ} \;&=\; \frac{1}{g_{\circ\circ}}, \\
    \widetilde{g}_{\mu\circ} \;&=\; \frac{B_{\mu\circ}}{g_{\circ\circ}}, \\
    \widetilde{g}_{\mu\nu} \;&=\; g_{\mu\nu} - \frac{1}{g_{\circ\circ}}\big(g_{\mu\circ}g_{\nu\circ}-B_{\mu\circ}B_{\nu\circ}\big), \\
    \widetilde{B}_{\mu\circ} \;&=\; \frac{g_{\mu\circ}}{g_{\circ\circ}}, \\
    \widetilde{B}_{\mu\nu} \;&=\; B_{\mu\nu} - \frac{1}{g_{\circ\circ}}\big(B_{\mu\circ}g_{\nu\circ}-g_{\mu\circ}B_{\nu\circ}\big).
    \end{aligned}
\end{equation}
Notice that, under T-duality, metric and Kalb-Ramond field are mixed. This is an intriguing suggestion that these two objects can be two faces of the same medal.

\subsection{T-duality on the torus}

Let us now consider a target space which is a torus $M=T^n$, with constant metric $g$ and constant Kalb-Ramond field $B$.
Now, we can pack the momentum $p_\mu$ and the winding numbers $w_\mu$ in a single $2n$-dimensional vector
\begin{equation}
\mathbbvar{p}^M \;:=\; \begin{pmatrix}
 w^\mu \\
 n_\mu 
 \end{pmatrix}.
\end{equation}
We can also define the following matrices:
\begin{equation}
\mathcal{G}_{MN}\;:=\; \begin{pmatrix}
 g_{\mu\nu}-B_{\mu\lambda}g^{\lambda\rho}B_{\rho\nu} & b_{\mu\lambda}g^{\lambda\nu} \\[0.1cm]
 -g^{\mu\lambda}B_{\lambda\nu} & g^{\mu\nu}
 \end{pmatrix}, \qquad \eta_{MN} \;:=\; \begin{pmatrix}
 0 & 1 \\
 1 & 0
 \end{pmatrix},
\end{equation}
where we arranged together the metric $g$ and the Kalb-Ramond field $B$.
The mass spectrum of our string and the Level Matching Condition can now be written by
\begin{equation}\label{eq:LMC}
    \begin{aligned}
    M^2 \;&=\; \mathbbvar{p}^M\mathcal{H}_{MN}\mathbbvar{p}^N + \frac{2}{\alpha'}(N+\widetilde{N}-2), \\
    N-\widetilde{N} \;&=\; \frac{1}{2}\mathbbvar{p}^M\eta_{MN}\mathbbvar{p}^N.
    \end{aligned}
\end{equation}
Let us define the group $O(n,n;\mathbb{Z})$ as the group of $(n\times n)$-matrices $\mathcal{O}$ which preserve the matrix $\eta_{MN}$, i.e. such that $\mathcal{O}^\mathrm{T}\eta\mathcal{O}=\eta$.
Notice that the mass spectrum is invariant under the following transformations:
\begin{equation}
\begin{aligned}
\mathcal{H}_{MN} \,&\mapsto\, \mathcal{O}^L_{\;M}\mathcal{O}^P_{\;N} \mathcal{H}_{LP},
    \mathbbvar{p}^N \,&\mapsto\, (\mathcal{O}^{-1})^M_{\;N}\mathbbvar{p}^N,
\end{aligned}
\end{equation}
where $\mathcal{O}\in O(n,n;\mathbb{Z})$.
Therefore, the T-duality symmetry group $\mathbb{Z}_2$ of the circle compactification is generalised to the bigger group $\mathrm{O}(n,n;\mathbb{Z})$ of a $n$-toroidal compactification.
Notice that the Buscher's rules \eqref{eq:introbusch} can be recovered by the $O(n,n;\mathbb{Z})$-transformation
\begin{equation}
    \mathcal{O}^M_{\;N} \;=\; \begin{pmatrix}
    \delta^\mu_\nu & 0 & 0 & 0 \\
    0 & 0 & 0 & 1 \\
    0 & 0 & \delta_\mu^\nu & 0 \\
    0 & 1 & 0 & 0 \\
 \end{pmatrix}.
\end{equation}

\subsection{Topological T-duality}
T-duality on the torus target space can be naturally generalised to a principal $T^n$-bundle target space. 
Crucially, T-duality is a global transformation of the string background, which exchanges the first Chern class of the $T^n$-bundle with the cohomology class of the Kalb-Ramond field. \vspace{0.2cm}

\noindent The idea of topological T-duality \cite{Bou03,Bou03x, Bou03xx, Bou04} is to disregard both the metric and the connection of the $T^n$-bundle and look at the underlying topological structure.
A \textit{topological T-duality} is, thus, a diagram of the following form:
\begin{equation}
    \begin{tikzcd}[row sep={11ex,between origins}, column sep={11ex,between origins}]
     & & M\times_{M_0}\widetilde{M}\arrow[dr, "\pi"']\arrow[dl, "\widetilde{\pi}"] & & \\
    T^n \arrow[r, hook] & M\arrow[dr, "\pi"'] & & \widetilde{M}\arrow[dl, "\widetilde{\pi}"] & \arrow[l, hook'] T^n\\
    & & M_0 & &
    \end{tikzcd}
\end{equation}
where $M$ and $\widetilde{M}$ are principal $T^n$-bundles on a common base manifold $M_0$ such that
\begin{equation}
    \mathrm{c}_1(\widetilde{M}) \;=\; \pi_\ast [H]\in H^2(M,\mathbb{Z}), \qquad \mathrm{c}_1(M) \;=\; \widetilde{\pi}_\ast [\widetilde{H}]\in H^2(M,\mathbb{Z}),
\end{equation}
where $\mathrm{c}_1(-)$ denotes the first Chern class of a bundle, and $\pi_\ast$ and $\widetilde{\pi}_\ast$ are the fibre integration respectively of $M$ and $\widetilde{M}$.\vspace{0.2cm}

\noindent For example, let us consider a $3$-sphere $S^3$, seen as a Hopf fibration $S^3\twoheadrightarrow S^2$ with $\mathrm{c}_1(S^3)=1$, with a trivial Kalb-Ramond charge $0\in H^3(S^3,\mathbb{Z})$. Its T-dual will immediately be an $S^1$-bundle with $\mathrm{c}_1(\widetilde{M})=0$, i.e. a trivial $\widetilde{M}=S^2\times S^1$, with Kalb-Ramond charge $1\in H^3(S^2\times S^1,\mathbb{Z})$.

\begin{figure}[ht!]\begin{center}
\tikzset{every picture/.style={line width=0.75pt}} 
\begin{tikzpicture}[x=0.75pt,y=0.75pt,yscale=-1,xscale=1]
\draw (204.93,70.83) node  {\includegraphics[width=84.39pt,height=68.25pt]{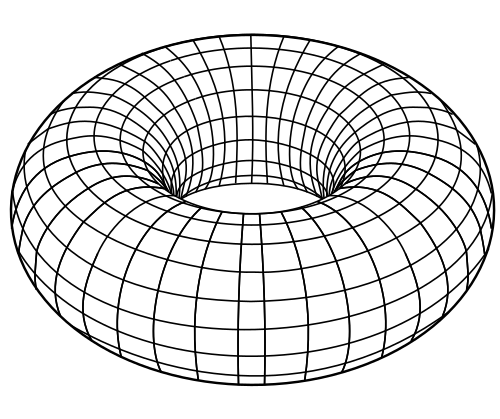}};
\draw (51.67,69.17) node  {\includegraphics[width=76pt,height=76pt]{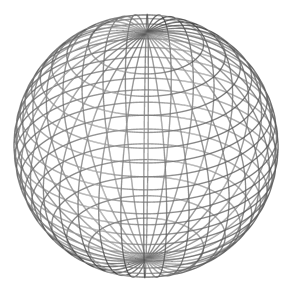}};
\draw  [fill={rgb, 255:red, 0; green, 0; blue, 0 }  ,fill opacity=1 ] (103.33,70.17) -- (109,65.33) -- (109,68.85) -- (139.33,68.85) -- (139.33,65.33) -- (145,70.17) -- (139.33,75) -- (139.33,71.48) -- (109,71.48) -- (109,75) -- cycle ;
\draw (205,129.33) node  [font=\small]  {$h=1$};
\draw (50.67,129.33) node  [font=\small]  {$h=0$};
\draw (52.67,10.33) node  [font=\small]  {$S^{3}$};
\draw (206.33,11) node  [font=\small]  {$S^{2} \times S^{1}$};
\end{tikzpicture}
\caption[Example of topology change under T-duality]{Example of topology change under T-duality, together with Kalb-Ramond charge.}
\end{center}\end{figure}
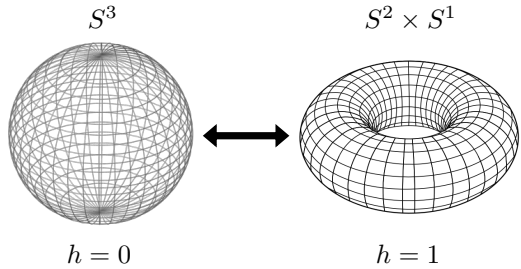

\subsection{Extended Field Theory}
An Extended Field Theory (ExFT) is a field theory which makes string dualities manifest symmetries. Such field theories do not live on spacetime, but on an extended spacetime, which is, at least locally, the space underlying the fundamental representation of the duality group. This framework includes Double Field Theory (DFT) and the Exceptional Field Theories, which are supergravity theories with respectively manifest T-duality and U-duality. \vspace{0.2cm}

\noindent Double Field Theory (DFT) was formulated by \cite{HulZwi09} and seminal work can be traced back to \cite{Siegel1, Siegel2}. DFT is the T-duality covariant formulation of the universal bosonic sector of the supergravity limit of closed String Theory. The T-duality covariant field content of DFT live on a doubled-dimensional coordinate chart, which one can interpret as the fibre product of a spacetime chart with its T-dual. However, the fields of DFT are constrained to depend only on half of the coordinates, to avoid unphysical degrees of freedom. This condition is usually known as strong constraint.\vspace{0.2cm}

\noindent Crucially, on doubled charts, DFT geometrises automatically not just T-duality but also the local gauge transformations of the Kalb-Ramond field, two distinct and apparently unrelated features of String Theory at once. 
The fact that doubling coordinates allows to describe the gauge transformations of the Kalb-Ramond field and T-duality at the same time is a strong hint of a bigger unification principle underlying. In this sense it has been suggested, e.g. see \cite{Park11,Hohm13space,BerBla14}, that the geometry underlying DFT should be thought as a new stringy geometry, which is for strings what Riemannian geometry is for usual point-particles. \vspace{0.2cm}

\noindent More recently, Exceptional Field Theory was defined by \cite{HohSam13} as a generalisation of the framework of DFT to the bosonic sector of $11$d Supergravity and by embedding the T-duality group in the larger U-duality group of M-theory.\vspace{0.2cm}

\noindent However, despite its importance in String Theory, the globalisation of Extended Geometry is still an open problem, both in Doubled Field Theory and in Exceptional Field Theory. Since T-duality is a global transformation, this poses a problem.

\section{Higher geometry and String Theory}
Since the discovery of General Relativity, geometry has become a privileged means to the conceptual understanding of the fundamental structures underlying Nature, where the use of physical intuition becomes difficult. 
Over the last 15 years, the geometry underlying the Kalb-Ramond field has been clarified and identified with a \textit{bundle gerbe}, a concept originally introduced by \cite{Murray,Murray2}, which generalises the idea of principal bundle. See \cite{Murray3} for an introduction. The idea of bundle gerbe was reformulated by \cite{Hit99} and recently generalised by \cite{Principal1} to the notion of principal $\infty$-bundle. \vspace{0.2cm}

\noindent A principal $\infty$-bundle is a structure which generalised a principal bundle to the case where the structure group is a Lie $\infty$-group,
Such an algebraic object encodes not only symmetries, but also symmetries of symmetries and so on. 
Thus, a Kalb-Ramond field can be globally formulated as the connection of a principal $\infty$-bundle. This means that the local $2$-forms $B_{(\alpha)}\in\Omega^2(U_\alpha)$ are patched by local $1$-form gauge transformations $\Lambda_{(\alpha\beta)}\in\Omega^1(U_\alpha\cap U_\beta)$ which are themselves patched by scalar gauge transformations $G_{(\alpha\beta\gamma)}\in\Coo(U_\alpha\cap U_\beta \cap U_\gamma)$ satisfying the cocycle condition on four-fold overlaps of patches. Therefore the patching conditions of the differential local data of the Kalb-Ramond field can be summed up by 
\begin{equation}\label{eq:introgerby}
    \begin{aligned}
    H \,&=\, \mathrm{d}B_{(\alpha)}, \\
    B_{(\beta)} - B_{(\alpha)} \,&=\, \mathrm{d}\Lambda_{(\alpha\beta)}, \\
    \Lambda_{(\alpha\beta)}+\Lambda_{(\beta\gamma)}+\Lambda_{(\gamma\alpha)} \,&=\, \mathrm{d}G_{(\alpha\beta\gamma)}, \\
    G_{(\alpha\beta\gamma)}-G_{(\beta\gamma\delta)}+G_{(\gamma\delta\alpha)}-G_{(\delta\alpha\beta)} \,&\in\, 2\pi\mathbb{Z}.
    \end{aligned}
\end{equation}
The bundle gerbes can be equipped with a natural generalisation of parallel transport which will be along surfaces, instead of curves. This corresponds exactly to the Wess-Zumino action term, which is given by the coupling of the worldsheet of the string with the Kalb-Ramond field in the target space. \vspace{0.2cm}

\noindent More generally, principal $\infty$-bundles are the natural framework to deal with \textit{higher gauge theories} \cite{Baez11}, which are the globally-defined theories whose fields are locally given by differential $n$-forms. For example, higher gauge theory has been used by \cite{Saem17, Saem19, Saem19x} to formulate a $6$d superconformal field theory which is a promising step in developing a M5-brane worldvolume theory.
\vspace{0.2cm}

\noindent Higher geometry has also been used to formulate \textit{higher prequantisation}: the generalisation of geometric quantisation from ordinary particles to string and branes. This field of research can be traced back to the idea of quantisation of $n$-plectic manifolds in \cite{Rog11, SaSza11, Rog13} and of loop spaces in \cite{SaSza11x}. The powerful formalism of higher stacks allowed to further generalise the theory in \cite{SaSza13, FSS13, Sch16, FSS16, BSS16, BS16, Sza19}.
Moreover, the $L_\infty$-algebras naturally appearing in higher geometry have been revealed to naturally encompass BV–BRST formalism for quantisation of field theories in \cite{Pau14, Saem18bv, Saem19bv, Doubek_2019, Jurco:2019yfd, Jurco:2020yyu}. Other properties of field theories related to higher Lie algebras have been explored by \cite{Hohm17, Hohm17x, Hohm19, Hohm19x}.
\vspace{0.2cm}

\noindent Higher geometry has been also successfully applied to the underlying geometry of M-theory by \cite{FSS12, FSS15x, FSS19x, FSS19xxx, BSS19, HSS19}. In these references the topological and differential structure of M-theory is investigated, until remarkably a proposal for the generalised cohomology theory that charge-quantises the supergravity $C$-field is made (known as \textit{Hypothesis H}) by \cite{FSS19coho}. This idea was further explored by \cite{BSS18,SS19,FSS19xx, Fiorenza:2020iax, Sati:2020nob, Sati:2021uhj}.
\vspace{0.2cm}

\noindent Significantly, the research in nonassociative physics, which emerges from open String Theory, has been linked not only to non-geometric fluxes \cite{SN1, SN2, SN4}, but also to higher geometry \cite{SN3, SN5, SN6, SN7, Sza18}. 
Finally, higher geometry provides a natural framework for Algebraic Quantum Field Theory \cite{Benini:2018oeh, Mathieu:2019lgi, Benini:2019hoc,Benini:2019uge, Benini:2020gbr}

\subsection{Higher geometry and T-duality}
Remarkably, higher geometry has been identified as the natural framework to study T-duality. This has been formalised by \cite{BunNik13,FSS16x,FSS17x,FSS18,FSS18x,NikWal18} in terms of an isomorphism of a pair bundle gerbes geometrically encoding a Kalb-Ramond field and its T-dual. Let us consider two principal $T^n$-bundle spacetimes $M\xrightarrow{\pi}M_0$ and $\widetilde{M}\xrightarrow{\widetilde{\pi}}M_0$ on a common base manifold $M_0$. In the references, a couple of bundle gerbes $\mathscr{G}\xrightarrow{\Pi}M$ and $\widetilde{\mathscr{G}}\xrightarrow{\widetilde{\Pi}}\widetilde{M}$, formalising two Kalb-Ramond fields respectively on $M$ and $\widetilde{M}$, are T-dual if the following isomorphism exists:
\begin{equation}
    \begin{tikzcd}[row sep={11ex,between origins}, column sep={11ex,between origins}]
    & \mathscr{G}\times_{M_0} \widetilde{M}\arrow[rr, "\cong"', "\text{T-duality}"]\arrow[dr, "\Pi"']\arrow[dl, "\widetilde{\pi}"] & & M\times_{M_0}\widetilde{\mathscr{G}}\arrow[dr, "\pi"']\arrow[dl, "\widetilde{\Pi}"] \\
    \mathscr{G}\arrow[dr, "\Pi"'] & & M\times_{M_0}\widetilde{M}\arrow[dr, "\pi"']\arrow[dl, "\widetilde{\pi}"] & & [-2.5em]\widetilde{\mathscr{G}}\arrow[dl, "\widetilde{\Pi}"] \\
    & M\arrow[dr, "\pi"'] & & \widetilde{M}\arrow[dl, "\widetilde{\pi}"] & \\
    & & M_0 & &
    \end{tikzcd}
\end{equation}
In the references it is shown that this induces an isomorphism between the twisted cohomology theory of D-branes of Type IIA and of Type IIB String Theory, which is closely connected with twisted K-theory.
\vspace{0.2cm}

\noindent Moreover, notice that the diagram above can be seen as the finite version of the T-duality diagram of Courant algebroids appearing in \cite{CavGua11}. In fact, the Courant algebroid appearing in generalised geometry \cite{Gua11} has been understood as a higher Atiyah algebroid for the bundle gerbe \cite{Col11,Rog13}. Supergravity can be naturally formulated in terms of generalised geometry by \cite{Hull07,Wald08E,Wald08,Wald11,Wald12} and T-duality, as shown by \cite{CavGua11}, can be easily formalised.
\vspace{0.2cm}

\noindent At this point, is not so surprising that research in DFT has been affected by these new geometric ideas. It was noticed by \cite{BCM14} that the doubled metric of DFT, to actually geometrise the Kalb-Ramond field, must carry a bundle gerbe structure and have non-trivial local data three-fold overlaps. These arguments lead to the idea that a finite well-defined DFT geometry must be constructed in the context of higher geometry. Moreover, \cite{DesSae18} proposed for DFT a formalism rooted in $L_\infty$-algebroids, which was generalised to Heterotic DFT by \cite{DesSae18x} and, then, applied to the particular case of nilmanifolds by \cite{DesSae19}. This successful idea was also translated to Exceptional Field Theory by \cite{Arv18,Arvanitakis:2021wkt}. Independently, \cite{Hohm19DFT, Hohm19x} showed that the gauge structure of the infinitesimal generalised diffeomorphisms of DFT has a $L_\infty$-algebra structure. Moreover, many interesting works relate DFT in the worldvolume perspective to $L_\infty$-algebras \cite{Chatzistavrakidis:2019rpp, Grewcoe:2020gka, Grewcoe:2020uih, Grewcoe:2020ren, Chatzistavrakidis:2019rpp}.

\section{Outline of this Thesis}

The main goal of this thesis is to develop a globally well-defined formalisation for Double Field Theory and to generalise it to the other Extended Field Theories. Secondly, we want to derive globally-defined compactifications of Double Field Theory from our formalisation. In particular, we will derive a global formulation of tensor hierarchies.\vspace{0.2cm}

\noindent Chapter \ref{ch:2} briefly reviews the main features of Double Field Theory and Exceptional Field Theory whose formalisation is the aim of this thesis. Particular focus will be given to the problem of the underlying global geometry. \vspace{0.2cm}

\noindent Chapter \ref{ch:3} will provide an introduction to higher geometry, which is aimed to make the thesis self-contained. In particular, we will introduce the notion of $\infty$-groupoid and its relation to the notion of $L_\infty$-algebra via $\infty$-Lie theory. We will also introduce stacks and a notion of atlas for stacks. Moreover, we will review the theory of principal $\infty$-bundles and twisted $\infty$-bundles, with particular focus on bundle gerbes. Finally, we will discuss some applications to String Theory. \vspace{0.2cm}

\noindent Chapter \ref{ch:4} we will review the main proposals of formalisation of the geometry underlying Double Field Theory and we will interpret them in the light of higher geometry. \vspace{0.2cm}

\noindent Chapter \ref{ch:5} and \ref{ch:6} are the core of the thesis.
In chapter \ref{ch:5} we develop a formalisation of Double Field Theory as a higher Kaluza-Klein theory, i.e. a generalisation of Kaluza-Klein theory which lives on a principal $\infty$-bundle, instead of an ordinary principal bundle. In the light of this formalisation, Double Field Theory can be seen as a field theory on the total space of a bundle gerbe. The usual coordinate description of Double Field Theory is naturally recovered by introducing an atlas for the bundle gerbe, which is naturally made up of doubled coordinate charts.
\vspace{0.2cm}

\noindent Chapter \ref{ch:6} will derive T-duality from the proposed formalisation of Double Field Theory. This will include abelian T-duality, both geometric and non-geometric, non-abelian T-duality and Poisson-Lie T-duality. We will also use this formalisation to propose a global definition of non-abelian T-fold and Poisson-Lie T-fold.
Moreover, we will derive a globally-defined notion of tensor hierarchy and we will discuss its topological properties.
\vspace{0.2cm}

\noindent Chapter \ref{ch:7} will make some steps in generalising or formalism for Double Field Theory to the general class of Extended Field Theories. In particular we will study the local chart description of the extended spaces of heterotic Double Field Theory, Type II super-Double Field Theory, Exceptional Field Theory and super-Exceptional Field Theory. \vspace{0.2cm}

\noindent In chapter \ref{ch:8} we will introduce a T-duality background-invariant geometric quantisation of Double Field Theory. Moreover, we will study the relation between the bundle gerbe and the phase space of a closed string.
\vspace{0.2cm}

\noindent This thesis is completed by three appendices which contain introductory material to facilitate the comprehension of the main body. In particular, appendix \ref{app:1} we will introduce generalised geometry and the theory of Courant algebroids, including their relation with Lie bialgebroids. Appendix \ref{app:2} will review the supergeometry underlying Supergravity, with particular focus on the higher geometric structures formalising the field content of the Supergravity limit of String Theory. Finally, appendix \ref{app:3} will provide a soft introduction to Hamiltonian mechanics and geometric quantisation.
\begin{savequote}[8cm]
The effort to understand the universe is one of the very few things which lifts human life a little above the level of farce and gives it some of the grace of tragedy.
  \qauthor{--- Steven Weinberg, \textit{The First Three Minutes}}
\end{savequote}

\chapter{\label{ch:2}Introduction to Extended Field Theories}

\minitoc

\noindent In this chapter we will provide an introduction to the formalism of local Double Field Theory and Exceptional Field Theory.

\section{Double Field Theory}

In this section we introduce the formalism of local Double Field Theory. Excellent reviews on the subject include \cite{Aldazabal:2013sca, Geissbuhler:2013uka, BerTho14}.

\subsection{Coordinate representation}

Let us consider an open simply connected $2d$-dimensional open patch $\mathcal{U}$. We can introduce coordinates $(x^\mu,\widetilde{x}_\mu):\mathcal{U}\rightarrow \mathbb{R}^{2d}$, which we will call collectively $x^M:=(x^\mu,\widetilde{x}_\mu)$. Now, we want to equip the vector space $\mathbb{R}^{2d}$ with the fundamental representation of the continuous T-duality group $O(d,d)$. Since the action of $O(d,d)$-matrices on $\mathbb{R}^{2d}$ preserves the matrix $\eta_{MN}:=\left(\begin{smallmatrix}0&1\\1&0\end{smallmatrix}\right)$, we can define a metric $\eta=\eta_{MN}\di x^M \otimes \di x^N \in \odot^2T^\ast\mathcal{U}$ with signature $(d,d)$.

\subsection{Generalised diffeomorphisms}

We want to define a \textit{generalised Lie derivative}
\begin{equation}
    \mathfrak{L}:\, \mathfrak{X}(\mathcal{U})\times \mathfrak{X}(\mathcal{U}) \;\longrightarrow\; \mathfrak{X}(\mathcal{U})
\end{equation}
which preserves the $\eta$-tensor, i.e. such that $\mathfrak{L}_X\eta=0$ for any vector $X\in\mathfrak{X}(\mathcal{U})$. Thus, for any couple of vectors $X,Y\in\mathfrak{X}(\mathcal{U})$ we can define
\begin{equation}
    \big(\mathfrak{L}_XY\big)^M \;:=\;  X^N\partial_N Y^M - (\partial\times_{\mathrm{ad}} X)^M_{\;\;\;N}Y^N.
\end{equation}
Here, we defined the adjoint projection 
\begin{equation}
    \times_{\mathrm{ad}}\,:\;T^\ast\mathcal{U}\times T\mathcal{U} \;\longrightarrow\; \mathrm{ad}(F\mathcal{U}),
\end{equation}
where $F\mathcal{U}$ is the frame bundle of the tangent bundle $T\mathcal{U}$ and it can be regarded as a principal $O(d,d)$-bundle. This projection can be given in coordinates by the tensor
\begin{equation}
    (\partial\times_{\mathrm{ad}} X)^M_{\;\;\;N} \;=\; \mathbb{P}^{MP}_{\mathrm{ad}\;\; LN} \partial_PX^L, \quad \mathbb{P}^{ML}_{\mathrm{ad}\;\; NP} \,:=\, \delta^M_P\delta^L_N  - \eta^{ML}\eta_{NP},
\end{equation}
which projects the $GL(2d)$-valued function $\partial_LX^N$ into an $\mathfrak{o}(d,d)$-valued one. The generalised Lie derivative is also known as D-bracket $\llbracket X,Y \rrbracket_{\mathrm{D}} := \mathfrak{L}_XY$.
The {C-bracket} is defined as the anti-symmetrisation of the D-bracket, i.e.
\begin{equation}
    \llbracket X,Y \rrbracket_{\mathrm{C}} \;:=\; \frac{1}{2}\big(\llbracket X,Y \rrbracket_{\mathrm{D}} - \llbracket Y,X \rrbracket_{\mathrm{D}}\big).
\end{equation}

\noindent Now, if we want to construct an algebra of generalised Lie derivatives, we immediately find out that it cannot be close, i.e. we generally have
\begin{equation}
    \big[\mathfrak{L}_X, \,\mathfrak{L}_Y\big] \;\neq\;\mathfrak{L}_{\llbracket X,Y \rrbracket_{\mathrm{C}}}
\end{equation}
Thus, to assure the closure, we need to impose extra conditions. The {weak} and the  {strong constraint} (also known collectively as  {section condition}) are respectively the conditions
\begin{equation}
    \eta^{MN}\partial_M\partial_N\phi_i =0, \qquad  \eta^{MN}\partial_M\phi_1\partial_N\phi_2 =0
\end{equation}
for any couple of fields or parameters $\phi_1,\phi_2$. The immediate solution to the section condition is obtained by considering only fields and parameters $\phi$ which satisfy the condition $\widetilde{\partial}^\mu\phi=0$. Therefore, upon application of the strong constraint, all the fields and parameters will depend on the $d$-dimensional submanifold $U := \mathcal{U}/\!\sim \,\,\subset \mathcal{U}$, where $\sim$ is the relation identifying points with the same physical coordinates $(x^\mu,\widetilde{x}_\mu)\sim(x^\mu,\widetilde{x}_\mu')$. In particular vectors $X\in\mathfrak{X}(\mathcal{U})$ satisfying the strong constraint can be identified with sections of the generalised tangent bundle $TU\otimes T^\ast U$ of generalised geometry. Moreover the C-bracket, when restricted to strong constrained vectors, reduces to the Courant bracket of generalised geometry, i.e. we have
\begin{equation}
    \llbracket -,- \rrbracket_{\mathrm{C}}\,\Big|_{\widetilde{\partial}^\mu=0} \;=\; [-,-]_{\mathrm{Cou}}
\end{equation}
In this sense, the geometry underlying Double Field Theory, when strong constrained, locally reduces to generalised geometry. 

\subsection{Generalised metric}

We can define the {generalised metric} $\mathcal{G}=\mathcal{G}_{MN}\di x^M\otimes \di x^N$ by requiring that it is symmetric and it satisfies the property $\mathcal{G}_{ML}\eta^{LP}\mathcal{G}_{PN}=\eta_{MN}$. Thus, the matrix $\mathcal{G}_{MN}$ can be parametrised as
\begin{equation}
    \mathcal{G}_{MN} \;=\; \begin{pmatrix}g_{\mu\nu}- B_{\mu\lambda}g^{\lambda\rho}B_{\rho\beta} & B_{\mu\lambda}g^{\lambda\nu} \\-g^{\mu\lambda}B_{\lambda\nu} & g^{\mu\nu} \end{pmatrix}.
\end{equation}
where $g_{\mu\nu}$ and $B_{\mu\nu}$ are respectively a symmetric and an anti-symmetric matrix. 
Finally, we must impose the strong constraint on $\mathcal{G}_{MN}$, so that its components are allowed to depend only on the $x^\mu$ coordinates, and not on the $\widetilde{x}_\mu$ ones. 
Now, $g:=g_{\mu\nu}\di x^\mu\otimes \di x^\nu$ is a symmetric tensor and $B:=\frac{1}{2}B_{\mu\nu}\di x^\mu\wedge \di x^\nu$ is an anti-symmetric tensor on the $d$-dimensional quotient manifold $U$. These can be respectively interpreted as a metric and a Kalb-Ramond field on the $d$-dimensional patch $U$. 
If we consider a strong constrained vector $V:=v+\widetilde{v}\in\mathfrak{X}(U)\oplus \Omega^1(U)$. The infinitesimal gauge transformation given by generalised Lie derivative $\delta \mathcal{G}_{MN}=\mathfrak{L}_V\mathcal{G}_{MN}$ is equivalent to the following gauge transformations:
\begin{equation}
    \delta g = \mathcal{L}_v g, \qquad \delta B = \mathcal{L}_v B + \di\widetilde{v}
\end{equation}
where $\mathcal{L}_v$ is the ordinary Lie derivative. This, then reproduces the gauge transformations of metric and Kalb-Ramond field. Therefore, the infinitesimal generalised diffeomorphisms of the $2d$-dimensional patch $\mathcal{U}$ unify the infinitesimal diffeomorphisms of the $d$-dimensional subpatch $U\subset \mathcal{U}$ with the infinitesimal gauge transformations of the Kalb-Ramond field, in analogy with Kaluza-Klein theory.

\subsection{Tensor hierarchy}

By following \cite{HohSam13KK}, we can consider an extended space in which only an internal fibre is doubled, i.e.
\begin{equation}
    \mathbb{R}^{d-n}\times\mathbb{R}^{2n},
\end{equation}
where $\{x^\mu,y^M\}$ are respectively the coordinates of the base space $\mathbb{R}^{d-n}$ and of the fibre space $\mathbb{R}^{2n}$.
The field content of Double Field Theory will be
\begin{equation}
\mathrm{Fields}:\;\{g_{\mu\nu},\,\mathcal{G}_{MN},\,\mathcal{A}^M_{\mu},\,\mathcal{B}_{\mu\nu}, \varphi\},
\end{equation}
where all the fields, in principle, can depend on all the coordinates $\{x^\mu,y^M\}$. 
First, the moduli field of the generalised metric will be a function valued in the following coset space
\begin{equation}
    \mathcal{G}_{MN} \;\in\; \frac{O(n,n)}{O(n)\times O(n)}.
\end{equation}
Secondly, the \textit{tensor hierarchy} is defined by the following subset of fields:
\begin{equation}
\mathrm{Tensor\,hierarchy}:\;\{\mathcal{A}^M_{\mu},\,\mathcal{B}_{\mu\nu}\}.
\end{equation}

\noindent Now, let $R_1\cong\mathbf{2n}$ be the fundamental representation of $O(n,n)$ and $R_2\cong\mathbf{1}$ be the singlet representation of $O(n,n)$. 
Notice that we can identify the former with the space of vectors
$R_1=\mathfrak{X}(\mathbb{R}^{2n})$ and the latter with the space of functions $R_2=\Coo(\mathbb{R}^{2n})$ on the fibre.
We can construct a chain complex
\begin{equation}
    R_1 \xleftarrow{\;\;\mathfrak{D}\;\;} R_2
\end{equation}
by defining the following differential
\begin{equation}
    (\mathfrak{D}f)^M \;:=\; \eta^{MN}\partial_{N}f
\end{equation}
for any $f\in R_2$, where $\partial_{N}$ is the partial derivative on the fibre $\mathbb{R}^{2n}$
At this point, we can naturally introduce the DFT \textit{exterior product}
\begin{equation}
    \langle-,-\rangle : R_1\otimes R_1 \longrightarrow R_2,
\end{equation}
which is defined by $\langle X,Y\rangle:=\eta_{MN}X^MY^N$ for any couple of doubled vectors $X,Y\in R_1$.
Finally, we obtain the expression known as the DFT \textit{magic formula}:
\begin{equation}
    \begin{aligned}
         \mathfrak{L}_XY \;&=\; \llbracket X, Y \rrbracket_\mathrm{C} + \mathfrak{D}\langle X,Y\rangle
    \end{aligned}
\end{equation}
for any given couple of doubled vectors $X,Y\in R_1$. This generalises the well-known Cartan's magic formula of ordinary differential geometry. \vspace{0.25cm}

\noindent Now, we can identify the fields of the tensor hierarchy with the following $R_i$-valued differential forms on the base space:
\begin{equation}
    \begin{aligned}
        \mathcal{A}^M \,&\in\; \Omega^1(\mathbb{R}^{d-n})\otimes R_1, \\
        \mathcal{B} \;&\in\; \Omega^2(\mathbb{R}^{d-n})\otimes R_2.
    \end{aligned}
\end{equation}
On such class of fields, we can naturally define the following operators:
\begin{equation}
     (\mathfrak{D}\mathcal{B})^M \;=\; \eta^{MN}\partial_{N}\mathcal{B}, \qquad \langle \mathcal{A}_1\,\overset{\wedge}{,}\,\mathcal{A}_2\rangle \;=\; \eta_{MN}\mathcal{A}^M_1\wedge \mathcal{A}^N_2,
\end{equation}
for any $\mathcal{B}\in \Omega^2(\mathbb{R}^{d-n})\otimes R_2$ and any couple $\mathcal{A}_1,\mathcal{A}_2\in \Omega^1(\mathbb{R}^{d-n})\otimes R_1$. 
The $O(n,n)$-covariant field strengths \cite{HohSam13KK} of the tensor hierarchy are then naturally given by
\begin{equation}
\begin{aligned}
    \mathcal{F}_{\mu\nu} \;&=\; 2\partial_{[\mu} \mathcal{A}_{\nu]} + \llbracket\mathcal{A}_\mu,\mathcal{A}_\nu\rrbracket_\mathrm{C} + \mathfrak{D}\mathcal{B} \\
    \mathcal{H}_{\mu\nu\lambda} \;&=\; 3\partial_{[\mu}\mathcal{B}_{\nu\lambda]} + \mathcal{A}^M_{[\mu}\partial_M\mathcal{B}_{\nu\lambda]} + \langle\mathcal{F}_{[\mu\nu},\mathcal{A}_{\nu]}\rangle + \langle\mathcal{A}_{[\mu},\llbracket\mathcal{A}_\nu,\mathcal{A}_{\lambda]}\rrbracket_\mathrm{C}\rangle. \\
\end{aligned}
\end{equation}
In a coordinate-invariant form, we can rewrite the previous equation as follows:
\begin{equation}
\begin{aligned}
    \mathcal{F} \;&=\; \di \mathcal{A} + \llbracket\mathcal{A}\,\overset{\wedge}{,}\,\mathcal{A}\rrbracket_\mathrm{C} + \mathfrak{D}\mathcal{B}, \\
    \mathcal{H} \;&=\; \di \mathcal{B} + \langle\mathcal{A}\,\overset{\wedge}{,}\,\mathfrak{D}\mathcal{B}\rangle + \frac{1}{2}\langle\mathcal{F}\,\overset{\wedge}{,}\,\mathcal{A}\rangle + \frac{1}{3!}\langle\mathcal{A}\,\overset{\wedge}{,}\,\llbracket\mathcal{A}\,\overset{\wedge}{,}\,\mathcal{A}\rrbracket_\mathrm{C}\rangle.
\end{aligned}
\end{equation}
In even a more compact fashion, we can rewrite the field strengths by
\begin{equation}
\begin{aligned}
    \mathcal{F} \;&=\; \mathrm{D} \mathcal{A} + \mathfrak{D}\mathcal{B}, \\
    \mathcal{H} \;&=\; \mathrm{D} \mathcal{B} + \frac{1}{2}\mathrm{CS}_3(\mathcal{A}), \\
\end{aligned}
\end{equation}
where the covariant derivatives are defined by $\mathrm{D}\mathcal{A}:=\di \mathcal{A} + \llbracket\mathcal{A}\,\overset{\wedge}{,}\,\mathcal{A}\rrbracket_\mathrm{C}$ for the $1$-form and $\mathrm{D}\mathcal{B}:=\di \mathcal{B} + \langle\mathcal{A}\,\overset{\wedge}{,}\,\mathfrak{D}\mathcal{B}\rangle$ for the $2$-form, and where the Chern-Simons differential $3$-form is defined by $\mathrm{CS}_3(\mathcal{A}):=\langle\mathcal{F}\,\overset{\wedge}{,}\,\mathcal{A}\rangle + \frac{1}{3}\langle\mathcal{A}\,\overset{\wedge}{,}\,\llbracket\mathcal{A}\,\overset{\wedge}{,}\,\mathcal{A}\rrbracket_\mathrm{C}\rangle$.
We can easily calculate that the Bianchi identities of the tensor hierarchy are
\begin{equation}
\begin{aligned}
    \mathrm{D}\mathcal{F} \;&=\; \mathfrak{D}\mathcal{H}, \\
    \mathrm{D}\mathcal{H} \;&=\; \frac{1}{2}\langle\mathcal{F}\,\overset{\wedge}{,}\,\mathcal{A}\rangle. \\
\end{aligned}
\end{equation}

\subsection{Generalised Scherk-Schwarz reductions}

Given a $2n$-dimensional Lie group $G$, we can consider an extended space with a doubled fibre of the following form:
\begin{equation}
    \mathbb{R}^{d-n}\times G,
\end{equation}
with split coordinates $\{x^\mu,y^M\}$.
The \textit{generalised Scherk-Schwarz reductions} \cite{Dabholkar:2002sy, Berman:2012uy, Berman:2013cli} are defined by the following ansatz on the fields of Double Field Theory:
\begin{equation}
    \begin{aligned}
        g_{\mu\nu}(x,y) \;&=\; g_{\mu\nu}(x), \\
        \mathcal{G}_{MN}(x,y) \;&=\; U^I_{\;\,M}(y)\,\mathcal{G}_{IJ}(x)\,U^J_{\;\,N}(y),\\
        \mathcal{A}^{M}_{\mu}(x,y) \;&=\; U^{M}_{\;\;\;I}(y)\,\mathcal{A}^I_{\mu}(x), \\
        \mathcal{B}_{\mu\nu}(x,y) \;&=\; \mathcal{B}_{\mu\nu}(x), \\
        \varphi(x,y) \;&=\; \varphi(x),
    \end{aligned}
\end{equation}
where $U^I_{\;\,M}$ is the inverse of $U^{M}_{\;\;\;I}$, and such that the vectors $U_I:=U^{M}_{\;\;\;I}\partial_M$can be seen as generalised frame fields for the $G$-fibre, so that they satisfy the generalised Lie derivative
\begin{equation}
    \mathfrak{L}_{U_I}U_J \;=\; C_{\;\;\,IJ}^K U_K,
\end{equation}
where $C_{\;\;\,IJ}^K$ are the structure constants of the Lie algebra $\mathfrak{g}=\mathrm{Lie}(G)$.
We also require that the gauge parameters are factorised similarly.
Now, the fields of the tensor hierarchy take the following form:
\begin{equation}
    \begin{aligned}
        \mathcal{A}^I \,&\in\; \Omega^1(\mathbb{R}^{d-n})\otimes \mathfrak{g}, \\
        \mathcal{B} \;&\in\; \Omega^2(\mathbb{R}^{d-n}).
    \end{aligned}
\end{equation}
It is easy to verify that the field strengths of the reduced tensor hierarchy becomes
\begin{equation}
\begin{aligned}
    \mathcal{F} \;&=\; \di \mathcal{A} + [\mathcal{A}\,\overset{\wedge}{,}\,\mathcal{A}]_\mathfrak{g}, \\
    \mathcal{H} \;&=\; \di \mathcal{B} + \frac{1}{2}\langle\mathcal{F}\,\overset{\wedge}{,}\,\mathcal{A}\rangle + \frac{1}{3!}\langle\mathcal{A}\,\overset{\wedge}{,}\,[\mathcal{A}\,\overset{\wedge}{,}\,\mathcal{A}]_\mathfrak{g}\rangle.
\end{aligned}
\end{equation}
Notice that $\mathcal{A}^I$ becomes a non-abelian gauge field with gauge group $G$.
The generalised Scherk-Schwarz reduced theory $\{g,\mathcal{G}_{IJ},\mathcal{A}^I,\mathcal{B},\varphi\}$ on the base space $\mathbb{R}^{d-n}$ is also known as \textit{gauged Double Field Theory}.

\section{Exceptional Field Theory}

In this section we will give a brief introduction to Exceptional Field Theory, which is the generalisation of Double Field Theory to U-duality. See table \ref{tab:egg} for a list of the U-duality groups $E_{n(n)}$ in any codimension $n$.
A wide review of Exceptional Field Theory can be found in \cite{BerBla20}.\vspace{0.2cm}

\noindent Let us split the $11$-dimensional Minkowski space as $\mathbb{R}^{1,10}=\mathbb{R}^{1,10-n}\times \mathbb{R}^n$.
We can consider an extended space in which only the internal fibre is extended, i.e.
\begin{equation}
    \mathbb{R}^{1,10-n}\times R_1,
\end{equation}
where $R_1$ is the fundamental representation of the U-duality group $E_{n(n)}$.
The generalised Lie derivative of vectors on $R_1$ is defined by
\begin{equation}
    \big(\mathfrak{L}_XY\big)^M \;:=\;  X^N\partial_N Y^M - \alpha(\partial\times_{\mathrm{ad}} X)^M_{\;\;\;N}Y^N + \lambda_Y\partial_NX^NY^M,
\end{equation}
where $\alpha=\alpha(n)$ is a coefficient which depends on the codimension $n$ and $\lambda_Y$ is the weight of the vector $Y$.
Now, we can define the tensor
\begin{equation}
    \mathrm{Y}^{MN}_{\quad\;\; QP} \;:=\; -\alpha \mathbb{P}^{MN}_{\mathrm{ad}\;\;\, QP} + \delta^M_{\;\;P}\delta^N_{\;\;Q} - \omega\delta^M_{\;\;Q}\delta^N_{\;\;P},
\end{equation}
where the tensor $\mathbb{P}^{MN}_{\mathrm{ad}\;\;\, QP}$ is the projection to the adjoint frame bundle and $\omega:=-1/(n-2)$ is a special weight. By using the $\mathrm{Y}$-tensor, we can recast the generalised Lie derivative as
\begin{equation}
    \big(\mathfrak{L}_XY\big)^M \;:=\;  [X,Y]_{\mathrm{Lie}}^M \mathrm{Y}^{MN}_{\quad\;\; PQ}\partial_NX^PY^Q + (\lambda_Y+\omega)\partial_NX^NY^M.
\end{equation}
Similarly to Double Field Theory, the $E$-bracket of Exceptional Field Theory is defined as the anti-symmetrisation of the generalised Lie derivative, i.e.
\begin{equation}
    \llbracket X,Y \rrbracket_{\mathrm{E}} \;:=\; \frac{1}{2}\big(\mathfrak{L}_XY - \mathfrak{L}_YX\big).
\end{equation}
Now, the field content of Exceptional Field Theory is
\begin{equation}
\mathrm{Fields}:\;\{g_{\mu\nu},\,\mathcal{G}_{MN},\,\mathcal{A}_{\mu},\,\mathcal{B}_{\mu\nu}, \,\mathcal{C}_{\mu\nu\lambda},\,\mathcal{D}_{\mu\nu\lambda\rho},\,\dots\}.
\end{equation}
First, the moduli field of the generalised metric of Exceptional Field Theory will be a function valued in the following coset space
\begin{equation}
    \mathcal{G}_{MN}\;\in\;E_{n(n)}/K_n
\end{equation}
where $K_n\subset E_{n(n)}$ is the maximal compact subgroup of the U-duality group.
Generalising Double Field Theory, we can construct a chain complex of representations of the U-duality group $E_{n(n)}$
\begin{equation}
    R_1 \,\xleftarrow{\;\;\mathfrak{D}\;\;}\, R_2 \,\xleftarrow{\;\;\mathfrak{D}\;\;}\, \cdots \,\xleftarrow{\;\;\mathfrak{D}\;\;}\, R_{8-n}\,\cong\,\overline{R}_1,
\end{equation}
where, by combining the derivatives $\partial_M$ on the extended fibre with the $\mathrm{Y}$-tensor, we defined the differential
\begin{equation}
    \mathfrak{D}\,: R_p \;\longrightarrow\; R_{p-1}.
\end{equation}
By contraction with the $\mathrm{Y}$-tensor, we can also introduce the EFT \textit{exterior product}
\begin{equation}
    \langle-,-\rangle : R_p\otimes R_q \longrightarrow R_{p+q},
\end{equation}
for any $p+q\leq 8-n$.
For a given vector $X\in R_1$, we have the EFT \textit{magic formula}:
\begin{equation}
    \begin{aligned}
         \mathfrak{L}_XY \;&=\; \llbracket X, Y \rrbracket_\mathrm{E} + \mathfrak{D}\langle X,Y\rangle & \text{for }Y\in R_1, \\
        \mathfrak{L}_XY \;&=\; \langle X, \mathfrak{D}Y\rangle + \mathfrak{D}\langle X,Y\rangle & \text{for }Y\in R_p,
    \end{aligned}
\end{equation}
with $1<p<8-n$. The tensor hierarchy is defined by the following subset of fields:
\begin{equation}
\mathrm{Tensor\,hierarchy}:\;\{\mathcal{A}_{\mu},\,\mathcal{B}_{\mu\nu}, \,\mathcal{C}_{\mu\nu\lambda},\,\mathcal{D}_{\mu\nu\lambda\rho},\,\dots\},
\end{equation}
where the fields can be identified with the following $R_i$-valued differential forms on the base space:
\begin{equation}
    \begin{aligned}
        \mathcal{A} \,&\in\; \Omega^1(\mathbb{R}^{d-n})\otimes R_1, \\
        \mathcal{B} \;&\in\; \Omega^2(\mathbb{R}^{d-n})\otimes R_2, \\
        \mathcal{C} \;&\in\; \Omega^3(\mathbb{R}^{d-n})\otimes R_3, \\
        \mathcal{D} \;&\in\; \Omega^4(\mathbb{R}^{d-n})\otimes R_4, \\
        &\vdots
    \end{aligned}
\end{equation}
Finally, we have field strengths of the tensor hierarchy of the following form:
\begin{equation}
\begin{aligned}
    \mathcal{F} \;&=\; \di \mathcal{A} + \llbracket\mathcal{A}\,\overset{\wedge}{,}\,\mathcal{A}\rrbracket_\mathrm{E} + \mathfrak{D}\mathcal{B}, \\
    \mathcal{H} \;&=\; \di \mathcal{B} + \langle\mathcal{A}\,\overset{\wedge}{,}\,\mathfrak{D}\mathcal{B}\rangle + \frac{1}{2}\langle\mathcal{F}\,\overset{\wedge}{,}\,\mathcal{A}\rangle + \frac{1}{3!}\langle\mathcal{A}\,\overset{\wedge}{,}\,\llbracket\mathcal{A}\,\overset{\wedge}{,}\,\mathcal{A}\rrbracket_\mathrm{E}\rangle + \mathfrak{D}\mathcal{C},\\
    \mathcal{J} \;&=\; \di \mathcal{C} + \langle\mathcal{A}\,\overset{\wedge}{,}\,\mathfrak{D}\mathcal{C}\rangle + \langle\mathcal{B}\,\overset{\wedge}{,}\,\mathfrak{D}\mathcal{B}\rangle + \frac{1}{2}\langle\mathcal{F}\,\overset{\wedge}{,}\,\mathcal{B}\rangle + \frac{1}{3!}\langle\mathcal{A}\,\overset{\wedge}{,}\,\langle\mathcal{A}\,\overset{\wedge}{,}\,\mathcal{F}\rangle\rangle + \mathfrak{D}\mathcal{D}, \\
    &\vdots
\end{aligned}
\end{equation}

\section{The globalisation problem}
However, the Kalb-Ramond field $B$ is geometrically the connection of a bundle gerbe and hence it is globalised by the patching conditions \eqref{eq:introgerby}. Thus, it is not obvious how the local patches $\big(\mathcal{U},\,\sim,\,\eta,\,\mathcal{G}\big)$, which we introduced here, can be consistently glued together? This is the substance of the globalisation problem of the doubled geometry underlying Double Field Theory. Seminal work in this direction was done by \cite{BCM14}. The geometrisation of Double Field Theory which is the object of this thesis is a candidate to answer this question. 
\vspace{0.2cm}

\noindent We know how to globalise the local geometry of Double Field Theory for particular classes of examples, where the gerby nature of the Kalb-Ramond field is not manifest. In particular, global Double Field Theory on group manifolds \cite{Hul09, DFTWZW15, DFTWZW15x, Hass18} is well-defined. Also, doubled torus bundles \cite{Hull06}, which are globally affine $T^{2n}$-bundles on an undoubled base manifold \cite{BelHulMin07}, are well-defined. However, there is no conclusive answer on how to globalise this geometry in the most general case. Moreover, it has been argued in \cite{HohSam13} that the doubled torus bundles should be recoverable by imposing a certain compactified topology to a general doubled space, whose geometry, however, remains an open problem. A problem which becomes even more obscure in the case of the geometry underlying Exceptional Field Theory.

\begin{landscape}
\begin{table}[h!]\vspace{4cm}\begin{center}
\begin{center}
 \begin{tabular}{||c||c|c|c||c|c|c|c|c|c||c||} 
 \hline
  dim & \multicolumn{3}{c||}{U-duality group} & \multicolumn{6}{c||}{Tensor hierarchy reps} & adjoint rep \\[0.5ex]
 \hline\hline
    $11-n$ & $E_{n(n)}$ & $K_n$ & $\mathrm{dim}\big(E_{n(n)}/K_n\big)$ & $R_1$ & $R_2$ & $R_3$ & $R_4$ & $R_5$ & $R_6$ & $R_\mathrm{ad}$ \\ [0.5ex] 
    \hline
    $9$ & $SL(2)\times\mathbb{R}^+$ & $SO(2)$ & $3$ & $\mathbf{2}_1\oplus\mathbf{1}_{-2}$ & $\mathbf{2}_0$ & $\mathbf{1}_1$ & $\mathbf{1}_0$ & $\mathbf{2}_1$ & $\mathbf{2}_0\oplus\mathbf{1}_2$ & $\mathbf{3}$ \\[0.8ex]  
    $8$ & $SL(3)\times SL(2)$ & $SO(3)\times SO(2)$ & $7$ & $(\mathbf{3},\mathbf{2})$ & $(\overline{\mathbf{3}},\mathbf{1})$ & $(\mathbf{1,2})$ & $\mathbf{(3,1)}$ & $(\overline{\mathbf{3}},\mathbf{2})$ & $\ast$ & $(\mathbf{8},\mathbf{1})\oplus(\mathbf{1},\mathbf{3})$ \\[0.8ex]  
    $7$ & $SL(5)$ & $SO(5)$ & $14$ & $\mathbf{10}$ & $\overline{\mathbf{5}}$ & $\mathbf{5}$ & $\overline{\mathbf{10}}$ & $\ast$ & & $\mathbf{24}$ \\[0.8ex]  
    $6$ & $SO(5,5)$ & $SO(5)\times SO(5)$ & $25$ & $\mathbf{16}$ & $\mathbf{10}$ & $\overline{\mathbf{16}}$ & $\ast$ & & & $\mathbf{45}$ \\[0.8ex] 
    $5$ & $E_{6(6)}$ & $USp(8)$ & $42$ & $\mathbf{27}$ & $\overline{\mathbf{27}}$ & $\ast$ & & & & $\mathbf{78}$ \\[0.8ex]
    $4$ & $E_{7(7)}$ & $SU(8)$ & $70$ & $\mathbf{56}$ & $\ast$ &  & & & & $\mathbf{133}$ \\[0.8ex]
 \hline
\end{tabular}
\end{center}
\caption[Classification of the U-duality groups]{\label{tab:egg}Classification of the U-duality groups $G$, their maximal compact subgroups $K_n\subset E_{n(n)}$, their $R_n$ representations and their adjoint representations $R_\mathrm{ad}\cong R_{9-n}$. The symbol $\ast$ denotes where the tensor hierarchy stops.}\vspace{-0.0cm}
\end{center}\end{table}
\end{landscape}
\begin{savequote}[6.7cm]
{\textgreekfont Mὴ εἶναι βασιλικὴν ἀτραπὸν ἐπί γεωμετρίαν.}

There is no royal road to geometry.
  \qauthor{--- Euclid}
\end{savequote}

\chapter{\label{ch:3}Elements of higher geometry}

\minitoc

\noindent In this chapter we will introduce all the needed mathematical notions for this thesis. The following introduction in higher geometry will make the thesis as self-contained as possible and it will be particularly directed to a reader in theoretical or mathematical physics. 
The general theory reviewed in this chapter can be found mostly in \cite{topos, lurie2009infinity2categories, DCCTv2, LurieHA, Jurco:2019woz}.
For a review of the latest development in the application of higher geometry to String Theory, see \cite{Jurco:2016qwv}. \vspace{0.25cm}

\noindent Higher geometry is, essentially, differential geometry where the notion of equality has been replaced by the weaker notion of equivalence. Thus the question whether two geometric objects are equal is replaced by the question whether they are equivalent. Two equivalences, in turn, can be related by an equivalence, and so on. 
\begin{equation}
\begin{tikzcd}[row sep=scriptsize, column sep=12ex]
    \phi \arrow[r, "\cong"]
    & \phi'
\end{tikzcd}, \quad\;
\begin{tikzcd}[row sep=scriptsize, column sep=14ex]
    \phi \arrow[r, bend left=60, ""{name=U, below}, "\,\cong"]
    \arrow[r, bend right=60, ""{name=D}, "\cong"']
    & \phi'
    \arrow[Rightarrow, from=U, to=D, "\cong", end anchor={[xshift=-0.18ex]}]
\end{tikzcd}, \quad\;
\begin{tikzcd}[row sep=scriptsize, column sep=16ex]
    \phi \arrow[r, bend left=60, ""{name=U, below}, "\cong"]
    \arrow[r, bend right=60, ""{name=D}, "\cong"']
    & \phi'
    \arrow[Rightarrow, from=U, to=D, bend left=55, ""{name=R, below}, "\cong"] \arrow[Rightarrow, from=U, to=D, bend right=55, ""{name=L}, "\cong"'] \tarrow[from=L, to=R, end anchor={[yshift=0.23ex]}]{r} \arrow[phantom, from=L, to=R, end anchor={[yshift=0.6ex]}, bend left=34, "{\footnotesize\cong}"]
\end{tikzcd}, \quad \cdots
\end{equation}
\noindent This way, higher geometry is a natural framework for theoretical physics, since equivalences can be interpreted as gauge transformations. Indeed, the natural and meaningful question in physics is whether two fields are gauge-equivalent.
In this specific sense, higher geometry can be seen as a generalisation of differential geometry which encompasses the gauge principle of physics.

\section{Category theory}

A category can be thought as a labeled directed graph, whose labelled vertices are known as objects and whose labelled directed edges are known as morphisms. A category is equipped with two fundamental properties. Firstly, the composition of morphisms is associative. Secondly, any object is canonically equipped with an identity morphism to itself. 
The point of category theory is studying the properties of classes of mathematical structures.

\begin{definition}[Category]
A category $\mathbf{C}$ is given by the following data:
\begin{itemize}
    \item a class of \textit{objects}, whose elements we will denote as $A\in\mathbf{C}$,
    \item a class of \textit{morphisms}, which are maps of the form $f:A\rightarrow B$ from an object $A\in\mathbf{C}$ (known as \textit{source}) to an object $B\in\mathbf{C}$ (known as \textit{target}). For any couple of objects $A,B\in\mathbf{C}$, we well denote as $\mathbf{C}(A,B)$ the class of all the morphisms from $A$ to $B$,
    \item a binary operation, known as \textit{composition}, $\circ: \mathbf{C}(A,B)\times \mathbf{C}(B,C)\rightarrow \mathbf{C}(A,C)$, which sends a couple of morphisms $f:A\rightarrow B$ and $g:B\rightarrow C$ to their composition $g\circ f:A\rightarrow C$. The composition $\circ$ satisfies the following properties:
    \begin{enumerate}
        \item \textit{associativity}, i.e. $h\circ (g\circ f) = (h\circ g)\circ f$,
        \item \textit{identity}, i.e. for any object $A\in\mathbf{C}$ there exists a morphism $\mathrm{id}_A:A\rightarrow A$, known as \textit{identity morphism}, satisfying $\mathrm{id}_B\circ f = f\circ \mathrm{id}_A$ for any $f:A\rightarrow B$.
    \end{enumerate}
\end{itemize}
\end{definition}

\begin{example}[Category of sets]
Let $\mathbf{Set}$ be the \textit{category of sets}, whose objects $S\in\mathbf{Set}$ are sets and whose morphisms $f:S\rightarrow T$ are functions between sets.
\end{example}

\begin{example}[Category of topological spaces]
Let $\mathbf{Top}$ be the \textit{category of topological spaces}, whose objects $X\in\mathbf{Top}$ are topological spaces and whose morphisms $f:X\rightarrow Y$ are continuous maps.
\end{example}

\begin{example}[Category of smooth manifolds]
Let $\mathbf{Diff}$ be the \textit{category of smooth manifold}, whose objects $M\in\mathbf{Diff}$ are smooth manifolds and whose morphisms $f:M\rightarrow N$ are smooth maps.
\end{example}

\begin{example}[Category of sheaves]
Let $\mathbf{Sh}$ be the \textit{category of sheaves}, whose objects $\mathscr{X}\in\mathbf{Sh}$ are sheaves over manifolds and whose morphisms $f:\mathscr{X}\rightarrow \mathscr{Y}$ are morphisms of sheaves over manifolds.
\end{example}

\begin{example}[Category of Lie algebras]
Let $\mathbf{LAlg}$ be the \textit{category of Lie algebras}, whose objects $\mathfrak{g}\in\mathbf{LAlg}$ are Lie algebras and whose morphisms $f:\mathfrak{g}\rightarrow \mathfrak{h}$ are homomorphisms of Lie algebras.
\end{example}

\begin{definition}[Opposite category]
Given a category $\mathbf{C}$, its \textit{opposite category} $\mathbf{C}^{\mathrm{op}}$ is a category whose objects are the same of $\mathbf{C}$ and whose morphisms interchange the source and target of the morphisms of $\mathbf{C}$. In other words, any morphism $f:A\rightarrow B$ in $\mathbf{C}$ gives a morphism $f^{\mathrm{op}}:B\rightarrow A$ in $\mathbf{C}^{\mathrm{op}}$.
\end{definition}

\begin{definition}[Functor]
Given two categories $\mathbf{C},\mathbf{D}$, a \textit{functor}
\begin{equation}
    F:\,\mathbf{C} \;\longrightarrow\; \mathbf{D}
\end{equation}
is a map which associates each object $X\in\mathbf{C}$ to an object $F(X)\in\mathbf{D}$ and each morphism $f:A\rightarrow B$ in $\mathbf{C}$ to a morphism $F(f):F(A)\rightarrow F(B)\in\mathbf{D}$, such that
\begin{itemize}
    \item $F(\mathrm{id}_X)=\mathrm{id}_{F(X)}$ for any object $X\in\mathbf{C}$,
    \item $F(g\circ f)=F(g)\circ F(f)$ for any couple of morphisms $f:A\rightarrow B$, $g:B\rightarrow C$ in $\mathbf{C}$.
\end{itemize}
\end{definition}

\begin{definition}[Natural transformation]
Given two functors $F,G:\mathbf{C}\rightarrow\mathbf{D}$, a \textit{natural transformation}
\begin{equation}
    \eta:\,F \;\longrightarrow\; G
\end{equation}
is a map which associates to any object $X\in \mathbf{C}$ a morphism $\eta_X:F(X)\rightarrow G(X)$ in $\mathbf{D}$ such that the following diagram commutes
\begin{equation}
\begin{tikzcd}[row sep=6.5ex, column sep=5ex]
    F(A) \arrow[r, "\eta_A"]\arrow[d, "F(f)"'] & G(A) \arrow[d, "G(f)"] \\
    F(B) \arrow[r, "\eta_B"] & G(B)
    \end{tikzcd}
\end{equation}
for any morphism $f:A\rightarrow B$ in $\mathbf{C}$.
\end{definition}

\begin{definition}[Category of functors]
We call $\Func(\mathbf{C},\mathbf{D})$ the \textit{category of functors}, whose objects are all the functors $F:\mathbf{C}\rightarrow \mathbf{D}$ and whose morphisms are all the natural transformations $\eta:F \rightarrow F'$ between such functors.
\end{definition}

\begin{definition}[Yoneda embedding]
The \textit{Yoneda embedding} is defined as the functor
\begin{equation}
    \begin{aligned}
        よ:\; \mathbf{C} \,\;&\longhookrightarrow\,\; \Func(\mathbf{C}^\mathrm{op},\mathbf{Set})\\[0.2ex]
        A\,\;&\longmapsto\,\; よ_{\!A}\,:=\,\mathbf{C}(-,A),
    \end{aligned}
\end{equation}
which is fully faithful.
\end{definition}

\begin{definition}[Terminal object]
A category $\mathbf{C}$ is equipped with a \textit{terminal object} $1\in\mathbf{C}$ if there exists exactly one morphism $X\rightarrow 1$ for any object $X\in\mathbf{C}$.
\end{definition}

\begin{definition}[Group object]
Given a category $\mathbf{C}$ equipped with a terminal object $1\in\mathbf{C}$ and finite products (i.e. for any $A,B\in\mathbf{C}$ there exists $A\times B\in\mathbf{C}$), we define a \textit{group object} $G\in\mathbf{C}$ as an object equipped with three morphisms
\begin{enumerate}
    \item $\mu:G\times G \rightarrow G$ (\textit{group multiplication}),
    \item $e:1\rightarrow G$ (\textit{inclusion of identity element}),
    \item $(-)^{-1}:G\rightarrow G$ (\textit{inversion operation}),
\end{enumerate}
such that they satisfy the following properties:
\begin{itemize}
    \item $\mu(\mu, \mathrm{id}_G)=\mu (\mathrm{id}_G, \mu)$ (i.e. \textit{associativity}),
    \item $\mu(\mathrm{id}_G, e) = \mathrm{id}_G$ and $\mu(e, \mathrm{id}_G) = \mathrm{id}_G$ (i.e. \textit{$e$ is a unit of $\mu$}),
    \item $\mu(\mathrm{id}_G, (-)^{-1})\circ \mathrm{diag}=e_G$ and $\mu((-)^{-1}, \mathrm{id}_G)\circ \mathrm{diag}=e_G$, where $\mathrm{diag}:G\rightarrow G\times G$ is the diagonal map and $e_G:G\rightarrow G$ is the composition of the unique morphism $G\rightarrow 1$ with $e$ (i.e. \textit{$(-)^{-1}$ is an inverse for $\mu$}).
\end{itemize}
We denote by $\mathrm{Grp}(\mathbf{C})$ the category of the group objects in $\mathbf{C}$.
\end{definition}

\begin{example}[Group]
A group object $G\in\mathrm{Grp}(\mathbf{Set})$ is a group.
\end{example}

\begin{example}[Topological group]
A group object $G\in\mathrm{Grp}(\mathbf{Top})$ is a topological group.
\end{example}

\begin{example}[Lie group]
A group object $G\in\mathrm{Grp}(\mathbf{Diff})$ is a Lie group.
\end{example}

\section{Smooth $\infty$-groupoids}

A topos is, very informally, a category equipped with a structure which provides a generalisation of topology and which allows us to apply to it much of the intuition about topological spaces. An $(\infty,1)$-topos \cite{topos} is a higher geometric generalisation of the notion of topos. We will not provide a definition of this idea, but we will limit ourselves to consider a particular example.
In fact, the smooth $\infty$-groupoids, together with the maps between them, make up an $(\infty,1)$-topos.
In a certain sense, this property makes smooth $\infty$-groupoid a suitable generalisation of smooth manifold.
Informally, this means that $\infty$-groupoids can be seen as "generalised spaces".\vspace{0.2cm}

\noindent This section will be devoted to the introduction of smooth $\infty$-groupoids and to the study of some of their properties. 

\subsection{Simplicial sets}

For any $n\in\mathbb{N}$, let us define the following linearly ordered set:
\begin{equation}
    [n]\;:=\; \{0,1,2,\dots,n\}.
\end{equation}
Such a set can be equivalently thought as an $n$-simplex. Let us explain how with some examples.
To begin with, we have
\begin{gather}\begin{aligned}
\begin{tikzpicture}[line join = round, line cap = round]
\draw (-1.5,0.3) node [anchor=north west][inner sep=0.75pt] {$[0]\;=\;$};
\coordinate [label=above:0] (0) at (0,0,0);
\begin{scope}[decoration={markings,mark=at position 0.5 with {\arrow{to}}}]
\fill(0) circle (1.5pt);
\end{scope}
\end{tikzpicture}
\end{aligned}\raisetag{10pt}
\end{gather}
which is just a point.
The next step is
\begin{gather}\begin{aligned}
\begin{tikzpicture}[line join = round, line cap = round]
\draw (-1.5,0.3) node [anchor=north west][inner sep=0.75pt] {$[1]\;=\;$};
\coordinate [label=above:1] (1) at (1,0,0);
\coordinate [label=above:0] (0) at (0,0,0);
\begin{scope}[decoration={markings,mark=at position 0.5 with {\arrow{to}}}]
\draw[] (0)--(1);
\fill(0) circle (1.5pt);
\fill(1) circle (1.5pt);
\end{scope}
\end{tikzpicture}
\end{aligned}\raisetag{20pt}
\end{gather}
which is a line segment. Subsequently, we have
\begin{gather}\begin{aligned}
\begin{tikzpicture}[line join = round, line cap = round]
\draw (-1.8,0.6) node [anchor=north west][inner sep=0.75pt] {$[2]\;=\;$};
\coordinate [label=left:2] (2) at (0,0,0);
\coordinate [label=right:1] (1) at (1,0,0);
\coordinate [label=above:0] (0) at (0.5,0.9,0);
\begin{scope}[decoration={markings,mark=at position 0.5 with {\arrow{to}}}]
\draw[fill=lightgray,fill opacity=.5] (1)--(0)--(2)--cycle;
\fill(0) circle (1.5pt);
\fill(1) circle (1.5pt);
\fill(2) circle (1.5pt);
\end{scope}
\end{tikzpicture}
\end{aligned}\raisetag{-10pt}
\end{gather}
which is a triangle. Finally, we have
\begin{gather}\begin{aligned}
\begin{tikzpicture}[line join = round, line cap = round]
\draw (-2.4,0.6) node [anchor=north west][inner sep=0.75pt] {$[3]\;=\;$};
\coordinate [label=above:3] (3) at (0,{sqrt(2)},0);
\coordinate [label=left:2] (2) at ({-.5*sqrt(3)},0,-.5);
\coordinate [label=below:1] (1) at (0,0,1);
\coordinate [label=right:0] (0) at ({.5*sqrt(3)},0,-.5);
\begin{scope}[decoration={markings,mark=at position 0.5 with {\arrow{to}}}]
\draw[densely dotted] (0)--(2);
\draw[fill=lightgray,fill opacity=.5] (1)--(0)--(3)--cycle;
\draw[fill=gray,fill opacity=.5] (2)--(1)--(3)--cycle;
\fill(0) circle (1.5pt);
\fill(1) circle (1.5pt);
\fill(2) circle (1.5pt);
\fill(3) circle (1.5pt);
\end{scope}
\end{tikzpicture}
\end{aligned}\raisetag{-10pt}
\end{gather}
which is a tetrahedron, and so on for $n>3$.
The $n$-simplices naturally make up a category.
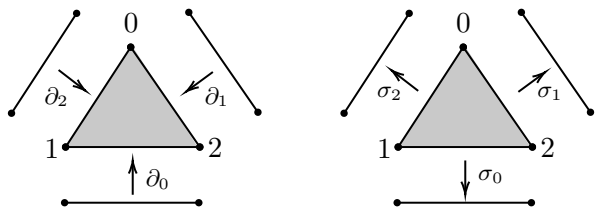
\begin{figure}[h]\begin{center}
\tikzset{every picture/.style={line width=0.75pt}} 
\begin{tikzpicture}[x=0.75pt,y=0.75pt,yscale=-1,xscale=1]
\draw  [draw opacity=0][fill={rgb, 255:red, 155; green, 155; blue, 155 }  ,fill opacity=0.54 ] (96,40.25) -- (129,91.5) -- (63,91.5) -- cycle ;
\draw    (63,91.5) -- (95.5,41.25) ;
\draw [shift={(95.5,41.25)}, rotate = 302.89] [color={rgb, 255:red, 0; green, 0; blue, 0 }  ][fill={rgb, 255:red, 0; green, 0; blue, 0 }  ][line width=0.75]      (0, 0) circle [x radius= 1.34, y radius= 1.34]   ;
\draw [shift={(63,91.5)}, rotate = 302.89] [color={rgb, 255:red, 0; green, 0; blue, 0 }  ][fill={rgb, 255:red, 0; green, 0; blue, 0 }  ][line width=0.75]      (0, 0) circle [x radius= 1.34, y radius= 1.34]   ;
\draw    (130,91.5) -- (95.5,41.25) ;
\draw [shift={(95.5,41.25)}, rotate = 235.53] [color={rgb, 255:red, 0; green, 0; blue, 0 }  ][fill={rgb, 255:red, 0; green, 0; blue, 0 }  ][line width=0.75]      (0, 0) circle [x radius= 1.34, y radius= 1.34]   ;
\draw [shift={(130,91.5)}, rotate = 235.53] [color={rgb, 255:red, 0; green, 0; blue, 0 }  ][fill={rgb, 255:red, 0; green, 0; blue, 0 }  ][line width=0.75]      (0, 0) circle [x radius= 1.34, y radius= 1.34]   ;
\draw    (130,91.5) -- (63,91.5) ;
\draw [shift={(63,91.5)}, rotate = 180] [color={rgb, 255:red, 0; green, 0; blue, 0 }  ][fill={rgb, 255:red, 0; green, 0; blue, 0 }  ][line width=0.75]      (0, 0) circle [x radius= 1.34, y radius= 1.34]   ;
\draw [shift={(130,91.5)}, rotate = 180] [color={rgb, 255:red, 0; green, 0; blue, 0 }  ][fill={rgb, 255:red, 0; green, 0; blue, 0 }  ][line width=0.75]      (0, 0) circle [x radius= 1.34, y radius= 1.34]   ;
\draw    (129.5,119.5) -- (62.5,119.5) ;
\draw [shift={(62.5,119.5)}, rotate = 180] [color={rgb, 255:red, 0; green, 0; blue, 0 }  ][fill={rgb, 255:red, 0; green, 0; blue, 0 }  ][line width=0.75]      (0, 0) circle [x radius= 1.34, y radius= 1.34]   ;
\draw [shift={(129.5,119.5)}, rotate = 180] [color={rgb, 255:red, 0; green, 0; blue, 0 }  ][fill={rgb, 255:red, 0; green, 0; blue, 0 }  ][line width=0.75]      (0, 0) circle [x radius= 1.34, y radius= 1.34]   ;
\draw    (159,74) -- (124.5,23.75) ;
\draw [shift={(124.5,23.75)}, rotate = 235.53] [color={rgb, 255:red, 0; green, 0; blue, 0 }  ][fill={rgb, 255:red, 0; green, 0; blue, 0 }  ][line width=0.75]      (0, 0) circle [x radius= 1.34, y radius= 1.34]   ;
\draw [shift={(159,74)}, rotate = 235.53] [color={rgb, 255:red, 0; green, 0; blue, 0 }  ][fill={rgb, 255:red, 0; green, 0; blue, 0 }  ][line width=0.75]      (0, 0) circle [x radius= 1.34, y radius= 1.34]   ;
\draw    (36,74.5) -- (68.5,24.25) ;
\draw [shift={(68.5,24.25)}, rotate = 302.89] [color={rgb, 255:red, 0; green, 0; blue, 0 }  ][fill={rgb, 255:red, 0; green, 0; blue, 0 }  ][line width=0.75]      (0, 0) circle [x radius= 1.34, y radius= 1.34]   ;
\draw [shift={(36,74.5)}, rotate = 302.89] [color={rgb, 255:red, 0; green, 0; blue, 0 }  ][fill={rgb, 255:red, 0; green, 0; blue, 0 }  ][line width=0.75]      (0, 0) circle [x radius= 1.34, y radius= 1.34]   ;
\draw    (96.5,115.5) -- (96.5,100.25) ;
\draw [shift={(96.5,98.25)}, rotate = 450] [color={rgb, 255:red, 0; green, 0; blue, 0 }  ][line width=0.75]    (6.56,-1.97) .. controls (4.17,-0.84) and (1.99,-0.18) .. (0,0) .. controls (1.99,0.18) and (4.17,0.84) .. (6.56,1.97)   ;
\draw    (136.5,53.75) -- (124.61,62.56) ;
\draw [shift={(123,63.75)}, rotate = 323.47] [color={rgb, 255:red, 0; green, 0; blue, 0 }  ][line width=0.75]    (6.56,-1.97) .. controls (4.17,-0.84) and (1.99,-0.18) .. (0,0) .. controls (1.99,0.18) and (4.17,0.84) .. (6.56,1.97)   ;
\draw    (59.5,52.75) -- (71.42,62.02) ;
\draw [shift={(73,63.25)}, rotate = 217.87] [color={rgb, 255:red, 0; green, 0; blue, 0 }  ][line width=0.75]    (6.56,-1.97) .. controls (4.17,-0.84) and (1.99,-0.18) .. (0,0) .. controls (1.99,0.18) and (4.17,0.84) .. (6.56,1.97)   ;
\draw  [draw opacity=0][fill={rgb, 255:red, 155; green, 155; blue, 155 }  ,fill opacity=0.54 ] (262,40.25) -- (295,91.5) -- (229,91.5) -- cycle ;
\draw    (229,91.5) -- (261.5,41.25) ;
\draw [shift={(261.5,41.25)}, rotate = 302.89] [color={rgb, 255:red, 0; green, 0; blue, 0 }  ][fill={rgb, 255:red, 0; green, 0; blue, 0 }  ][line width=0.75]      (0, 0) circle [x radius= 1.34, y radius= 1.34]   ;
\draw [shift={(229,91.5)}, rotate = 302.89] [color={rgb, 255:red, 0; green, 0; blue, 0 }  ][fill={rgb, 255:red, 0; green, 0; blue, 0 }  ][line width=0.75]      (0, 0) circle [x radius= 1.34, y radius= 1.34]   ;
\draw    (296,91.5) -- (261.5,41.25) ;
\draw [shift={(261.5,41.25)}, rotate = 235.53] [color={rgb, 255:red, 0; green, 0; blue, 0 }  ][fill={rgb, 255:red, 0; green, 0; blue, 0 }  ][line width=0.75]      (0, 0) circle [x radius= 1.34, y radius= 1.34]   ;
\draw [shift={(296,91.5)}, rotate = 235.53] [color={rgb, 255:red, 0; green, 0; blue, 0 }  ][fill={rgb, 255:red, 0; green, 0; blue, 0 }  ][line width=0.75]      (0, 0) circle [x radius= 1.34, y radius= 1.34]   ;
\draw    (296,91.5) -- (229,91.5) ;
\draw [shift={(229,91.5)}, rotate = 180] [color={rgb, 255:red, 0; green, 0; blue, 0 }  ][fill={rgb, 255:red, 0; green, 0; blue, 0 }  ][line width=0.75]      (0, 0) circle [x radius= 1.34, y radius= 1.34]   ;
\draw [shift={(296,91.5)}, rotate = 180] [color={rgb, 255:red, 0; green, 0; blue, 0 }  ][fill={rgb, 255:red, 0; green, 0; blue, 0 }  ][line width=0.75]      (0, 0) circle [x radius= 1.34, y radius= 1.34]   ;
\draw    (295.5,119.5) -- (228.5,119.5) ;
\draw [shift={(228.5,119.5)}, rotate = 180] [color={rgb, 255:red, 0; green, 0; blue, 0 }  ][fill={rgb, 255:red, 0; green, 0; blue, 0 }  ][line width=0.75]      (0, 0) circle [x radius= 1.34, y radius= 1.34]   ;
\draw [shift={(295.5,119.5)}, rotate = 180] [color={rgb, 255:red, 0; green, 0; blue, 0 }  ][fill={rgb, 255:red, 0; green, 0; blue, 0 }  ][line width=0.75]      (0, 0) circle [x radius= 1.34, y radius= 1.34]   ;
\draw    (325,74) -- (290.5,23.75) ;
\draw [shift={(290.5,23.75)}, rotate = 235.53] [color={rgb, 255:red, 0; green, 0; blue, 0 }  ][fill={rgb, 255:red, 0; green, 0; blue, 0 }  ][line width=0.75]      (0, 0) circle [x radius= 1.34, y radius= 1.34]   ;
\draw [shift={(325,74)}, rotate = 235.53] [color={rgb, 255:red, 0; green, 0; blue, 0 }  ][fill={rgb, 255:red, 0; green, 0; blue, 0 }  ][line width=0.75]      (0, 0) circle [x radius= 1.34, y radius= 1.34]   ;
\draw    (202,74.5) -- (234.5,24.25) ;
\draw [shift={(234.5,24.25)}, rotate = 302.89] [color={rgb, 255:red, 0; green, 0; blue, 0 }  ][fill={rgb, 255:red, 0; green, 0; blue, 0 }  ][line width=0.75]      (0, 0) circle [x radius= 1.34, y radius= 1.34]   ;
\draw [shift={(202,74.5)}, rotate = 302.89] [color={rgb, 255:red, 0; green, 0; blue, 0 }  ][fill={rgb, 255:red, 0; green, 0; blue, 0 }  ][line width=0.75]      (0, 0) circle [x radius= 1.34, y radius= 1.34]   ;
\draw    (262.5,113.5) -- (262.5,98.25) ;
\draw [shift={(262.5,115.5)}, rotate = 270] [color={rgb, 255:red, 0; green, 0; blue, 0 }  ][line width=0.75]    (6.56,-1.97) .. controls (4.17,-0.84) and (1.99,-0.18) .. (0,0) .. controls (1.99,0.18) and (4.17,0.84) .. (6.56,1.97)   ;
\draw    (300.89,54.94) -- (289,63.75) ;
\draw [shift={(302.5,53.75)}, rotate = 143.47] [color={rgb, 255:red, 0; green, 0; blue, 0 }  ][line width=0.75]    (6.56,-1.97) .. controls (4.17,-0.84) and (1.99,-0.18) .. (0,0) .. controls (1.99,0.18) and (4.17,0.84) .. (6.56,1.97)   ;
\draw    (227.08,53.98) -- (239,63.25) ;
\draw [shift={(225.5,52.75)}, rotate = 37.87] [color={rgb, 255:red, 0; green, 0; blue, 0 }  ][line width=0.75]    (6.56,-1.97) .. controls (4.17,-0.84) and (1.99,-0.18) .. (0,0) .. controls (1.99,0.18) and (4.17,0.84) .. (6.56,1.97)   ;
\draw (101.5,101.4) node [anchor=north west][inner sep=0.75pt]  [font=\footnotesize]  {$\partial _{0}$};
\draw (131,59.9) node [anchor=north west][inner sep=0.75pt]  [font=\footnotesize]  {$\partial _{1}$};
\draw (51,59.9) node [anchor=north west][inner sep=0.75pt]  [font=\footnotesize]  {$\partial _{2}$};
\draw (90.5,22.9) node [anchor=north west][inner sep=0.75pt]    {$0$};
\draw (51,86.4) node [anchor=north west][inner sep=0.75pt]    {$1$};
\draw (132.5,85.9) node [anchor=north west][inner sep=0.75pt]    {$2$};
\draw (267.5,101.4) node [anchor=north west][inner sep=0.75pt]  [font=\footnotesize]  {$\sigma _{0}$};
\draw (297,59.9) node [anchor=north west][inner sep=0.75pt]  [font=\footnotesize]  {$\sigma _{1}$};
\draw (217,59.9) node [anchor=north west][inner sep=0.75pt]  [font=\footnotesize]  {$\sigma _{2}$};
\draw (256.5,22.9) node [anchor=north west][inner sep=0.75pt]    {$0$};
\draw (217,86.4) node [anchor=north west][inner sep=0.75pt]    {$1$};
\draw (298.5,85.9) node [anchor=north west][inner sep=0.75pt]    {$2$};
\end{tikzpicture}
\caption{Example of face inclusion $\partial_i$ and degeneracy projection $\sigma_i$ maps for $[2]$.}
\end{center}\end{figure}

\begin{definition}[Simplex category]
The \textit{simplex category} $\mathbf{\Delta}$ is the category such that
\begin{itemize}
    \item objects are $n$-simplices $[n]$ for any $n\in\mathbb{N}$,
    \item morphisms are as follows:\begin{enumerate}
        \item  \textit{face inclusion} $\partial_i:[n-1] \longhookrightarrow [n]$ for $i=0,\dots,n$, is the only (order-preserving) injection that "misses" $i$,
        \item  \textit{degenerate projection} $\sigma_i:[n+1]\longtwoheadrightarrow [n]$ for $i=0,\dots,n$, is the only (order-preserving) surjection  that "hits" $i$ twice,
    \end{enumerate}
    which satisfy the following relations, also known as \textit{simplicial identities}:
    \begin{equation}
        \begin{aligned}
            \partial_i\circ\partial_j \;&=\; \partial_{j+1}\circ\partial_i && (i<j),\\
            \sigma_i\circ \partial_j \;&=\; \partial_{j-1}\circ\sigma_i && (i<j), \\
            \sigma_i\circ \partial_i \;&=\; \mathrm{id}_{[n]} && (0\leq i \leq n), \\
            \sigma_{i+1}\circ \partial_i \;&=\; \mathrm{id}_{[n]} && (0\leq i \leq n), \\
            \sigma_i\circ\partial_j  \;&=\; \partial_j \circ \sigma_{i-1} &&(i>j+1), \\
            \sigma_i\circ\sigma_j \;&=\; \sigma_{j}\circ\sigma_{i+1} &&(i\leq j).
        \end{aligned}
    \end{equation}
\end{itemize}
\end{definition}

\noindent Notice that, by using the face inclusions maps $\partial_i$, we obtain a diagram of the following form:
\vspace{-0.2cm}
\begin{equation}
    \begin{tikzcd}[row sep=scriptsize, column sep=5ex]
    \; \cdots\; \arrow[r, yshift=1.4ex, leftarrow] \arrow[r, yshift=2.8ex, leftarrow] \arrow[r, leftarrow] \arrow[r, yshift=-1.4ex, leftarrow]\arrow[r, yshift=-2.8ex, leftarrow] & \;\begin{tikzpicture}[line join = round, line cap = round]
\coordinate [label=above:3] (3) at (0,{sqrt(2)},0);
\coordinate [label=left:2] (2) at ({-.5*sqrt(3)},0,-.5);
\coordinate [label=below:1] (1) at (0,0,1);
\coordinate [label=right:0] (0) at ({.5*sqrt(3)},0,-.5);
\begin{scope}[decoration={markings,mark=at position 0.5 with {\arrow{to}}}]
\draw[densely dotted] (0)--(2);
\draw[fill=lightgray,fill opacity=.5] (1)--(0)--(3)--cycle;
\draw[fill=gray,fill opacity=.5] (2)--(1)--(3)--cycle;
\fill(0) circle (1.5pt);
\fill(1) circle (1.5pt);
\fill(2) circle (1.5pt);
\fill(3) circle (1.5pt);
\end{scope}
\end{tikzpicture}\; \arrow[r, yshift=1.8ex, leftarrow]\arrow[r, yshift=0.6ex, leftarrow]\arrow[r, yshift=-1.8ex, leftarrow]\arrow[r, yshift=-0.6ex, leftarrow]& \;\begin{tikzpicture}[line join = round, line cap = round]
\coordinate [label=left:2] (2) at (0,-0.2,0);
\coordinate [label=right:1] (1) at (1,-0.2,0);
\coordinate [label=above:0] (0) at (0.5,0.7,0);
\begin{scope}[decoration={markings,mark=at position 0.5 with {\arrow{to}}}]
\draw[fill=lightgray,fill opacity=.5] (1)--(0)--(2)--cycle;
\fill(0) circle (1.5pt);
\fill(1) circle (1.5pt);
\fill(2) circle (1.5pt);
\end{scope}
\end{tikzpicture}\;
    \arrow[r, yshift=1.4ex, leftarrow] \arrow[r, leftarrow] \arrow[r, yshift=-1.4ex, leftarrow] & \;\begin{tikzpicture}[line join = round, line cap = round]
\coordinate [label=above:1] (1) at (1,0,0);
\coordinate [label=above:0] (0) at (0,0,0);
\begin{scope}[decoration={markings,mark=at position 0.5 with {\arrow{to}}}]
\draw[] (0)--(1);
\fill(0) circle (1.5pt);
\fill(1) circle (1.5pt);
\end{scope}
\end{tikzpicture} \; \arrow[r, yshift=0.7ex, leftarrow] \arrow[r, yshift=-0.7ex, leftarrow] &\; \begin{tikzpicture}[line join = round, line cap = round]
\coordinate [label=above:0] (0) at (0,0,0);
\begin{scope}[decoration={markings,mark=at position 0.5 with {\arrow{to}}}]
\fill(0) circle (1.5pt);
\end{scope}
\end{tikzpicture}
    \end{tikzcd}    
\end{equation}

\begin{definition}[Simplicial set]
Let $\mathbf{Set}$ be the category of sets and $\mathbf{\Delta}$ the simplex category. A \textit{simplicial set} is defined as a functor
\begin{equation}
    K :\, \mathbf{\Delta}^\mathrm{op}\;\longrightarrow \; \mathbf{Set}.
\end{equation}
In other words, a simplicial set $K$ is given by the following sets:
\begin{itemize}
    \item $K_0 := K([0])$ is the set of objects,
    \item $K_n:= K([n])$ is the set of $n$-morphisms for any natural $n>0$,
\end{itemize} 
which are equipped with the following maps:
\begin{enumerate}
    \item \textit{face maps} $\di_i:=K(\partial_i):\,K_n\,\longtwoheadrightarrow\,K_{n-1}$ send $n$-morphisms to its $i$-th boundary $(n-1)$-morphisms,
    \item \textit{degeneracy maps} $s_i:=K(\sigma_i):\,K_n\,\longhookrightarrow\,K_{n+1}$ send $n$-morphisms to the identity $(n+1)$-morphisms on them,
\end{enumerate} 
which satisfy the following relations, also known as \textit{simplicial identities}:
 \begin{equation}
        \begin{aligned}
            \di_i\circ\di_j \;&=\; \di_{j-1}\circ\di_i && (i<j),\\
            \di_i\circ s_j \;&=\;  s_{j-1}\circ\di_i && (i<j), \\
            \di_i\circ  s_i \;&=\; \mathrm{id} && (0\leq i \leq n), \\
            \di_{i+1}\circ  s_i \;&=\; \mathrm{id} && (0\leq i \leq n), \\
            \di_i\circ  s_j \;&=\;  s_j \circ \di_{i-1} &&(i>j+1), \\
             s_i\circ s_j \;&=\;  s_{j+1}\circ s_i &&(i\leq j).
        \end{aligned}
    \end{equation}
\noindent Notice that the collection of face maps $\di_i$ defines a diagram of the following form:
\begin{equation}
    \begin{tikzcd}[row sep=scriptsize, column sep=5ex]
    \; \cdots\; \arrow[r, yshift=1.4ex] \arrow[r, yshift=2.8ex] \arrow[r] \arrow[r, yshift=-1.4ex]\arrow[r, yshift=-2.8ex] & K_3 \arrow[r, yshift=1.8ex]\arrow[r, yshift=0.6ex]\arrow[r, yshift=-1.8ex]\arrow[r, yshift=-0.6ex]& K_2
    \arrow[r, yshift=1.4ex] \arrow[r] \arrow[r, yshift=-1.4ex] & K_1  \arrow[r, yshift=0.7ex] \arrow[r, yshift=-0.7ex] & K_0 .
    \end{tikzcd}    
\end{equation}
\end{definition}

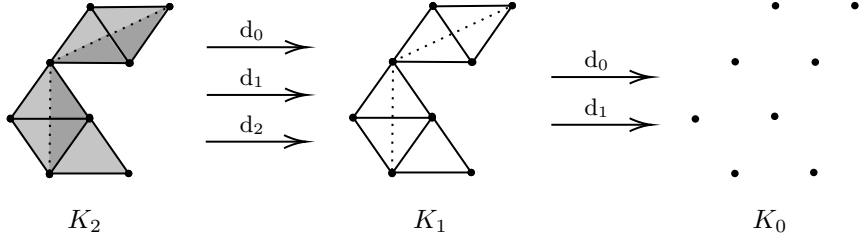
\begin{figure}[h]\begin{center}
\tikzset{every picture/.style={line width=0.75pt}} 
\begin{tikzpicture}[x=0.75pt,y=0.75pt,yscale=-1,xscale=1]
\draw    (197.5,38.5) -- (217,66.75) ;
\draw [shift={(217,66.75)}, rotate = 55.38] [color={rgb, 255:red, 0; green, 0; blue, 0 }  ][fill={rgb, 255:red, 0; green, 0; blue, 0 }  ][line width=0.75]      (0, 0) circle [x radius= 1.34, y radius= 1.34]   ;
\draw [shift={(197.5,38.5)}, rotate = 55.38] [color={rgb, 255:red, 0; green, 0; blue, 0 }  ][fill={rgb, 255:red, 0; green, 0; blue, 0 }  ][line width=0.75]      (0, 0) circle [x radius= 1.34, y radius= 1.34]   ;
\draw    (177.5,66.75) -- (197.5,38.5) ;
\draw [shift={(197.5,38.5)}, rotate = 305.3] [color={rgb, 255:red, 0; green, 0; blue, 0 }  ][fill={rgb, 255:red, 0; green, 0; blue, 0 }  ][line width=0.75]      (0, 0) circle [x radius= 1.34, y radius= 1.34]   ;
\draw [shift={(177.5,66.75)}, rotate = 305.3] [color={rgb, 255:red, 0; green, 0; blue, 0 }  ][fill={rgb, 255:red, 0; green, 0; blue, 0 }  ][line width=0.75]      (0, 0) circle [x radius= 1.34, y radius= 1.34]   ;
\draw    (177.5,66.75) -- (217,66.75) ;
\draw [shift={(217,66.75)}, rotate = 0] [color={rgb, 255:red, 0; green, 0; blue, 0 }  ][fill={rgb, 255:red, 0; green, 0; blue, 0 }  ][line width=0.75]      (0, 0) circle [x radius= 1.34, y radius= 1.34]   ;
\draw [shift={(177.5,66.75)}, rotate = 0] [color={rgb, 255:red, 0; green, 0; blue, 0 }  ][fill={rgb, 255:red, 0; green, 0; blue, 0 }  ][line width=0.75]      (0, 0) circle [x radius= 1.34, y radius= 1.34]   ;
\draw    (197,94.75) -- (177.5,66.49) ;
\draw [shift={(177.5,66.49)}, rotate = 235.4] [color={rgb, 255:red, 0; green, 0; blue, 0 }  ][fill={rgb, 255:red, 0; green, 0; blue, 0 }  ][line width=0.75]      (0, 0) circle [x radius= 1.34, y radius= 1.34]   ;
\draw [shift={(197,94.75)}, rotate = 235.4] [color={rgb, 255:red, 0; green, 0; blue, 0 }  ][fill={rgb, 255:red, 0; green, 0; blue, 0 }  ][line width=0.75]      (0, 0) circle [x radius= 1.34, y radius= 1.34]   ;
\draw    (217,66.51) -- (197,94.75) ;
\draw [shift={(197,94.75)}, rotate = 125.32] [color={rgb, 255:red, 0; green, 0; blue, 0 }  ][fill={rgb, 255:red, 0; green, 0; blue, 0 }  ][line width=0.75]      (0, 0) circle [x radius= 1.34, y radius= 1.34]   ;
\draw [shift={(217,66.51)}, rotate = 125.32] [color={rgb, 255:red, 0; green, 0; blue, 0 }  ][fill={rgb, 255:red, 0; green, 0; blue, 0 }  ][line width=0.75]      (0, 0) circle [x radius= 1.34, y radius= 1.34]   ;
\draw    (217,66) -- (236.5,94.25) ;
\draw [shift={(236.5,94.25)}, rotate = 55.38] [color={rgb, 255:red, 0; green, 0; blue, 0 }  ][fill={rgb, 255:red, 0; green, 0; blue, 0 }  ][line width=0.75]      (0, 0) circle [x radius= 1.34, y radius= 1.34]   ;
\draw [shift={(217,66)}, rotate = 55.38] [color={rgb, 255:red, 0; green, 0; blue, 0 }  ][fill={rgb, 255:red, 0; green, 0; blue, 0 }  ][line width=0.75]      (0, 0) circle [x radius= 1.34, y radius= 1.34]   ;
\draw    (197,94.25) -- (236.5,94.25) ;
\draw [shift={(236.5,94.25)}, rotate = 0] [color={rgb, 255:red, 0; green, 0; blue, 0 }  ][fill={rgb, 255:red, 0; green, 0; blue, 0 }  ][line width=0.75]      (0, 0) circle [x radius= 1.34, y radius= 1.34]   ;
\draw [shift={(197,94.25)}, rotate = 0] [color={rgb, 255:red, 0; green, 0; blue, 0 }  ][fill={rgb, 255:red, 0; green, 0; blue, 0 }  ][line width=0.75]      (0, 0) circle [x radius= 1.34, y radius= 1.34]   ;
\draw  [dash pattern={on 0.84pt off 2.51pt}]  (197.5,38.5) -- (197,94.25) ;
\draw    (217.4,10.38) -- (236.72,38.75) ;
\draw [shift={(236.72,38.75)}, rotate = 55.73] [color={rgb, 255:red, 0; green, 0; blue, 0 }  ][fill={rgb, 255:red, 0; green, 0; blue, 0 }  ][line width=0.75]      (0, 0) circle [x radius= 1.34, y radius= 1.34]   ;
\draw [shift={(217.4,10.38)}, rotate = 55.73] [color={rgb, 255:red, 0; green, 0; blue, 0 }  ][fill={rgb, 255:red, 0; green, 0; blue, 0 }  ][line width=0.75]      (0, 0) circle [x radius= 1.34, y radius= 1.34]   ;
\draw    (197.23,38.51) -- (236.72,38.75) ;
\draw [shift={(236.72,38.75)}, rotate = 0.35] [color={rgb, 255:red, 0; green, 0; blue, 0 }  ][fill={rgb, 255:red, 0; green, 0; blue, 0 }  ][line width=0.75]      (0, 0) circle [x radius= 1.34, y radius= 1.34]   ;
\draw [shift={(197.23,38.51)}, rotate = 0.35] [color={rgb, 255:red, 0; green, 0; blue, 0 }  ][fill={rgb, 255:red, 0; green, 0; blue, 0 }  ][line width=0.75]      (0, 0) circle [x radius= 1.34, y radius= 1.34]   ;
\draw    (197.23,38.51) -- (217.4,10.38) ;
\draw [shift={(217.4,10.38)}, rotate = 305.65] [color={rgb, 255:red, 0; green, 0; blue, 0 }  ][fill={rgb, 255:red, 0; green, 0; blue, 0 }  ][line width=0.75]      (0, 0) circle [x radius= 1.34, y radius= 1.34]   ;
\draw [shift={(197.23,38.51)}, rotate = 305.65] [color={rgb, 255:red, 0; green, 0; blue, 0 }  ][fill={rgb, 255:red, 0; green, 0; blue, 0 }  ][line width=0.75]      (0, 0) circle [x radius= 1.34, y radius= 1.34]   ;
\draw    (257.01,10.32) -- (217.51,10.68) ;
\draw [shift={(217.51,10.68)}, rotate = 179.48] [color={rgb, 255:red, 0; green, 0; blue, 0 }  ][fill={rgb, 255:red, 0; green, 0; blue, 0 }  ][line width=0.75]      (0, 0) circle [x radius= 1.34, y radius= 1.34]   ;
\draw [shift={(257.01,10.32)}, rotate = 179.48] [color={rgb, 255:red, 0; green, 0; blue, 0 }  ][fill={rgb, 255:red, 0; green, 0; blue, 0 }  ][line width=0.75]      (0, 0) circle [x radius= 1.34, y radius= 1.34]   ;
\draw    (257.01,10.32) -- (237.27,38.75) ;
\draw [shift={(237.27,38.75)}, rotate = 124.78] [color={rgb, 255:red, 0; green, 0; blue, 0 }  ][fill={rgb, 255:red, 0; green, 0; blue, 0 }  ][line width=0.75]      (0, 0) circle [x radius= 1.34, y radius= 1.34]   ;
\draw [shift={(257.01,10.32)}, rotate = 124.78] [color={rgb, 255:red, 0; green, 0; blue, 0 }  ][fill={rgb, 255:red, 0; green, 0; blue, 0 }  ][line width=0.75]      (0, 0) circle [x radius= 1.34, y radius= 1.34]   ;
\draw  [dash pattern={on 0.84pt off 2.51pt}]  (197.23,38.51) -- (257.01,10.32) ;
\draw    (26.5,39) -- (46,67.25) ;
\draw [shift={(46,67.25)}, rotate = 55.38] [color={rgb, 255:red, 0; green, 0; blue, 0 }  ][fill={rgb, 255:red, 0; green, 0; blue, 0 }  ][line width=0.75]      (0, 0) circle [x radius= 1.34, y radius= 1.34]   ;
\draw [shift={(26.5,39)}, rotate = 55.38] [color={rgb, 255:red, 0; green, 0; blue, 0 }  ][fill={rgb, 255:red, 0; green, 0; blue, 0 }  ][line width=0.75]      (0, 0) circle [x radius= 1.34, y radius= 1.34]   ;
\draw    (6.5,67.25) -- (26.5,39) ;
\draw [shift={(26.5,39)}, rotate = 305.3] [color={rgb, 255:red, 0; green, 0; blue, 0 }  ][fill={rgb, 255:red, 0; green, 0; blue, 0 }  ][line width=0.75]      (0, 0) circle [x radius= 1.34, y radius= 1.34]   ;
\draw [shift={(6.5,67.25)}, rotate = 305.3] [color={rgb, 255:red, 0; green, 0; blue, 0 }  ][fill={rgb, 255:red, 0; green, 0; blue, 0 }  ][line width=0.75]      (0, 0) circle [x radius= 1.34, y radius= 1.34]   ;
\draw    (6.5,67.25) -- (46,67.25) ;
\draw [shift={(46,67.25)}, rotate = 0] [color={rgb, 255:red, 0; green, 0; blue, 0 }  ][fill={rgb, 255:red, 0; green, 0; blue, 0 }  ][line width=0.75]      (0, 0) circle [x radius= 1.34, y radius= 1.34]   ;
\draw [shift={(6.5,67.25)}, rotate = 0] [color={rgb, 255:red, 0; green, 0; blue, 0 }  ][fill={rgb, 255:red, 0; green, 0; blue, 0 }  ][line width=0.75]      (0, 0) circle [x radius= 1.34, y radius= 1.34]   ;
\draw    (26,95.25) -- (6.5,66.99) ;
\draw [shift={(6.5,66.99)}, rotate = 235.4] [color={rgb, 255:red, 0; green, 0; blue, 0 }  ][fill={rgb, 255:red, 0; green, 0; blue, 0 }  ][line width=0.75]      (0, 0) circle [x radius= 1.34, y radius= 1.34]   ;
\draw [shift={(26,95.25)}, rotate = 235.4] [color={rgb, 255:red, 0; green, 0; blue, 0 }  ][fill={rgb, 255:red, 0; green, 0; blue, 0 }  ][line width=0.75]      (0, 0) circle [x radius= 1.34, y radius= 1.34]   ;
\draw    (46,67.01) -- (26,95.25) ;
\draw [shift={(26,95.25)}, rotate = 125.32] [color={rgb, 255:red, 0; green, 0; blue, 0 }  ][fill={rgb, 255:red, 0; green, 0; blue, 0 }  ][line width=0.75]      (0, 0) circle [x radius= 1.34, y radius= 1.34]   ;
\draw [shift={(46,67.01)}, rotate = 125.32] [color={rgb, 255:red, 0; green, 0; blue, 0 }  ][fill={rgb, 255:red, 0; green, 0; blue, 0 }  ][line width=0.75]      (0, 0) circle [x radius= 1.34, y radius= 1.34]   ;
\draw    (46,66.5) -- (65.5,94.75) ;
\draw [shift={(65.5,94.75)}, rotate = 55.38] [color={rgb, 255:red, 0; green, 0; blue, 0 }  ][fill={rgb, 255:red, 0; green, 0; blue, 0 }  ][line width=0.75]      (0, 0) circle [x radius= 1.34, y radius= 1.34]   ;
\draw [shift={(46,66.5)}, rotate = 55.38] [color={rgb, 255:red, 0; green, 0; blue, 0 }  ][fill={rgb, 255:red, 0; green, 0; blue, 0 }  ][line width=0.75]      (0, 0) circle [x radius= 1.34, y radius= 1.34]   ;
\draw    (26,94.75) -- (65.5,94.75) ;
\draw [shift={(65.5,94.75)}, rotate = 0] [color={rgb, 255:red, 0; green, 0; blue, 0 }  ][fill={rgb, 255:red, 0; green, 0; blue, 0 }  ][line width=0.75]      (0, 0) circle [x radius= 1.34, y radius= 1.34]   ;
\draw [shift={(26,94.75)}, rotate = 0] [color={rgb, 255:red, 0; green, 0; blue, 0 }  ][fill={rgb, 255:red, 0; green, 0; blue, 0 }  ][line width=0.75]      (0, 0) circle [x radius= 1.34, y radius= 1.34]   ;
\draw  [dash pattern={on 0.84pt off 2.51pt}]  (26.5,39) -- (26,94.75) ;
\draw    (46.4,10.88) -- (65.72,39.25) ;
\draw [shift={(65.72,39.25)}, rotate = 55.73] [color={rgb, 255:red, 0; green, 0; blue, 0 }  ][fill={rgb, 255:red, 0; green, 0; blue, 0 }  ][line width=0.75]      (0, 0) circle [x radius= 1.34, y radius= 1.34]   ;
\draw [shift={(46.4,10.88)}, rotate = 55.73] [color={rgb, 255:red, 0; green, 0; blue, 0 }  ][fill={rgb, 255:red, 0; green, 0; blue, 0 }  ][line width=0.75]      (0, 0) circle [x radius= 1.34, y radius= 1.34]   ;
\draw    (26.23,39.01) -- (65.72,39.25) ;
\draw [shift={(65.72,39.25)}, rotate = 0.35] [color={rgb, 255:red, 0; green, 0; blue, 0 }  ][fill={rgb, 255:red, 0; green, 0; blue, 0 }  ][line width=0.75]      (0, 0) circle [x radius= 1.34, y radius= 1.34]   ;
\draw [shift={(26.23,39.01)}, rotate = 0.35] [color={rgb, 255:red, 0; green, 0; blue, 0 }  ][fill={rgb, 255:red, 0; green, 0; blue, 0 }  ][line width=0.75]      (0, 0) circle [x radius= 1.34, y radius= 1.34]   ;
\draw    (26.23,39.01) -- (46.4,10.88) ;
\draw [shift={(46.4,10.88)}, rotate = 305.65] [color={rgb, 255:red, 0; green, 0; blue, 0 }  ][fill={rgb, 255:red, 0; green, 0; blue, 0 }  ][line width=0.75]      (0, 0) circle [x radius= 1.34, y radius= 1.34]   ;
\draw [shift={(26.23,39.01)}, rotate = 305.65] [color={rgb, 255:red, 0; green, 0; blue, 0 }  ][fill={rgb, 255:red, 0; green, 0; blue, 0 }  ][line width=0.75]      (0, 0) circle [x radius= 1.34, y radius= 1.34]   ;
\draw    (86.01,10.82) -- (46.51,11.18) ;
\draw [shift={(46.51,11.18)}, rotate = 179.48] [color={rgb, 255:red, 0; green, 0; blue, 0 }  ][fill={rgb, 255:red, 0; green, 0; blue, 0 }  ][line width=0.75]      (0, 0) circle [x radius= 1.34, y radius= 1.34]   ;
\draw [shift={(86.01,10.82)}, rotate = 179.48] [color={rgb, 255:red, 0; green, 0; blue, 0 }  ][fill={rgb, 255:red, 0; green, 0; blue, 0 }  ][line width=0.75]      (0, 0) circle [x radius= 1.34, y radius= 1.34]   ;
\draw    (86.01,10.82) -- (66.27,39.25) ;
\draw [shift={(66.27,39.25)}, rotate = 124.78] [color={rgb, 255:red, 0; green, 0; blue, 0 }  ][fill={rgb, 255:red, 0; green, 0; blue, 0 }  ][line width=0.75]      (0, 0) circle [x radius= 1.34, y radius= 1.34]   ;
\draw [shift={(86.01,10.82)}, rotate = 124.78] [color={rgb, 255:red, 0; green, 0; blue, 0 }  ][fill={rgb, 255:red, 0; green, 0; blue, 0 }  ][line width=0.75]      (0, 0) circle [x radius= 1.34, y radius= 1.34]   ;
\draw  [dash pattern={on 0.84pt off 2.51pt}]  (26.23,39.01) -- (86.01,10.82) ;
\draw  [draw opacity=0][fill={rgb, 255:red, 0; green, 0; blue, 0 }  ,fill opacity=0.23 ] (46,66.25) -- (66,94.75) -- (26,94.75) -- cycle ;
\draw  [draw opacity=0][fill={rgb, 255:red, 0; green, 0; blue, 0 }  ,fill opacity=0.23 ] (26.5,39) -- (46.5,67.5) -- (6.5,67.5) -- cycle ;
\draw  [draw opacity=0][fill={rgb, 255:red, 0; green, 0; blue, 0 }  ,fill opacity=0.23 ] (46.4,10.88) -- (66.4,39.38) -- (26.4,39.38) -- cycle ;
\draw  [draw opacity=0][fill={rgb, 255:red, 0; green, 0; blue, 0 }  ,fill opacity=0.23 ] (26,95.75) -- (6,67.25) -- (46,67.25) -- cycle ;
\draw  [draw opacity=0][fill={rgb, 255:red, 0; green, 0; blue, 0 }  ,fill opacity=0.23 ] (66.01,39.32) -- (46.01,10.82) -- (86.01,10.82) -- cycle ;
\draw  [draw opacity=0][fill={rgb, 255:red, 0; green, 0; blue, 0 }  ,fill opacity=1 ] (366.25,94.75) .. controls (366.25,93.71) and (367.08,92.87) .. (368.12,92.87) .. controls (369.16,92.87) and (370,93.71) .. (370,94.75) .. controls (370,95.79) and (369.16,96.62) .. (368.12,96.62) .. controls (367.08,96.62) and (366.25,95.79) .. (366.25,94.75) -- cycle ;
\draw  [draw opacity=0][fill={rgb, 255:red, 0; green, 0; blue, 0 }  ,fill opacity=1 ] (386.13,66) .. controls (386.13,64.96) and (386.96,64.13) .. (388,64.13) .. controls (389.04,64.13) and (389.88,64.96) .. (389.88,66) .. controls (389.88,67.04) and (389.04,67.88) .. (388,67.88) .. controls (386.96,67.88) and (386.13,67.04) .. (386.13,66) -- cycle ;
\draw  [draw opacity=0][fill={rgb, 255:red, 0; green, 0; blue, 0 }  ,fill opacity=1 ] (405.63,94.25) .. controls (405.63,93.21) and (406.46,92.38) .. (407.5,92.38) .. controls (408.54,92.38) and (409.38,93.21) .. (409.38,94.25) .. controls (409.38,95.29) and (408.54,96.13) .. (407.5,96.13) .. controls (406.46,96.13) and (405.63,95.29) .. (405.63,94.25) -- cycle ;
\draw  [draw opacity=0][fill={rgb, 255:red, 0; green, 0; blue, 0 }  ,fill opacity=1 ] (346.53,67.32) .. controls (346.53,66.29) and (347.37,65.45) .. (348.41,65.45) .. controls (349.44,65.45) and (350.28,66.29) .. (350.28,67.32) .. controls (350.28,68.36) and (349.44,69.2) .. (348.41,69.2) .. controls (347.37,69.2) and (346.53,68.36) .. (346.53,67.32) -- cycle ;
\draw  [draw opacity=0][fill={rgb, 255:red, 0; green, 0; blue, 0 }  ,fill opacity=1 ] (366.41,38.57) .. controls (366.41,37.54) and (367.25,36.7) .. (368.29,36.7) .. controls (369.32,36.7) and (370.16,37.54) .. (370.16,38.57) .. controls (370.16,39.61) and (369.32,40.45) .. (368.29,40.45) .. controls (367.25,40.45) and (366.41,39.61) .. (366.41,38.57) -- cycle ;
\draw  [draw opacity=0][fill={rgb, 255:red, 0; green, 0; blue, 0 }  ,fill opacity=1 ] (386.52,10.38) .. controls (386.52,9.34) and (387.36,8.5) .. (388.4,8.5) .. controls (389.43,8.5) and (390.27,9.34) .. (390.27,10.38) .. controls (390.27,11.42) and (389.43,12.25) .. (388.4,12.25) .. controls (387.36,12.25) and (386.52,11.42) .. (386.52,10.38) -- cycle ;
\draw  [draw opacity=0][fill={rgb, 255:red, 0; green, 0; blue, 0 }  ,fill opacity=1 ] (406.39,38.75) .. controls (406.39,37.72) and (407.23,36.88) .. (408.27,36.88) .. controls (409.3,36.88) and (410.14,37.72) .. (410.14,38.75) .. controls (410.14,39.79) and (409.3,40.63) .. (408.27,40.63) .. controls (407.23,40.63) and (406.39,39.79) .. (406.39,38.75) -- cycle ;
\draw  [draw opacity=0][fill={rgb, 255:red, 0; green, 0; blue, 0 }  ,fill opacity=1 ] (426.14,10.32) .. controls (426.14,9.29) and (426.98,8.45) .. (428.01,8.45) .. controls (429.05,8.45) and (429.89,9.29) .. (429.89,10.32) .. controls (429.89,11.36) and (429.05,12.2) .. (428.01,12.2) .. controls (426.98,12.2) and (426.14,11.36) .. (426.14,10.32) -- cycle ;
\draw  [draw opacity=0][fill={rgb, 255:red, 0; green, 0; blue, 0 }  ,fill opacity=0.18 ] (26.4,39.38) -- (46.5,66.75) -- (26.4,66.75) -- cycle ;
\draw  [draw opacity=0][fill={rgb, 255:red, 0; green, 0; blue, 0 }  ,fill opacity=0.18 ] (25.9,95.75) -- (46,66.25) -- (25.9,66.25) -- cycle ;
\draw  [draw opacity=0][fill={rgb, 255:red, 0; green, 0; blue, 0 }  ,fill opacity=0.18 ] (84.83,10.41) -- (66.23,40.9) -- (56.06,25.07) -- cycle ;
\draw  [draw opacity=0][fill={rgb, 255:red, 0; green, 0; blue, 0 }  ,fill opacity=0.18 ] (25.55,39.14) -- (64.89,39.33) -- (54.72,24.81) -- cycle ;
\draw    (276.5,70.25) -- (324.5,70.25) ;
\draw [shift={(326.5,70.25)}, rotate = 180] [color={rgb, 255:red, 0; green, 0; blue, 0 }  ][line width=0.75]    (10.93,-3.29) .. controls (6.95,-1.4) and (3.31,-0.3) .. (0,0) .. controls (3.31,0.3) and (6.95,1.4) .. (10.93,3.29)   ;
\draw    (276.5,46.25) -- (324.5,46.25) ;
\draw [shift={(326.5,46.25)}, rotate = 180] [color={rgb, 255:red, 0; green, 0; blue, 0 }  ][line width=0.75]    (10.93,-3.29) .. controls (6.95,-1.4) and (3.31,-0.3) .. (0,0) .. controls (3.31,0.3) and (6.95,1.4) .. (10.93,3.29)   ;
\draw    (104.5,55.25) -- (152.5,55.25) ;
\draw [shift={(154.5,55.25)}, rotate = 180] [color={rgb, 255:red, 0; green, 0; blue, 0 }  ][line width=0.75]    (10.93,-3.29) .. controls (6.95,-1.4) and (3.31,-0.3) .. (0,0) .. controls (3.31,0.3) and (6.95,1.4) .. (10.93,3.29)   ;
\draw    (104.5,31.25) -- (152.5,31.25) ;
\draw [shift={(154.5,31.25)}, rotate = 180] [color={rgb, 255:red, 0; green, 0; blue, 0 }  ][line width=0.75]    (10.93,-3.29) .. controls (6.95,-1.4) and (3.31,-0.3) .. (0,0) .. controls (3.31,0.3) and (6.95,1.4) .. (10.93,3.29)   ;
\draw    (104,78.75) -- (152,78.75) ;
\draw [shift={(154,78.75)}, rotate = 180] [color={rgb, 255:red, 0; green, 0; blue, 0 }  ][line width=0.75]    (10.93,-3.29) .. controls (6.95,-1.4) and (3.31,-0.3) .. (0,0) .. controls (3.31,0.3) and (6.95,1.4) .. (10.93,3.29)   ;
\draw (375.5,111.4) node [anchor=north west][inner sep=0.75pt]  [font=\small]  {$K_{0}$};
\draw (206,111.4) node [anchor=north west][inner sep=0.75pt]  [font=\small]  {$K_{1}$};
\draw (33.5,111.4) node [anchor=north west][inner sep=0.75pt]  [font=\small]  {$K_{2}$};
\draw (291.5,31.4) node [anchor=north west][inner sep=0.75pt]  [font=\footnotesize]  {$\di_{0}$};
\draw (291.5,55.4) node [anchor=north west][inner sep=0.75pt]  [font=\footnotesize]  {$\di_{1}$};
\draw (119.5,16.4) node [anchor=north west][inner sep=0.75pt]  [font=\footnotesize]  {$\di_{0}$};
\draw (119.5,40.4) node [anchor=north west][inner sep=0.75pt]  [font=\footnotesize]  {$\di_{1}$};
\draw (119.5,63.9) node [anchor=north west][inner sep=0.75pt]  [font=\footnotesize]  {$\di_{2}$};
\end{tikzpicture}
\caption[Simplicial set $K$]{Simplicial set $K$ as a sequence of sets $K_n$, together with face maps $\di_i: K_n\twoheadrightarrow K_{n-1}$ and degeneracy maps.\label{fig:simplicialset}}
\end{center}\end{figure}

\noindent See figure \ref{fig:simplicialset} for an intuitive picture of an example of simplicial set. \vspace{0.3cm}

\begin{definition}[Category of simplicial sets]
We define the \textit{category of simplicial sets},
\begin{equation}
    \mathbf{sSet} \;:=\; \Func(\mathbf{\Delta}^{\mathrm{op}},\mathbf{Set}),
\end{equation}
whose objects $K\in\mathbf{sSet}$ are simplicial sets $K : \mathbf{\Delta}^\mathrm{op}\rightarrow \mathbf{Set}$ and whose morphisms are natural transformations $\alpha:K\Rightarrow K'$.
\end{definition}

\begin{definition}[$n$-simplex as simplicial set]
For any $n\in\mathbb{N}$, we define the \textit{$n$-simplex simplicial set} $\Delta^n\in\mathbf{sSet}$ as the following Yoneda embedding of the $n$-simplex $[n]$ into the category of simplicial sets:
\begin{equation}
    \Delta^n \;:=\; \mathbf{\Delta}(-,[n]).
\end{equation}
\end{definition}

\begin{definition}[$n$-horns]
For any $0\leq i \leq n$ and $n\in\mathbb{N}^+$, we define the $i$-th $n$\textit{-horn} $\Lambda^n_i\in\mathbf{sSet}$ as the sub-simplicial set
\begin{equation}
    \Lambda^n_i \;\longhookrightarrow\; \Delta^n
\end{equation}
which is the union of all faces of $\Delta^n$ except the $i$-th one.
\end{definition}

\begin{figure}[h]\begin{center}
\tikzset{every picture/.style={line width=0.75pt}} 
\begin{tikzpicture}[x=0.75pt,y=0.75pt,yscale=-1,xscale=1]
\draw  [draw opacity=0][fill={rgb, 255:red, 155; green, 155; blue, 155 }  ,fill opacity=0.54 ] (238.5,31.25) -- (271.5,82.5) -- (205.5,82.5) -- cycle ;
\draw    (205.5,82.5) -- (238,32.25) ;
\draw [shift={(238,32.25)}, rotate = 302.89] [color={rgb, 255:red, 0; green, 0; blue, 0 }  ][fill={rgb, 255:red, 0; green, 0; blue, 0 }  ][line width=0.75]      (0, 0) circle [x radius= 1.34, y radius= 1.34]   ;
\draw [shift={(205.5,82.5)}, rotate = 302.89] [color={rgb, 255:red, 0; green, 0; blue, 0 }  ][fill={rgb, 255:red, 0; green, 0; blue, 0 }  ][line width=0.75]      (0, 0) circle [x radius= 1.34, y radius= 1.34]   ;
\draw    (272.5,82.5) -- (238,32.25) ;
\draw [shift={(238,32.25)}, rotate = 235.53] [color={rgb, 255:red, 0; green, 0; blue, 0 }  ][fill={rgb, 255:red, 0; green, 0; blue, 0 }  ][line width=0.75]      (0, 0) circle [x radius= 1.34, y radius= 1.34]   ;
\draw [shift={(272.5,82.5)}, rotate = 235.53] [color={rgb, 255:red, 0; green, 0; blue, 0 }  ][fill={rgb, 255:red, 0; green, 0; blue, 0 }  ][line width=0.75]      (0, 0) circle [x radius= 1.34, y radius= 1.34]   ;
\draw    (272.5,82.5) -- (205.5,82.5) ;
\draw [shift={(205.5,82.5)}, rotate = 180] [color={rgb, 255:red, 0; green, 0; blue, 0 }  ][fill={rgb, 255:red, 0; green, 0; blue, 0 }  ][line width=0.75]      (0, 0) circle [x radius= 1.34, y radius= 1.34]   ;
\draw [shift={(272.5,82.5)}, rotate = 180] [color={rgb, 255:red, 0; green, 0; blue, 0 }  ][fill={rgb, 255:red, 0; green, 0; blue, 0 }  ][line width=0.75]      (0, 0) circle [x radius= 1.34, y radius= 1.34]   ;
\draw    (105.5,82.5) -- (71,32.25) ;
\draw [shift={(71,32.25)}, rotate = 235.53] [color={rgb, 255:red, 0; green, 0; blue, 0 }  ][fill={rgb, 255:red, 0; green, 0; blue, 0 }  ][line width=0.75]      (0, 0) circle [x radius= 1.34, y radius= 1.34]   ;
\draw [shift={(105.5,82.5)}, rotate = 235.53] [color={rgb, 255:red, 0; green, 0; blue, 0 }  ][fill={rgb, 255:red, 0; green, 0; blue, 0 }  ][line width=0.75]      (0, 0) circle [x radius= 1.34, y radius= 1.34]   ;
\draw    (105.5,82.5) -- (38.5,82.5) ;
\draw [shift={(38.5,82.5)}, rotate = 180] [color={rgb, 255:red, 0; green, 0; blue, 0 }  ][fill={rgb, 255:red, 0; green, 0; blue, 0 }  ][line width=0.75]      (0, 0) circle [x radius= 1.34, y radius= 1.34]   ;
\draw [shift={(105.5,82.5)}, rotate = 180] [color={rgb, 255:red, 0; green, 0; blue, 0 }  ][fill={rgb, 255:red, 0; green, 0; blue, 0 }  ][line width=0.75]      (0, 0) circle [x radius= 1.34, y radius= 1.34]   ;
\draw    (127,57) -- (184.5,57.24) ;
\draw [shift={(186.5,57.25)}, rotate = 180.24] [color={rgb, 255:red, 0; green, 0; blue, 0 }  ][line width=0.75]    (7.65,-2.3) .. controls (4.86,-0.97) and (2.31,-0.21) .. (0,0) .. controls (2.31,0.21) and (4.86,0.98) .. (7.65,2.3)   ;
\draw [shift={(127,57)}, rotate = 0.24] [color={rgb, 255:red, 0; green, 0; blue, 0 }  ][line width=0.75]      (0,-7.83) .. controls (-2.16,-7.83) and (-3.91,-6.07) .. (-3.91,-3.91) .. controls (-3.91,-1.75) and (-2.16,0) .. (0,0) ;
\draw  [draw opacity=0][fill={rgb, 255:red, 155; green, 155; blue, 155 }  ,fill opacity=0.54 ] (239,137.75) -- (272,189) -- (206,189) -- cycle ;
\draw    (206,189) -- (238.5,138.75) ;
\draw [shift={(238.5,138.75)}, rotate = 302.89] [color={rgb, 255:red, 0; green, 0; blue, 0 }  ][fill={rgb, 255:red, 0; green, 0; blue, 0 }  ][line width=0.75]      (0, 0) circle [x radius= 1.34, y radius= 1.34]   ;
\draw [shift={(206,189)}, rotate = 302.89] [color={rgb, 255:red, 0; green, 0; blue, 0 }  ][fill={rgb, 255:red, 0; green, 0; blue, 0 }  ][line width=0.75]      (0, 0) circle [x radius= 1.34, y radius= 1.34]   ;
\draw    (273,189) -- (238.5,138.75) ;
\draw [shift={(238.5,138.75)}, rotate = 235.53] [color={rgb, 255:red, 0; green, 0; blue, 0 }  ][fill={rgb, 255:red, 0; green, 0; blue, 0 }  ][line width=0.75]      (0, 0) circle [x radius= 1.34, y radius= 1.34]   ;
\draw [shift={(273,189)}, rotate = 235.53] [color={rgb, 255:red, 0; green, 0; blue, 0 }  ][fill={rgb, 255:red, 0; green, 0; blue, 0 }  ][line width=0.75]      (0, 0) circle [x radius= 1.34, y radius= 1.34]   ;
\draw    (273,189) -- (206,189) ;
\draw [shift={(206,189)}, rotate = 180] [color={rgb, 255:red, 0; green, 0; blue, 0 }  ][fill={rgb, 255:red, 0; green, 0; blue, 0 }  ][line width=0.75]      (0, 0) circle [x radius= 1.34, y radius= 1.34]   ;
\draw [shift={(273,189)}, rotate = 180] [color={rgb, 255:red, 0; green, 0; blue, 0 }  ][fill={rgb, 255:red, 0; green, 0; blue, 0 }  ][line width=0.75]      (0, 0) circle [x radius= 1.34, y radius= 1.34]   ;
\draw    (106,189) -- (71.5,138.75) ;
\draw [shift={(71.5,138.75)}, rotate = 235.53] [color={rgb, 255:red, 0; green, 0; blue, 0 }  ][fill={rgb, 255:red, 0; green, 0; blue, 0 }  ][line width=0.75]      (0, 0) circle [x radius= 1.34, y radius= 1.34]   ;
\draw [shift={(106,189)}, rotate = 235.53] [color={rgb, 255:red, 0; green, 0; blue, 0 }  ][fill={rgb, 255:red, 0; green, 0; blue, 0 }  ][line width=0.75]      (0, 0) circle [x radius= 1.34, y radius= 1.34]   ;
\draw    (71.5,138.75) -- (39,189) ;
\draw [shift={(39,189)}, rotate = 122.89] [color={rgb, 255:red, 0; green, 0; blue, 0 }  ][fill={rgb, 255:red, 0; green, 0; blue, 0 }  ][line width=0.75]      (0, 0) circle [x radius= 1.34, y radius= 1.34]   ;
\draw [shift={(71.5,138.75)}, rotate = 122.89] [color={rgb, 255:red, 0; green, 0; blue, 0 }  ][fill={rgb, 255:red, 0; green, 0; blue, 0 }  ][line width=0.75]      (0, 0) circle [x radius= 1.34, y radius= 1.34]   ;
\draw    (127.5,163.5) -- (185,163.74) ;
\draw [shift={(187,163.75)}, rotate = 180.24] [color={rgb, 255:red, 0; green, 0; blue, 0 }  ][line width=0.75]    (7.65,-2.3) .. controls (4.86,-0.97) and (2.31,-0.21) .. (0,0) .. controls (2.31,0.21) and (4.86,0.98) .. (7.65,2.3)   ;
\draw [shift={(127.5,163.5)}, rotate = 0.24] [color={rgb, 255:red, 0; green, 0; blue, 0 }  ][line width=0.75]      (0,-7.83) .. controls (-2.16,-7.83) and (-3.91,-6.07) .. (-3.91,-3.91) .. controls (-3.91,-1.75) and (-2.16,0) .. (0,0) ;
\draw  [draw opacity=0][fill={rgb, 255:red, 155; green, 155; blue, 155 }  ,fill opacity=0.54 ] (238.5,244.25) -- (271.5,295.5) -- (205.5,295.5) -- cycle ;
\draw    (205.5,295.5) -- (238,245.25) ;
\draw [shift={(238,245.25)}, rotate = 302.89] [color={rgb, 255:red, 0; green, 0; blue, 0 }  ][fill={rgb, 255:red, 0; green, 0; blue, 0 }  ][line width=0.75]      (0, 0) circle [x radius= 1.34, y radius= 1.34]   ;
\draw [shift={(205.5,295.5)}, rotate = 302.89] [color={rgb, 255:red, 0; green, 0; blue, 0 }  ][fill={rgb, 255:red, 0; green, 0; blue, 0 }  ][line width=0.75]      (0, 0) circle [x radius= 1.34, y radius= 1.34]   ;
\draw    (272.5,295.5) -- (238,245.25) ;
\draw [shift={(238,245.25)}, rotate = 235.53] [color={rgb, 255:red, 0; green, 0; blue, 0 }  ][fill={rgb, 255:red, 0; green, 0; blue, 0 }  ][line width=0.75]      (0, 0) circle [x radius= 1.34, y radius= 1.34]   ;
\draw [shift={(272.5,295.5)}, rotate = 235.53] [color={rgb, 255:red, 0; green, 0; blue, 0 }  ][fill={rgb, 255:red, 0; green, 0; blue, 0 }  ][line width=0.75]      (0, 0) circle [x radius= 1.34, y radius= 1.34]   ;
\draw    (272.5,295.5) -- (205.5,295.5) ;
\draw [shift={(205.5,295.5)}, rotate = 180] [color={rgb, 255:red, 0; green, 0; blue, 0 }  ][fill={rgb, 255:red, 0; green, 0; blue, 0 }  ][line width=0.75]      (0, 0) circle [x radius= 1.34, y radius= 1.34]   ;
\draw [shift={(272.5,295.5)}, rotate = 180] [color={rgb, 255:red, 0; green, 0; blue, 0 }  ][fill={rgb, 255:red, 0; green, 0; blue, 0 }  ][line width=0.75]      (0, 0) circle [x radius= 1.34, y radius= 1.34]   ;
\draw    (105.5,295.5) -- (38.5,295.5) ;
\draw [shift={(38.5,295.5)}, rotate = 180] [color={rgb, 255:red, 0; green, 0; blue, 0 }  ][fill={rgb, 255:red, 0; green, 0; blue, 0 }  ][line width=0.75]      (0, 0) circle [x radius= 1.34, y radius= 1.34]   ;
\draw [shift={(105.5,295.5)}, rotate = 180] [color={rgb, 255:red, 0; green, 0; blue, 0 }  ][fill={rgb, 255:red, 0; green, 0; blue, 0 }  ][line width=0.75]      (0, 0) circle [x radius= 1.34, y radius= 1.34]   ;
\draw    (71,245.25) -- (38.5,295.5) ;
\draw [shift={(38.5,295.5)}, rotate = 122.89] [color={rgb, 255:red, 0; green, 0; blue, 0 }  ][fill={rgb, 255:red, 0; green, 0; blue, 0 }  ][line width=0.75]      (0, 0) circle [x radius= 1.34, y radius= 1.34]   ;
\draw [shift={(71,245.25)}, rotate = 122.89] [color={rgb, 255:red, 0; green, 0; blue, 0 }  ][fill={rgb, 255:red, 0; green, 0; blue, 0 }  ][line width=0.75]      (0, 0) circle [x radius= 1.34, y radius= 1.34]   ;
\draw    (127,270) -- (184.5,270.24) ;
\draw [shift={(186.5,270.25)}, rotate = 180.24] [color={rgb, 255:red, 0; green, 0; blue, 0 }  ][line width=0.75]    (7.65,-2.3) .. controls (4.86,-0.97) and (2.31,-0.21) .. (0,0) .. controls (2.31,0.21) and (4.86,0.98) .. (7.65,2.3)   ;
\draw [shift={(127,270)}, rotate = 0.24] [color={rgb, 255:red, 0; green, 0; blue, 0 }  ][line width=0.75]      (0,-7.83) .. controls (-2.16,-7.83) and (-3.91,-6.07) .. (-3.91,-3.91) .. controls (-3.91,-1.75) and (-2.16,0) .. (0,0) ;
\draw (233,13.9) node [anchor=north west][inner sep=0.75pt]    {$0$};
\draw (193.5,77.4) node [anchor=north west][inner sep=0.75pt]    {$1$};
\draw (275,76.9) node [anchor=north west][inner sep=0.75pt]    {$2$};
\draw (66,13.9) node [anchor=north west][inner sep=0.75pt]    {$0$};
\draw (26.5,77.4) node [anchor=north west][inner sep=0.75pt]    {$1$};
\draw (108,76.9) node [anchor=north west][inner sep=0.75pt]    {$2$};
\draw (233.5,120.4) node [anchor=north west][inner sep=0.75pt]    {$0$};
\draw (194,183.9) node [anchor=north west][inner sep=0.75pt]    {$1$};
\draw (275.5,183.4) node [anchor=north west][inner sep=0.75pt]    {$2$};
\draw (66.5,120.4) node [anchor=north west][inner sep=0.75pt]    {$0$};
\draw (27,183.9) node [anchor=north west][inner sep=0.75pt]    {$1$};
\draw (108.5,183.4) node [anchor=north west][inner sep=0.75pt]    {$2$};
\draw (233,226.9) node [anchor=north west][inner sep=0.75pt]    {$0$};
\draw (193.5,290.4) node [anchor=north west][inner sep=0.75pt]    {$1$};
\draw (275,289.9) node [anchor=north west][inner sep=0.75pt]    {$2$};
\draw (66,226.9) node [anchor=north west][inner sep=0.75pt]    {$0$};
\draw (26.5,290.4) node [anchor=north west][inner sep=0.75pt]    {$1$};
\draw (108,289.9) node [anchor=north west][inner sep=0.75pt]    {$2$};
\draw (320.5,43.9) node [anchor=north west][inner sep=0.75pt]    {$\Lambda ^{2}_{0} \hookrightarrow \Delta ^{2}$};
\draw (320.5,150.9) node [anchor=north west][inner sep=0.75pt]    {$\Lambda ^{2}_{1} \hookrightarrow \Delta ^{2}$};
\draw (320.5,256.9) node [anchor=north west][inner sep=0.75pt]    {$\Lambda ^{2}_{2} \hookrightarrow \Delta ^{2}$};
\end{tikzpicture}
\caption{All the $2$-horns $\Lambda_i^2$ of a $2$-simplex $\Delta^2$.}
\end{center}\end{figure}
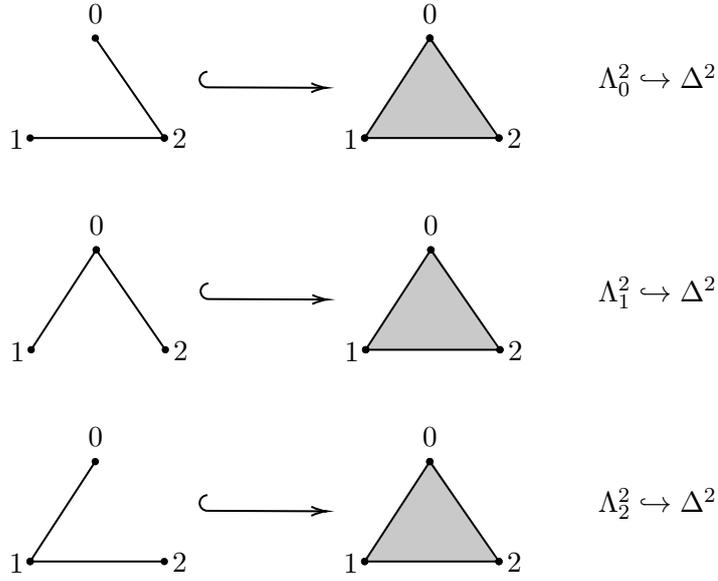

\begin{definition}[Kan complex]
A \textit{Kan complex} is a simplicial set $K\in\mathbf{sSet}$ that satisfies the \textit{Kan condition}: for any image of a horn $f:\Lambda^n_i\rightarrow K$ with $0\leq i\leq n$ in $K$, the missing $i$-th face must be in $K$ too, i.e. there must exist a map $f'$ given as follows:
\begin{equation}\label{eq:kan}
    \begin{tikzcd}[row sep=5.5ex, column sep=6.5ex]
    \Lambda^n_i \arrow[r, "f"]\arrow[d, hook]& K \\
    \Delta^n \arrow[ur, "f'"']
    \end{tikzcd}  .  
\end{equation}
\end{definition}

\begin{remark}[Kan condition]
Notice that we can equivalently reformulate the Kan condition \eqref{eq:kan} by requiring that the map
\begin{equation}\label{eq:kan2}
    K_n \,\cong\, \mathbf{sSet}(\Delta^n,K)\,\longrightarrow\,\mathbf{sSet}(\Lambda^n_i,K)
\end{equation}
is surjective.
\end{remark}

\begin{definition}[Weak Kan complex]
A \textit{weak Kan complex} is a simplicial set $K\in\mathbf{sSet}$ that satisfies the Kan condition only for $n$-horns $\Lambda_i^n$ with $0<i<n$.
\end{definition}

\subsection{$\infty$-groupoids}

The Kan complex is a geometric model for the $\infty$-groupoid.
Similarly, the weak Kan complex is a geometric model for the $(\infty,1)$-category.

\begin{definition}[$\infty$-groupoid]
An $\infty$\textit{-groupoid} $G$ is a Kan complex, i.e.
\begin{equation}
    X\;=\; \left( \begin{tikzcd}[row sep=scriptsize, column sep=5ex]
    \; \cdots\; \arrow[r, yshift=1.4ex] \arrow[r, yshift=2.8ex] \arrow[r] \arrow[r, yshift=-1.4ex]\arrow[r, yshift=-2.8ex] & X_3 \arrow[r, yshift=1.8ex]\arrow[r, yshift=0.6ex]\arrow[r, yshift=-1.8ex]\arrow[r, yshift=-0.6ex]& X_2
    \arrow[r, yshift=1.4ex] \arrow[r] \arrow[r, yshift=-1.4ex] & X_1  \arrow[r, yshift=0.7ex] \arrow[r, yshift=-0.7ex] & X_0 
    \end{tikzcd}    \right),
\end{equation}
where we call the elements of $X_0$ the \textit{objects} of the $\infty$-groupoid and the elements of $X_n$ with $n>1$ the $n$\textit{-morphisms} of the $\infty$-groupoid.
\end{definition}

\begin{definition}[Morphisms of $\infty$-groupoids]
A \textit{morphism of $\infty$-groupoids} $f:X\rightarrow Y$ is a map of simplicial sets which respect all the face maps $\di_i$ and the degeneracy maps $s_i$.
\end{definition}

\begin{definition}[Product of $\infty$-groupoids]
The \textit{product }$X\times Y$ of two $\infty$-groupoids $X$ and $Y$ is an $\infty$-groupoid defined by the Kan complex
\begin{equation}
    \begin{aligned}
    (X\times Y)_n \;&:=\; X_n\times Y_n, \\
    \di_i^{X\times Y}(x,y) \;&:=\; (\di_i^{X}x, \di_i^{Y}y), \\
    s_i^{X\times Y}(x,y) \;&:=\; (s_i^{X}x, s_i^{Y}y),
    \end{aligned}
\end{equation}
where we called $\di_i^{X}$ and $s_i^{X}$ the face and degeneracy maps of $X$ and so on.
\end{definition}

\begin{definition}[Internal hom of $\infty$-groupoids]
The \textit{internal hom} $[X,Y]$ of two $\infty$-groupoids $X$ and $Y$ is an $\infty$-groupoid defined by the Kan complex
\begin{equation}
    [X,Y]\,: [n] \;\longrightarrow\; \mathbf{sSet}(X\times \Delta^n, \,Y),
\end{equation}
where $\Delta^n\in\mathbf{sSet}$ is the $n$-simplex simplicial set.
\end{definition}

\noindent Notice that the internal hom of $\infty$-groupoids satisfies the fundamental property:
\begin{equation}
    \mathbf{sSet}(X\times Y,\,Z)\;\cong\;\mathbf{sSet}(X,\,[Y,Z]).
\end{equation}

\begin{definition}[Homotopy of $\infty$-groupoids]
Let $f,g:X\longrightarrow Y$ be morphisms of $\infty$-groupoids. A \textit{homotopy} $\alpha:f\Rightarrow g$ is defined as a morphism of $\infty$-groupoids $\alpha: X \longrightarrow [\Delta^1,Y]$ such that the following diagram commutes:
\begin{equation}
    \begin{tikzcd}[row sep=6ex, column sep=5ex]
    & Y \\
    X \arrow[ur, "f"]\arrow[dr, "g"']\arrow[r, "\alpha"] & {[\Delta^1,Y]} \arrow[u, "{[\di_0,Y]}"']\arrow[d, "{[\di_1,Y]}"] \\
    & Y
    \end{tikzcd}
\end{equation}
\end{definition}

\noindent Therefore, there exists a simplicial set (in particular, a Kan complex) of morphisms between any two $\infty$-groupoids. Equivalently, for any couple of $\infty$-groupoids, there exists an $\infty$-groupoid of morphisms between them. This means that $\infty$-groupoids make up a well-defined $(\infty,1)$-category, which we will call $\mathbf{\infty Grpd}$.

\begin{definition}[Set of connected components of an $\infty$-groupoid]
Given an $\infty$-groupoid $X\in\mathbf{\infty Grpd}$, we define its \textit{set of connected components} $\pi_0(X)\in\mathbf{Set}$ by
\begin{equation}
    \pi_0(X) \;:=\; X_0/X_1,
\end{equation}
i.e. by the set of equivalence classes of elements of $X_0$ modulo the equivalence relation $\di_0(x)\sim\di_1(x)\in X_0$ for any $x\in X_1$.
\end{definition}

\begin{definition}[Geometric realisation of an $\infty$-groupoid]
The \textit{geometric realisation} is the functor
\begin{equation}
    |-|\,:\; \mathbf{\infty Grpd} \;\longrightarrow\; \mathbf{Top}
\end{equation}
which is defined, for any $\infty$-groupoid $X\in\mathbf{\infty Grpd}$, by the topological space
\begin{equation}
    |X|\;:=\; \bigsqcup_{k\in\mathbb{N}} \!\left(X_k\times\Delta^k_{\mathrm{top}}\right)/\sim\,,
\end{equation}
where the equivalence relation glues points $\big(x,\partial_i(t)\big)\sim\big(\di_i(x),t\big)$ and $\big(x,\sigma_i(t)\big)\sim\big(s_i(x),t\big)$.
\end{definition}

\subsection{Smooth stacks}

Since the dawn of geometry, with Euclid's Elements, the point of studying geometry has always been the investigation of the relations between geometric objects. The problems of synthetic geometry are usually about whether a point lies on a certain line, or whether a pair of lines meet, etc..
In this section we will introduce the notion of smooth stack, which can be intuitively understood as a geometric object which captures this fundamental relational aspect of geometry in the context of higher geometry.
\vspace{0.2cm}

\noindent A smooth stack (or smooth $\infty$-groupoid) can be understood as an $\infty$-groupoid equipped with a smooth structure, similarly to how an ordinary Lie group is a group with manifold structure. Such a smooth structure is given by specifying how any manifold can be mapped onto our $\infty$-groupoid. This concept originated in \cite{Sat09}.

\begin{definition}[Smooth stack]
Let $\mathbf{Diff}$ be the ordinary category of smooth manifolds and $\Grpd$ the $(\infty,1)$-category of Lie $\infty$-groupoids. A \textit{smooth stack} (or \textit{smooth }$\infty$\textit{-groupoid}) $\mathscr{X}$ is defined as an $\infty$-functor
\begin{equation}
    \mathscr{X}:\mathbf{Diff}^{\mathrm{op}}\longrightarrow \Grpd
\end{equation}
which satisfies some higher gluing properties, known as descent. This can be though as a generalisation of the notion of sheaf which takes value in $\infty$-groupoids.
\end{definition}

\noindent Morally speaking an higher smooth stack $\mathscr{X}$ is a sheaf of $\infty$-groupoids over manifolds. Given a local patch $U$, we can picture it by an assignment of the form
\begin{equation*}
    U\,\mapsto\, \mathscr{X}(U)\,=\, \left( \begin{tikzcd}[row sep=scriptsize, column sep=5ex]
    \; \cdots\; \arrow[r, yshift=1.4ex] \arrow[r, yshift=2.8ex] \arrow[r] \arrow[r, yshift=-1.4ex]\arrow[r, yshift=-2.8ex] & \mathscr{X}_3(U) \arrow[r, yshift=1.8ex]\arrow[r, yshift=0.6ex]\arrow[r, yshift=-1.8ex]\arrow[r, yshift=-0.6ex]& \mathscr{X}_2(U)
    \arrow[r, yshift=1.4ex] \arrow[r] \arrow[r, yshift=-1.4ex] & \mathscr{X}_1(U)  \arrow[r, yshift=0.7ex] \arrow[r, yshift=-0.7ex] & \mathscr{X}_0(U) 
    \end{tikzcd}    \right).
\end{equation*}
This means that for a manifold $M$ with good cover $\mathcal{U}=\bigsqcup_{\alpha\in I} U_\alpha$, the higher groupoid $\mathscr{X}(M)$ can be described in local data by a collection of higher groupoids $\mathscr{X}(U_{\alpha_1}\cap\cdots\cap U_{\alpha_k})$ on any $k$-fold overlap of patches, which are glued by their groupoid morphisms. For a formal exposition see \cite{Principal2}.

\begin{definition}[$(\infty,1)$-category of smooth stacks]
We call $\mathbf{H}$ the $(\infty,1)$\textit{-category of smooth stacks} on manifolds, such that
\begin{itemize}
    \item objects are smooth stacks,
    \item $k$-morphisms for any $k\in\mathbb{N}^+$ are $k$-morphisms of smooth stacks.
\end{itemize} 
\end{definition}

\begin{example}[Manifolds as smooth stacks]
Given any smooth manifold $M\in\mathbf{Diff}$, we can easily construct a sheaf $\Coo(-,M)\in\mathbf{H}$ of smooth functions to $M$, which is in particular a stack. This is nothing but a Yoneda embedding $\mathbf{Diff}\hookrightarrow\mathbf{H}$ of the smooth manifolds into the $(\infty,1)$-category of stacks.
\end{example}

\begin{remark}[Smooth $\infty$-groupoids as generalised spaces]
Notice that we can naturally generalise the argument from the previous subsection and show that there exists a smooth $\infty$-groupoid of morphisms between any couple of smooth $\infty$-groupoids.
Thus, smooth $\infty$-groupoids make up a well-defined $(\infty,1)$-category, which we will call $\mathbf{H}$.
We can regard such an $(\infty,1)$-category as a generalisation of the category of smooth manifolds and, thus, a smooth $\infty$-groupoid as a generalisation of a smooth manifold.
Notice that we have the embeddings:
\begin{equation}
    \mathbf{H}\; \hookleftarrow \; \mathbf{Diff}\; \hookleftarrow \; \mathbf{Top} \; \hookleftarrow \; \mathbf{Set}.
\end{equation}
\end{remark}

\begin{remark}[$0$-truncation of a smooth $\infty$-groupoid]
The inclusion $\mathbf{Sh}\hookrightarrow \mathbf{H}$ of sheaves into smooth $\infty$-groupoids has a left adjoint $\tau_0:\mathbf{H}\rightarrow \mathbf{Sh}$, which is called \textit{$0$-truncation}. This functor sends a smooth $\infty$-groupoid $X\in\mathbf{H}$ to its underlying sheaf of objects $X_0$.
\end{remark}

\noindent Let us now give simple examples of $0$-truncated stacks, which are just ordinary sheaves.

\begin{example}[Some useful $0$-stacks]
Interestingly $\mathrm{Diff}(M)$ can be thought as a $0$-stack sending a manifold $M$ to its group of diffeomorphisms, while $\Omega^{n}(M)$ is a $0$-stack sending $M$ to the vector space of its $n$-forms. Analogously $\Omega^{n}_{\mathrm{cl}}(M)$ is the $0$-stack of closed $n$-forms. However we remark that a $0$-stack of exact forms $\Omega^{n}_{\mathrm{ex}}(M)$ \textit{does not} exist, because it would not satisfy the gluing conditions on overlaps of patches $\Omega^n_{\mathrm{ex}}(U_\alpha\cap U_\beta)$.
\end{example}

\noindent Let us now introduce some useful notions in the context of smooth stacks.

\begin{definition}[Hom $\infty$-groupoid]
Given any couple of smooth stacks $\mathscr{X},\mathscr{Y}\in\mathbf{H}$ we can define the \textit{hom }$\infty$\textit{-groupoid} 
\begin{equation}
    \mathbf{H}(\mathscr{X},\mathscr{Y}) \;\in\; \mathbf{\infty Grpd}
\end{equation}
as the $\infty$-groupoid of morphisms from $\mathscr{X}$ to $\mathscr{Y}$, i.e. such that
\begin{itemize}
    \item objects are $1$-morphisms $f:\mathscr{X}\rightarrow\mathscr{Y}$ in $\mathbf{H}$,
    \item $k$-morphisms are $(k+1)$-morphisms of stacks $f\Mapsto f'$ in $\mathbf{H}$.
\end{itemize}
\end{definition}

\noindent Notice that, given an higher smooth stack $\mathscr{X}$ over smooth manifolds, we have the natural equivalence $\mathbf{H}(M,\,\mathscr{X})\cong\mathscr{X}(M)$ for any smooth manifold $M$ (regarded here as a smooth $0$-stack).

\begin{definition}[Internal hom stack]\label{def:inthom}
Given any couple of smooth stacks $\mathscr{X},\mathscr{Y}\in\mathbf{H}$, we can define the \textit{internal hom stack} $[\mathscr{X},\mathscr{Y}]\in\mathbf{H}$ as the smooth stack which satisfies the following equivalence:
\begin{equation}
    \mathbf{H}\big(\mathscr{S},\,[\mathscr{X},\mathscr{Y}]\big) \;\cong\; \mathbf{H}(\mathscr{S}\times\mathscr{X},\, \mathscr{Y}),
\end{equation}
for any fixed smooth stack $\mathscr{S}\in\mathbf{H}$.
\end{definition}

\begin{theorem}[Explicit form of internal hom stack]
Given any couple of smooth stacks $\mathscr{X},\mathscr{Y}\in\mathbf{H}$, the internal hom stack $[\mathscr{X},\mathscr{Y}]\in\mathbf{H}$ is given by
\begin{equation}
    [\mathscr{X},\mathscr{Y}]: \, U\longrightarrow\mathbf{H}(U\times \mathscr{X},\,\mathscr{Y}),
\end{equation}
for any smooth manifold $U\in\mathbf{Diff}$.
\end{theorem}

\begin{proof}
Given a smooth manifold $U\in\mathbf{Diff}$, we have $[\mathscr{X},\mathscr{Y}](U) = \mathbf{H}(U,[\mathscr{X},\mathscr{Y}])$. By using definition \ref{def:inthom}, we immediately have $\mathbf{H}(U,[\mathscr{X},\mathscr{Y}])\cong\mathbf{H}(\mathscr{X}\times U,\mathscr{Y})$.
\end{proof}

\noindent Notice that, if $\,\ast\,\in\mathbf{H}$ is a point, we have the natural equivalence $[\,\ast\,,\,\mathscr{X}\,]\,\cong\,\mathscr{X}\,$ for any stack $\,\mathscr{X}\in\mathbf{H}$.

\begin{example}[Loop space of a manifold]\label{ex:loopspace}
The \textit{loop space} of a manifold $M\in\mathbf{Diff}$ is the Fr\'{e}chet manifold $\mathcal{L}M:=[S^1,M]$.
\end{example}

\begin{definition}[Slice $\infty$-category]\label{def:slice}
For any given object $\mathscr{X}\in\mathbf{H}$, according to \cite{DCCTv2} we can define the \textit{slice }$\infty$\textit{-category} $\mathbf{H}_{/\mathscr{X}}$ as the $\infty$-category such that
\begin{itemize}
    \item objects are $1$-morphisms $f:\mathscr{Z}\rightarrow\mathscr{X}$ in $\mathbf{H}$, 
    \item $1$-morphisms $F:f\mapsto f'$ are homotopy commutative diagrams of the following form
    \begin{equation}
        \begin{tikzcd}[row sep=scriptsize, column sep=8ex]
        \mathscr{Z} \arrow[rr, "F"{name=D}]\arrow[rd, "f"'] & & \mathscr{Z}'\arrow[ld, "f'"] \\
        & \mathscr{X}\arrow[Rightarrow, to=D] &
    \end{tikzcd}
    \end{equation}
    \item and so on for $k$-morphisms with $k>1$.
\end{itemize}
\end{definition}

\begin{definition}[Loop space object of an $\infty$-category]
For any object $X\in\mathbf{C}$ in an $\infty$-category $\mathbf{C}$, we can define the \textit{loop space object} $\Omega_{X}\mathbf{C}$ as the $\infty$-category such that
\begin{itemize}
    \item objects are $1$-morphisms $f:X\rightarrow X$ in $\mathbf{C}$, 
    \item $k$-morphisms are $(k+1)$-morphisms $f_1\Mapsto f_2$ in $\mathbf{C}$.
\end{itemize}
This category must not be confused with the loop space of a manifold of example \ref{ex:loopspace}.
\end{definition}

\begin{example}[Path $\infty$-groupoid]
The \textit{path $\infty$-groupoid} is the $\infty$-groupoid whose objects are points $x\in M$, $1$-morphisms $\gamma:x\mapsto y$ are paths, $2$-morphisms are homotopies of paths, $3$-morphisms are homotopies of homotopies, etc... In other words it is the $\infty$-groupoid given by the following simplicial object:
\begin{equation*}
    \mathscr{P}(M) \,=\, \left(\begin{tikzcd}[row sep=scriptsize, column sep=4ex]
    \; \cdots\; \arrow[r, yshift=1.8ex]\arrow[r, yshift=0.6ex]\arrow[r, yshift=-1.8ex]\arrow[r, yshift=-0.6ex]& {\mathbf{Top}}(\Delta^2_{\mathrm{top}},M)
    \arrow[r, yshift=1.4ex] \arrow[r] \arrow[r, yshift=-1.4ex] & {\mathbf{Top}}(\Delta^1_{\mathrm{top}},M)  \arrow[r, yshift=0.7ex] \arrow[r, yshift=-0.7ex] & {\mathbf{Top}}(\Delta^0_{\mathrm{top}},M)
    \end{tikzcd}    \right),
\end{equation*}
where, here, $\Delta^n_{\mathrm{top}}$ is the standard $n$-simplex for $n\in\mathbb{N}$, regarded as a topological space.
\end{example}

\begin{figure}[h]\begin{center}
\tikzset {_rkw01qkrw/.code = {\pgfsetadditionalshadetransform{ \pgftransformshift{\pgfpoint{83.16 bp } { -104.94 bp }  }  \pgftransformscale{1.32 }  }}}
\pgfdeclareradialshading{_oe83dzfyc}{\pgfpoint{-72bp}{88bp}}{rgb(0bp)=(1,1,1);
rgb(0bp)=(1,1,1);
rgb(25bp)=(0.61,0.61,0.61);
rgb(400bp)=(0.61,0.61,0.61)}
\tikzset{every picture/.style={line width=0.75pt}} 
\begin{tikzpicture}[x=0.75pt,y=0.75pt,yscale=-1,xscale=1]
\path  [shading=_oe83dzfyc,_rkw01qkrw] (1.8,75.2) .. controls (1.8,34.66) and (41.02,1.8) .. (89.4,1.8) .. controls (137.78,1.8) and (177,34.66) .. (177,75.2) .. controls (177,115.74) and (137.78,148.6) .. (89.4,148.6) .. controls (41.02,148.6) and (1.8,115.74) .. (1.8,75.2) -- cycle ; 
 \draw  [color={rgb, 255:red, 0; green, 0; blue, 0 }  ,draw opacity=1 ] (1.8,75.2) .. controls (1.8,34.66) and (41.02,1.8) .. (89.4,1.8) .. controls (137.78,1.8) and (177,34.66) .. (177,75.2) .. controls (177,115.74) and (137.78,148.6) .. (89.4,148.6) .. controls (41.02,148.6) and (1.8,115.74) .. (1.8,75.2) -- cycle ; 
\draw    (36.4,62.8) .. controls (70.6,3.4) and (95.4,46.2) .. (139.4,61.4) ;
\draw [shift={(139.4,61.4)}, rotate = 19.06] [color={rgb, 255:red, 0; green, 0; blue, 0 }  ][fill={rgb, 255:red, 0; green, 0; blue, 0 }  ][line width=0.75]      (0, 0) circle [x radius= 1.34, y radius= 1.34]   ;
\draw [shift={(36.4,62.8)}, rotate = 299.93] [color={rgb, 255:red, 0; green, 0; blue, 0 }  ][fill={rgb, 255:red, 0; green, 0; blue, 0 }  ][line width=0.75]      (0, 0) circle [x radius= 1.34, y radius= 1.34]   ;
\draw    (36.4,62.8) .. controls (88.2,144.2) and (127.8,113.4) .. (139.4,61.4) ;
\draw    (84.25,37.83) .. controls (74.41,44.63) and (68.42,54.96) .. (68.42,67.62) .. controls (68.42,78.76) and (73.03,91.64) .. (80.13,100.75)(82.55,35.37) .. controls (71.89,42.73) and (65.42,53.92) .. (65.42,67.62) .. controls (65.42,79.27) and (70.13,92.79) .. (77.72,102.55) ;
\draw [shift={(83.4,107.8)}, rotate = 231.97] [color={rgb, 255:red, 0; green, 0; blue, 0 }  ][line width=0.75]    (10.93,-3.29) .. controls (6.95,-1.4) and (3.31,-0.3) .. (0,0) .. controls (3.31,0.3) and (6.95,1.4) .. (10.93,3.29)   ;
\draw    (93.39,38.89) .. controls (103.84,52.55) and (108.71,64.65) .. (108.71,76.21) .. controls (108.71,81.87) and (107.54,87.41) .. (105.27,92.94) .. controls (102.89,98.76) and (99.28,104.58) .. (98.05,105.83)(91.01,40.71) .. controls (100.95,53.71) and (105.71,65.19) .. (105.71,76.21) .. controls (105.71,81.49) and (104.61,86.65) .. (102.5,91.81) .. controls (100.21,97.38) and (96.73,102.96) .. (95.74,103.9) ;
\draw [shift={(92.2,111)}, rotate = 309.61] [color={rgb, 255:red, 0; green, 0; blue, 0 }  ][line width=0.75]    (10.93,-3.29) .. controls (6.95,-1.4) and (3.31,-0.3) .. (0,0) .. controls (3.31,0.3) and (6.95,1.4) .. (10.93,3.29)   ;
\draw    (72.22,71.5) -- (98.22,71.9)(72.18,74.5) -- (98.18,74.9) ;
\draw    (72.62,68.5) -- (98.62,68.9)(72.58,71.5) -- (98.58,71.9) ;
\draw    (91.4,71.8) -- (103.8,71.63) ;
\draw [shift={(105.8,71.6)}, rotate = 539.2] [color={rgb, 255:red, 0; green, 0; blue, 0 }  ][line width=0.75]    (12.02,-5.39) .. controls (7.65,-2.53) and (3.64,-0.73) .. (0,0) .. controls (3.64,0.73) and (7.65,2.53) .. (12.02,5.39)   ;
\draw (26,48.2) node [anchor=north west][inner sep=0.75pt]  [font=\footnotesize]  {$x$};
\draw (142.8,46.6) node [anchor=north west][inner sep=0.75pt]  [font=\footnotesize]  {$y$};
\draw (-12.4,0.4) node [anchor=north west][inner sep=0.75pt]    {$\mathscr{P}( M)$};
\end{tikzpicture}
\caption{Path $\infty$-groupoid of a smooth manifold $M$.}
\end{center}\end{figure}

\begin{definition}[Sheaf of connected components of a smooth $\infty$-groupoid]
Given a smooth $\infty$-groupoid $\mathscr{X}\in\mathbf{H}$, we define its \textit{sheaf of connected components} $\pi_0(\mathscr{X})\in\mathbf{Sh}$ by
\begin{equation}
    \pi_0(\mathscr{X}) \;:=\; \mathscr{X}_0/\mathscr{X}_1
\end{equation}
i.e. by the set of equivalence classes of elements of $X_0$ modulo the equivalence relation $\di_0(x)\sim\di_1(x)\in \mathscr{X}_0(U)$ for any $x\in \mathscr{X}_1(U)$ and $U\in\mathbf{Diff}$.
\end{definition}

\begin{example}[Set of connected components of the path $\infty$-groupoid]
Given a smooth manifold $M$, the set of connected components of the path $\infty$-groupoid $\mathscr{P}(M)$ is $\pi_0\big(\mathscr{P}(M)\big)=\pi_0(M)$, i.e. the set of connected components of the manifold $M$.
\end{example}

\begin{example}[\v{C}ech groupoid]
Given a good open cover $\mathcal{U}:=\{U_\alpha\}$ of a smooth manifold $M$, its \textit{\v{C}ech groupoid} is defined by the following simplicial set
\begin{equation}\label{eq:chechgroupoidex}
    \check{C}(\mathcal{U}) \;:=\; \left(\begin{tikzcd}[row sep=scriptsize, column sep=3ex]\; \cdots\; \arrow[r, yshift=1.8ex]\arrow[r, yshift=0.6ex]\arrow[r, yshift=-1.8ex]\arrow[r, yshift=-0.6ex]& \underset{\alpha\beta\gamma}{\bigsqcup} U_{\alpha}\cap U_\beta\cap U_\gamma \;
    \arrow[r, yshift=1.4ex] \arrow[r] \arrow[r, yshift=-1.4ex] & \; \underset{\alpha\beta}{\bigsqcup}\; U_{\alpha}\cap U_\beta  \arrow[r, yshift=0.7ex] \arrow[r, yshift=-0.7ex] & \; \underset{\alpha}{\bigsqcup}\; U_{\alpha}\;
    \end{tikzcd}\right)
\end{equation}
where all the arrows are embedding of open sets. 
\end{example}

\noindent Notice that in the \v{C}ech groupoid the gluing conditions of the manifold $M$ between its patches $\{U_\alpha\}$ have been promoted to morphisms.

\subsection{Lie $\infty$-groups}

Now, our $(\infty,1)$-topos of generalised spaces is $\mathbf{H}$. Just like objects of $\mathrm{Grp}(\mathbf{Top})$ are topological groups and of $\mathrm{Grp}(\mathbf{Diff})$ are Lie groups, we may wonder if there exist a new notion of group $\mathrm{Grp}(\mathbf{H})$ on such generalised spaces.
The definition of a $G\in\mathrm{Grp}(\mathbf{H})$ provided by \cite{LurieHA} is the following.

\begin{definition}[Lie $\infty$-group]
A \textit{Lie $\infty$-group} $G$ is an $A_\infty$-algebra in $\mathbf{H}$ such that the sheaf of its connected components $\pi_0(G)$ is a group object in the category of sheaves, i.e. $\pi_0(G)\in\mathrm{Grp}(\mathbf{Sh})$.
\end{definition}

\noindent See \cite{Principal1, Principal2} for details. 
Since this definition is technically difficult, we will now provide an equivalent way to think about Lie $\infty$-groups as simplicial objects.

\begin{remark}[Lie $\infty$-groups as Lie $\infty$-groupoids]
There exists an embedding of $(\infty,1)$-categories
\begin{equation}
    \mathrm{Grp}(\mathbf{H}) \; \xhookrightarrow{\quad\mathbf{B}\quad} \; \mathbf{H}.
\end{equation}
which identifies Lie $\infty$-groups $G$ with smooth $\infty$-groupoids $\mathbf{B}G$ whose space of objects is a point, i.e.
\begin{equation}
    \mathbf{B}G\;=\; \left( \begin{tikzcd}[row sep=scriptsize, column sep=5ex]
    \; \cdots\; \arrow[r, yshift=1.4ex] \arrow[r, yshift=2.8ex] \arrow[r] \arrow[r, yshift=-1.4ex]\arrow[r, yshift=-2.8ex] & G_3 \arrow[r, yshift=1.8ex]\arrow[r, yshift=0.6ex]\arrow[r, yshift=-1.8ex]\arrow[r, yshift=-0.6ex]& G_2
    \arrow[r, yshift=1.4ex] \arrow[r] \arrow[r, yshift=-1.4ex] & G_1  \arrow[r, yshift=0.7ex] \arrow[r, yshift=-0.7ex] & \ast 
    \end{tikzcd}    \right).
\end{equation}
We will refer to the functor $\mathbf{B}$ as \textit{delooping}.
\end{remark}

\begin{example}[Ordinary Lie group]
Let $G$ be an ordinary Lie group. The simplicial set corresponding to $G$ is given by $[n]\mapsto G^{\times n}$. More precisely, this is given by the following diagram:
\begin{equation*}
    \mathbf{B}G\;=\; \left( \begin{tikzcd}[row sep=scriptsize, column sep=14ex]
    \; \cdots\;\; G\times G\times G \arrow[r, yshift=3.8ex, "{(\pi_1,\pi_2)}"]\arrow[r, yshift=1.4ex, "{((-)\cdot(-),\,(-))}"]\arrow[r, yshift=-1.4ex, "{((-),\,(-)\cdot(-))}"]\arrow[r, yshift=-3.8ex, "{(\pi_2,\pi_3)}"] & G\times G
    \arrow[r, yshift=3.2ex, "\pi_1"] \arrow[r, "(-)\cdot(-)"] \arrow[r, yshift=-3.2ex, "\pi_2"] & G  \arrow[r, yshift=1.4ex, "0"] \arrow[r, yshift=-1.4ex, "0"] & \ast 
    \end{tikzcd}    \right),
\end{equation*}
where $\pi_i$ is the projection of the $i$-th copy of $G$ in a product $G^{\times n}$ and $(-)\cdot(-)$ is the group product of $G$.
\end{example}

\begin{example}[Classifying space of a Lie group]
The geometric realisation of the delooping $\mathbf{B}G$ of an ordinary Lie group $G$ is exactly the usual \textit{classifying space}
\begin{equation}
    |\mathbf{B}G| \; =\; BG.
\end{equation}
\end{example}

\begin{example}[Strict Lie $2$-group]
A \textit{strict Lie $2$-group} is defined as a Lie $2$-group whose corresponding Kan complex is of the following form:
\begin{equation}
    \mathbf{B}G\;=\; \left( \begin{tikzcd}[row sep=scriptsize, column sep=5ex]
    \; \cdots\; \arrow[r, yshift=1.4ex] \arrow[r, yshift=2.8ex] \arrow[r] \arrow[r, yshift=-1.4ex]\arrow[r, yshift=-2.8ex] & K^{\times 3} \times H^{\times 2} \arrow[r, yshift=1.8ex]\arrow[r, yshift=0.6ex]\arrow[r, yshift=-1.8ex]\arrow[r, yshift=-0.6ex]& K^{\times 2} \times H
    \arrow[r, yshift=1.4ex] \arrow[r] \arrow[r, yshift=-1.4ex] & K  \arrow[r, yshift=0.7ex] \arrow[r, yshift=-0.7ex] & \ast 
    \end{tikzcd}    \right),
\end{equation}
where $K,H$ are two ordinary Lie groups. Such a Kan complex is also known as \textit{Duskin nerve} \cite{Duskin} of a Lie $2$-group.
This implies that the horizontal and vertical composition of $2$-morphisms is given by the following rules:
\begin{equation}
\begin{aligned}
\begin{tikzcd}[row sep=scriptsize, column sep=12ex]
    \ast \arrow[r, bend left=50, ""{name=U, below}, "g_1"]
    \arrow[r, bend right=50, "h_1"', ""{name=D}]
    & \ast
    \arrow[Rightarrow, from=U, to=D, "\epsilon_1"] \arrow[r, bend left=50, ""{name=I, below}, "g_2"]
    \arrow[r, bend right=50, "h_2"', ""{name=F}] & \ast \arrow[Rightarrow, from=I, to=F, "\epsilon_2"]
\end{tikzcd} \;\;&=\;\; \begin{tikzcd}[row sep=scriptsize, column sep=12ex]
    \ast \arrow[r, bend left=50, ""{name=U, below}, "g_2\circ g_1"]
    \arrow[r, bend right=50, "h_2\circ h_1"', ""{name=D}]
    & \ast
    \arrow[Rightarrow, from=U, to=D, "\epsilon_2\circ\epsilon_1"]
\end{tikzcd} \\[1ex]
\begin{tikzcd}[row sep=scriptsize, column sep=12ex]
    \ast \arrow[r, bend left=50, ""{name=U, below}, "g"]
    \arrow[r, ""{name=D}, ""'{name=I}] \arrow[r, bend right=50, "h"', ""{name=F}] & \ast
    \arrow[Rightarrow, from=U, to=D, "\epsilon_1"] \arrow[Rightarrow, from=I, to=F, "\epsilon_2"] 
\end{tikzcd} \;\;&=\;\; \begin{tikzcd}[row sep=scriptsize, column sep=12ex]
    \ast \arrow[r, bend left=50, ""{name=U, below}, "g"]
    \arrow[r, bend right=50, "h"', ""{name=D}]
    & \ast
    \arrow[Rightarrow, from=U, to=D, "\epsilon_2\circ\epsilon_1"]
\end{tikzcd}
\end{aligned}
\end{equation}
where $\epsilon_i\in H$ and $g_i,h_i\in K$.
\end{example}

\subsection{Dold-Kan correspondence}

Now we will briefly present a correspondence which allows us to write abelian stacks in a very simple and immediate fashion. See \cite[ch.$\,$3]{Goess:1999jar} for a detailed discussion.

\begin{remark}[Dold-Kan correspondence]\label{rem:dkc}
\textit{Dold-Kan correspondence} exhibits an equivalence between abelian smooth higher stacks and chain complexes of abelian sheaves over manifolds. In our notation, an abelian smooth stack $\mathscr{A}\in\mathbf{H}$ will correspond a chain complex $(\mathcal{A}_\bullet,\mathrm{d}_\bullet)$ of abelian sheaves, i.e. explicitly
\begin{equation}
    \mathscr{A} \; = \;  \mathrm{DK}\!\left(
    \begin{tikzcd}[row sep=scriptsize, column sep=3ex]\cdots\arrow[r, "\mathrm{d}_{3}"]& \mathcal{A}_2 \arrow[r, "\mathrm{d}_{2}"]& \mathcal{A}_1\arrow[r, "\mathrm{d}_{1}"]& \mathcal{A}_0
    \end{tikzcd}
    \right),
\end{equation}
where the $\mathcal{A}_i$ are all abelian sheaves.
The stack $\mathbf{B}\mathscr{A}$, which is called \textit{delooping} of $\mathscr{A}$, is exactly the stack corresponding to the shifted chain $\mathcal{A}_\bullet[1]$ of smooth sheaves, i.e. explicitly
\begin{equation}
    \mathbf{B}\mathscr{A} \; = \;  \mathrm{DK}\!\left(
    \begin{tikzcd}[row sep=scriptsize, column sep=3ex]\cdots \arrow[r, "\mathrm{d}_{2}"]& \mathcal{A}_1\arrow[r, "\mathrm{d}_{1}"]& \mathcal{A}_0 \arrow[r, "0"]& \,0\,
    \end{tikzcd}
    \right).
\end{equation}
Let $M$ be a smooth manifold. An element of the $\infty$-groupoid $\mathbf{H}(M,\mathscr{A})$ of sections of $\mathscr{A}$ is given under Dold-Kan correspondence by a cocycle $\big(a^0_{(\alpha)},a^1_{(\alpha\beta)},a^2_{(\alpha\beta\gamma)},\dots\big)$, such that
\begin{equation}
    a^0_{(\alpha)}\in\mathcal{A}_0(U_\alpha), \quad a^1_{(\alpha\beta)}\in\mathcal{A}_1(U_\alpha\cap U_\beta), \quad a^2_{(\alpha\beta\gamma)}\in\mathcal{A}_2(U_\alpha\cap U_\beta\cap U_\gamma), \quad \dots
\end{equation}
and whose patching conditions are given, on any $k$-fold overlap of patches $U_\alpha$ of $M$, by
\begin{equation}
\begin{aligned}
    a^0_{(\beta)}-a^0_{(\alpha)} \;&=\; \di_1 a^1_{(\alpha\beta)},\\[0.1cm]
    a^1_{(\alpha\beta)}+a^1_{(\beta\gamma)}+a^1_{(\gamma\alpha)} \;&=\; \di_2 a^2_{(\alpha\beta\gamma)}, \\[0.1cm]
    a^2_{(\alpha\beta\gamma)}-a^2_{(\beta\gamma\delta)}+a^2_{(\gamma\delta\alpha)}-a^2_{(\delta\alpha\beta)} \;&=\; \di_3 a^3_{(\alpha\beta\gamma\delta)}, \\[0.1cm]
    \;&\;\;\vdots\;
\end{aligned}
\end{equation}
\end{remark}

\noindent The following physically relevant applications of the Dold-Kan correspondence were introduced by \cite{Fiorenza:2012ec}.

\begin{example}[Abelian $1$-stacks and $2$-stacks]\label{ex:stacks}
The following are the relevant examples of abelian $1$-stacks and $2$-stacks we are going to use in the next discussion. They are presented through Dold-Kan correspondence (remark \ref{rem:dkc}) as chain complexes of abelian sheaves. Notice that in this form they are Deligne complexes:
\begin{equation}
    \begin{aligned}
    \mathbf{B}U(1) \;&=\;  \mathrm{DK}\!\left(
    \begin{tikzcd}[row sep=scriptsize, column sep=8ex]\Coo\left(-,U(1)\right)\;\arrow[r,"0"]&\; 0
    \end{tikzcd}
    \right), \\
    \mathbf{B}U(1)_{\mathrm{conn}} \;&=\;   \mathrm{DK}\!\left(
    \begin{tikzcd}[row sep=scriptsize, column sep=8ex]\Coo\left(-,U(1)\right)\;\arrow[r,"\frac{1}{2\pi i}\mathrm{d}\cdot\log"]&\;\Omega^1(-)
    \end{tikzcd}
    \right), \\
    \mathbf{B}^2U(1)  \;&=\; \mathrm{DK}\!\left(
    \begin{tikzcd}[row sep=scriptsize, column sep=8ex]\Coo\left(-,U(1)\right)\;\arrow[r,"0"]&\; 0\arrow[r,"0"]&\; 0
    \end{tikzcd}
    \right), \\
    \mathbf{B}(\mathbf{B}U(1)_{\mathrm{conn}})\;&=\;   \mathrm{DK}\!\left(
    \begin{tikzcd}[row sep=scriptsize, column sep=8ex]\Coo\left(-,U(1)\right)\;\arrow[r,"\frac{1}{2\pi i}\mathrm{d}\cdot\log"]&\;\Omega^1(-)\arrow[r,"0"]&\; 0
    \end{tikzcd}
    \right), \\
    \mathbf{B}^2U(1)_{\mathrm{conn}}  \;&=\;  \mathrm{DK}\!\left(
    \begin{tikzcd}[row sep=scriptsize, column sep=8ex]\Coo\left(-,U(1)\right)\;\arrow[r,"\frac{1}{2\pi i}\mathrm{d}\cdot\log"]&\;\Omega^1(-)\arrow[r,"\mathrm{d}"]&\; \Omega^2(-)
    \end{tikzcd}
    \right).
        \end{aligned}
\end{equation}
More generally, we can write the following abelian $k$-stack for any $k\in\mathbb{N}^+$ by using the Dold-Kan correspondence:
\begin{equation*}
    \mathbf{B}^kU(1)_{\mathrm{conn}}  \,=\,  \mathrm{DK}\!\left(
    \begin{tikzcd}[row sep=scriptsize, column sep=7ex]\!\!\Coo\left(-,U(1)\right)\arrow[r,"\frac{1}{2\pi i}\mathrm{d}\cdot\log"]&\Omega^1(-)\arrow[r,"\mathrm{d}"]& \cdots\arrow[r,"\mathrm{d}"]& \Omega^k(-)\!
    \end{tikzcd}\right).
\end{equation*}
\end{example}

\begin{remark}[Forgetful functor]
Notice we can naturally introduce a \textit{forgetful functor} which forgets the $1$-degree $1$-form part of the chain complex and retains only the $0$-degree sheaf for
\begin{equation}
    \begin{tikzcd}[row sep=scriptsize, column sep=7ex]
    \mathbf{B}U(1)_{\mathrm{conn}} \arrow[r, "\frgt"] & \mathbf{B}U(1).
\end{tikzcd}
\end{equation}
Analogously we can define natural forgetful functors for the $2$,$1$-degree sheaves of the chains
\begin{equation}
    \begin{tikzcd}[row sep=scriptsize, column sep=7ex]
    \mathbf{B}^2U(1)_{\mathrm{conn}} \arrow[r, "\frgt"] & \mathbf{B}(\BU) \arrow[r, "\frgt"] &  \mathbf{B}^2U(1).
\end{tikzcd}
\end{equation}
\end{remark}

\begin{remark}[$\BU$ is a group-stack]\label{rem:gs}
The stack $\BU$ of circle bundles with connection is a \textit{group-stack}, which means that it satisfies the ordinary defining properties of a group up to an isomorphism. First of all $\BU$ is naturally equipped with a tensor product
\begin{equation}
    \otimes:\;\BU \,\times\, \BU\; \longrightarrow\; \BU,
\end{equation}
which maps a couple of circle bundles $P_1\rightarrow M$ and $P_2\rightarrow M$ to a new one $P_1\otimes P_2\rightarrow M$. Moreover, the dual bundle $P^\ast\rightarrow M$ of any circle bundle $P\rightarrow M$ plays the role of its \textit{inverse element}, while the trivial circle bundle $M\times U(1)\rightarrow M$ with trivial connection plays the role of the \textit{identity element} $\mathrm{id}$. It is easy to verify that ordinary group properties
\begin{equation}
    \begin{aligned}
    P^\ast\otimes P\cong\mathrm{id}, \quad P\otimes P^\ast\cong \mathrm{id}, \\
    P_1\otimes(P_2\otimes P_3)\cong (P_1\otimes P_2)\otimes P_3
    \end{aligned}
\end{equation}
are satisfied only up to gauge transformation of circle bundles. In local \v{C}ech data on a manifold $M$ we have  $(\eta_{(\alpha)},\eta_{(\alpha\beta)})\otimes(\eta'_{(\alpha)},\eta'_{(\alpha\beta)})= (\eta_{(\alpha)}+\eta'_{(\alpha)},\eta_{(\alpha\beta)}+\eta'_{(\alpha\beta)})$ and  $(\eta_{(\alpha)},\eta_{(\alpha\beta)})^\ast:=(-\eta_{(\alpha)},-\eta_{(\alpha\beta)})$.
\end{remark}

\subsection{Geometric stacks}

Intuitively, a geometric stack is a smooth stack which is represented by a suitable notion of Kan complex in the category of smooth manifolds.
Such stacks are particularly important in the context of $\infty$-Lie theory (see section \ref{sec:lietheory}), when we deal with Lie integration of $L_\infty$-algebroids.

\begin{definition}[Kan simplicial manifold]
A \textit{Kan simplicial manifold} is defined as a simplicial manifold, i.e. an object of the category
\begin{equation}
    \mathbf{sDiff} \;:=\; \Func(\mathbf{\Delta}^{\mathrm{op}},\mathbf{Diff}),
\end{equation}
such that it satisfies a differential version of the Kan condition \eqref{eq:kan2}, i.e. the map
\begin{equation}
    X_n \,\cong\, \mathbf{sDiff}(\Delta^n,K)\,\longrightarrow\,\mathbf{sDiff}(\Lambda^n_i,K)
\end{equation}
is a surjective submersion for any $0\leq i\leq n$.
\end{definition}

\begin{definition}[Geometric stack]
A \textit{geometric stack} is defined as a smooth stack $\mathscr{X}\in\mathbf{H}$ represented by a Kan simplicial manifold \cite{Pridham_2013}, i.e. such that it is given by $\mathscr{X}(U)=\mathbf{sDiff}(U,X)$ for some fixed Kan simplicial manifold $X$, on any $U\in\mathbf{Diff}$.
\end{definition}

\begin{remark}[Technicalities on Kan simplicial manifolds]
We remark that a smooth stack is not represented, in general, by a Kan simplicial manifold. Moreover, geometric stacks do not constitute an $(\infty,1)$-topos, For more details see \cite{Principal1, Principal2}.
\end{remark}

\section{$\infty$-Lie theory}\label{sec:lietheory}

$\infty$-Lie theory, or higher Lie theory, is the refinement of Lie theory to higher geometry. It studies $L_\infty$-algebras and, more generally, $L_\infty$-algebroids and their relation to Lie $\infty$-groupoids by Lie integration.

\subsection{$L_\infty$-algebras}

The reader will be probably familiar with the notion of $L_\infty$-algebra $\mathfrak{g}$. Roughly, this is a generalisation of Lie algebra whose underlying vector space is graded and which is equipped with a potentially infinite number of $n$-ary brackets $[-,-,\dots,-]_n:\mathfrak{g}^{\otimes n}\rightarrow\mathfrak{g}$, whose Jacobi identities only holds up to a homotopy given by the $(n+1)$-ary brackets. 

\begin{definition}[Unshuffle]
An $(i,n-i)$-unshuffle with $0<i<n$ is a permutation $\sigma\in\mathsf{Perm}(n)$ such that 
\begin{equation}
    \sigma(1)<\cdots<\sigma(i) \quad \text{and}\quad \sigma(i+1)<\cdots<\sigma(n).
\end{equation}
We will denote as $\mathsf{Unsh}(i,n-i)\subset \mathsf{Perm}(n)$ the set of $(i,n-i)$-unshuffles.
\end{definition}

\begin{definition}[$L_\infty$-algebra] An $L_\infty$-algebra $\mathfrak{g}=\left(V,[-,\,\cdots\,,-]_n\right)$ is a positively graded vector space $V$ equipped with a collection of graded anti-symmetric brackets of degree $n-2$
\begin{equation}
    [-,\,\cdots\,,-]_n\!:\, V^{\otimes n} \rightarrow V
\end{equation}
for any positive integer $n\in\mathbb{N}^+$, such that they satisfy the following conditions
\begin{enumerate}
    \item \textit{graded skew symmetry}: for any $n$-tuple $(x_1,\dots,x_n)$ of homogeneously graded elements, we have
    \begin{equation}
        [x_{1},\cdots,x_{i},x_j\cdots,x_n]_n \,=\, -(-1)^{\deg(x_i)\deg(x_j)}[x_{1},\cdots,x_{j},x_i\cdots,x_n]_n,
    \end{equation}
    where $\deg(x_i)$ is the degree of the element $x_i\in V$.
    \item \textit{strong homotopy Jacobi identity}: for any $n$-tuple $(x_1,\dots,x_n)$ of homogeneously graded elements, we have
\begin{equation}
    \mathrm{Jac}_n(x_1,\cdots,x_n) = 0 \qquad (n>1), 
\end{equation}
where the $n$-th \textit{Jacobiator} $\mathrm{Jac}_n(-,\cdots,-)$ is defined by
\begin{equation*}
    \!\!\!\!\!\!\mathrm{Jac}_n(x_1,\cdots,x_n):=\!\!\!\!\sum_{i+j=n+1}\!\!\!\!(-1)^{i(j-1)}\!\!\!\!\!\!\!\!\!\!\sum_{\sigma\in\mathsf{Unsh}(i,j-1)}\!\!\!\!\!\!\!\!\chi(\sigma)\!\left[\big[x_{\sigma(1)},\cdots,x_{\sigma(i)}\big]_i,x_{\sigma(i+1)},\cdots,x_{\sigma(n)}\right]_{j+1},
\end{equation*}
where $\chi(\sigma)=\{+1,-1\}$, known as \textit{Koszul-signature}, is the sign obtained by multiplying the sign $(-1)^{|\sigma|}$ of the permutation with the sign $(-1)^{\deg(x_i)\deg(x_j)}$ that arises from the degrees of the permuted elements.
\end{enumerate}
\end{definition}

\noindent We redirect to the reference \cite{Lada:1992wc} for details.

\begin{example}[Ordinary Lie algebra]
An \textit{ordinary Lie algebra} is a $L_\infty$-algebra whose underlying graded vector space $V$ is trivial also at degree $>1$.
\end{example}

\begin{example}[Lie $n$-algebra]
A \textit{Lie $n$-algebra} is a $L_\infty$-algebra whose underlying graded vector space $V$ is trivial also at degree $>n$.
\end{example}

\subsection{$L_\infty$-algebroids}

\noindent A $L_\infty$-algebroid on a smooth manifold $M$ is a generalisation of $L_\infty$-algebra, defined by a graded vector bundle $\mathfrak{a}\twoheadrightarrow M$ whose graded space of its sections $\Gamma(M,\mathfrak{a})$ is equipped with a $L_\infty$-algebra structure. This concept was introduced by \cite{Sat09}.

\begin{definition}[$L_\infty$-algebroid] An $L_\infty$ algebroid $\mathfrak{a}=\left(E,[-,\,\cdots\,,-]_n,\varrho\right)$ on a smooth manifold $M$ is a graded vector bundle $E\twoheadrightarrow M$ equipped with an $L_\infty$-algebra structure on the graded $\Coo(M)$-module $\Gamma(M,E)$ of its sections and with a morphism of graded vector bundles $\varrho:E\rightarrow TM$, called \textit{anchor map}, such that
\begin{enumerate}[label=(\alph*)$\;$]
    \item the anchor induces a homomorphism of $L_\infty$-algebras (see remark \ref{def:homlinfty})
    \begin{equation}
        \varrho:(\Gamma(M,E),[-,\,\cdots\,,-]_n)\;\longtwoheadrightarrow\;(\mathfrak{X}(M),[-,-]_{\mathrm{Lie}})
    \end{equation}
    \item the brackets $[-,\,\cdots\,,-]_n$ satisfy the following Leibniz rule
    \begin{equation}
        \begin{aligned}
        \left[x_1, fx_2\right]_2 \,&=\, \varrho(x_1)(f)x_2 + (-1)^{\deg(x_1)}f[x_1,x_2]_2, \\
        \left[x_1, \cdots, fx_n\right]_n \,&=\, (-1)^{n+\sum_{i=1}^{n-1}\!\deg(x_i)}f[x_1,\cdots,x_n]_n \qquad\;\; (n>2),
    \end{aligned}
\end{equation}
for any smooth function $f\in\Coo(M)$ and sections $x_i\in\Gamma(M,E)$.
\end{enumerate}
\end{definition}

\begin{remark}[$L_\infty$-algebra as $L_\infty$-algebroid]
Let $\mathbf{L_\infty Alg}$ and $\mathbf{L_\infty Algbd}$ be the $(\infty,1)$-categories respectively of $L_\infty$-algebras ans $L_\infty$-algebroids. We have an embedding
\begin{equation}
\begin{aligned}
    \mathbf{L_\infty Alg} \;&\xhookrightarrow{\;\;\;\mathbf{b}\;\;\;}\; \mathbf{L_\infty Algbd} \\
    \mathfrak{g} \;&\xmapsto{\qquad}\; \mathbf{b}\mathfrak{g}.
\end{aligned}
\end{equation}
which maps a $L_\infty$-algebra $\mathfrak{g}$ to a $L_\infty$-algebroid $\mathbf{b}\mathfrak{g}\twoheadrightarrow \ast$ whose base manifold is a point $\ast$.
\end{remark}

\subsection{Chevalley-Eilenberg dg-algebras}

For simplicity, let us denote, from now on, the underlying graded vector space of a $L_\infty$-algebra $\mathfrak{g}$ not anymore as $V$, but just as $\mathfrak{g}$.

\begin{theorem}[Chevalley-Eilenberg dg-algebra of an $L_\infty$-algebra]
An $L_\infty$-algebra structure on $\mathfrak{g}$ is equivalently a dg-algebra structure on $\wedge^\bullet\mathfrak{g}^\ast$, which we will call \textit{Chevalley-Eilenberg dg-algebra} of $\mathfrak{g}$.
\end{theorem}
\begin{proof}
Given an $L_\infty$-algebra $\mathfrak{g}$, let us define the following differential-graded algebra:
\begin{equation}
    \mathrm{CE}(\mathfrak{g}) \;:=\; \big(\wedge^\bullet\mathfrak{g}^\ast,\,\di\,\big),
\end{equation}
where the underlying graded vector space is
\begin{equation}
    \wedge^\bullet\mathfrak{g}^\ast \;=\; \mathbb{R}\,\oplus\, \mathfrak{g}_0^\ast \,\oplus\, (\mathfrak{g}_1^\ast \,\oplus\, \mathfrak{g}_0^\ast\wedge \mathfrak{g}_0^\ast)\,\oplus\, (\mathfrak{g}_2^\ast\,\oplus\, \mathfrak{g}_0^\ast\wedge \mathfrak{g}_1^\ast\,\oplus\, \mathfrak{g}_0^\ast\wedge\mathfrak{g}_0^\ast\wedge\mathfrak{g}_0^\ast )\,\oplus\,\cdots
\end{equation}
and the $+1$-degree differential is defined by
\begin{equation}\label{eq:cobracket}
    \di: \; t^a \quad \longmapsto \quad \di t^a = -\sum_{n\in\mathbb{N}^+}\frac{1}{n!}[t_{a_1}, t_{a_2},\cdots,t_{a_n}]_n^a \, t^{a_1}\wedge t^{a_2} \wedge\cdots\wedge t^{a_n},
\end{equation}
where $\{t^a\}$ is a basis of $\mathfrak{g}$ and $\{t_a\}$ is its dual basis of $\mathfrak{g}^\ast$.
Thus, the differential encodes the $n$-ary brackets of the $L_\infty$ algebra $\mathfrak{g}$.
Now, we will show that the differential condition $\di^2=0$ on the dg-algebra $\mathrm{CE}(\mathfrak{g})$ is equivalent to the condition $\mathrm{Jac}_n=0$ for the Jacobiator for any $n\in\mathbb{N}^+$ on the $L_\infty$-algebra $\mathfrak{g}$.
Thus, we can directly calculate
\begin{equation*}
    \di^2t^a \,= \!\!\!\sum_{n,m\in\mathbb{N}^+}\!\frac{1}{n!(m-1)!}\big[[t_{a_1},\cdots,t_{a_n}]_n, t_{b_2},\cdots,t_{b_m} \big]^a_{m} \, t^{a_1}\wedge\cdots\wedge t^{a_n}\wedge t^{b_1}\wedge\cdots\wedge t^{b_m}.
\end{equation*}
This can be produced by summing over all the unshuffles $\sigma\in\mathsf{Unsh}(n,m-1)$ weighted by the Koszul-sign of the permutation, i.e.
\begin{equation*}
    0\,=\!\!\!\sum_{m+n=k+1}\!\!\!(-1)^{n(m-1)} \!\!\!\!\!\!\!\!\sum_{\sigma\in\mathsf{Unsh}(n,m+1)}\!\!\!\!\!\!\chi(\sigma)\big[[t_{\sigma(1)},\cdots,t_{\sigma(n)}]_n,t_{\sigma(n+1)},\cdots,t_{\sigma(m)}\big]_{m},
\end{equation*}
which is exactly the Jacobiator $\mathrm{Jac}_k$ of any $k$-ary bracket of the $L_\infty$-algebra $\mathfrak{g}$. 
\end{proof}

\begin{example}[Chevalley-Eilenberg dg-algebra of an ordinary Lie algebra]
Let us explicitly consider the Chevalley-Eilenberg dg-algebra $\mathrm{CE}(\mathfrak{g})$ of an ordinary Lie algebra $\mathfrak{g}$.
Let $\{t^a\}$ be the $1$-degree generators of $\mathfrak{g}^\ast$. Then, $\mathrm{CE}(\mathfrak{g})$ is identified by 
\begin{equation}
\begin{aligned}
        \di t^a \;&=\; -\frac{1}{2}C^a_{\;\;bc}t^b\wedge t^c
\end{aligned}
\end{equation}
where $ C^a_{\;\;bc}$ are the structure constants of the Lie algebra $\mathfrak{g}$.
\end{example}

\begin{example}[Chevalley-Eilenberg dg-algebra of a Lie $2$-algebra]
Let us explicitly consider the Chevalley-Eilenberg dg-algebra $\mathrm{CE}(\mathfrak{g})$ of a Lie $2$-algebra $\mathfrak{g}$.
Let $\{t^a,b^i\}$ be respectively the $1$-degree and $2$-degree generators of $\mathfrak{g}^\ast$. Then, $\mathrm{CE}(\mathfrak{g})$ is given by 
\begin{equation}
\begin{aligned}
        \di t^a \;&=\; -\frac{1}{2}C^a_{\;\;bc}t^b\wedge t^c - C^a_{\;i}b^i \\
        \di b^i \;&=\; -C^i_{\;aj}t^a\wedge b^j - C_{abc}^i t^a\wedge t^b\wedge t^c
\end{aligned}
\end{equation}
where $\{C^a_{\;\;bc},C^a_{\;i},C^i_{\;aj},C_{abc}^i\}$ are the structure constants corresponding to the bracket structure $\{[-]_1,\,[-,-]_2,\,[-,-,-]_3\}$ of the Lie $2$-algebra $\mathfrak{g}$.
\end{example}

\begin{example}[String $2$-algebra $\mathfrak{string}(\mathfrak{g})$]
Following \cite{FSS12}, we define the string $2$-algebra $\mathfrak{string}(\mathfrak{g})$ of an ordinary Lie algebra $\mathfrak{g}$ by dually defining the dg-algebra $\mathrm{CE}(\mathfrak{string}(\mathfrak{g}))$ as follows:
\begin{equation}
\begin{aligned}
        \di t^a \;&=\; -\frac{1}{2}C^a_{\;\;bc}t^b\wedge t^c \\
        \di b \;&=\; - k_{aa'}C^{a'}_{\;\,bc} t^a\wedge t^b\wedge t^c,
\end{aligned}
\end{equation}
where $C^a_{\;\;bc}$ are the structure constants of the ordinary Lie algebra $\mathfrak{g}$, where $k_{ab}$ is a Killing form on $\mathfrak{g}$ and where $\{t^a,b\}$ are respectively the $1$-degree generators of $\mathfrak{g}^\ast$ and a $2$-degree generator.
\end{example}

\begin{remark}[Homomorphisms of $L_\infty$-algebras]\label{def:homlinfty}
Given a graded vector space $V$, the shift isomorphism of graded vector spaces $V\xrightarrow{\;\cong\;} V[1]$ induces an isomorphism of dg-algebras $\wedge^\bullet V \xrightarrow{\cong} \bigodot^\bullet V[1]$. Thus, we can equivalently write $ \mathrm{CE}(\mathfrak{g}) = \big(\bigodot^\bullet\mathfrak{g}^\ast[1],\,\di\,\big)$.
Therefore, there is an embedding (i.e. a fully faithful functor)
\begin{equation}
\begin{aligned}
    \mathbf{L_\infty Alg} \;&\xhookrightarrow{\;\; \mathrm{CE}(-)\;\;}\; \mathbf{dgcAlg}^\mathrm{op} \\
    \mathfrak{g} \;&\xmapsto{\qquad\quad\,}\; \mathrm{CE}(\mathfrak{g})
\end{aligned}
\end{equation}
where $\mathbf{dgcAlg}$ is the $\infty$-category of the differential-graded commutative algebras. 
Thus, the homomorphisms $\mathfrak{g}_1\rightarrow\mathfrak{g}_2$ of $L_\infty$-algebras can be defined by
\begin{equation}
    \mathbf{L_\infty Alg}(\mathfrak{g}_1,\mathfrak{g}_2) \;:=\; \mathbf{dgcAlg}\big(\mathrm{CE}(\mathfrak{g}_2),\,\mathrm{CE}(\mathfrak{g}_1)\big) .
\end{equation}
\end{remark}

\noindent The definition of the Chevalley-Eilenberg dg-algebra of an $L_\infty$-algebra can be easily generalised to $L_\infty$-algebroids.

\begin{definition}[Chevalley-Eilenberg dg-algebra of an $L_\infty$-algebroid]
The \textit{Chevalley-Eilenberg dg-algebra of an $L_\infty$-algebroid} $\mathfrak{a}\twoheadrightarrow M$ is defined by
\begin{equation}
    \mathrm{CE}(\mathfrak{a}) \;:=\; \big(\wedge^\bullet\Gamma(M,\mathfrak{a}^\ast),\,\di\,\big),
\end{equation}
where the underlying graded space is
\begin{equation}
    \wedge^\bullet\Gamma(M,\mathfrak{a}^\ast) \;=\; \Coo(M)\,\oplus\, \Gamma(M,\mathfrak{a}^\ast_0) \,\oplus\, \big(\Gamma(M,\mathfrak{a}^\ast_1) \,\oplus\, \Gamma(M,\mathfrak{a}^\ast_0\big)\wedge \Gamma(M,\mathfrak{a}_0^\ast))\,\oplus\, \cdots
\end{equation}
and where the Chevalley-Eilenberg differential, now, acts also on smooth functions $f\in\Coo(M)$ as follows:
\begin{equation}
    \di f \;=\; \varrho^\mu_{\;\,a}t^a \wedge \partial_\mu f \;\;\,\in\,\Gamma(M,\mathfrak{a}^\ast_0),
\end{equation}
where $\varrho^\mu_{\;\,a}$ is the anchor map of the $L_\infty$-algebroid, in components. The differential of the generators in the higher degrees is given in analogy with the differential \eqref{eq:cobracket}.
\end{definition}

\begin{remark}[Homomorphisms of $L_\infty$-algebroids]
In analogy with $L_\infty$-algebra in remark \ref{def:homlinfty}, we can rewrite $\mathrm{CE}(\mathfrak{a}) = \big(\bigodot^\bullet\Gamma(M,\mathfrak{a}^\ast)[1],\,\di\,\big)$.
Thus, we have an embedding (i.e. a fully faithful functor)
\begin{equation}
\begin{aligned}
    \mathbf{L_\infty Algbd} \;&\xhookrightarrow{\;\; \mathrm{CE}(-)\;\;}\; \mathbf{dgcAlg}^\mathrm{op} \\
    \mathfrak{a} \;&\xmapsto{\qquad\quad\,}\; \mathrm{CE}(\mathfrak{a})
\end{aligned}
\end{equation}
where $\mathbf{dgcAlg}$ is the $\infty$-category of the differential-graded commutative algebras. 
Now, the homomorphisms $\mathfrak{a}_1\rightarrow\mathfrak{a}_2$ of $L_\infty$-algebroids can be defined by
\begin{equation}
    \mathbf{L_\infty Algbd}(\mathfrak{a}_1,\mathfrak{a}_2) \;:=\; \mathbf{dgcAlg}\big(\mathrm{CE}(\mathfrak{a}_2),\,\mathrm{CE}(\mathfrak{a}_1)\big) .
\end{equation}
Notice that $\mathbf{dgcAlg}\big(\mathrm{CE}(\mathfrak{a}_2),\,\mathrm{CE}(\mathfrak{a}_1)\big)$ is itself a differential-graded algebra. This dg-algebra can be regarded as the $L_\infty$-algebroid of the homomorphisms from $\mathfrak{a}_1$ to $\mathfrak{a}_2$, which we can denote as $\mathbf{L_\infty Algbd}(\mathfrak{a}_1,\mathfrak{a}_2)$.
\end{remark}

\begin{definition}[dg-algebra of flat $L_\infty$-algebra-valued differential forms]
For a given $L_\infty$-algebra $\mathfrak{g}$, the dg-algebra of \textit{flat $\mathfrak{g}$-valued differential forms} is defined by
\begin{equation}
    \big(\Omega_\mathrm{flat}^\bullet(M,\mathfrak{g}),\nabla\big) \,\;:=\,\; \mathbf{dgcAlg}\big(\mathrm{CE}(\mathfrak{g}),\,(\Omega^\bullet,\di)\big),
\end{equation}
where $\nabla$ is the covariant derivative of any $1$-degree differential form.
\end{definition}

\begin{example}[dg-algebra of flat Lie algebra-valued differential forms]
Given an ordinary Lie algebra $\mathfrak{g}$, a $1$-degree element $A\in\Omega_\mathrm{flat}^1(M,\mathfrak{g})$ is a $\mathfrak{g}$-valued differential form $A\in\Omega^1(M)\otimes\mathfrak{g}$ which satisfies the differential equation
\begin{equation}
\begin{aligned}
        \di A^a + \frac{1}{2}C^a_{\;\;bc}A^b\wedge A^c \;&=\; 0,
\end{aligned}
\end{equation}
where $ C^a_{\;\;bc}$ are the structure constants of $\mathfrak{g}$.
\end{example}

\noindent Interestingly, Chevalley-Eilenberg dg-algebra is also useful to define a cohomology theory for Lie algebras.

\begin{definition}[Lie algebra cohomology]
The \textit{Lie algebra cohomology} of $\mathfrak{g}$ is the cohomology of the differential $\di$ of the Chevalley-Eilenberg dg-algebra $\mathrm{CE}(\mathfrak{g})$, i.e.
\begin{equation}
H_{\mathrm{Lie}}^n(\mathfrak{g}) \;=\; \frac{\mathrm{Ker}\,\big(\di:\wedge^{n}\mathfrak{g}^\ast\rightarrow\wedge^{n+1}\mathfrak{g}^\ast\big)} {\mathrm{Im}\,\big(\di:\wedge^{n-1}\mathfrak{g}^\ast\rightarrow\wedge^{n}\mathfrak{g}^\ast\big)}.
\end{equation}
In other words, we have the identification
\begin{equation}
H_{\mathrm{Lie}}^n(\mathfrak{g}) \;=\; H^n\big(\mathrm{CE}(\mathfrak{g})\big).
\end{equation}
\end{definition}

\subsection{Weil dg-algebra}

\begin{definition}[Tangent $L_\infty$-algebra]
We define the \textit{tangent $L_\infty$-algebra} $T\mathfrak{g}$ of an $L_\infty$-algebra $\mathfrak{g}$ is defined by the following Chevalley-Eilenberg algebra:
\begin{equation}
        \mathrm{CE}(T\mathfrak{g}) \;:=\; \big(\wedge^\bullet(\mathfrak{g}^\ast \oplus \mathfrak{g}^\ast[1]),\;\di:=\di_{\mathrm{CE}(\mathfrak{g})} + \delta\, \big)
\end{equation}
where $\mathfrak{g}^\ast[1]$ is a degree-shifted copy of $\mathfrak{g}^\ast$, the differential $\di_{\mathrm{CE}(\mathfrak{g})}$ is the differential of ${\mathrm{CE}(\mathfrak{g})}$ and the differential $\delta:\mathfrak{g}^\ast\xrightarrow{\;\cong\;}\mathfrak{g}^\ast[1]$ is an isomorphism of graded spaces such that $[\di_{\mathrm{CE}(\mathfrak{g})},\delta]=0$.
\end{definition}

\noindent The tangent $L_\infty$-algebra $T\mathfrak{g}$ is also known as $L_\infty$\textit{-algebra of inner derivations} $\mathrm{inn}(\mathfrak{g})$.

\begin{definition}[Weil dg-algebra]\label{def:weildgalgebra}
The \textit{Weil dg-algebra} $\mathrm{W}(\mathfrak{g})$ of an $L_\infty$-algebra $\mathfrak{g}$ is
\begin{equation}
    \mathrm{W}(\mathfrak{g}) \;:=\; \mathrm{CE}(T\mathfrak{g}).
\end{equation}
\end{definition}

\begin{example}[Weil dg-algebra of an ordinary Lie algebra]
Let us explicitly consider the Weil dg-algebra $\mathrm{W}(\mathfrak{g})$ of an ordinary Lie algebra $\mathfrak{g}$.
Let $\{t^a\}$ be the $1$-degree generators of $\mathfrak{g}^\ast$ and $\{r^a\}$ the $2$-degree ones of $\mathfrak{g}^\ast[1]$. Then, the Weil dg-algebra $\mathrm{W}(\mathfrak{g})$ is identified by 
\begin{equation}
\begin{aligned}
        \di t^a \;&=\; -\frac{1}{2}C^a_{\;\;bc}t^b\wedge t^c + r^a, \\
        \di r^a \;&=\; -C^a_{\;\;bc}t^b\wedge r^c,
\end{aligned}
\end{equation}
where $ C^a_{\;\;bc}$ are the structure constants of the Lie algebra $\mathfrak{g}$.
\end{example}

\begin{example}[Weil dg-algebra of the string $2$-algebra $\mathfrak{string}(\mathfrak{g})$]\label{ex:weilstring}
The Weil dg-algebra of the string $2$-algebra $\mathfrak{string}(\mathfrak{g})$ is given by
\begin{equation}
\begin{aligned}
        \di t^a \;&=\; -\frac{1}{2}C^a_{\;\;bc}t^b\wedge t^c + r^a, \\
        \di b \;&=\; - k_{aa'}C^{a'}_{\;\,bc} t^a\wedge t^b\wedge t^c + h, \\
        \di r^a \;&=\; -C^a_{\;\;bc}t^b\wedge r^c, \\
        \di h \;&=\; - k_{aa'}C^{a'}_{\;\,bc} r^a\wedge t^b\wedge t^c
\end{aligned}
\end{equation}
where $C^a_{\;\;bc}$ are the structure constants of the ordinary Lie algebra $\mathfrak{g}$, where $k_{ab}$ is a Killing form on $\mathfrak{g}$ and where $\{t^a,b,r^a,h\}$ are respectively the $1$-degree generators $\{t^a\}$, the $2$-degree generators $\{b,r^a\}$ and the $3$-degree generator $\{h\}$.
\end{example}

\begin{definition}[dg-algebra of $L_\infty$-algebra-valued differential forms]\label{def:gvaluedforms}
For a given $L_\infty$-algebra $\mathfrak{g}$, the dg-algebra of \textit{$\mathfrak{g}$-valued differential forms} is defined by
\begin{equation}
     \big(\Omega^\bullet(M,\mathfrak{g}),\nabla\big) \,\;:=\,\; \mathbf{dgcAlg}\big(\mathrm{W}(\mathfrak{g}),\,(\Omega^\bullet,\di)\big).
\end{equation}
\end{definition}

\begin{remark}[Higher gauge theory]
The geometric meaning of this dg-algebra and its physical interpretation of as higher gauge theory will be explored in subsection \ref{HGFandLinfty}. For the moment, notice that we can physically interpret the underlying graded space on some open set $U$ as follows:
\begin{equation*}
    \cdots \;\longrightarrow \!\!\!\!\!\! \underbrace{\Omega^{-1}(U,\mathfrak{g})}_{\text{gauge of gauge param.}}\!\!\!\! \longrightarrow \!\! \,\underbrace{\Omega^0(U,\mathfrak{g})}_{\text{gauge parameters}}\!\! \longrightarrow\; \, \underbrace{\Omega^1(U,\mathfrak{g})}_{\text{gauge fields}}\;\longrightarrow\;\underbrace{\Omega^2(U,\mathfrak{g})}_{\text{curvatures}}\;\longrightarrow\!\!\underbrace{\Omega^3(U,\mathfrak{g})}_{\text{Bianchi identities}}\!\!\!,
\end{equation*}
where $1$-degree elements $A\in\Omega^1(U,\mathfrak{g})$ are local \textit{higher gauge fields} (given by a collection of local higher form fields) and  $0$-degree elements are gauge parameters of the form
\begin{equation}
    \delta_\lambda A \;=\; [\lambda]_1 + [\lambda\,\overset{\wedge}{,}\,A]_2 + [\lambda\,\overset{\wedge}{,}\,A\,\overset{\wedge}{,}\,A]_3 + \dots.
\end{equation}
Moreover, $2$-degree elements are curvatures of the form
\begin{equation}
    F \;=\; [A]_1 +\frac{1}{2}[A\,\overset{\wedge}{,}\,A]_2 + \frac{1}{3!}[A\,\overset{\wedge}{,}\,A\,\overset{\wedge}{,}\,A]_3 + \dots
\end{equation}
and $3$-degree elements are Bianchi identities for higher gauge fields, i.e.
\begin{equation}
    0 \;=\; [F]_1 +[F\,\overset{\wedge}{,}\,A]_2 + [F\,\overset{\wedge}{,}\,A\,\overset{\wedge}{,}\,A]_3 + \dots.
\end{equation}
\end{remark} 

\begin{example}[dg-algebra of Lie algebra-valued differential forms]
Given an ordinary Lie algebra $\mathfrak{g}$, a $1$-degree element $A\in\Omega^1(M,\mathfrak{g})$ is a $\mathfrak{g}$-valued differential form $A\in\Omega^1(M)\otimes\mathfrak{g}$ which satisfy the differential equations
\begin{equation}
\begin{aligned}
        \di A^a + \frac{1}{2}C^a_{\;\;bc}A^b\wedge A^c \;&=\; F^a, \\
        \di F^a + C^a_{\;\;bc}A^b\wedge F^c\;&=\; 0,
\end{aligned}
\end{equation}
where $F\in\Omega^2(M)\otimes\mathfrak{g}$ is a $2$-degree element which encodes the curvature.
Since $A$ is not constrained to be flat, we can regard $\Omega^1(M,\mathfrak{g})$ as the algebroid of general $\mathfrak{g}$-valued $1$-forms $A\in\Omega^1(M)\otimes \mathfrak{g}$.
\end{example}

\begin{example}[dg-algebra of $\mathfrak{string}(\mathfrak{g})$-valued differential forms]
Given an ordinary Lie algebra $\mathfrak{g}$, a $1$-degree element $(A,B)\in\Omega^1\big(M,\mathfrak{string}(\mathfrak{g})\big)$ is a couple of differential forms $A\in\Omega^1(M)\otimes\mathfrak{g}$ and $B\in\Omega^2(M)$ which satisfy the differential equations
\begin{equation}
\begin{aligned}
        \di A^a + \frac{1}{2}C^a_{\;\;bc}A^b\wedge A^c \;&=\; F^a, \\
        \di B + k_{aa'}C^{a'}_{\;\,bc} A^a\wedge A^b\wedge A^c \;&=\; H, \\[0.5ex]
        \di F^a + C^a_{\;\;bc}A^b\wedge F^c\;&=\; 0, \\[0.5ex]
        \di H + k_{aa'}C^{a'}_{\;\,bc} F^a\wedge A^b\wedge A^c \;&=\; 0,
\end{aligned}
\end{equation}
where $(F,H)\in\Omega^2\big(M,\mathfrak{string}(\mathfrak{g})\big)$ is a $2$-degree element which encodes the curvature.
\end{example}

\subsection{NQ-manifolds}

The first definition of Q-manifold is due to \cite{Alexandrov:1995kv}. See \cite{Fairon_2017} for a review.

\begin{definition}[NQ-manifold]
An \textit{NQ-manifold}, also known as differential graded manifold or \textit{dg-manifold}, is a couple $\mathcal{M}=(|\mathcal{M}|,\Coo)$ of a topological space $|\mathcal{M}|$ and a sheaf of differential graded algebras $\Coo$ on $|\mathcal{M}|$ such that, for any open set $U\subset |\mathcal{M}|$,
\begin{equation}
    \Coo(U) \;\cong\; \big( \Coo(\mathbb{R}^n)\otimes \wedge^\bullet V, \;Q\big)
\end{equation}
where $V$ is a graded vector space and $Q$ is a differential.
\end{definition}

\begin{remark}[NQ-manifolds are $L_\infty$-algebroids]
there exists an $L_\infty$-algebroid $\mathfrak{a}$ satisfying 
\begin{equation}
    \Coo(U) \;\cong\; \mathrm{CE}(\mathfrak{a}|_U) 
\end{equation}
\begin{equation}
    \text{dg-manifold}\;\;\Longleftrightarrow\;\;L_\infty\text{-algebroid}
\end{equation}
\end{remark}

\noindent See \cite{Arvanitakis:2021wkt} for applications in String Theory.

\begin{example}[Shifted tangent bundle]
The NQ-manifold corresponding to the tangent algebroid $TM$ is the \textit{shifted tangent bundle} $T[1]M$, whose coordinates $\{x^\mu,\di x^\mu\}$ are respectively of degree $0$ and $1$, and whose differential is
\begin{equation}
    Q\;=\; \di x^\mu\frac{\partial}{\partial x^\mu}.
\end{equation}
Immediately, we have $\mathrm{CE}(TM)=\Omega^\bullet(M)=\Coo(T[1]M)$.
\end{example}

\begin{example}[Lie algebra]
The NQ-manifold corresponding to an ordinary Lie algebra $\mathfrak{g}$ is the dg-manifold $\mathcal{M}=\mathfrak{g}[1]$, whose differential is given by
\begin{equation}
    Q\;=\; \frac{1}{2}C^i_{\;jk} \xi^j\xi^k\frac{\partial}{\partial \xi^i} \quad\Longrightarrow\quad Q\xi^i \;=\; \frac{1}{2}C^i_{\;jk} \xi^j\xi^k,
\end{equation}
where $\{\xi^i\}$ are the $1$-degree coordinates of $\mathfrak{g}[1]$ and $C^i_{\;jk}$ are the structure constants of the Lie algebra. Clearly, we have $\mathrm{CE}(\mathfrak{g})=\Coo(\mathfrak{g}[1])$.
\end{example}

\begin{remark}[Shifted tangent bundle and tangent $L_\infty$-algebra]
Given the NQ-manifold $\mathcal{M}=\mathfrak{g}[1]$ corresponding to an $L_\infty$-algebra $\mathfrak{g}$, its shifted tangent bundle $T[1]\mathcal{M}=T[1]\mathfrak{g}[1]$ will correspond to the tangent $L_\infty$-algebra $T\mathfrak{g}$. Thus, we will have the identity $\mathrm{W}(\mathfrak{g})=\mathrm{CE}(T\mathfrak{g})=\Coo(T[1]\mathfrak{g}[1])$.
\end{remark}

\subsection{Lie integration}

\begin{definition}[Lie integration]
Given a $L_\infty$-algebroid $\mathfrak{a}$, we define its \textit{Lie integration} as the Lie $\infty$-groupoid $\exp(\mathfrak{a})\in\mathbf{H}$ given by the simplicial presheaf
\begin{equation}
    \exp(\mathfrak{a}) \;:\; (U,[k]) \;\longmapsto \; {\mathbf{dgcAlg}}\Big(\mathrm{CE}(\mathfrak{a}),\, \big(\Omega^\bullet(U\times \Delta^k)^{\mathrm{si}}_{\mathrm{vert}},\di\big)\Big) ,
\end{equation}
where $\Delta^k$ is, for any $k\in\mathbb{N}$, the standard $k$-simplex regarded as a smooth manifold and $(\Omega^\bullet(U\times \Delta^k)^{\mathrm{si}}_{\mathrm{vert}},\di)$ is the dg-algebra of differential forms such that
\begin{enumerate}
    \item they are vertical respect to the trivial bundle $U\times \Delta^k\twoheadrightarrow U$,
    \item they have \textit{sitting instants} along the boundary, i.e. for every $p$-face $\Delta^p$ in $\Delta^k$ with $p<k$ there is an open neighbourhood $U_{\Delta^p}\subset \Delta^k$ of the face $\Delta^p$ such that the forms restricted to $U_{\Delta^p}$ are constant in the directions perpendicular to the $p$-face on its value restricted to that face.
\end{enumerate}
\end{definition}

\noindent For more details, see \cite{Fiorenza:2010mh, Severa2015integration}.

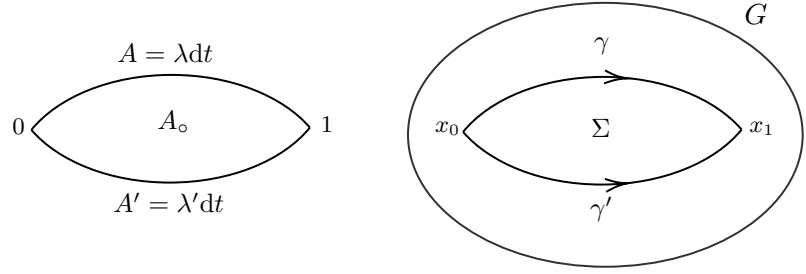
\begin{figure}[h]\begin{center}
\tikzset {_hgbr0lbm8/.code = {\pgfsetadditionalshadetransform{ \pgftransformshift{\pgfpoint{83.16 bp } { -109.56 bp }  }  \pgftransformscale{1.32 }  }}}
\pgfdeclareradialshading{_l0l7ohths}{\pgfpoint{-72bp}{88bp}}{rgb(0bp)=(1,1,1);
rgb(0bp)=(1,1,1);
rgb(25bp)=(0,0,0);
rgb(400bp)=(0,0,0)}
\tikzset{_dr9zug6ru/.code = {\pgfsetadditionalshadetransform{\pgftransformshift{\pgfpoint{83.16 bp } { -109.56 bp }  }  \pgftransformscale{1.32 } }}}
\pgfdeclareradialshading{_4hqeei4k3} { \pgfpoint{-72bp} {88bp}} {color(0bp)=(transparent!0);
color(0bp)=(transparent!0);
color(25bp)=(transparent!64);
color(400bp)=(transparent!64)} 
\pgfdeclarefading{_2k14fzarl}{\tikz \fill[shading=_4hqeei4k3,_dr9zug6ru] (0,0) rectangle (50bp,50bp); } 
\tikzset{every picture/.style={line width=0.75pt}} 
\begin{tikzpicture}[x=0.75pt,y=0.75pt,yscale=-1,xscale=1]
\path  [shading=_l0l7ohths,_hgbr0lbm8,path fading= _2k14fzarl ,fading transform={xshift=2}] (202.5,51.38) .. controls (202.5,14.72) and (246.6,-15) .. (301,-15) .. controls (355.4,-15) and (399.5,14.72) .. (399.5,51.38) .. controls (399.5,88.03) and (355.4,117.75) .. (301,117.75) .. controls (246.6,117.75) and (202.5,88.03) .. (202.5,51.38) -- cycle ; 
 \draw  [color={rgb, 255:red, 0; green, 0; blue, 0 }  ,draw opacity=0.81 ][line width=0.75]  (202.5,51.38) .. controls (202.5,14.72) and (246.6,-15) .. (301,-15) .. controls (355.4,-15) and (399.5,14.72) .. (399.5,51.38) .. controls (399.5,88.03) and (355.4,117.75) .. (301,117.75) .. controls (246.6,117.75) and (202.5,88.03) .. (202.5,51.38) -- cycle ; 
\draw  [draw opacity=0] (14.5,48.92) .. controls (27.71,32.47) and (54.1,21.25) .. (84.5,21.25) .. controls (114.17,21.25) and (140.01,31.94) .. (153.52,47.74) -- (84.5,72.88) -- cycle ; \draw   (14.5,48.92) .. controls (27.71,32.47) and (54.1,21.25) .. (84.5,21.25) .. controls (114.17,21.25) and (140.01,31.94) .. (153.52,47.74) ;
\draw  [draw opacity=0] (153.52,47.74) .. controls (140.31,64.19) and (113.92,75.41) .. (83.52,75.41) .. controls (53.85,75.41) and (28.01,64.73) .. (14.5,48.92) -- (83.52,23.79) -- cycle ; \draw   (153.52,47.74) .. controls (140.31,64.19) and (113.92,75.41) .. (83.52,75.41) .. controls (53.85,75.41) and (28.01,64.73) .. (14.5,48.92) ;
\draw  [draw opacity=0] (230,49.92) .. controls (243.21,33.47) and (269.6,22.25) .. (300,22.25) .. controls (329.67,22.25) and (355.51,32.94) .. (369.02,48.74) -- (300,73.88) -- cycle ; \draw   (230,49.92) .. controls (243.21,33.47) and (269.6,22.25) .. (300,22.25) .. controls (329.67,22.25) and (355.51,32.94) .. (369.02,48.74) ;
\draw  [draw opacity=0] (369.02,48.74) .. controls (355.81,65.19) and (329.42,76.41) .. (299.02,76.41) .. controls (269.35,76.41) and (243.51,65.73) .. (230,49.92) -- (299.02,24.79) -- cycle ; \draw   (369.02,48.74) .. controls (355.81,65.19) and (329.42,76.41) .. (299.02,76.41) .. controls (269.35,76.41) and (243.51,65.73) .. (230,49.92) ;
\draw    (308.02,22.79) -- (309.53,22.89) ;
\draw [shift={(311.52,23.04)}, rotate = 184.09] [color={rgb, 255:red, 0; green, 0; blue, 0 }  ][line width=0.75]    (10.93,-3.29) .. controls (6.95,-1.4) and (3.31,-0.3) .. (0,0) .. controls (3.31,0.3) and (6.95,1.4) .. (10.93,3.29)   ;
\draw    (306.52,76.29) -- (310.01,75.95) ;
\draw [shift={(312,75.75)}, rotate = 534.4] [color={rgb, 255:red, 0; green, 0; blue, 0 }  ][line width=0.75]    (10.93,-3.29) .. controls (6.95,-1.4) and (3.31,-0.3) .. (0,0) .. controls (3.31,0.3) and (6.95,1.4) .. (10.93,3.29)   ;
\draw (3.5,41.9) node [anchor=north west][inner sep=0.75pt]  [font=\footnotesize]  {$0$};
\draw (157.5,40.9) node [anchor=north west][inner sep=0.75pt]  [font=\footnotesize]  {$1$};
\draw (56,3.9) node [anchor=north west][inner sep=0.75pt]  [font=\small]  {$A=\lambda \mathrm{d} t$};
\draw (54,78.9) node [anchor=north west][inner sep=0.75pt]  [font=\small]  {$A'=\lambda '\mathrm{d} t$};
\draw (76,37.9) node [anchor=north west][inner sep=0.75pt]  [font=\small]  {$A_{\circ }$};
\draw (214.5,43.4) node [anchor=north west][inner sep=0.75pt]  [font=\footnotesize]  {$x_{0}$};
\draw (371.5,43.4) node [anchor=north west][inner sep=0.75pt]  [font=\footnotesize]  {$x_{1}$};
\draw (294,0.9) node [anchor=north west][inner sep=0.75pt]  [font=\small]  {$\gamma $};
\draw (292,80.9) node [anchor=north west][inner sep=0.75pt]  [font=\small]  {$\gamma '$};
\draw (292.5,41.9) node [anchor=north west][inner sep=0.75pt]  [font=\small]  {$\Sigma $};
\draw (369,-15.6) node [anchor=north west][inner sep=0.75pt]    {$G$};
\end{tikzpicture}
\caption{Lie-integration of a Lie algebra.}\label{fig:integration}
\end{center}\end{figure}

\begin{example}[Moduli stack $\mathbf{B}G$]
Let $G$ be an ordinary Lie group, whose Lie algebra is $\mathfrak{g}$. Then, we have the Lie integration $\exp(\mathfrak{g}) \cong \mathbf{B}G$. Let us apply the definition:
\begin{itemize}
    \item the set of objects is trivial, i.e. $\{0\}$,
    \item $1$-morphisms are $\{A\in\Omega^1_{\mathrm{flat}}(U\times[0,1], \mathfrak{g})^{\mathrm{si}}_{\mathrm{vert}}\}$, i.e. vertical flat $\mathfrak{g}$-valued $1$-forms $A=\lambda\di t$ on the interval, where $t\in[0,1]$,
    \item $2$-morphisms are $\{A_{\circ}\in\Omega^1_{\mathrm{flat}}(U\times D^2,\mathfrak{g})^{\mathrm{si}}_{\mathrm{vert}}\}$, i.e. vertical flat $\mathfrak{g}$-valued $1$-forms $A_\circ$ on the disk $D^2$, which interpolates between two 1-morphisms $A=\lambda\di t$ and $A'=\lambda'\di t$, which are defined on the two semicircles (see figure \ref{fig:integration}).
\end{itemize}
Now, $1$-morpshisms can be interpreted as pullbacks $A=\gamma^\ast\omega$ and $A'=\gamma^{\prime\ast}\omega$, where $\gamma,\gamma':[0,1]\rightarrow G$ are paths on the Lie group $G$, with $\gamma(0)=\gamma'(0)$ and $\gamma(1)=\gamma'(1)$, and the $1$-form $\omega\in\Omega^1(G,\mathfrak{g})$ is the Maurer-Cartan form. Similarly, a $2$-morphism can be regarded as a homotopy $\Sigma:\gamma\Rightarrow\gamma'$ between two paths on $G$. This is equivalent to $\mathbf{B}G$.
\end{example}

\begin{remark}[$\infty$-Lie theory]
If we consider Lie integration
\begin{equation}
    \exp:\,\mathbf{L_\infty Algbd}\;\longhookrightarrow\; \mathbf{H}
\end{equation}
and its natural inverse, Lie differentiation,
\begin{equation}
    \mathrm{Lie}:\,\mathbf{H}\;\longtwoheadrightarrow\; \mathbf{L_\infty Algbd}
\end{equation}
we have all the ingredients which make up $\infty$-Lie theory.
The Lie differentiation of Lie $\infty$-groups was introduced in \cite{Severa2006}.
\end{remark}

\section{Principal $\infty$-bundles}

In this section we will give a simple introduction to the theory of principal $\infty$-bundles developed by \cite{Principal1,Principal2}. Moreover, from the general theory, we will recover the local differential data of abelian bundle gerbes as presented by \cite{Hit99, GERBE03}.\vspace{0.2cm}

\noindent In ordinary differential geometry a principal $G$-bundle on a manifold $M$ is defined by an element of the first non-abelian $G$-cohomology group $H^1(M,G) \,\cong\, G\mathrm{Bund}(M)_{/\cong}$. These are equivalence classes $\big[f_{(\alpha\beta)}\big]$ where the representatives are given by \v{C}ech $G$-cocycles $f_{(\alpha\beta)}\in\Coo(U_\alpha\cap U_\beta,G)$ on $M$ and the equivalence relation is given by \v{C}ech coboundaries $\eta_{(\alpha)}\in\Coo(U_\alpha,G)$ by $f_{(\alpha\beta)}\,\sim\,\eta_{(\alpha)} f_{(\alpha\beta)}\eta_{(\beta)}^{-1}$. We would like to refine this formalism to a stack description, where we consider $G$-bundles without slashing out gauge transformations.

\begin{remark}[Ordinary principal bundle]\label{modulirem}
For a given manifold $M$ and ordinary Lie group $G$, the groupoid $\mathbf{H}(M,\mathbf{B}G)$ has as objects all the nonabelian \v{C}ech $G$-cocycles $f_{(\alpha\beta)}$ on $M$ and as morphisms all the couboundaries $f_{(\alpha\beta)}\mapsto\eta_{(\alpha)} f_{(\alpha\beta)}\eta_\beta^{-1}$ between them. Schematically, we have:
\begin{equation}
    \mathbf{H}(M,\mathbf{B}G)\,\cong\,\left\{\begin{tikzcd}[row sep=scriptsize, column sep=12ex]
    M \arrow[r, bend left=50, ""{name=U, below}, "f_{(\alpha\beta)}"]
    \arrow[r, bend right=50, "\eta_{(\alpha)} f_{(\alpha\beta)}\eta_{(\beta)}^{-1}"', ""{name=D}]
    & \mathbf{B}G
    \arrow[Rightarrow, from=U, to=D, "\eta_{(\alpha)}"]
\end{tikzcd} \right\}
\end{equation}
In geometric terms the objects are all the principal $G$-bundles over $M$ and the morphisms are all the isomorphisms (i.e. gauge transformations) between them. Thus we will operatively define a principal $G$-bundle as just an object of groupoid $\mathbf{H}(M,\mathbf{B}G)$.
\end{remark}

\noindent To recover the topological picture we only need to take the set of connected components of the groupoid of ordinary principal bundles:
\begin{equation}\label{eq:pathconnected}
    H^1(M,G) \,=\, \pi_0\mathbf{H}(M,\mathbf{B}G).
\end{equation}

\noindent To any cocycle $ M \rightarrow \mathbf{B}G$ is canonically associated a principal bundle $\pi:P\twoheadrightarrow M$ given by
\begin{equation}\label{eq:firstbundle}
    \begin{tikzcd}[row sep=7.5ex, column sep=8ex]
    P \arrow[d, "\mathrm{hofib}(f)"', two heads]\arrow[r] & \ast \arrow[d] \\
    M \arrow[r, "f"] & \mathbf{B}G
\end{tikzcd}
\end{equation}
where we called $\pi=:\mathrm{hofib}(f)$ is the fibre projection of the bundle. \vspace{0.25cm}

\noindent The fundamental idea for defining principal $\infty$-bundles is letting the formalism \eqref{modulirem} and \eqref{eq:firstbundle} work for Lie $\infty$-groups too. 

\begin{definition}[Principal $\infty$-bundles]
A \textit{principal $\infty$-bundle} $P\twoheadrightarrow M$ is a fibration
\begin{equation}
    \begin{tikzcd}[row sep=7.5ex, column sep=8ex]
    P \arrow[d, "\mathrm{hofib}(f)"', two heads]\arrow[r] & \ast \arrow[d] \\
    M \arrow[r, "f"] & \mathbf{B}G
\end{tikzcd}
\end{equation}
corresponding to a cocycle $f\in\mathbf{H}(M,\mathbf{B}G)$, where $G\in\mathrm{Grp}(\mathbf{H})$ is a Lie $\infty$-group in $\mathbf{H}$.
Thus, we can define the $\infty$-groupoid of principal $G$-bundles on $M$ by
\begin{equation}\label{eq:groupoidofprinc}
    G\mathrm{Bund}(M) \;:=\; \mathbf{H}(M,\mathbf{B}G).
\end{equation}
\end{definition}

\begin{definition}[Bundle $n$-gerbe]
A \textit{bundle $n$-gerbe} $\mathscr{G}\twoheadrightarrow M$ is defined as a principal $\mathbf{B}^nU(1)$-bundle on a base manifold $M$. 
The $\infty$-groupoid of bundle $n$-gerbes is, thus, 
\begin{equation}
    n\text{-}\mathrm{Gerb}(M) \;:=\; \mathbf{H}(M,\mathbf{B}^{n+1}G).
\end{equation}
\end{definition}

\begin{remark}[Cohomology and bundle $n$-gerbes]
Notice that bundle $n$-gerbes on a base manifold $M$ are topologically classified exactly by the $n$-th cohomology group of $M$, i.e.
\begin{equation}
    H^{n+2}(M,\mathbb{Z}) \;\cong\; H^{n+1}(M,U(1)) \;\cong\; \pi_0\mathbf{H}\big(M,\mathbf{B}^{n+1}U(1)\big).
\end{equation}
\end{remark}

\noindent In all the rest of this section we will explore several properties of the bundle gerbes.

\begin{remark}[Principal action]
Every principal $\infty$-bundle $P$ comes naturally equipped with a \textit{principal action}
\begin{equation}
    \rho:\;G\times P\;\longrightarrow \;P,
\end{equation}
where $G$ is the structure $\infty$-group of $P$.
\end{remark}

\begin{definition}[Sections of an $\infty$-bundle]\label{def:secgroupoid}
Given any bundle $\pi:P\twoheadrightarrow M$ with $P,M\in\mathbf{H}$, according to \cite{Principal1} we can define its $\infty$-groupoid of its \textit{sections} on $M$ by 
\begin{equation}
    \Gamma(M,P) \;:=\; \mathbf{H}_{/M}(\mathrm{id}_M,\pi)
\end{equation}
where $\mathbf{H}_{/M}(-,-)$ is the internal hom $\infty$-groupoid (definition \ref{def:inthom}) of the slice $\infty$-category $\mathbf{H}_{/M}$ (definition \ref{def:slice}).
\end{definition}

\subsection{Action $\infty$-groupoid}

\begin{definition}[Action $\infty$-groupoid]\label{def:actgroupoid}
Given a Lie group $G\in\mathrm{Grp}(\mathbf{H})$ and a smooth stack $\mathscr{V}\in\mathbf{H}$, an \textit{action $\infty$-groupoid} (also known as \textit{quotient stack}) $\mathscr{V}/\!/_{\!\rho} G\in\mathbf{H}$ is defined by a principal bundle of the following form:
\begin{equation}
    \begin{tikzcd}[row sep=8.5ex, column sep=8ex]
    \mathscr{V} \arrow[d, "\mathrm{hofib}(f)"', two heads]\arrow[r] & \ast \arrow[d] \\
    \mathscr{V}/\!/_{\!\rho} G \arrow[r, "f"] & \mathbf{B}G,
\end{tikzcd}
\end{equation}
where we called $\rho:G\times \mathscr{V}\rightarrow \mathscr{V}$ the principal action. 
\end{definition}

\noindent See \cite{Sharpe:2001bs} for more physical applications.

\begin{example}[Universal principal $\infty$-bundle]
We define the \textit{universal principal $\infty$-bundle} $\mathbf{E}G$ over a Lie $\infty$-group $\mathbf{B}G$ by the following action $\infty$-groupoid
\begin{equation}
    \mathbf{E}G \;:=\; G/\!/_{\!\rho}G,
\end{equation}
where $\rho$ is the natural action of $G$ on itself.
\end{example}

\begin{remark}[$\infty$-group of inner automorphisms]\label{rem:inng}
The universal principal $\infty$-bundle $\mathbf{E}G$ has a natural Lie $\infty$-group structure. As a Lie $\infty$-group, the universal principal $\infty$-bundle is also known as $\mathrm{Inn}(G)$, i.e. as $\infty$\textit{-group of inner automorphisms} of $G$.
\end{remark}

\begin{remark}[Borel construction]
Given an action $\infty$-groupoid $M/\!/_{\!\rho} G\in\mathbf{H}$ where $M$ is a smooth manifold, we have the equivalence $M/\!/_{\!\rho} G \cong \mathbf{E}G\times_G M$ and, thus, the geometric realisation
\begin{equation}
    \big|M/\!/_{\!\rho} G\big| \;=\; EG\times_G M,
\end{equation}
where $EG$ is the (topological) universal principal $G$-bundle and where the fiber product identifies couples $(a\Psi,x)\sim(a,\Psi x)$ where $a\in EG,\; x\in M$ and $\Psi\in G$.
\end{remark}

\noindent Geometrically $EG\times_G M$ is the bundle over the classifying space $BG$ associated with the universal bundle $EG$ and with fibre $M$.

\begin{remark}[Equivariant cohomology]
Usually, the equivariant cohomology of a $G$-space $M$ is defined as the ordinary cohomology of the space $EG\times_G M$, i.e.
\begin{equation}\label{equiv}
H_G^n(M,\mathbb{Z}) \;:=\; H^n(EG\times_G M,\mathbb{Z}).
\end{equation}
Notice that this can be equivalently given by
\begin{equation}
H_G^n(M,\mathbb{Z}) \;\cong\; H_G^{n-1}(M,U(1)) \;\cong\; \pi_0\mathbf{H}\Big(M/\!/_{\!\rho} G,\,\,\mathbf{B}^{n-1}U(1)\Big),
\end{equation}
i.e. by a bundle $(n-2)$-gerbe on the action $\infty$-groupoid $M/\!/_{\!\rho} G$.
\end{remark}

\subsection{Bundle gerbes}

he abelian bundle gerbe is a categorification of the principal $U(1)$-bundle introduced by \cite{Murray, Murray2}. More recently, in \cite{Principal1}, the bundle gerbe has been reformalised as a special case of principal $\infty$-bundle, where the structure Lie $2$-group is $G=\mathbf{B}U(1)$, i.e. the circle $2$-group. 
For an introductory self-contained review, see \cite{Bunk:2021quu}.

\begin{example}[Bundle gerbe]
An abelian bundle gerbe is a principal $\mathbf{B}U(1)$-bundle (i.e. a circle $2$-bundle).
\end{example}

\begin{remark}[Dixmier-Douady class]
By taking the group of path-connected components of the groupoid $\mathbf{H}(M,\mathbf{B}^2U(1))$ of the abelian bundle gerbes we obtain the $3$rd cohomology group
\begin{equation}
    \pi_0\mathbf{H}(M,\mathbf{B}^2U(1)) \;=\; H^2(M,U(1)) \;\cong\; H^3(M,\mathbb{Z}).
\end{equation}
Hence, bundle gerbes $P\rightarrow M$ over a base manifold $M$ are topologically classified by their Dixmier-Douady class, i.e. by an element $\mathrm{dd}(P)\in H^3(M,\mathbb{Z})$ of the third integer cohomology group of the base manifold. This is totally analogous to how the first Chern class $\mathrm{c}_1(P)\in H^2(M,\mathbb{Z})$ classifies ordinary circle bundles $P\rightarrow M$. In general we have a sequence of circle $n$-bundles:
\begin{equation*}
    H^1(M,\mathbb{Z})\cong \Coo(M,S^1), \;\; H^2(M,\mathbb{Z})\cong S^1\mathrm{Bund}(M)_{/\sim}, \;\; H^3(M,\mathbb{Z})\cong \mathrm{Gerb}(M)_{/\sim}, \;\;\dots
\end{equation*}
where $S^1\mathrm{Bund}(M)_{/\sim}$ and $\mathrm{Gerb}(M)_{/\sim}$ are respectively the group of isomorphism classes of circle bundles and abelian bundle gerbes over the base manifold $M$. Therefore, in this context, a global map in $\Coo(M,S^1)$ can be seen as a circle $0$-bundle.
\end{remark}

\begin{remark}[Bundle gerbe in \v{C}ech picture]
An object of $\mathbf{H}(M,\mathbf{B}^2U(1))$ is given in \v{C}ech data for a good cover $\mathcal{U}=\{U_\alpha\}$ of $M$ by a collection $(G_{(\alpha\beta\gamma)})$ of local scalars on $U_\alpha\cap U_\beta\cap U_\gamma$ satisfying
\begin{equation}
    \begin{aligned}
        G_{(\alpha\beta\gamma)}-G_{(\beta\gamma\delta)}+G_{(\gamma\delta\alpha)}-G_{(\delta\alpha\beta)}\;\in\; 2\pi\mathbb{Z},
    \end{aligned}
\end{equation}
i.e. an abelian bundle gerbe in \v{C}ech data. The $1$-morphisms between these objects are \v{C}ech coboundaries (in physical words the gauge transformations of the bundle gerbe) given by collections $(\eta_{(\alpha\beta)})$ of local scalars on overlaps $U_\alpha\cap U_\beta$ so that
\begin{equation}
    \begin{aligned}
        G_{(\alpha\beta\gamma)} \;\mapsto\; G_{(\alpha\beta\gamma)} + \eta_{(\alpha\beta)}+\eta_{(\beta\gamma)}+\eta_{(\gamma\alpha)}
    \end{aligned}
\end{equation}
The $2$-morphisms between $1$-morphisms (in physical words the gauge-of-gauge transformations of the bundle gerbe) are given by collections $(\epsilon_{(\alpha)})$ of local scalars on each $U_\alpha$ so that
\begin{equation}
    \begin{aligned}
        \eta_{(\alpha\beta)} \;\Mapsto\; \eta_{(\alpha\beta)}+\epsilon_{(\alpha)}-\epsilon_{(\beta)} .
    \end{aligned}
\end{equation}
In terms of diagrams we can write this $2$-groupoid of abelian bundle gerbes as follows:
\begin{equation}
    \mathbf{H}\big(M,\mathbf{B}^2U(1)\big) \,\cong\, \left\{\; \begin{tikzcd}[row sep=scriptsize, column sep=26ex]
    \;\;\; M \arrow[r, bend left=60, ""{name=U, below}, "(G_{(\alpha\beta\gamma)})"]
    \arrow[r, bend right=60, "(G'_{(\alpha\beta\gamma)})"', ""{name=D}]
    & \qquad\,\mathbf{B}^2U(1)
    \arrow[Rightarrow, from=U, to=D, bend left=55, "(\eta'_{(\alpha\beta)})", ""{name=R, below}] \arrow[Rightarrow, from=U, to=D, bend right=55, "(\eta_{(\alpha\beta)})"', ""{name=L}] \tarrow[from=L, to=R, end anchor={[yshift=0.6ex]}]{r} \arrow["(\epsilon_{(\alpha)})", phantom, from=L, to=R, end anchor={[yshift=0.6ex]}, bend left=34]
\end{tikzcd}\right\}
\end{equation}
\end{remark}

\begin{remark}[Bundle gerbe in Murray picture]
There is an alternative but equivalent way to geometrically describe a bundle gerbe: the \textit{Murray} description by \cite{Murray, Murray2, Murray3}.
A bundle gerbe will be given by a circle bundle $P_{\alpha\beta}\in\mathbf{H}(U_\alpha\cap U_\beta,\,\mathbf{B}U(1))$ on each overlap of patches and an isomorphism between each tensor product $P_{\alpha\beta}\otimes P_{\beta\gamma}$ and $P_{\alpha\gamma}$ on every three-fold overlap of patches. The latter is a gauge transformation $G_{(\alpha\beta\gamma)}\in\Coo(U_\alpha\cap U_\beta\cap U_\gamma)$, so that
\begin{equation}
    P_{\alpha\beta}\otimes P_{\beta\gamma} \xrightarrow[\;G_{(\alpha\beta\gamma)}\;]{\cong} P_{\alpha\gamma}
\end{equation}
and which satisfies the cocycle condition on four-fold overlaps of patches. This notation is reminiscent of the transition functions $(G_{(\alpha\beta)})$ of an ordinary circle bundle $P\rightarrow M$, which are indeed bundle $0$-gerbes $G_{(\alpha\beta)}\in\Coo(U_\alpha\cap U_\beta,\,U(1))$ and which satisfy exactly $G_{(\alpha\beta)}\cdot G_{(\beta\gamma)} = G_{\alpha\gamma}$.
\end{remark}

\begin{remark}[Bundle gerbe in local data]
Let $\mathcal{U}:=\{U_\alpha\}$ be any good cover for the base manifold $M$. The \v{C}ech groupoid $\check{C}(\mathcal{U})$ is defined as the $\infty$-groupoid corresponding to the following simplicial object
\begin{equation}\label{eq:cechgroupoidsimplicial}
    \begin{tikzcd}[row sep=scriptsize, column sep=6ex] \cdots\arrow[r, yshift=1.8ex]\arrow[r, yshift=0.6ex]\arrow[r, yshift=-0.6ex]\arrow[r, yshift=-1.8ex] & \bigsqcup_{\alpha\beta\gamma}U_{\alpha}\cap U_\beta\cap U_\gamma\arrow[r, yshift=1.4ex]\arrow[r]\arrow[r, yshift=-1.4ex]& \bigsqcup_{\alpha\beta}U_{\alpha}\cap U_\beta  \arrow[r, yshift=0.7ex] \arrow[r, yshift=-0.7ex] & \; \bigsqcup_{\alpha} U_{\alpha} .
    \end{tikzcd}
\end{equation}
Now, by using the natural equivalence between the \v{C}ech groupoid $\check{C}(\mathcal{U})$ and the manifold $M$ in the $(\infty,1)$-category of stacks, we can express the map between $M$ and the moduli stack $\mathbf{B}^2U(1)$ as a functor of the form
\begin{equation}
\begin{tikzcd}[row sep=scriptsize, column sep=8ex] 
    M \;\cong\; \check{C}(\mathcal{U}) \arrow[r, "f"] & \mathbf{B}^2U(1).
\end{tikzcd}
\end{equation}
By using the definition of the \v{C}ech groupoid, such a map can be presented as a collection of cocycles $\bigsqcup_{\alpha\beta}U_{\alpha}\cap U_\beta\rightarrow \mathbf{B}U(1)$ which are glued by isomorphisms on three-fold overlaps of patches $\bigsqcup_{\alpha\beta\gamma}U_{\alpha}\cap U_\beta\cap U_\gamma$. Since, as we have seen, any map $U\rightarrow\mathbf{B}U(1)$ from an open set $U$ is equivalently a $U(1)$-bundle $P\twoheadrightarrow U$, we obtain the following diagram:
\begin{equation}
    \begin{tikzcd}[row sep={11ex,between origins}, column sep={6ex}]
    \big\{\mu_{(\alpha\beta\gamma)}:P_{\alpha\beta}\otimes P_{\beta\gamma}\xrightarrow{\;\cong\;}P_{\alpha\gamma}\big\} \arrow[d, two heads] & \bigsqcup_{\alpha\beta}P_{\alpha\beta} \arrow[d, two heads] & \\
    \bigsqcup_{\alpha\beta\gamma}U_{\alpha}\cap U_\beta\cap U_\gamma \arrow[r, yshift=1.4ex, two heads]\arrow[r, two heads]\arrow[r, yshift=-1.4ex, two heads] & \bigsqcup_{\alpha\beta}U_{\alpha}\cap U_\beta  \arrow[r, yshift=0.7ex, two heads] \arrow[r, yshift=-0.7ex, two heads] & \; \bigsqcup_{\alpha} U_{\alpha} \arrow[d, two heads] \\
    & & M
    \end{tikzcd}
\end{equation}
More in detail, we have a collection of circle bundles $\{P_{\alpha\beta}\twoheadrightarrow U_\alpha\cap U_\beta\}$ on each overlap of patches $U_\alpha\cap U_\beta\subset M$ such that:
\begin{itemize}
    \item there exists a bundle isomorphism $P_{\alpha\beta}\cong P^{-1}_{\beta\alpha}$ on any two-fold overlap of patches $U_\alpha\cap U_\beta$,
    \item there exists a bundle isomorphism $\mu_{(\alpha\beta\gamma)}:P_{\alpha\beta}\otimes P_{\beta\gamma}\xrightarrow{\;\cong\;} P_{\alpha\gamma}$ on any three-fold overlap of patches $U_\alpha\cap U_\beta\cap U_\gamma$,
    \item this isomorphism satisfies $\mu_{(\alpha\beta\gamma)}\circ\mu_{(\beta\gamma\delta)}^{-1}\circ\mu_{(\gamma\delta\alpha)}^{-1}\circ\mu_{(\delta\alpha\beta)}= 1$ on any four-fold overlaps of patches $U_\alpha\cap U_\beta\cap U_\gamma \cap U_\delta$.
\end{itemize}
We, thus, recovered the Murray formulation \cite{Murray, Murray2, Murray3} of the bundle gerbe $\Pi:\mathscr{G}\longtwoheadrightarrow M$. 
\end{remark}

\subsection{Principal $\infty$-bundles with connection}\label{subsectionconnection}

In this subsection we want to introduce the moduli stack of principal bundle with connection $\mathbf{B}G_{\mathrm{conn}}$, which refines the moduli stack of principal bundles $\mathbf{B}G$. We will have the following diagram:
\begin{equation}
    \begin{tikzcd}[row sep={11ex,between origins}, column sep={6ex}]
    & \mathbf{B}G_{\mathrm{conn}} \arrow[d, two heads, "\frgt"] \\
    M \arrow[r, "f"]\arrow[ur, "{(A,f)}"] & \mathbf{B}G.
    \end{tikzcd}
\end{equation}
Just like a cocycle $f:M\rightarrow \mathbf{B}G$ encodes the global geometric data of a principal bundle, a cocycle $(A,f):M\rightarrow \mathbf{B}G_{\mathrm{conn}}$ will encode both the global geometric data of a principal bundle and the global differential data of a principal connection. \vspace{0.25cm}

\noindent Before providing a full definition for the stack $\mathbf{B}G_\mathrm{conn}$ for any given Lie $\infty$-group, let us first construct a simple example.

\begin{example}[Ordinary principal $G$-bundle]
Let $G$ be an ordinary Lie group and $M$ a smooth manifold. We can explicitly construct the stack $\mathbf{B}G_\mathrm{conn}\in\mathbf{H}$ such that a cocycle $\big(A_{(\alpha)},f_{(\alpha\beta)}\big)\in\mathbf{H}(M,\mathbf{B}G_{\mathrm{conn}})$ encodes the global differential data of a principal $G$-bundle with connection on $M$ as follows: $A_{(\alpha)}\in\Omega(U_\alpha,\mathfrak{g})$ is a local $1$-form, which is globally patched by
\begin{equation}
    A_{(\beta)} \;=\; f_{(\beta\alpha)}^{-1}\big(A_{(\alpha)}+\di\big)f_{(\beta\alpha)},
\end{equation}
and $f_{(\alpha\beta)}:M\longrightarrow \mathbf{B}G$ is the \v{C}ech cocycle of a principal $G$-bundle, i.e. satisfying
\begin{equation}
    f_{(\alpha\beta)}f_{(\beta\gamma)} \,=\, f_{(\alpha\gamma)}.
\end{equation}
A coboundary $\big(A_{(\alpha)},f_{(\alpha\beta)}\big) \Rightarrow \big(A_{(\alpha)}',f'_{(\alpha\beta)}\big)$ is given by a local scalar $\eta_{(\alpha)}\in\Coo(U_\alpha,G)$ such that
\begin{equation}
    \begin{aligned}
        A_{(\alpha)}' \;&=\; \eta_{(\alpha)}^{-1}\big(A_{(\alpha)}+\di\big)\eta_{(\alpha)},\\
        f'_{(\alpha\beta)} \;&=\; \eta_{(\beta)}^{-1}f_{(\alpha\beta)}\eta_{(\alpha)}.
    \end{aligned}
\end{equation}
\end{example}

\begin{remark}[Underlying principal $G$-bundle]
In general, there is a forgetful functor
\begin{equation}
    \mathbf{B}G_{\mathrm{conn}}\; \xrightarrow{\quad\frgt\quad}\; \mathbf{B}G,
\end{equation}
which forgets the connection of the $G$-bundles.
Thus, it is important to remark that a cocycle $M\rightarrow \mathbf{B}G_{\mathrm{conn}}$ does not contain only local connection data, but it remembers the underlying bundle structure $M\rightarrow \mathbf{B}G$.
For example, if $G$ is an ordinary Lie group as in the example above, then cocycles are mapped as
\begin{equation}
   \mathbf{H}(M,\mathbf{B}G_{\mathrm{conn}}) \,\ni\;\, \big(A_{(\alpha)},f_{(\alpha\beta)}\big) \;\xmapsto{\;\;\;\frgt\;\;\;}\; (f_{(\alpha\beta)}) \;\,\in\, \mathbf{H}(M,\mathbf{B}G),
\end{equation}
so that the functor forgets the connection data, but retains the global geometric data.
\end{remark}

\noindent Once we have defined the stack $\mathbf{B}G_{\mathrm{conn}}$, we immediately have the following definition.

\begin{definition}[$\infty$-groupoid of $G$-bundles with connection]
Given a smooth manifold $M$ and a Lie $\infty$-group $G$, the $\infty$\textit{-groupoid of $G$-bundles with connection} on $M$ is defined as follows:
\begin{equation}
    G\mathrm{Bund}_\mathrm{conn}(M) \;:=\; \mathbf{H}(M,\mathbf{B}G_{\mathrm{conn}}).
\end{equation}
\end{definition}

\begin{remark}[Differential cohomology]
Notice that we have the isomorphism
\begin{equation}
    \pi_0\mathbf{H}(M,\mathbf{B}G_{\mathrm{conn}}) \;\cong\; \widehat{H}^1(M,G),
\end{equation}
where $\widehat{H}^\bullet(-,G)$ is the differential cohomology.
\end{remark}

\noindent At this point, we must discuss how it is possible to construct, in general, the stack $\mathbf{B}G_{\mathrm{conn}}$ for a given general Lie $\infty$-group $G$. This issue was discussed e.g. in \cite{schreiber2013connections, 2017arXiv170408542W}.
Such a construction will have to generalise ordinary parallel transport to principal $\infty$-bundles.
Na\"{i}vely, we could expect to define our wanted stack by $\mathbf{B}G_{\mathrm{conn}}^{ff}: U \, \longmapsto \, \mathbf{H}\big(\mathscr{P}(U),\,\mathbf{B}G\big)$, where $\mathscr{P}(U)$ is the path $\infty$-groupoid of the smooth manifold $U$. 
However, this definition works only for ordinary Lie groups. For a general Lie $\infty$-group $G$, the stack such a definition makes the cocycles valued in $\mathbf{B}G_{\mathrm{conn}}^{ff}$ \textit{fake-flat} connections, i.e. connections whose $2$-form component of the curvature is forced to vanish. This issue was discussed already in \cite{baez2005higher} and more recently in \cite{Saem19x}. We could attempt a definition of the form $\mathbf{B}G_{\mathrm{conn}}^{ff}: U \, \longmapsto \, \mathbf{H}\big(\mathscr{P}(U),\,\mathbf{B}\mathrm{Inn}(G)\big)$, where $\mathrm{Inn}(G)$ is the $\infty$-group of inner automorphisms of $G$ (see remark \ref{rem:inng}), but this still leads to fake-flatness for higher Lie groups.
A proposed solution has been developed by \cite{Saem19x, Kim:2019owc, Borsten:2021ljb}. To understand it, we must introduce the notion of adjusted tangent $L_\infty$-algebras \cite{Borsten:2021ljb} first.

\begin{definition}[Adjusted tangent $L_\infty$-algebra]
Given an $L_\infty$-algebra $\mathfrak{g}$, we can define the \textit{(firmly) adjusted tangent $L_\infty$-algebra} $T_{\mathrm{adj}}\mathfrak{g}$ as deformation of the $L_\infty$-algebra structure of the tangent $L_\infty$-algebra $T\mathfrak{g}$, given as follows.
Let $\{t^i\}$ and $\{r^i\}$ be respectively the generators of $\mathfrak{g}^\ast[2]$ and $\mathfrak{g}^\ast[1]$. We can obtain $\mathrm{CE}(T_{\mathrm{adj}}\mathfrak{g})$ by a coordinate transformation of $\mathrm{CE}(T\mathfrak{g})$ the form
\begin{equation}
\begin{aligned}
    t^i \;&\mapsto\; t^i, \\
    r^i \;&\mapsto\; r^i + p^i_{j_1\cdots j_m k_1 \cdots k_n} r^{j_1}\cdots r^{j_m} t^{k_1}\cdots t^{k_n},
\end{aligned}
\end{equation}
such that the image of the resulting differential $\di_{\mathrm{CE}}$ on generators in $\mathfrak{g}^\ast[2]$ contains no generator in $\mathfrak{g}^\ast[1]$, except for at most one of degree $1$.
\end{definition}

\noindent In analogy with definition \ref{def:weildgalgebra}, we can define the \textit{(firmly) adjusted Weil dg-algebra} \cite{Kim:2019owc} of an $L_\infty$-algebra $\mathfrak{g}$ by $\mathrm{W}_{\mathrm{adj}}(\mathfrak{g}):= \mathrm{CE}(T_{\mathrm{adj}}\mathfrak{g})$. Let us now look at an important example of such dg-algebras.

\begin{example}[Adjusted Weil dg-algebra of the string $2$-algebra $\mathfrak{string}(\mathfrak{g})$]\label{ex:adjweilstring}
The (firmly) adjusted Weil dg-algebra \cite{Kim:2019owc} of the string $2$-algebra $\mathfrak{string}(\mathfrak{g})$ is given by
\begin{equation}
\begin{aligned}
        \di t^a \;&=\; -\frac{1}{2}C^a_{\;\;bc}t^b\wedge t^c + r^a, \\
        \di b \;&=\; - k_{aa'}C^{a'}_{\;\,bc} t^a\wedge t^b\wedge t^c - k_{ab}t^a\wedge r^b + h, \\
        \di r^a \;&=\; -C^a_{\;\;bc}t^b\wedge r^c, \\
        \di h \;&=\; - k_{ab}r^a\wedge r^b 
\end{aligned}
\end{equation}
where $C^a_{\;\;bc}$ are the structure constants of the ordinary Lie algebra $\mathfrak{g}$, where $k_{ab}$ is a Killing form on $\mathfrak{g}$ and where $\{t^a,b,r^a,h\}$ are respectively the $1$-degree generators $\{t^a\}$, the $2$-degree generators $\{b,r^a\}$ and the $3$-degree generator $\{h\}$. It is not hard to check that this dg-algebra can be obtained from the usual Weil dg-algebra in example \ref{ex:weilstring} by the change of coordinates $h \mapsto h' = h + k_{ab}r^a t^b$.
\end{example}

\noindent Now, we can construct the stack $\mathbf{B}G_{\mathrm{conn}}$ by defining an "adjusted" notion of parallel transport which makes use of the idea of adjusted tangent $L_\infty$-algebra.

\begin{definition}[Moduli stack of $G$-bundles with connection]
Given a Lie $\infty$-group $G$, the \textit{moduli stack of $G$-bundles with connection} $\mathbf{B}G_\mathrm{conn}\in\mathbf{H}$ is defined by the functor
\begin{equation}
    \begin{aligned}
        \mathbf{B}G_{\mathrm{conn}}:\, \mathbf{Diff}^\mathrm{op} \; &\longrightarrow \; \mathbf{\infty Grpd} \\
        M \; &\longmapsto \; \mathbf{H}\big(\mathscr{P}(M),\,\mathbf{B}\mathrm{Inn}_{\mathrm{adj}}(G)\big),
    \end{aligned}
\end{equation}
where $\mathscr{P}(M)$ is the path $\infty$-groupoid of the smooth manifold $M$ and $\mathrm{Inn}_{\mathrm{adj}}(G)$ is the \textit{adjusted $\infty$-group of inner automorphisms} of $G$, which can be defined by Lie-integration of the adjusted tangent $L_\infty$-algebra $T_\mathrm{adj}\mathfrak{g}$, where $\mathfrak{g}:=\mathrm{Lie}(G)$. \vspace{0.25cm}

\noindent For more details and discussion, we redirect to \cite{Kim:2019owc, Borsten:2021ljb}.
\end{definition}


\begin{figure}[h]\begin{center}
\tikzset {_rkw01qkrw/.code = {\pgfsetadditionalshadetransform{ \pgftransformshift{\pgfpoint{83.16 bp } { -104.94 bp }  }  \pgftransformscale{1.32 }  }}}
\pgfdeclareradialshading{_oe83dzfyc}{\pgfpoint{-72bp}{88bp}}{rgb(0bp)=(1,1,1);
rgb(0bp)=(1,1,1);
rgb(25bp)=(0.61,0.61,0.61);
rgb(400bp)=(0.61,0.61,0.61)}
\tikzset{every picture/.style={line width=0.75pt}} 
\begin{tikzpicture}[x=0.75pt,y=0.75pt,yscale=-1,xscale=1]
\path  [shading=_oe83dzfyc,_rkw01qkrw] (1.8,75.2) .. controls (1.8,34.66) and (41.02,1.8) .. (89.4,1.8) .. controls (137.78,1.8) and (177,34.66) .. (177,75.2) .. controls (177,115.74) and (137.78,148.6) .. (89.4,148.6) .. controls (41.02,148.6) and (1.8,115.74) .. (1.8,75.2) -- cycle ; 
 \draw  [color={rgb, 255:red, 0; green, 0; blue, 0 }  ,draw opacity=1 ] (1.8,75.2) .. controls (1.8,34.66) and (41.02,1.8) .. (89.4,1.8) .. controls (137.78,1.8) and (177,34.66) .. (177,75.2) .. controls (177,115.74) and (137.78,148.6) .. (89.4,148.6) .. controls (41.02,148.6) and (1.8,115.74) .. (1.8,75.2) -- cycle ; 
\draw    (36.4,62.8) .. controls (70.6,3.4) and (95.4,46.2) .. (139.4,61.4) ;
\draw [shift={(139.4,61.4)}, rotate = 19.06] [color={rgb, 255:red, 0; green, 0; blue, 0 }  ][fill={rgb, 255:red, 0; green, 0; blue, 0 }  ][line width=0.75]      (0, 0) circle [x radius= 1.34, y radius= 1.34]   ;
\draw [shift={(36.4,62.8)}, rotate = 299.93] [color={rgb, 255:red, 0; green, 0; blue, 0 }  ][fill={rgb, 255:red, 0; green, 0; blue, 0 }  ][line width=0.75]      (0, 0) circle [x radius= 1.34, y radius= 1.34]   ;
\draw    (36.4,62.8) .. controls (88.2,144.2) and (127.8,113.4) .. (139.4,61.4) ;
\draw    (84.25,37.83) .. controls (74.41,44.63) and (68.42,54.96) .. (68.42,67.62) .. controls (68.42,78.76) and (73.03,91.64) .. (80.13,100.75)(82.55,35.37) .. controls (71.89,42.73) and (65.42,53.92) .. (65.42,67.62) .. controls (65.42,79.27) and (70.13,92.79) .. (77.72,102.55) ;
\draw [shift={(83.4,107.8)}, rotate = 231.97] [color={rgb, 255:red, 0; green, 0; blue, 0 }  ][line width=0.75]    (10.93,-3.29) .. controls (6.95,-1.4) and (3.31,-0.3) .. (0,0) .. controls (3.31,0.3) and (6.95,1.4) .. (10.93,3.29)   ;
\draw    (93.39,38.89) .. controls (103.84,52.55) and (108.71,64.65) .. (108.71,76.21) .. controls (108.71,81.87) and (107.54,87.41) .. (105.27,92.94) .. controls (102.89,98.76) and (99.28,104.58) .. (98.05,105.83)(91.01,40.71) .. controls (100.95,53.71) and (105.71,65.19) .. (105.71,76.21) .. controls (105.71,81.49) and (104.61,86.65) .. (102.5,91.81) .. controls (100.21,97.38) and (96.73,102.96) .. (95.74,103.9) ;
\draw [shift={(92.2,111)}, rotate = 309.61] [color={rgb, 255:red, 0; green, 0; blue, 0 }  ][line width=0.75]    (10.93,-3.29) .. controls (6.95,-1.4) and (3.31,-0.3) .. (0,0) .. controls (3.31,0.3) and (6.95,1.4) .. (10.93,3.29)   ;
\draw    (72.22,71.5) -- (98.22,71.9)(72.18,74.5) -- (98.18,74.9) ;
\draw    (72.62,68.5) -- (98.62,68.9)(72.58,71.5) -- (98.58,71.9) ;
\draw    (91.4,71.8) -- (103.8,71.63) ;
\draw [shift={(105.8,71.6)}, rotate = 539.2] [color={rgb, 255:red, 0; green, 0; blue, 0 }  ][line width=0.75]    (12.02,-5.39) .. controls (7.65,-2.53) and (3.64,-0.73) .. (0,0) .. controls (3.64,0.73) and (7.65,2.53) .. (12.02,5.39)   ;
\draw (26,48.2) node [anchor=north west][inner sep=0.75pt]  [font=\footnotesize]  {$x$};
\draw (142.8,46.6) node [anchor=north west][inner sep=0.75pt]  [font=\footnotesize]  {$y$};
\draw (-12.4,0.4) node [anchor=north west][inner sep=0.75pt]    {$\mathscr{P}(M)$};
\draw (300,7) node [anchor=north west][inner sep=0.75pt]    {$\mathbf{B}\mathrm{Inn}_{\mathrm{adj}}(G)$};
\draw (200,15) node [anchor=north west][inner sep=0.75pt]    {$\xrightarrow{\quad\mathrm{tra}_A\quad}\quad\begin{tikzcd}[row sep=scriptsize, column sep=22ex]
    \ast \arrow[r, bend left=60, ""{name=U, below}]
    \arrow[r, bend right=60, ""{name=D}]
    & \ast
    \arrow[Rightarrow, from=U, to=D, bend left=55, ""{name=R, below}] \arrow[Rightarrow, from=U, to=D, bend right=55, ""{name=L}] \tarrow[from=L, to=R, end anchor={[yshift=0.4ex]}]{r} \arrow[phantom, from=L, to=R, end anchor={[yshift=0.6ex]}, bend left=34]
\end{tikzcd}$};
\end{tikzpicture} 
\caption[Parallel transport on a principal $G$-bundle]{Parallel transport on a principal $G$-bundle is given by a functor between the path $\infty$-groupoid $\mathscr{P}(M)$ of the smooth manifold $M$ and the stack $\mathbf{B}\mathrm{Inn}_{\mathrm{adj}}(G)$.}
\end{center}\end{figure}

\noindent For an ordinary Lie group $G$, a functor $\mathscr{P}(M)\longrightarrow \mathbf{B}\mathrm{Inn}_{\mathrm{adj}}(G)$ can be equivalently given by a functor $\mathscr{P}(M)\longrightarrow \mathbf{B}G$. For more details see \cite{Kim:2019owc}.

\begin{example}[Ordinary parallel transport]
Given an ordinary Lie group $G$ and a smooth manifold $M$, a functor $\mathrm{tra}_A:\mathscr{P}(M)\longrightarrow \mathbf{B}G$ is called \textit{parallel transport} and it is given by the map
\begin{equation}
    \mathrm{tra}_A :\, \gamma \;\longmapsto\; \mathcal{P}\exp\!\left(\int_\gamma A\right) \,\in G,
\end{equation}
where $\gamma$ is any path on the base manifold $M$ and $A_{(\alpha)}\in\Omega(U_\alpha,\mathfrak{g})$ is a local $1$-form, which is globally patched by
\begin{equation}
    A_{(\beta)} \;=\; f_{(\beta\alpha)}^{-1}\big(A_{(\alpha)}+\di\big)f_{(\beta\alpha)},
\end{equation}
where $f_{(\alpha\beta)}:M\longrightarrow \mathbf{B}G$ is the \v{C}ech cocycle of a principal $G$-bundle, i.e. satisfying
\begin{equation}
    f_{(\alpha\beta)}f_{(\beta\gamma)} \,=\, f_{(\alpha\gamma)}.
\end{equation}
Thus, a map $\mathrm{tra}_A$ is equivalently a cocycle $\big(A_{(\alpha)},f_{(\alpha\beta)}\big)\in\mathbf{H}(M,\mathbf{B}G_{\mathrm{conn}})$ which encodes the global differential data of a principal $G$-bundle with connection.
The functorial nature of the parallel transport is clear from
\begin{equation}
    \mathrm{tra}_A(\gamma')\cdot\mathrm{tra}_A(\gamma) \;=\; \mathrm{tra}_A(\gamma'\circ\gamma),
\end{equation}
where $\circ$ is the composition of paths.
\end{example}

\subsection{Bundle gerbes with connection}

\noindent Let us now construct a bundle gerbe equipped with connection, which globally formalises a Kalb-Ramond field. This is mostly based on \cite{Murray, Murray2, Murray3, DCCTv2}. For bundle gerbes, the construction of the stack $\mathbf{B}^2 U(1)_{\mathrm{conn}}$ is particularly simple.

\begin{definition}[Bundle gerbe with connection]
An abelian \textit{bundle gerbe with connection} is given by a cocycle $M\rightarrow\mathbf{B}^2U(1)_{\mathrm{conn}}$ where we defined the stack $\mathbf{B}^2U(1)_{\mathrm{conn}}$ in example \ref{ex:stacks}.
\end{definition}

\begin{remark}[Bundle gerbe with connection in \v{C}ech-Deligne picture]
An object of the $\infty$-groupoid $\mathbf{H}(M,\mathbf{B}^2U(1)_{\mathrm{conn}})$ is given in \v{C}ech-Deligne data for a good cover $\mathcal{U}=\{U_\alpha\}$ of $M$ by a collection $(B_{(\alpha)},\Lambda_{(\alpha\beta)},G_{(\alpha\beta\gamma)})$ of $2$-forms $B_{(\alpha)}\in\Omega^2(U_\alpha)$, $1$-forms $\Lambda_{(\alpha\beta)}\in\Omega^2(U_\alpha\cap U_\beta)$ and scalars $G_{(\alpha\beta\gamma)}\in\Coo(U_\alpha\cap U_\beta\cap U_\gamma)$, patched by
\begin{equation}
    \begin{aligned}
        B_{(\beta)}-B_{(\alpha)} &= \mathrm{d}\Lambda_{(\alpha\beta)}, \\
        \Lambda_{(\alpha\beta)}+\Lambda_{(\beta\gamma)}+\Lambda_{(\gamma\alpha)} &= \mathrm{d}G_{(\alpha\beta\gamma)} \\
        G_{(\alpha\beta\gamma)}-G_{(\beta\gamma\delta)}+G_{(\gamma\delta\alpha)}-G_{(\delta\alpha\beta)}&\in 2\pi\mathbb{Z}
    \end{aligned}
\end{equation}
i.e. an abelian bundle gerbe with connection in \v{C}ech-Deligne data. The $1$-morphisms between these objects are \v{C}ech-Deligne coboundaries (in physical words the gauge transformations of the bundle gerbe), given by collections $(\eta_{\alpha},\eta_{(\alpha\beta)})$ of local $1$-forms $\eta_{(\alpha)}\in\Omega^1(U_\alpha)$ and local scalars $\eta_{(\alpha\beta)}\in\Coo(U_\alpha\cap U_\beta)$, so that
\begin{equation}\label{eq:coboundaries}
    \begin{aligned}
        B_{(\alpha)} &\mapsto B_{(\alpha)} + \mathrm{d}\eta_{\alpha}, \\
        \Lambda_{(\alpha\beta)} &\mapsto \Lambda_{(\alpha\beta)}+\eta_{\alpha}-\eta_{\beta}+\mathrm{d}\eta_{(\alpha\beta)} \\
        G_{(\alpha\beta\gamma)} &\mapsto G_{(\alpha\beta\gamma)} + \eta_{(\alpha\beta)}+\eta_{(\beta\gamma)}+\eta_{(\gamma\alpha)}
    \end{aligned}
\end{equation}
The $2$-morphisms between $1$-morphisms (in physical words the gauge-of-gauge transformations of the bundle gerbe) are given by collections $(\epsilon_{(\alpha)})$ of local scalars on each $U_\alpha$ so that
\begin{equation}
    \begin{aligned}
        \eta_{(\alpha)} &\Mapsto \eta_{(\alpha)} + \mathrm{d}\epsilon_{(\alpha)}, \\
        \eta_{(\alpha\beta)} &\Mapsto \eta_{(\alpha\beta)}+\epsilon_{(\alpha)}-\epsilon_{(\beta)} .
    \end{aligned}
\end{equation}
In terms of diagrams we can write this $2$-groupoid of abelian bundle gerbes with connection as follows:
\begin{equation}
    \mathbf{H}\big(M,\mathbf{B}^2U(1)_{\mathrm{conn}}\big) \,\cong\, \left\{\; \begin{tikzcd}[row sep=scriptsize, column sep=26ex]
    \, M \arrow[r, bend left=60, ""{name=U, below}, "(B_{(\alpha)}\text{,}\,\Lambda_{(\alpha\beta)}\text{,}\,G_{(\alpha\beta\gamma)})"]
    \arrow[r, bend right=60, "(B'_{(\alpha)}\text{,}\,\Lambda'_{(\alpha\beta)}\text{,}\,G'_{(\alpha\beta\gamma)})"', ""{name=D}]
    & \qquad\quad\,\mathbf{B}^2U(1)_{\mathrm{conn}}
    \arrow[Rightarrow, from=U, to=D, bend left=55, "(\eta'_{(\alpha)}\text{,}\,\eta'_{(\alpha\beta)})", ""{name=R, below}] \arrow[Rightarrow, from=U, to=D, bend right=55, "(\eta_{(\alpha)}\text{,}\,\eta_{(\alpha\beta)})"', ""{name=L}] \tarrow[from=L, to=R, end anchor={[yshift=0.6ex]}]{r} \arrow["(\epsilon_{(\alpha)})", phantom, from=L, to=R, end anchor={[yshift=0.6ex]}, bend left=34]
\end{tikzcd}\right\}
\end{equation}
\end{remark}

\begin{definition}[Flat and trivial bundle gerbe]
A \textit{flat bundle gerbe} is defined as a bundle gerbe $(B_{(\alpha)},\Lambda_{(\alpha\beta)},G_{(\alpha\beta\gamma)})$ with vanishing curvature $\mathrm{d}B_{(\alpha)}=0$. We use the symbol $\flat\mathbf{B}^2U(1)_{\mathrm{conn}}$ for the moduli stack of flat bundle gerbes with connection. A \textit{trivial bundle gerbe} is defined as a bundle gerbe with trivial Dixmier-Douady class.
\end{definition}

\begin{remark}[Flat and trivial bundle gerbe in \v{C}ech-Deligne picture]
Let us express in local data a flat bundle gerbe $(B_{(\alpha)},\Lambda_{(\alpha\beta)},G_{(\alpha\beta\gamma)})\in\mathbf{H}(M,\flat\mathbf{B}^2U(1)_{\mathrm{conn}})$. Since $B_{(\alpha)}$ is closed on each patch $U_\alpha$ we can rewrite
\begin{equation}\label{eq:flatgerbe}
    \begin{aligned}
        B_{(\alpha)} \,=&\; \mathrm{d}\eta_{(\alpha)}, \\
        \Lambda_{(\alpha\beta)} \,=&\; \eta_{(\beta)}-\eta_{(\alpha)} + \mathrm{d}\eta_{(\alpha\beta)}, \\
        G_{(\alpha\beta\gamma)} \,=&\; \eta_{(\alpha\beta)} + \eta_{(\beta\gamma)} + \eta_{(\gamma\alpha)} +c_{(\alpha\beta\gamma)}, \\
        &\; c_{(\alpha\beta\gamma)} - c_{(\beta\gamma\delta)} + c_{(\gamma\delta\alpha)} - c_{(\delta\alpha\beta)} \in 2\pi\mathbb{Z}
    \end{aligned}
\end{equation}
Hence flat bundle gerbes are classified by \textit{holonomy classes} $[c_{(\alpha\beta\gamma)}]\in H^2(M,U(1)_{\mathrm{discr}})$. The \v{C}ech-Deligne local data of a trivial bundle gerbe will be exactly \eqref{eq:flatgerbe}, but with trivial constants $c_{(\alpha\beta\gamma)}=0$.
\end{remark}

\begin{definition}[Flat holonomy class]\label{eq:holonomyclass}
Flat bundle gerbes are classified by elements of the cohomology group $H^2(M,U(1)_{\mathrm{discr}})\cong \mathrm{Hom}\big(H_2(M),U(1)_{\mathrm{discr}}\big)$, where $U(1)_{\mathrm{discr}}$ is the circle equipped with discrete topology. Such class is called \textit{flat holonomy class} of the bundle gerbe.
\end{definition}

\noindent Hence a class $[c_{(\alpha\beta\gamma)}]\in H^2(M,U(1)_{\mathrm{discr}})$ encode the holonomy of the bundle gerbe, meaning that to any surface $[\Sigma]\in H_2(M)$ of the base manifold will be associated an angle $\mathrm{hol}(\Sigma,B_{(\alpha)})\in U(1)$.

\begin{remark}[Flat bundle gerbe has torsion Dixmier-Douady class]
There exists a natural map $H^2(M,U(1)_{\mathrm{discr}})\rightarrow H^2(M,U(1))\cong H^3(M,\mathbb{Z})$ sending a flat bundle gerbe to its Dixmier-Douady class. The Dixmier-Douady class of the flat bundle gerbe has not to be zero, but its image in the de Rham cohomology $H^3(M,\mathbb{Z})\rightarrow H^3(M,\mathbb{R})\cong H^3_{\mathrm{dR}}(M)$ must be zero, since $\mathrm{d}B_{(\alpha)}=0$. This implies that the Dixmier-Douady class is, in general, torsion.
\end{remark}

\begin{remark}[Sections of the bundle gerbe]
We can construct the $2$-groupoid $\Gamma(M,\mathscr{G}):=\mathbf{H}_{/M}(\mathrm{id}_M,\Pi)$ of sections of the bundle gerbe $\mathscr{G}\xrightarrow{\Pi}M$ according to definition \ref{def:secgroupoid}. They will be given by a collection $(\widetilde{x}_{(\alpha)},\phi_{(\alpha\beta)})\in\Gamma(M,\mathcal{M})$ where
\begin{equation}
    \widetilde{x}_{(\alpha)}\in\Omega^1(U_\alpha), \;\quad
    \phi_{(\alpha\beta)}\in\Coo(U_\alpha\cap U_\beta)
\end{equation}
are local $1$-forms and scalars, such that they are patched on two-fold and three-fold overlaps by using $(\Lambda_{(\alpha\beta)},G_{(\alpha\beta\gamma)})$ as transition functions by
\begin{equation}\label{eq:coords}
    \begin{aligned}
        \widetilde{x}_{(\alpha)} - \widetilde{x}_{(\beta)}  \,&=\, -\Lambda_{(\alpha\beta)} + \mathrm{d}\phi_{(\alpha\beta)},\\
        \phi_{(\alpha\beta)}+\phi_{(\beta\gamma)}+\phi_{(\gamma\alpha)} \,&=\, G_{(\alpha\beta\gamma)} \quad\! \mathrm{mod}\,2\pi\mathbb{Z}.
    \end{aligned}
\end{equation}
Gauge transformations between global sections are given by a collection of local functions on each patch $\epsilon_\alpha\in\Coo(U_\alpha)$ so that $(\widetilde{x}_{(\alpha)},\phi_{(\alpha\beta)})\mapsto(\widetilde{x}_{(\alpha)}+\mathrm{d}\epsilon_{(\alpha)},\, \phi_{(\alpha\beta)}+\epsilon_{(\alpha)}-\epsilon_{(\beta)})$.
Global sections $(\widetilde{x}_{(\alpha)},\phi_{(\alpha\beta)})$ and gauge transformations $(\epsilon_{(\alpha)})$ are respectively the objects and the morphisms of the groupoid $\Gamma(M,\mathscr{G})$ of sections of the bundle gerbe.
\end{remark}

\begin{remark}[Sections of the bundle gerbe are twisted circle bundles]
As explained by \cite{Principal1}, sections of a principal $2$-bundle $f:M\rightarrow\mathbf{B}G$ can be interpreted as ordinary principal bundles on $M$ twisted by the cocycle $f$. Coherently with this, in the case of the bundle gerbe we can immediately interpret sections $(\widetilde{x}_{(\alpha)},\phi_{(\alpha\beta)})\in\Gamma(M,\mathscr{G})$, which are patched according to \eqref{eq:coords}, as $U(1)$-bundles with connection on $M$ twisted by the \v{C}ech-Deligne cocycle $(\Lambda_{(\alpha\beta)},G_{(\alpha\beta\gamma)})$.
\end{remark}

\begin{remark}[Principal action on the bundle gerbe in \v{C}ech-Deligne picture]
This principal action  on the bundle gerbe is described in local \v{C}ech-Deligne data on sections $(\widetilde{x}_{(\alpha)},\phi_{(\alpha\beta)})\in\Gamma(M,\mathscr{G})$ as follows:
\begin{itemize}
    \item a circle bundle on $M$, given by the \v{C}ech-Deligne cocycle $(\eta_\alpha,\eta_{(\alpha\beta)})\in\mathbf{H}(M,\BU)$ acting by
    \begin{equation}\label{eqgauge}
        \begin{aligned}
            x_{(\alpha)} \,&\mapsto\, x_{(\alpha)} \\
            \widetilde{x}_{(\alpha)} \,&\mapsto\, \widetilde{x}_{(\alpha)} + \eta_{(\alpha)} \\
            \phi_{(\alpha\beta)} \,&\mapsto\, \phi_{(\alpha\beta)} + \eta_{(\alpha\beta)},
        \end{aligned}
    \end{equation}
    \item gauge transformations $(\eta_{(\alpha)},\eta_{(\alpha\beta)})\Mapsto(\eta'_\alpha,\eta'_{(\alpha\beta)})$ are just gauge transformations between circle bundles and are given in local data by a collection of functions $\epsilon_{(\alpha)}\in\Coo(U_\alpha)$ acting by
    \begin{equation}\label{eqgaugeofgauge}
        \begin{aligned}
            \eta_{(\alpha)} \,&\Mapsto\, \eta_{(\alpha)} + \mathrm{d}\epsilon_{(\alpha)} \\
            \eta_{(\alpha\beta)} \,&\Mapsto\, \eta_{(\alpha\beta)}+\epsilon_{(\alpha)}-\epsilon_{(\beta)}.
        \end{aligned}
    \end{equation}
\end{itemize}
In terms of diagrams we can rewrite the principal action on the bundle gerbe as the groupoid
\begin{equation}
   \mathbf{H}(M,\,\BU) \;\simeq\; \left\{\begin{tikzcd}[row sep=scriptsize, column sep=15ex]
    \mathscr{G} \arrow[r, bend left=50, ""{name=U, below}, "(\eta_\alpha\text{, }\eta_{(\alpha\beta)})"]
    \arrow[r, bend right=50, "(\eta_{(\alpha)} + \mathrm{d}\epsilon_{(\alpha)}\text{, }\eta_{(\alpha\beta)}+\epsilon_{(\alpha)}-\epsilon_{(\beta)})"', ""{name=D}]
    & \mathscr{G}
    \arrow[Rightarrow, from=U, to=D, "(\epsilon_{(\alpha)})"]
\end{tikzcd}\right\}
\end{equation}
\end{remark}

\begin{remark}[Principal action gives global gauge transformations]\label{rem:globgauge}
Let us consider a \v{C}ech-Deligne cocycle  $(B_{(\alpha)},\Lambda_{(\alpha\beta)},G_{(\alpha\beta\gamma)})\in\mathbf{H}(M,\mathbf{B}^2U(1)_{\mathrm{conn}})$. Then the principal action $\rho$ (remark \ref{principalaction}) gives global gauge transformations $(\eta_{(\alpha)},\eta_{(\alpha\beta)})\in\mathbf{H}\big(M,\BU\big)$ of the connection of the bundle gerbe. From the expression of coboundaries \eqref{eq:coboundaries} we find
\begin{equation}
    \begin{aligned}
        B_{(\alpha)} \;&\mapsto\; B_{(\alpha)} + \mathrm{d}\eta_{(\alpha)} \\
        \Lambda_{(\alpha\beta)} \;&\mapsto\; \Lambda_{(\alpha\beta)} + \eta_{(\alpha)}- \eta_{(\beta)} +\mathrm{d}\eta_{(\alpha\beta)} &=&  \;\Lambda_{(\alpha\beta)}\\
        G_{(\alpha\beta\gamma)} \;&\mapsto\; G_{(\alpha\beta\gamma)} + \eta_{(\alpha\beta)}  + \eta_{(\beta\gamma)}  + \eta_{(\gamma\alpha)} &=& \;G_{(\alpha\beta\gamma)}.
    \end{aligned}
\end{equation}
and the \eqref{eqgaugeofgauge} are the gauge transformations of these gauge transformations. The gerbe curvature $H=\mathrm{d}B_{(\alpha)}$ is clearly unaffected.
Transformation \eqref{eqgauge} and \eqref{eqgaugeofgauge} can be also understood as a change of local trivialisation (e.g. like in \cite{Hit99}) for the cocycle $(\Lambda_{(\alpha\beta)},G_{(\alpha\beta\gamma)})\in\mathbf{H}\big(M,\mathbf{B}(\BU)\big)$. Where, for a circle bundle, a change of local trivialisation is a global $U(1)$-valued function, for a bundle gerbe this is a global $U(1)$-bundle.
By using the curvature functor $\mathrm{curv}:\BU\longrightarrow\Omega^2_{\mathrm{cl}}$, which sends a $U(1)$-bundle $(\eta_{(\alpha)},\eta_{(\alpha\beta)})\in\mathbf{H}\big(M,\BU\big)$ to the closed global $2$-form $b\in\Omega^2_{\mathrm{cl}}(M)$ defined by $b|_{U_\alpha} = \di\eta_{(\alpha)}$, we can rewrite this transformation as a global $B$-shift $B_{(\alpha)}\mapsto B_{(\alpha)} + b$.
\end{remark}

\noindent This generalises ordinary Kaluza-Klein, where the principal $U(1)$-action on the circle bundle is given by global shifts in the angular coordinates $\theta_{(\alpha)}\mapsto\theta_{(\alpha)}+\eta_{(\alpha)}$ with $\eta_{(\alpha)}=\eta_{(\beta)}$ and it encodes global gauge transformations $A_{(\alpha)} \mapsto A_{(\alpha)}+\mathrm{d}\eta_{(\alpha)}$ and $G_{(\alpha\beta)}\mapsto G_{(\alpha\beta)} + \eta_{(\alpha)} - \eta_{(\beta)} = G_{(\alpha\beta)} $.

\subsection{Global higher gauge fields and $L_\infty$-algebras}\label{HGFandLinfty}

In this subsection we will discuss the relation between the definition of a higher gauge field as a cocycle $M\rightarrow \mathbf{B}G_{\mathrm{conn}}$ and the one that is more common in the literature, i.e. as a map $T[1]M\rightarrow T[1]\mathfrak{g}[1]$ of NQ-manifolds. In particular we will discuss their global aspects. \vspace{0.35cm}

\noindent In subsection \ref{subsectionconnection} we learnt that a \v{C}ech cocycle $\big(A_{(\alpha)},f_{(\alpha\beta)}\big)$ encoding the global data of a principal $G$-bundle with connection, where $G$ is an ordinary Lie group, can be expressed as a map
\begin{equation}\label{eq:mapconnectionex}
    \big(A_{(\alpha)},\,f_{(\alpha\beta)}\big):\;\mathscr{P}(M) \;\longrightarrow\; \mathbf{B}G.
\end{equation}
The morphisms between such maps are nothing but global gauge transformations, i.e.
\begin{equation}
    \big(A_{(\alpha)},\,f_{(\alpha\beta)}\big) \;\xmapsto{\quad \lambda_{(\alpha)} \quad}\; \Big(\lambda_{(\alpha)}^{-1}(A_{(\alpha)}+\di)\lambda_{(\alpha)},\, \lambda_{(\alpha)}f_{(\alpha\beta)}\lambda_{(\beta)}^{-1}\Big) ,
\end{equation}
where the gauge parameter $\lambda_{(\alpha)}\in\Coo(U_\alpha,G)$ is a local $G$-valued function. Such cocycles and such gauge transformations are exactly the objects and the morphisms of the groupoid $\mathbf{H}(\mathscr{P}(M),\mathbf{B}G)$. \vspace{0.35cm}

\noindent The infinitesimal version of the map \eqref{eq:mapconnectionex} is a homomorphisms of $L_\infty$-algebroids
\begin{equation}
    A:\;TM \;\longrightarrow\; \mathbf{b}(T\mathfrak{g}),
\end{equation}
where $TM$ is regarded as an algebroid with Lie bracket $[-,-]_\mathrm{Lie}$ and $\mathbf{b}(T\mathfrak{g})$ is just the tangent $L_\infty$-algebra $T\mathfrak{g}$ regarded as an $L_\infty$-algebroid.
Let us now consider its dual map of Chevalley-Eilenberg dg-algebras. First of all, recall that $\mathrm{CE}(TM)=\big(\Omega^\bullet(M),\di\big)$. Thus, we have the following map of dg-algebras
\begin{equation}
    \big(\Omega^\bullet(M),\di\big) \;\longleftarrow\; \mathrm{CE}(T\mathfrak{g})=\mathrm{W}(\mathfrak{g}) \;:A.
\end{equation}
Such a map is nothing but a $\mathfrak{g}$-valued differential form
\begin{equation}
    A\in\Omega^1(M,\mathfrak{g})
\end{equation}
with curvature $F:=\di A + [A\,\overset{\wedge}{,}\,A]\in\Omega^2(M,\mathfrak{g})$, which can be non-vanishing.
Notice that, equivalently, in NQ-manifold notation, this map can be rewritten as
\begin{equation}
    A:\; T[1]M \;\longrightarrow\; T[1]\mathfrak{g}[1],
\end{equation}
which is a definition commonly adopted of gauge field in literature. The dg-algebra of such maps is nothing but $\mathrm{Maps}(T[1]M,\,T[1]\mathfrak{g}[1])=\mathbf{dgcAlg}\big(\mathrm{W}(\mathfrak{g}),\,\mathrm{CE}(TM)\big)\cong\Omega^\bullet(M,\mathfrak{g})$ from definition \ref{def:gvaluedforms}. Its underlying graded space will be given by the cochain complex
\begin{equation}
    \underbrace{\Coo(M,\mathfrak{g})}_{\text{gauge parameters}}\!\! \longrightarrow\; \, \underbrace{\Omega^1(M,\mathfrak{g})}_{\text{gauge fields}}\;\longrightarrow\;\underbrace{\Omega^2(M,\mathfrak{g})}_{\text{curvatures}}\;\longrightarrow\!\!\underbrace{\Omega^3(M,\mathfrak{g})}_{\text{Bianchi identities}}.
\end{equation}
It is important to notice that the space of such maps is itself an algebroid, whose objects are $A\in\Omega^1(M,\mathfrak{g})$ and whose morphisms are infinitesimal gauge transformation
\begin{equation}
    \delta_\lambda A \;=\; \di\lambda +[\lambda,A]_\mathfrak{g},
\end{equation}
where the gauge parameter is a global $\mathfrak{g}$-valued function $\lambda\in\Coo(M,\mathfrak{g})$.
For any fixed object $A\in\Omega^1(M,\mathfrak{g})$ we have a Lie algebra of gauge parameters, i.e. a \textit{gauge algebra}: this Lie algebra structure is given by the Lie bracket
\begin{equation}
    \lambda_{12} \;=\; [\lambda_1,\lambda_2]_\mathfrak{g}
\end{equation}
for any couple of gauge parameters $\lambda_1,\lambda_2\in\Coo(M,\mathfrak{g})$.
It is intuitively clear that such an algebroid is nothing but the infinitesimal version of the groupoid $\mathbf{H}\big(M,\mathbf{B}G_{\mathrm{conn}}\big) = \mathbf{H}(\mathscr{P}(M),\mathbf{B}G)$ of $G$-bundles with connection. \vspace{0.2cm}

\begin{table}[h!]\begin{center}
\begin{center}
 \begin{tabular}{|| c | c |  c ||} 
 \hline
 \multicolumn{3}{||c||}{Gauge fields} \\
 \hline\hline
  & Groupoid & Algebroid \\ [0.8ex] 
 & $\mathbf{H}(\mathscr{P}(M),\mathbf{B}G)$ &$\mathrm{Maps}(T[1]M,\,T[1]\mathfrak{g}[1])$   \\[0.8ex]  
 \hline
 Objects & \v{C}ech cocycles $\big(A_{(\alpha)},\,f_{(\alpha\beta)}\big)$ & $\mathfrak{g}$-valued $1$-forms   \\[0.8ex] 
  & $A_{(\alpha)}\in\Omega^1(U_\alpha,\mathfrak{g})$ & $A\in\Omega^1(M,\mathfrak{g})$   \\[0.5ex]  
    & $f_{(\alpha\beta)}\in\Coo(U_\alpha\cap U_\beta,G)$ &    \\[0.5ex]  
 \hline
 Morphisms & Global gauge transformations & Infinitesimal gauge transformations   \\[0.8ex] 
  & $\lambda_{(\alpha)}\in\Coo(U_\alpha,G)$ & $\lambda\in\Coo(M,\mathfrak{g})$   \\[0.8ex] 
  \hline
\end{tabular}
\end{center}
\caption[Global and local gauge fields in higher geometry]{\label{tab:gaugefields}Global aspects of the definition of ordinary gauge fields in higher geometry.}\vspace{-0.2cm}
\end{center}\end{table}

\noindent However, notice that the infinitesimal definition of a gauge field as a globally-defined $1$-form $A\in\Omega^1(M,\mathfrak{g})$ does not capture the global geometry of a general gauge field. In fact, we know that a topologically non-trivial gauge field (e.g. in presence of non-zero charges) cannot be defined as a global $1$-form, but it must be given by a cocycle $\big(A_{(\alpha)},f_{(\alpha\beta)}\big)\in\mathbf{H}(\mathscr{P}(M),\mathbf{B}G)$, which allows non-trivial patchings.
\vspace{0.25cm}

\noindent Everything we said in this subsection can be immediately generalised to higher gauge fields by replacing the ordinary Lie group $G$ with a Lie $\infty$-group.

\begin{remark}[Global vs infinitesimal higher gauge theory]
Given a Lie $\infty$-group $G$, often in the literature a higher gauge field is defined as a map $A:T[1]M\rightarrow T[1]\mathfrak{g}[1]$. Since $T[1]\mathfrak{g}[1]$ is the NQ-manifold corresponding to the tangent $L_\infty$-algebra $T\mathfrak{g}$, this map is equivalently a map of dg-algebras $A:\mathrm{W}(\mathfrak{g})\rightarrow \mathrm{CE}(TM)$. However, in subsection \ref{subsectionconnection} we learnt that we should replace the Weil dg-algebra with the adjusted Weil algebra $\mathrm{W}_{\mathrm{adj}}(\mathfrak{g})$ to avoid fake-flatness.
The dg-algebra $\mathbf{dgcAlg}\big(\mathrm{W}_{\mathrm{adj}}(\mathfrak{g}),\,\mathrm{CE}(TM)\big)\cong\big(\Omega^\bullet(M,\mathfrak{g}),\nabla\big)$ of such maps has the following underlying cochain complex
\begin{equation*}
    \cdots \;\longrightarrow \!\!\!\!\!\! \underbrace{\Omega^{-1}(M,\mathfrak{g})}_{\text{gauge of gauge param.}}\!\!\!\! \longrightarrow \!\! \,\underbrace{\Omega^0(M,\mathfrak{g})}_{\text{gauge parameters}}\!\! \longrightarrow\; \, \underbrace{\Omega^1(M,\mathfrak{g})}_{\text{gauge fields}}\;\longrightarrow\;\underbrace{\Omega^2(M,\mathfrak{g})}_{\text{curvatures}}\;\longrightarrow\!\!\underbrace{\Omega^3(M,\mathfrak{g})}_{\text{Bianchi identities}}\!\!\!,
\end{equation*}
where $1$-degree elements $A\in\Omega^1(M,\mathfrak{g})$ are local higher gauge fields;  $0$-degree elements are gauge parameters given by
\begin{equation}
    \delta_\lambda A \;=\; [\lambda]_1 +[\lambda\,\overset{\wedge}{,}\,A]_2 + [\lambda\,\overset{\wedge}{,}\,A\,\overset{\wedge}{,}\,A]_3 + \dots,
\end{equation}
$2$-degree elements are curvatures given by
\begin{equation}
    F \;=\; [A]_1 +[A\,\overset{\wedge}{,}\,A]_2 + [A\,\overset{\wedge}{,}\,A\,\overset{\wedge}{,}\,A]_3 + \dots
\end{equation}
and $3$-degree elements are Bianchi identities for higher gauge fields, given by
\begin{equation}
    0 \;=\; [F]_1 +[F\,\overset{\wedge}{,}\,A]_2 + [F\,\overset{\wedge}{,}\,A\,\overset{\wedge}{,}\,A]_3 + \dots.
\end{equation}
Similarly to the case of the ordinary gauge field we previously discussed, by fixing an object in $\mathbf{dgcAlg}\big(\mathrm{W}_{\mathrm{adj}}(\mathfrak{g}),\,\mathrm{CE}(TM)\big)$ regarded as a $L_\infty$-algebroid, we obtain the \textit{gauge} $L_\infty$\textit{-algebra} of the higher gauge field. \vspace{0.25cm}

\noindent However, as we explicitly worked out for the particular example of an ordinary Lie algebra, this definition does not capture at all the global geometry of the higher gauge field (e.g. topologically non-trivial higher gauge fields), which are encoded by $\mathbf{H}\big(M,\mathbf{B}G_{\mathrm{conn}}\big) = \mathbf{H}\big(\mathscr{P}(M),\mathbf{B}\mathrm{Inn}_{\mathrm{adj}}(G)\big)$  of $G$-bundles with connection.
\end{remark}

\section{Automorphisms of principal $\infty$-bundles}

In this section we will apply the definition of automorphism $\infty$-groupoid from \cite{FSS16} to obtain the $\infty$-groupoid of finite symmetries of a principal $\infty$-bundle. The Lie differentiation of such an $\infty$-groupoid will be closely related to generalised geometry. This is closely related to \cite{BMS20}.

\begin{definition}[Automorphism $\infty$-groupoid]
Given any object $X\in\mathbf{C}$ of an $(\infty,1)$-category $\mathbf{C}$, we define its \textit{automorphism $\infty$-groupoid} $\Aut_\mathbf{C}(X)$ as the sub-$\infty$-groupoid of $\mathbf{C}(X,X)$ of invertible morphisms. 
For a given morphism $f:X\rightarrow Y$, the automorphism groupoid $\Aut_{\mathbf{C}_/}(f)$ is analogously defined as the sub-$\infty$-groupoid of $\mathbf{C}_{/}(f,f)$ of invertible morphisms.
\end{definition}

\begin{example}[Automorphisms of principal $\infty$-bundles]
Let $P\rightarrow M$ be a principal $\infty$-bundle given by $f:M\rightarrow\mathbf{B}G$. The automorphism $\infty$-group of $f$ (i.e. the $\infty$-group of automorphisms of $P$ preserving the principal structure) will sit at the center of a short exact sequence of $\infty$-groups
\begin{equation}\label{eq:autdef}
    1\longrightarrow\Omega_f\mathbf{H}(M,\mathbf{B}G)\longrightarrow \Aut_{\mathbf{H}_/}(f) \longrightarrow \mathrm{Diff}(M)\longrightarrow 1.
\end{equation}
We will also, equivalently, use the semidirect product notation 
\begin{equation}
    \Aut_{\mathbf{H}_/}(f) \;=\; \mathrm{Diff}(M)\ltimes\Omega_f\mathbf{H}(M,\mathbf{B}G).
\end{equation}
\end{example}

\begin{example}[Automorphisms of ordinary $G$-bundles]\label{ex:nonabelianaut}
Let $G$ be an ordinary Lie group and let $P$ be an ordinary principal $G$-bundle given by the cocycle $f:M\rightarrow\mathbf{B}G$. Hence we have the isomorphism $\Omega_f\mathbf{H}(M,\mathbf{B}G)\cong \Gamma\big(M,\mathrm{Ad}(P)\big)$, where the associated bundle $\mathrm{Ad}(P) := P\times_G G$ with the adjoint action $\mathrm{Ad}:G\times G\rightarrow G$ is just the non-linear \textit{adjoint bundle} of $P$. So we have the usual automorphism group of a principal $G$-bundle
\begin{equation}
    1\longrightarrow\Gamma\big(M,\mathrm{Ad}(P)\big)\longrightarrow \Aut_{\mathbf{H}_/}(f) \longrightarrow \mathrm{Diff}(M)\longrightarrow 1.
\end{equation}
In other words, we have
\begin{equation}
    \Aut_{\mathbf{H}_/}(f) \;=\; \mathrm{Diff}(M)\ltimes \Gamma\big(M,\mathrm{Ad}(P)\big).
\end{equation}
\end{example}

\begin{example}[Automorphisms of circle bundles]
For the case of the ordinary structure group $G=U(1)$, we have $\Omega_f\mathbf{H}(M,\mathbf{B}U(1))\cong \Coo(M,U(1))$ and hence the usual automorphism $1$-group of a circle bundle
\begin{equation}
    1\longrightarrow\Coo(M,U(1))\longrightarrow \Aut_{\mathbf{H}_/}(f) \longrightarrow \mathrm{Diff}(M)\longrightarrow 1.
\end{equation}
In other words, we have
\begin{equation}
    \Aut_{\mathbf{H}_/}(f) \;=\; \mathrm{Diff}(M)\ltimes \Coo\big(M,U(1)\big).
\end{equation}
\end{example}

\begin{example}[Automorphisms of bundle gerbes]\label{ex:aut}
It is possible to prove there exists an equivalence of $2$-groups $\Omega_f\mathbf{H}\big(M,\mathbf{B}(\BU)\big) \cong \mathbf{H}\big(M,\BU\big)$ for any bundle gerbe $f:M\rightarrow\mathbf{B}(\BU)$. Therefore global gauge transformations of this bundle gerbe are global circle bundles with connection on $M$. Thus the $2$-group of automorphisms will sit at the center of the exact sequence of $2$-groups 
\begin{equation}\label{eq:finiteautomorphismseq}
    1\longrightarrow\mathbf{H}\big(M,\BU\big)\longrightarrow \Aut_{\mathbf{H}_/}(f) \longrightarrow \mathrm{Diff}(M)\longrightarrow 1.
\end{equation}
Let us introduce the curvature map of stacks
\begin{equation}
    \mathrm{curv}:\BU\longrightarrow\Omega^2_{\mathrm{cl}}
\end{equation}
which maps a circle bundle $(\eta_{(\alpha)},\eta_{(\alpha\beta)})$ over a manifold $M$ into a global closed $2$-form $b\in\Omega_{\mathrm{cl}}^2(M)$ such that $b|_{U_\alpha} = \mathrm{d}\eta_{(\alpha)}$. Then gauge transformations can be expressed as global $B$-shifts of the form $B_{(\alpha)}\mapsto B_{(\alpha)}+b$.
Notice $\mathrm{Diff}(M)\ltimes\Omega_{\mathrm{cl}}^2(M)$ is the gauge group proposed by \cite{Hull14} for DFT.
\end{example}

\noindent From remark \ref{ex:aut} we know $1$-morphisms between bundle gerbes over $M$ are circle bundles over $M$ and $2$-morphisms are gauge transformations between these circle bundles. This corresponds, in general, to the idea that global gauge transformations of bundle $n$-gerbes are bundle $(n-1)$-gerbes and so on. This is a clear categorical feature of these geometrical objects.

\subsection{Atiyah $L_\infty$-algebroids and generalised geometry}

In this section we will deal with the infinitesimal automorphisms of a principal $\infty$-bundles and we will show how they are related to the more familiar generalised geometry (see \cite{Gua11}). For an introduction to generalised geometry and Courant algebroids, see appendix \ref{app:1}.

\begin{definition}[Atiyah $L_\infty$-algebroids]
Let $P\rightarrow M$ be a principal $\infty$-bundle corresponding to a map $f:M\rightarrow\mathbf{B}G$. The Atiyah $L_\infty$-algebroid of this principal $\infty$-bundle was defined in \cite{FSS16} as the Lie differentiation of its automorphism $\infty$-groupoid
\begin{equation}
    \mathfrak{at}(P) \;:=\; \mathrm{Lie}\big(\Aut_{\mathbf{H}_/}(f)\big).
\end{equation}
\end{definition}

\noindent This $L_\infty$-algebra encodes the infinitesimal symmetries of the principal structure. By differentiating sequence \eqref{eq:autdef} we have that it will sit at the center of the short exact sequence of $L_\infty$-algebras
\begin{equation}
    0\longrightarrow\mathrm{Lie}\big(\Omega_f\mathbf{H}(M,\mathbf{B}G)\big)\longrightarrow \mathfrak{at}(P) \longrightarrow \mathfrak{X}(M)\longrightarrow 0.
\end{equation}

\begin{example}[Ordinary Atiyah algebroid]
If $P\rightarrow M$ is a principal $G$-bundle for some ordinary Lie group $G$ we get the short exact sequence of ordinary algebras
\begin{equation}
    0\longrightarrow\Gamma\big(M,\,\mathrm{ad}(P)\big)\longrightarrow \mathfrak{at}(P) \longrightarrow \mathfrak{X}(M)\longrightarrow 0.
\end{equation}
Here, the symbol $\mathrm{ad}(P):=P\times_G\mathfrak{g}$, where the $G$-action is the adjoint action $\mathrm{ad}:G\rightarrow\mathfrak{g}$, denotes the linear \textit{adjoint bundle} of $P$.
\end{example}

\begin{example}[Ordinary Atiyah algebroid of a circle bundle]
If $P\rightarrow M$ is a circle bundle we get the familiar short exact sequence of ordinary algebras
\begin{equation}
    0\longrightarrow\Coo(M,\mathbb{R})\longrightarrow \mathfrak{at}(P) \longrightarrow \mathfrak{X}(M)\longrightarrow 0.
\end{equation}
Locally, on any patch $U\subset M$, this reduces to the familiar algebra $\mathfrak{at}(P)|_U = \mathfrak{X}(U)\oplus\Coo(U)$ of infinitesimal gauge transformation of an abelian gauge field.
\end{example}

\begin{example}[Courant $2$-algebroid]
If $P\rightarrow M$ is a bundle gerbe with connection data corresponding to a map $M\rightarrow\mathbf{B}(\BU)$, as explained in \cite{Col11, Rog11}, we get that the Atiyah $2$-algebroid is the so-called \textit{Courant }$2$\textit{-algebra} sitting in the short exact sequence of $2$-algebras
\begin{equation}
    0\longrightarrow\mathbf{H}(M,\mathbf{B}\mathbb{R}_{\mathrm{conn}})\longrightarrow \mathfrak{at}(P) \longrightarrow \mathfrak{X}(M)\longrightarrow 0.
\end{equation}
Locally, on a patch $U\subset M$, this reduces the familiar Courant $2$-algebra of infinitesimal gauge transformations of the bundle gerbe, whose underlying complex is just
\begin{equation}
    \mathfrak{at}(P)|_U \,\cong\, \Big( \Coo(U)\,\xrightarrow{\;\mathrm{d}\;}\,\mathfrak{X}(U)\oplus\Omega^1(U) \Big).
\end{equation}
Notice that the definition of Courant $2$-algebroid recovers the formalisation of Courant algebroids in terms of symplectic differential-graded geometry by \cite{Roy02}.
\end{example}

\section{Twisted $\infty$-bundles and $G$-structures}

Given a smooth manifold $M\in\mathbf{Diff}$, a smooth stack $\mathscr{X}\in\mathbf{H}$ and an $\infty$-group $K\in\mathbf{Grp}(\mathbf{H})$, we are interested in cocycles of the form $\hat{f}:M\longrightarrow \mathscr{X}$, where the new stack $\mathscr{V}$ is a principal $G$-bundle on the original stack $\mathscr{X}$. In other words, we are interested in diagrams of the form
\begin{equation}\label{diag:generaltwisted}
    \begin{tikzcd}[row sep=11.5ex, column sep=10.5ex]
P\arrow[d, "\mathrm{hofib}(g)"']\arrow[r] & \ast \arrow[d] & \\
Q\arrow[d, "\mathrm{hofib}(f)"']\arrow[r, "g"] & \mathscr{V}\arrow[r]\arrow[d, two heads, ""']&\ast\arrow[d] \\ 
M\arrow[rr, bend right, "f"']\arrow[r, "\hat{f}"] & \mathscr{V}/\!/K \arrow[r, "c"] & \mathbf{B}K.
\end{tikzcd}
\end{equation}
The new cocycle $\hat{f}:M\longrightarrow \mathscr{V}/\!/K$ can be thought as an object of the $\infty$-groupoid
\begin{equation}
   \hat{f}\;\in\; \mathbf{H}_{/\mathbf{B}K}(f,c).
\end{equation}
As derived by \cite{Principal1}, there exists an isomorphism of $\infty$-groupoids
\begin{equation}
   \mathbf{H}_{/\mathbf{B}K}(f,c) \;\cong\; \Gamma(M,\,Q\times_K\mathscr{V}),
\end{equation}
which identifies such cocycles with sections of the $\mathscr{V}$-associated bundle of the principal $K$-bundle $Q\twoheadrightarrow M$. This equivalence will be useful to simplify calculations.

\begin{remark}[Equivariance of the twisted $\infty$-bundle]
Notice that the following subdiagram of the diagram \eqref{diag:generaltwisted} 
\begin{equation}
\begin{tikzcd}[row sep=8.5ex, column sep=7ex]
    \mathscr{V}\arrow[r]\arrow[d, two heads, ""']&\ast\arrow[d] \\ 
    \mathscr{V}/\!/K \arrow[r, "c"] & \mathbf{B}K
\end{tikzcd}
\end{equation}
is exactly an action $\infty$-groupoid, as defined in definition \ref{def:actgroupoid}.
Notice, then, that in diagram \eqref{diag:generaltwisted} both $Q$ and $\mathscr{V}$ are both principal $G$-bundles and the morphism $g:Q\longrightarrow \mathscr{V}$ is $G$-equivariant.
\end{remark}

\subsection{Twisted $\infty$-bundles}

A first interesting case of the diagram \eqref{diag:generaltwisted} is the one of twisted $\infty$-bundles. This idea was introduced by \cite{Sat09} and further developed by \cite{Principal1, DCCTv2}.

\begin{definition}[Twisted $\infty$-bundles]
We define a \textit{twisted $G$-bundle} $P\twoheadrightarrow M$ as a diagram of the following form
\begin{equation}
    \begin{tikzcd}[row sep=11.5ex, column sep=10ex]
P\arrow[d, "\mathrm{hofib}(g)"']\arrow[r] & \ast \arrow[d] & \\
Q\arrow[d, "\mathrm{hofib}(f)"']\arrow[r, "g"] & \mathbf{B}G\arrow[r]\arrow[d, two heads, "\mathrm{hofib}(c)"']&\ast\arrow[d] \\ 
M\arrow[rr, bend right, "f"']\arrow[r, "\hat{f}"] & \mathbf{B}G/\!/K \arrow[r, "c"] & \mathbf{B}K
\end{tikzcd}
\end{equation}
where $Q\twoheadrightarrow M$ is its \textit{twisting principal $K$-bundle.}
In analogy with \eqref{eq:groupoidofprinc}, we define the $\infty$-groupoid of twisted $G$-bundles with fixed twisting principal $K$-bundle $f: M\rightarrow\mathbf{B}G$ by
\begin{equation}
    G\mathrm{Bund}^{[f]}(M) \;:=\; \Gamma(M,\,Q\times_K\mathbf{B}G).
\end{equation}
\end{definition}

\begin{remark}[Trivial twisting bundle]
From the definition, we immediately obtain that, if the twisting $K$-bundle $M\rightarrow\mathbf{B}G$ is trivial, then a twisted $G$-bundle reduces to a principal $G$-bundle, i.e.
\begin{equation}
    G\mathrm{Bund}^{[0]}(M) \;\cong\; G\mathrm{Bund}(M).
\end{equation}
\end{remark}

\begin{example}[Ramond-Ramond fields]
Let us define the stack 
\begin{equation}
    \mathbf{K}U \; := \; \prod_{k\in\mathbb{N}}\mathbf{B}^{2k+1}U(1),
\end{equation}
which is a moduli stack of a tower of abelian bundle $2k$-gerbes for any $k\in\mathbb{N}$. 
Now, let us consider the twisted bundle $\mathscr{D}\twoheadrightarrow M$ given as follows:
\begin{equation}
     \begin{tikzcd}[row sep={13ex,between origins}, column sep={15ex,between origins}]
      \mathscr{D}\arrow[r]\arrow[d,two heads] & \ast\arrow[d] & \\
      \mathscr{G}\arrow[r]\arrow[d,two heads] & \mathbf{K}U\arrow[r]\arrow[d] & \ast\arrow[d]\\
      M\arrow[r]\arrow[rr, bend right] & \mathbf{K}U/\!/\mathbf{B}U(1)\arrow[r] & \mathbf{B}^2U(1),
    \end{tikzcd}
\end{equation}
where the twisting principal bundle is a bundle gerbe $\mathscr{G}\twoheadrightarrow M$. 
The $L_\infty$-algebroid corresponding to the smooth $\infty$-groupoid $\mathbf{K}U/\!/\mathbf{B}U(1)$ will be
$\mathrm{Lie}(\mathbf{K}U/\!/\mathbf{B}U(1))=\mathfrak{ku}/\!/\mathbf{B}U(1)$ and its Chevalley-Eilenberg dg-algebra will be 
\begin{equation}
    \mathrm{CE}\big(\mathfrak{ku}/\!/\mathbf{B}U(1)\big) \;=\; \mathbb{R}[f_{2},f_{4},f_{6},\dots,h_3]/\!\left(\begin{array}{l}\di h_3=0, \\[0.5ex] \di f_{2(k+1)}=h_3\wedge f_{2k} \;\,\forall k\text{ even}  \end{array}\right).
\end{equation}
The twisted cocycle $M\longrightarrow\mathbf{K}U/\!/\mathbf{B}U(1)$ is then the \v{C}ech cocycle of a tower of bundle $2k$-gerbes on $M$, twisted by the cocycle of the Kalb-Ramond field. Its curvature will be then the usual curvature of the Ramond-Ramond fields in Type IIA String Theory, i.e.
\begin{equation}
    \begin{aligned}
        \mathrm{d}H =0, \qquad\qquad\qquad \mathrm{d}F_{D0} =0, \quad \mathrm{d}F_{D2} + F_{D0}\wedge H=0, \\
        \mathrm{d}F_{D4} + F_{D2}\wedge H=0, \quad \mathrm{d}F_{D6} + F_{D4}\wedge H=0, \quad \mathrm{d}F_{D8} + F_{D6}\wedge H=0,
\end{aligned}
\end{equation}
when truncated at $k=4$. For more details, see \cite[p.22]{FSS18x}.
\end{example}

\begin{example}[Twisted K-theory]
Consider twisted $U(n)$-bundles for a given twisting principal $\mathbf{B}U(1)$-bundle (i.e. a bundle gerbe) with Dixmier-Douady class $[H]\in H^3(M,\mathbb{Z})$. 
\begin{equation}
     \begin{tikzcd}[row sep={12ex,between origins}, column sep={13ex,between origins}]
      \mathscr{K}\arrow[r]\arrow[d,two heads] & \ast\arrow[d] & \\
      \mathscr{G}\arrow[r]\arrow[d,two heads] & \mathbf{B}U(n)\arrow[r]\arrow[d] & \ast\arrow[d]\\
      M\arrow[r]\arrow[rr, bend right] & \mathbf{B}PU(n) \arrow[r] & \mathbf{B}^2U(1),
    \end{tikzcd}
\end{equation}
Notice that equivalence classes of such bundles are nothing but the objects studied by twisted K-theory. 
\end{example}

\subsection{Principal $\String^{\mathbf{a}}(G)$-bundle}

$\String^{\mathbf{a}}(G)$-bundles are extremely relevant in String Theory, since they formalise the geometry underlying the gauge structure of the Supergravity limit of heterotic String Theory. This structure was introduced by \cite{Sat09} and further developed by \cite{Fiorenza:2012mr}. For an introductory review of $\String^{\mathbf{a}}(G)$-bundles, see \cite{nikolaus2013equivalent}.

\begin{definition}[$\String^{\mathbf{a}}(G)$ $2$-group]
Let $G$ be an ordinary Lie group whose Lie algebra $\mathfrak{g}=\mathrm{Lie}(G)$ is equipped with a Killing form $\langle -,-\rangle:\mathfrak{g}\otimes\mathfrak{g}\longrightarrow \mathbb{R}$.
The $2$-group $\String^{\mathbf{a}}(G)$ is defined by the following diagram: 
\begin{equation}
\begin{tikzcd}[row sep=10ex, column sep=5ex]
\mathbf{B}\String^{\mathbf{a}}(G) \arrow[r]\arrow[d, "\mathrm{hofib}(\mathbf{a})"']&\ast\arrow[d] \\ 
\mathbf{B}G \arrow[r, "{\mathbf{a}}"] & \mathbf{B}^3U(1),
\end{tikzcd}
\end{equation}
where the map $\mathbf{a}=\exp(a)$ is given by Lie exponentiating the map $a:\mathfrak{g}\rightarrow \mathbf{b}^2\mathfrak{u}(1)$ of $L_\infty$-algebras which is dually defined by the map
\begin{equation}
    \mathbb{R}[t_3]/\langle \di t_3=0\rangle \;\ni\, t_3 \;\xmapsto{\;\;\;a^\ast\;\;\;}\; \langle -, [-,-]_\mathfrak{g}\rangle \;\in\,\mathrm{CE}(\mathfrak{g}).
\end{equation}
\end{definition}

\noindent From de definition, follows that the Chevalley-Eilenberg dg-algebra of the Lie $2$-algebra $\mathfrak{string}^\mathbf{a}(\mathfrak{g})$ of the Lie $2$-group $\String^{\mathbf{a}}(G)$ is
\begin{equation}
    \mathrm{CE}(\mathfrak{string}^\mathbf{a}(\mathfrak{g})) \;=\; \mathbb{R}\!\left[e^\mu,B\right]/ \!\left(\begin{array}{l} \di e^\mu = 0, \\[0.5ex] \di B = \kappa_{\mu\mu'}C^{\mu'}_{\;\;\nu\lambda}e^\mu\wedge e^\nu\wedge e^\lambda \end{array}\right),
\end{equation}
where $\kappa_{\mu\nu}$ is the Killing form and $C^{\mu}_{\;\;\nu\lambda}$ are the structure constants of $\mathfrak{g}$. \vspace{0.25cm}

\noindent Its definition implies that the $\String^{\mathbf{a}}(G)$ $2$-group can also be thought as a Lie $\infty$-group extension of the form $\mathbf{B}U(1) \xhookrightarrow{\;\;i\;\;} \String^{\mathbf{a}}(G) \xtwoheadrightarrow{\;\;\pi\;\;} G$.
Let us now consider a principal $\infty$-bundle $\mathscr{G}_{\mathrm{het}}\twoheadrightarrow M$ whose structure group is the Lie $2$-group $\String^{\mathbf{a}}(G)$.
From the definition of the $2$-group $\String^{\mathbf{a}}(G)$, we can obtain a diagram of the following form:
\begin{equation}
\begin{tikzcd}[row sep=14ex, column sep=9ex]
\mathscr{G}_{\mathrm{het}}\arrow[d, "\mathrm{hofib}(g)"']\arrow[dd, bend right=55, "\mathrm{hofib}(\hat{f})"']\arrow[r] & \ast \arrow[d] & \\
P\arrow[d,"\mathrm{hofib}(f)"']\arrow[r, "g"] & \mathbf{B}^2U(1) \arrow[r]\arrow[d, "\mathbf{B}i", hook]&\ast\arrow[d] \\ 
M\arrow[r, "\hat{f}"]\arrow[rr, bend right=25, "f"'] & \mathbf{B}\String^{\mathbf{a}}(G) \arrow[r, "\mathbf{B}\pi", two heads] & \mathbf{B}G,
\end{tikzcd}
\end{equation}
where $P\twoheadrightarrow M$ is a principal $G$-bundle.
We just proved the following corollary.

\begin{theorem}[$\String^{\mathbf{a}}(G)$-bundle as bundle gerbe on a $G$-bundle]
A $\String^{\mathbf{a}}(G)$-bundle $\mathscr{G}_{\mathrm{het}}\twoheadrightarrow M$ is a bundle gerbe $\mathscr{G}_{\mathrm{het}}\twoheadrightarrow P$ on a principal $G$-bundle $P\twoheadrightarrow M$ such that its Dixmier-Douady class $\mathrm{dd}(\mathscr{G}_{\mathrm{het}})\in H^3(P,\mathbb{Z})$ restricted to the fibre $G$ of $P$ is a generator of $H^3(G,\mathbb{Z})\cong\mathbb{Z}$.
\end{theorem}

\begin{remark}[Connection of a $\String^{\mathbf{a}}(G)$-bundle]
Locally, the connection of a $\String^{\mathbf{a}}(G)$-bundle is given by a couple $\big(A_{(\alpha)},B_{(\alpha)}\big)$, where $A_{(\alpha)}\in\Omega^1(U_\alpha,\mathfrak{g})$ and $B_{(\alpha)}\in\Omega^2(U_\alpha)$ for every open set $U_\alpha\subset M$.
The curvature of a $\String^{\mathbf{a}}(G)$-bundle is locally given by
\begin{equation}
\begin{aligned}
    F_{(\alpha)} \;&=\; \di A_{(\alpha)} + [A_{(\alpha)}\,\overset{\wedge}{,}\,A_{(\alpha)}]_\mathfrak{g}, \\
    H_{(\alpha)} \;&=\; \di B_{(\alpha)} + \frac{1}{2}\langle F_{(\alpha)}\,\overset{\wedge}{,}\,A_{(\alpha)}\rangle + \frac{1}{3!}\langle A_{(\alpha)}\,\overset{\wedge}{,}\,[A_{(\alpha)}\,\overset{\wedge}{,}\,A_{(\alpha)}]_\mathfrak{g}\rangle,
\end{aligned}
\end{equation}
where $\big(F_{(\alpha)},H_{(\alpha)}\big)$ where $F_{(\alpha)}\in\Omega^2(U_\alpha,\mathfrak{g})$ and $H_{(\alpha)}\in\Omega^3(U_\alpha)$.
On every open set, by taking the differential of the curvature, we obtain the Bianchi identities
\begin{equation}
\begin{aligned}
    \mathrm{D}_A F_{(\alpha)} \;&=\; 0, \\
    \di H_{(\alpha)} \;&=\; \frac{1}{2}\langle F_{(\alpha)}\,\overset{\wedge}{,}\,F_{(\alpha)}\rangle.
\end{aligned}
\end{equation}
Notice that the bosonic fields of heterotic Supergravity are encoded by the connection of a $\String^{\mathbf{a}}(G)$-bundle with Lie group $G=\mathrm{Spin}(d)\times SO(32)$ or $G=\mathrm{Spin}(d)\times E_8\times E_8$.
\end{remark}

\subsection{11d Supergravity}

Let us now briefly discuss the twisted $\infty$-bundle underlying the global geometry of 11d Supergravity.
See appendix \ref{app:2} for a review of supergeometry.\vspace{0.15cm}

\noindent We can define the super-Minkowski space of 11d Supergravity as a super-algebra $\mathbb{R}^{1,10|\mathbf{32}}$ whose Chevalley-Eilenberg dg-algebra is
\begin{equation}
    \mathrm{CE}(\mathbb{R}^{1,10|\mathbf{32}}) \;=\; \mathbb{R}\!\left[e^\mu,\psi^\upalpha\right]/ \!\left(\begin{array}{l} \di e^\mu = \bar{\psi}\Gamma^\mu\psi, \\[0.5ex] \di \psi^\upalpha = 0\end{array}\right)
\end{equation}
where the degree of the generators are $\deg(e^\mu)=(1,\mathrm{even})$ and $\deg(\psi^\upalpha)=(1,\mathrm{odd})$.

\begin{definition}[$\mathfrak{m2brane}$ $3$-algebra]
Let us define the Lie $3$-algebra $\mathfrak{m2brane}$ as the following fibration:
\begin{equation}
     \begin{tikzcd}[row sep=2cm, column sep=1.5cm]
        \mathfrak{m2brane}\arrow[r]\arrow[d,two heads, "\mathrm{hofib}(g_4)"'] & \ast\arrow[d]  \\
        \mathbb{R}^{1,10|\mathbf{32}}\arrow[r,"g_4"] & \mathbf{b}^3\mathfrak{u}(1)
    \end{tikzcd}
\end{equation}
where $g_4$ is a map which can be dually defined by a map of Chevalley-Eilenberg dg-algebras $g_4^\ast:\mathbb{R}[t_4]/\langle \di t_4=0\rangle\longrightarrow \mathrm{CE}(\mathbb{R}^{1,10|\mathbf{32}})$ which sends the generator $t_3$ to the Lie algebra cocycle
\begin{equation}
g_4 \;=\; \bar{\psi}\Gamma_{\mu\nu}\psi\wedge e^\mu \wedge e^\nu \;\, \in\; \mathrm{CE}(\mathbb{R}^{1,10|\mathbf{32}}),
\end{equation}
which satisfies the equation $\di g_4 =0$.
This implies that the Lie $3$-algebra $\mathfrak{m2brane}$ is given by
\begin{equation}
    \mathrm{CE}(\mathfrak{m2brane}) \;=\; \mathbb{R}\!\left[e^\mu,\psi^\upalpha, c_3\right]/ \!\left(\begin{array}{l} \di e^\mu = \bar{\psi}\Gamma^\mu\psi, \\[0.5ex] \di \psi^\upalpha = 0, \\[0.5ex] \di c_3 = g_4\end{array}\right).
\end{equation}
\end{definition}

\begin{definition}[$\mathfrak{m5brane}$ $6$-algebra]
Let us define the Lie $6$-algebra $\mathfrak{m5brane}$ as the following fibration:
\begin{equation}
     \begin{tikzcd}[row sep=2cm, column sep=2cm]
        \mathfrak{m5brane}\arrow[r]\arrow[d,two heads, "\mathrm{hofib}(c_3\wedge g_4+\frac{1}{2}g_7)"'] & \ast\arrow[d]  \\
        \mathfrak{m2brane}\arrow[r,"c_3\wedge g_4+\frac{1}{2}g_7"] & \mathbf{b}^6\mathfrak{u}(1)
    \end{tikzcd}
\end{equation}
where $c_3\wedge g_4+\frac{1}{2}g_7$ is a map which can be dually defined by a map of Chevalley-Eilenberg dg-algebras $\big(c_3\wedge g_4+\frac{1}{2}g_7\big)^\ast:\mathbb{R}[t_7]/\langle \di t_7=0\rangle\longrightarrow \mathrm{CE}(\mathfrak{m2brane})$ which sends the generator $t_7$ to the Lie algebra cocycle
\begin{equation}
c_3\wedge g_4+\frac{1}{2}g_7\,\in\,\mathrm{CE}(\mathfrak{m2brane}) \,\;\;\text{ with }\,\;\; g_7 \;:=\; \bar{\psi}\Gamma_{\mu_1\cdots \mu_5}\psi\wedge e^{\mu_1} \wedge \cdots \wedge e^{\mu_5},
\end{equation}
which satisfies the equation $\di \big(c_3\wedge g_4+\frac{1}{2}g_7\big) =0$.
This implies that the Lie $6$-algebra $\mathfrak{m5brane}$ is given by
\begin{equation}\label{eq:m5branelinftyalgebra}
    \mathrm{CE}(\mathfrak{m5brane}) \;=\; \mathbb{R}\!\left[e^\mu,\psi^\upalpha, c_3, c_6\right]/ \!\left(\begin{array}{l} \di e^\mu = \bar{\psi}\Gamma^\mu\psi, \\[0.5ex] \di \psi^\upalpha = 0, \\[0.5ex] \di c_3 = g_4 , \\[0.5ex] \di c_6 = c_3\wedge g_4 + \frac{1}{2}g_7\end{array}\right).
\end{equation}
\end{definition}

\noindent By putting everything together, we obtain the following fibration of super $L_\infty$-algebras:
\begin{equation}\label{diag:m25branealg}
\begin{tikzcd}[row sep=13ex, column sep=9ex]
\mathfrak{m5brane}\arrow[d, "\mathrm{hofib}\big(c_3\wedge g_4 + \frac{1}{2}g_7\big)"']\arrow[r] & \ast \arrow[d] & \\
\mathfrak{m2brane}\arrow[d,"\mathrm{hofib}(g_4)"']\arrow[r, "c_3\wedge g_4 + \frac{1}{2}g_7"] & \mathbf{b}^6\mathfrak{u}(1) \arrow[r]\arrow[d]&\ast\arrow[d] \\ 
\mathbb{R}^{1,10|\mathbf{32}}\arrow[r]\arrow[rr, bend right=25, "g_4"'] & \mathbf{b}^6\mathfrak{u}(1)/\!/\mathbf{b}^2\mathfrak{u}(1) \arrow[r] & \mathbf{b}^3\mathfrak{u}(1).
\end{tikzcd}
\end{equation}

\begin{remark}[11d Supergravity]
Let $M$ be a $(1,10|\mathbf{32})$-dimensional super-manifold.
We can, then, integrate the diagram \eqref{diag:m25branealg} to the following twisted $\infty$-bundle:
\begin{equation}
\begin{tikzcd}[row sep=14ex, column sep=9ex]
\mathscr{G}_{\mathrm{M5}}\arrow[d, "\mathrm{hofib}(g)"']\arrow[dd, bend right=55, "\mathrm{hofib}(\hat{f})"']\arrow[r] & \ast \arrow[d] & \\
\mathscr{G}_{\mathrm{M2}}\arrow[d,"\mathrm{hofib}(f)"']\arrow[r, "g"] & \mathbf{B}^6U(1) \arrow[r]\arrow[d]&\ast\arrow[d] \\ 
M\arrow[r, "\hat{f}"]\arrow[rr, bend right=25, "f"'] & \mathbf{B}^6U(1)/\!/\mathbf{B}^2U(1) \arrow[r] & \mathbf{B}^3U(1).
\end{tikzcd}
\end{equation}
By looking at equation \eqref{eq:m5branelinftyalgebra}, it is not hard to see that the curvature of this twisted $\infty$-bundle recovers the bosonic fields of 11d Supergravity. In other words, we have 
\begin{equation}
    \begin{aligned}
        G_4 \;&=\; \di C_{3(\alpha)},  \\
        G_7 \;&=\; \di C_{6(\alpha)} + C_{3(\alpha)}\wedge G_4.
    \end{aligned}
\end{equation}
By taking the differential on any patch of the base manifold, we recover the Bianchi identities of 11d Supergravity, i.e.
\begin{equation}
    \begin{aligned}
        \di G_4 \;&=\; 0,  \\
        \di G_7 \;&=\; G_4\wedge G_4.
    \end{aligned}
\end{equation}
\end{remark}
\noindent This geometric structure underlying $11$d Supergravity was introduced by \cite{Fiorenza:2013nha, FSS15x}.

\subsection{$G$-structures}

One can ask if a principal $K$-bundle with structure group $K$ "comes from" a subgroup $G$ of $K$. This is called reduction of the structure group to $G$. If we regard the frame bundle $FM\twoheadrightarrow M$ as a $GL(d,\mathbb{R})$-bundle with $d=\mathrm{dim}(M)$, then a $G$-structure can be seen as a reduction of the structure group $G\xhookrightarrow{\;i\;} GL(d,\mathbb{R})$ of $FM$.

\begin{definition}[$G$-structure]\label{ex:reductionofstructuregroup}
Let $M$ be a $d$-dimensional smooth manifold.
A $G$\textit{-structure} on $M$ is a twisted bundle of the following form:
\begin{equation}
    \begin{tikzcd}[row sep={16ex,between origins}, column sep={17ex,between origins}]
P\arrow[d]\arrow[r] & \ast \arrow[d] & \\
FM\arrow[d, "\mathrm{hofib}(N)"']\arrow[r] & GL(d)/\!/G\arrow[r]\arrow[d, two heads, "\mathrm{hofib}(\mathbf{B}i)"']&\ast\arrow[d] \\ 
M\arrow[rr, bend right, "N"']\arrow[r, "\hat{N}"] & \mathbf{B}G \arrow[r, "\mathbf{B}i", hook] & \mathbf{B}GL(d),
\end{tikzcd}
\end{equation}
where $M\xrightarrow{\;\;N\;\;}\mathbf{B}GL(d)$ is the \v{C}ech cocycle corresponding to the frame bundle of $M$.
The groupoid of all the $G$-structures on a smooth manifold $M$ is defined by
\begin{equation}
    G\Struc(M) \;:=\; \Gamma\Big(M,\, FM\times_{GL(d)}GL(d)/\!/G\Big).
\end{equation}
\end{definition}

\begin{figure}[h]\begin{center}
\tikzset {_9lyqfs408/.code = {\pgfsetadditionalshadetransform{ \pgftransformshift{\pgfpoint{83.16 bp } { -104.94 bp }  }  \pgftransformscale{1.32 }  }}}
\pgfdeclareradialshading{_96t8aag1b}{\pgfpoint{-72bp}{88bp}}{rgb(0bp)=(1,1,1);
rgb(0bp)=(1,1,1);
rgb(25bp)=(0.61,0.61,0.61);
rgb(400bp)=(0.61,0.61,0.61)}
\tikzset{every picture/.style={line width=0.75pt}} 
\begin{tikzpicture}[x=0.75pt,y=0.75pt,yscale=-1,xscale=1]
\path  [shading=_96t8aag1b,_9lyqfs408] (24.3,137.52) .. controls (24.3,90.46) and (64.42,52.3) .. (113.9,52.3) .. controls (163.38,52.3) and (203.5,90.46) .. (203.5,137.52) .. controls (203.5,184.59) and (163.38,222.75) .. (113.9,222.75) .. controls (64.42,222.75) and (24.3,184.59) .. (24.3,137.52) -- cycle ; 
 \draw  [color={rgb, 255:red, 0; green, 0; blue, 0 }  ,draw opacity=1 ] (24.3,137.52) .. controls (24.3,90.46) and (64.42,52.3) .. (113.9,52.3) .. controls (163.38,52.3) and (203.5,90.46) .. (203.5,137.52) .. controls (203.5,184.59) and (163.38,222.75) .. (113.9,222.75) .. controls (64.42,222.75) and (24.3,184.59) .. (24.3,137.52) -- cycle ; 
\draw  [fill={rgb, 255:red, 255; green, 0; blue, 31 }  ,fill opacity=0.17 ] (42.34,143.36) .. controls (33.61,126.6) and (47.52,102.07) .. (73.42,88.58) .. controls (99.32,75.09) and (127.4,77.74) .. (136.13,94.5) .. controls (144.86,111.26) and (130.94,135.78) .. (105.05,149.28) .. controls (79.15,162.77) and (51.07,160.12) .. (42.34,143.36) -- cycle ;
\draw  [fill={rgb, 255:red, 0; green, 24; blue, 255 }  ,fill opacity=0.17 ] (186.96,145.76) .. controls (177.46,162.1) and (149.3,163.45) .. (124.05,148.78) .. controls (98.8,134.1) and (86.02,108.96) .. (95.52,92.62) .. controls (105.01,76.28) and (133.18,74.93) .. (158.43,89.6) .. controls (183.68,104.27) and (196.45,129.41) .. (186.96,145.76) -- cycle ;
\draw [color={rgb, 255:red, 208; green, 2; blue, 27 }  ,draw opacity=1 ] (39.69,55.9) -- (68.83,40.94)(29.13,28.18) -- (44.1,57.32) (60.32,39.69) -- (68.83,40.94) -- (64.88,48.59) (27.88,36.69) -- (29.13,28.18) -- (36.78,32.12)  ;
\draw [color={rgb, 255:red, 0; green, 94; blue, 205 }  ,draw opacity=1 ] (177.74,50.58) -- (206.98,65.35)(193.96,25.74) -- (179.19,54.98) (202.99,57.73) -- (206.98,65.35) -- (198.48,66.65) (186.34,29.73) -- (193.96,25.74) -- (195.27,34.24)  ;
\draw [color={rgb, 255:red, 208; green, 2; blue, 27 }  ,draw opacity=1 ] (110.69,29.4) -- (139.83,14.44)(100.13,1.68) -- (115.1,30.82) (131.32,13.19) -- (139.83,14.44) -- (135.88,22.09) (98.88,10.19) -- (100.13,1.68) -- (107.78,5.62)  ;
\draw [color={rgb, 255:red, 0; green, 94; blue, 205 }  ,draw opacity=1 ] (110.73,26.34) -- (139.47,42.05)(127.74,2.04) -- (112.03,30.78) (135.73,34.31) -- (139.47,42.05) -- (130.93,43.08) (120,5.78) -- (127.74,2.04) -- (128.77,10.58)  ;
\draw  [dash pattern={on 0.84pt off 2.51pt}]  (42.6,54.41) -- (77.5,118.75) ;
\draw  [dash pattern={on 0.84pt off 2.51pt}]  (180.49,51.48) -- (150.5,118.25) ;
\draw  [dash pattern={on 0.84pt off 2.51pt}]  (113.6,27.91) -- (114.5,112.75) ;
\draw  [draw opacity=0][dash pattern={on 4.5pt off 4.5pt}] (100.39,1.32) .. controls (104.19,-2.48) and (109.01,-4.75) .. (114.25,-4.75) .. controls (118.93,-4.75) and (123.28,-2.94) .. (126.86,0.16) -- (114.25,23.13) -- cycle ; \draw  [color={rgb, 255:red, 144; green, 19; blue, 254 }  ,draw opacity=1 ][dash pattern={on 4.5pt off 4.5pt}] (100.39,1.32) .. controls (104.19,-2.48) and (109.01,-4.75) .. (114.25,-4.75) .. controls (118.93,-4.75) and (123.28,-2.94) .. (126.86,0.16) ;
\draw  [draw opacity=0][dash pattern={on 4.5pt off 4.5pt}] (141.13,16.12) .. controls (144.6,20.23) and (146.46,25.22) .. (146.03,30.44) .. controls (145.64,35.11) and (143.47,39.29) .. (140.09,42.6) -- (118.25,28.13) -- cycle ; \draw  [color={rgb, 255:red, 144; green, 19; blue, 254 }  ,draw opacity=1 ][dash pattern={on 4.5pt off 4.5pt}] (141.13,16.12) .. controls (144.6,20.23) and (146.46,25.22) .. (146.03,30.44) .. controls (145.64,35.11) and (143.47,39.29) .. (140.09,42.6) ;
\draw (34,198.9) node [anchor=north west][inner sep=0.75pt]  [font=\footnotesize]  {$M$};
\draw (48.5,131.9) node [anchor=north west][inner sep=0.75pt]  [font=\footnotesize]  {$U_{\alpha }$};
\draw (159,134.4) node [anchor=north west][inner sep=0.75pt]  [font=\footnotesize]  {$U_{\beta }$};
\draw (96,113.4) node [anchor=north west][inner sep=0.75pt]  [font=\tiny]  {$U_{\alpha }\!\cap\!U_{\beta }$};
\draw (82.5,-22.6) node [anchor=north west][inner sep=0.75pt]  [font=\footnotesize,color={rgb, 255:red, 144; green, 19; blue, 254 }  ,opacity=1 ]  {$G\subset GL( d,\mathbb{R})$};
\draw (39,28.9) node [anchor=north west][inner sep=0.75pt]  [font=\scriptsize,color={rgb, 255:red, 208; green, 2; blue, 27 }  ,opacity=1 ]  {$e^{\mu }_{( \alpha )}$};
\draw (190.5,38.4) node [anchor=north west][inner sep=0.75pt]  [font=\scriptsize,color={rgb, 255:red, 0; green, 76; blue, 166 }  ,opacity=1 ]  {$e^{\mu }_{( \beta )}$};
\end{tikzpicture}
\caption[$G$-structure]{$G$-structure as a reduction of the structure group $GL(d,\mathbb{R})$ of the frame bundle $FM\twoheadrightarrow M$, on a $d$-dimensional base manifold $M$.}
\end{center}\end{figure}
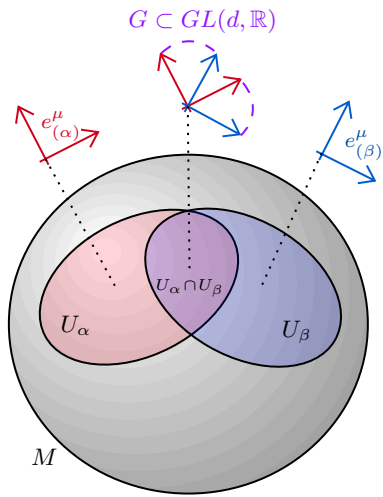

\begin{table}[h!]\begin{center}
\begin{center}
 \begin{tabular}{||c|c|||} 
 \hline
 \multicolumn{2}{||c||}{Examples of ordinary $G$-structures} \\[0.5ex]
 \hline\hline
    Structure & $G \subset GL(d,\mathbb{R})$ \\ [0.5ex] 
    \hline
    Orientation structure & $SL(d,\mathbb{R})$ \\[0.2ex]  
    Orthogonal structure & $O(d)$ \\[0.2ex]   
    (almost) symplectic structure & $Sp(d,\mathbb{R})$ \\[0.2ex]  
    (almost) complex structure & $GL(\nicefrac{d}{2},\mathbb{C})$ \\[0.2ex]  
    (almost) Hermitian structure & $U(\nicefrac{d}{2})$ \\[0.2ex]
    (almost) para-complex structure & $GL(\nicefrac{d}{2},\mathbb{R})\times GL(\nicefrac{d}{2},\mathbb{R})$ \\[0.2ex]  
    (almost) para-Hermitian structure & $GL(\nicefrac{d}{2},\mathbb{R})$ \\[0.2ex]  
 \hline
\end{tabular}
\end{center}
\caption[$G$-structures]{\label{tab:structures}Some relevant ordinary $G$-structures on a smooth manifold $M$ with $\mathrm{dim}(M)=d$.}\vspace{-0.0cm}
\end{center}\end{table}

\subsection{Orthogonal structure}\label{ex:orth}

The \textit{orthogonal structure moduli stack} is the stack which encodes a Riemannian metric structure on a smooth manifold $M$. 
First, notice that, given a smooth manifold $M\in\mathbf{Diff}$, the frame bundle $FM\twoheadrightarrow M$ of its tangent bundle can be seen as a principal $GL(d)$-bundle with coefficients $M\xrightarrow{\;\;N\;\;}\mathbf{B}GL(d)$ given by its transition functions $N_{(\alpha\beta)}\in\Coo\big(U_\alpha\cap U_\beta,\, GL(d)\big)$. 
Now, the existence of a Riemannian metric $g$ on $M$ is equivalent to the fact that the structure group of the frame bundle $FM\twoheadrightarrow M$ can be reduced to $O(d)$, along the inclusion of Lie groups $i:O(d)\hookrightarrow GL(d)$.
Therefore, we are exactly in the situation of example \ref{ex:reductionofstructuregroup}. In particular, we have
\begin{equation}
\begin{tikzcd}[row sep=12.5ex, column sep=9.2ex]
GL(d)/\!/O(d)\arrow[r]\arrow[d, "\mathrm{hofib}(\mathbf{B}i)"', two heads]&\ast\arrow[d] \\ 
\mathbf{B}O(d) \arrow[r, "\mathbf{B}i", hook] & \mathbf{B}GL(d).
\end{tikzcd}
\end{equation}
We want to determine the $\infty$-groupoid $O(d)\Struc(M)$ of orthogonal structures on a $d$-dimensional smooth manifold $M$. Firstly, a map
\begin{equation}
    M \xrightarrow{\quad{(e_{(\alpha)},h_{(\alpha\beta)})}\quad} O(d)\Struc
\end{equation}
is a collection $(e,h)$ of local $GL(d)$-functions $e_{(\alpha)}\in\Coo\big(U_\alpha,\,GL(d)\big)$ on patches and of local $O(d)$-functions $h_{(\alpha\beta)}\in\Coo\big(U_\alpha\cap U_\beta,\,O(d)\big)$ on overlaps of patches, such that they are patched by
\begin{equation}
    \begin{aligned}
        e_{(\alpha)} &= h_{(\alpha\beta)}\cdot e_{(\beta)}\cdot N_{(\alpha\beta)} \\
        h_{(\alpha\gamma)} &= h_{(\alpha\beta)}\cdot h_{(\beta\gamma)}
    \end{aligned}
\end{equation}
on two-fold and on three-fold overlaps. The morphisms $\eta:\big(e_{(\alpha)},h_{(\alpha\beta)}\big)\Mapsto\big(e_{(\alpha)}',h_{(\alpha\beta)}'\big)$ between these maps 
\begin{equation}
    \begin{tikzcd}[row sep=6ex, column sep=14ex]
         M\cong \check{C}(\mathcal{U}) \arrow[r, bend left=50, ""{name=U, below}, "{(e_{(\alpha)},h_{(\alpha\beta)})}"]
        \arrow[r, bend right=50, "{(e_{(\alpha)}',h_{(\alpha\beta)}')}"', ""{name=D}]
        &  O(d)\Struc
        \arrow[Rightarrow, from=U, to=D, "(\eta_{(\alpha)})"]
    \end{tikzcd}
\end{equation}
are collections of local $O(d)$-functions $\eta_{(\alpha)}\in\Coo\big(U_\alpha,\,O(d)\big)$ on each patch, such that they give
\begin{equation}
    \begin{aligned}
        e'_{(\alpha)} &= \eta_{(\alpha)} \cdot e_{(\alpha)} \\
        h'_{(\alpha\beta)} &= \eta_{(\alpha)}\cdot h_{(\alpha\beta)}\cdot\eta_\beta^{-1}.
    \end{aligned}
\end{equation}
Notice that the $e_{(\alpha)}\in\Coo\big(U_\alpha,\,GL(d)\big)$ are the vielbein matrices of the Riemannian metric.

\begin{remark}[Trivialisable tangent bundle]
If the tangent bundle $TM$ of $M$ is trivialisable, an orthogonal structure is not twisted by the frame bundle and it can be described by a simpler cocycle $M\longrightarrow GL(d)/\!/O(d)$. Indeed, for a trivial frame bundle we have $\Gamma\big(M,\,FM\times_{GL(d)}GL(d)/\!/O(d)\big)\cong \mathbf{H}(M,\,GL(d)/\!/O(d))$.
\end{remark}

\begin{remark}[Moduli space of the orthogonal structure]
Notice the moduli space of an orthogonal structure is locally given by $\Coo\big(U_\alpha,\,GL(d)/O(d)\big)$ and globally by non-trivially gluing these spaces by involving the transition functions $N_{(\alpha\beta)}$ of the frame bundle $FM$.
\end{remark}

\subsection{Cartan geometry}

In this subsection we will formalise Cartan geometry with gauge group of the form $G = H \ltimes \mathbb{R}^{1,d}$, so that the Klein coset space $G/H \cong \mathbb{R}^{1,d}$ is a $(1+d)$-dimensional Minkowski space. In particular, we will consider Cartan geometries with $H:=SO(1,d)$. This approach will specialise some of the geometric ideas delineated by \cite{HSS19}.
We will formalise a Cartan geometry on $M$ as an $H$-structure on $M$, whose underlying principal $H$-bundle is equipped with a principal connection. Thus we will define the groupoid $\Cartan(M)$ of Cartan connections of the smooth manifold $M$ as a refinement of the groupoid $H\Struc(M)$ of $H$-structures on $M$. \vspace{0.25cm}

\noindent Let $\mathcal{U}:=\{U_{\alpha}\}$ be a good open cover of a smooth manifold $M$.
A Cartan connection, i.e. an object of $\Cartan(M)$, is a map
\begin{equation}
    \begin{tikzcd}[row sep=7ex, column sep=18ex]
         M \arrow[r, "{(\omega_{(\alpha)},\,e_{(\alpha)},\,h_{(\alpha\beta)})}"]
        & \Cartan,
    \end{tikzcd}
\end{equation}
and it is locally given by the following differential data:
\begin{equation}
    \begin{aligned}
     \omega_{(\alpha)} \,&\in\, \Omega^1\big( U_{\alpha},\, \mathfrak{so}(1,d) \big) &&\;\text{(\textit{spin connection})}, \\
    e_{(\alpha)} \,&\in\, \Omega^1\big( U_{\alpha},\, \mathbb{R}^{1,d} \big) &&\;\text{(\textit{vielbein})}, \\
    h_{(\alpha\beta)} \,&\in\, \Coo\big(U_{\alpha}\cap U_{\beta},\, SO(1,d) \big),
    \end{aligned}
\end{equation}
which satisfy the patching following conditions:
\begin{equation}
    \begin{aligned}
    \omega_{(\beta)} \,&=\,  \mathrm{Ad}_{h_{(\alpha\beta)}^{-1}}\omega_{(\alpha)} + h_{(\alpha\beta)}^{-1}\di h_{(\alpha\beta)}, \\
    e_{(\beta)} \,&=\,  h_{(\alpha\beta)}^{-1} \cdot e_{(\alpha)}, \\
    h_{(\alpha\gamma)}\,&=\, h_{(\alpha\beta)}\cdot h_{(\beta\gamma)},
    \end{aligned}
\end{equation}
where $\cdot$ is the group multiplication of $\mathrm{ISO}(1,d)$.
A gauge transformation
\begin{equation}
    \begin{tikzcd}[row sep=7ex, column sep=14ex]
         M\cong \check{C}(\mathcal{U}) \arrow[r, bend left=50, ""{name=U, below}, "{(\omega_{(\alpha)},\,e_{(\alpha)},\,h_{(\alpha\beta)})}"]
        \arrow[r, bend right=50, "{(\omega_{(\alpha)}',\,e_{(\alpha)}',\,h_{(\alpha\beta)}')}"', ""{name=D}]
        & \Cartan
        \arrow[Rightarrow, from=U, to=D, "(\eta_{(\alpha)})"]
    \end{tikzcd}
\end{equation}
between two Cartan connections is given by a coboundary
\begin{equation}
    \begin{aligned}
    \omega_{(\alpha)}' \,&=\,  \mathrm{Ad}_{\eta_{(\alpha)}^{-1}}\omega_{(\alpha)} + \eta_{(\alpha)}^{-1}\di \eta_{(\alpha)}, \\
    e_{(\alpha)}' \,&=\,  \eta_{(\alpha)}^{-1} \cdot e_{(\alpha)}, \\
    h_{(\alpha\beta)}' \,&=\, \eta_{(\alpha)}^{-1}\cdot h_{(\alpha\beta)}\cdot \eta_{(\beta)}.
    \end{aligned}
\end{equation}

\noindent Notice that the vielbein and the spin connection are \textit{not} globally defined $1$-forms on the base manifold $M$.

\section{Atlases and charts}

In this section we define a notion of atlas for geometric stacks, by generalising the atlas of a smooth manifold.

\begin{remark}[$0$-truncation of stacks]
Let $\mathbf{H}_0$ be the ordinary category of sheaves on manifolds. Then, the inclusion $\mathbf{H}_0\hookrightarrow \mathbf{H}$ has a left adjoint $\tau_0:\mathbf{H}\rightarrow \mathbf{H}_0$ which is called $0$-truncation and which sends a higher stack $\mathscr{X}\in\mathbf{H}$ to its restricted sheaf $\tau_0\mathscr{X}\in\mathbf{H}_0$ at the $0$-degree.
\end{remark}

\begin{definition}[Atlas of a smooth stack]
The atlas of a smooth stack $\mathscr{X}\in\mathbf{H}$ is defined by a smooth manifold $\mathcal{U}\in\mathbf{Diff}$ equipped with a morphism of smooth stacks
\begin{equation}
    \Phi:\,\mathcal{U }\; \longrightarrow\; \mathscr{X}
\end{equation}
which is, in particular, an effective epimorphism, i.e. whose $0$-truncation $\tau_0\Phi:\mathcal{U }\longtwoheadrightarrow \tau_0\mathscr{X}$ is an epimorphism of sheaves. See \cite{Hei05} and \cite{topos} for more detail.
\end{definition}

\noindent This formalizes the idea that to any geometric stack $\mathscr{X}\in\mathbf{H}$ we can associate an atlas which is made up of ordinary manifolds $\mathcal{U }\in\mathbf{Diff}$. This provides a remarkably handy tool to deal with higher geometric objects. Moreover, the notion of atlas will be a pivotal in establishing a correspondence between doubled and higher geometry.

\begin{example}[Atlas for a smooth manifold]
If our geometric stack is an ordinary smooth manifold $\mathscr{G}:=M$, we can choose an atlas given by $\mathcal{U}:=\bigsqcup_{\alpha\in I}\mathbb{R}^{d}$ and by a surjective map $\phi:\bigsqcup_{\alpha\in I}\mathbb{R}^{d}\xtwoheadrightarrow{\;\;\{\phi_\alpha\}_{\alpha\in I}\;\;} M$ given by the local charts $\{\phi_\alpha: \mathbb{R}^{d}\twoheadrightarrow U_\alpha\}_{\alpha\in I}$ of any cover $\{U\}_{\alpha\in I}$ of the manifold $M$. This formalizes the intuitive idea that any smooth manifold looks locally like a Cartesian space $\mathbb{R}^d$. 
\end{example}

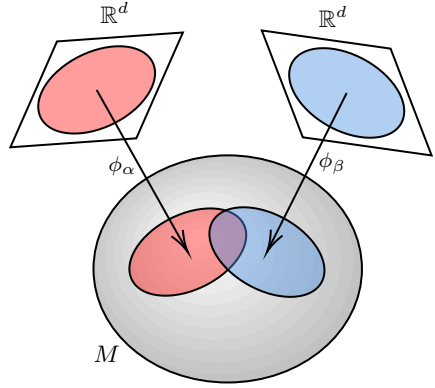
\begin{figure}[h]\begin{center}
\tikzset {_5fwgb8f9k/.code = {\pgfsetadditionalshadetransform{ \pgftransformshift{\pgfpoint{89.1 bp } { -108.9 bp }  }  \pgftransformscale{1.32 }  }}}
\pgfdeclareradialshading{_5zqvwi5cg}{\pgfpoint{-72bp}{88bp}}{rgb(0bp)=(1,1,1);
rgb(0bp)=(1,1,1);
rgb(25bp)=(0.68,0.68,0.68);
rgb(400bp)=(0.68,0.68,0.68)}
\tikzset{every picture/.style={line width=0.75pt}} 
\begin{tikzpicture}[x=0.75pt,y=0.75pt,yscale=-1,xscale=1]
\path  [shading=_5zqvwi5cg,_5fwgb8f9k] (53.5,149.13) .. controls (53.5,117.58) and (83.5,92) .. (120.5,92) .. controls (157.5,92) and (187.5,117.58) .. (187.5,149.13) .. controls (187.5,180.67) and (157.5,206.25) .. (120.5,206.25) .. controls (83.5,206.25) and (53.5,180.67) .. (53.5,149.13) -- cycle ; 
 \draw   (53.5,149.13) .. controls (53.5,117.58) and (83.5,92) .. (120.5,92) .. controls (157.5,92) and (187.5,117.58) .. (187.5,149.13) .. controls (187.5,180.67) and (157.5,206.25) .. (120.5,206.25) .. controls (83.5,206.25) and (53.5,180.67) .. (53.5,149.13) -- cycle ; 
\draw   (35.28,34.71) -- (98.14,30.21) -- (74.87,83.45) -- (12.02,87.95) -- cycle ;
\draw   (201.8,38.32) -- (137.51,29.18) -- (157.98,82.92) -- (222.28,92.05) -- cycle ;
\draw  [fill={rgb, 255:red, 255; green, 0; blue, 0 }  ,fill opacity=0.39 ] (72.68,154.62) .. controls (68,145.27) and (76.69,131.44) .. (92.1,123.72) .. controls (107.51,116) and (123.8,117.33) .. (128.48,126.68) .. controls (133.16,136.03) and (124.47,149.87) .. (109.06,157.58) .. controls (93.65,165.3) and (77.36,163.97) .. (72.68,154.62) -- cycle ;
\draw  [fill={rgb, 255:red, 74; green, 144; blue, 226 }  ,fill opacity=0.43 ] (112.92,125.29) .. controls (118.07,116.19) and (134.41,115.69) .. (149.4,124.17) .. controls (164.4,132.66) and (172.39,146.91) .. (167.24,156.02) .. controls (162.09,165.12) and (145.75,165.62) .. (130.76,157.13) .. controls (115.76,148.65) and (107.77,134.39) .. (112.92,125.29) -- cycle ;
\draw  [fill={rgb, 255:red, 255; green, 0; blue, 0 }  ,fill opacity=0.39 ] (27.18,73.05) .. controls (22.5,63.7) and (31.19,49.86) .. (46.6,42.15) .. controls (62.01,34.43) and (78.29,35.76) .. (82.98,45.11) .. controls (87.66,54.46) and (78.96,68.3) .. (63.56,76.01) .. controls (48.15,83.73) and (31.86,82.4) .. (27.18,73.05) -- cycle ;
\draw  [fill={rgb, 255:red, 74; green, 144; blue, 226 }  ,fill opacity=0.43 ] (152.74,45.25) .. controls (157.89,36.15) and (174.22,35.65) .. (189.22,44.14) .. controls (204.21,52.62) and (212.2,66.88) .. (207.05,75.98) .. controls (201.9,85.08) and (185.57,85.58) .. (170.57,77.1) .. controls (155.57,68.61) and (147.59,54.36) .. (152.74,45.25) -- cycle ;
\draw    (55.08,59.08) -- (99.61,138.91) ;
\draw [shift={(100.58,140.65)}, rotate = 240.85] [color={rgb, 255:red, 0; green, 0; blue, 0 }  ][line width=0.75]    (10.93,-3.29) .. controls (6.95,-1.4) and (3.31,-0.3) .. (0,0) .. controls (3.31,0.3) and (6.95,1.4) .. (10.93,3.29)   ;
\draw    (179.89,60.62) -- (140.97,138.86) ;
\draw [shift={(140.08,140.65)}, rotate = 296.45] [color={rgb, 255:red, 0; green, 0; blue, 0 }  ][line width=0.75]    (10.93,-3.29) .. controls (6.95,-1.4) and (3.31,-0.3) .. (0,0) .. controls (3.31,0.3) and (6.95,1.4) .. (10.93,3.29)   ;
\draw (55.5,13.4) node [anchor=north west][inner sep=0.75pt]  [font=\footnotesize]  {$\mathbb{R}^{d}$};
\draw (164.5,15.4) node [anchor=north west][inner sep=0.75pt]  [font=\footnotesize]  {$\mathbb{R}^{d}$};
\draw (52,186.4) node [anchor=north west][inner sep=0.75pt]  [font=\footnotesize]  {$M$};
\draw (59.5,90.9) node [anchor=north west][inner sep=0.75pt]  [font=\scriptsize]  {$\phi _{\alpha }$};
\draw (164.5,88.9) node [anchor=north west][inner sep=0.75pt]  [font=\scriptsize]  {$\phi _{\beta }$};
\end{tikzpicture}
\caption[Atlas of a smooth manifold]{Atlas of a smooth manifold.}
\end{center}\end{figure}

\noindent We physicists, in fact, work not directly on a manifold $M$, but on local charts of the form $\mathbb{R}^d$.

\begin{example}[Atlas for ordinary Kaluza-Klein theory]
In ordinary Kaluza-Klein Theory the space to consider is a circle bundle $P$ on the base manifold $M$. On an open cover $\{U_\alpha\}_{\alpha\in I}$ of the base manifold $M$ this is locally trivialised by a collection of local trivial bundles $\{U_\alpha\times U(1)\}_{\alpha\in I}$. From ordinary differential geometry we know that the total space $P$ of the bundle can be covered by local charts $\{\phi_\alpha:\,\mathbb{R}^{d+1}\twoheadrightarrow U_\alpha\times U(1)\subset P\}_{\alpha\in I}$. These charts define an atlas for the total space of the bundle $P$ of the form $\bigsqcup_{\alpha\in I}\mathbb{R}^{d+1}\xtwoheadrightarrow{\;\;\{\phi_\alpha\}_{\alpha\in I}\;\;} P$.
This corresponds to the well-known idea in differential geometry that the total space $P$ locally looks like a Cartesian space $\mathbb{R}^{d+1}$.
\noindent Any such map $\phi_\alpha$ uniquely factorizes as $\phi_\alpha:\,\mathbb{R}^{1,d+1}\xtwoheadrightarrow{\;F\;} \mathbb{R}^{1,d}\times U(1) \xtwoheadrightarrow{\;\varphi_\alpha\;} U_\alpha\times U(1)$, where the first map is just the surjection
\begin{equation}
    F:\,\mathbb{R}^{d+1}\; \longtwoheadrightarrow\; \mathbb{R}^{d}\times U(1),
\end{equation}
which is the identity on $\mathbb{R}^{1,d}$ and the quotient map $\mathbb{R}\twoheadrightarrow U(1)=\mathbb{R}/2\pi\mathbb{Z}$. Crucially, the map $F$ is an atlas of ordinary Lie groups.

\noindent The surjective map $\bigsqcup_{\alpha\in I}\mathbb{R}^{d}\times U(1)\xtwoheadrightarrow{\;\;\{\varphi_\alpha\}_{\alpha\in I}\;\;} P$ is an atlas for the total space $P$, in the stacky sense of the word. This corresponds to the intuitive idea that the total space of a circle bundle locally looks like the Lie group $\mathbb{R}^{d}\times U(1)$.
\end{example}

\begin{remark}[The \v{C}ech nerve of an atlas]
When we defined the atlas $\phi:\,\mathcal{U}\; \longtwoheadrightarrow\; \mathscr{G}$ for the stack $\mathscr{G}$, we said that it must be an \textit{effective epimorphism}. An effective epimorphism is defined as the colimit of a certain \textit{simplicial object} which is called \textit{\v{C}ech nerve}. In other words we have
\begin{equation}
\Big(\begin{tikzcd} \mathcal{U}  \arrow[r, two heads, "\phi"] & \mathscr{G} \end{tikzcd}\Big) \,\;=\;\, \varinjlim\bigg( \begin{tikzcd} \cdots \;\arrow[r, yshift=1.7ex, two heads] \arrow[r, yshift=-0.6ex, two heads] \arrow[r, yshift=0.6ex, two heads] \arrow[r, yshift=-1.7ex, two heads] & \mathcal{U}\times_{\mathscr{G}}\mathcal{U}\times_{\mathscr{G}}\mathcal{U} \arrow[r, two heads] \arrow[r, yshift=1.2ex, two heads] \arrow[r, yshift=-1.2ex, two heads] & \mathcal{U}\times_{\mathscr{G}}\mathcal{U} \arrow[r, yshift=0.7ex, two heads] \arrow[r, yshift=-0.7ex, two heads] & \mathcal{U}
\end{tikzcd} \bigg).
\end{equation}
The \v{C}ech nerve of an atlas can be interpreted as a $\infty$-groupoid, which we will call \textit{\v{C}ech groupoid}. This groupoid encodes the global geometry of the stack in terms of the smooth manifold $\mathcal{U}$, which makes it easier to deal with. Besides the original stack can always be recovered by the colimit of the nerve.
\end{remark}

\noindent How do we construct such a simplicial object? Let us firstly consider the \textit{kernel pair} of the map $\phi$, which is defined as the pullback (in the category theory meaning) of two copies of the map $\phi$. The \textit{coequalizer} diagram of this kernel pair will thus be of the following form:
\begin{equation}\label{eq:kernelpairdef}
\begin{tikzcd}
\mathcal{U}\times_{\mathscr{G}}\mathcal{U} \arrow[r, yshift=0.7ex, two heads] \arrow[r, yshift=-0.7ex, two heads] & \mathcal{U}  \arrow[r, two heads, "\phi"] & \mathscr{G}.
\end{tikzcd}
\end{equation}
This was the first step. By iterating this construction we obtain all the rest of the \v{C}ech nerve.

\begin{example}[\v{C}ech nerve of the atlas of a smooth manifold]
Let us consider, like in the first example, the case where our stack $\mathscr{G}=M$ is just a smooth manifold with an atlas $\bigsqcup_{\alpha\in I}\mathbb{R}^{d}\xtwoheadrightarrow{\;\;\{\phi_\alpha\}_{\alpha\in I}\;\;} M$. In this case the kernel pair, defined in \eqref{eq:kernelpairdef}, is the following:
\begin{equation}
\begin{tikzcd}[column sep=9ex]
\displaystyle\bigsqcup_{\alpha,\beta\in I}\mathbb{R}^d \cap_{\phi} \mathbb{R}^d \arrow[r, yshift=0.7ex, two heads] \arrow[r, yshift=-0.7ex, two heads] & \displaystyle\bigsqcup_{\alpha\in I}\mathbb{R}^d  \arrow[r, two heads, "\{\phi_\alpha\}_{\alpha\in I}"] & M,
\end{tikzcd}
\end{equation}
where we called $\mathbb{R}^d \cap_{\phi} \mathbb{R}^d := \big\{(x,y)\in\mathbb{R}^d\times\mathbb{R}^d|\,  \phi_{\alpha}(x)=\phi_\beta(y)\big\}$. We notice that the kernel pair of the atlas encodes nothing but the information about how the charts are glued together over the manifold $M$. We intuitively have that the global geometry of a smooth manifold $M$ is entirely encoded in its \v{C}ech groupoid $\big(\!\begin{tikzcd}
\bigsqcup_{\alpha,\beta\in I}\mathbb{R}^d \cap_{\phi} \mathbb{R}^d \arrow[r, yshift=0.7ex, two heads] \arrow[r, yshift=-0.7ex, two heads] & \bigsqcup_{\alpha\in I}\mathbb{R}^d
\end{tikzcd}\!\big)$. 
\end{example}

\noindent We physicists are actually very familiar with this perspective: in fact we usually describe our fields as functions on the local charts $\mathbb{R}^d$ of a manifold $M$ and, if we want to describe how they behave globally, we simply write how they transform on the overlaps $\mathbb{R}^d \cap_{\phi} \mathbb{R}^d$ of these charts.
In the next paragraph we will formalize exactly this perspective on fields.

\begin{remark}[Gluing morphisms of stacks]
Given a geometric stack $\mathscr{X}\in\mathbf{H}$ equipped with an atlas $\Phi:\mathcal{U} \longrightarrow \mathscr{X}$, we can write the \v{C}ech nerve of $\Phi$ as the following simplicial object
\begin{equation}
\begin{tikzcd}[column sep=6ex]
\dots \arrow[r, yshift=2ex, two heads]\arrow[r, yshift=0.7ex, two heads]\arrow[r, yshift=-0.7ex, two heads] \arrow[r, yshift=-2ex, two heads] &\mathcal{U } \times_\mathscr{X} \mathcal{U } \times_\mathscr{X} \mathcal{U } \arrow[r, yshift=1.4ex, two heads]\arrow[r, two heads] \arrow[r, yshift=-1.4ex, two heads] &\mathcal{U } \times_\mathscr{X} \mathcal{U } \arrow[r, yshift=0.7ex, two heads] \arrow[r, yshift=-0.7ex, two heads] & \mathcal{U }  \arrow[r, two heads, "\Phi"] & \mathscr{X} .
\end{tikzcd}
\end{equation}
For simplicity, let us now consider just a geometric $1$-stack $\mathscr{X}\in\mathbf{H}$. A complicated object such as a morphism of stacks $\underline{\sigma}:\mathscr{X}\longrightarrow\mathscr{S}$, for some $\mathscr{S}\in\mathbf{H}$, can be equivalently expressed on the atlas $\mathcal{U }$ of the stack $\mathscr{X}$. This can be done as the map induced by the atlas
\begin{equation}
\begin{tikzcd}[column sep=7ex, row sep=7ex]
\mathcal{U } \arrow[r, "\Phi"]\arrow[rr, bend right=35, "\sigma"]& \mathscr{X} \arrow[r, "\underline{\sigma}"] & \mathscr{S}
\end{tikzcd}
\end{equation}
together with an isomorphism of the two maps induced by the kernel pair of the atlas
\begin{equation}
\begin{tikzcd}[column sep=7ex, row sep=7ex]
\mathcal{U } \times_\mathscr{X} \mathcal{U } \arrow[bend left=40, "\sigma"]{r}[name=U,below]{}
\arrow[bend right=40, "\sigma'"']{r}[name=D]{} &
\mathscr{S} \arrow[Rightarrow, to path=(U) -- (D)]{}
\end{tikzcd}
\end{equation}
such that it satisfies the cocycle condition on $\mathcal{U } \times_\mathscr{X} \mathcal{U } \times_\mathscr{X} \mathcal{U }$. For more details, see \cite{Hei05}.
\end{remark}

\noindent The idea of gluing morphisms of stacks on the atlas will be useful in this section, when we will have to consider geometric structures on a bundle gerbe.

\begin{remark}[Gluing a field on a stack]
Let $\mathcal{U}\xtwoheadrightarrow{\;\phi\;}\mathscr{G}$ be an atlas for the stack $\mathscr{G}$ and let $\mathscr{F}$ be another stack, which we will interpret as the moduli-stack of some physical field. Now let $A:\mathscr{G}\rightarrow\mathscr{F}$ be a morphism of stacks (i.e. a physical field on $\mathscr{G}$). We obtain an induced morphism $A\circ\phi:\mathcal{U}\rightarrow \mathscr{F}$ together with an isomorphism between the two induced morphisms $\begin{tikzcd}\mathcal{U}\times_{\mathscr{G}}\mathcal{U} \arrow[r, yshift=0.7ex] \arrow[r, yshift=-0.7ex] &\mathscr{F}\end{tikzcd}$ which satisfies the cocycle condition on $\mathcal{U}\times_{\mathscr{G}}\mathcal{U}\times_{\mathscr{G}}\mathcal{U}$.
\end{remark}

\begin{example}[Gluing a gauge field on a smooth manifold.]
Let $\bigsqcup_{\alpha\in I}\mathbb{R}^{d}\xtwoheadrightarrow{\,\{\phi_\alpha\}_{\alpha\in I}\,}M$ be an atlas for the smooth manifold $M$ and let $\mathscr{F}:=\mathbf{B}G_{\mathrm{conn}}$ be the moduli-stack of Yang-Mills fields with gauge group $G$. Then a gauge field $A:M\rightarrow\mathbf{B}G_{\mathrm{conn}}$ on the smooth manifold induces a local $1$-form $A_{(\alpha)}:=A\circ\phi_{\alpha}\in\Omega^1(\mathbb{R}^d,\mathfrak{g})$ on each chart of the atlas. Notice that these $1$-forms $A_{(\alpha)}(x)$ depends on local coordinates $x\in\mathbb{R}^d$, like we physicists are used. On overlaps of charts we must also have an isomorphism between $A_{(\alpha)}$ and $A_{(\beta)}$ given by a gauge transformation $A_{(\alpha)}= h_{(\alpha\beta)}^{-1}(A_{(\beta)}+\di)h_{(\alpha\beta)}$ with $h_{(\alpha\beta)}\in\Coo(\mathbb{R}^d \cap_{\phi} \mathbb{R}^d,\,G)$. Again, these $h_{(\alpha\beta)}(x)$ are not $G$-valued functions directly on the manifold, but on the atlas. Finally, these isomorphisms must satisfy the cocycle condition $h_{(\alpha\beta)}h_{(\beta\gamma)}h_{(\gamma\alpha)}=1$.
\end{example}

\noindent In this subsection we explained how geometric structures on smooth manifolds become the familiar and more treatable objects on local $\mathbb{R}^d$ coordinates we physicists use. We will see in the next subsection that these intermediate steps become much less trivial if we want to glue local charts for DFT.

\section{Higher geometric quantisation}

Higher geometric quantisation generalises the method of canonical quantisation of ordinary particles to $(n+1)$-dimensional branes, by replacing the ordinary prequantum $U(1)$-bundle with a prequantum bundle $n$-gerbe on the phase space. See \cite{Bunk:2021quu} for an introduction.
For some physical systems, e.g. topological branes, the phase space is itself categorified from a symplectic manifold to a symplectic Lie $\infty$-groupoid \cite{Sev01}. Interestingly, this second form of higher geometric quantisation will be crucial for deriving the BV-BRST quantisation, which can be thought as its infinitesimal version.

\subsection{Prequantum $\infty$-bundles}

A prequantum bundle on a symplectic manifold $(M,\omega)$ is just a principal $U(1)$-bundle $P\twoheadrightarrow M$ whose curvature is $\mathrm{curv}(P)=\omega\in\Omega^2_{\mathrm{cl}}(M)$.
The prequantum Hilbert space is then defined by the vector space
\begin{equation}
    \mathfrak{H}_{\mathrm{pre}} \;:=\; \Gamma(M, P\times_{U(1)}\mathbb{C})
\end{equation}
where we defined the $\mathbb{C}$-associated bundle
\begin{equation}
    E \;:=\; P\times_{U(1)}\mathbb{C}.
\end{equation}
The elements $\Ket{\psi}\in\mathfrak{H}_{\mathrm{pre}}$ of the prequantum Hilbert space are nothing but local wave-functions $\psi_{(\alpha)}\in\Coo(U_\alpha,\mathbb{C})$ on $U_\alpha\subset M$ which are globally patched by
\begin{equation}
    \psi_{(\alpha)} \;=\; e^{if_{(\alpha\beta)}}\psi_{(\beta)},
\end{equation}
where $f_{(\alpha\beta)}:M\rightarrow \mathbf{B}U(1)$ are the transition functions of the prequantum bundle $P\twoheadrightarrow M$.

\begin{definition}[Prequantum $\infty$-Hilbert space]
Let $(M,\omega)$ be an $n$\textit{-plectic manifold} \cite{Rog11,Rog13}, i.e. a smooth manifold $M$ equipped with a closed $(n+1)$-form $\omega\in\Omega^{n+1}_{\mathrm{cl}}(M)$.
A prequantum $\infty$-bundle on an $n$-plectic manifold $(M,\omega)$ is a bundle $n$-gerbe $\mathscr{G}\twoheadrightarrow M$ whose curvature is $\mathrm{curv}(\mathscr{G})=\omega\in\Omega^{n+1}_{\mathrm{cl}}(M)$.
The prequantum $n$-Hilbert space is then defined by the stack
\begin{equation}
    \mathfrak{H}_{\mathrm{pre}} \;:=\; \Gamma\big(M,\,\mathscr{G}\times_{\mathbf{B}^nU(1)}\mathscr{V}\big)
\end{equation}
where we defined the $\mathscr{V}$-associated $\infty$-bundle
\begin{equation}
    \mathscr{E} \;:=\; \mathscr{G}\times_{\mathbf{B}^nU(1)}\mathscr{V}
\end{equation}
where $\mathscr{V}\in\mathbf{H}$ is any stack equipped with a $\mathbf{B}^nU(1)$-action. 
\end{definition}

\begin{example}[Prequantum $2$-Hilbert space]
Notice that there exist a canonical inclusion of unitary groups
\begin{equation}
    U(1) \,\hookrightarrow\, U(2)  \,\hookrightarrow\, \cdots  \,\hookrightarrow\, U(N) \,\hookrightarrow\, U(N+1)  \,\hookrightarrow\, \cdots.
\end{equation}
Let us define the moduli stack $\mathbf{B}U\in\mathbf{H}$ of principal $U(N)$-bundles for any $N\in\mathbb{N}^+$ by
\begin{equation}
    \mathbf{B}U \;:=\; \lim_{N\rightarrow \infty}\mathbf{B}U(N).
\end{equation}
Given a prequantum bundle gerbe $\mathscr{G}$, we can define its $\mathbf{B}U$-associated bundle
\begin{equation}
    \mathscr{E} \;=\; \mathscr{G}\times_{\mathbf{B}U(1)}\mathbf{B}U.
\end{equation}
The prequantum $2$-Hilbert space can thus be identified with the groupoid
\begin{equation}
    \mathfrak{H}_{\mathrm{pre}} \;=\; \Gamma\big(M,\, \mathscr{G}\times_{\mathbf{B}U(1)}\mathbf{B}U \big),
\end{equation}
whose objects are $U(N)$-bundles twisted by $\mathscr{G}$ on the base manifold $M$ and whose morphisms are coboundaries between them.
For a detailed construction of such $2$-Hilbert space (including a categorified notion of Hilbert product), we redirect to the seminal work by \cite{BSS16,BS16} in higher geometric quantisation.
\end{example}

\begin{remark}[Transgression]
Given a prequantum bundle $n$-gerbe $\mathscr{G}\twoheadrightarrow M$ defined by the cocycle
\begin{equation}
    M \;\xrightarrow{\;\;f\;\;}\; \mathbf{B}^{n+1}U(1),
\end{equation}
can be transgressed to a cocycle on the mapping space $[\Sigma_d,M]$, i.e.
\begin{equation}
    [\Sigma_d,M] \;\xrightarrow{\;\;[\Sigma_d,f]\;\;}\; \mathbf{B}^{n-d+1}U(1),
\end{equation}
where $\Sigma_d$ is a $d$-dimensional smooth manifold.
\end{remark}

\begin{example}[Transgression of a prequantum bundle gerbe]
Let a prequantum bundle gerbe $\mathscr{G}\twoheadrightarrow M$ be given by a cocycle $M\xrightarrow{\;f\;} \mathbf{B}^2U(1)$.
We can transgress this cocycle to a $U(1)$-bundle $P$ on the loop space by 
\begin{equation}
    \mathcal{L}M \,=\, [S^1,M] \;\xrightarrow{\;[S^1,f]\;}\; [S^1,\mathbf{B}^2U(1)] \,\cong\, \mathbf{B}U(1).
\end{equation}
Consequently, the $\mathbf{B}U$-associated bundle $\mathscr{G}\times_{\mathbf{B}U(1)}\mathbf{B}U \twoheadrightarrow\ M$ given by a cocycle
\begin{equation}
    M\;\longrightarrow \; \mathbf{B}U/\!/\mathbf{B}U(1)
\end{equation}
is transgressed exactly to an ordinary $\mathbb{C}$-associated bundle $P\times_{U(1)}\mathbb{C}\twoheadrightarrow\mathcal{L}M$ given by the transgressed cocycle
\begin{equation}
     [S^1,M]\;\longrightarrow \; \mathbb{C}/\!/U(1)
\end{equation}
on the loop space $\mathcal{L}M=[S^1,M]$ of the smooth manifold $M$. 
See \cite{SaSza11x,SaSza13} for discussion of higher geometric quantisation and loop space.
\end{example}

\begin{example}[Higher quantisation of stacks]
In many applications, e.g. \cite{Sev01, Fiorenza:2013jz, FSS16}, it is useful to consider a prequantum $\infty$-bundle on a stack $\mathscr{X}\in\mathbf{H}$, i.e. 
\begin{equation}
    \mathscr{X} \;\longrightarrow\; \mathbf{B}^{n+1}U(1).
\end{equation}
In particular, in quantum field theory, it is useful to consider prequantum bundle $n$-gerbes on the moduli stack of $G$-bundles \cite{FRS18, Fiorenza:2013jz, FSS16}, i.e.
\begin{equation}
    \mathbf{B}G \;\longrightarrow\; \mathbf{B}^{n+1}U(1)
\end{equation}
For example, for $n=2$, we can consider the smooth refinement of the $2$nd Chern class, i.e. the map which sends a $G$-bundle with curvature $F$ to a bundle $2$-gerbe whose curvature is the $4$-form $\mathrm{tr}(F\wedge F)$.
Recall that we can transgress the prequantum $\infty$-bundle to an ordinary prequantum bundle of the form
\begin{equation}
    [\Sigma_2,\mathbf{B}G] \;\longrightarrow\; \mathbf{B}U(1),
\end{equation}
where $\Sigma_2$ is a smooth surface.
This picture is closely related to the quantisation of $3$d Chern-Simons theory. In fact, if we transgress the prequantum $\infty$-bundle by using a $3$-dimensional manifold $\Sigma_3$, we obtain the map
\begin{equation}
    [\Sigma_3,\mathbf{B}G] \;\longrightarrow\; U(1),
\end{equation}
which sends a gauge field $A$ on $\Sigma_3$ to the element $\exp\big(i\int_{\Sigma_3}\!\mathrm{cs}_3(A)\big)\in U(1)$.
\end{example}

\subsection{BRST complex}

Let $M$ be a smooth manifold. Let $\mathcal{F} := [M,\mathbf{B}G_{\mathrm{conn}}]$ be the groupoid of gauge fields with structure Lie group $G$ on $M$.
In particular, let us consider a base manifold $M=U$, which is just an open set, so that we can rewrite the $\infty$-groupoid of fields $\mathcal{F}$ as an action $\infty$-groupoid
\begin{equation}
    \mathcal{F}|_U \;\cong\; \Omega^1(U,\mathfrak{g})/\!/_{\!\rho\,}\Coo(U,G),
\end{equation}
where the action $\rho:\Coo(U,G)\times\Omega^1(U,\mathfrak{g})\rightarrow \Omega^1(U,\mathfrak{g})$ is given by a gauge transformation $(\lambda,\,A)\longmapsto \lambda^{-1}(A + \di)\lambda$ with gauge parameter $\lambda\in\Coo(U,G)$.
Thus, its Lie differentiation $\mathfrak{a}_{\mathrm{BRST}} \;:=\; \mathrm{Lie}(\mathcal{F}|_U)$ is an action $L_\infty$-algebroid of the form
\begin{equation}
    \mathfrak{a}_{\mathrm{BRST}} \;\cong\; \Omega^1(U,\mathfrak{g})/\!/_{\!\rho\,}\Coo(U,\mathfrak{g}),
\end{equation}
where objects are gauge fields $A\in\Omega^1(U)$ and the infinitesimal action is given by infinitesimal gauge transformations $\delta_\lambda A = \di\lambda + [\lambda, A]_\mathfrak{g}$ with infinitesimal parameter $\lambda\in\Coo(\mathfrak{g})$.
The Chevalley-Eilenberg dg-algebra of the $L_\infty$-algebroid $\mathfrak{a}_{\mathrm{BRST}}$ is exactly the BRST complex
\begin{equation}
    \mathrm{CE}(\mathfrak{a}_{\mathrm{BRST}}) \;=\; \mathbb{R}\!\left[A,c\right]/ \!\left(\begin{array}{l} QA \,=\, \mathrm{D}_Ac,\\[0.8ex] Qc \;=\, -\frac{1}{2}[c,c]_\mathfrak{g}\end{array}\right).
\end{equation}
Thus, we are interested in the cohomology $H^n_{\mathrm{Lie}}(\mathfrak{a}_{\mathrm{BRST}})= H^n\big(\mathrm{CE}(\mathfrak{a}_{\mathrm{BRST}})\big)$.
Now, let us rename the Lie algebras $V:=\Omega^1(U,\mathfrak{g})$ and $\hat{\mathfrak{g}}:=\Coo(U,\mathfrak{g})$. We have the isomorphism $H^n_{\mathrm{Lie}}(\mathfrak{a}_{\mathrm{BRST}})\cong H^n_{\mathrm{Lie}}(V/\!/_{\!\rho\,}\hat{\mathfrak{g}}) \cong H^n_{\mathrm{Lie}}(\hat{\mathfrak{g}};V)$, where in the last cohomology group $V$ is regarded as a $\hat{\mathfrak{g}}$-module. This way, we make contact with the more common definition in terms of cohomology of $\hat{\mathfrak{g}}$ valued in the $\hat{\mathfrak{g}}$-module $V$.

\subsection{Batalin–Vilkovisky quantisation}

Let us define the NQ-manifold $\mathcal{M}_{\mathrm{BRST}}$ corresponding to the BRST-algebroid, i.e.
\begin{equation}
    \Coo(\mathcal{M}_{\mathrm{BRST}}) \;:=\; \mathrm{CE}\big(\mathfrak{a}_{\mathrm{BRST}}\big).
\end{equation}
We can encode the field equations by a closed $1$-form $\varepsilon\in\Omega^1_{\mathrm{cl}}(\mathcal{M}_{\mathrm{BRST}})$. Indeed, we can locally express the differential of this $1$-form by
\begin{equation}
    \varepsilon \;=\; \delta S \;=\; \frac{\delta S}{\delta \Phi^i} \,\delta\Phi^i,
\end{equation}
where $\delta$ is the functional derivative, $S[\Phi]$ is the action and the fields $\{\Phi^i(x)\}$ are the coordinates of $\mathcal{M}_{\mathrm{BRST}}$.
We can also define $\mathcal{F}_{\mathrm{shell}}:=\mathrm{ker}(\varepsilon)\subset \mathcal{F}$, which is exactly the sub-$\infty$-groupoid of fields which satisfy the Euler-Lagrange equations $\frac{\delta S}{\delta \Phi^i}=0$. \vspace{0.2cm}

\noindent Now, we can define the phase space of the BV-BRST quantisation by
\begin{equation}
    \mathcal{M}_{\mathrm{BV}} \;:=\; T^\ast[-1] \mathcal{M}_{\mathrm{BRST}},
\end{equation}
i.e. as a $(-1)$-shifted cotangent bundle of $\mathcal{M}_{\mathrm{BRST}}$.
Let $\{\Phi(x),\Phi^+(x)\}$ be the Darboux coordinates of the phase space $\mathcal{M}_{\mathrm{BV}}$, where $\Phi^+(x)$ can be interpreted as antifields.
We can introduce a vector field
\begin{equation}
    Q_\varepsilon \;:=\; \frac{\delta S}{\delta \Phi^i}\, \frac{\delta}{\delta \Phi_i^+} \;\,\in\,\mathfrak{X}(\mathcal{M}_{\mathrm{BV}}),
\end{equation}
which maps the antifields to
\begin{equation}
    Q_\varepsilon \Phi^+_i \;=\; \frac{\delta S}{\delta \Phi^i}.
\end{equation}
The dg-manifold $\mathcal{M}_\mathrm{BV}$ comes canonically equipped with a symplectic form
\begin{equation}
    \omega_{\mathrm{BV}} \;:=\; \int_U\di^nx\,\Big(\delta\Phi_i^+(x)\wedge \delta\Phi^i(x)\Big) \;\,\in\; \Omega^2_{\mathrm{cl}}(\mathcal{M}_{\mathrm{BV}}),
\end{equation}
which satisfies the equation $\iota_{Q_\varepsilon}\omega_{\mathrm{BV}} = \delta S$.
Now, we can define the vector
\begin{equation}
    Q_{\mathrm{BV}} \;:=\; Q_\varepsilon + Q_{\mathrm{BRST}},
\end{equation}
which satisfies the Hamilton equation
\begin{equation}
    \iota_{Q_{\mathrm{BV}}}\omega_{\mathrm{BV}} \;=\; \delta S_{\mathrm{BV}},
\end{equation}
where $Q_{\mathrm{BV}}$ plays the role of the Hamiltonian vector field and the Batalin–Vilkovisky action $S_{\mathrm{BV}}$ the one of the Hamiltonian. This is a functional $S_{\mathrm{BV}}[\Phi,\Phi^+]\in\Coo(\mathcal{M}_{\mathrm{BV}})$ of the form
\begin{equation}
    S_{\mathrm{BV}}[\Phi,\Phi^+] \;=\; \sum_{k\in\mathbb{N}}\frac{1}{(n+1)!}\langle \Phi^+,[\Phi,\cdots,\Phi]_k \rangle.
\end{equation}
Let us now focus on the concrete example of a Yang-Mills theory, so that we have fields $\{\Phi,\Phi^+\}:=\{A,c,A^+,c^+\}$ with the $0$-degree $A\in\Omega^1(U,\mathfrak{g})$, $1$-degree ghost $c\in\Coo(U,\mathfrak{g})$, $(-1)$-degree antifield $A^+\in\Omega^{n-1}(U,\mathfrak{g})$ and $(-2)$-degree antighost $c^+\in\Omega^n(U,\mathfrak{g})$.
In this context, the symplectic form becomes
\begin{equation}
    \omega_{\mathrm{BV}} \;=\; \int_U\di^nx\,\Big(\delta A^+_i \wedge \delta A^i + \delta c^+_i \wedge \delta c^i\Big).
\end{equation}
The Batalin–Vilkovisky action will be of the form
\begin{equation}
    S_{\mathrm{BV}}[A,c,A^+,c^+] \;=\; \int_U\left(\frac{1}{2}\kappa_{ij} F^i \wedge \star F^j - A^+_i \wedge \mathrm{D}_Ac^i + \frac{1}{2} c^+_i [c,c]^i \right).
\end{equation}
Finally, notice that the dg-algebra of function on the phase space is given by
\begin{equation}
    \Coo(\mathcal{M}_{\mathrm{BV}}) \;=\; \mathbb{R}\!\left[A,c,A^+,c^+\right]/ \!\left(\begin{array}{l} Q_{\mathrm{BV}}A \,=\, \mathrm{D}_Ac,\\[0.6ex] 
Q_{\mathrm{BV}}c \,=\, -\frac{1}{2}[c,c],\\[0.6ex] 
Q_{\mathrm{BV}}A^+ \,=\, -\mathrm{D}_A \star F - [c,A^+],\\[0.8ex] 
Q_{\mathrm{BV}}c^+ \,=\, -\mathrm{D}_A A^+ - [c,c^+]\end{array}\right).
\end{equation}
For a detailed discussion of BV-BRST quantisation in the context of higher geometry, we redirect to \cite{Benini:2019hoc}.
\begin{savequote}[8cm]
The history of science is the history of mankind’s unity, of its sublime purpose, of its gradual redemption.
  \qauthor{--- George Sarton, \textit{Introduction to the History of Science}}
\end{savequote}

\chapter{\label{ch:4}Review of proposals of doubled geometry}

\minitoc

\noindent In this chapter we will give a brief overview on the main proposals for a formulation of the doubled geometry underlying Double Field Theory. We will underline the relations between the approaches and we will discuss some open problems.

\section{Non-associative proposal}

The non-associative proposal was presented by \cite{HohZwi12} and further developed by \cite{Hohm13space}. Its aim is to realise the group of gauge transformations of DFT by diffeomorphisms of the doubled space. However, since the C-bracket structure on doubled vectors does not satisfy the Jacobi identity, its exponentiation will not give us a Lie group, but a geometric object which does not satisfy the associativity property.\vspace{0.2cm}

\noindent In the proposal by \cite{Hohm13space} the doubled space $\mathcal{M}$ is just a $2d$-dimensional smooth manifold. This means that we can consider a cover $\{\mathcal{U}_\alpha\}$, so that $\bigcup_\alpha \mathcal{U}_\alpha = \mathcal{M}$, and glue the coordinate patches on each two-fold overlap $\mathcal{U}_\alpha \cap \mathcal{U}_\beta$ of the doubled space by diffeomorphisms $x_{(\beta)} = f_{(\alpha\beta)}\big(x_{(\alpha)}\big)$. Vectors of the tangent bundle $T\mathcal{M}$ will be then glued on each $T(\mathcal{U}_\alpha \cap \mathcal{U}_\beta)$ by the $GL(2d)$-valued Jacobian matrix $J_{(\alpha\beta)} := {\partial x_{(\alpha)}}/{\partial x_{(\beta)}}$. However these transformations do not work for doubled vectors from DFT, thus \cite{Hohm13space} proposed that the doubled vectors should transform by the $O(d,d)$-valued matrix given by
\begin{equation}\label{eq:fmatrix}
    \mathcal{F}_{(\alpha\beta)} \;:=\; \frac{1}{2}\Big(J_{(\alpha\beta)}J^{-\mathrm{T}}_{(\alpha\beta)}+J^{-\mathrm{T}}_{(\alpha\beta)}J_{(\alpha\beta)}\Big),
\end{equation}
which indeed preserves the $O(d,d)$-metric $\eta:=\eta_{MN}\di x^M \otimes \di x^N$. Now, if we go to the three-fold overlaps of patches $\mathcal{U}_\alpha \cap \mathcal{U}_\beta \cap \mathcal{U}_\gamma$ we realise that these transition functions do not satisfy the expected cocycle condition. In other words we generally have
\begin{equation}\label{eq:noncocycle}
    \mathcal{F}_{(\alpha\beta)}\,\mathcal{F}_{(\beta\gamma)}\,\mathcal{F}_{(\gamma\alpha)} \,\neq\, 1. 
\end{equation}
Notice that for the first time we see something resembling a gerbe-like structure spontaneously emerging in DFT geometry.

\subsection{Modified exponential map}
The solution proposed by \cite{Hohm13space} consists, first of all, in a modified exponential map ${\exp}\Theta:\mathfrak{X}(\mathcal{U})\rightarrow \Diff(\mathcal{U})$. This will map any vector $X\in\mathfrak{X}(\mathcal{U})$ in the diffeomorphism given by
\begin{equation}\label{eq:thetaexp}
    x'= e^{\Theta(X)} x \;\quad\; \text{with} \;\quad\; \Theta(X)^M \,:=\, X^M + \underbrace{\sum_i \rho_i \partial^M \chi_i}_{\mathcal{O}(X^3)},
\end{equation} \vspace{-0.46cm}

\noindent where $\rho_i$ and $\chi_i$ are functions on $x$ depending on the vector $X$ in a way which guarantees that $\Theta(X)^M\partial_M = X^M\partial_M$ when applied to any field satisfying the strong constraint. This modified diffeomorphism crucially agrees with the gauge transformation $V'(x)=e^{\mathfrak{L}_X}V(x)$ of DFT, where $\mathfrak{L}_X$ is the generalised Lie derivative defined by the D-bracket.

\subsection{$\star$-product and non-associativity}
In ordinary differential geometry the exponential map $\exp:\big(\mathfrak{X}(\mathcal{U}),\,[-,-]\big)\rightarrow \big(\Diff(\mathcal{U}),\,\circ\,\big)$ maps a vector $X\mapsto e^X$ into the diffeomorphism that it generates. The usual exponential map notoriously satisfies the property $e^{X}\circ e^{Y} = e^{Z}$ with $Z\in\mathfrak{X}(\mathcal{U})$ given by the Baker-Campbell-Hausdorff series $Z=X+Y+[X,Y]/2+\dots$ for any couple of vectors $X,Y\in\mathfrak{X}(\mathcal{U})$.
The idea by \cite{Hohm13space} consists in equipping the space of vector fields $\mathfrak{X}(\mathcal{U})$ with another bracket structure $\big(\mathfrak{X}(\mathcal{U}),\,\llbracket -,-\rrbracket_{\mathrm{C}}\big)$, where $\llbracket -,-\rrbracket_{\mathrm{C}}$ is the C-bracket of DFT. Now this algebra can be integrated by using the modified exponential map ${\exp}\Theta$ defined in \eqref{eq:thetaexp} to a quasigroup $\big(\Diff(\mathcal{U}),\,\star\,\big)$ that satisfies
\begin{equation}
    e^{\Theta(X)} \star e^{\Theta(Y)} = e^{\Theta(Z)} \;\quad\; \text{with} \;\quad\; Z=X+Y+\frac{1}{2}\llbracket X,Y\rrbracket_{\mathrm{C}}+\dots
\end{equation}
It is possible to check that this $\star$-product is \textit{not associative}: in other words the inequality
\begin{equation}
    (f\star g) \star h \,\neq\, f\star (g \star h),
\end{equation}
where $f,g,h\in\Diff(\mathcal{U})$ are diffeomorphisms, generally holds. Now let us call the diffeomorphisms $f:=e^{\Theta(X)}$, $g:=e^{\Theta(Y)}$ and $h:=e^{\Theta(Z)}$ obtained by exponentiating three vectors $X,Y,Z\in\mathfrak{X}(\mathcal{U})$. Then the obstruction of the $\star$-product from being associative is controlled by an element $W$ which satisfies the equation
\begin{equation}
    (f \star g)\star h \,=\, W \star \big( f \star (g\star h) \big)
\end{equation}
and which is given by $W=\exp\Theta\!\left(-\frac{1}{6}\mathcal{J}(X,Y,Z)+\dots\right)$, where $\mathcal{J}(-,-,-)$ is the Jacobiator of the C-bracket. Even if it is well-known that the Jacobiator is of the form $\mathcal{J}^M=\partial^M\mathcal{N}$ for a function $\mathcal{N}\in\Coo(\mathcal{U})$, notice that the transformation $W$ is non-trivial. Also if we consider diffeomorphisms on doubled space which satisfy $f_{(\alpha\beta)}\star f_{(\beta\gamma)}=f_{(\alpha\gamma)}$, we re-obtain the desired property $\mathcal{F}_{(\alpha\beta)} \mathcal{F}_{(\beta\gamma)}=\mathcal{F}_{(\alpha\gamma)}$ for doubled vectors.
\vspace{0.25cm}

\noindent We know that the diffeomorphisms group of the doubled space is not homeomorphic to the group $ G_{\mathrm{DFT}}$ of DFT gauge transformations $e^{\mathfrak{L}_X}$. But now, by replacing the composition of diffeomorphisms with the $\star$-product, we can define a homomorphism 
\begin{equation}
    \varphi:\big(\Diff(\mathcal{U}),\,\star\,\big) \,\longrightarrow\, G_{\mathrm{DFT}},
\end{equation}
which therefore satisfies the property
\begin{equation}
    \varphi(f\star g) \,=\, \varphi(f)\,\varphi(g).
\end{equation}
This property determines the $\star$-product up to trivial gauge transformation. In the logic of \cite{Hohm13space} this will allow to geometrically realise DFT gauge transformation as diffeomorphisms of the doubled space. 

\section{Proposal with gerbe-like local transformations}

The first paper in the literature explicitly recognising the higher geometrical property of DFT is \cite{BCM14}. In the reference it is argued that we can overcome many of the difficulties of the non-associative proposal by describing the geometry of DFT modulo local $O(d,d)$-transformations.\vspace{0.2cm}

\noindent Their proposal starts from the same problem \eqref{eq:noncocycle}, but proposes a different solution. 
We can rewrite the C-bracket of doubled vectors by $\llbracket X,Y \rrbracket_{\mathrm{C}}= [X,Y]_{\mathrm{Lie}}+ \lambda(X,Y)$ where we called $\lambda^M(X,Y) := X^N\partial^MY_N$. This means that we can rewrite the algebra of DFT gauge transformations as $[\mathfrak{L}_X,\mathfrak{L}_Y] = \mathfrak{L}_{[X,Y]_{\mathrm{Lie}}}+\Delta(X,Y)$ where we defined $\Delta(X,Y) := \mathfrak{L}_{\lambda(X,Y)}$. In \cite{BCM14} it is noticed that the extra $\Delta$-transformation appearing in the DFT gauge algebra is \textit{non-translating}, i.e. it involves no translation term if acting on tensors satisfying the strong constraint. Thus the diffeomorphism $e^X$ and the gauge transformation $e^{\mathfrak{L}_X}$ agree up to a local transformation $e^{\Delta}=1+\Delta$.
In fact, if we impose the strong constraint on fields and parameters 
\begin{equation}
    \Delta_M^{\;\;N} \;=\; \begin{pmatrix}0 & 0 \\ \partial_{[\mu} \tilde{\lambda}_{\nu]} & 0 \end{pmatrix},
\end{equation}
where $\tilde{\lambda}_\mu = X^N\partial_\mu Y_N$ depends only on the $d$-dimensional physical subset $U\subset\mathcal{U}$ of our doubled space patch. Hence the local $\Delta$-transformation is just an infinitesimal gauge transformation $\mathfrak{L}_{\tilde{\lambda}} B=\di\tilde{\lambda}$ of the Kalb-Ramond field. 

\paragraph{Further discussion.} As noticed by \cite{Hull14}, $\Delta$-transformations are integrated on a patch $\mathcal{U}$ to the group $\Omega^1(U)$ of finite gauge transformations of the Kalb-Ramond field, while full gauge transformations generated by a strong constrained doubled vectors are integrated to $\Diff(U)\ltimes\Omega^1(U)\subset \Diff(\mathcal{U})$. Now we notice that the group of DFT gauge transformations effectively becomes the homotopy quotient $G_{\mathrm{DFT}}=\big(\Diff(U)\ltimes\Omega^1(U)\big)/\!/\Omega^1(U)$, thus a $2$-group.
\vspace{0.25cm}

\noindent The doubled space is still a $2d$-dimensional manifold $\mathcal{M}$ and then its coordinate patches on each two-fold overlap $\mathcal{U}_\alpha \cap \mathcal{U}_\beta$ are still glued by by diffeomorphisms $x_{(\beta)} = f_{(\alpha\beta)}\big(x_{(\alpha)}\big)$. The doubled vectors are still glued by the $O(d,d)$-valued matrix $\mathcal{F}_{(\alpha\beta)}$ defined in \eqref{eq:fmatrix}, like in the non-associative proposal. Now, according to \cite{BCM14}, on three-fold overlaps of patches $\mathcal{U}_\alpha\cap \mathcal{U}_\beta \cap \mathcal{U}_\gamma$ the transition functions of doubled vectors satisfy 
\begin{equation}\label{eq:bbb}
    \mathcal{F}_{(\alpha\beta)}\,\mathcal{F}_{(\beta\gamma)} \,=\, \mathcal{F}_{(\alpha\gamma)}e^{\Delta_{(\alpha\beta\gamma)}},
\end{equation}
i.e. they satisfy the desired transitive property up to a local $\Delta$-transformation.
In a more mathematical language we can say that \textit{doubled vectors would be sections of a stack} on the $2d$-dimensional manifold $\mathcal{M}$. This is not surprising since the algebra of $G_{\mathrm{DFT}}$ is of the form $\big(\mathfrak{X}(U)\oplus\Omega^1(U)\big)/\!/\Omega^1(U)$, which then must be glued on overlaps of patches by $B$-shifts $\di \tilde{\lambda}$. Thus we could replace the concept of non-associative transformations with a gerbe-like structure.

\section{Doubled-yet-gauged space proposal}

The idea of \textit{doubled-yet-gauged space} was proposed by \cite{Park13} as a solution for the discrepancy between finite gauge transformations and diffeomorphisms of the doubled space, in alternative to the non-associative proposal. Then it was further explored in \cite{Par13x}, where a covariant action was obtained, and in \cite{Par16x}, where it was generalised to the super-string case. Very intriguingly this formalism led to novel non-Riemannian backgrounds in \cite{Par17xx}, \cite{Par18xx} and \cite{Par19xx}. Recently a BRST formulation for the action of a particle on the doubled-yet-gauged space has been proposed by \cite{Par19x} and related to the NQP-geometry involved by other proposals.

\subsection{The coordinate gauge symmetry}
In doubled-yet-gauged space proposal the doubled space $\mathcal{M}$ is, at least locally, a smooth manifold. A local $2d$-dimensional coordinate patch $\mathcal{U}$ is characterised by \textit{coordinate symmetry}, i.e. there exists a canonical gauge action on its local coordinates expressed by
\begin{equation}
    x^M \;\sim\; x^M + \sum_i \rho_i \partial^M \chi_i(x)
\end{equation}
for any choice of functions $\rho_i,\chi_i\in\Coo(\mathcal{U})$.
This observation is motivated by the fact that any strong constrained tensor satisfies the identity $T_{A_1\dots A_n}\big(x+\lambda(x)\big) = T_{A_1\dots A_n}(x)$ where we called $\lambda^M := \sum_i \rho_i \partial^M \chi_i$ at any point $x\in\mathcal{U}$. \vspace{0.25cm}

\noindent Let us choose coordinates for our doubled patch $\mathcal{U}$ such that the strong constraint is solved by letting all the fields and parameters depend only on the $d$-dimensional subpatch $U\subset\mathcal{U}$. Then the coordinate symmetry on the doubled space reduces to
\begin{equation}
    \big(x^\mu,\,\tilde{x}_\mu\big) \;\sim\; \big(x^\mu,\,\tilde{x}_\mu+\tilde{\lambda}_\mu(x)\big),
\end{equation}
where $\tilde{\lambda}_\mu = \sum_i \rho_i \partial_\mu \chi_i$. This coordinate symmetry, similarly to the $\Delta$-transformations in the previous proposal, can be identified with the local gauge symmetry of the Kalb-Ramond field by $\delta_{\tilde{\lambda}}B=\di\tilde{\lambda}$, where the parameter is exactly $\tilde{\lambda}:=\tilde{\lambda}_\mu\di x^\mu$. We can thus identify the physical $d$-dimensional patches with the quotients $U_\alpha\,\cong\,\mathcal{U}_\alpha/\sim$. Thus, as argued by \cite{Park13}, physical spacetime points must be identified with gauge orbits of the doubled-yet-gauged space.\vspace{0.25cm}

\noindent The coordinate gauge symmetry is also the key to solve the discrepancy between DFT gauge transformations $e^{\mathfrak{L}_V}$ and diffeomorphisms $e^V$. Indeed, as argued by \cite{Park13}, the two exponentials induce two finite coordinate transformations $x^M\mapsto x^{\prime M}$ and $x^M\mapsto x^{\prime\prime M}$ whose ending points are coordinate gauge equivalent, i.e. $x^{\prime M}\sim x^{\prime\prime M}$. Therefore, upon section constraint, they differ just by a Kalb-Ramond field gauge transformation. 

\paragraph{Further discussion: how can we globalise?}
Now, the doubled-yet-gauged formalism encompasses the local geometry of the doubled space. However in this review we are interested in the global aspects of DFT, so we may try to understand how these doubled patches can be glued together. Let us first try a na\"{i}ve approach, for pedagogical reasons: we will try to glue our doubled patches by diffeomorphisms that respect the section condition, i.e. on two-fold overlaps of patches $\mathcal{U}_\alpha \cap \mathcal{U}_\beta$ we will have
\begin{equation}\label{eq:patchingpa}
    x_{(\beta)}\,=\,f_{(\alpha\beta)}\big(x_{(\alpha)}\big), \qquad \tilde{x}_{(\beta)} \,=\, \tilde{x}_{(\alpha)} + \Lambda_{(\alpha\beta)}\big(x_{(\alpha)}\big).
\end{equation}
This would imply the patching conditions $B_{(\beta)} = f^\ast_{(\alpha\beta)}B_{(\alpha)} + \di\Lambda_{(\alpha\beta)}$ where the local $1$-forms $\Lambda_{(\alpha\beta)}:=\Lambda_{(\alpha\beta)\mu} \di x_{(\beta)}^\mu$ are given by the gluing conditions \eqref{eq:patchingpa}. But then, with these assumptions, the doubled space $\mathcal{M}$ would become just the total space $(\mathbb{R}^{d})^\ast$-bundle on the physical $d$-dimensional spacetime $M$. If we compose the transformations \eqref{eq:patchingpa} on three-fold overlaps of patches $\mathcal{U}_\alpha \cap \mathcal{U}_\beta \cap \mathcal{U}_\gamma$ we immediately obtain the cocycle condition $\Lambda_{(\alpha\beta)} + \Lambda_{(\beta\gamma)} + \Lambda_{(\gamma\alpha)} = 0$, which is the cocycle describing a topologically trivial gerbe bundle with $[H]=0\in H^3(M,\mathbb{Z})$ and not a general string background. Therefore this na\"{i}ve attempt at gluing by using the coordinate gauge symmetry is not enough. \vspace{0.2cm}

\noindent The doubled-yet-gauged formalism gives us an unprecedented interpretation of the coordinates of DFT. Upon choice of coordinates which are compatible with the section constraint, indeed, the coordinate gauge symmetry can be identified with the gauge transformations of the Kalb-Ramond field. This is a fundamental link between the geometry of the bundle gerbe formalising the Kalb-Ramond field and the geometry of the doubled space. This also provides an interesting link with ordinary Kaluza-Klein geometry, where the points of the base manifold of a $G$-bundle are in bijection with the gauge $G$-orbits of the bundle. As we will see in chapter \ref{ch:5}, the local coordinate gauge symmetry which was discovered by \cite{Park13} will be also recovered as fundamental property of the double space which arises from the Higher Kaluza-Klein perspective. We will see that the Higher Kaluza-Klein formalism recovers a globalised version of the doubled-yet-gauged space with gluing conditions which are a gerby version of the na\"{i}ve patching conditions \eqref{eq:patchingpa}. Therefore the Higher Kaluza-Klein proposal can be seen also as a proposal of globalization of the doubled-yet-gauged space approach.

\section{Finite gauge transformations proposal}

In \cite[pag.$\,$23]{Hull14} it was proposed that, given a geometric background $M$, the group of gauge transformations of DFT should be just
\begin{equation}\label{eq:ghull}
    G_{\mathrm{DFT}}\;=\;\Diff(M)\ltimes \Omega^2_{\mathrm{cl}}(M),
\end{equation}
i.e. diffeomorphisms of the manifold $M$ and $B$-shifts. 
In particular it was argued that any try of realising the group of gauge transformations of DFT as diffeomorphisms of a $2d$-dimensional space should fail, because it is not homomorphic to the group of diffeomorphisms.\vspace{0.2cm}

\noindent In \cite[pag.$\,$20]{Hull14} it was then proposed that \textit{double vectors on a geometric background $M$ are just sections of a Courant algebroid $E\twoheadrightarrow M$ twisted by a bundle gerbe}. In other words, on any patch $U_\alpha$ of the manifold $M$, a doubled vector would be of the form
\begin{equation}\label{eq:vec}
    V_{(\alpha)} \;=\; \begin{pmatrix}1 & 0 \\ -B_{(\alpha)} & 1\end{pmatrix}  \begin{pmatrix} v_{(\alpha)} \\ \tilde{v}_{(\alpha)} \end{pmatrix} \;=\; \begin{pmatrix} v^\mu_{(\alpha)} \\[0.3em] \tilde{v}_{(\alpha)\mu} + B_{(\alpha)\mu\nu}v^\nu_{(\alpha)}\end{pmatrix}.
\end{equation}
It was also shown by \cite[pag.$\,$23]{Hull14} that the $O(d,d)$-matrix \eqref{eq:fmatrix} transforming double vectors under a finite gauge transformation of DFT, i.e. a diffeomorphism $x_{(\beta)}=f_{(\alpha\beta)}\big(x_{(\alpha)}\big)$ and a $B$-shift $\di\lambda_{(\alpha\beta)}$, reduces to 
\begin{equation}
    \mathcal{F}_{(\alpha\beta)} \;=\; \begin{pmatrix}j_{(\alpha\beta)} & 0 \\ 0 & j^{-\mathrm{T}}_{(\alpha\beta)}\end{pmatrix} \begin{pmatrix}1 & 0 \\ -B_{(\alpha)} & 1\end{pmatrix}\begin{pmatrix}1 & 0 \\ \di\lambda_{(\alpha\beta)} & 1\end{pmatrix},
\end{equation}
where we called $j_{(\alpha\beta)}:= \partial x_{(\beta)}/\partial x_{(\alpha)}$ the Jacobian matrix of the diffeomorphism. This way it is natural to recover equation \eqref{eq:bbb}, i.e.
\begin{equation}
    \mathcal{F}_{(\alpha\beta)}\mathcal{F}_{(\beta\gamma)}\mathcal{F}_{(\gamma\alpha)} \;=\; e^{\Delta_{(\alpha\beta\gamma)}},
\end{equation}
where $e^{\Delta_{(\alpha\beta\gamma)}}$ will generally be a non-trivial local $B$-shift.

\paragraph{Further discussion.} 
This proposal clarifies the previous ones by prescribing that, whenever the strong constraint can be globally solved by letting the fields depend on a $d$-dimensional submanifold $M$, doubled vectors (which encode the infinitesimal symmetries of Double Field Theory) must be seen as sections of a Courant algebroid twisted by a bundle gerbe on $M$. For non-geometric backgrounds, however, this picture holds only locally.

\section{C-space proposal}

The idea of C-spaces was born in \cite{Pap13}, matured in \cite{Pap14} and further explored in \cite{HowPap17} in relation to topological T-duality. This was the first proposal to suggest that a global double space should consist of the total space of a bundle gerbe, equipped with a particular notion of coordinates, which was renamed \textit{C-space}.\vspace{0.2cm}

\noindent The notation $\mathscr{C}_M^{[H]}$ for a C-space makes explicit that it is topologically classified only by the base manifold $M$ and by the Dixmier-Douady class $[H]\in H^3(M,\mathbb{Z})$, i.e. the H-flux.
According to \cite{Pap14} we can introduce two sets of coordinates for a C-space $\mathscr{C}_M^{[H]}\longtwoheadrightarrow M$ on some base manifold $M$. According to \cite{Pap13,Pap14,HowPap17}, we must consider new coordinates $y^1_{(\alpha)}$ on each patch $U_\alpha$ and $\theta_{(\alpha\beta)}$ on each two-fold overlap of patches $U_\alpha\cap U_\beta$ of $M$. Let us now recall that the differential data of a bundle gerbe on $M$ is specified by a \v{C}ech cocycle $\big(B_{(\alpha)},\Lambda_{(\alpha\beta)},G_{(\alpha\beta\gamma)}\big)$, with $B_{(\alpha)}\in\Omega^2(U_\alpha)$, $\Lambda_{(\alpha\beta)}\in\Omega^1(U_\alpha\cap U_\beta)$ and $G_{(\alpha\beta\gamma)}\in\Coo(U_\alpha\cap U_\beta\cap U_\gamma)$ which satisfy
\begin{equation}\label{eq:cgerbe}
    \begin{aligned}
        B_{(\beta)}-B_{(\alpha)} &= \mathrm{d}\Lambda_{(\alpha\beta)} ,\\
        \Lambda_{(\alpha\beta)}+\Lambda_{(\beta\gamma)}+\Lambda_{(\gamma\alpha)} &= \mathrm{d}G_{(\alpha\beta\gamma)} ,\\
        G_{(\alpha\beta\gamma)}-G_{(\beta\gamma\delta)}+G_{(\gamma\delta\alpha)}-G_{(\delta\alpha\beta)}&\in 2\pi\mathbb{Z}.
    \end{aligned}
\end{equation}
Then, the extra coordinates $\big(y^1_{(\alpha)},\, \theta_{(\alpha\beta)}\big)$, according to \cite{HowPap17}, have "the degree of a $1$-form" and of a scalar, and must be then glued on two-fold and three-fold overlaps of patches of $M$ by using the transition functions of the gerbe, i.e. by
\begin{equation}\label{eq:papapatch}
    \begin{aligned}
        - y^1_{(\alpha)} + y^1_{(\beta)} +\di\theta_{(\alpha\beta)} \,&=\, \Lambda_{(\alpha\beta)}, \\
        \theta_{(\alpha\beta)} + \theta_{(\beta\gamma)} + \theta_{(\gamma\alpha)} \,&=\, G_{(\alpha\beta\gamma)}\;\mathrm{mod}\,2\pi\mathbb{Z}.
    \end{aligned}
\end{equation}
With this identification, a change of coordinates $\big(y^1_{(\alpha)},\, \theta_{(\alpha\beta)}\big) \mapsto\big(y^1_{(\alpha)}+\eta_{(\alpha)},\, \theta_{(\alpha\beta)}+\eta_{(\alpha\beta)}\big)$ induces a gauge transformation for the Kalb-Ramond field given by
\begin{equation}
    \begin{aligned}
        B_{(\alpha)} &\mapsto B_{(\alpha)} + \mathrm{d}\eta_{(\alpha)}, \\
        \Lambda_{(\alpha\beta)} &\mapsto \Lambda_{(\alpha\beta)}+\eta_{(\alpha)}-\eta_{(\beta)}+\mathrm{d}\eta_{(\alpha\beta)} ,\\
        G_{(\alpha\beta\gamma)} &\mapsto G_{(\alpha\beta\gamma)} + \eta_{(\alpha\beta)}+\eta_{(\beta\gamma)}+\eta_{(\gamma\alpha)},
    \end{aligned}
\end{equation}
in analogy with the extra coordinate of ordinary Kaluza-Klein Theory.\vspace{0.25cm}

\noindent Moreover, if we take the differential of the first patching condition in \eqref{eq:papapatch}, we obtain the condition $ - \di y^1_{(\alpha)} + \di y^1_{(\beta)} = \di\Lambda_{(\alpha\beta)}$ for the differentials. This means that if we rewrite in components $y^1_{(\alpha)}=y^1_{(\alpha)\mu}\di x^\mu $, we can also rewrite $ - \di y^1_{(\alpha)\mu} + \di y^1_{(\alpha)\mu} = \di\Lambda_{(\alpha\beta)\mu}$. If we define the dual vectors $\partial/\partial y^1_{(\alpha)\mu}$ to the $1$-forms $\di y^1_{(\alpha)\mu}$ as vectors satisfying $\big\langle \partial/\partial y^1_{(\alpha)\mu},\, \di y^1_{(\alpha)\nu} \big\rangle =\delta^\mu_{\;\nu}$, we obtain doubled vector of the following form:
\begin{equation}
    V_{(\alpha)} \,=\, v^\mu_{(\alpha)} \frac{\partial}{\partial x^\mu_{(\alpha)}} + \bigg( \tilde{v}_{(\alpha)\mu} + B_{(\alpha)\mu\nu}\,v^\nu_{(\alpha)} \bigg)\frac{\partial}{\partial y^1_{(\alpha)\mu}},
\end{equation}
which are exactly the same as the ones in \eqref{eq:vec}.
Therefore the analogue of \textit{the tangent bundle of the C-space can be identified with a Courant algebroid $E\twoheadrightarrow M$ twisted by the gerbe \eqref{eq:cgerbe}}.

\paragraph{Further discussion.} 
The proposal seems to capture something quite fundamental of the geometry of DFT, by suggesting that the doubled space should be the total space of the gerbe itself. This looks consistent with the existing idea that doubled vectors should belong to a Courant algebroid twisted by a gerbe, which is the analogous to the tangent bundle for a gerbe. However this intuition is still waiting for a proper formalization: for example it is not clear how to construct coordinates that are $1$-forms on $M$. Moreover it is still not clear what is the relation with the new extra coordinates and the T-dual spacetime.

\section{Pre-NQP manifold proposal}

The pre-NQP manifold proposal was developed by \cite{DesSae18}, generalised to Heterotic DFT by \cite{DesSae18x} and then applied to the particular example of nilmanifolds by \cite{DesSae19}. This approach to DFT is based on the fact that $n$-algebroids can be equivalently described by differential-graded manifolds, including the Courant algebroid, which describes the local symmetries of the bundle gerbe of the Kalb-Ramond field. The idea is thus that we can describe the geometry of DFT by considering the differential graded manifold which geometrises the Courant algebroid and by relaxing some of the conditions.

\subsection{Symplectic $L_\infty$-algebroids as NQP-manifolds}
Given a $L_\infty$-algebroid $\mathfrak{a}\twoheadrightarrow M$ on some base manifold $M$, we can always associate to $\mathfrak{a}$ its Chevalley-Eilenberg algebra $\mathrm{CE}(\mathfrak{a})$, which is essentially the differential graded algebra of its sections. This is defined by
\begin{equation}
    \mathrm{CE}(\mathfrak{a})\;:=\; \Big(\wedge^\bullet\Gamma(M,\mathfrak{a}_\bullet^\ast),\,\di_{\mathrm{CE}}\Big),
\end{equation}
where the underlying complex is defined by 
\begin{equation}
    \wedge^\bullet\Gamma(M,\mathfrak{a}_\bullet^\ast) \;\,:=\;\, \underbrace{\Coo(M)}_{\text{degree 0}} \,\oplus\, \underbrace{\Gamma(M,\mathfrak{a}_0^\ast)}_{\text{degree 1}} \,\oplus\, \underbrace{\Gamma\big(M,\mathfrak{a}_1^\ast \oplus(\mathfrak{a}_0^\ast\wedge\mathfrak{a}_0^\ast)\big)}_{\text{degree 2}} \,\oplus\; \dots
\end{equation}
where the $\mathfrak{a}_k$ for any $k\in\mathbb{N}$ are the ordinary vector bundles underlying the $L_\infty$-algebroid. In the definition $\di_{\mathrm{CE}}$ is a degree $1$ differential operator on the graded complex $\wedge^\bullet\Gamma(M,\mathfrak{a}_\bullet^\ast)$ which encodes the $L_\infty$-bracket structure of the original $L_\infty$-algebroid $\mathfrak{a}$.
\vspace{0.25cm}

\noindent Now a NQ-manifold is defined as a graded manifold $\mathcal{M}$ equipped with a degree $1$ vector field $Q$ satisfying $Q^2=0$. The fundamental feature of NQ-manifolds is that the algebra of functions of any NQ-manifold $\mathcal{M}$ is itself a differential graded algebra $\left(\Coo(\mathcal{M}),\,Q\right)$ where the role of the differential operator is played by the vector $Q$, which is thus called \textit{cohomological}. This terminology was introduced in \cite{Alexandrov:1995kv}.
\vspace{0.25cm}

\noindent Crucially, there exists an equivalence between $L_\infty$-algebroids and NQ-manifolds given by
\begin{equation}
    \mathrm{CE}(\mathfrak{m})\;=\; \Big(\Coo(\mathcal{M}),\,Q\Big),
\end{equation}
so that any $L_\infty$-algebroid $\mathfrak{m}$ can be equivalently seen as a NQ-manifold $\mathcal{M}$. In the particular case which is relevant for DFT we consider the $2$-algebroid $\mathfrak{at}({\mathcal{G}})$ of infinitesimal gauge transformation of the bundle gerbe of the Kalb-Ramond field on  a manifold $M$. This is notoriously given by a NQ-manifold $T^\ast[2]T[1]M$ by the usual identification
\begin{equation}\label{eq:CEat}
    \mathrm{CE}\big(\mathfrak{at}({\mathcal{G}})\big)\;=\; \Big(\Coo\big(T^\ast[2]T[1]M\big),\,Q_H\Big), 
\end{equation}
where $Q_H$ is the cohomological vector twisted by the curvature $H\in\Omega^3_{\mathrm{cl}}(M)$ of the gerbe. To show this, notice first that in this case the differential graded algebra of functions on our NQ-manifold will be truncated at degree $< 2$. The degree $1$ sections will be sums of a vector and a $1$-form $X+\xi\in\Gamma(M,TM\oplus T^\ast M)$ and the degree $0$ sections will be just functions $f\in\Coo(M)$ on the base manifold. Now we can explicitly rewrite the underlying chain complexes of the two differential graded algebras \eqref{eq:CEat} by
\begin{equation}
    \begin{aligned}
    \mathrm{CE}\big(\mathfrak{at}({\mathcal{G}})\big) \;&=\;  \Big(\Coo(M) \xrightarrow{\;\,\di\,\;}\Gamma(M,\,TM\oplus T^\ast M)\Big),\\[0.25em]
    \Big(\Coo\big(T^\ast[2]T[1]M\big),\,Q_H\Big) \;&=\; \Big(\Coo(M) \xrightarrow{\;\,\di\,\;}\Gamma(M,\,TM\oplus T^\ast M)\Big),
    \end{aligned}
\end{equation}
moreover the derived bracket structure (see \cite{Roy02, DesSae18} for details) defined by the cohomological vector $Q_H$ on $\Coo\big(T^\ast[2]T[1]M\big)$ is exactly the bracket structure of the Courant $2$-algebroid, i.e.
\begin{gather}
    \begin{aligned}
    \ell_1(f) \;&=\; \di f, \\
    \ell_2(X+\xi,\,Y+\eta) \;&=\; [X,Y] + \mathcal{L}_X\eta-\mathcal{L}_Y\xi - \frac{1}{2}\di\langle X+\xi,\,Y+\eta \rangle+ \iota_X\iota_YH ,\\
    \;&=\; [X+\xi,\,Y+\eta]_{\mathrm{Cou}} ,\\
    \ell_2(X+\xi,\,f) \;&=\; \mathcal{L}_Xf ,\\
    \ell_2(X+\xi,\,Y+\eta,\,Z+\zeta) \;&=\; \frac{1}{3!}\big\langle X+\xi,\,[Y+\eta,\,Z+\zeta]_{\mathrm{Cou}} \big\rangle+\mathrm{cycl.}\,,
    \end{aligned}
    \raisetag{0.6cm}
\end{gather}
where $[-,-]_{\mathrm{Cou}}$ is the Courant bracket and $\langle-,-\rangle$ is the bundle metric defined by the contraction $\langle X+\xi,\,Y+\eta\rangle= \iota_X\eta+\iota_Y\xi$ for every sections $X+\xi,Y+\eta\in\Gamma(M,TM\oplus T^\ast M)$. \vspace{0.2cm}

\noindent The Courant $2$-algebroid is canonically a symplectic $2$-algebroid (see \cite{Roy02} for details), i.e. it can be equipped with a canonical symplectic form $\omega$, which can be easily expressed in local coordinates on the corresponding NQ-manifold. 
On each local patch of the NQ-manifold $T^\ast[2]T[1]M$ we can choose local coordinates $(x^\mu,\,  e^\mu,\, \bar{e}_\mu,\, p_\mu)$ where the $x^\mu$ are in degree $0$, while the $(e^\mu,\, \bar{e}_\mu)$ are both in degree $1$ and the $p_\mu$ are in degree $2$. On the local patches we can express the symplectic form $\omega\in\Omega^2(T^\ast[2]T[1]M)$ in local coordinates by
\begin{equation}
    \omega = \di x^\mu \wedge \di p_\mu + \di e^\mu \wedge \di \bar{e}_\mu.
\end{equation}
We can also use Hamilton's equations $\iota_{Q_H}\omega=\mathcal{Q}_H$ to express the vector $Q_H$ by an Hamiltonian function $\mathcal{Q}_H$. We find
\begin{equation}
    \mathcal{Q}_H = e^\mu p_\mu + H_{\mu\nu\lambda} e^\mu e^\nu e^\lambda,
\end{equation}
where $H\in\Omega^3_{\mathrm{cl}}(M)$ is a representative of the Dixmier-Douady class $[H]\in H^3(M,\mathbb{Z})$ of the original bundle gerbe $\mathcal{G}\twoheadrightarrow M$. This class in the literature of differential graded manifolds changes name in \textit{\v{S}evera class} \cite{Sev01}.

\subsection{A pre-NQP-manifold for Double Field Theory}
By following \cite{DesSae18}, we choose as $2d$-dimensional base manifold $M=T^\ast U$ the cotangent bundle of some $d$-dimensional local patch. This is because we are interested in the local geometry of the doubled space and we have still no information about how to patch together these local $2d$-dimensional $T^\ast U$ manifolds. Thus the Courant algebroid on $T^\ast U$ will be given by the NQP-manifold $T^\ast[2]T[1](T^\ast U)$, as we have seen.
This will have coordinates $(x^M,\,  e^M,\, \bar{e}_M,\, p_M)$ still respectively in degrees $0$, $1$, $1$ and $2$, but with $M=1,\dots,2d$. We must then think the local coordinates $x^M = (x^\mu, \tilde{x}_\mu)$ to be the doubled coordinates of DFT.
\vspace{0.2cm}

\noindent Since $T^\ast U$ is canonically equipped with the tensor $\eta_{MN}$, we can make a change of degree $1$ coordinates by
\begin{equation}
    E^M \,:=\, \frac{1}{\sqrt{2}}(e^M + \eta^{MN}\bar{e}_N), \qquad  \bar{E}_M \,:=\, \frac{1}{\sqrt{2}}(\bar{e}_M - \eta_{MN}e^N).
\end{equation}
Now we must restrict ourselves to the submanifold $\mathcal{M}:=\{\bar{E}_M=0\}$ of the original manifold $T^\ast[2]T[1](T^\ast U)$. It is not hard to check that this submanifold will be $\mathcal{M}=(T^\ast[2]\oplus T[1])(T^\ast U)$. The degree $1$ functions on $\mathcal{M}$ will then be doubled vectors of the form
\begin{equation}
    X^\mu(x,\tilde{x})\left(\frac{\partial}{\partial x^\mu}+\di \tilde{x}_\mu\right) + \xi_\mu(x,\tilde{x})\left(\frac{\partial}{\partial \tilde{x}_\mu}+\di x^\mu\right)
\end{equation}
and the degree $0$ functions will be just ordinary functions of the form $f\in\Coo(T^\ast U)$. The symplectic form restricted to the submanifold $\mathcal{M}$ will now be
\begin{equation}
    \omega|_{\mathcal{M}}\,=\, \di x^M \wedge \di p_M + \frac{1}{2}\eta_{MN}\di E^M \wedge \di E^N.
\end{equation}
The new Hamiltonian function will be $\mathcal{Q}|_{\mathcal{M}}=E^M p_M + H_{MNL}E^ME^NE^L$, where $H_{MNL}$ now is the curvature of a bundle gerbe on the $2d$-dimensional base $T^\ast U$, which we should think as the extended fluxes of DFT. Crucially our $\mathcal{M}$ will still be a symplectic graded manifold, however it will not be a NQP-manifold since the new restricted vector $Q$ is not nilpotent on $\mathcal{M}$, i.e. we have that $Q^2\neq 0$. This is exactly the reason why \cite{DesSae18} named $\mathcal{M}$ \textit{pre-NQP manifold} and therefore this cannot be seen an $L_\infty$-algebroid.
\vspace{0.2cm}

\noindent However this pre-NQP manifold satisfies a very interesting property: the pre-NQP-manifold has a number of sub-manifold which are proper NQP-manifolds and thus well-defined sub-$2$-algebroids. Schematically we have
\begin{equation}
    \mathrm{CE}(\mathfrak{a}) \;\subset\; \big(\Coo(\mathcal{M}),\,Q\big),
\end{equation}
where $\mathfrak{a}$ is one of these sub-$2$-algebroids.
On any of these, the bracket of doubled vectors in degree $1$ will be exactly the D-bracket of DFT, which will be given by $\llbracket X,Y \rrbracket_{\mathrm{D}} := \{QX,Y\}$. For instance we can choose the differential graded algebra of functions which are pullbacks from the submanifold $\mathcal{N}:=\{\tilde{x}_\mu=\tilde{p}^\mu=0\}\subset\mathcal{M}$, which is exactly the Courant $2$-algebroid $\mathcal{N}= T^\ast[2]T[1]U$. This corresponds to choosing a sub-$2$-algebroid which satisfies the strong constraint and therefore this restriction reduces the pre-NQP-geometry to bare Generalised Geometry on the manifold $U$. Any other solution of the strong constraint will correspond to a viable choice of sub-$2$-algebroid. \vspace{0.25cm}

\noindent We can also introduce tensors of the form $\mathcal{G}_{MN}E^M\otimes E^N$ on $\mathcal{M}$ and use the Poisson bracket to define a natural notion of D- and C-bracket on tensors. This allows to define a notion of generalised metric, curvature and torsion in analogy with Riemannian geometry.

\subsection{An example of global pre-NQP manifold}
It is well-known that higher geometry is the natural framework for geometric T-duality, see the formalization by \cite{BunNik13, FSS16x, FSS17x, FSS18, FSS18x, NikWal18}. Assume that we have two $T^n$-bundle spacetimes $M\xrightarrow{\pi}M_0$ and $\widetilde{M}\xrightarrow{\tilde{\pi}}M_0$ over a common $(d-n)$-dimensional base manifold $M_0$. A couple of bundle gerbes $\mathscr{G}\xrightarrow{\Pi}M$ and $\widetilde{\mathscr{G}}\xrightarrow{\tilde{\Pi}}\widetilde{M}$, formalising two Kalb-Ramond fields respectively on $M$ and $\widetilde{M}$, are geometric T-dual if the following isomorphism exists
\begin{equation}
    \begin{tikzcd}[row sep={12ex,between origins}, column sep={12ex,between origins}]
    & \mathscr{G}\times_{M_0} \widetilde{M}\arrow[rr, "\cong"', "\text{T-duality}"]\arrow[dr, "\Pi"']\arrow[dl, "\tilde{\pi}"] & & M\times_{M_0}\widetilde{\mathscr{G}}\arrow[dr, "\pi"']\arrow[dl, "\tilde{\Pi}"] \\
    \mathscr{G}\arrow[dr, "\Pi"'] & & M\times_{M_0}\widetilde{M}\arrow[dr, "\pi"']\arrow[dl, "\tilde{\pi}"] & & [-2.5em]\widetilde{P}\arrow[dl, "\tilde{\Pi}"] \\
    & M\arrow[dr, "\pi"'] & & \widetilde{M}\arrow[dl, "\tilde{\pi}"] & \\
    & & M_0 & &
    \end{tikzcd}
\end{equation}
This picture is nothing but the finite version of T-duality between Courant algebroids illustrated by \cite{CavGua11}. Now, in \cite{DesSae19} it is proposed that we should consider the fiber product of the pull-back of both the gerbes $\mathscr{G}$ and $\widetilde{\mathscr{G}}$ to the correspondence space $M\times_{M_0}\widetilde{M}$ of the T-duality, which will be itself a gerbe of the form
\begin{equation}\label{eq:dsgerbe}
    \Pi\otimes\tilde{\Pi}:\; \mathscr{G}\otimes\widetilde{\mathscr{G}}\;\longtwoheadrightarrow\; M\times_{M_0}\widetilde{M}.
\end{equation}
Now, as previously explained, we can take the algebroid of infinitesimal gauge transformations of this gerbe $\mathfrak{at}({\mathscr{G}\otimes\widetilde{\mathscr{G}}})\longtwoheadrightarrow M\times_{M_0}\widetilde{M}$ and express it as a differential graded manifold $\big( T^\ast[2]T[1](M\times_{M_0}\widetilde{M}),\,Q\big)$ with local coordinates $(x^\mu,x^I,e^\mu, e^I, \bar{e}_\mu, \bar{e}_I, p_\mu, p_I)$ with indices $\mu=1,\dots,d-n$ and $I=1,\dots,n$. Now, as we explained for the local doubled space, we can change coordinates to $E^I:=(e^I+\eta^{IJ}\bar{e}_J)/\sqrt{2}$ and $\bar{E}_I:=(\bar{e}_I+\eta_{IJ}e^J)/\sqrt{2}$ and set $\bar{E}_I=0$ to zero so that we obtain a new differential graded manifold $\mathcal{M}$. This new manifold will be locally isomorphic to $(T^\ast[2]\oplus T[1])T^{2n} \oplus T^\ast[2]T[1]U$ on each patch $U\subset M_0$ of the base manifold, but which is globally well-defined.
In \cite{DesSae19} this machinery is applied for fiber dimension $n=1$ to the particular case where $M$ and $\widetilde{M}$ are nilmanifolds on a common base torus $M_0=T^2$.

\paragraph{Further discussion.}
This is the first proposal to interpret strong constrained doubled vectors as sections of the $2$-algebroid of the local symmetries of a gerbe: the Courant $2$-algebroid. This suggests that it could be a complementary approach to the ones attempting to realise the doubled space as a geometrization the bundle gerbe itself.
\vspace{0.25cm}

\noindent However there are still some open problems. 
The only non-trivial global case that was constructed in this framework was, as we saw, on the correspondence space $M\times_{M_0}\widetilde{M}$ equipped with the pullback of both the gerbe $\mathscr{G}$ and its dual $\widetilde{\mathscr{G}}$. But, for this construction, the correspondence space of the T-duality is not derived from the pre-NQP manifold theory, but it must be assumed and prepared by using the machinery of topological T-duality. 
Besides, the total gerbe \eqref{eq:dsgerbe} has "repeated" information: for example, if we start from a gerbe $\mathscr{G}_{i,j}$ with Dixmier-Douady number $i$ on a nilmanifold with $1$st Chern number $j$, its dual will be a gerbe $\widetilde{\mathscr{G}}_{j,i}$ on a nilmanifold with inverted Dixmier-Douady and $1$st Chern number. Now the total gerbe $\mathscr{G}_{i,j}\otimes\widetilde{\mathscr{G}}_{j,i}$ contain each number twice: as $1$st Chern number and as Dixmier-Douady number. Moreover, in literature, a globally defined pre-NQP manifold for a non-trivially fibrated spacetime $M$ was proposed only for the case of geometric T-duality. Recently \cite{Crow-Watamura:2018liw} applied pre-NQP geometry to the case of DFT on group manifolds. However the extension of this formalism to general T-dualizable backgrounds is not immediate.

\section{Tensor hierarchies proposal}

The idea of tensor hierarchy was introduced in \cite{HohSam13KK} in the context of the dimensional reduction of DFT, then further formalised in \cite{Hohm19DFT}, \cite{Hohm19} and \cite{Hohm19x} as a higher gauge structure. See also work by \cite{Ced20a} and \cite{Ced20b}. 

\subsection{Embedding tensor and Leibniz-Loday algebra}
Let $\rho:\mathfrak{o}(d,d)\otimes R \rightarrow R$ be the fundamental representation of the Lie algebra $\mathfrak{o}(d,d)$ of the Lie group $O(d,d)$. The vector space underlying the fundamental representation of $O(d,d)$ is nothing but $R \cong \mathbb{R}^{2d}$. Let us use the notation $\mathbf{x}\otimes Y\mapsto \rho_\mathbf{x}Y \in R$. The \textit{embedding tensor} of DFT is defined as a linear map $\Theta: \mathfrak{X}(\mathbb{R}^{2d}) \,\hookrightarrow\, \Coo(\mathbb{R}^{2d},\,\mathfrak{o}(d,d))$ which satisfies the following compatibility condition, usually called \textit{quadratic constraint}:
\begin{equation}
    [\Theta(X),\Theta(Y)] \,=\, \Theta(\rho_{\Theta(X)}Y),
\end{equation}
where $[-,-]$ are the Lie bracket of the Lie algebra $\mathfrak{o}(d,d)$.
Concretely the embedding tensor maps a vector field by $X^M\mapsto (X^M\!,\,\partial_{[M}X_{N]})\in\Coo(\mathbb{R}^{2d},\,\mathfrak{o}(d,d))$. Now the embedding tensor defines a natural action of $\mathfrak{X}(\mathbb{R}^{2d})$ on itself by
\begin{align}
    \circ:\,\mathfrak{X}(\mathbb{R}^{2d}) \otimes \mathfrak{X}(\mathbb{R}^{2d}) \;&\longrightarrow\; \mathfrak{X}(\mathbb{R}^{2d})\\
    (X,Y) \;&\longmapsto\; X \circ Y \,:=\, \rho_{\Theta(X)}Y.
\end{align}
This is exactly the D-bracket of DFT. Thus the anti-symmetric part will be the C-bracket
\begin{equation}
\begin{aligned}
    \frac{1}{2}(X\circ Y - Y\circ X) \,=\, \llbracket X,Y\rrbracket_{\mathrm{C}}.
\end{aligned}
\end{equation}
On the other hand the symmetric part of the D-bracket is given by $X\circ Y + Y\circ X \,=\, \mathfrak{D}\langle X,Y \rangle$, where $\mathfrak{D}:\Coo(\mathbb{R}^{2d}) \longrightarrow \mathfrak{X}(\mathbb{R}^{2d})$ is defined by $f\mapsto \partial^M\!{f}$ and the metric is defined by the contraction $\langle X,Y\rangle:=\eta_{MN}X^MY^N$. Therefore the D-bracket can be expressed in terms of these operators by
\begin{equation}
    X\circ Y \,=\, \llbracket X,Y\rrbracket_{\mathrm{C}} + \frac{1}{2}\mathfrak{D}\langle X,Y \rangle.
\end{equation}
An interesting consequence is that the couple $\big(\mathfrak{X}(\mathbb{R}^{2n}),\,\circ\;\big)$ is not a Lie algebra, since the D-bracket is not anti-symmetric, but it is a Leibniz-Loday algebra, since it satisfies the Leibniz property $X\circ (Y\circ Z) = (X \circ Y)\circ Z + Y \circ (X \circ Z)$ for any triple of vectors $X,Y,Z\in\mathfrak{X}(\mathbb{R}^{2n})$. \vspace{0.25cm}

\noindent Now something remarkable happens: the Leibniz-Loday algebra $\big(\mathfrak{X}(\mathbb{R}^{2n}),\,\circ\;\big)$ of infinitesimal DFT gauge transformations naturally defines a Lie $2$-algebra $\big(\mathscr{D}(\mathbb{R}^{2n}),\,\ell_i\big)$ of infinitesimal DFT gauge transformations. This is given by the underlying cochain complex
\begin{equation}\label{eq:dofd}
    \mathscr{D}(\mathbb{R}^{2n}) \;:=\; \Big( \Coo(\mathbb{R}^{2n}) \xrightarrow{\;\;\mathfrak{D}\;\;} \mathfrak{X}(\mathbb{R}^{2n})\Big),
\end{equation}
equipped with the following $L_\infty$-bracket structure:
\begin{align}
    \ell_1(f) \;&=\; \mathfrak{D}f ,\\
    \ell_2(X,Y) \;&=\; \llbracket X,Y\rrbracket_{\mathrm{C}},\\
    \ell_2(X,f) \;&=\; \langle X,\mathfrak{D}f\rangle ,\\
    \ell_3(X,Y,Z) \;&=\; -\frac{1}{2}\langle\llbracket X,Y\rrbracket_{\mathrm{C}},Z\rangle + \mathrm{cycl.}\,,
\end{align}
for any $f\in\Coo(\mathbb{R}^{2d})$ and $X,Y,Z\in\mathfrak{X}(\mathbb{R}^{2n})$.
Now notice that the quadratic constraint, which is the condition controlling the closure of the Leibniz bracket $X\circ Y$, requires to impose an additional constraint: this condition is nothing but the strong constraint. This makes the underlying complex of sheaves reduce to the one of sections of the standard Courant $2$-algebroid
\begin{equation}
    \mathscr{D}_{\mathrm{sc}}(\mathbb{R}^{2n}) \;=\; \Big( \Coo(\mathbb{R}^{n}) \xrightarrow{\;\;\di\;\;} \mathfrak{X}(\mathbb{R}^{n}) \oplus \Omega^1(\mathbb{R}^{d})\Big).
\end{equation}
Hence if we want $(\mathfrak{X}(\mathbb{R}^{2n}),\,\circ\,)$ to be a well-defined Leibniz-Loday algebra we need to restrict to Generalised Geometry and the D-bracket $\circ$ must reduce to the Dorfman bracket of Generalised Geometry, not twisted by any flux. At the present time no ways to generalise this construction beyond the strong constraint have been found, despite the community expecting such generalisation to exist.

\subsection{Tensor hierarchies}
Now that we have our well-defined $L_\infty$-algebra $\big(\mathscr{D}(\mathbb{R}^{2n}),\,\ell_n\big)$, we can ask ourselves what happens if we use it to construct an \textit{higher gauge field theory} on a $(d-n)$-dimensional manifold $M$. The answer is that the theory resulting from this gauging process is exactly a tensor hierarchy, which is supposed to describe DFT truncated at codimension $n$.
\vspace{0.25cm}

\noindent Luckily for our gauging purposes, there exists a well-defined notion of the tensor product of a commutative differential graded algebra with an $L_\infty$-algebra. Thus we can define the prestack of local tensor hierarchies $\Omega\!\left(U,\,\mathscr{D}(\mathbb{R}^{2n})\right)$ by the tensor product of the differential graded algebra of the de Rham complex $(\Omega^\bullet(U),\,\di)$ with the $L_\infty$-algebra $\big(\mathscr{D}(\mathbb{R}^{2n}),\,\ell_i\big)$. In other words we define $\Omega\!\left(U,\,\mathscr{D}(\mathbb{R}^{2n})\right) := \Omega^\bullet(U) \otimes \mathscr{D}(\mathbb{R}^{2n})$ for any contractible open set $U\subset M$. Its underlying complex of sheaves of this prestack will be
\begin{equation*}
    \Omega\big(U,\,\mathscr{D}(\mathbb{R}^{2n})\big) \,=\, \underbrace{\Coo(U\times \mathbb{R}^{2n})}_{\text{degree }0} \,\oplus\, \bigoplus_{k>0} \bigg(\underbrace{ \Omega^k(U)\otimes\Coo(\mathbb{R}^{2n}) \,\oplus\, \Omega^{k-1}(U)\otimes\mathfrak{X}(\mathbb{R}^{2n})}_{\text{degree }k}\bigg)
\end{equation*}
and the bracket structure is found by applying the definition by \cite{JRSW19}. Explicitly, for any elements $\mathcal{A}_p\in\Omega^\bullet(U)\otimes \mathfrak{X}(\mathbb{R}^{2n})$ and $\mathcal{B}_p\in\Omega^\bullet(U)\otimes \Coo(\mathbb{R}^{2n})$, we have the following bracket structure:
\begin{equation}\begin{aligned}
   \ell_1(\mathcal{A}+\mathcal{B}) \;&=\; (\di \mathcal{A} + \mathfrak{D}\mathcal{B}) + \di \mathcal{B},\\
   \ell_2(\mathcal{A}_1,\mathcal{A}_2) \;&=\; -\llbracket \mathcal{A}_1\,\overset{\wedge}{,}\, \mathcal{A}_2 \rrbracket_{\mathrm{C}},\\
   \ell_2(\mathcal{A}_1,\mathcal{B}_2) \;&=\; \langle \mathcal{A}\,\overset{\wedge}{,}\, \mathfrak{D}\mathcal{B} \rangle ,\\
   \ell_2(\mathcal{B}_1,\mathcal{B}_2) \;&=\; 0 ,\\
   \ell_3(\mathcal{A}_1,\mathcal{A}_2,\mathcal{A}_3) \;&=\; -\frac{1}{2}\langle\llbracket \mathcal{A}_1\,\overset{\wedge}{,}\, \mathcal{A}_2 \rrbracket_{\mathrm{C}}\,\overset{\wedge}{,}\, \mathcal{A}_3 \rangle + \mathrm{cycl.} ,\\
   \ell_3(\mathcal{A}_1,\mathcal{A}_2,\mathcal{B}_3) \;&=\;  \ell_3(\mathcal{A}_1,\mathcal{B}_2,\mathcal{B}_3)\;=\;  \ell_3(\mathcal{B}_1,\mathcal{B}_2,\mathcal{B}_3) \;=\; 0 , 
\end{aligned}\end{equation}
where we introduced the following compact notation for $\mathscr{D}(\mathbb{R}^{2n})$-valued differential forms:
\begin{itemize}
    \item $\llbracket - \,\overset{\wedge}{,}\, -\rrbracket_{\mathrm{C}}$ is a wedge product on $\Omega^\bullet(U)$ and a C-bracket on $\mathfrak{X}(\mathbb{R}^{2n})$,
    \item $\langle - \,\overset{\wedge}{,}\, - \rangle$ is a wedge product on $\Omega^\bullet(U)$ and a contraction $\langle -,-\rangle$ on $\mathfrak{X}(\mathbb{R}^{2n})$.
\end{itemize}
The prestack $\Omega\!\left(U,\,\mathscr{D}(\mathbb{R}^{2n})\right)$ encodes the local fields of a tensor hierarchy on a local doubled space of the form $U\times \mathbb{R}^{2n}$ with base manifold $\mathrm{dim}(U)=d-n$.
In our degree convention the connection data of a tensor hierarchy is given by a degree $2$ multiplet
\begin{equation}
    \begin{aligned}
    \mathcal{A}^I_\mu \;&\in\; \Omega^1(U)\otimes \mathfrak{X}(\mathbb{R}^{2n}), \\
    \mathcal{B}_{\mu\nu} \;&\in\; \Omega^2(U)\otimes \Coo(\mathbb{R}^{2n}).
    \end{aligned}
\end{equation}
If we consider differential forms valued in a suitable adjusted Weil algebra $\mathrm{W}_{\mathrm{adj}}\big(\mathscr{D}(\mathbb{R}^{2n})\big)$, as in example \ref{ex:adjweilstring}, we can also introduce the curvature of the tensor hierarchy, which will be given by the degree $3$ multiplet 
\begin{equation}
    \begin{aligned}
    \mathcal{F}^I_{\mu\nu} \;&\in\; \Omega^2(U)\otimes \mathfrak{X}(\mathbb{R}^{2n}) ,\\
    \mathcal{H}_{\mu\nu\lambda} \;&\in\; \Omega^3(U)\otimes \Coo(\mathbb{R}^{2n}) .
    \end{aligned}
\end{equation}
Notice that all the fields of the hierarchy depend not just on the coordinates $x$ of the base manifold $U$, but also on the coordinates $(y, \tilde{y})$ of the vector space $\mathbb{R}^{2n}$. The curvature of the tensor hierarchy can be expressed in terms of the connection, as it is found in \cite{HohSam13KK}, by
\begin{equation}\label{eq:tensorhi}
    \begin{aligned}
    \mathcal{F} \;&=\; \di \mathcal{A} - \llbracket \mathcal{A} \,\overset{\wedge}{,}\, \mathcal{A}\rrbracket_{\mathrm{C}} + \mathfrak{D}\mathcal{B} ,\\[0.2em]
    \mathcal{H} \;&=\;  \mathrm{D}\mathcal{B} + \frac{1}{2}\langle \mathcal{A} \,\overset{\wedge}{,}\, \di A\rangle - \frac{1}{3!} \langle \mathcal{A} \,\overset{\wedge}{,}\,   \llbracket \mathcal{A} \,\overset{\wedge}{,}\, \mathcal{A}\rrbracket_{\mathrm{C}} \rangle, 
    \end{aligned}
\end{equation}
where we introduced  the covariant derivative $\mathrm{D} := \di -  \mathcal{A}\circ\wedge$ defined by the $1$-form connection $A$, which acts explicitly by $\mathrm{D}\mathcal{A} = \di \mathcal{A} + \llbracket \mathcal{A} \,\overset{\wedge}{,}\, \mathcal{A}\rrbracket_{\mathrm{C}}$ and $\mathrm{D}\mathcal{B} = \di \mathcal{B} + \langle A\,\overset{\wedge}{,}\, \mathfrak{D} \mathcal{B}\rangle$.
Notice the characteristic C-bracket Chern-Simons term in the expression of $3$-form curvature. We will call $\mathrm{CS}_3(\mathcal{A})$, so we will be able to write the curvature of the tensor hierarchy in a compact fashion:
\begin{equation}
    \begin{aligned}
    \mathcal{F} \;&=\; \mathrm{D} \mathcal{A} + \mathfrak{D}\mathcal{B}, \\
    \mathcal{H} \;&=\;  \mathrm{D}\mathcal{B} + \frac{1}{2}\mathrm{CS}_3(\mathcal{A}) .
    \end{aligned}
\end{equation}
By calculating the differential of the field curvature multiplet, this immediately gives the Bianchi identities of the tensor hierarchy:
\begin{equation}
    \begin{aligned}
    \mathrm{D}\mathcal{F} +\mathfrak{D}\mathcal{H}\;&=\; 0 \quad&\in\; \Omega^3(U)\otimes \mathfrak{X}(\mathbb{R}^{2n}),\\
    \mathrm{D}\mathcal{H} - \frac{1}{2}\langle \mathcal{F} \;\overset{\wedge}{,}\; \mathcal{F} \rangle\;&=\;  0  \quad&\in\; \Omega^4(U)\otimes \Coo(\mathbb{R}^{2n}).
    \end{aligned}
\end{equation}
The infinitesimal gauge transformations of a tensor hierarchy are given by $1$-degree multiplets of the form
\begin{equation}
    \begin{aligned}
    \lambda^I \;&\in\; \Coo(U)\otimes \mathfrak{X}(\mathbb{R}^{2n}), \\
    \Xi_{\mu} \;&\in\; \Omega^1(U)\otimes \Coo(\mathbb{R}^{2n}) ,
    \end{aligned}
\end{equation}
so that
\begin{equation}
    \begin{aligned}
    \mathcal{A} \;&\longmapsto\; \mathcal{A} + \mathrm{D}\lambda + \mathfrak{D}\Xi, \\
    \mathcal{B} \;&\longmapsto\; \mathcal{B} + \mathrm{D}\Xi - \langle\lambda,\mathcal{F}\rangle,
    \end{aligned}
\end{equation}
where the covariant derivative acts by $\mathrm{D}\lambda \;=\; \di \lambda + \llbracket \mathcal{A} , \lambda \rrbracket_{\mathrm{C}}$ and $\mathrm{D}\Xi \;=\; \di \Xi + \langle \mathcal{A}\,\overset{\wedge}{,}\, \mathfrak{D} \Xi\rangle$.
Notice the extraordinary similarity of these equations to the ones defining a principal $\mathrm{String}$-bundle. This similarity will be discussed in chapter \ref{ch:6}.

\begin{example}[Topological T-duality]
Notice that, in the particular case of a tensor hierarchy where none of the fields depend on the internal space $\mathbb{R}^{2n}$, the curvature reduces to the familiar equations of a doubled torus bundle, i.e.
\begin{equation}
    \begin{aligned}
    \mathcal{F} \;&=\; \di \mathcal{A} && \in\;\Omega^2_{\mathrm{cl}}(U, \mathbb{R}^{2n}), \\
    \mathcal{H} \;&=\;  \di \mathcal{B} + \frac{1}{2}\langle \mathcal{A} \,\overset{\wedge}{,}\, \di \mathcal{A}\rangle && \in\;\Omega^3_{\mathrm{cl}}(U),
    \end{aligned}
\end{equation}
which is exactly the curvature of the $\mathrm{String}(T^n\times T^n)$-bundle arising in the case of a globally geometric T-duality. Also the gauge transformations reduce to
\begin{equation}
    \begin{aligned}
    \mathcal{A}^{ I} \;&\mapsto\; \mathcal{A}^{I} + \di\lambda^I, \\
    \mathcal{B} \;&\mapsto\; \mathcal{B} + \di\Xi - \langle\lambda,\mathcal{F}\rangle.
    \end{aligned}
\end{equation}
The local field $\mathcal{F}\in\Omega^2_{\mathrm{cl}}(U, \mathbb{R}^{2n})$ can thus be globalised to the curvature of a doubled torus bundle with $1$st Chern class $[\mathcal{F}]\in H^2(M,\mathbb{Z}^{2n})$. At this point topological T-duality is immediately encompassed by the $O(n,n;\mathbb{Z})$-rotation $[\widetilde{\mathcal{F}}]_I := \eta_{IJ}[\mathcal{F}]^J$ of the $1$st Chern class of the doubled torus bundle.
\end{example}

\noindent This particular example of tensor hierarchy allows a globalization to a principal $2$-bundle with gauge $2$-group $\mathrm{String}(T^n\times T^n)$. Moreover, if we forget the higher form field, we stay with a well-defined $T^{2n}$-bundle on the $(d-n)$-dimensional base manifold $M$. This leads to the question about how to geometrically globalise and interpret general tensor hierarchies. 

\subsection{The puzzle of the global tensor hierarchy}
This proposal is the first to understand that the doubled connections $\mathcal{A}^I_\mu$, which we have also for the doubled torus bundles, are just a part of the full connection of the prestack $\Omega(-,\mathscr{D}(\mathbb{R}^{2n}))$, including also $\mathcal{B}_{\mu\nu}$. Thus the doubled space is intrinsically a higher geometric object.\vspace{0.2cm}

\noindent In \cite{Hohm19} it was proposed that the global higher gauge theory of tensor hierarchies on a $(d-n)$-dimensional manifold $M$ should consist in the $L_\infty$-algebra of $\mathrm{W}_{\mathrm{adj}}\big(\mathscr{D}(\mathbb{R}^{2n})\big)$-valued differential forms on $M$. However this must be taken as a local statement, since we know that gauge and $p$-form fields are not generally global differential forms on $M$, unless their underlying principal bundles are topologically trivial. 
Exactly like gauge fields, the global stack of tensor hierarchies can be constructed by using the notion of \textit{parallel transport}, as we have seen in chapters \ref{ch:3}. This is true, at least, if we want to formalise tensor hierarchies as higher gauge theories. In chapter \ref{ch:6} we will discuss a different perspective. Let $\exp \mathscr{D}(\mathbb{R}^{2n})\in\mathbf{H}$ be the Lie integration of the Lie $2$-algebra $\mathscr{D}(\mathbb{R}^{2n})$. Now we can define the higher gauge theory
\begin{equation}
    \mathscr{T\!\!H}:\, M \, \longmapsto \, \mathbf{H}\Big( \mathscr{P}(M),\, \mathbf{B}\mathrm{Inn}_{\mathrm{adj}}\big(\exp \mathscr{D}(\mathbb{R}^{2n})\big) \Big),
\end{equation}
where $\mathscr{P}(M)$ is the path $\infty$-groupoid of the smooth manifold $M$.
By construction this means that on any set $U\subset M$ of a good cover of our $(d-n)$-dimensional manifold $M$ we will have the isomorphism
\begin{equation}
    \tenhie(U) \;\cong\; \Omega^\bullet(U) \otimes \mathrm{W}_{\mathrm{adj}}\big(\mathscr{D}(\mathbb{R}^{2n})\big),
\end{equation}
This conveys the intuition that $\tenhie(-)$ is a globalization of the prestack of $ \mathrm{W}_{\mathrm{adj}}\big(\mathscr{D}(\mathbb{R}^{2n})\big)$-valued forms, but not necessarily a topologically-trivial one. By construction $\tenhie(-)$ maps any $(d-n)$-dimensional smooth manifold $M$ to the $2$-groupoid $\tenhie(M)$ whose objects are tensor hierarchies and whose morphisms are gauge transformations of tensor hierarchies on $M$. 
\vspace{0.25cm}

\noindent In more concrete terms a global tensor hierarchy, which is an object of the $2$-groupoid $\tenhie(M)$, can be expressed in a local trivialization by a \v{C}ech cocycle. Given any good cover $\{U_\alpha\}$ for the $(d-n)$-dimensional manifold $M$, such a cocycle will be of the form
\begin{equation}
    \big(\mathcal{F}^I_{(\alpha)},\,\mathcal{H},\,\mathcal{A}_{(\alpha)}^I,\,\mathcal{B}_{(\alpha)},\,\lambda^I_{(\alpha\beta)},\, \Xi_{(\alpha\beta)},\,g_{(\alpha\beta\gamma)}\big) \;\in\; \tenhie(M),
\end{equation}
where the fields are of the following differential forms:
\begin{equation}
    \begin{aligned}
    \mathcal{F}^I_{(\alpha)\mu\mu} \;&\in\; \Omega^2(U_{\alpha})\otimes \mathfrak{X}(\mathbb{R}^{2n}) ,\\
    \mathcal{H}_{\mu\nu\lambda} \;&\in\; \Omega^3(M)\otimes \Coo(\mathbb{R}^{2n}) ,\\[0.5em]
    A^I_{(\alpha)\mu} \;&\in\; \Omega^1(U_{\alpha})\otimes \mathfrak{X}(\mathbb{R}^{2n}) ,\\
    B_{(\alpha)\mu\nu} \;&\in\; \Omega^2(U_{\alpha})\otimes \Coo(\mathbb{R}^{2n}) ,\\[0.5em]
    \lambda^I_{(\alpha\beta)} \;&\in\; \Coo(U_\alpha\cap U_\beta)\otimes \mathfrak{X}(\mathbb{R}^{2n}) ,\\
    \Xi_{(\alpha\beta)\mu} \;&\in\; \Omega^1(U_\alpha\cap U_\beta)\otimes \Coo(\mathbb{R}^{2n}) ,\\[0.5em]
    g_{(\alpha\beta\gamma)}  \;&\in\; \Coo(U_\alpha\cap U_\beta\cap U_\gamma \times \mathbb{R}^{2n}),
    \end{aligned}
\end{equation}
and they are glued on two-fold, three-fold and four-fold overlap of patches as follows:  
\begin{equation}\label{eq:tensorhipatched}
    \begin{aligned}
        \mathcal{F}_{(\alpha)} \;&=\; \mathrm{D} \mathcal{A}_{(\alpha)} + \mathfrak{D}\mathcal{B}_{(\alpha)} ,\\
    \mathcal{H} \;&=\;  \mathrm{D}\mathcal{B}_{(\alpha)} + \frac{1}{2}\mathrm{CS}_3(\mathcal{A}_{(\alpha)}) ,\\[0.8em]
    \mathcal{A}_{(\alpha)} \;&=\;  e^{-\lambda_{(\alpha\beta)}} (\mathcal{A}_{(\beta)}+\di)e^{\lambda_{(\alpha\beta)}} + \mathfrak{D}\Xi_{(\alpha\beta)} ,\\
    \mathcal{B}_{(\alpha)} - \mathcal{B}_{(\beta)} \;&=\; \mathrm{D}\Xi_{(\alpha\beta)} - \langle\lambda_{(\alpha\beta)},\mathcal{F}_{(\alpha)}\rangle ,\\[0.8em]
    e^{\lambda_{(\alpha\beta)}} e^{\lambda_{(\beta\gamma)}} e^{\lambda_{(\gamma\alpha)}} \;&=\; e^{\mathfrak{D} g_{(\alpha\beta\gamma)}} ,\\
    \Xi_{(\alpha\beta)} + \Xi_{(\beta\gamma)} + \Xi_{(\gamma\alpha)} \;&=\; \di g_{(\alpha\beta\gamma)} ,\\[0.8em]
    g_{(\alpha\beta\gamma)} - g_{(\beta\gamma\delta)} +  g_{(\gamma\delta\alpha)} -  g_{(\delta\alpha\beta)}\;&\in\;2\pi\mathbb{Z},
    \end{aligned}
\end{equation}
where the covariant derivatives are $\mathrm{D}=\di- \mathcal{A}_{(\alpha)} 
\circ\wedge$. Notice the similarity, at least locally, of the potential $\mathcal{A}_{(\alpha)}\in\Omega^1\big(U_\alpha\times\mathbb{R}^{2n},\,U_\alpha\times T\mathbb{R}^{2n}\big)$ with the non-principal connection defined for instance in \cite[pag.$\,$77]{Nat93} for a general bundle. \vspace{0.25cm}

\noindent In chapter \ref{ch:6} we discuss the possibility of a more general definition of global tensor hierarchy, which can be obtained by directly dimensionally reducing the bundle gerbe and not as a higher gauge theory.\vspace{0.20cm}

\noindent However, if we accept that the global picture of tensor hierarchies is a higher gauge algebra, we would still have some open questions. From \cite{HohSam13KK} we know that a tensor hierarchy is supposed to be a split version of DFT with a $(d-n)$-dimensional base manifold $M$ and $2n$-dimensional fibers for an arbitrary $n$. But since tensor hierarchies are higher gauge theories, this hints that the full $2d$-dimensional doubled space should carry a bundle gerbe structure. Such structure, as we have seen for previous proposals, still needs to be clarified.

\section{Born Geometry}

The first proposal of formalisation of the geometry underlying Double Field Theory as a para-K\"{a}hler manifold was developed by \cite{Vais12} and then generalised to a para-Hermitian manifold by \cite{Vai13}. The para-Hermitian program was further developed by \cite{Svo17, Svo18, MarSza18, Svo19, MarSza19, Shiozawa:2019jul, DFTWZW19, BPV20, Ba20, Ikeda:2020lxz, Svo20}.

\subsection{Para-Hermitian geometry}
An almost para-complex manifold $(\M,J)$ is a $2d$-dimensional smooth manifold $\M$ which is equipped with a $(1,1)$-tensor field $J\in\mathrm{End}(T\M)$, called almost para-complex structure, such that $J^2=\mathrm{id}_{T\mathcal{M}}$ and that the $\pm 1$-eigenbundles $L_\pm\subset T\mathcal{M}$ of $J$ have both $\mathrm{rank}(L_\pm)=d$. A para-complex structure is, then, equivalently given by a splitting of the form
\begin{equation}
    T\mathcal{M}\;=\;L_+\oplus L_-
\end{equation}
Therefore, the structure group of the tangent bundle $T\mathcal{M}$ of the almost para-complex manifold is reduced to $GL(d,\mathbb{R})\times GL(d,\mathbb{R})\subset GL(2d,\mathbb{R})$. The para-complex structure also canonically defines the following projectors to its eigenbundles:
\begin{equation}
    \Pi_\pm \,:=\, \frac{1}{2}(1\pm J): \,T\M \,\longtwoheadrightarrow\, L_\pm.
\end{equation}
An almost para-complex structure $J$ is said to be, respectively, $\pm$-integrable if $L_\pm$ is closed under Lie bracket, i.e. if it satisfies the property
\begin{equation}
    \big[\Gamma(\mathcal{M},L_\pm),\,\Gamma(\mathcal{M},L_\pm)\big]_{\mathrm{Lie}} \,\subseteq\, \Gamma(\mathcal{M},L_\pm).
\end{equation}
The $\pm$-integrability of $J$ implies the existence a foliation $\mathcal{F}_\pm$ of the manifold $\mathcal{M}$ such that $L_\pm = T \mathcal{F}_\pm$. An almost para-complex manifold $(\mathcal{M},J)$ is a para-complex manifold if and only if $J$ is both $+$-integrable and $-$-integrable at the same time.
\vspace{0.25cm}

\noindent An almost para-Hermitian manifold $(\mathcal{M},J,\eta)$ is an almost para-complex manifold $(\mathcal{M},J)$ equipped with a metric $\eta \in \bigodot^2T^\ast\mathcal{M}$ of Lorentzian signature $(d,d)$ which is compatible with the almost para-complex structure as follows:
\begin{equation}
    \eta(J-,J-) \,=\, - \eta(-,-).
\end{equation}
A para-Hermitian structure $(J,\eta)$ canonically defines an almost symplectic structure $\omega\in\Omega^2(\mathcal{M})$, called fundamental $2$-form, by $\omega(-,-) := \eta(J-,-)$. An almost para-Hermitian manifold can be equivalently expressed as $(\mathcal{M},J,\omega)$, since the para-Hermitian metric can be uniquely determined by $\eta(-,-) = \omega(J-,-)$. Notice that the subbundles $L_\pm$ are both maximal isotropic subbundles respect to $\eta$ and Lagrangian subbundles respect to $\omega$. 

\subsection{Recovering generalised geometry}
The para-Hermitian metric immediately induces an isomorphism $\eta^\sharp:L_{\pm}\xrightarrow{\;\cong\;}L_{\mp}^\ast$. In the case of a $+$-integrable para-Hermitian manifold, this implies the existence of an isomorphism
\begin{equation}
    T\mathcal{M}\;\cong\;T\mathcal{F}_+\oplus T^\ast\mathcal{F}_+
\end{equation}
given by $X\mapsto \Pi_+(X) + \eta^\sharp(\Pi_-(X))$, for any vector $X\in T\M$.
As shown by \cite{Svo17,MarSza18}, it is possible to define a bracket structure $\llbracket-,-\rrbracket_{\mathrm{D}}: \mathfrak{X}(\mathcal{M})\times\mathfrak{X}(\mathcal{M})\rightarrow\mathfrak{X}(\mathcal{M})$ which is compatible with the para-Hermitian metric, so that $(T\M,\llbracket-,-\rrbracket_{\mathrm{D}},\eta)$ is a metric algebroid, and which makes a generalised version of the Nijenhuis tensor of $J$ vanish \cite[p.$\,$13]{MarSza18}.
If we consider any couple of sections $X+\xi,Y+\zeta\in\Gamma(\M,T\mathcal{F}_+\oplus T^\ast\mathcal{F}_+)$, the bracket can be rewritten as
\begin{equation}
    \llbracket X+\xi, Y+\zeta\rrbracket_{\mathrm{D}} \;=\!\! \underbrace{\big([X,Y] + \mathcal{L}_X\zeta - \iota_Y\di\xi  \big)}_{\text{Dorfman bracket on }T\mathcal{F}_+\oplus T^\ast\mathcal{F}_+}\!\!\! + \;\big( [\xi,\zeta]^\ast + \mathcal{L}_\xi^\ast Y - \iota_\zeta\di^\ast X \big)
\end{equation}
where $[-.-]^\ast$, $\mathcal{L}^\ast_{(-)}$ and $\di^\ast$ are operators induced by the Lie bracket of $T\M$. Therefore, if we restrict ourselves to couples of strongly foliated vectors, i.e. $X+\xi,Y+\zeta\in\mathfrak{X}(\mathcal{F}_+)\oplus\Omega^1(\mathcal{F}_+)$, we recover the usual Dorfman bracket
\begin{equation}
    \llbracket X+\xi, Y+\zeta\rrbracket_{\mathrm{D}} \;=\; [X,Y] + \mathcal{L}_X\zeta - \iota_Y\di\xi,
\end{equation}
i.e. we recover generalised geometry.
\vspace{0.25cm}

\noindent An almost para-Hermitian manifold $(\mathcal{M},J,\eta)$ is, in particular, a para-K\"{a}hler manifold if the fundamental $2$-form is symplectic, i.e. $\di\omega=0$. In the general case, the closed $3$-form $\mathcal{K}\in\Omega^3_{\mathrm{cl}}(\mathcal{M})$ defined by $\mathcal{K}:=\di\omega$, which embodies the obstruction of $\omega$ from being symplectic, is interpreted as the generalised fluxes of Double Field Theory.

\subsection{Born geometry}
A Born geometry is the datum of an almost para-Hermitian manifold $(\mathcal{M},J,\omega)$ equipped with a Riemannian metric $\mathcal{G}\in \bigodot^2T^\ast\mathcal{M}$ which is compatible with both the metric $\eta$ and the fundamental $2$-form $\omega$ as follows:
\begin{equation}
    \eta^{-1}\mathcal{G} \,=\, \mathcal{G}^{-1}\eta \quad\;\text{and}\;\quad \omega^{-1}\mathcal{G} \,=\, -\mathcal{G}^{-1}\omega.
\end{equation}
Such a Riemannian metric can be naturally identified with the generalised metric of Double Field Theory.\vspace{0.2cm}

\noindent The generalised diffeomorphisms of Double Field Theory can now be identified with diffeomorphisms of $\mathcal{M}$ which preserve the para-Hermitian metric $\eta$, i.e isometries $\mathrm{Iso}(\mathcal{M},\eta)$. The push-forward of a generalised diffeomorphism $f\in\mathrm{Iso}(\mathcal{M},\eta)$ is nothing but an $O(d,d)$-valued function $f_\ast\in\Coo(\mathcal{M},O(d,d))$. This group of symmetries can be further extended to the group of general bundle automorphisms of $T\mathcal{M}$ preserving the para-Hermitian metric $\eta$.
A generalised diffeomorphism induces a morphism of Born geometries
\begin{equation}
    (\mathcal{M},\,J,\,\omega,\,\mathcal{G})\;\longmapsto\;(\mathcal{M},\, f^\ast J,\,f^\ast\omega,\, f^\ast\mathcal{G}),
\end{equation}
which is an isometry of the para-Hermitian metric, i.e. such that it preserves $\eta = f^\ast\eta$. 
\vspace{0.2cm}

\noindent Particularly interesting is the case of $b$-shifts, which can be seen as a bundle morphisms $e^{b}:T\mathcal{M}\rightarrow T\mathcal{M}$ covering the identity $\mathrm{id}_{\mathcal{M}}$ of the base manifold. This transforms the para-complex structure by $J\mapsto J+b$, which also implies $\omega\mapsto \omega+b$. Therefore, a $b$-shift maps the splitting $T\mathcal{M} = L_+\oplus L_-$ to a new one $T\mathcal{M} =L_+^\prime\oplus L_-$, preserving the eigenbundle $L_-$, but not $L_+$. Therefore, it does not preserve $+$-integrability. \vspace{0.25cm}

\paragraph{Further discussion.} We can notice that Born Geometry is not (at least immediately) related to bundle gerbes, even if theory of foliations is closely related to higher structures as seen by \cite{Vit14}. In the next subsection we will mostly discuss the relation between Born Geometry and the bundle gerbe of the Kalb-Ramond field, trying to clarify it.

\section{Can DFT actually recover bosonic supergravity?}

\subsection{Recovering a string background}
We will now try to recover a general bosonic string background, consisting in a pseudo-Riemannian manifold $(M,g)$ equipped with a Kalb-Ramond field whose H-flux $[H]\in H^3(M,\mathbb{Z})$ is generally non-trivial, from Born Geometry as prescribed by \cite{MarSza19} and \cite{Svo20}.
\vspace{0.25cm}

\noindent Let us start from the almost para-Hermitian manifold $(\mathcal{M},K,\eta)$. The para-complex structure $K$ splits the tangent bundle $T\mathcal{M}= L_+ \oplus L_-$ where $L_\pm$ are its $\pm 1$-eigenbundles. Since we want to recover a conventional supergravity background let us firstly assume that $L_-$ is integrable (physically this corresponds to set the $R$-flux to zero, see \cite{MarSza19}). This implies that there exists a foliation $\mathcal{F}_-$ of $\mathcal{M}$ such that $L_- = T\mathcal{F}_-$. Secondly, since we want to recover a conventional supergravity background, let us require that the leaf space $M:=\mathcal{M}/\mathcal{F}_-$ of this foliation is a smooth manifold. Indeed, according to \cite{MarSza19} and \cite{Svo20}, physical spacetime must be identified with the leaf space $M$. Thus the foliation $\mathcal{F}_-$ is simple and the canonical quotient map $\pi:\mathcal{M}\twoheadrightarrow M= \mathcal{M}/\mathcal{F}_-$ is a surjective submersion, making $\mathcal{M}$ a fibered manifold.
\vspace{0.25cm}

\noindent Now we can use adapted (or fibered) coordinates $(x_{(\alpha)},\tilde{x}_{(\alpha)})$ on each patch $\mathcal{U}_{\alpha}$ of a good cover of the manifold $\mathcal{M}=\bigcup_\alpha \mathcal{U}_{\alpha}$. Thus there exist a frame $\{Z_\mu,\tilde{Z}^\mu\}$ and a dual coframe $\{e^\mu,\tilde{e}_\mu\}$, given on local patches $\mathcal{U}_{\alpha}$ as follows
\begin{equation}
    \begin{aligned}
    Z_{(\alpha)\mu} \,&=\, \frac{\partial}{\partial x^\mu_{(\alpha)}} + N_{(\alpha)\mu\nu}\frac{\partial}{\partial \tilde{x}_{(\alpha)\nu}}, \quad & \tilde{Z}^\mu_{(\alpha)} \,&=\, \frac{\partial}{\partial \tilde{x}_{(\alpha)\mu}}, \\[0.1cm]
    e^\mu_{(\alpha)} \,&=\, \di x^\mu_{(\alpha)}, \quad & \tilde{e}_{(\alpha)\mu} \,&=\, \di \tilde{x}_{(\alpha)\mu} + N_{(\alpha)\mu\nu}\di x^\nu_{(\alpha)},
    \end{aligned}
\end{equation}
such that they diagonalise the tensor $K$ and such that $\{Z_\mu\}$ is a local completion of the holonomic frame for $\Gamma(L_-)$. Notice that the $N_{(\alpha)\mu\nu}\in\Coo\!\left(\mathcal{U}_{\alpha}\right)$ are local functions. In this frame we can express the global $O(d,d)$-metric $\eta = \eta^\mu_\nu \tilde{e}_\mu \odot e^\nu$ and the fundamental $2$-form $\omega = \eta^\mu_\nu \tilde{e}_\mu \wedge e^\nu$. In local coordinates $(x_{(\alpha)},\tilde{x}_{(\alpha)})$ the latter can be written on each patch $\mathcal{U}_{\alpha}$ as
\begin{equation}
    \omega|_{\mathcal{U}_{\alpha}} \,=\, \eta^\mu_\nu \di \tilde{x}_{\mu(\alpha)} \wedge \di x^\nu_{(\alpha)} + \eta^\mu_\nu N_{(\alpha)\mu\lambda}\di x^\lambda_{(\alpha)}\wedge \di x^\nu_{(\alpha)}.
\end{equation}
Now, by following \cite{MarSza19}, we can define a local $2$-form $B_{(\alpha)}\in\Omega^2\!\left(\mathcal{U}_{\alpha}\right)$ by the second term of the $2$-form $\omega|_{\mathcal{U}_{\alpha}}$, i.e.
\begin{equation}
    B_{(\alpha)} \,:=\, \eta^\mu_\nu N_{(\alpha)\mu\lambda}\di x^\lambda_{(\alpha)}\wedge \di x^\nu_{(\alpha)}.
\end{equation}
Now we must ask: what is the condition to make the local $2$-form $B_{(\alpha)}$ on $\mathcal{U}_\alpha$ descend to a proper local $2$-form on the leaf space $M=\mathcal{M}/\mathcal{F}_-$ (which is the physical spacetime)? By following \cite{MarSza19} we can impose the condition that $N_{(\alpha)\mu\lambda}$ are basic functions, i.e.
\begin{equation}
    \mathcal{L}_{X_-}N_{(\alpha)\mu\lambda} =0 \quad \forall X_-\in\Gamma(L_-),
\end{equation}
which assures exactly this. In local coordinates on $\mathcal{U}_{\alpha}$ this condition can be rewritten as
\begin{equation}
    \frac{\partial N_{(\alpha)\mu\nu}}{\partial \tilde{x}_{(\alpha)\lambda}} =0,
\end{equation}
which is solved by $N_{(\alpha)\mu\nu}=N_{(\alpha)\mu\nu}\big(x_{(\alpha)}\big)$, i.e. by asking that the $N_{(\alpha)\mu\nu}$ are local functions only of the $x_{(\alpha)}$-coordinates on each patch $\mathcal{U}_{\alpha}$. Therefore, if our local $2$-form is of the form $B_{(\alpha)}= B_{(\alpha)\mu\nu}\big(x_{(\alpha)}\big)\di x^\mu_{(\alpha)}\wedge \di x^\nu_{(\alpha)}$ it will descend to a local $2$-form $\pi_{\ast} B_{(\alpha)}\in\Omega^2(U_{\alpha})$ on $M$.

\subsection{Papadopoulos' puzzle revised}\label{papadopoulospuzzle}
In the adapted (or fibered) coordinates the transition functions of $\mathcal{M}$ on two-fold overlaps of patches $\mathcal{U}_{\alpha}\cap\mathcal{U}_{\beta}$ will have the simple following form:
\begin{equation}\label{eq:introadapted}
    x_{(\alpha)} = f_{(\alpha\beta)}\big(x_{(\beta)}\big), \quad \tilde{x}_{(\alpha)} = \tilde{f}_{(\alpha\beta)}\big(x_{(\beta)},\tilde{x}_{(\beta)}\big).
\end{equation}
An adapted atlas will be also provided with the property that the sets $U_{\alpha}:=\pi\!\left(\mathcal{U}_{\alpha}\right)$, where $\pi:\mathcal{M}\twoheadrightarrow M= \mathcal{M}/\mathcal{F}_-$ is the quotient map, are patches of the leaf space $M$ with local coordinates $(q_{(\alpha)})$ defined by the equation $x_{(\alpha)}^\mu = q_{(\alpha)}^\mu\circ \pi$. These charts $(U_{\alpha},q_{(\alpha)})$ are uniquely defined. The local $2$-form $B_{(\alpha)}$ will then descend to the local $2$-form $\pi_{\ast} B_{(\alpha)} = B_{(\alpha)\mu\nu}\big(q_{(\alpha)}\big)\di q^\mu_{(\alpha)}\wedge \di q^\nu_{(\alpha)}$.
\vspace{0.25cm}

\noindent Since $\omega\in\Omega^2(\mathcal{M})$ is a global $2$-form, on each two-fold overlap of patches $\mathcal{U}_{\alpha}\cap\mathcal{U}_{\beta}$ we have
\begin{equation}\label{eq:introdiff}
    \begin{aligned}
         \di \tilde{x}_{\mu(\alpha)} \wedge \di x^\mu + B_{(\alpha)} \,=\, \di \tilde{x}_{\mu(\beta)} \wedge \di x^\mu + B_{(\beta)},
    \end{aligned}
\end{equation}
where we suppressed the patch indices on the $1$-forms $\{\di x^\mu\}$: this is because $\{e^\mu\}$ are global $1$-forms on $\mathcal{M}$ and thus we can slightly abuse the notation by calling $e^\mu\equiv \di x^\mu$. Thus we have
\begin{equation}\label{eq:introdx}
(\di \tilde{x}_{(\alpha)\mu} - \di \tilde{x}_{(\beta)\mu}) \wedge \di x^\mu  =  B_{(\beta)} - B_{(\alpha)}.
\end{equation}
Since the local $2$-forms $B_{(\alpha)}$ descend to local $2$-forms on patches $U_{\alpha}\subset M=\mathcal{M}/\mathcal{F}_-$ of the leaf space and these, according to \cite{MarSza19} and \cite{Svo20}, must be physically identified with the local data of the Kalb-Ramond Field, we must have bundle gerbe local data of the following form: 
\begin{equation}\label{eq:introgerbe}
    \begin{aligned}
    \pi_{\ast} B_{(\beta)} - \pi_{\ast} B_{(\alpha)} &= \di \Lambda_{(\alpha\beta)} &&\text{ on } U_{\alpha}\cap U_{\beta}, \\
    \Lambda_{(\alpha\beta)} + \Lambda_{(\beta\gamma)} + \Lambda_{(\gamma\alpha)} &=\di G_{(\alpha\beta\gamma)}  &&\text{ on }U_{\alpha}\cap U_{\beta}\cap U_{\gamma},\\
    G_{(\alpha\beta\gamma)} + G_{(\beta\alpha\delta)} + G_{(\gamma\beta\delta)} + G_{(\delta\alpha\gamma)} &\in 2\pi\mathbb{Z} &&\text{ on }U_{\alpha}\cap U_{\beta}\cap U_{\gamma}\cap U_{\delta}.
    \end{aligned}
\end{equation}
Now, from the patching relations \eqref{eq:introadapted} of the adapted coordinates we obtain
\begin{equation}
    \di \tilde{x}_{(\alpha)\mu} \,=\, \frac{\partial \tilde{f}_{(\alpha\beta)\mu}}{\partial x_{(\beta)}^\nu} \,\di x_{(\beta)}^\nu \,+\, \frac{\partial \tilde{f}_{(\alpha\beta)\mu}}{\partial \tilde{x}_{(\beta)\nu}} \,\di \tilde{x}_{(\beta)\nu}.
\end{equation}
This, combined with \eqref{eq:introdx} and \eqref{eq:introgerbe}, implies the following equations
\begin{equation}
    \frac{\partial \tilde{f}_{(\alpha\beta)\mu}}{\partial \tilde{x}_{(\beta)\nu}} = \delta_{\mu}^{\;\,\nu}, \qquad
    \frac{\partial \tilde{f}_{(\alpha\beta)\mu}}{\partial x_{(\beta)}^\nu} = \left(\di\Lambda_{(\alpha\beta)}\right)\!_{\mu\nu}.
\end{equation}
We can immediately solve the first equation by decomposing $\tilde{f}_{(\alpha\beta)}\!\left(x_{(\beta)},\tilde{x}_{(\beta)}\right) = \tilde{x}_{(\beta)} + \tilde{f}'_{(\alpha\beta)}(x_{(\beta)})$, where $\tilde{f}'_{(\alpha\beta)}(x_{(\beta)})$ is a new basic function of the $x_{(\beta)}$-coordinates only. Now the second equation is equivalent to the new equation $\di (\tilde{f}'_{(\alpha\beta)\mu}\di x^\mu)=-\di\Lambda_{(\alpha\beta)}$ on $U_\alpha\cap U_\beta$, which is solved by
\begin{equation}
    \tilde{f}_{(\alpha\beta)\mu}'\di x^\mu \,=\, -\Lambda_{(\alpha\beta)\mu}\di x^\mu + \di\eta_{(\alpha\beta)},
\end{equation}
where $\di\eta_{(\alpha\beta)}\in\pi^\ast\Omega^1_{\mathrm{ex}}(U_{\alpha}\cap U_{\beta})$ are local exact basic $1$-forms on overlaps of patches. The cocycle condition for transition functions of a manifold on three-fold overlaps of patches implies then
\begin{equation}\label{eq:introlambda}
    \di(\eta_{(\alpha\beta)} + \eta_{(\beta\gamma)} + \eta_{(\gamma\alpha)}) \,=\, \Lambda_{(\alpha\beta)} + \Lambda_{(\beta\gamma)} + \Lambda_{(\gamma\alpha)}.
\end{equation}
Since $\Lambda_{(\alpha\beta)} + \Lambda_{(\beta\gamma)} + \Lambda_{(\gamma\alpha)} = \di G_{\alpha\beta\gamma}$ from \eqref{eq:introgerbe}, then we must have the trivialization
\begin{equation}\label{eq:introtriv}
    G_{(\alpha\beta\gamma)} \,=\, \left(\eta_{(\alpha\beta)} + \eta_{(\beta\gamma)} + \eta_{(\gamma\alpha)}\right) + c_{(\alpha\beta\gamma)},
\end{equation}
where $c_{(\alpha\beta\gamma)}\in\mathbb{R}$ are local constants which must satisfy the following cocycle condition
\begin{equation}
    c_{(\alpha\beta\gamma)} + c_{(\beta\alpha\delta)} + c_{(\gamma\beta\delta)} + c_{(\delta\alpha\gamma)} \in 2\pi\mathbb{Z}
\end{equation}
on each four-fold overlaps of patches of the leaf space. 
\vspace{0.25cm}

\noindent This implies that the gerbe $(\pi_{\ast} B_{(\alpha)},\Lambda_{(\alpha\beta)},G_{(\alpha\beta\gamma)})$ on our physical spacetime $M=\mathcal{M}/\mathcal{F}_-$ is \textit{flat}, i.e. that the curvature of the Kalb-Ramond field $H\in\Omega^3_{\mathrm{ex}}(M)$ is exact and that the topological Dixmier-Douady class $[H]\in H^3(M,\mathbb{Z})$ of the gerbe is torsion. To see this, it is enough to check that the equation \eqref{eq:introlambda} implies that $\Lambda_{(\alpha\beta)}=\di\eta_{(\alpha\beta)}+\tau_{(\alpha)}-\tau_{(\beta)}$ for some local $1$-forms $\tau_{(\alpha)}\in \Omega^1\!\left(U_{\alpha}\right)$. But this implies that $\pi_{\ast} B_{(\alpha)}-\pi_{\ast} B_{(\beta)}=\tau_{(\beta)}-\tau_{(\alpha)}$ so that, in other words, there exist a global $2$-form on the leaf space $M=\mathcal{M}/\mathcal{F}_-$ given by gauge transformations of the Kalb-Ramond field and expressed on overlaps of patches by
\begin{equation}
    \pi_{\ast} B_{(\alpha)} + \di\tau_{(\alpha)} \,=\, \pi_{\ast} B_{(\beta)} + \di\tau_{(\beta)}.
\end{equation}
If we call $B'|_{U_{\alpha}} := \pi_{\ast} B_{(\alpha)} + \di\tau_{(\alpha)}$ the gauge transformed Kalb-Ramond field, we immediately see that it satisfies $H=\di B'$ globally on $M=\mathcal{M}/\mathcal{F}_-$. Therefore in the de Rham cohomology we have the class $[H]=0\in H^3_{\mathrm{dR}}(M)$, which is mapped to a torsion element of the integral cohomology $H^3(M,\mathbb{Z})$. In this context the constants $c_{(\alpha\beta\gamma)}$ are interpreted as a representative of the flat holonomy class $[c_{(\alpha\beta\gamma)}]\in H^2(M,\mathbb{R}/2\pi\mathbb{Z})$ of the flat bundle gerbe.

\paragraph{Open problem.} 
Therefore it does not seem possible to recover a general geometric string background made of a smooth manifold $M$ equipped with a non-trivial Kalb-Ramond field $[H]\in H^3(M,\mathbb{Z})$. And, since DFT was introduced to extend supergravity, the impossibility of recovering supergravity poses a problem. This means that the original argument by \cite{Pap13} is still relevant whenever we try to construct the doubled space as a manifold.\vspace{0.25cm}

\noindent However, as we will see, Born geometry is still extremely efficient in dealing with doubled group manifold and, in particular, Drinfel'd doubles. Remarkable results from the application of para-Hermitian geometry to group manifolds can be found in \cite{MarSza18}, \cite{DFTWZW19} and \cite{MarSza19}. These groups, where fluxes are constant, allow a simple geometrization of the gerbe with a group manifold which is not possible in the general case. Notice that a link between Drinfel'd doubles and bundle gerbes was firstly found by \cite{Wil08}.\vspace{0.2cm}

\noindent The Higher Kaluza-Klein proposal is an attempt to attack this problem and allow the geometrization of general bundle gerbes. In the Higher Kaluza-Klein perspective the doubled space would be identified with the total space of a gerbe $\mathscr{G}\twoheadrightarrow M$. Thus the quotient $\pi:\mathcal{M}\twoheadrightarrow M=\mathcal{M}/\mathcal{F}_-$ is reinterpreted as a local version of the projector of the gerbe as a principal $\infty$-bundle, i.e.
\begin{equation}
    \Pi:\,\mathscr{G}\; \longtwoheadrightarrow \; M \,\cong\, \mathscr{G}/\!/\mathbf{B}U(1),
\end{equation}
which is the higher geometric version of the statement $\pi:P\twoheadrightarrow M\cong P/G$ for any $G$-bundle $P$. See the next section for an introduction.

\begin{savequote}[8cm]
Definitio ut dicatur perfecta, debebit intimam essentiam rei explicare, et cavere, ne eius loco propria quaedam usurpemus.

A definition, if it is to be called perfect, must explain the inmost essence of a thing, and must take care not to substitute for this any of its properties. 
  \qauthor{--- Spinoza, \textit{Tractatus de Intellectus Emendatione}}
\end{savequote}

\chapter{\label{ch:5}Global Double Field Theory as higher Kaluza-Klein theory}

\minitoc


\noindent In this chapter we will give a formal definition of higher Kaluza-Klein Theory and we will explain how this can be interpreted as a global version of Double Field Theory (DFT). 
In this chapter we will construct the correspondence between the doubled geometry of Double Field Theory and the higher geometry of bundle gerbes. We will define the atlas of a bundle gerbe and we will show that it can be identified with the doubled space of Double Field Theory. This will have the consequence that Double Field Theory can be globally interpreted as a field theory on the total space of a bundle gerbe, just like ordinary Kaluza-Klein theory lives on the total space of a principal bundle. This chapter is based on \cite{Alf19,Alfonsi:2021uwh}

\section{Doubled/higher correspondence}

The aim of this section will be to prove the existence of a correspondence between doubled and higher geometry in a linearised form. 

\begin{definition}[$\mathfrak{string}$ Lie $2$-algebra]
Let us call $\mathfrak{string}:=\mathbb{R}^{d}\oplus\mathbf{b}\mathfrak{u}(1)$ the $2$-algebra of the abelian Lie $2$-group $\mathbb{R}^{d}\times\mathbf{B}U(1)$. It is well-understood that any $L_\infty$-algebra $\mathfrak{g}$ is equivalently described in terms of its Chevalley-Eilenberg dg-algebra $\mathrm{CE}(\mathfrak{g})$. In our particular case this is
\begin{equation}
    \mathrm{CE}\!\left(\mathfrak{string}\right) \;=\; \mathbb{R}[e^a,B]/\langle\di e^a=0,\;\di B = 0\rangle,
\end{equation}
where $\{e^a\}_{a=0,\dots,d-1}$ are generators in degree $1$ and $B$ is a generator in degree $2$.
\end{definition}

\noindent The Lie $2$-algebra $\mathfrak{string}=\mathbb{R}^{d}\oplus\mathbf{b}\mathfrak{u}(1)\twoheadrightarrow \mathbb{R}^{d}$ can be interpreted as a linearisation of a bundle gerbe, in the sense proposed by \cite{Fiorenza:2013nha}. Thus, such a Lie $2$-algebra can be thought as trivially made up of a flat Minkowski space and a trivial Kalb-Ramond field. 
Now, we want to introduce a notion of atlas for this $2$-algebra.

\begin{remark}[Atlas of an $L_\infty$-algebra]\label{def:atlasAlg}
By linearising the notion of atlas of a smooth stack \cite{khavkine2017synthetic}, we obtain that the atlas of an $L_\infty$-algebra $\mathfrak{g}$ can be defined by an ordinary Lie algebra $\mathfrak{atlas}$ equipped with a homomorphism of $L_\infty$-algebras $\phi:\mathfrak{atlas} \longrightarrow \mathfrak{g}$ that is surjective onto the $0$-truncation of $\mathfrak{g}$.
In this dissertation, we will also require the dual homomorphism $\phi^\ast:\mathrm{CE}(\mathfrak{g})\hookrightarrow\mathrm{CE}(\mathfrak{atlas})$ of dg-algebras to be injective.
\end{remark}

\noindent We will also need the following slight specialisation of the notion of atlas of an $L_\infty$-algebra, which will be useful to deal with our physically motivated examples.

\begin{definition}[Lorentz-compatible atlas]\label{def:lorentzcom}
Let $\mathfrak{g}\twoheadrightarrow \mathbb{R}^{d}$ be an $L_\infty$-algebra fibrated on a Minkowski space and equipped with an atlas $\phi:\mathfrak{atlas} \longtwoheadrightarrow \mathfrak{g}$. We say that the atlas is \textit{Lorentz-compatible} if $\mathfrak{atlas}$ comes equipped with a $SO(1,d-1)$-action such that
\begin{enumerate}
    \item it non-trivially extends the natural $SO(1,d-1)$-action on $\mathbb{R}^d$,
    \item $\phi$ is $SO(1,d-1)$-equivariant,
\end{enumerate}
and if $\mathrm{dim}(\mathfrak{atlas})$ is the minimal dimension for which the above conditions are satisfied.
\end{definition}

\noindent Notice that, in a Lorentz-compatible atlas, the images of the higher generators of $\mathfrak{g}$ through the map $\phi^\ast:\mathrm{CE}(\mathfrak{g})\hookrightarrow\mathrm{CE}(\mathfrak{atlas})$ are non-zero elements which are invariant under Lorentz action.
We are now ready to present the main result of this section.

\begin{theorem}[$\mathfrak{double}/\mathfrak{string}$ correspondence]
The Lorentz-compatible atlas of the Lie $2$-algebra $\mathfrak{string}$ is the para-K\"{a}hler vector space $\big(\mathbb{R}^{d}\oplus (\mathbb{R}^{d})^\ast,\,J,\,\omega\big)$, where 
\begin{itemize}
    \item $J$ is the para-complex structure corresponding to the canonical splitting $\mathbb{R}^{d}\oplus (\mathbb{R}^{d})^\ast$,
    \item $\omega$ is the symplectic structure given by the transgression of the higher generator of $\mathfrak{string}$ to the space of the atlas $\mathbb{R}^{d}\oplus (\mathbb{R}^{d})^\ast$.
\end{itemize}
\end{theorem}
\begin{proof}
Recall the definition \ref{def:atlasAlg} of atlas $\phi:\mathfrak{atlas} \longtwoheadrightarrow \mathfrak{string}$ for an $L_\infty$-algebra. The map $\phi$ can be dually given as an embedding $\phi^\ast:\mathrm{CE}\!\left(\mathfrak{string}\right)\longhookrightarrow \mathrm{CE}(\mathfrak{atlas})$ between their Chevalley-Eilenberg dg-algebras.
Thus, we want to identify an ordinary Lie algebra $\mathfrak{atlas}$ such that its Chevalley-Eilenberg dg-algebra contains a $2$-degree element 
\begin{equation}
    \omega \,:=\, \phi^\ast (B)\;\in\,\mathrm{CE}(\mathfrak{atlas})
\end{equation}
which is the image of the $2$-degree generator of $\mathrm{CE}\!\left(\mathfrak{string}\right)$ and which must satisfy the equation
\begin{equation}
    \di \omega \,=\, 0,
\end{equation}
given by the fact that a homomorphism of dg-algebras maps $\phi^\ast(0)=0$.
Recall that $\mathfrak{atlas}$ must be an ordinary Lie algebra, so its Chevalley-Eilenberg dg-algebra $\mathrm{CE}(\mathfrak{atlas})$ will only have $1$-degree generators. 
Since we want a Lorentz-compatible atlas, $\omega$ must also be a singlet under Lorentz transformations. Thus the generators of the atlas must consist not only in the images $\{e^a:=\phi^\ast(e^a)\}_{a=0,\dots,d-1}$, but also in an extra set $\{\widetilde{e}_a\}_{a=0,\dots,d-1}$ which generates $(\mathbb{R}^d)^\ast$. This way the image of the generator $b\in\mathrm{CE}(\mathfrak{string})$ is
\begin{equation}
    \omega \,=\, \widetilde{e}_a \wedge e^a,
\end{equation}
which is Lorentz-invariant.
Indeed, the generators $\{e^a\}$ of $\mathbb{R}^d$ transform by $e^a\mapsto N^a_{\;b}e^b$ for $N\in SO(1,d-1)$, while the generators $\{\widetilde{e}_a\}$ of $(\mathbb{R}^d)^\ast$ transform by $\widetilde{e}_a\mapsto (N^{-1})^{\;b}_{a}\widetilde{e}_b$.
Now, the equation $\di \omega =0$, combined with the equation $\di e^a = 0$, implies that the differential of the new generator is zero, i.e. $\di \widetilde{e}_a=0$. Therefore, we found the dg-algebra
\begin{equation}
    \mathrm{CE}(\mathfrak{double}) \;=\; \mathbb{R}[e^a,\widetilde{e}_a]/\langle\di e^a=0,\;\di \widetilde{e}_a = 0\rangle
\end{equation}
where we renamed the Lie algebra $\mathfrak{atlas}$ to $\mathfrak{double}$. This ordinary Lie algebra is immediately $\mathfrak{double}=\big(\mathbb{R}^{d}\oplus (\mathbb{R}^{d})^\ast,\,[-,-]=0\big)$, i.e. the abelian Lie algebra whose underlying $2d$-dimensional vector space is $\mathbb{R}^{d}\oplus (\mathbb{R}^{d})^\ast$.
Now, recall that the Chevalley-Eilenberg dg-algebra $\mathrm{CE}(\mathfrak{g})$ of any ordinary Lie algebra $\mathfrak{g}$ is isomorphic to the dg-algebra $\big(\Omega_{\mathrm{li}}^\bullet (G),\,\di\big)$ of left-invariant differential forms on the corresponding Lie group $G=\exp(\mathfrak{g})$. Therefore, we have the isomorphism
\begin{equation}
    \mathrm{CE}(\mathfrak{double}) \;\cong\; \big(\Omega_{\mathrm{li}}^\bullet (\mathbb{R}^{d,d}),\,\di\big)
\end{equation}
where we called $\mathbb{R}^{d,d}$ the abelian Lie group integrating $\mathfrak{double}$ whose underlying smooth manifold is still the linear space $\mathbb{R}^d\times(\mathbb{R}^d)^\ast$. Thus, the smooth functions $\Coo(\mathbb{R}^{d,d})$ are generated by coordinate functions $x^a$ and $\widetilde{x}_a$ and the basis of left-invariant $1$-forms on $\mathbb{R}^{d,d}$ is simply given by
\begin{equation}
    \begin{aligned}
    e^a \,=\, \di x^a,\quad \widetilde{e}_a \,=\, \di \widetilde{x}_a
    \end{aligned}
\end{equation}
Thus, the transgressed element $\omega\in\mathrm{CE}(\mathfrak{double})$ is equivalently the symplectic form $\omega = \di \widetilde{x}_a \wedge \di x^a$.
Moreover, the canonical splitting $\mathbb{R}^d\oplus(\mathbb{R}^d)^\ast$ induces a canonical para-complex structure $J$, which is compatible with the symplectic form $\omega$. Therefore, the atlas of $\mathfrak{string}$ is equivalently a para-K\"{a}hler vector space $\big(\mathbb{R}^d\oplus(\mathbb{R}^d)^\ast,\,J,\,\omega\big)$.
\end{proof}

\begin{remark}[Emergence of para-Hermitian geometry]
On one side of the correspondence, the Lie $2$-algebra $\mathfrak{string}=\mathbb{R}^d\oplus\mathbf{b}\mathfrak{u}(1)$ is the linearisation of a bundle gerbe and, on the other side, the para-K\"{a}hler vector space $\big(\mathbb{R}^d\oplus(\mathbb{R}^d)^\ast,\,J,\,\omega\big)$ is the linearisation of a para-Hermitian manifold. The latter is the atlas of the former.
\end{remark}

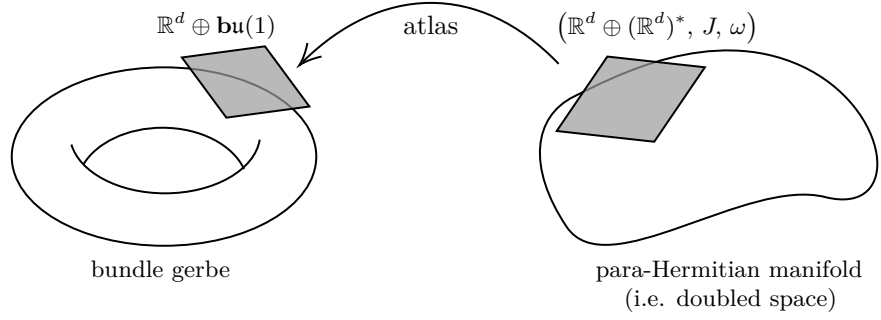
\begin{figure}[h]\begin{center}
\tikzset{every picture/.style={line width=0.75pt}} 
\begin{tikzpicture}[x=0.75pt,y=0.75pt,yscale=-1,xscale=1]
\draw   (10,74.38) .. controls (10,49.59) and (44.03,29.5) .. (86,29.5) .. controls (127.97,29.5) and (162,49.59) .. (162,74.38) .. controls (162,99.16) and (127.97,119.25) .. (86,119.25) .. controls (44.03,119.25) and (10,99.16) .. (10,74.38) -- cycle ;
\draw   (290.5,44.25) .. controls (310.5,34.25) and (359,8) .. (404,28.75) .. controls (449,49.5) and (455.5,104.75) .. (416,94.75) .. controls (376.5,84.75) and (307,139.75) .. (287,109.75) .. controls (267,79.75) and (270.5,54.25) .. (290.5,44.25) -- cycle ;
\draw  [draw opacity=0] (133.77,65.87) .. controls (131.39,81.03) and (111.14,92.87) .. (86.5,92.87) .. controls (63.22,92.87) and (43.84,82.29) .. (39.78,68.33) -- (86.5,62.87) -- cycle ; \draw   (133.77,65.87) .. controls (131.39,81.03) and (111.14,92.87) .. (86.5,92.87) .. controls (63.22,92.87) and (43.84,82.29) .. (39.78,68.33) ;
\draw  [draw opacity=0] (45.78,78.33) .. controls (52.22,67.49) and (67.47,59.88) .. (85.25,59.88) .. controls (104.28,59.88) and (120.4,68.6) .. (125.94,80.65) -- (85.25,89.88) -- cycle ; \draw   (45.78,78.33) .. controls (52.22,67.49) and (67.47,59.88) .. (85.25,59.88) .. controls (104.28,59.88) and (120.4,68.6) .. (125.94,80.65) ;
\draw  [fill={rgb, 255:red, 155; green, 155; blue, 155 }  ,fill opacity=0.7 ] (307.29,24.08) -- (355.99,29.47) -- (330.71,66.92) -- (282.01,61.53) -- cycle ;
\draw  [fill={rgb, 255:red, 155; green, 155; blue, 155 }  ,fill opacity=0.7 ] (136.39,18.93) -- (94.94,24.54) -- (117.12,54.81) -- (158.57,49.2) -- cycle ;
\draw    (155.26,26.71) .. controls (194,-11.83) and (243.11,-14.11) .. (283,27.75) ;
\draw [shift={(153.5,28.5)}, rotate = 314.1] [color={rgb, 255:red, 0; green, 0; blue, 0 }  ][line width=0.75]    (10.93,-4.9) .. controls (6.95,-2.3) and (3.31,-0.67) .. (0,0) .. controls (3.31,0.67) and (6.95,2.3) .. (10.93,4.9)   ;
\draw (204,2) node [anchor=north west][inner sep=0.75pt]  [font=\small] [align=left] {atlas};
\draw (280,0) node [anchor=north west][inner sep=0.75pt]  [font=\footnotesize]  {$\big(\mathbb{R}^{d} \oplus (\mathbb{R}^{d})^{\ast},\, J, \,\omega \big)$};
\draw (81,0) node [anchor=north west][inner sep=0.75pt]  [font=\footnotesize]  {$\mathbb{R}^{d} \oplus \mathbf{b}\mathfrak{u}( 1)$};
\draw (48,125) node [anchor=north west][inner sep=0.75pt]  [font=\footnotesize] [align=left] {bundle gerbe};
\draw (300,125) node [anchor=north west][inner sep=0.75pt]  [font=\footnotesize] [align=left] {\begin{minipage}[lt]{100pt}\setlength\topsep{0pt}
para-Hermitian manifold
\begin{center}
(i.e. doubled space)
\end{center}
\end{minipage}};
\end{tikzpicture}
\caption[Para-Hermitian geometry of the atlas of a bundle gerbe]{para-Hermitian geometry (i.e. the geometry of doubled spaces) as the atlas description of bundle gerbes.}
\end{center}\vspace{-0.5cm}\end{figure}

\noindent Notice that the subgroup of linear transformations $GL(2d)$ preserving the para-complex structure $J$ is given by $GL(d)\times GL(d)$. The subgroup preserving the fundamental $2$-form $\omega$ is $Sp(2d,\mathbb{R})$, while the one preserving para-hermitian metric $\eta$ is $O(d,d)$. The subgroup of linear transformations $GL(2d)$ preserving the whole para-K\"{a}hler structure $(J,\omega,\eta)$ is just the group of linear transformations of the base $GL(d)$. Hence, we have the usual identities
\begin{equation}\begin{aligned}
    O(d,d) \,\cap\, Sp(2d,\mathbb{R}) \,&=\, GL(d), \\
    Sp(2d,\mathbb{R}) \,\cap\, \big(GL(d)\times GL(d)\big) \,&=\, GL(d), \\
    \big(GL(d)\times GL(d)\big) \,\cap\, O(d,d) \,&=\, GL(d).
\end{aligned}\end{equation}

\begin{remark}[Kernel pair of the atlas of $\mathfrak{string}$]
Now let us discuss the kernel pair of the atlas $\phi:\mathfrak{atlas} \longtwoheadrightarrow \mathfrak{string}$. This is defined as the pullback (in the category theory sense) of two copies of the map $\phi$ of the atlas. The coequalizer diagram of these maps is
\begin{equation}\label{eq:kernelpair}
\begin{tikzcd}
\mathfrak{double}\times_{\mathfrak{string}}\mathfrak{double} \arrow[r, yshift=0.7ex, two heads] \arrow[r, yshift=-0.7ex, two heads] & \mathfrak{double}  \arrow[r, two heads, "\phi"] & \mathfrak{string}.
\end{tikzcd}
\end{equation}
To deal with it, we can consider the Chevalley-Eilenberg algebras of all the involved $L_\infty$-algebras and look at the equalizer diagram of the cokernel pair which is dual to the starting kernel pair \eqref{eq:kernelpair}. This will be given by the following maps of differential graded algebras:
\begin{equation}
\begin{tikzcd}
\mathrm{CE}(\mathfrak{double})\sqcup_{\mathrm{CE}\left(\mathfrak{string}\right)}\mathrm{CE}(\mathfrak{double})  & \mathrm{CE}(\mathfrak{double}) \arrow[l, yshift=0.7ex, hook'] \arrow[l, yshift=-0.7ex, hook']  & \mathrm{CE}(\mathfrak{string}) \arrow[l, hook', "\phi^\ast"'].
\end{tikzcd}
\end{equation}
Let us describe this in more detail. When composed with $\phi^\ast$, the two maps at the centre of the diagram both send the generators $e^a\in \mathrm{CE}(\mathfrak{string})$ to $e^a \in \mathrm{CE}(\mathfrak{double})\sqcup_{\mathrm{CE}\left(\mathfrak{string}\right)}\mathrm{CE}(\mathfrak{double})$. However, they map the generator $B\in \mathrm{CE}(\mathfrak{string})$ to two different elements $\omega=\widetilde{e}_a\wedge e^a$ and $\omega'=\widetilde{e}'_a\wedge e^a$, where $\widetilde{e}_a$ and $\widetilde{e}'_a$ are such that they both satisfy the same equation $\di \widetilde{e}_a' = \di \widetilde{e}_a$. This implies that they are related by a gauge transformation $\widetilde{e}'_a = \widetilde{e}_a + \di\lambda_a$. This can be seen as a consequence of the gauge transformations $B'=B+\di\lambda$ with parameter $\lambda:=\lambda_ae^a$.
\end{remark}

\begin{remark}[T-duality on the $\mathfrak{double}$ algebra]\label{rem:algduality}
The ordinary Lie algebra $\mathfrak{double}$ is not the atlas only of the Lie $2$-algebra $\mathfrak{string}$, but of an entire class of Lie $2$-algebras. For example, we have
\begin{equation}
    \begin{tikzcd}[row sep=5ex, column sep=2ex] 
    & \mathfrak{double} \arrow[dr, two heads, "\widetilde{\phi}"]\arrow[dl, two heads, "\phi"']& \\
    \mathfrak{string} & & \widetilde{\mathfrak{string}}  
\end{tikzcd}
\end{equation}
where we called $\widetilde{\mathfrak{string}}$ the Lie $2$-algebra whose Chevalley-Eilenberg dg-algebra is given by $\mathrm{CE}\big(\widetilde{\mathfrak{string}}\big)= \mathbb{R}\big[\widetilde{e}_a,\widetilde{B}\big]/\langle\di \widetilde{e}_a = 0,\;\di\widetilde{B}=0\rangle$ and where $\widetilde{\phi}$ is the atlas mapping the generators by $\widetilde{e}_a\mapsto\widetilde{e}_a$ and $\widetilde{B}\mapsto e^a\wedge \widetilde{e}_a$. 
The Lie $2$-algebra $\widetilde{\mathfrak{string}}=(\mathbb{R}^d)^\ast\oplus \mathbf{b}\mathfrak{u}(1)$ can be immediately seen as the T-dualisation of $\mathfrak{string}$ along all the $d$ directions of the underlying spacetime. More generally, $\mathfrak{double}$ will be the atlas of any T-dual of the Lie $2$-algebra $\mathfrak{string}$: this is nothing but a linearised version of T-duality of bundle gerbes. This remark can look quite trivial, but it will be important both in the next section and in chapter \ref{ch:7}.
\end{remark}

\section{Doubled space as atlas of a bundle gerbe}

In the previous section, we established a correspondence between linearised doubled geometries and $L_\infty$-algebras, which interprets the former as the atlas description of the latter. In this section we will globalise this relation to construct a correspondence between doubled spaces and bundle gerbes based on the notion of atlas of a smooth stack.

\begin{remark}[On the nature of the extra coordinates]
The $2d$-dimensional atlas of the bundle gerbe is the natural candidate for being an atlas for the doubled space where Double Field Theory lives.
This way, we can completely avoid the conceptual issue of postulating many new extra dimensions in extended geometry, because the extra coordinates which appears in the extended charts describe the degrees of freedom of a bundle gerbe. In this sense, a flat doubled space $\mathbb{R}^{d,d}$ can be seen as a coordinate description of a trivial bundle gerbe. 
\end{remark}

\begin{remark}[Atlas for the Lie $2$-group]
Let $\mathbb{R}^{d}\times \mathbf{B}U(1)$ be the Lie $2$-group which integrates the Lie $2$-algebra $\mathfrak{string}:=\mathbb{R}^{d}\oplus\mathbf{b}\mathfrak{u}(1)$. Let us call again $\mathbb{R}^{d,d}$ the ordinary Lie group which integrates the ordinary abelian Lie algebra $\mathbb{R}^{d}\oplus (\mathbb{R}^{d})^\ast$. Therefore, we have a homomorphism of Lie groups
\begin{equation}\label{eq:groupatlas}
   \exp(\phi):\,\mathbb{R}^{d,d}\; \longtwoheadrightarrow\; \mathbb{R}^{d}\times\mathbf{B}U(1),
\end{equation}
which exponentiates the homomorphism of Lie algebras $\phi:\mathfrak{atlas} \longtwoheadrightarrow \mathfrak{string}$ from the previous section. Consequently, this is also a well defined atlas for $\mathbb{R}^{d}\times\mathbf{B}U(1)$, seen as a smooth stack. 
\end{remark}

\begin{definition}[Lorentz-compatible atlas of an $\infty$-bundle]
Let us define a \textit{Lorentz-compatible atlas} for an $\infty$-bundle $\mathscr{P}\xtwoheadrightarrow{\;\Pi\;} M$ as an atlas whose charts are Lorentz-compatible in the sense of definition \ref{def:lorentzcom}.
\end{definition}

\begin{theorem}[Doubled space/bundle gerbe correspondence]
The Lorentz-compatible atlas of a bundle gerbe $\mathscr{G}\xtwoheadrightarrow{\;\Pi\;} M$ is a para-Hermitian manifold $\left(\M,\,J,\,\omega\right)$, where
\begin{itemize}
    \item $J$ is the para-complex structure corresponding to the splitting of $T\M$ into horizontal and vertical bundle induced by the connection of the bundle gerbe,
    \item $\omega$ is the fundamental $2$-form given by the transgression of the connection of the bundle gerbe, i.e. which satisfies $\pi^\ast H=-\di\omega$ with $H\in\Omega^{3}_{\mathrm{cl}}(M)$ curvature of the bundle gerbe.
\end{itemize}
\end{theorem}

\begin{proof}
Let $\mathscr{G}\twoheadrightarrow M$ be a bundle gerbe on a base manifold $M$. Thus $\mathscr{G}$ can be locally trivialised as a collection of local trivial gerbes $\{U_\alpha \times \mathbf{B}U(1)\}_{\alpha\in I}$ on a given open cover $\{U_\alpha\}_{\alpha\in I}$ of the base manifold $M$. 
Thus, we have an effective epimorphism $\varphi_\alpha: \mathbb{R}^{d}\times \mathbf{B}U(1)\twoheadrightarrow U_\alpha \times \mathbf{B}U(1)$ for any chart. These can be combined in a single effective epimorphism $\bigsqcup_{\alpha\in I}\mathbb{R}^{d}\times \mathbf{B}U(1)\xtwoheadrightarrow{\;\;\{\varphi_\alpha\}_{\alpha\in I}\;\;} \mathscr{G}$. Thus, we can cover the bundle gerbe with copies of the Lie $2$-group $\mathbb{R}^d\times\mathbf{B}U(1)$. 
Since this Lie $2$-group comes equipped with the natural atlas \eqref{eq:groupatlas}, we can define the composition maps
$\Phi_\alpha:\mathbb{R}^{d,d} \xtwoheadrightarrow{\;\;\exp(\phi)\;\;} \mathbb{R}^d\times\mathbf{B}U(1) \xtwoheadrightarrow{\;\;\varphi_\alpha\;\;} U_\alpha \times\mathbf{B}U(1)$. By combining them we can construct an effective epimorphism
\begin{equation}\label{eq:atlas}
    \Phi:\; \bigsqcup_{\alpha\in I}\mathbb{R}^{d,d}\; \xtwoheadrightarrow{\;\; \{\Phi_\alpha\}_{\alpha\in I} \;\;}\; \mathscr{G} 
\end{equation}
From now on, let us call the total space of the atlas $\M:=\bigsqcup_{\alpha\in I}\mathbb{R}^{d,d}$. Notice that, in general, this is a disjoint union of $\mathbb{R}^{d,d}$-charts.
We can now use the map \eqref{eq:atlas} to explicitly construct the \v{C}ech nerve of the atlas. What we obtain is the following simplicial object:
\begin{equation*}
\begin{tikzcd}[column sep=9ex]
\displaystyle\bigsqcup_{\alpha,\beta,\gamma\in I}\!\!\mathbb{R}^{d,d} \times_\mathscr{G} \mathbb{R}^{d,d} \times_\mathscr{G} \mathbb{R}^{d,d} \arrow[r, yshift=1.4ex, two heads]\arrow[r, two heads] \arrow[r, yshift=-1.4ex, two heads] & \displaystyle\bigsqcup_{\alpha,\beta\in I}\!\mathbb{R}^{d,d} \times_\mathscr{G} \mathbb{R}^{d,d} \arrow[r, yshift=0.7ex, two heads] \arrow[r, yshift=-0.7ex, two heads] & \displaystyle\bigsqcup_{\alpha\in I}\mathbb{R}^{d,d}  \arrow[r, two heads, "\{\Phi_\alpha\}_{\alpha\in I} "] & \mathscr{G},
\end{tikzcd}
\end{equation*}
which tells us how the charts of the atlas are glued by morphisms.
Let us describe this diagram in more detail in terms of its dual diagram of Chevalley-Eilenberg dg-algebras. Let us also call $\big(B_{(\alpha)},\,\Lambda_{(\alpha\beta)},\,G_{(\alpha\beta\gamma)}\big)$ the \v{C}ech cocycle of the bundle gerbe. The two maps of the kernel pair send the local $1$-degree generator to $\di x^\mu$ and the local $2$-degree generator to a couple of local $2$-forms $\omega_{(\alpha)}^{\mathrm{triv}}=\di\widetilde{x}_{(\alpha)\mu}\wedge \di x^\mu$ and $\omega_{(\beta)}^{\mathrm{triv}}=\di\widetilde{x}_{(\beta)\mu}\wedge \di x^\mu$ on the fiber product of the $\alpha$-th and $\beta$-th charts. Now the local $1$-forms $\di\widetilde{x}_{(\alpha)\mu}$ and $\di\widetilde{x}_{(\beta)\mu}$ are required to be related by a gauge transformation $\di\widetilde{x}_{(\alpha)\mu} = \di\widetilde{x}_{(\beta)\mu} + \di\Lambda_{(\alpha\beta)\mu}$ where the gauge parameters $\Lambda_{(\alpha\beta)\mu}$ are given by the cocycle of the bundle gerbe. Equivalently, the two $2$-forms must be related by a gauge transformation $\omega_{(\alpha)}^{\mathrm{triv}}=\omega_{(\beta)}^{\mathrm{triv}}+\di\Lambda_{(\alpha\beta)}$  with gauge parameter $\Lambda_{(\alpha\beta)}:=\Lambda_{(\alpha\beta)\mu}\di x^\mu$. The gauge parameters are required to satisfy the cocycle condition $\Lambda_{(\alpha\beta)}+\Lambda_{(\beta\gamma)}+\Lambda_{(\gamma\alpha)}=\di G_{(\alpha\beta\gamma)}$ on three-fold overlaps of charts.\vspace{0.2cm}

\noindent On the atlas $\M$ of the bundle gerbe, we can define a $2$-form $\omega\in\Omega^2(\M)$ by taking the difference $\omega_{(\alpha)} := \omega^{\mathrm{triv}}_{(\alpha)} - \pi^\ast B_{(\alpha)}$ of the local $2$-form $\omega^{\mathrm{triv}}_{(\alpha)}$ and the pullback of the local connection $2$-form $B_{(\alpha)}$ of the bundle gerbe from the base manifold on each chart $\mathbb{R}^{d,d}$. This definition assures that $\omega_{(\alpha)} =  \omega_{(\beta)}$ on overlaps of charts $\mathbb{R}^{d,d}\times_\mathscr{G}\mathbb{R}^{d,d}$. Therefore, this $2$-form is globally well-defined and we can write it simply as $\omega$, by removing the $\alpha$-index. In local coordinates we can write
\begin{equation}\label{eq:omegarev}
    \omega \;=\; \big(\di\widetilde{x}_{(\alpha)\mu}+B_{(\alpha)\mu\nu}\di x^\nu \big)\wedge \di x^\mu
\end{equation}
Notice that the form $\omega$ is, more generally, invariant under gauge transformations of the bundle gerbe. From the definition of $\omega$, we obtain the relation with curvature of the bundle gerbe:
\begin{equation}
    \pi^\ast H\,=\,-\di\omega, \quad \text{with} \quad H\in\Omega^3_{\mathrm{cl}}(M).
\end{equation}
Now, we want to show that $\M$ is canonically para-Hermitian with fundamental $2$-form $\omega$. The projection $\pi: \M \twoheadrightarrow M$ induces a short exact sequence of vector bundles: 
\begin{equation}\label{eq1}
    0 \longhookrightarrow \mathrm{Ker}(\pi_\ast) \longhookrightarrow T\M \xtwoheadrightarrow{\;\pi_\ast\;} \pi^\ast TM \xtwoheadrightarrow{\quad} 0
\end{equation}
From the definition of the $2$-form $\omega$, we can see that it is a projector to the vertical bundle
\begin{equation}
    \omega:\, T\M \,\longtwoheadrightarrow\, \mathrm{Ker}(\pi_\ast)
\end{equation}
Therefore, the $2$-form $\omega$ defines the splitting $\pi_\ast\oplus\omega$ into horizontal and vertical bundle
\begin{equation}\label{eq2}
    T\M \; \cong\; \pi^\ast TM \,\oplus\,  \mathrm{Ker}(\pi_\ast).
\end{equation}
This splitting canonically defines a para-complex structure $J\in\mathrm{Aut}(T\M)$. If we split any vector in horizontal and vertical projection $X=X_H+X_V$, the para-complex structure $J$ is defined such that $J(X)= X_H-X_V$. Notice that, since $J$ defines a splitting $T\M =L_+ \oplus L_-$ of the tangent bundle of $\M$, as seen in chapter \ref{ch:4}, this identifies $L_+ \equiv \pi^\ast TM$ and $L_-\equiv\mathrm{Ker}(\pi_\ast)$.
Therefore, the atlas of a bundle gerbe is a para-Hermitian manifold $(\M,\,J,\,\omega)$ with para-complex structure $J$ and fundamental $2$-form $\omega$, defined above. Thus, we have the conclusion of the lemma.
\end{proof}

\noindent Therefore, we can now make the following identification.

\begin{post}[Doubled space]\label{post1}
A \textit{doubled space} $\mathcal{M}$ is the atlas of a bundle gerbe $\mathscr{G}\xrightarrow{\;\bbpi\;}M$ with connective structure (i.e. a principal $\mathbf{B}U(1)$-bundle on a smooth manifold $M$).
\end{post}

\begin{remark}[Topological classification of doubled spaces]\label{rem:ddds}
Notice that the trivialisation we introduced defines \v{C}ech cocycle $\big[G_{(\alpha\beta\gamma)}\big]\in H^{2}\big(M,\,\Coo(-,S^1)\big)$, where $\Coo(-,S^1)$ is the sheaf of maps to the circle. It is a well-established result (e.g. see see \cite{Hit99} for details) that there exists an isomorphism $H^{2}\big(M,\,\Coo(-,S^1)\big)\cong H^{3}(M,\mathbb{Z})$, induced by the short exact sequence of sheaves $0\rightarrow \mathbb{Z} \rightarrow \Coo(-,\mathbb{R}) \rightarrow \Coo(-,S^1) \rightarrow 0$. The image of $\big[G_{(\alpha\beta\gamma)}\big]$ along such an isomorphism will be an element of the $3$rd cohomology group of the base manifold $M$, which we will call Dixmier-Douady class $\mathrm{dd}(\mathscr{G})\in H^{3}(M,\mathbb{Z})$ of the bundle gerbe. Thus, bundle gerbes $\mathscr{G}\in\mathbf{H}$ are topologically classified by a Dixmier-Douady class $\mathrm{dd}(\mathscr{G})\in H^{3}(M,\mathbb{Z})$.
\end{remark}

\begin{remark}[Principal connection of the bundle gerbe]\label{remarkrev}
Let $\Omega^2\in\mathbf{H}$ be the usual sheaf of differential $2$-forms over smooth manifolds.
We can define a differential $2$-form $\underline{\omega}$ on the bundle gerbe $\mathscr{G}$ as a map  $\mathscr{G}\xrightarrow{\;\underline{\omega}\;}\Omega^2$. Notice that, given the fundamental $2$-form $\omega$ in \eqref{eq:omegarev}, we can construct a $2$-form $\mathscr{G}\xrightarrow{\;\underline{\omega}\;}\Omega^2$ given as follows:
\begin{equation*}
\begin{tikzcd}[column sep=7ex, row sep=7ex]
\M \arrow[r, "\Phi_{\alpha}"]\arrow[rr, bend right=35, "\omega_{(\alpha)}"]& \mathscr{G} \arrow[r, "\underline{\omega}"] & \Omega^2
\end{tikzcd}, \qquad
\begin{tikzcd}[column sep=7ex, row sep=7ex]
\M \times_\mathscr{G} \M \arrow[bend left=40, "\omega_{(\alpha)}"]{r}[name=U,below]{}
\arrow[bend right=40, "\omega_{(\beta)}"']{r}[name=D]{} &
\Omega^2 \arrow[Rightarrow, to path=(U) -- (D)]{}
\end{tikzcd}.
\end{equation*}
Since $\Omega^2$ is a $0$-truncated stack, the $2$-morphism in the second diagram is just an identity. In other words we obtain that $\underline{\omega}\in\Omega^2(\mathscr{G})$ is given on the atlas by a collection of local $2$-forms $\omega_{(\alpha)} =\big(\di\widetilde{x}_{(\alpha)\mu}+B_{(\alpha)\mu\nu}\di x^\nu \big)\wedge \di x^\mu$ on any chart, which satisfy $\omega_{(\alpha)}=\omega_{(\beta)}$ on any overlap of charts. Thus, the fundamental $2$-form $\omega$ on the atlas $\mathcal{M}$ from the previous theorem can be interpreted as a $2$-form $\underline{\omega}$ on the bundle gerbe $\mathscr{G}$.
\end{remark}

\begin{remark}[Analogy with a principal $U(1)$-bundle]
The way of expressing the fundamental $2$-form on our atlas as in remark \ref{remarkrev} is, despite of the appearance, very natural and familiar. When we write the connection of a $U(1)$-bundle in local coordinates, we are exactly writing a $1$-form $\omega_{(\alpha)} := \di\theta_{(\alpha)}+A_{(\alpha)\mu}(x_{(\alpha)})\di x^\mu_{(\alpha)}\in\Omega^1(\mathbb{R}^{d+1})$ on the local chart $\mathbb{R}^{d+1}$, where $\big\{\di\theta_{(\alpha)},\di x^\mu_{(\alpha)}\big\}$ is the coordinate basis of $\Omega^1(\mathbb{R}^{d+1})$. On the overlaps of charts we have $\omega_{(\alpha)}=\omega_{(\beta)}$, which assures that the $1$-form we are writing in local coordinates is equivalently the pullback $\omega_{(\alpha)}=\phi_\alpha^\ast\underline{\omega}$ of a well-defined $1$-form $\underline{\omega}$ on the total space of the $U(1)$-bundle.
Notice that this is in perfect analogy with remark \ref{remarkrev}.
\end{remark}

\begin{example}[Topologically trivial doubled space]
Let us consider a topologically trivial bundle gerbe $\mathscr{G}=M\times \mathbf{B}U(1)$ with connective structure.
The corresponding doubled space is a para-K\"{a}hler manifold $(\M,J,\omega)$, where $\M= T^\ast M$ is just the cotangent bundle of the base manifold, the para-complex structure $J$ corresponds to the canonical splitting $T\M \cong TM\oplus T^\ast M$ and the connection $\omega = \di\widetilde{x}_\mu\wedge \di x^\mu$ is the canonical symplectic form on $T^\ast M$ with $\{x^\mu,\widetilde{x}_\mu\}$ Darboux coordinates.
\end{example}

\begin{example}[Doubled Minkowski space]
If, in the previous example, we choose as base manifold the Minkowski space $M=\mathbb{R}^d$, the corresponding doubled space will be the para-K\"{a}hler vector space $(\mathbb{R}^{d,d},J,\omega)$.
\end{example}

\begin{remark}[Correspondence between sections of the bundle gerbe and the doubled space]
Let us consider again a topologically trivial bundle gerbe $\mathscr{G}=M\times \mathbf{B}U(1)$ with connective structure. Any section $M\xhookrightarrow{\;I\;}\mathscr{G}$ will be a $U(1)$-bundle $I\twoheadrightarrow M$ with connection, while any section $M\xhookrightarrow{\;\iota\;}\M$ will be an embedding $\widetilde{x}=\widetilde{x}(x)$. These two objects are immediately related by
\begin{equation}
    \iota^\ast\omega \;=\; \mathrm{curv}(I)
\end{equation}
where $\mathrm{curv}(-)$ is the curvature $2$-form of a $U(1)$-bundle. Since any bundle gerbe can be locally trivialised, it is possible to generalise this relation to the general topologically non-trivial case. The correspondence between bundle gerbes and doubled spaces was firstly presented and studied in \cite{Alf19} by using this observation.
\end{remark}

\begin{remark}[Principal action on the bundle gerbe]\label{principalaction}
By definition (see postulate \ref{post1}) the doubled space is nothing but the atlas of a bundle gerbe $\mathscr{G}$ with connective structure, which, therefore, will be canonically equipped with a higher principal action $\rho:\mathbf{B}U(1)_{\mathrm{conn}}\times\mathscr{G}\longrightarrow\mathscr{G}$. This means that we will have not only transformations, but also isomorphisms between them. On the base manifold this is a functor
\begin{equation}
    \mathbf{H}\big(M,\mathbf{B}U(1)_{\mathrm{conn}}\big)\,\times\, \Gamma(M,\mathscr{G}) \;\longrightarrow\; \Gamma(M,\mathscr{G}).
\end{equation}
Since local sections of $\mathcal{M}$ are local circle bundles $P_\alpha$, for any $Q\in\mathbf{H}(M,\mathbf{B}U(1)_{\mathrm{conn}})$, this action is locally given by the tensor product of $\mathbf{B}U(1)_{\mathrm{conn}}$ from remark \ref{rem:gs}, i.e. by $(Q,\,P_\alpha)\mapsto Q\otimes P_\alpha$.
\end{remark}

\noindent Notice this generalises the principal circle action of ordinary Kaluza-Klein Theory. Indeed a section of a circle bundle is in local data a collection of $U(1)$-functions $\theta_{(\alpha)}$ and the $U(1)$-action is given by a global shift $(g,\,\theta_{(\alpha)})\mapsto g\cdot\theta_{(\alpha)}$, where $g$ is a global $U(1)$-function.

\begin{remark}[Basis of global forms]\label{rem:globalforms}
In general it is also possible to express the principal connection $\omega=\widetilde{e}_{a}\wedge e^a$ in terms of the globally defined $1$-forms $\widetilde{e}_{a}=\di\widetilde{x}_{(\alpha)a} + B_{(\alpha)a\nu}\di x^\nu$ and $e^a=\di x^a$ on the atlas.
We pack both in a single global $1$-form $E^A$ with index $A=1,\dots,2d$ which is defined by $E^a:=e^a$ and $E_{a}:=\widetilde{e}_{a}$. In this basis, we have that the connection can be expressed by $\omega \;=\; \omega_{AB}\,E^A\wedge E^B$, where $\omega_{AB}$ is the $2d$-dimensional standard symplectic matrix.
\end{remark}

\section{Global generalised metric}

In this subsection we will give a global definition of a generalised metric on our doubled space. This will clarify the fundamental intuition by \cite{BCM14}, that firstly discussed the intrinsic higher geometric nature of a generalised metric structure. Here, we will introduce a generalised metric structure as a structure reduction of the tangent bundle of the doubled space, in the spirit of \cite{DCCTv2}.

\begin{definition}[Generalised metric]\label{defdoubledmetric}
A global \textit{generalised metric} can be defined, in analogy with a Riemannian metric from example \ref{ex:orth}, as an orthogonal structure
\begin{equation}
    \mathscr{G} \xrightarrow{\;\;\;\underline{\mathcal{G}}\;\;\;} O(2d)\Struc,
\end{equation}
on the bundle gerbe.
\end{definition}

\noindent On the atlas $(\M,J,\omega)$, this will be given by a collection of Riemannian metrics $\mathcal{G}_{(\alpha)}$ with the following patching conditions:
\begin{equation*}
\begin{tikzcd}[column sep=7ex, row sep=7ex]
\M \arrow[r, "\Phi_{\alpha}"]\arrow[rr, bend right=35, "\mathcal{G}_{(\alpha)}"]& \mathscr{G} \arrow[r, "\underline{\mathcal{G}}"] &   O(2d)\Struc
\end{tikzcd}, \qquad
\begin{tikzcd}[column sep=7ex, row sep=7ex]
\M \times_\mathscr{G} \M \arrow[bend left=40, "\mathcal{G}_{(\alpha)}"]{r}[name=U,below]{}
\arrow[bend right=40, "\mathcal{G}_{(\beta)}"']{r}[name=D]{} &
  O(2d)\Struc \arrow[Rightarrow, to path=(U) -- (D)]{}
\end{tikzcd}.
\end{equation*}
Now, the tangent stack of the bundle gerbe $\mathscr{G}$ is given by the short exact sequence
\begin{equation}
    0 \longhookrightarrow \mathscr{G}\ltimes\mathbf{b}\mathfrak{u}(1)_{\mathrm{conn}} \longhookrightarrow T\mathscr{G} \xtwoheadrightarrow{\;\pi_\ast\;} \pi^\ast TM \xtwoheadrightarrow{\quad} 0.
\end{equation}
This is nothing but the stack version of the short exact sequence \eqref{eq1}.
The connection of the bundle gerbe induces the isomorphism of stacks
\begin{equation}
    T\mathscr{G} \; \cong\; \pi^\ast TM \,\oplus\, \mathscr{G}\ltimes\mathbf{b}\mathfrak{u}(1)_{\mathrm{conn}}.
\end{equation}
The bundle $T\mathscr{G}$ naturally corresponds to a cocycle valued in $\mathbf{B}GL(d)\ltimes\mathbf{b}\mathfrak{u}(1)_\mathrm{conn}$. Notice that this can be embedded in a cocycle valued in $\mathbf{B}O(d,d)$, i.e.
\begin{equation}
     \mathbf{B}GL(d)\ltimes\mathbf{b}\mathfrak{u}(1)_\mathrm{conn} \;\hookrightarrow\;\mathbf{B}O(d,d)\;\hookrightarrow\;\mathbf{B}GL(2d).
\end{equation}
Such a cocycle is given by the following $O(d,d)$-valued matrices on each overlap of patches:
\begin{equation}
    \mathcal{N}_{(\alpha\beta)} \;=\;   \begin{pmatrix}
 N_{(\alpha\beta)} & 0 \\
 \mathrm{d}\Lambda_{(\alpha\beta)} & N_{(\alpha\beta)}^{-\mathrm{T}}
 \end{pmatrix},
\end{equation}
where $N_{(\alpha\beta)}$ are the transition corresponding to $TM$ and $(\Lambda_{(\alpha\beta)},G_{(\alpha\beta\gamma)})$ is the cocycle corresponding to the bundle gerbe.
The cocycle $\mathcal{N}_{(\alpha\beta)}$ can be seen as the cocycle corresponding to $T\M$ appearing in the short exact sequence \eqref{eq1}. 
Moreover, notice that $O(d,d)\cap O(2d)\cong O(d)\times O(d)$. Therefore the inclusion $\mathbf{B}O(2d)\hookrightarrow \mathbf{B}GL(2d)$ which defines a general orthogonal structure reduces to $\mathbf{B}\big(O(d)\times O(d)\big)\hookrightarrow \mathbf{B}O(d,d)$.

\noindent Thus, in local data, a generalised metric can be seen as locally given by matrices $E_{(\alpha)}\in\Coo\big(\mathbb{R}^{d,d},\,GL(2d)\big)$ which are patched by 
\begin{equation}
    \begin{aligned}
    E_{(\alpha)} \;&=\; \mathcal{O}_{(\alpha\beta)}\cdot E_{(\beta)}\cdot\mathcal{N}_{(\alpha\beta)}, \\[0.1cm]
    \mathcal{O}_{(\alpha\gamma)}  \;&=\; \mathcal{O}_{(\alpha\beta)}\cdot \mathcal{O}_{(\beta\gamma)},
    \end{aligned}
\end{equation}
where $\mathcal{O}_{(\alpha\beta)}$ is a $\big(O(d)\times O(d)\big)$-cocycle and $\mathcal{N}_{(\alpha\beta)}$ are the transition functions of the tangent bundle $T\mathcal{M}$.
The generalised metric from definition \ref{defdoubledmetric} can be recovered by defining 
\begin{equation}
    \mathcal{G}_{(\alpha)} :=  E_{(\alpha)}^\mathrm{T} E_{(\alpha)}
\end{equation}
and patched by the condition $\mathcal{G}_{(\beta)}=\mathcal{N}_{(\alpha\beta)}^{\mathrm{T}}\mathcal{G}_{(\alpha)}\mathcal{N}_{(\alpha\beta)}$.\vspace{0.2cm}

\noindent Hence, the moduli space of a general generalised metric is locally given by a mapping space $\Coo\big(\mathbb{R}^{d,d},\,O(d,d)/(O(d)\times O(d))\big)$, which is what we expect from a generalised metric on $\mathcal{M}$, in the spirit of definition \ref{defdoubledmetric}.

\begin{remark}[Field content of Double Field Theory]\label{def:metricdb}
Thus, the field content of Double Field Theory is given by a bundle gerbe $\mathscr{G}$ (postulate \ref{post1}) equipped with a generalised metric $\mathcal{G}$ (definition \ref{defdoubledmetric}) on it. 
\end{remark}

\section{Global strong constraint as higher cylindricity}

In this subsection we will give a global definition of the strong constraint and we will explain how it can be used to formulate a higher Kaluza-Klein reduction of the generalised metric.

\begin{post}[Global strong constraint]\label{post3}
The bundle gerbe $\mathscr{G}$ with generalised metric $\mathcal{G}$ must satisfy the \textit{strong constraint}, i.e. the generalised metric structure $\mathscr{G}\xrightarrow{\mathcal{G}} O(2d)\Struc$ must be equivariant under the principal action of $\mathscr{G}$ (remark \ref{principalaction}).
\end{post}

\noindent This is exactly an higher version of the \textit{cylindricity condition} in Kaluza-Klein Theory, which forbids the dependence of the bundle metric on the extra coordinate by asking it is invariant under the principal $U(1)$-action. Since for a principal circle bundle $P$ we have $P/U(1)\cong M$, the bundle metric will be actually a structure on the base manifold $M$. Analogously for the doubled space we have the following.

\begin{remark}[Geometry of the global strong constraint]\label{rem:geomsc}
From postulate \ref{post3} we can easily derived the familiar formulation of the strong constraint we know from DFT literature. Recall the principal action $\rho$ from remark \ref{principalaction} on the doubled space. The equivariance implies that the generalised metric is actually not a structure on $\mathscr{G}$, but on the (homotopy) quotient induced by the principal action
\begin{equation}
    \mathscr{G} /\!/\!_\rho \,\mathbf{B}U(1) \,\cong\, M,
\end{equation}
i.e. on the base manifold. Thus, the generalised metric locally depends only on physical $\{x^\mu\}$ coordinates of the smooth manifold $M$.
\end{remark}

\begin{remark}[A doubled-yet-gauged space]\label{d-y-g}
The principal action of the bundle gerbe is transgressed to the atlas by a shift $(x^\mu,\widetilde{x}_\mu)\mapsto(x^\mu,\widetilde{x}_\mu+\lambda_\mu(x))$ in the unphysical coordinates, which can be identified with a gauge transformation $B\mapsto B + \di(\lambda_\mu\di x^\mu)$ of the Kalb-Ramond field.
Moreover, the property $\mathscr{G}/\!/\mathbf{B}U(1)\cong M$ of bundle gerbes, when transgressed to the atlas, can be identified with the idea that physical points correspond to gauge orbits of the doubled space \cite{Park13}. This was further explored in \cite{Par13x, Par16x, Par17x, Par18x}. Remarkably, this will provide a global geometric interpretation of the strong constraint of Double Field Theory in the next section. 
Therefore, the atlas $\mathcal{M}$ of the bundle gerbe is naturally a doubled-yet-gauged space, according to the definition given by \cite{Park13}.
\end{remark}

\noindent Now we will introduce the powerful notion of \textit{higher Kaluza-Klein reduction} \cite{FSS18x, BSS19, DCCTv2}.

\begin{definition}[Higher Kaluza-Klein reduction]\label{def:dimred}
For any principal $\infty$-bundle $P\xrightarrow{\;\pi\;\,}M$ defined by the cocycle $M\xrightarrow{\;f\;\,}\mathbf{B}G$ and any stack $\mathscr{S}\in\mathbf{H}$, there is an equivalence
\begin{equation}
    \begin{tikzcd}[row sep=5ex, column sep=2.7ex]
    \mathbf{H}\big(P,\, \mathscr{S}\big) \arrow[rr, "\text{reduction}", yshift=1.6ex] & \cong & \mathbf{H}_{/\mathbf{B}G}\big(M,\, [G,\mathscr{S}]/\!/G\big), \arrow[ll, "\text{oxidation}", yshift=-1.5ex]
    \end{tikzcd}
\end{equation}
where the \textit{reduction} (which we will also call \textit{Kaluza-Klein reduction}) is given by the map
\begin{equation}
    \left(\begin{tikzcd}[row sep=5ex, column sep=4ex]
    P \arrow[r, "s"] & \mathscr{S}
    \end{tikzcd}\right) \quad\mapsto\quad
    \left(\begin{tikzcd}[row sep=7ex, column sep=5ex]
    & \left[G,\mathscr{S}\right]/\!/G \arrow[d]\\
    M \arrow[ru, "\hat{s}"]\arrow[r, "f"] & \mathbf{B}G
    \end{tikzcd}\right)
\end{equation}
This equivalence is also called \textit{double dimensional reduction} in the reference.
\end{definition}

\begin{definition}[Equivariant structure]
In this thesis we call $M\xrightarrow{\;s\;}\mathscr{S}$ a $G$-\textit{equivariant structure} if the morphism $s$ is $G$-equivariant with respect to a $G$-action on $M$.
\end{definition}

\begin{theorem}[Higher Kaluza-Klein reduction of equivariant structure]
When the structure $P\xrightarrow{s}\mathscr{S}$ on a principal $G$-bundle $P$ is $G$-equivariant, its higher Kaluza-Klein reduction (definition \ref{def:dimred}) will be a structure $M\xrightarrow{s/G}\mathscr{S}/\!/G$ on the base manifold.
\end{theorem}
\begin{proof}
Structure $P\xrightarrow{s}\mathscr{S}$ is equivariant if there exists a map $P/\!/G\cong M\xrightarrow{s/G}\mathscr{S}/\!/G$ such that $s/G \circ \pi \cong \pi'\circ s$, where $\pi:P\twoheadrightarrow M$ and $\pi':\mathscr{S}\twoheadrightarrow\mathscr{S}/\!/G$ are the bundle projections. This means that the reduction decomposes as $\hat{s}:M\xrightarrow{\;s/G\;} \mathscr{S}/\!/G \longhookrightarrow[G,\mathscr{S}]/\!/G$, where the last embedding comes immediately from the embedding $M\cong [\,\ast\,,M]\longhookrightarrow [G,M]$.
\end{proof}

\noindent In the following discussion we will apply this abstract definition to the concrete cases of ordinary Kaluza-Klein theory and, finally, of the bundle gerbe with generalised metric.

\begin{example}[Ordinary Kaluza-Klein reduction of the circle bundle]
Before looking at what happens to doubled space let us reformulate ordinary Kaluza-Klein reduction. Let us consider a circle bundle $P\rightarrow M$ and an $U(1)$-invariant Riemannian metric $\mathcal{G}$ on $P$. We have the following reduction:
\begin{equation}
    \left(\begin{tikzcd}[row sep=5ex, column sep=4ex]
    P \arrow[r, "\mathcal{G}"] &  O(2d)\Struc
    \end{tikzcd}\right) \quad\mapsto\quad
    \left(\begin{tikzcd}[row sep=8ex, column sep=3ex]
    &  O(d)\Struc\times\mathbb{R}\times\BU \arrow[d, "\ast\,\times\,\mathrm{frgt}"]\\
    M \arrow[ru, "(g{,}\varphi\text{,}A_{(\alpha)}\text{,}f_{(\alpha\beta)})"]\arrow[r, "f_{(\alpha\beta)}"'] & \mathbf{B}U(1)
    \end{tikzcd}\right).
\end{equation}
Let us now explain how this reduction is obtained.
Firstly, let us recall that $\mathfrak{at}(P)=TP/U(1)$ is locally given by sections of $TU_\alpha\times\mathbb{R}$ patched on overlaps $U_\alpha\cap U_\beta$ by
\begin{equation}
    \mathcal{N}_{(\alpha\beta)} =   \begin{pmatrix}
 N_{(\alpha\beta)} & 0 \\
 \mathrm{d}f_{(\alpha\beta)} & 1
 \end{pmatrix}.
\end{equation}
Hence to have a $U(1)$-equivariant orthogonal structure we need to restrict the structure group $GL(d)\ltimes\mathbb{R}^d\times\mathbb{R}\subset GL(d+1)$. Since $\Omega^1(U_\alpha)\cong\Coo(U_\alpha,\mathbb{R}^d)$ are isomorphic we can write the vielbein as
\begin{equation}
     E_{(\alpha)} =   \begin{pmatrix}
 e_{(\alpha)} & 0 \\
 \varphi_{(\alpha)}A_{(\alpha)} & \varphi_{(\alpha)}
 \end{pmatrix}.
\end{equation}
But also $O(d)\subset O(d+1)$. Therefore we have $e_{(\alpha)} = h_{(\alpha\beta)}\cdot e_{(\beta)}\cdot N_{(\alpha\beta)}$ and $\varphi_{(\alpha)}=\varphi_{(\beta)}$, but also $A_{(\alpha)}-A_{(\beta)} = \mathrm{d}f_{(\alpha\beta)}$. This means that the cocycle $P\rightarrow O(d+1)\Struc$ is Kaluza-Klein reduced to a cocycle $M\rightarrow O(d)\Struc\times\mathbb{R}\times\BU$ and the map $\mathrm{frgt}$ is just the forgetful functor $\BU\longrightarrow\mathbf{B}U(1)$ which forgets the connection data. Hence, if we call $g_{(\alpha)}:=e_{(\alpha)}^{\mathrm{T}}e_{(\alpha)}$, we can rewrite our metric on each local patch as
\begin{equation}
    \mathcal{G}_{(\alpha)}= E^\mathrm{T}_{(\alpha)} E_{(\alpha)} =   \begin{pmatrix}
 g_{(\alpha)} + \varphi^2A^{\mathrm{T}}_{(\alpha)} A_{(\alpha)} & \varphi^2A^{\mathrm{T}}_{(\alpha)} \\
 \varphi^2A_{(\alpha)} & \varphi^2
 \end{pmatrix}
\end{equation}
satisfying $\mathcal{G}_{(\alpha)}=\mathcal{N}_{(\alpha\beta)}^{\mathrm{T}}\mathcal{G}_{(\beta)}\mathcal{N}_{(\alpha\beta)}$ on two-fold overlaps of patches $U_\alpha\cap U_\beta$.
This assures that $g$ is a global Riemannian metric on $M$ and the $1$-form is a $U(1)$-gauge field patched by $A_{(\beta)}-A_{(\alpha)} = \mathrm{d}f_{(\alpha\beta)}$.
Thus, globally, we have
\begin{equation}
    \mathcal{G} \;=\; g\oplus \varphi^2(\di\theta_{(\alpha)} + A_{(\alpha)})\otimes(\di\theta_{(\alpha)} + A_{(\alpha)}).
\end{equation}
\end{example}

\noindent Let us now higher Kaluza-Klein reduce our generalised metric structure in a totally analogous way.

\begin{theorem}[Higher Kaluza-Klein reduction of the doubled space]\label{bosfieldsfromgeom}
A bundle gerbe $\mathscr{G}$ equipped with generalised metric $\mathcal{G}$ which satisfies global strong constraint (postulate \ref{post3}) reduces to a bosonic supergravity background $M$ with a Riemannian metric $g$ and a bundle gerbe structure with connection $(B_{(\alpha)},\Lambda_{(\alpha\beta)},G_{(\alpha\beta\gamma)})$. In diagrams we have the following reduction
\begin{equation}
    \left(\begin{tikzcd}[row sep=5ex, column sep=4ex]
    \mathscr{G} \arrow[r, "\mathcal{G}"] &  O(2d)\Struc
    \end{tikzcd}\right) \;\mapsto\;
    \left(\begin{tikzcd}[row sep=8ex, column sep=5ex]
    &  O(d)\Struc\times\mathbf{B}^2U(1)_{\mathrm{conn}} \arrow[d, "\ast\,\times\,\mathrm{frgt}"]\\
    M \arrow[ru, "\big(g\text{,}B_{(\alpha)}\text{,}\Lambda_{(\alpha\beta)}\text{,}G_{(\alpha\beta\gamma)}\big)"]\arrow[r, "G_{(\alpha\beta\gamma)}"'] & \mathbf{B}^2U(1)
    \end{tikzcd}\right)
\end{equation}
\end{theorem}

\begin{proof}
As seen in lemma \ref{thm:dtbild}, the local data of a Courant $2$-algebroid $\mathfrak{at}(\mathscr{G})\cong T\mathscr{G}/\!/\mathbf{B}U(1)_{\mathrm{conn}}$ is given by a cocycle $N_{(\alpha\beta)}:M\longrightarrow\mathbf{B}GL(d)$ which describes the tangent bundle $TM$ and the cocycle $\mathrm{d}\Lambda_{(\alpha\beta)}:M\longrightarrow\mathbf{B}\wedge^2\mathbb{R}^d$. Hence, its sections $\Gamma\big(M,\mathfrak{at}(\mathscr{G})\big)\cong\mathfrak{X}(M)\ltimes\mathbf{H}(M,\mathbf{b}\mathfrak{u}(1)_{\mathrm{conn}})$ are patched by transition functions in $\Coo\big(U_\alpha\cap U_\beta,\, GL(d)\ltimes\wedge^2\mathbb{R}^d\big)$ where $GL(d)\ltimes\wedge^2\mathbb{R}^d \subset O(d,d)$ is exactly the so-called geometric group. Thus, we can rewrite the transition functions as $O(d,d)$-valued functions on overlaps of patches, up to gauge transformations of the cocycle $(\Lambda_{(\alpha\beta)},G_{(\alpha\beta\gamma)})$, as
\begin{equation}
    \mathcal{N}_{(\alpha\beta)} =   \begin{pmatrix}
 N_{(\alpha\beta)} & 0 \\
 \mathrm{d}\Lambda_{(\alpha\beta)} & N_{(\alpha\beta)}^{-\mathrm{T}}
 \end{pmatrix},
\end{equation}
Consequently, we can write the vielbein as a local $O(d,d)$-valued function on each patch $U_\alpha$ as
\begin{equation}
     E_{(\alpha)} =   \begin{pmatrix}
 e_{(\alpha)} & 0 \\[0.1cm]
 -e^{-\mathrm{T}}_{(\alpha)} B_{(\alpha)} & e^{-\mathrm{T}}_{(\alpha)}
 \end{pmatrix}
\end{equation}
where $e_{(\alpha)}$ and $B_{(\alpha)}$ are respectively a $GL(d)$-valued and a $\wedge^2\mathbb{R}^d$-valued function on patches $U_\alpha$. Since $O(d,d)\cap O(2d) = O(d)\times O(d)$, the local $O(2d)$ symmetry of the vielbein breaks to $O(d)\times O(d)$. Hence we can write the generalised metric as
\begin{equation}\label{eq:doubledmetric0}
    \mathcal{G}_{(\alpha)} \;=\;  E^\mathrm{T}_{(\alpha)} E_{(\alpha)} =   \begin{pmatrix}
 g_{(\alpha)} - B^{ }_{(\alpha)} g^{-1}_{(\alpha)} B_{(\alpha)} & B^{ }_{(\alpha)} g^{-1}_{(\alpha)} \\[0.1cm]
 -g^{-1}_{(\alpha)} B_{(\alpha)} & g^{-1}_{(\alpha)}
 \end{pmatrix}
\end{equation}
where we called the symmetric matrix $g_{(\alpha)}:=e_{(\alpha)}^\mathrm{T}e_{(\alpha)}$. On two-fold overlaps of patches $U_\alpha\cap U_\beta$ the generalised metric is hence patched by $\mathcal{G}_{(\alpha)} = \mathcal{N}^{\mathrm{T}}_{(\alpha\beta)}\mathcal{G}_{(\beta)} \mathcal{N}_{(\alpha\beta)}$, which assures that $g$ is a globally defined tensor on $M$ and that $B_{(\alpha)}$ is patched by $B_{(\beta)}-B_{(\alpha)} = \mathrm{d}\Lambda_{(\alpha\beta)}$ and hence it is a gerbe connection. 
Therefore, the equivariant cocycle $M\rightarrow O(2d)\Struc$ "breaks" to a cocycle $M\longrightarrow O(d)\Struc\times\mathbf{B}^2U(1)_{\mathrm{conn}}$ and the map $\frgt$ is just the forgetful functor $\mathbf{B}^2U(1)_{\mathrm{conn}}\twoheadrightarrow\mathbf{B}^2U(1)$ which forgets the connection.
\end{proof}

\noindent Thus, if we require the generalised metric structure to be invariant under the principal $\mathbf{B}U(1)$-action of the bundle gerbe, by using the differential forms of remark \ref{rem:globalforms}, this will have to be of the form
\begin{equation}
\begin{aligned}
    \mathcal{G} \;=\; \mathcal{G}_{AB}\,E^A \odot E^B \;=\; g_{ab}\,e^a\odot e^b \,+\, g^{ab}\,\widetilde{e}_{a} \odot \widetilde{e}_{b} 
    \end{aligned}
\end{equation}
where we called the matrix $\mathcal{G}_{AB}:=(g\oplus g^{-1})_{AB}$ and where $g\in\odot^2\Omega^1(M)$ is a Riemannian metric on the base manifold. In the coordinate basis $\{\di x^\mu_{(\alpha)},\di\widetilde{x}_{(\alpha)\mu}\}$ we find the expression
\begin{equation}\label{eq:doubledmetric}
    \mathcal{G} \;=\; g \oplus g^{\mu\nu}(\mathrm{d}\widetilde{x}_{(\alpha)\mu} + B_{(\alpha)\mu\lambda}\mathrm{d}x^\lambda)\otimes(\mathrm{d}\widetilde{x}_{(\alpha)\nu} + B_{(\alpha)\nu\lambda}\mathrm{d}x^\lambda).
\end{equation}
Notice the gerbe connection $\omega = \di\widetilde{x}_{(\alpha)\mu}\wedge \di x_{(\alpha)}^\mu-B_{(\alpha)}$ plays an analogous role of the connection $1$-form of a circle bundle in a Kaluza-Klein metric.
Therefore, we recover the usual matrix expression
\begin{equation}
    \mathcal{G}_{(\alpha)MN} \;=\; \begin{pmatrix}g_{\mu\nu}- B_{(\alpha)\mu\lambda}g^{\lambda\rho}B_{(\alpha)\rho\beta} & B_{(\alpha)\mu\lambda}g^{\lambda\nu} \\[0.1cm] -g^{\mu\lambda}B_{(\alpha)\lambda\nu} & g^{\mu\nu} \end{pmatrix},
\end{equation}
where $B_{(\alpha)}$ is the connection of the bundle gerbe. \vspace{0.2cm}

\noindent The moduli space, which is locally given by $\Coo\big(U_\alpha,\,GL(2d)/O(2d)\big)$, is therefore broken to $\Coo\big(U_\alpha,\,O(d,d)/\big(O(d)\times O(d)\big)\big)$ and glued by the transition functions $\mathcal{N}_{(\alpha\beta)}$ of the Courant $2$-algebroid $\mathfrak{at}(\mathscr{G})$. Hence we recovered generalised geometry by higher Kaluza-Klein reduction.

\begin{remark}[Check: invariance of generalised metric]
For any gauge transformation, i.e. circle bundle $(\eta_{(\alpha)},\eta_{(\alpha\beta)})\in\mathbf{H}(M,\BU)$, sections transform according to $\widetilde{x}_{(\alpha)}\mapsto\widetilde{x}_{(\alpha)}+\eta_{(\alpha)}$ on each patch, while the connection to $B_{(\alpha)} \mapsto B_{(\alpha)} + \eta_{(\alpha)}$. Hence generalised metric \eqref{eq:doubledmetric} is invariant.
\end{remark}

\begin{digression}[Recovering Born geometry]
The generalised metric we defined (see definition \ref{defdoubledmetric}) can be reduced to the one introduced in para-Hermitian geometry by \cite{Svo18} and further clarified by \cite{MarSza18,MarSza19}. Recall that the atlas of the bundle gerbe is a para-Hermitian manifold. Our generalised metric $\mathcal{G}$ on each patch $\mathbb{R}^{d,d}$ of the atlas $\M$ automatically satisfy the equations
\begin{equation}
    \eta^{-1}\mathcal{G} \,=\, \mathcal{G}^{-1}\eta, \qquad \omega^{-1}\mathcal{G} \,=\, -\mathcal{G}^{-1}\omega.
\end{equation}
It is immediate to check it by using the basis of remark \ref{rem:genvielbein}, so that we obtain
\begin{equation}
    \eta_{{MN}} = \begin{pmatrix}0 & 1 \\1& 0 \end{pmatrix}, \quad \omega_{{MN}} = \begin{pmatrix}0 & 1 \\-1 & 0 \end{pmatrix}, \quad \mathcal{G}_{{MN}}= \begin{pmatrix}g & 0 \\0 & g^{-1} \end{pmatrix}.
\end{equation}
A linear transformation of $\mathbb{R}^{d,d}$ which preserves both the para-Hermitian structure and the generalised metric is then a linear isometry of the Riemannian metric $g$ on $\mathbb{R}^{d}$. In terms of structure groups we have the familiar expression
\begin{equation}
O(d,d) \,\cap\, Sp(2d,\mathbb{R}) \,\cap\, O(2d) \;=\; O(d).
\end{equation}
\end{digression}

\begin{remark}[Isometry $2$-group of the generalised metric]\label{rem:isometry}
Given a bundle gerbe $\mathscr{G}$ with generalised metric $\mathcal{G}$ which satisfies global strong constraint (postulate \ref{post3}) there is a sub-$2$-group $\mathbf{Iso}(\mathscr{G},\mathcal{G})\subset \mathbf{Aut}_{\mathbf{H}_/}(\mathscr{G}) = \mathrm{Diff}(M)\ltimes\mathbf{H}(M,\,\BU)$ of automorphisms of the bundle gerbe which preserve the generalised metric structure:
\begin{equation}
\begin{tikzcd}[row sep=scriptsize, column sep=5ex]
   1 \arrow[r] & \mathbf{H}(M,\,\BU)\, \arrow[r, hook] & \,\mathbf{Iso}(\mathscr{G},\mathcal{G})\, \arrow[r, two heads] & \, \mathrm{Iso}(M,g)\arrow[r] & 1.
\end{tikzcd}
\end{equation}
i.e. the $2$-group of automorphisms covering the group of isometries of the Riemannian base manifold $(M,g)$.
\end{remark}

\begin{remark}[Generalised vielbein]\label{rem:genvielbein}
There are $2d$ global sections of $\mathfrak{at}(\mathscr{G})_{\mathrm{hor}}$ (see remark \ref{rem:horizontalsec}) giving the generalised vielbein $E_{(\alpha)}$ in terms of horizontal generalised vectors
\begin{equation}
    \begin{aligned}
    E_{M} :=\begin{cases}(e_\mu + \iota_{e_\mu}B_{(\alpha)}, \, 0), & {M} = \mu\\(e^\mu, \, 0), & {M} = d+\mu\end{cases}.
    \end{aligned}
\end{equation}
where $\{e_\mu\}\subset\Omega^1(M)$ are the $d$ vielbein $1$-forms of the Riemannian metric $g=\delta_{\mu\nu}\,e^\mu\otimes e^\nu$ on the base manifold $M$, while $\{e_\mu\}\subset\mathfrak{X}(M)$ are their dual vielbein vectors.
\end{remark}

\noindent Let us now conclude this subsection by giving a quick example of a relevant higher Kaluza-Klein reduction of some structure which is not the generalised metric.

\begin{example}[D-branes as objects in the doubled space]
Recall from chapter \ref{ch:3} the definition of the stack $\mathbf{K}U\in\mathbf{H}$, i.e.
\begin{equation}
    \mathbf{K}U \; := \; \prod_{k\in\mathbb{N}}\mathbf{B}^{2k+1}U(1),
\end{equation}
which is the moduli stack of an abelian bundle $2k$-gerbes for each $k\in\mathbb{N}$. 
Let us consider a morphism $\mathscr{G}\rightarrow\mathbf{K}U$ from a bundle gerbe to this stack such that it fits in a twisted $\infty$-bundle diagram of the form
\begin{equation}
     \begin{tikzcd}[row sep={12ex,between origins}, column sep={15ex,between origins}]
      \mathscr{D}\arrow[r]\arrow[d,two heads] & \ast\arrow[d] & \\
      \mathscr{G}\arrow[r]\arrow[d,two heads] & \mathbf{K}U\arrow[r]\arrow[d] & \ast\arrow[d]\\
      M\arrow[r] & \mathbf{K}U/\!/\mathbf{B}U(1)\arrow[r,two heads] & \mathbf{B}^2U(1).
    \end{tikzcd}
\end{equation}
Now, such morphism to this stack is immediately higher Kaluza-Klein reduced as follows:
\begin{equation}
    \left(\begin{tikzcd}[row sep=5ex, column sep=4ex]
    \mathscr{G} \arrow[r, " "] & \mathbf{K}U
    \end{tikzcd}\right) \quad\longmapsto\quad
    \left(\begin{tikzcd}[row sep=5ex, column sep=5ex]
    & \qquad\;\, \mathbf{K}U/\!/\mathbf{B}U(1) \arrow[d, " "]\\
    M \arrow[ru, "{(H,\,F_{Dp})}"]\arrow[r, "G_{(\alpha\beta\gamma)}"'] & \mathbf{B}^2U(1)
    \end{tikzcd}\right)
\end{equation}
The reduced map $M\rightarrow\mathbf{K}U/\!/\mathbf{B}U(1)$ is then the \v{C}ech cocycle of bundle $2k$-gerbes on $M$ twisted by the cocycle $(G_{(\alpha\beta)})$ of the Kalb-Ramond field.
Recall that we have
\begin{equation}
    \mathrm{CE}(\mathfrak{ku}/\!/\mathbf{B}U(1)) = \mathbb{R}[\omega_{2},\omega_{4},\omega_{6},\dots,h_3]/\!\left(\begin{array}{l}\di h_3=0, \\[0.5ex] \di\omega_{2(k+1)}=h_3\wedge \omega_{2k} \;\,\forall k\text{ even}  \end{array}\right).
\end{equation}
It is easy to see that the curvature of the structure $M\rightarrow\mathbf{K}U/\!/\mathbf{B}U(1)$ will be given by the usual expression for the RR fields for Type IIA String Theory:
\begin{equation}
    \begin{aligned}
        \mathrm{d}H =0, \qquad\qquad\quad\;\;\;\, \mathrm{d}F_{D0} =0, \quad \mathrm{d}F_{D2} + F_{D0}\wedge H=0, \\
        \mathrm{d}F_{D4} + F_{D2}\wedge H=0, \quad \mathrm{d}F_{D6} + F_{D4}\wedge H=0, \quad \mathrm{d}F_{D8} + F_{D6}\wedge H=0,
\end{aligned}
\end{equation}
as explained in a more general setting and formally by \cite[p.22]{FSS18x}. This means that RR fields are simpler if they are thought as fields on the bundle gerbe rather than on the base manifold. This appears consistent with the unified algebra description of spacetime and branes by Lie algebra extensions in \cite{FSS18}.
\end{example}

\section{Recovering generalised geometry}

Here we will show that generalised geometry is naturally recovered from the bundle gerbe perspective upon imposition of the strong constraint, i.e. invariance under the principal $\mathbf{B}U(1)$-action.

\subsection{Generalised tangent bundle on the atlas}

\begin{remark}[Generalised geometry on the atlas]\label{vecatlas}
Let $\{\partial_M\}=\{\partial_\mu, \widetilde{\partial}^\mu\}$ be the local coordinate basis of $T\M$.
A vector on the atlas $(\M,J,\omega)$ can be written in local coordinates as $V=V^M_{(\alpha)}\partial_M = v^\mu_{(\alpha)} \partial_\mu + \widetilde{v}_{(\alpha)\mu}\widetilde{\partial}^\mu$, where the components $V^M_{(\alpha)}$ are locally defined. The fundamental $2$-form $\omega$ will project this into a vertical vector $\omega(V)=(\widetilde{v}_{(\alpha)\mu}+B_{(\alpha)\mu\nu}v^\nu_{(\alpha)})\widetilde{\partial}^\mu$. 
Now, if we call $\{D_A\}$ the basis of globally defined vectors on $\M$ dual to the global $1$-forms $\{E^A\}$, we can write a vector on the atlas by $V=V^AD_A$, where now the components $V^A$ are globally defined.
We can now express the isomorphism $\pi_\ast\oplus\omega$ in \eqref{eq2} by
\begin{equation}
    V^AD_A\;=\; v^\mu_{(\alpha)} \partial_\mu + \big(\widetilde{v}_{{(\alpha)}\mu}+B_{(\alpha)\mu\nu}v^\nu_{(\alpha)}\big)\widetilde{\partial}^\mu
\end{equation}
Notice that, if we restrict ourselves to strong constrained vectors, i.e. vectors whose components $V^M_{(\alpha)}$ only depend on the coordinates of the base manifold $M$, these are immediately sections of a Courant algebroid twisted by the bundle gerbe $\mathscr{G}\twoheadrightarrow M$ with local potential $B_{(\alpha)}$.
\end{remark}
 
\noindent We have already shown that strong constrained vectors on the atlas reduce to sections of a Courant algebroid. Now, we want to show that the bracket structure of the Courant algebroid also comes from the bundle gerbe. To show this, we will need to start from the $2$-group of automorphisms of bundle gerbe and Lie differentiate it.

\subsection{Finite symmetries}

In this subsection we will deal with finite symmetries of the bundle gerbe and we will prove they are the gauge transformations we expect for Double Field Theory.

\begin{remark}[$2$-group of gauge transformations]\label{subgroupoidgauge}
The automorphisms $2$-group of the principal structure (example \ref{ex:aut}) of the bundle gerbe $\mathscr{G}\xrightarrow{\bbpi}M$ with connective structure is exactly the $2$-group extending the diffeomorphisms of the base $M$ through gauge transformations of the gerbe
\begin{equation}\label{eq:DFTgroup}
    \begin{tikzcd}[row sep=scriptsize, column sep=6ex]
    \mathbf{Aut}_{\mathbf{H}_/}(\mathscr{G}) \,=\, \Diff(M)\ltimes\mathbf{H}(M,\BU).
    \end{tikzcd}
\end{equation}
This $2$-group is the refined version of the gauge group of DFT proposed by \cite{Hull14}, i.e.
\begin{equation}\label{eq:hull}
   G_{\mathrm{NS}} \,=\, \Diff(M)\ltimes\Omega^2_{\mathrm{cl}}(M),
\end{equation}
which is obtained by taking the curvature of the circle bundle, as seen in example \ref{ex:aut}.
\end{remark}

\noindent Notice \eqref{eq:DFTgroup} is exactly the analogous to the familiar $\Diff(M)\ltimes\Coo(M,U(1))\subset\Diff(P)$ of gauge transformations in ordinary Kaluza-Klein theory, where $P$ is a circle bundle.
\vspace{0.2cm}

\noindent The map between \eqref{eq:DFTgroup} and \eqref{eq:hull} is just the curvature $$\mathrm{curv}:\mathbf{H}(M,\BU)\rightarrow\Omega^2_{\mathrm{cl}}(M),$$ 
which maps $(\eta_{(\alpha)},\eta_{(\alpha\beta)})\mapsto b$ where $b|_{U_\alpha}:=\mathrm{d}\eta_{(\alpha)}$ on each patch, in accord with remark \ref{rem:globgauge}. This global closed $2$-form $b\in\Omega^2_{\mathrm{cl}}(M)$ is usually called $B$-shift in DFT literature.

\begin{remark}[$2$-group of gauge transformations of DFT in \v{C}ech data]\label{diffgaugeconc}
The $2$-group of gauge transformations $\mathbf{Aut}_{\mathbf{H}_/}(\mathscr{G})=\Diff(M)\ltimes\mathbf{H}(M,\BU)$ from remark \ref{subgroupoidgauge} will naturally define an action on the groupoid $\Gamma(M,\mathcal{M})$ of sections of the bundle gerbe by the functor
\begin{equation}
    \mathbf{Aut}_{\mathbf{H}_/}(\mathscr{G})\,\times\, \Gamma(M,\mathcal{M}) \;\longrightarrow\; \Gamma(M,\mathcal{M}).
\end{equation}
In local \v{C}ech data on the base manifold $M$ this action will be given by the following.
\begin{itemize}
\item objects of $\mathbf{Aut}_{\mathbf{H}_/}(\mathscr{G})$ are triples $(f,\eta_{(\alpha)},\eta_{(\alpha\beta)})$, made up of a diffeomorphism $f\in\Diff(M)$ and a circle bundle $(\eta_{(\alpha)},\eta_{(\alpha\beta)})\in\mathbf{H}(M,\BU)$ and acting on sections $(\widetilde{x}_{(\alpha)},\phi_{(\alpha\beta)})\in\Gamma(M,\mathscr{G})$ by
\begin{equation}
\begin{aligned}
    (f,\eta_{(\alpha)},\,\eta_{(\alpha\beta)})\, :\, (\widetilde{x}_{(\alpha)},\phi_{(\alpha\beta)}) \,&\mapsto\, \big(f_\ast\widetilde{x}_{(\alpha)}+\eta_{(\alpha)}, \,\phi_{(\alpha\beta)}+\eta_{(\alpha\beta)}\big) \\
\end{aligned}
\end{equation}
\item isomorphisms of $\mathbf{Aut}_{\mathbf{H}_/}(\mathscr{G})$ between these objects are just ordinary gauge transformations of circle bundles, made up of local functions $\epsilon_{(\alpha)}\in\Coo(U_\alpha)$acting by 
\begin{equation}
   (\epsilon_{(\alpha)}):\, (f,\eta_{(\alpha)},\,\eta_{(\alpha\beta)}) \,\Mapsto\, (f,\eta_{(\alpha)}+\mathrm{d}\epsilon_{(\alpha)},\,\eta_{(\alpha\beta)}+\epsilon_{(\alpha)}-\epsilon_\beta).
\end{equation}
\end{itemize}
In terms of diagrams, we can rewrite the $2$-group of automorphisms of the bundle gerbe $\mathscr{G}$ as follows:
\begin{equation}
    \mathbf{Aut}_{\mathbf{H}_/}(\mathscr{G}) \,\cong\, \left\{ \begin{tikzcd}[row sep=scriptsize, column sep=18ex]
    \mathscr{G} \arrow[r, bend left=50, ""{name=U, below}, "(f\text{,}\,\eta_{(\alpha)}\text{,}\,\eta_{(\alpha\beta)})"]
    \arrow[r, bend right=50, "(f\text{,}\,\eta_{(\alpha)}+\mathrm{d}\epsilon_{(\alpha)}\text{,}\,\eta_{(\alpha\beta)}+\epsilon_{(\alpha)}-\epsilon_\beta)"', ""{name=D}]
    & \mathscr{G}
    \arrow[Rightarrow, from=U, to=D, "\,(\epsilon_{(\alpha)})"]
\end{tikzcd} \right\}.
\end{equation}
Hence the action of the sub-$2$-group $\mathbf{H}(M,\BU)\subset\mathbf{Aut}_{\mathbf{H}_/}(\mathscr{G})$ of automorphisms which cover the identity $\mathrm{id}_M\in\mathrm{Diff}(M)$ is exactly the principal action of the bundle gerbe $\mathscr{G}$ from remark \ref{principalaction}. This is directly analogous to Kaluza-Klein theory, where the translation along the compactified dimension coincides with the principal circle action.
\end{remark}

\begin{digression}[Generalised diffeomorphisms]
In the DFT literature the automorphisms $\mathbf{Aut}_{\mathbf{H}_/}(\mathscr{G})$ from remark \ref{diffgaugeconc} are usually called \textit{generalised diffeomorphisms} (or \textit{large gauge transformations}). Notice that the research by \cite{DesSae18} and by \cite{Hohm19DFT} are independently already pointing in the direction of generalised diffeomorphisms having a higher group structure. 
\end{digression}

\noindent Now we can explain how automorphisms of definition \ref{subgroupoidgauge} can be used to glue the doubled space in a way that is not affected by Papadopolous' puzzle (section \ref{papadopoulospuzzle}).

\begin{remark}[Doubled space is glued in a $(\infty,1)$-category]
Gluing local patches of the doubled space has always been a puzzle in DFT (which becomes even more problematic in Exceptional Field Theory). In our proposal the solution to this puzzle is given by the fact that our doubled space is not a manifold, but the atlas of a bundle gerbe: therefore it is glued not in the category of smooth manifolds, but in the $(\infty,1)$-category of smooth stacks.
Let us call $\mathscr{G}_\alpha:= \mathscr{G}\big|_{U_\alpha}$. These are trivial bundle gerbes on each patch $U_\alpha$ and their groupoid of sections $\Gamma(U_\alpha,\mathscr{G}_\alpha)\cong\mathbf{H}(U_\alpha,\mathbf{B}U(1))$ are just the groupoids of local circle bundles $P_\alpha$.
The sub-spaces $\mathscr{G}_\alpha$, $\mathscr{G}_\beta$ and $\mathscr{G}_\gamma$. They are then patched by automorphisms $e_{(\alpha\beta)}\in\mathrm{Diff}(U_\alpha\cap U_\beta)\ltimes\mathbf{H}(U_\alpha\cap U_\beta,\,\BU)$ on two-fold overlaps $\mathscr{G}_\alpha|_{U_\alpha\cap U_\beta}\cong\mathscr{G}_\beta|_{U_\alpha\cap U_\beta}$. On three-fold overlaps we have gauge transformations of automorphisms (see remark \ref{diffgaugeconc}) which are given by $e_{(\alpha\beta)}\circ e_{(\beta\gamma)}\xRightarrow{G_{(\alpha\beta\gamma)}\;} e_{\alpha\gamma}$, or equivalently by the $2$-commuting diagram
\begin{equation}
    \begin{tikzcd}[row sep=7ex, column sep=0ex]
    & \mathscr{G}_\beta|_{U_\alpha\cap U_\beta\cap U_\gamma}\arrow[rd, "\text{}e_{(\beta\gamma)}"] \arrow[Rightarrow, from=1-2, to=D,  "\,G_{(\alpha\beta\gamma)} "] & \\
    \mathscr{G}_\alpha|_{U_\alpha\cap U_\beta\cap U_\gamma}\arrow[rr, ""{name=D}, "\text{}e_{\alpha\gamma}"']{} \arrow[ru, "\text{}e_{(\alpha\beta)}"] & & \mathscr{G}_\gamma|_{U_\alpha\cap U_\beta\cap U_\gamma}
    \end{tikzcd}
\end{equation}
Therefore we can glue the local $\{\mathscr{G}_\alpha\}$ by automorphisms to get the global bundle gerbe $\mathscr{G}$.
\end{remark}

\noindent This idea of higher gluing the doubled space in a $(2,1)$-category, contrary to appearances, is not totally unprecedented. Let us mention some of its relevant progenitors in the literature.

\begin{digression}[Hints of higher gluing]
Notice that \cite{BCM14} proposed for the first time non-trivial patching conditions on three-fold overlaps of patches of the doubled space. More recently the local differential-graded patches $T^\ast[2]T[1]U_\alpha$ with $U_\alpha\subset M$ proposed by \cite{DesSae18} would need to be glued together in the $2$-category of derived spaces to give a global picture. This is because differential-graded manifolds are a simple model for derived spaces (see \cite{dman}). But this is consistent with our proposal: we will see in chapter \ref{ch:6} that the formalism by \cite{DesSae18} can be related to an infinitesimal version of our doubled space when there is T-duality.
\end{digression}

\begin{digression}[Recovering Papadopolous' C-space]\label{diga}
Recall that sections $(\widetilde{x}_{(\alpha}),\phi_{(\alpha\beta)})\in\Gamma(M,\mathscr{G})$ of the bundle gerbe are patched by condition \eqref{eq:coords}. Notice that these are exactly of the same form of the coordinates of a \textit{C-space}, defined by Papadopolous in \cite{Pap13}, \cite{Pap14} and further developed by \cite{HowPap17}, which was prescribed in the references to accommodate DFT geometry. Hence our construction of $\mathscr{G}$ may be seen also as a formalisation of that intuition.
\end{digression}

\subsection{Courant algebroid as Atiyah $L_\infty$-algebroid}

In this subsection we will deal with infinitesimal symmetries of the bundle gerbe and we will prove they locally reduce to the one expected from DFT. Indeed we will show that the Courant $2$-algebroid formalism can be recovered as infinitesimal descriptions of the geometry of the bundle gerbe. Finally, from the Courant $2$-algebroid, we will explicitly recover ordinary generalised geometry. Recall that generalised geometry has been revealed by \cite{Wald08, Wald11, Wald12} to be the natural language to express Type II supergravity.

\begin{definition}[Infinitesimally thickened point]
An \textit{infinitesimally thickened point} is defined (see \cite{DCCTv2}) as the locally ringed space given by the spectrum of the ring of dual numbers
\begin{equation}
    \mathbb{D}^1 \;:=\; \mathrm{Spec}\left(\frac{\mathbb{R}[\epsilon]}{\langle\epsilon^2\rangle}\right).
\end{equation}
Hence, its underlying topological space is just a single point $\{\ast\}$, but its smooth algebra of functions is $\mathcal{O}_{\mathbb{D}^1}(\{\ast\}) = \mathbb{R}[\epsilon]/\langle\epsilon^2\rangle \cong \mathbb{R}\oplus\epsilon\mathbb{R}$ with $\epsilon^2=0$, i.e. it is the ring of dual numbers.
\end{definition}

\begin{remark}[Technicalities about the infinitesimally thickened point]
Notice that $\mathbb{D}^1$ is not a smooth stack, as we defined it. However, it is possible to define a new $(\infty,1)$-category $\mathbf{H}_{\mathrm{formal}}$ by enlarging the category $\mathbf{Diff}$ of smooth manifolds (on which the objects of $\mathbf{H}$ are sheaves) to the category $\mathbf{Diff}_\mathrm{formal}$ of formal smooth manifolds, i.e. smooth manifolds equipped with infinitesimal extension \cite{khavkine2017synthetic}. In this dissertation, we will commit a slight abuse of notation and we will denote $\mathbf{H}_{\mathrm{formal}}$ just by $\mathbf{H}$.
\end{remark}

\begin{example}[Tangent bundle of a manifold]
This idea is widely used in algebraic geometry to define tangent spaces. For example a map $\mathbb{D}^1\xrightarrow{X} M$ is given on the underlying topological spaces by sending $\{\ast\}$ to a point $x$ in a manifold $M$ and on the algebras of smooth functions by a map $\Coo(M)\xrightarrow{X^\sharp}\mathbb{R}\oplus\epsilon\mathbb{R}$, which is given by $f \mapsto f(x)+\epsilon X^\mu\partial_\mu f(x)$ for some $X\in T_xM$. Thus vectors $X$ on $M$ can equivalently be seen as maps $\mathbb{D}^1\xrightarrow{X} M$ and therefore $TM\cong[\mathbb{D}^1,\,M]$.
\end{example}

\noindent This motivates the following definition for the tangent stack of a bundle gerbe $\mathscr{G}$.

\begin{definition}[Tangent stack of a bundle gerbe]\label{def:doubledtangentbundle}
We define the tangent bundle of a bundle gerbe $\mathscr{G}$ as the internal hom stack (definition \ref{def:inthom}) of maps from the infinitesimally thickened point $\mathbb{D}^1$ to $\mathscr{G}$
\begin{equation}
    T\mathscr{G} \;:=\; \big[\mathbb{D}^1,\, \mathscr{G}\big]
\end{equation}
\end{definition}

\begin{remark}[Atiyah sequence of the bundle gerbe]
A calculation shows that $[\mathbb{D}^1,M]=TM$ and $[\mathbb{D}^1,\mathbf{B}U(1)_{\mathrm{conn}}]=\mathbf{B}U(1)_{\mathrm{conn}}\ltimes \mathbf{b}\mathfrak{u}(1)_{\mathrm{conn}}$, where $\mathbf{b}\mathfrak{u}(1)_{\mathrm{conn}}$ is the Lie $2$-algebra of the Lie $2$-group $\mathbf{B}U(1)_{\mathrm{conn}}$ and it can be seen as the stack of real line bundles with connection.
From this result, given a bundle gerbe $\mathscr{G}$ with connective structure, we obtain the short exact sequence
\begin{equation}
    0 \longhookrightarrow \mathscr{G}\ltimes\mathbf{b}\mathfrak{u}(1)_{\mathrm{conn}} \longhookrightarrow T\mathscr{G} \xtwoheadrightarrow{\;\,\pi_\ast\,\;} \pi^\ast TM \xtwoheadrightarrow{\quad} 0,
\end{equation}
where $\pi:\mathscr{G}\twoheadrightarrow M$ is the bundle projection.
This sequence is nothing but the stack version of the short exact sequence \eqref{eq1}.
The connection of the bundle gerbe induces the isomorphism of stacks
\begin{equation}\label{eqa}
    T\mathscr{G} \; \cong\; \pi^\ast TM \,\oplus\, \mathscr{G}\ltimes\mathbf{b}\mathfrak{u}(1)_{\mathrm{conn}},
\end{equation}
which is the stack version of the isomorphism \eqref{eq2}.
\end{remark}

\begin{theorem}[Tangent stack of a bundle gerbe in local data]\label{thm:dtbild}
Let us consider the bundle gerbe $\mathscr{G}$ equipped with connection structure $(\Lambda_{(\alpha\beta)},G_{(\alpha\beta\gamma)})\in\mathbf{H}\big(M,\mathbf{B}(\BU)\big)$.
On a patch $U\subset M$ of the base manifold we have the isomorphism
\begin{equation}\label{eq:DTBs}
    \Gamma(U,\; T\mathscr{G}) \;\cong\; \Gamma(U,\,\mathscr{G})\ltimes\Big(\mathfrak{X}(U)\ltimes\mathbf{H}(U,\mathbf{b}\mathfrak{u}(1)_{\mathrm{conn}})\Big).
\end{equation}
These local sections will be non-trivially glued on the whole smooth base manifold $M$ and the patching conditions are the following:
\begin{itemize}
    \item local vectors $X\in\mathfrak{X}(U)$ will be patched to global vectors $X\in\mathfrak{X}(M)$ on $M$,
    \item local $\mathfrak{u}(1)$-bundles $(\xi_{(\alpha)},\eta_{(\alpha\beta)})\in\mathbf{H}(U,\mathbf{b}\mathfrak{u}(1)_{\mathrm{conn}})$ will be patched by the Lie derivatives of the transition functions of the bundle gerbe $(\mathcal{L}_{X}\Lambda_{(\alpha\beta)},\,\mathcal{L}_{X}G_{(\alpha\beta\gamma)})$, i.e.
    \begin{equation}\label{eq:localsymdata}
        \begin{aligned}
        \xi_{(\alpha)} - \xi_{(\beta)} \;&=\; -\mathcal{L}_{X}\Lambda_{(\alpha\beta)} +\mathrm{d}\eta_{(\alpha\beta)}, \\
       \eta_{(\alpha\beta)}+ \eta_{(\beta\gamma)} + \eta_{(\gamma\alpha)} \;&=\; \mathcal{L}_{X}G_{(\alpha\beta\gamma)}.
        \end{aligned}
    \end{equation}
\end{itemize}
Notice that we can reparametrize the scalars $f_{(\alpha\beta)} :=  \eta_{(\alpha\beta)} - \iota_X\Lambda_{(\alpha\beta)}$ in \eqref{eq:localsymdata} and obtain sections $\mathbbvar{X}:=(X+\xi_\alpha,f_{(\alpha\beta)})$, which are now patched by the familiar condition
    \begin{equation}
        \begin{aligned}
        \xi_{(\alpha)} - \xi_{(\beta)} \;&=\; -\iota_{X}\mathrm{d}\Lambda_{(\alpha\beta)} +\mathrm{d}f_{(\alpha\beta)}, \\
        f_{(\alpha\beta)}+f_{(\beta\gamma)} + f_{(\gamma\alpha)} \;&=\; 0,
        \end{aligned}
    \end{equation}
\end{theorem}

\begin{proof}
Since local sections of $\mathscr{G}$ are local circle bundles $\Gamma(U,\mathscr{G})\cong\mathbf{H}(U,\BU)$ equipped with connection, we can find the $2$-groupoid of sections of the tangent stack $T\mathscr{G}|_U \cong \left[\mathbb{D}^1,\,U\times\BU\right]$ by calculating
\begin{equation}
    \begin{aligned}
    \mathbf{H}(U\times\mathbb{D}^1, U) \,&\cong\, \mathrm{Diff}(U)\times\mathfrak{X}(U), \\
    \mathbf{H}(U\times\mathbb{D}^1, \BU) \,&\cong\, \mathbf{H}(U, \BU) \times \mathbf{H}(U, \mathbf{b}\mathfrak{u}(1)_{\mathrm{conn}}),
    \end{aligned}
\end{equation}
and then by considering only the subgroupoid of maps covering the identity $\mathrm{id}_U\in\mathrm{Diff}(U)$. Hence a local section in $\Gamma(U,\, T\mathscr{G}) \cong \Gamma(U\times\mathbb{D}^1,\, \mathscr{G})$ must be given by \eqref{eq:DTBs}. The transition functions of the tangent stack $T\mathscr{G}\twoheadrightarrow\mathscr{G}$ are pullbacks of functions on the base manifold $M$, because the transition functions $(\Lambda_{(\alpha\beta)},G_{(\alpha\beta\gamma)})$ of $\mathscr{G}$ are functions on $M$. Since the bundle gerbe is patched by bundle automorphisms $\mathrm{Diff}(U)\ltimes\mathbf{H}(U,\BU)$, its tangent stack will be patched by their linearised version $\Coo\big(U,\,GL(d)\big)\ltimes\mathbf{H}(U,\,\mathbf{b}\mathfrak{u}(1)_{\mathrm{conn}})$. Hence $T\mathscr{G}$ naturally induces a bundle with transition functions $M\rightarrow\mathbf{B}(GL(d)\ltimes\mathbf{b}\mathfrak{u}(1)_{\mathrm{conn}})$. The transition functions $M\rightarrow\mathbf{B}GL(d)$ are just the ones of the frame bundle $FM$, while the cocycle $M\rightarrow\mathbf{b}\mathfrak{u}(1)_{\mathrm{conn}}$ is just the collection $(\mathcal{L}_{X}\Lambda_{(\alpha\beta)},\,\mathcal{L}_{X}G_{(\alpha\beta\gamma)})$. Thus, sections are patched as in equation \eqref{eq:localsymdata}.
\end{proof} 

\noindent Notice \eqref{eq:DTBs} is analogous to the familiar idea that the tangent bundle $TP$ of a circle bundle $P$ is locally of the form $TU\times U(1)\times\mathbb{R}$ and patched by using the transition functions of $P$.
\vspace{0.2cm}

\noindent Hence, the \v{C}ech data of a doubled vector $\mathbbvar{X}=(X+\xi_{(\alpha)},f_{(\alpha\beta)})$ are the data of an infinitesimal gauge transformation given by $x\mapsto x+\epsilon X$ on the base manifold and by $(\widetilde{x}_\alpha,\,\phi_{(\alpha\beta)})\mapsto(\widetilde{x}_\alpha+\epsilon\xi_\alpha,\,\phi_{(\alpha\beta)}+\epsilon f_{(\alpha\beta)}+\epsilon\iota_X\Lambda_{(\alpha\beta)})$ on sections of the bundle gerbe. 
\vspace{0.2cm}

\noindent The only doubled vectors we are actually interested in are the ones which are invariant under the principal action of the bundle gerbe $\mathscr{G}$, i.e. the doubled vectors which satisfies the strong constraint. These are also called \textit{strongly foliated} by \cite{Vai12}.

\begin{remark}[Atiyah $L_\infty$-algebroid of the bundle gerbe]
We can define the Atiyah $L_\infty$-algebroid of our bundle gerbe with connective structure by $\mathfrak{at}(\mathscr{G}) := T\mathscr{G}/\!/\mathbf{B}U(1)_{\mathrm{conn}} \twoheadrightarrow M$, in perfect analogy with the Atiyah algebroid of a principal bundle.
\end{remark}

\begin{definition}[Courant $2$-algebroid]\label{def:courantalg}
The $2$-algebra of sections of the \textit{Courant }$2$\textit{-algebroid} over the base manifold $M$ sits in the center of the following short exact sequence of $2$-algebras:
\begin{equation}\label{eq:courasection}
\begin{tikzcd}[row sep=scriptsize, column sep=6.5ex]
   0 \arrow[r] & \mathbf{H}\big(M,\mathbf{b}\mathfrak{u}(1)_{\mathrm{conn}}\big)\; \arrow[r, hook, "\mathrm{injection}"] & \;\Gamma\big(M,\mathfrak{at}(\mathscr{G})\big)\; \arrow[r, two heads, "\mathrm{anchor}"] & \; \mathfrak{X}(M) \arrow[r] & 0,
\end{tikzcd}
\end{equation}
where $\mathfrak{X}(M)$ is the algebra of vector fields on $M$ and $\mathbf{H}\big(M,\mathbf{b}\mathfrak{u}(1)_{\mathrm{conn}}\big)$ is the abelian $2$-algebra of line bundles with connection on $M$, i.e. of infinitesimal gauge transformations of the gerbe. This is obtained by differentiating the finite automorphisms sequence \eqref{eq:finiteautomorphismseq}.
\end{definition}

\begin{remark}[Analogy with Atiyah $1$-algebroid]
Definition \ref{def:courantalg} is analogous to Atiyah algebroid in Kaluza-Klein, which encodes vectors on a circle bundle $P$ invariant under principal action. As explained in \cite{Col11} and \cite{Rog13} exact sequence \eqref{courantseq} is the higher version of the ordinary Atiyah sequence
\begin{equation}
\begin{tikzcd}[row sep=scriptsize, column sep=6.5ex]
   0 \arrow[r] & \Coo(M,\mathbb{R})\; \arrow[r, hook,"\mathrm{injection}"] & \;\Gamma\big(M,\mathfrak{at}(P)\big)\; \arrow[r, two heads, "\mathrm{anchor}"] & \; \mathfrak{X}(M) \arrow[r] & 0.
\end{tikzcd}
\end{equation}
\end{remark}

\begin{definition}[Standard Courant $2$-algebroid]
We define the \textit{standard Courant }$2$\textit{-algebroid} by the semidirect sum of $2$-algebroids
\begin{equation}
    \mathfrak{at}(\mathscr{G}) \; \cong\; TM \,\oplus_\mathrm{s}\, M\times\mathbf{b}\mathfrak{u}(1)_{\mathrm{conn}}.
\end{equation}
Thus, the $2$-algebra of its sections will be semidirect sum of $2$-algebras
\begin{equation}
    \Gamma\big(M,\mathfrak{at}(\mathscr{G})\big) \; \cong\; \mathfrak{X}(M) \,\oplus_\mathrm{s}\, \mathbf{H}\big(M,\mathbf{b}\mathfrak{u}(1)_{\mathrm{conn}}\big).
\end{equation}
This means its sections will be of the form $(X+\xi_\alpha, f_{(\alpha\beta)})$ where $X\in\mathfrak{X}(M)$ is a global vector and $(\xi_\alpha, f_{(\alpha\beta)})$ is a \v{C}ech cocycle
\begin{equation}
\begin{aligned}
    \begin{aligned}
        \xi_{(\alpha)} - \xi_{(\beta)} \;&=\; \mathrm{d}f_{(\alpha\beta)}, \\
        f_{(\alpha\beta)}+f_{(\beta\gamma)}+f_{(\gamma\alpha)} \;&=\; 0.
    \end{aligned}
\end{aligned}
\end{equation}
The morphisms will be gauge transformations $(\varepsilon_{(\alpha)}):(X+\xi_{(\alpha)}, f_{(\alpha\beta)})\mapsto(X+\xi_{(\alpha)}+\mathrm{d}\varepsilon_{(\alpha)},\, f_{(\alpha\beta)}+\varepsilon_{(\alpha)}-\varepsilon_{(\beta)})$ between line bundles, as usual. By slightly extending \cite{Col11}, the $2$-algebra structure of this semidirect sum is isomorphic to the bracket structure given by the following bracket structure
\begin{gather}\label{eq:brac}
\begin{aligned}
    \big\llbracket (\varepsilon_{(\alpha)}) \big\rrbracket_{\mathrm{std}} &= \big(\mathrm{d}\varepsilon_{(\alpha)}, \, \varepsilon_{(\alpha)}-\varepsilon_{(\beta)}) \\
    \!\!\!\!\!\big\llbracket (X+\xi_{(\alpha)}, f_{(\alpha\beta)}), (Y+\eta_{(\alpha)}, g_{(\alpha\beta)})\big\rrbracket_{\mathrm{std}}  &= \bigg([X,Y]+\mathcal{L}_X\eta_{(\alpha)}-\mathcal{L}_Y\xi_{(\alpha)} - \frac{1}{2}\mathrm{d}(\iota_X\eta_{(\alpha)}-\iota_Y\xi_{(\alpha)}), \\
    & \hspace{4.8cm} \frac{1}{2}X(g_{(\alpha\beta)})-\frac{1}{2}Y(f_{(\alpha\beta)})\bigg) \\
    \big\llbracket (X+\xi_{(\alpha)}, \, f_{(\alpha\beta)}), (\varepsilon_{(\alpha)})\big\rrbracket_{\mathrm{std}}  &= (\mathcal{L}_X\varepsilon_{(\alpha)}) \\
    \big\llbracket (X+\xi_{(\alpha)}, f_{(\alpha\beta)}), (Y+\eta_{(\alpha)}, g_{(\alpha\beta)}), (Z+&\zeta_{(\alpha)}, h_{(\alpha\beta)})\big\rrbracket_{\mathrm{std}}  = \frac{1}{3!}\bigg( \iota_X\iota_Y\mathrm{d}\zeta_{(\alpha)} + \frac{3}{2}\iota_X\mathrm{d}\iota_Y\zeta_{(\alpha)} + \text{perm.}\!\bigg)
\end{aligned}\raisetag{4.4cm}
\end{gather}
Let us notice that the underlying groupoid of sections $\mathfrak{X}(M)\oplus\mathbf{H}\big(M,\mathbf{b}\mathfrak{u}(1)_{\mathrm{conn}}\big)$ of the standard Courant $2$-algebroid is nothing but the stackification (in other words the globalisation) of the familiar local $\Coo(U,\mathbb{R})\,\xrightarrow{\;\mathrm{d}\;}\,\mathfrak{X}(U)\oplus\Omega^1(U)$ on patches $U\subset M$ of the base manifold.
\end{definition}

\begin{theorem}[General Courant $2$-algebroid]\label{gengeom}
Given the higher Atiyah sequence \eqref{eq:courasection}, for any choice of splitting homomorphism
\begin{equation}\label{courantseq}
\begin{tikzcd}[row sep=scriptsize, column sep=8ex]
   0 \arrow[r] &\mathbf{H}\big(M,\mathbf{b}\mathfrak{u}(1)_{\mathrm{conn}}\big)\; \arrow[r, hook, "\mathrm{injection}"] & \;\Gamma\big(M,\mathfrak{at}(\mathscr{G})\big)\; \arrow[r, two heads, "\mathrm{anchor}"] & \; \mathfrak{X}(M) \arrow[l, bend right=50, hook', "\mathrm{splitting}"'] \arrow[r] & 0.
\end{tikzcd}
\end{equation}
we obtain the following results.
\begin{itemize}
\item A section $\mathbbvar{X}:=(X+\xi_{(\alpha)}, \, f_{(\alpha\beta)})$ consists of a global vector field $X\in\mathfrak{X}(M)$, a collection of $1$-forms $\xi_{(\alpha)}\in\Omega^1(U_\alpha)$ on each patch $U_\alpha$ of $M$ and a collection of functions $f_{(\alpha\beta)}\in\Coo(U_\alpha\cap U_\beta)$ on each overlap $U_\alpha\cap U_\beta$ of $M$,
such that they are glued according to
\begin{equation}\label{eq:pathcthevector}
\begin{aligned}
    \begin{aligned}
        \xi_{(\alpha)} - \xi_{(\beta)} \;&=\; -\iota_{X}\mathrm{d}\Lambda_{(\alpha\beta)}+\mathrm{d}f_{(\alpha\beta)}, \\
        f_{(\alpha\beta)}+f_{(\beta\gamma)}+f_{(\gamma\alpha)} \;&=\; 0.
    \end{aligned}
\end{aligned}
\end{equation}
\item A morphism between two sections is a gauge transformation $(\varepsilon_{(\alpha)})$ of line bundles, given in local data by a collection of local functions $\varepsilon_{(\alpha)}\in\Coo(U_\alpha)$ so that
\begin{equation}
    (\varepsilon_{(\alpha)}):\,(X+\xi_{(\alpha)},\, f_{(\alpha\beta)}) \;\;\mapsto\;\; (X + \xi_{(\alpha)}+\mathrm{d}\varepsilon_{(\alpha)},\, f_{(\alpha\beta)}+\varepsilon_{(\alpha)}-\varepsilon_{(\beta)})
\end{equation}
\item The brackets of $\mathfrak{at}(\mathscr{G})$ are induced by the one of the standard Courant algebroid by
\begin{equation}\label{eq:bracketfromstd}
    \big\llbracket (s(X)+\xi_{(\alpha)},\,f_{(\alpha\beta)}),\,(s(Y)+\eta_{(\alpha)},\,f_{(\alpha\beta)})\big\rrbracket \,:=\, \big\llbracket (X+\xi_{(\alpha)},\,f_{(\alpha\beta)}),\,(Y+\eta_{(\alpha)},\,f_{(\alpha\beta)})\big\rrbracket_{\mathrm{std}}
\end{equation}
and analogously for the other brackets by using the remaining \eqref{eq:brac}.
\end{itemize}
\end{theorem}

\begin{proof}
By using the splitting $s$ and the injection $i$ in \eqref{courantseq} we can construct an isomorphism of $2$-algebras $s\oplus i:\,\mathfrak{X}(M)\oplus\mathbf{H}\big(M,\mathbf{b}\mathfrak{u}(1)_{\mathrm{conn}}\big)\xrightarrow{\;\cong\;}\mathfrak{at}(\mathscr{G})$. As explained in \cite{Rog13}, we have a splitting for any connection local data $B_{(\alpha)}\in\Omega^2(U_\alpha)$ which satisfy $B_{(\beta)} - B_{(\alpha)} = \mathrm{d}\Lambda_{(\alpha\beta)}$.
The isomorphism $s\oplus i$ is then given by the map $(X+\xi'_{(\alpha)},\,f_{(\alpha\beta)}) \mapsto (X+ \iota_XB_{(\alpha)} +\xi'_{(\alpha)},\, f_{(\alpha\beta)})$ for objects and by the identity for gauge transformations. Recall that $(\xi'_{(\alpha)},f_{(\alpha\beta)})$ is patched by $\xi'_{(\alpha)}-\xi'_\beta = \mathrm{d}f_{(\alpha\beta)}$ and $f_{(\alpha\beta)}+f_{(\beta\gamma)}+f_{(\gamma\alpha)}=0$. Now we only have to redefine $\xi_{(\alpha)} := \iota_XB_{(\alpha)}+\xi'_{(\alpha)}$ to get the wanted patching conditions.
\end{proof}

\begin{digression}[Twisted and untwisted generalised vectors]
The isomorphism of $L_\infty$-algebras
\begin{equation}\label{eq:isotw}
    \underbrace{\mathfrak{X}(M)\oplus\mathbf{H}\big(M,\mathbf{b}\mathfrak{u}(1)_{\mathrm{conn}}\big)}_\text{untwisted gen. \!vectors}\;\;\cong\underbrace{\Gamma\big(M,\mathfrak{at}(\mathscr{G})\big)}_\text{twisted gen. \!vectors}
\end{equation}
we used in lemma \ref{gengeom} is exactly the isomorphism locally presented by \cite{Hull14} between \textit{twisted} and \textit{untwisted} generalised vectors, but globally defined. Indeed, in the reference, on a given patch $U_\alpha\subset M$, if $X+\xi'_{(\alpha)}$ is an untwisted generalised vector, then $X+\iota_XB_{(\alpha)}+\xi'_{(\alpha)}$ is its twisted form. This perfectly matches with our construction. Moreover, our construction gives to the notion of twisted generalised vectors a precise higher geometrical interpretation: the connection $B_{(\alpha)}$ splits the Courant $2$-algebroid in a \textit{horizontal bundle} $TM$ and \textit{vertical bundle} $M\times\mathbf{b}\mathfrak{u}(1)_{\mathrm{conn}}$.
\end{digression}

\noindent This is analogous to how the tangent bundle of an ordinary circle bundle $P\rightarrow M$ is split in horizontal and vertical bundle $TP \cong TM\oplus\mathbb{R}$ by a connection $A_{(\alpha)}$.

\begin{remark}[Twisted bracket]
By explicitly writing the bracket structure \eqref{eq:bracketfromstd}, we recover the $H$-\textit{twisted bracket} of generalised geometry:
\begin{equation*}
\begin{aligned}
    &\big\llbracket (X+\xi_{(\alpha)}, f_{(\alpha\beta)}), (Y+\eta_{(\alpha)}, g_{(\alpha\beta)})\big\rrbracket =\\
    &=\!\bigg([X,Y]+\mathcal{L}_X\eta_{(\alpha)\!}-\mathcal{L}_Y\xi_{(\alpha)\!} - \frac{1}{2}\mathrm{d}(\iota_X\eta_{(\alpha)\!}-\iota_Y\xi_{(\alpha)\!}) + \iota_X\iota_YH,\,  \frac{1}{2}X(g_{(\alpha\beta)})-\frac{1}{2}Y(f_{(\alpha\beta)})\bigg).
\end{aligned}
\end{equation*}
\end{remark}

\begin{remark}[Pushforward of automorphisms as local $O(d,d)$ transformations]
We will now see how the Courant $2$-algebroid transforms under automorphisms of the bundle gerbe. Given an automorphism $(\varphi,\eta_{(\alpha)},\eta_{(\alpha\beta)})\in\mathrm{Diff}(M)\ltimes\mathbf{H}(M,\BU)$, any section $(X+\xi_{(\alpha)},f_{(\alpha\beta)})\in\Gamma\big(M,\mathfrak{at}(\mathscr{G})\big)$ will transform under its pushforward, which is its infinitesimal version, by
\begin{equation}
    \big(X+\xi_{(\alpha)},\;f_{(\alpha\beta)} \big) \;\,\mapsto\;\, \big(\varphi_\ast X+(\varphi^\ast)^{-1}\xi_{(\alpha)} +\iota_X\mathrm{d}\eta_{(\alpha)},\;f_{(\alpha\beta)}\circ\varphi\big).
\end{equation}
Therefore on each local patch $U_\alpha\subset M$ we recover the following transformations
\begin{equation}
    \begin{pmatrix}X\\\xi_{(\alpha)}\end{pmatrix} \;\,\mapsto\;\, \begin{pmatrix}\varphi_\ast & 0 \\\mathrm{d}\eta_{(\alpha)} & (\varphi^\ast)^{-1} \end{pmatrix}\begin{pmatrix}X\\\xi_{(\alpha)}\end{pmatrix}
\end{equation}
Since we have $\varphi_\ast\in\Coo(M,GL(d))$ and $\mathrm{d}\eta_{(\alpha)}\in\Omega_{\mathrm{cl}}^2(M)$, we can interpret this as a local $GL(d)\ltimes\wedge^2\mathbb{R}^d\subset O(d,d)$ transformation. Hence, we recover the local geometric $O(d,d)$ transformations of a Courant algebroid.
\end{remark}

\begin{remark}[Recovering ordinary generalised geometry]\label{rem:horizontalsec}
The $2$-algebra $\Gamma\big(M,\mathfrak{at}(\mathscr{G})\big)$ immediately reduces to the one appearing in \cite{DesSae18} and \cite{Col11} for sections of the form $(X+\xi_{(\alpha)}):=(X+\xi_{(\alpha)},0)$ with $f_{(\alpha\beta)}=0$. Therefore these sections satisfy the patching condition $\xi_{(\alpha)} - \xi_{(\beta)} =- \iota_{X}\mathrm{d}\Lambda_{(\alpha\beta)}$ on overlaps of patches and their morphisms are gauge transformations given by global functions $\varepsilon\in\Coo(M)$. These are exactly the sections which \cite{Col11} calls "horizontal lifts" of a vector $X$. Moreover their brackets \eqref{eq:brac} reduce to the $2$-algebra structure which appears in \cite{DesSae18}:
\begin{gather}\label{eq:generalised geometry}
\begin{aligned}
    \big\llbracket \,\varepsilon\, \big\rrbracket &= \,\mathrm{d}\varepsilon \\
    \big\llbracket (X+\xi_{(\alpha)}), (Y+\eta_{(\alpha)})\big\rrbracket &= \bigg([X,Y]+\mathcal{L}_X\eta_{(\alpha)}-\mathcal{L}_Y\xi_{(\alpha)} - \frac{1}{2}\mathrm{d}(\iota_X\eta_{(\alpha)}-\iota_Y\xi_{(\alpha)})+\iota_X\iota_YH\bigg) \\
    \big\llbracket (X+\xi_{(\alpha)}), \,\varepsilon\, \big\rrbracket &= \,\mathcal{L}_X\varepsilon  \\
    \big\llbracket (X+\xi_{(\alpha)}), (Y+\eta_{(\alpha)}), &(Z+\zeta_{(\alpha)})\big\rrbracket = \frac{1}{3!}\bigg( \iota_X\iota_Y\mathrm{d}\zeta_{(\alpha)} + \frac{3}{2}\iota_X\mathrm{d}\iota_Y\zeta_{(\alpha)} + \text{perm.}\bigg)
\end{aligned}\raisetag{20pt}
\end{gather}
We will denote this $2$-algebra with the symbol $\Gamma\big(M,\mathfrak{at}(\mathscr{G})\big)_{\mathrm{hor}}$. Notice that these horizontal sections $(X+\xi_{(\alpha)})\in\Gamma\big(M,\mathfrak{at}(\mathscr{G})\big)_{\mathrm{hor}}$ equipped with the bracket $\llbracket -,-\rrbracket$ from \eqref{eq:generalised geometry} can be also seen as sections $(X+\xi_{(\alpha)})\in\Gamma(M,E)$ of an ordinary Courant algebroid $E$ appearing in generalised geometry at the center of a short exact sequence $T^\ast M\rightarrow E \rightarrow TM$. In other words the underlying chain complex of the $2$-algebra $\Gamma\big(M,\mathfrak{at}(\mathscr{G})\big)_{\mathrm{hor}}$ will be $\Coo(M)\xrightarrow{\mathrm{d}}\Gamma(M,E)$. Hence if we restrict to horizontal sections we recover explicitly ordinary generalised geometry (see \cite{Gua11} for details).
\end{remark}

\noindent Notice from remark \ref{rem:horizontalsec} that $\mathrm{d}\Lambda_{(\alpha\beta)}$ satisfies the cocycle condition, even if $\Lambda_{(\alpha\beta)}$ does not. This is why the transition functions of the Courant $2$-algebroid define a global vector bundle, i.e. the vector bundle underlying the ordinary Courant algebroid. Now the reader may wonder why we considered sections from lemma \ref{gengeom} with non-vanishing gauge transformations on two-fold overlaps of patches instead of just horizontal ones. The answer is that there are applications where these data cannot be neglected, such as in geometry of T-duality in the next section.
\vspace{0.2cm}

\noindent Let us conclude this section by mentioning the relation between this stack perspective on generalised geometry and symplectic dg-geometry.

\begin{digression}[Relation with NQP-manifolds]
Let us recall that a differential-graded manifold (or NQ-manifold) is a locally ringed space $(N,\Coo)$ where $N$ is a topological space and $\Coo$ a sheaf of differential-graded algebras on $N$ satisfying some extra properties. It is well-understood that, given a $L_\infty$-algebroid $\mathfrak{a}\twoheadrightarrow M$, its Chevalley-Eilenberg dg-algebra $\mathrm{CE}(\mathfrak{a})$ can be seen as the dg-algebra of functions on a dg-manifold, also known as NQ-manifold.
The differential graded manifold $T^\ast[2]T[1]U$ with $U\subset M$ and local coordinates $\{x^\mu,\zeta_\mu,\chi^\mu,p_\mu\}$ respectively in degree $0$, $1$, $1$, $2$
and a sheaf of functions $\Coo(-)$ with differential $Q_H$. As shown by \cite{Roy02}, remarkably, its differential-graded algebra of functions is exactly the Chevalley-Eilenberg dg-algebra of the Atiyah $L_\infty$-algebroid $\mathfrak{at}(\mathscr{G}|_U)$, i.e.
\begin{equation}
    \Big(  \Coo(T^\ast[2]T[1]U),\,Q_H\Big) \,=\, \mathrm{CE}\big(\mathfrak{at}(\mathscr{G}|_U)\big),
\end{equation}
where the dg-manifold $T^\ast[2]T[1]M$, called Vinogradov algebroid, is canonically symplectic, i.e. it is canonically a NQP-manifold. 
In other words $T^\ast[2]T[1]U$ is just an alternative way to express $\mathfrak{at}(\mathscr{G}|_U)$. Notice that this recovers Extended Riemannian Geometry by \cite{DesSae18} in the simple case of generalised geometry. We will explain how we can to recover their geometry of doubled torus bundles in the section 4.
Inspired by this relation, a purely dg-geometric approach to Double Field Theory was developed by \cite{DesSae18, DesSae18x, Crow-Watamura:2018liw, DesSae19}.
\end{digression}

\subsection{A prequantum interpretation for the bundle gerbe}

\noindent Let us now conclude this section with an interesting digression. In fact, higher Kaluza-Klein Theory is also able to explicitly link DFT with a distinct field of research: higher geometric quantisation for String Theory. See the following digression for a brief discussion.

\begin{digression}[A relation with higher geometric prequantisation]\label{dig:preq}
Notice our higher Kaluza-Klein is as closely related to higher geometric prequantisation as ordinary Kaluza-Klein is to ordinary geometric prequantisation. The parallel transport of a section $(\theta_{(\alpha)})\in\Gamma(M,P)$ along a vector flow $\ell(t,x)$ with $\ell(0,x)=x$ of some Hamiltonian vector field $X$ is given by
\begin{equation}\label{eq:circlepreq}
    \theta_{(\alpha)}\big(\ell(t,x)\big) \,=\, \exp 2\pi i\Bigg(\sum_{\ell_\alpha}\int_{\ell_\alpha} \!A_{(\alpha)} + \sum_{x_{\alpha\beta}}f_{(\alpha\beta)}(x_{\alpha\beta})\Bigg)\cdot\theta_{(\alpha)}(x)
\end{equation}
which is a global gauge transformation in $\Coo(M,U(1))$ at any $t\in\mathbb{R}$. Recall that the underlying vector space of an ordinary prequantisation Hilbert space is $\Gamma(M,\,P\times_{U(1)}\mathbb{C})$, i.e. the space of sections of the associated bundle $P\times_{U(1)}\mathbb{C}$. Hence parallel transport \eqref{eq:circlepreq} can be immediately generalised to prequantum states $(\psi_{(\alpha)})\in\Gamma(M,\,P\times_{U(1)}\mathbb{C})$. Analogously in higher Geometric Prequantisation we can define a parallel transport of a section $(\widetilde{x}_{(\alpha)},\phi_{(\alpha\beta)})\in\Gamma(M,\mathcal{M})$ of the bundle gerbe along a vector flow $\ell(\tau,x)$ with $\ell(0,x)=x$ of a Hamiltonian vector field $X$ by
\begin{equation}\label{eq:higherpreq}
\begin{aligned}
    \ell(\tau,-)_\ast\big(\widetilde{x}_{(\alpha)},\phi_{(\alpha\beta)}\big) \;=&\; \Bigg(\sum_{l_\alpha}\int_{l_\alpha} \! B_{(\alpha)} + \sum_{x_{\alpha\beta}}\Lambda_{(\alpha\beta)}(x_{\alpha\beta}),\\
    &\;\;\;\, \sum_{l_\alpha}\int_{l_\alpha} \! \Lambda_{(\alpha\beta)} + \!\!\!\sum_{x_{\alpha\beta\gamma}}\!G_{(\alpha\beta\gamma)}(x_{\alpha\beta\gamma}) \Bigg)\otimes \big(\widetilde{x}_{(\alpha)},\phi_{(\alpha\beta)}\big)
\end{aligned}
\end{equation}
which is a global gauge transformation in $\mathbf{H}(M,\BU)$ at any $\tau\in\mathbb{R}$. In \cite{FSS13,FSS16,Sza19} is explained, and reviewed in chapter \ref{ch:3}, that the \textit{prequantum }$2$\textit{-Hilbert space} is defined by $\mathfrak{H}_\mathrm{pre}:=\Gamma(M,\,\mathscr{G}\times_{\mathbf{B}U(1)}\!\mathbf{B}U)$, i.e. by the groupoid of sections of the associated $2$-bundle $\mathscr{G}\times_{\mathbf{B}U(1)}\!\mathbf{B}U$ where the fiber stack is the direct limit $\mathbf{B}U := \Lim{ N \to \infty}\mathbf{B}U(N)$. Hence parallel transport \eqref{eq:higherpreq} can be immediately generalised to \textit{prequantum }$2$\textit{-states} of the form $(\psi_{(\alpha)},\psi_{(\alpha\beta)})\in\Gamma(M,\,\mathscr{G}\times_{\mathbf{B}U(1)}\!\mathbf{B}U)$. These are principal $U(N)$-bundles on the base manifold $M$ for any $N\in\mathbb{N}^+$, twisted by the bundle gerbe $\mathscr{G}\xrightarrow{\bbpi}M$ with cocycle $(B_{(\alpha)},\Lambda_{(\alpha\beta)},G_{(\alpha\beta\gamma)})$. These twisted $U(N)$-bundles are given in \v{C}ech data by $\psi_{(\alpha)}\in\Omega^1\big(U_\alpha,\mathfrak{u}(N)\big)$ and $\psi_{(\alpha\beta)}\in\Coo\big(U_\alpha\cap U_\beta,U(N)\big)$ patched by
\begin{equation*}
    \begin{aligned}
    \psi_{(\alpha)} - \psi_{(\alpha\beta)}^{-1}\big(\psi_\beta + \mathrm{d} \big)\psi_{(\alpha\beta)} \,&=\, - \Lambda_{(\alpha\beta)} \\
    \psi_{(\alpha\beta)}\cdot\psi_{(\beta\gamma)}\cdot\psi_{(\gamma\alpha)} \,&=\, \exp{i2\pi G_{(\alpha\beta\gamma)}}
    \end{aligned}
\end{equation*}
and they can be interpreted as states of $N$ coincident D-branes in a Kalb-Ramond field background.
In \cite{Sza19} this was explicitly calculated, in an infinitesimal fashion, for $M=\mathbb{R}^d$ and its non-associative behaviour was pointed out.
As explained by \cite{Rog13} and \cite{FSS13} we recover an ordinary prequantisation on the loop space $\mathcal{L}M:=[S^1,M]$. Indeed, given any loop $S_0^1\subset M$, its evolution $S^1_\tau := \ell(\tau,\,S^1_0)$ with $\tau\in\mathbb{R}$ can be seen both as a surface $\Sigma\subset M$ with $\partial\Sigma=S^1_0\sqcup S^1_\tau$ or as a path in the loop space $\mathcal{L}M$. By integrating along surface $\Sigma$ and taking the trace we get
\begin{equation*}
\begin{aligned}
    \mathrm{Hol}_{\ell(\tau,-)^\ast(\psi_{(\alpha)},\psi_{(\alpha\beta)})}(S^1_\tau)\,=\, \mathrm{Hol}_{(B_{\alpha},\Lambda_{(\alpha\beta)},G_{(\alpha\beta\gamma)})}(\Sigma)\,\cdot\, \mathrm{Hol}_{(\psi_{(\alpha)},\psi_{(\alpha\beta)})}(S^1_0) \\
\end{aligned}
\end{equation*}
where the holonomy of a twisted $U(N)$-bundle $(\psi_{(\alpha)},\psi_{(\alpha\beta)})$ along a loop $S^1\subset M$ and the holonomy of a gerbe $(B_{\alpha},\Lambda_{(\alpha\beta)},G_{(\alpha\beta\gamma)})$ along a surface $\Sigma\subset M$ are respectively given by the usual expressions
\begin{equation*}\begin{aligned}
    \mathrm{Hol}_{(\psi_{(\alpha)},\psi_{(\alpha\beta)})}(S^1) \,&:=\, \mathrm{Tr}\,\mathcal{P}\bigg(\exp{ 2\pi i\bigg(\int_{l_\alpha}\!\psi_{(\alpha)}}\bigg)\cdot\prod_{x_{\alpha\beta}}\psi_{(\alpha\beta)}(x_{\alpha\beta})\bigg), \\
    \mathrm{Hol}_{(B_{(\alpha)},\Lambda_{(\alpha\beta)},G_{(\alpha\beta\gamma)})}(\Sigma) \,&:=\, \exp 2\pi i\Bigg(\!\sum_\alpha\int_{\Sigma_\alpha} \!\!\!B_{(\alpha)} + \sum_{l_{(\alpha\beta)}}\int_{l_{(\alpha\beta)}} \!\!\!\Lambda_{(\alpha\beta)} + \!\!\sum_{x_{\alpha\beta\gamma}}\!G_{(\alpha\beta\gamma)}(x_{\alpha\beta\gamma})\Bigg).
\end{aligned}\end{equation*}
More generally, for an open surface $\Sigma\subset M$ such that the boundary $\partial\Sigma=\sqcup_iS_i^1$ is a disjoint union of loops of any orientation, we recover Bohr-Sommerfeld condition for a surface with boundary 
\begin{equation}
    \mathrm{Hol}_{(B_{(\alpha)},\Lambda_{(\alpha\beta)},G_{(\alpha\beta\gamma)})}(\Sigma) \,\cdot\, \prod_i \mathrm{Hol}_{(\psi_{(\alpha)},\psi_{(\alpha\beta)})}(S^1_i) \,=\,1
\end{equation}
Notice that this is nothing but the bosonic part of \textit{Freed-Witten anomaly} cancellation of a string.

\begin{figure}[!ht]\centering
\includegraphics[scale=0.35]{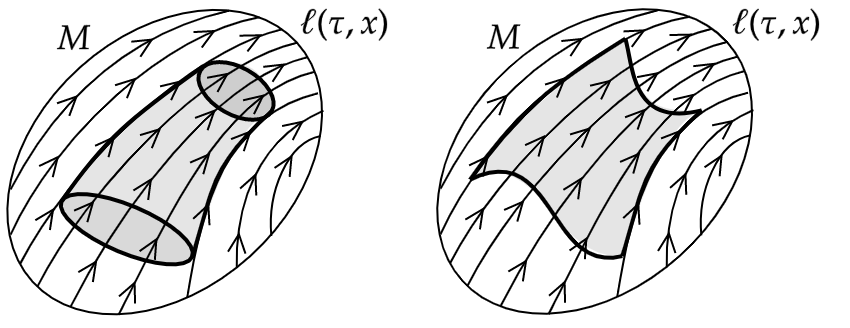}
\caption{Parallel transport along surfaces $\Sigma$, open and closed.}\label{FIGstring}\end{figure}
\end{digression}

\begin{remark}[A prequantum interpretation of the sections]
It is suggestive to notice that in higher prequantisation (see digression \ref{dig:preq}) a section of the bundle gerbe $(\widetilde{x}_{(\alpha)},\phi_{(\alpha\beta)})$ on the world-volume of $N$ coincident D-branes plays the role of a "higher phase" for the $U(N)$-field $(\psi_{(\alpha)},\psi_{(\alpha\beta)})$. This is analogous to the well-known fact that a section $(\theta_{(\alpha)})$ of the prequantum circle bundle plays the role of the phase of a wave-function $(\psi_{(\alpha)})$. This provides an evocative interpretation of the bundle gerbe underlying DFT in the context of prequantisation.
\end{remark}

\begin{table}[ht!]\begin{center}\vspace{0.5cm}\begin{tabular}{ c | c c }
 & $\;\;\qquad$Prequantisation$\qquad\quad$ & $\qquad\qquad$Higher Prequantisation$\qquad\quad$ \\[0.8ex] \hline \\[-1.5ex]
Phase$\,$ & $\theta_{(\alpha)} -\theta_{(\beta)} = G_{(\alpha\beta)} \;\mathrm{mod}\,2\pi\mathbb{Z}$ & \makecell{$\widetilde{x}_{(\alpha)} - \widetilde{x}_{(\beta)} -\mathrm{d}\phi_{(\alpha\beta)}  \,=\, -\Lambda_{(\alpha\beta)}$, \\ $\phi_{(\alpha\beta)}+\phi_{(\beta\gamma)}+\phi_{(\gamma\alpha)} \,=\, G_{(\alpha\beta\gamma)} \; \mathrm{mod}\,2\pi\mathbb{Z}$} \\[6.2ex]
Matter$\,$ & $\psi_{(\alpha)}\cdot\psi_{(\beta)}^{-1} = \exp i2\pi G_{(\alpha\beta)}$ & \makecell{$\psi_{(\alpha)} - \psi_{(\alpha\beta)}^{-1}\big(\psi_{(\beta)} + \mathrm{d} \big)\psi_{(\alpha\beta)} \,=\, - \Lambda_{(\alpha\beta)}$ \\ $\psi_{(\alpha\beta)}\cdot\psi_{(\beta\gamma)}\cdot\psi_{(\gamma\alpha)} \,=\, \exp{i2\pi G_{(\alpha\beta\gamma)}}$ }
\end{tabular}\end{center}\caption[Phases and states in ordinary and higher prequantisation.]{A comparison of phases and states between ordinary and higher prequantisation.}\end{table}

\section{NS5-brane as higher Kaluza-Klein monopole}
In this section we will present a new, globally defined monopole for higher Kaluza-Klein Theory, by directly generalising the ordinary Kaluza-Klein monopole by \cite{GrosPer83}. This can be interpreted as a globally defined monopole for DFT which does not need compactified dimensions to be well-defined. We will show that this monopole is an NS5-brane with non-trivial $H$-charge by higher Kaluza-Klein reduction. Finally we will prove that by smearing it we recover the familiar Berman-Rudolph DFT monopole.

\subsection{Higher Dirac monopole of the Kalb-Ramond field}

Let us give a quick review of the Dirac monopole in classical electromagnetism in this subsection. Then we will directly generalize this notion to a Kalb-Ramond field monopole.

\begin{definition}[Dirac monopole]
A Dirac monopole is a circle bundle of the form
\begin{equation}
    \mathbb{R}^{1}\times\left(\mathbb{R}^{3}-\{0\}\right) \longrightarrow \mathbf{B}U(1).
\end{equation}
with non-trivial first Chern class $[F]\in H^2\big(\mathbb{R}^{3}-\{0\},\,\mathbb{Z}\big)$ on the transverse space $\mathbb{R}^{3}-\{0\}$ and trivial on the time line $\mathbb{R}^{1}$. Here $\mathbb{R}^{1}$ can be seen as a magnetically charged world-line.
\end{definition}

\begin{remark}[Dirac charge-quantisation]
This spacetime can be alternatively written as
\begin{equation}
    \mathbb{R}^{1}\times\left(\mathbb{R}^{3}-\{0\}\right) \,\simeq\, \mathbb{R}^{1} \times \mathbb{R}^+ \times S^2
\end{equation}
where $\mathbb{R}^+$ is the radial direction in the transversal space and $S^2$ embodies the angular directions. Since $\mathbb{R}^+ \times S^2$ is homotopy equivalent to the $2$-sphere, its cohomology groups will be clearly isomorphic to the ones of the $2$-sphere. The underlying topological space of the stack $\mathbf{B}U(1)$ is the classifying space $BU(1)$ of circle bundles, i.e. the second Eilenberg–MacLane space
\begin{equation}
    \left|\mathbf{B}U(1)\right| = BU(1) = K(\mathbb{Z},2),
\end{equation}
where $|-|$ gives the geometric realisation of an $\infty$-groupoid. Circle bundles over the $2$-sphere are then classified by maps $S^2\rightarrow K(\mathbb{Z},2)$ whose group is just $\pi_2\big(K(\mathbb{Z},2)\big) \cong \mathbb{Z}$. Dually the second cohomology group of the $2$-sphere is $H^2(S^2,\mathbb{Z})\cong \mathbb{Z}$. Hence the first Chern number of any such bundle will be an integer
\begin{equation}
    \frac{1}{2\pi}\int_{S^2} F = m\in\mathbb{Z}
\end{equation}
and the curvature of the bundle will be a closed non-exact form $F=m\,\mathrm{Vol}(S^2)/2$. The trivial fibration $S^2\times S^1\rightarrow S^2$ corresponds to $m=0$ and the Hopf fibration $S^3\rightarrow S^2$ to $m=1$, while in general we will have a Lens space fibration $L(1,m)\rightarrow S^2$ for any $m\in\mathbb{Z}$.
\end{remark}

\begin{remark}[Local description of Dirac monopole]\label{rem:dirac}
Let us quickly look at what this means in terms of gauge fields. As very well known we can cover the $2$-sphere with just two open sets $\mathcal{U}=\{U,U'\}$ such that in spherical coordinates $(\phi,\theta)$ they are overlapping spherical caps $U=[0,2\pi)\times[0,\pi/2+u)$ and $U'=[0,2\pi)\times(\pi/2-u,\pi]$ for some $u\ll\pi/2$. The curvature can be explicitly written as $F=\sin{\theta}\,\mathrm{d}\theta\wedge\mathrm{d}\phi$. On the two charts the connection of the bundle $S^2\rightarrow K(\mathbb{Z},2)$ will then be given respectively by
\begin{equation}
    \begin{aligned}
        A = \frac{m}{2}(1-\cos{\theta})\mathrm{d}\phi, \quad A' = -\frac{m}{2}(1+\cos{\theta})\mathrm{d}\phi.
    \end{aligned}
\end{equation}
We can see that on the overlap $U\cap U'=(\pi/2-u,\pi/2+u)\times S^1$ we have $A-A'=m\,\mathrm{d}\phi$, which, integrated along the equator, gives $4\pi m$ and equivalently the Dirac quantisation condition
\begin{equation}
    \begin{aligned}
        \frac{1}{2\pi}\int_{S^1}A-A' = m.
    \end{aligned}
\end{equation}
\end{remark}

\begin{definition}[Higher Dirac monopole]\label{def:hdm}
A \textit{higher Dirac monopole} is a bundle gerbe of the form
\begin{equation}
    \mathbb{R}^{1,5} \times \left(\mathbb{R}^4-\{0\}\right) \longrightarrow \mathbf{B}^2U(1),
\end{equation}
with non-trivial Dixmier-Douady class $[H]$ on the transverse space $\mathbb{R}^4-\{0\}$ and trivial over $\mathbb{R}^{1,5}$. Here $\mathbb{R}^{1,5}$ can be seen as a magnetically $H$-charged world-volume.
\end{definition}

\begin{remark}[Higher Dirac charge-quantisation]
This spacetime can be also written as
\begin{equation}
    \mathbb{R}^{1,5} \times \left(\mathbb{R}^4-\{0\}\right) \,\simeq\, \mathbb{R}^{1,5} \times \mathbb{R}^+ \times S^3,
\end{equation}
where $\mathbb{R}^+$ is the radial direction in the transversal space, while $S^3$ embodies the angular directions. Since $\mathbb{R}^+ \times S^3$ is homotopy equivalent to the $3$-sphere, the cohomology groups of the transversal space will be immediately isomorphic to the ones of the $3$-sphere. The classifying space of abelian gerbes is the third Eilenberg–MacLane space 
\begin{equation}
    \left|\mathbf{B}^2U(1)\right| = B^2U(1) = K(\mathbb{Z},3)
\end{equation}
and the gerbes on the $3$-spheres are given by maps $S^3\rightarrow K(\mathbb{Z},3)$. The group of these maps is just the third homotopy group of the Eilenberg–MacLane space 
\begin{equation}
    \pi_3\big(K(\mathbb{Z},3)\big) \cong \mathbb{Z}
\end{equation}
which is isomorphic to the integers. Dually the third cohomology group of the $3$-sphere is $H^3(S^3,\mathbb{Z})\cong\mathbb{Z}$. Hence the Dixmier-Douady number of any such bundle will be an integer
\begin{equation}
    \frac{1}{4\pi^2}\int_{S^3} H = m\in\mathbb{Z}
\end{equation}
that we may call \textit{higher magnetic charge} or just $H$-charge. Then the curvature of the gerbe will be in general a non-exact $3$-form
\begin{equation}
    H = \frac{m}{2}\,\mathrm{Vol}(S^3),
\end{equation}
in direct analogy with the ordinary Dirac monopole.
\end{remark}

\begin{remark}[Atlas for the $3$-sphere]
A $3$-sphere $S^3$ can be seen as the submanifold of $\mathbb{C}^2$ defined by the condition $w_1^\ast w_1+w_2^\ast w_2=1$ on the complex coordinates $(w_1,w_2)\in\mathbb{C}^2$. This condition can be solved by
\begin{equation}
    w_1 = e^{i(\psi_1+\psi_2)}\sin{\chi}, \quad  w_2 = e^{i(\psi_1-\psi_2)}\cos{\chi},
\end{equation}
where $(\chi,\psi_1,\psi_2)$ with ranges $\chi\in[0,\pi/2],\psi_1\in[0,2\pi)$ and $\psi_2\in[0,\pi)$, are called \textit{Hopf coordinates}. Topologically $S^3-\{\ast\}\simeq \mathbb{R}^3$, thus we can give $S^3$ an open cover $\mathcal{U}=\{U,U'\}$ of just two open sets, for example the ones of $S^3$ deprived respectively of north and south pole.
\end{remark}

\begin{remark}[Local description of higher Dirac monopole]
We can cover the $3$-sphere with two open spherical caps $\mathcal{U}=\{U,U'\}$ and solve the gerbe curvature $H=m\mathrm{Vol}(S^3)/2$ by
\begin{equation}
    \begin{aligned}
        B = \frac{m}{2}(1-\cos{2\chi})\,\mathrm{d}\psi_1\wedge\mathrm{d}\psi_2, \quad B' = -\frac{m}{2}(1+\cos{2\chi})\,\mathrm{d}\psi_1\wedge\mathrm{d}\psi_2.
    \end{aligned}
\end{equation}
The overlap of the patches $U\cap U'\simeq (u,\pi/2-u)\times T^2$ for some $u\ll\pi/2$ is homotopy equivalent to $T^2$, so that $H^2(T^2,\mathbb{Z})\cong\mathbb{Z}$ and hence we have 
\begin{equation}
    \begin{aligned}
        \frac{1}{4\pi^2}\int_{T^2}B-B' = m,
    \end{aligned}
\end{equation}
which is exactly the charge-quantisation condition for String Theory with the constant $\alpha'=1$.
\end{remark}

\subsection{Higher Kaluza-Klein monopole in 10d}

In this subsection we will define the higher Kaluza-Klein monopole and we will look at its properties. Moreover we will show that it physically reduces to the NS5-brane. First of all let us give a quick review of the ordinary Kaluza-Klein monopole that we are going to generalize

\begin{digression}[Kaluza-Klein monopole by \cite{GrosPer83}]
A Kaluza-Klein monopole is a spacetime $(M,g)$ such that $M=\mathbb{R}^{1}\times\mathbb{R}^+\times L(1,m)$, where $L(1,m)$ is a Lens space, and the metric is
\begin{equation}
    \begin{aligned}
        g &= -\mathrm{d}t^2 + h(r)\delta_{ij}\mathrm{d}y^i\mathrm{d}y^j + \frac{1}{h(r)}(\mathrm{d}\widetilde{y}+A_i\mathrm{d}y^i)^2
    \end{aligned}
\end{equation}
where we called $r^2:=\delta_{ij}y^iy^j$ the radius in the transverse space, $t$ the coordinate on $\mathbb{R}^{1}$, $\{y^i\}_{i=1,2,3}$ the coordinate of the transverse space and $\widetilde{y}$ the coordinate of the fiber $S^1$. This spacetime encompass a Dirac monopole on the base manifold $\mathbb{R}^{1}\times\left(\mathbb{R}^3-\{0\}\right) \simeq \mathbb{R}^{1}\times\mathbb{R}^+\times S^2$
 The gauge field $A$ is required to satisfy the following conditions
\begin{equation}
    F = \star_{\mathbb{R}^3}\mathrm{d}h, \quad h(r)=1+\frac{m}{r}
\end{equation}
for $m\in\mathbb{Z}$. In other words the curvature of the bundle will be $F=m\mathrm{Vol}(S^2)/2$ with first Chern number $m$, representing the magnetic charge.
If we cover the $2$-sphere with two charts we can rewrite this metric in polar coordinates $(r,\theta,\phi)$ and we obtain on the first one
\begin{equation}
    g = -\mathrm{d}t^2 + h(r)(\mathrm{d}r^2 +r^2\mathrm{d}\theta^2 + r^2 \sin^2{\theta}\,\mathrm{d}\phi^2) + \frac{1}{h(r)}\bigg(\mathrm{d}\widetilde{y}+\frac{m}{2}(1-\cos{\theta})\mathrm{d}\phi\bigg)^2  \\
\end{equation}
while on the second one the gauge field is $A'=-m(1+\cos{\theta})\mathrm{d}\phi/2$, in agreement with remark \ref{rem:dirac}.
\end{digression}

\noindent Now we can give a precise definition of a new monopole, which directly generalises the ordinary Kaluza-Klein monopole and which geometrises the higher Dirac monopole of definition \ref{def:hdm}.

\begin{definition}[Higher Kaluza-Klein monopole]\label{def:hkkm}
A higher Kaluza-Klein monopole is a bundle gerbe $\mathscr{G}$ with generalised metric $\mathcal{G}$ such that 
\begin{itemize}
    \item $\mathscr{G}$ is a bundle gerbe on the base manifold $M=\mathbb{R}^{1,5}\times\left(\mathbb{R}^4-\{0\}\right) \simeq \mathbb{R}^{1,5}\times\mathbb{R}^+\times S^3$ which is non-trivial only on $S^3$,
    \item $\mathcal{G}$ is a (global) generalised metric, which is given on the atlas $\mathcal{M}$ of the bundle gerbe $\mathscr{G}$ by
\end{itemize}
\begin{equation}\label{eq:hkkmonopole}
    \begin{aligned}
        \mathcal{G} &= \eta_{\mu\nu}\mathrm{d}x^\mu\mathrm{d}x^\nu + \eta^{\mu\nu}\mathrm{d}\widetilde{x}_\mu\mathrm{d}\widetilde{x}_\nu + h(r)\delta_{ij}\mathrm{d}y^i\mathrm{d}y^j + \frac{\delta^{ij}}{h(r)}(\mathrm{d}\widetilde{y}_i+B_{ik}\mathrm{d}y^k)(\mathrm{d}\widetilde{y}_j+B_{jk}\mathrm{d}y^k),
    \end{aligned}
\end{equation}
where $\{x^\mu\}$ and $\{y^i\}$ are respectively the coordinates of $\mathbb{R}^{1,5}$ and $\mathbb{R}^{+}\times S^3$ and where the curvature of the gerbe and the harmonic function are respectively constraint to
\begin{equation}\label{eq:transversalcond}
    H = \star_{\mathbb{R}^4}\mathrm{d}h, \quad h(r)=1+\frac{m}{r^2}
\end{equation}
for any $m\in\mathbb{Z}$, with $r^2:=\delta_{ij}y^iy^j$ radius in the four-dimensional transverse space. 
\end{definition}

\noindent This generalised metric encompasses a higher Dirac monopole from definition \ref{def:hdm} on the base manifold, just as the Kaluza-Klein monopole does with an ordinary Dirac monopole. In other words the curvature of the gerbe will be $H=m\mathrm{Vol}(S^3)/2$ with non-trivial $H$-charge $m\in\mathbb{Z}$.

\begin{remark}[Correspondence space of higher Kaluza-Klein monopole]
By using the fact that $S^3$ is a $U(1)$-bundle (Hopf fibration) on $S^2$ we can apply the result of lemma \ref{thm:corrspace}. Thus, our bundle gerbe $\mathscr{G}|_{S^3}\twoheadrightarrow S^3$, restricted to the $3$-sphere, is Kaluza-Klein reduced to a $\String$-bundle $S^2\longrightarrow\mathbf{B}\String(S^1\times S^1)$, whose underlying $(S^1\times S^1)$-bundle is the correspondence space
\begin{equation}
    \begin{tikzcd}[row sep={11ex,between origins}, column sep={11ex,between origins}]
     & & S^3\times_{S^2}L(1,m)\arrow[dr, "\pi"']\arrow[dl, "\widetilde{\pi}"] & & \\
    S^1 \arrow[r, hook] & S^3\arrow[dr, "\pi"'] & & L(1,m)\arrow[dl, "\widetilde{\pi}"] & \arrow[l, hook'] S^1\\
    & & S^2. & &
    \end{tikzcd}
\end{equation}
\end{remark}

\begin{remark}[NS5-brane is higher Kaluza-Klein monopole]
By higher Kaluza-Klein reduction of the generalised metric \eqref{eq:hkkmonopole} to the base manifold $M=\mathbb{R}^{1,5}\times\mathbb{R}^+\times S^3$ we get the following metric and gerbe connection:
\begin{equation}
    \begin{aligned}
        g &= \eta_{\mu\nu}\mathrm{d}x^\mu\mathrm{d}x^\nu + h(r)\delta_{ij}\mathrm{d}y^i\mathrm{d}y^j,\\
        B &= B_{ij}\,\mathrm{d}y^i\wedge\mathrm{d}y^j
    \end{aligned}
\end{equation}
which satisfy the conditions \eqref{eq:transversalcond} on the transversal space.
If we rewrite this metric and $B$-field on a chart in spherical coordinates $(r,\chi,\psi_1,\psi_2)$ we obtain
\begin{equation}
    \begin{aligned}
         g &=   \eta_{\mu\nu}\mathrm{d}x^\mu\mathrm{d}x^\nu + h(r)\mathrm{d}r^2 + h(r)r^2\left(\mathrm{d}\chi^2+\mathrm{d}\psi_1^2+\mathrm{d}\psi_2^2-2\cos{2\chi}\,\mathrm{d}\psi_1\mathrm{d}\psi_2\right) \\
         B &= -\frac{m}{2}\cos{2\chi}\,\mathrm{d}\psi_1\wedge\mathrm{d}\psi_2
    \end{aligned}
\end{equation}
These are exactly the metric and Kalb-Ramond field of an NS5-brane with non-trivial $H$-charge $m$ in $10d$ spacetime $M$. Hence in our higher Kaluza-Klein framework encompassing NS5-branes is as natural as considering the direct higher version of a Kaluza-Klein monopole.
\end{remark}

\begin{remark}[Geometric interpretation of the NS5-brane]
We know that the Kaluza-Klein brane appears when spacetime $P\rightarrow M$ is a non-trivial circle bundle with some first Chern class $[F]\in H^2(M,\mathbb{Z})$. In perfect analogy, the NS5-brane appears when the bundle gerbe $\mathscr{G}\rightarrow M$ underlying the doubled space is a non-trivial bundle gerbe with Dixmier-Douady class $[H]\in H^3(M,\mathbb{Z})$.
\end{remark}

\begin{remark}[Angular T-dual of the NS5-brane]
The $3$-sphere is nothing but the Lens space $L(1,1)=S^3$ corresponding to the Hopf fibration. As we have seen the transverse space of the NS5 brane with $H$-charge $m$ is $\mathbb{R}^+\times S^3$. Let us perform a T-duality along the $\psi_1$ circle fiber
\begin{equation*}
    \begin{aligned}
         \widetilde{g} &=  \eta_{\mu\nu}\mathrm{d}x^\mu\mathrm{d}x^\nu + h(r)\mathrm{d}r^2 + h(r)r^2\!\left(\mathrm{d}\chi^2+\sin^2{2\chi}\,\mathrm{d}\psi_2^2\right) +\frac{1}{h(r)r^2}\bigg(\frac{\mathrm{d}\widetilde{\psi}_1}{2}-\frac{m}{2}\cos{2\chi}\,\mathrm{d}\psi_2\bigg)^{\! 2}, \\
         \widetilde{B} &= -\frac{1}{2}\cos{2\chi}\,\mathrm{d}\widetilde{\psi}_1\wedge\mathrm{d}\psi_2.
    \end{aligned}
\end{equation*}
This is again a supergravity solution, but it is not asymptotically flat: the $\widetilde{\psi}_1$ circle is not Hopf-fibered over the $2$-sphere, but it is be fibered with first Chern number $m$, generally making the whole bundle a Lens space $L(1,m)$. In other words the $H$-charge of the NS5-brane is mapped to a NUT charge $m$ under T-duality. However this background is not actually Taub-NUT, because the harmonic function is $h(r)=1+1/r^2$. 
\end{remark}

\begin{digression}[Recovering angular T-dualities]
The previous is the Plauschinn-Camell solution appearing in \cite{Pla18}. The authors perform angular T-dualities of NS5-brane backgrounds and speculate about implementing them in DFT. In our formulation this is totally natural as long as the angular directions are isometries of the generalised metric $\mathcal{G}$.
\end{digression}

\begin{remark}[Array of higher Kaluza-Klein monopoles]
Since higher Kaluza-Klein monopoles do not interact, we can construct a multi-monopole solution. This will be a bundle gerbe $\mathscr{G}$ on the base base manifold $\mathbb{R}^{1,5}\times\left(\mathbb{R}^4-\{y_p\}\right)$, but non-trivial only on the transverse space $\mathbb{R}^4-\{y_p\}$, which is equipped with the generalised metric
$\mathcal{G}$ given by
\begin{equation*}
    \begin{aligned}
        \mathcal{G} &= \eta_{\mu\nu}\mathrm{d}x^\mu\mathrm{d}x^\nu + \eta^{\mu\nu}\mathrm{d}\widetilde{x}_\mu\mathrm{d}\widetilde{x}_\nu + h(r)\delta_{ij}\mathrm{d}y^i\mathrm{d}y^j + \frac{\delta^{ij}}{h(r)}(\mathrm{d}\widetilde{y}_i+B_{ik}\mathrm{d}y^k)(\mathrm{d}\widetilde{y}_j+B_{jk}\mathrm{d}y^k)
    \end{aligned}
\end{equation*}
satisfying the conditions
\begin{equation}
    H = \star_{\mathbb{R}^4}\mathrm{d}h, \quad h(y)=1+\sum_p\frac{m_p}{|y-y_p|^2},
\end{equation}
where $y_p$ are the positions of the monopoles in the transverse space and $m_p$ are their $H$-charges.
\end{remark}

\subsection{Berman-Rudolph DFT monopole in 9d}
In this subsection we will give a global definition of the usual DFT monopole in our formalism. See \cite{BR14} for its original definition and \cite{Jen11} for seminal work. Then we will show that it is immediately related to our higher Kaluza-Klein monopole.

\begin{definition}[DFT monopole]\label{def:dftm}
The Berman-Rudolph DFT monopole \cite{BR14} is a bundle gerbe $\mathscr{G}$ equipped with generalised metric $\mathcal{G}$ such that 
\begin{itemize}
    \item $\mathscr{G}$ is a bundle gerbe on the base manifold $M=\mathbb{R}^{1,5}\times\left(\mathbb{R}^3-\{0\}\right)\times S^1 \,\simeq\, \mathbb{R}^{1,5}\times\mathbb{R}^+\times S^2\times S^1$ which is non-trivial only on $S^2\times S^1$,
    \item $\mathcal{G}$ is the (global) generalised metric given by
\end{itemize}
\begin{equation}\label{eq:smearedhkkmonopole}
    \begin{aligned}
        \mathcal{G} \,&=\, \eta_{\mu\nu}\mathrm{d}x^\mu\mathrm{d}x^\nu + \eta^{\mu\nu}\mathrm{d}\widetilde{x}_\mu\mathrm{d}\widetilde{x}_\nu + h(r)\delta_{ij}\mathrm{d}y^i\mathrm{d}y^j + \frac{\delta^{ij}}{h(r)}\mathrm{d}\widetilde{y}_i\mathrm{d}\widetilde{y}_j\\
        &\hspace{4cm} +h(r)\mathrm{d}z^{2} + \frac{1}{h(r)}(\mathrm{d}\widetilde{z}+{A}_{k}\mathrm{d}y^k)^{2},
    \end{aligned}
\end{equation}
where $\{x^\mu\}$, $\{z\}$ and $\{y^i\}$ are respectively coordinates for $\mathbb{R}^{1,5}$, $\mathbb{R}^{+}\times S^2$ and $S^1$, where we decomposed the connection by $B_{(\alpha)} = A_{(\alpha)i}\wedge\di z^i$, and where the curvature of the connection and the harmonic function are respectively constraint to
\begin{equation}\label{eq:transversalcond2}
    \mathrm{d}A = \star_{\mathbb{R}^3}\mathrm{d}h, \quad h(r)=1+\frac{m'}{r}.
\end{equation}
\end{definition}

\begin{remark}[Doubled space of the DFT monopole]
The global doubled space of the DFT monopole of definition \ref{def:dftm} is similar, but simpler respect to the one of a general higher Kaluza-Klein monopole. The circle coordinate $z\in\mathbb{R}/\mathbb{Z}$ is trivially fibered over the $2$-sphere, while its dual $\widetilde{z}_{(\alpha)}$ is in general non-trivially fibered by a connection $A_{(\alpha)}$ over it. Hence, by lemma \ref{thm:corrspace}, our bundle gerbe $\mathscr{G}|_{S^3}\twoheadrightarrow S^3$, is Kaluza-Klein reduced to a $\String$-bundle $S^2\longrightarrow\mathbf{B}\String(S^1\times S^1)$, whose underlying $(S^1\times S^1)$-bundle is the following correspondence space:
\begin{equation}
    \begin{tikzcd}[row sep={11ex,between origins}, column sep={11ex,between origins}]
     & & S^1\times L(1,m)\arrow[dr, "\pi"']\arrow[dl, "\widetilde{\pi}"] & & \\
    S^1 \arrow[r, hook] & S^2\times S^1\arrow[dr, "\pi"'] & & L(1,m')\arrow[dl, "\widetilde{\pi}"] & \arrow[l, hook'] S^1\\
    & & S^2. & &
    \end{tikzcd}
\end{equation}
In this case the original connection $\omega_B$ of the bundle gerbe has no $\omega_B^{(2)}$ component, but, on the other hand, the component $\omega_B^{(1)} = \mathrm{d}\widetilde{z}_{(\alpha)} +A_{(\alpha)}$ is exactly the connection of $L(1,m')$ on $S^2$.
\end{remark}

\begin{theorem}[Recovering the DFT monopole]
A Berman-Rudolph DFT monopole (definition \ref{def:dftm}) is a smeared higher Kaluza-Klein monopole (definition \ref{def:hkkm}).
\end{theorem}
\begin{proof}
The transverse space is a trivial circle fibration $(\mathbb{R}^3-\{0\})\times S^1$ with trivial connection $\mathrm{d}z$. As usual can decompose $B_{(\alpha)} = B_{(\alpha)}^{(2)} + B_{(\alpha)}^{(1)} \wedge \mathrm{d}z$, where $B_{(\alpha)}^{(2)}$ and $B_{(\alpha)}^{(1)}$ are respectively a gerbe connection and a circle connection on $\mathbb{R}^3-\{0\}$. Now we can gauge away the component $B_{(\alpha)}^{(2)}$ on $\mathbb{R}^3-\{0\}$ since this space is homotopy equivalent to a $2$-sphere and $H^3(S^2,\mathbb{Z})=0$ implies that any gerbe over $S^2$ is trivial. Thus the only non-trivial contribution to the $H$-flux will come from the connection ${A}_{(\alpha)}:=B_{(\alpha)}^{(1)}$ of the dual circle bundle and the gerbe curvature is $\mathrm{d}B_{(\alpha)} = \mathrm{d}A_{(\alpha)}\wedge\mathrm{d}z$.
\begin{equation}
    h(r,z)=1+\sum_{p\in\mathbb{Z}}\frac{m}{r^2 + (z-2\pi p)^2} \;\,\xrightarrow{\;\;\; r\gg 1 \;\;\;}\;\, 1 + \frac{m'}{r}
\end{equation}
with modified charge $m':=m/2$. Moreover we have $\star_{\mathbb{R}^4}\mathrm{d}h = (\star_{\mathbb{R}^3}\mathrm{d}h)\wedge\mathrm{d}z$ and thus the condition $\mathrm{d}B_{(\alpha)}=\star_{\mathbb{R}^4}\mathrm{d}h$ becomes the equation \eqref{eq:transversalcond2}.
\end{proof}

\begin{remark}[DFT monopole is smeared NS5-brane]
By higher Kaluza-Klein reduction of \eqref{eq:smearedhkkmonopole} to $M=\mathbb{R}^{1,5}\times \mathbb{R}^+\times S^2\times S^1$ we get the metric and gerbe connection
\begin{equation}
    \begin{aligned}
        g &= \eta_{\mu\nu}\mathrm{d}x^\mu\mathrm{d}x^\nu + h(r)\big(\delta_{ij}\mathrm{d}y^i\mathrm{d}y^j+\mathrm{d}z^{2}\big), \\
        B &= {A}_{k}\mathrm{d}y^k\wedge\mathrm{d}z.
    \end{aligned}
\end{equation}
This solution is unsurprisingly the smeared NS5-brane background. Notice the asymptotic geometry is the trivial Lens space $L(1,0)=S^2\times S^1$.
\end{remark}

\begin{remark}[KK5-brane is the T-dual of NS5-brane]
By higher Kaluza-Klein reduction of the generalised metric \eqref{eq:smearedhkkmonopole} to the dual bundle $\widetilde{M}=\mathbb{R}^{1,5}\times\mathbb{R}^+\times L(1,m)$ we get the metric and bundle gerbe connection
\begin{equation}
    \begin{aligned}
        \widetilde{g} &=  \eta_{\mu\nu}\mathrm{d}x^\mu\mathrm{d}x^\nu + h(r)\delta_{ij}\mathrm{d}y^i\mathrm{d}y^j + \frac{1}{h(r)}(\mathrm{d}\widetilde{z}+{A}_{k}\mathrm{d}y^k)^{2} \\
        \widetilde{B} &= 0.
    \end{aligned}
\end{equation}
The transverse space is a Taub-NUT space with asymptotic geometry $L(1,m)$ and it has zero $H$-charge. This solution is exactly the KK5-brane with isometry along the $\widetilde{z}_{(\alpha)}$ circle.
\end{remark}

\begin{digression}[Localisation of DFT monopole in the previous literature]
In \cite{BR14} we have a (local) definition of the DFT monopole and then a bottom-up generalisation to its non-smeared version on $\mathbb{R}^{1,5}\times(\mathbb{R}^3-\{0\})\times S^1$. The winding mode corrections of this process are studied in \cite{Kim13}. The resulting generalised metric is a (locally defined) version of higher Kaluza-Klein monopole on this particular background. In \cite{Ber19} it is argued that, in the case of an torus compactified spacetime we can write higher Dirac monopole in terms of an ordinary Dirac monopole by $H^3(S^2\times S^1,\mathbb{Z}) \cong H^2(S^2,\mathbb{Z})\otimes_\mathbb{Z} H^1(S^1,\mathbb{Z})$. But also that a full DFT monopole should require a geometrisation of the gerbe which is impossible to achieve with just manifolds: higher Kaluza-Klein geometry is hopefully an answer to this.
\end{digression}

\subsection{Berman-Rudolph DFT monopole in 8d}

In this subsection we will take a quick look to a further dimensional reduction of our monopole.

\begin{remark}[Reduction to $8d$ and $5_2^2$-brane]
If we compactify again spacetime to a trivial torus bundle $M=\mathbb{R}^{1,5}\times(\mathbb{R}^2-\{0\})\times T^2$, we can further smear and Kaluza-Klein reduce our higher Kaluza-Klein monopole to recover the zoo of exotic branes by \cite{Bak16}. By explicitly writing the T-dualities along the two directions we have the following commutative diagram
\begin{equation}
    \begin{tikzcd}[row sep=21ex, column sep=21ex]
    \mathrm{NS5}_{12} \arrow[r, leftrightarrow, "\mathcal{T}_1"] \arrow[d, leftrightarrow, "\mathcal{T}_2"] & \mathrm{KK5}^{\;2}_{1} \arrow[d, leftrightarrow, "\mathcal{T}_2"'] \\
    \mathrm{NS5}^1_{\;2} \arrow[r, leftrightarrow, "\mathcal{T}_1"] & (5^2_2)^{12} 
\end{tikzcd}
\end{equation}
where $\mathrm{NS5}_{12}$ is the NS5-brane smeared along both the directions of the $T^2$ fiber, while $\mathrm{NS5}^a_{\;b}$ is the KK-brane with isometry along the $a$-th direction and smeared along the $b$-th direction, while $(5^2_2)^{12}$ is the $5^2_2$-brane with isometry along both the directions. 
\end{remark}

\noindent Let $\{z^a\}$ be coordinates of $T^2$ with $a=1,2$ and $\{y^i\}$ be coordinates of $\mathbb{R}^2-\{0\}$ with $i=3,4$. The generalised correspondence space $K$ of the transverse space will be a non-trivial $T^2$-bundle on $(\mathbb{R}^2-\{0\})\times T^2$ with curvature $\widetilde{F}_a = H^{(1)}_{ab}\wedge\mathrm{d}z^b$. 

\begin{remark}[Geometric interpretation of $5_2^2$-brane]
The NS5-brane is associated to a non-trivial Dixmier-Douady class $[H]\in H^3(\mathbb{R}^4-\{0\},\mathbb{Z})$ on the transverse space. If spacetime is a trivial $T^2$-fibration and the NS5-brane is smeared along these directions, there will be a non-trivial flux compactification $[H^{(1)}_{12}]\in H^1(\mathbb{R}^2-\{0\},\mathbb{Z})\cong\mathbb{Z}$ on the reduced transverse space. A T-duality along the $1$st direction gives a $[F^{(1)1}_{\quad \;\;2}]\in H^1(\mathbb{R}^2-\{0\},\mathbb{Z})$. Indeed notice that $H^{>1}(\mathbb{R}^2-\{0\},\mathbb{Z})=0$ so there are no non-trivial abelian gauge field on the transverse space. Now by performing a T-duality in the $2$nd direction we get a flux $[Q^{(1)12}]\in H^1(\mathbb{R}^2-\{0\},\mathbb{Z})$.
Just like the Kaluza-Klein brane generally appears when spacetime is a circle fibration with non-trivial first Chern class, a $5_2^2$ brane appears when spacetime is $T^2$-fibration with nontrivial $Q$-flux given by the cohomology class $[Q^{(1)12}]\in H^1(\mathbb{R}^2-\{0\},\mathbb{Z})\cong\mathbb{Z}$. Notice that this class corresponds to an integer number which we can name $Q$-charge of the $5^2_2$-brane.
\end{remark}

\begin{remark}[Reduction to $8d$ and $5_2^3$-brane]
Now recall that our spacetime manifold is $M=\mathbb{R}^{1,5}\times(\mathbb{R}^2-\{0\})\times T^2$. Now translations along the two circles of the torus are legit isometries, while translations along any fixed direction in $\langle y^3\rangle\subset\mathbb{R}^2-\{0\}$ are not. As explained by \cite{Bak16}, we can still perform a local $\mathcal{T}_3\in O(10,10)$ transformation along this direction on the patches $\mathbb{R}^{10,10}$ of the doubled space $\mathcal{M}$, even if it will not be a global T-duality. Hence we will recover the usual picture of exotic branes by the diagram
\begin{equation}
   \begin{tikzcd}[row sep=38,column sep=28]
    & \mathrm{NS5}_{123}  \arrow[rr, "\mathcal{T}_1"] \arrow[dd, "\mathcal{T}_2" near start] \arrow[dl, "\mathcal{T}_3"'] & &   \mathrm{KK5}^{1}_{\;23}  \arrow[dd, "\mathcal{T}_2"] \arrow[dl, "\mathcal{T}_3"] \\
    \mathrm{KK5}^{\;\;\,3}_{12} \ar[crossing over, "\mathcal{T}_1" near start]{rr} \arrow[dd, "\mathcal{T}_2"'] & & (5^2_2)^{1\; 3}_{\; 2} \\
      &  \mathrm{KK5}^{\;2}_{1\; 3}  \arrow[rr, "\mathcal{T}_1" near start] \arrow[dl, "\mathcal{T}_3"'] & &  (5^2_2)^{12}_{\;\;\, 3}  \arrow[dl, "\mathcal{T}_3"] \\
    (5^2_2)^{\;23}_{1} \arrow[rr, "\mathcal{T}_1"'] && (5_2^3)^{123} \arrow[from=uu,crossing over, "\mathcal{T}_2" near start]
 \end{tikzcd} 
\end{equation}
where the superscript $3$ and the subscript $3$ this time do not mean isometry/smearing like for $1$ and $2$, but they respectively mean dependence either on the coordinate $y^3$ or its dual $\widetilde{y}_3$.
\end{remark}

\begin{remark}[Geometric interpretation of $5_2^3$-brane]
A geometric interpretation can be adopted also for the hypothetical $5_2^3$-brane, which will correspond to a $T^2$-compactification carrying a non-trivial $R$-flux class $[R\,]\in H^1(\mathbb{R}^2-\{0\},\mathbb{Z})\cong\mathbb{Z}$ which we may call $R$-charge.
\end{remark}
\begin{savequote}[8cm]
\chinese{此兩者同出而異名，}\\
\chinese{同謂之玄。}\\
\chinese{玄之又玄，眾妙之門。}

These two are one in origin, but they differ in name;\\
their unity is said to be the mystery.\\
Mystery of mysteries, the gateway to all understanding.
  \qauthor{--- Laozi, \textit{Tao Te Ching}}
\end{savequote}

\chapter{\label{ch:6}Global T-duality from higher geometry}

\minitoc


\noindent This chapter will be devoted to the study of T-duality covariant string compactifications with non-trivial fluxes. The global moduli stack of such non-trivial compactifications will be derived by Kaluza-Klein reducing the total space of the bundle gerbe we introduced in chapter \ref{ch:5}. This way, we will retain global geometric data on the base manifold, which will give the global geometry of the string compactification. Crucially, we will provide e a unifying framework for defining tensor hierarchies and determining their topology and global properties. This chapter is based on \cite{Alf19,Alf20,Alfonsi:2021uwh}.\vspace{0.2cm}

\noindent In section \ref{td2} and \ref{td3}, from the reduction of the bundle gerbe we will recover the usual global geometry of abelian T-duality developed by \cite{Bou03, Bou03x, Bou03xx, Bou04, Bou08}, both geometric and non-geometric. In addition we will lift such T-duality to the doubled space, i.e. the atlas of the underlying bundle gerbe, and show it provides the familiar picture in a global fashion. Moreover, we will focus on the global geometry of the tensor hierarchies which the reduction provides.
In section \ref{td4} we will generalise these results to the most general case of abelian T-duality. 
In section \ref{td5} we will widen the discussion to non-abelian T-duality. In particular, we will provide a global definition of non-abelian T-fold and we will discuss the topology of the corresponding global tensor hierarchy.
In section \ref{td6} we will generalise the previous results to Poisson-Lie T-duality. Thus, we will provide a global definition of Poisson-Lie T-fold and we will discuss the global geometry of generalised Scherk-Schwarz reductions.

\section{Moduli stack of global string compactifications}

The aim of this first section will be the introduction of the concept of moduli stack of global DFT compactifications, by starting from the idea of flux compactification. \vspace{0.2cm}

\noindent Let us consider a spacetime smooth manifold $M$. We require that the spacetime $M$ is \textit{compactified}, i.e. that it is the total space of a $G$-bundle for some compact Lie group $G$. We have a diagram of the form
\begin{equation}
\begin{tikzcd}[row sep=5.5ex, column sep=scriptsize]
   G \arrow[r,hook, "\iota"] & M \arrow[d,two heads, "\pi"] \\
   & M_0,
\end{tikzcd}
\end{equation}
where $\iota$ is the embedding of the fibre and $\pi$ is the principal projection.
Recall from chapter \ref{ch:3}, that a $G$-bundle can be expressed by the pullback diagram
\begin{equation}
\begin{tikzcd}[row sep=6.5ex, column sep=6ex]
   M \arrow[r]\arrow[d,two heads, "\pi"] & \ast \arrow[d] \\
   M_0 \arrow[r,"g_{(\alpha\beta)}"]& \mathbf{B}G,
\end{tikzcd}
\end{equation}
where $g_{(\alpha\beta)}:M_0\rightarrow \mathbf{B}G$ is the \v{C}ech cocycle which encodes the transition functions of the $G$-bundle $M$. \vspace{0.2cm}

\noindent Now let $\mathscr{S}\in\mathbf{H}$ be a stack encoding some higher structure. Let us consider a cocycle $s:M\rightarrow \mathscr{S}$ on the total space $M$ of our fibration.
As seen in chapter \ref{ch:5}, there is an isomorphism, known as Kaluza-Klein reduction,
\begin{equation}\label{eq:KKreduction6}
\begin{aligned}
    \mathbf{H}(M,\mathscr{S}) \;\,&\cong\,\; \mathbf{H}_{/\mathbf{B}G}\big(M,\, [G,\mathscr{S}]/\!/G\big)\\[0.15cm]
    \left(\begin{tikzcd}[row sep=7ex, column sep=3ex]
    M \arrow[r, "s"] & \mathscr{S}
    \end{tikzcd}\right) \;\,&\mapsto\,\; \left(\begin{tikzcd}[row sep=7ex, column sep=5ex]
    & \left[G,\mathscr{S}\right]\!/\!/G \arrow[d, two heads]\\
    M_0 \arrow[ru, "\hat{s}"]\arrow[r, "g_{(\alpha\beta)}"] & \mathbf{B}G
    \end{tikzcd}\right),  
\end{aligned}
\end{equation}
which sends a cocycle $M\rightarrow \mathscr{S}$ on the total space $M$ to some cocycle $M_0\rightarrow \left[G,\mathscr{S}\right]\!/\!/G$ on the base manifold $M_0$. This resulting cocycle can be interpreted as the Kaluza-Klein reduction of the original higher structure. \vspace{0.2cm}

\noindent We stress that this definition of Kaluza-Klein reduction is global. Let $\mathscr{S}$ be the moduli stack of some field. By reducing a cocycle $M\rightarrow \mathscr{S}$ to a cocycle $M_0\rightarrow \left[G,\mathscr{S}\right]\!/\!/G$, we retain both the global geometric data of the structure $M\rightarrow \mathscr{S}$ and of $M\twoheadrightarrow M_0$ as a generally non-trivial principal bundle on $M_0$.
In more concrete words, we must start from the local sections $s_{(\alpha)}:U_\alpha \times G\longrightarrow \mathscr{S}$ on a local trivialisation of $M$ and reduce them to sections $\hat{s}_{(\alpha)}:U_\alpha \longrightarrow [G,\mathscr{S}]/\!/G$, then glue all the sections together by taking into account the transition functions $g_{(\alpha\beta)}$ of $M$.

\begin{figure}[!ht]\centering
\vspace{0.2cm}
\tikzset {_kaehlo7n5/.code = {\pgfsetadditionalshadetransform{ \pgftransformshift{\pgfpoint{0 bp } { 0 bp }  }  \pgftransformrotate{-74 }  \pgftransformscale{2 }  }}}
\pgfdeclarehorizontalshading{_qxybxumn1}{150bp}{rgb(0bp)=(0.49,0.49,0.49);
rgb(37.5bp)=(0.49,0.49,0.49);
rgb(62.5bp)=(0.05,0.05,0.05);
rgb(100bp)=(0.05,0.05,0.05)}
\tikzset{_a2z2iqf77/.code = {\pgfsetadditionalshadetransform{\pgftransformshift{\pgfpoint{0 bp } { 0 bp }  }  \pgftransformrotate{-74 }  \pgftransformscale{2 } }}}
\pgfdeclarehorizontalshading{_iuisqx6xc} {150bp} {color(0bp)=(transparent!89);
color(37.5bp)=(transparent!89);
color(62.5bp)=(transparent!80);
color(100bp)=(transparent!80) } 
\pgfdeclarefading{_06gwwgem5}{\tikz \fill[shading=_iuisqx6xc,_a2z2iqf77] (0,0) rectangle (50bp,50bp); } 
\tikzset{every picture/.style={line width=0.75pt}} 

\begin{tikzpicture}[x=0.75pt,y=0.75pt,yscale=-1,xscale=1]

\draw (312,69.03) node  {\includegraphics[width=127.5pt,height=98.58pt]{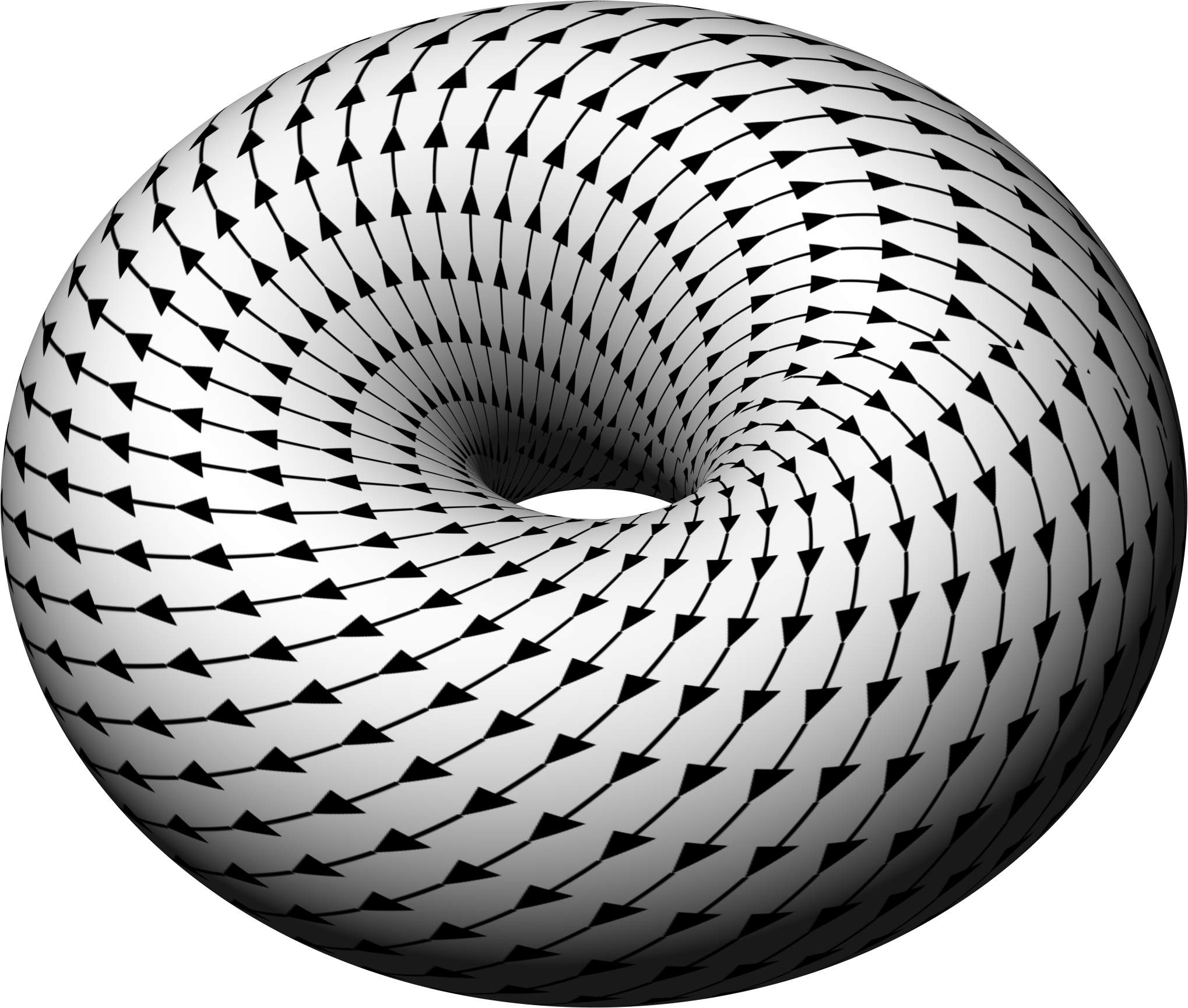}};
\draw (80.13,74.5) node  {\includegraphics[width=139.42pt,height=81.38pt]{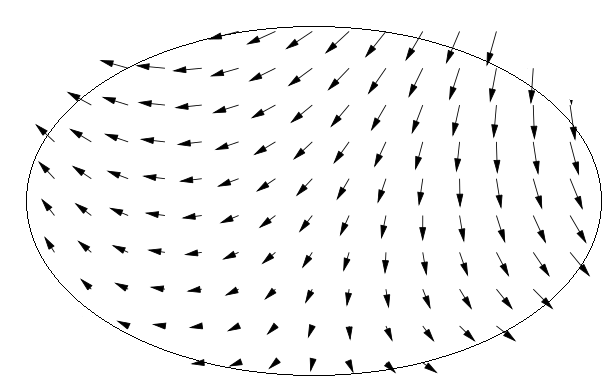}};
\path  [shading=_qxybxumn1,_kaehlo7n5,path fading= _06gwwgem5 ,fading transform={xshift=2}] (-4.82,77.36) .. controls (-4.82,50.12) and (34.58,28.03) .. (83.18,28.03) .. controls (131.78,28.03) and (171.18,50.12) .. (171.18,77.36) .. controls (171.18,104.61) and (131.78,126.69) .. (83.18,126.69) .. controls (34.58,126.69) and (-4.82,104.61) .. (-4.82,77.36) -- cycle ; 
 \draw   (-4.82,77.36) .. controls (-4.82,50.12) and (34.58,28.03) .. (83.18,28.03) .. controls (131.78,28.03) and (171.18,50.12) .. (171.18,77.36) .. controls (171.18,104.61) and (131.78,126.69) .. (83.18,126.69) .. controls (34.58,126.69) and (-4.82,104.61) .. (-4.82,77.36) -- cycle ; 

\draw   (227.62,68.72) .. controls (227.62,32.25) and (265.32,2.69) .. (311.83,2.69) .. controls (358.34,2.69) and (396.04,32.25) .. (396.04,68.72) .. controls (396.04,105.19) and (358.34,134.75) .. (311.83,134.75) .. controls (265.32,134.75) and (227.62,105.19) .. (227.62,68.72) -- cycle ;

\draw (82,144.4) node [anchor=north west][inner sep=0.75pt]  [font=\small]  {$U_{\alpha }$};
\draw (295.5,144.4) node [anchor=north west][inner sep=0.75pt]    {$\;G$};
\draw (189,60.4) node [anchor=north west][inner sep=0.75pt]  [font=\LARGE]  {$\times $};
\end{tikzpicture}\caption[Field on a locally trivialised bundle]{Field (represented as a vector field) on the local trivialisation $U_\alpha\times G$ of a compactified spacetime with non-trivial flux on the fibre.}\end{figure}
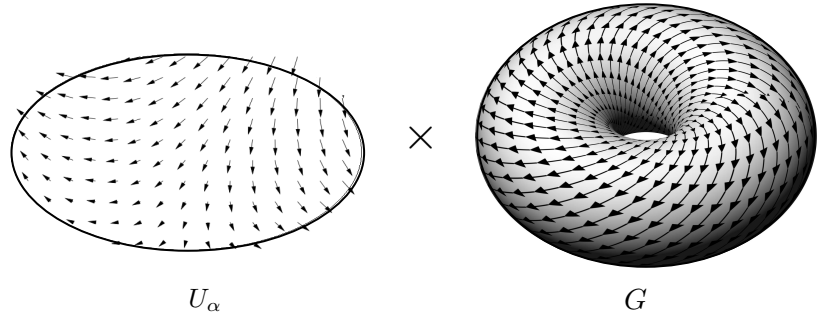

\noindent Let us provide an algebraic topological sense of why taking into account these global data matters. It is relevant to notice that, if $M\twoheadrightarrow M_0$ is a non-trivial $G$-bundle, K\"{u}nneth theorem does not hold, i.e.
\begin{equation}
    H^n(M,\mathbb{Z}) \,\;\neq\, \bigoplus_{i=0}^nH^{n-i}(M_0,\mathbb{Z})\otimes_\mathbb{Z} H^{i}(G,\mathbb{Z}),
\end{equation}
and, thus, the identification of the fluxes, or charges, of the reduced theory on the base manifold $M_0$ becomes complicated, even for abelian fields. Dimensionally reducing the cocycles as in \eqref{eq:KKreduction6} provides us all the global information of the reduced theory on $M_0$.

\subsection{Moduli stack of generalised Scherk-Schwarz reductions}

We are interested in globally-defined generalised Scherk-Schwarz reductions of the field content of Double Field Theory of the following form:
\begin{equation}\label{eq:gssred}
\big\{\underbrace{\mathcal{G}_{MN}}_{\substack{\text{generalised}\\\text{metric}}},\,\underbrace{\varphi}_{\text{dilaton}}\big\} \,\;\xrightarrow[\;\;\text{gSS reduction}\;\;]{\cong}\;\, \big\{\underbrace{g_{\mu\nu}}_{\text{metric}},\,\underbrace{\mathcal{G}_{IJ}}_{\substack{\text{moduli}\\\text{generalised}\\\text{metric}}},\,\underbrace{\mathcal{A}^I_{\mu},\,\mathcal{B}_{\mu\nu}}_{\substack{\text{tensor}\\\text{hierarchy}}}, \underbrace{\varphi}_{\text{dilaton}}\big\}.
\end{equation}
The aim of this chapter will be exactly the global study and the classification of these reductions, which are known as gauged Supergravities \cite{Geissbuhler:2011mx, Geissbuhler:2013uka}. \vspace{0.2cm}

\begin{figure}[!ht]\centering
\tikzset{every picture/.style={line width=0.75pt}} 
\begin{tikzpicture}[x=0.75pt,y=0.75pt,yscale=-1,xscale=1]
\draw    (85,31) -- (85,98.25) ;
\draw [shift={(85,100.25)}, rotate = 270] [color={rgb, 255:red, 0; green, 0; blue, 0 }  ][line width=0.75]    (6.56,-1.97) .. controls (4.17,-0.84) and (1.99,-0.18) .. (0,0) .. controls (1.99,0.18) and (4.17,0.84) .. (6.56,1.97)   ;
\draw    (85,132) -- (85,199.25) ;
\draw [shift={(85,201.25)}, rotate = 270] [color={rgb, 255:red, 0; green, 0; blue, 0 }  ][line width=0.75]    (6.56,-1.97) .. controls (4.17,-0.84) and (1.99,-0.18) .. (0,0) .. controls (1.99,0.18) and (4.17,0.84) .. (6.56,1.97)   ;
\draw    (140,16.5) -- (208.81,60.18) ;
\draw [shift={(210.5,61.25)}, rotate = 212.41] [color={rgb, 255:red, 0; green, 0; blue, 0 }  ][line width=0.75]    (6.56,-1.97) .. controls (4.17,-0.84) and (1.99,-0.18) .. (0,0) .. controls (1.99,0.18) and (4.17,0.84) .. (6.56,1.97)   ;
\draw    (140,217.5) -- (209,217.26) ;
\draw [shift={(211,217.25)}, rotate = 539.8] [color={rgb, 255:red, 0; green, 0; blue, 0 }  ][line width=0.75]    (6.56,-1.97) .. controls (4.17,-0.84) and (1.99,-0.18) .. (0,0) .. controls (1.99,0.18) and (4.17,0.84) .. (6.56,1.97)   ;
\draw    (265,76.5) -- (265,189.75) ;
\draw [shift={(265,191.75)}, rotate = 270] [color={rgb, 255:red, 0; green, 0; blue, 0 }  ][line width=0.75]    (6.56,-1.97) .. controls (4.17,-0.84) and (1.99,-0.18) .. (0,0) .. controls (1.99,0.18) and (4.17,0.84) .. (6.56,1.97)   ;
\draw    (210.5,61.25) -- (141.57,115.51) ;
\draw [shift={(140,116.75)}, rotate = 321.78999999999996] [color={rgb, 255:red, 0; green, 0; blue, 0 }  ][line width=0.75]    (6.56,-1.97) .. controls (4.17,-0.84) and (1.99,-0.18) .. (0,0) .. controls (1.99,0.18) and (4.17,0.84) .. (6.56,1.97)   ;
\draw    (140,116.75) -- (209.85,215.62) ;
\draw [shift={(211,217.25)}, rotate = 234.76] [color={rgb, 255:red, 0; green, 0; blue, 0 }  ][line width=0.75]    (6.56,-1.97) .. controls (4.17,-0.84) and (1.99,-0.18) .. (0,0) .. controls (1.99,0.18) and (4.17,0.84) .. (6.56,1.97)   ;
\draw    (30.5,-0.12) -- (139.5,-0.12) -- (139.5,30.88) -- (30.5,30.88) -- cycle  ;
\draw (85,15.38) node   [align=left] {\begin{minipage}[lt]{71.4pt}\setlength\topsep{0pt}
\begin{center}
String Theory
\end{center}
\end{minipage}};
\draw    (30.5,100.88) -- (139.5,100.88) -- (139.5,131.88) -- (30.5,131.88) -- cycle  ;
\draw (85,116.38) node   [align=left] {\begin{minipage}[lt]{71.4pt}\setlength\topsep{0pt}
\begin{center}
10d Sugra
\end{center}
\end{minipage}};
\draw    (30.5,201.88) -- (139.5,201.88) -- (139.5,232.88) -- (30.5,232.88) -- cycle  ;
\draw (85,217.38) node   [align=left] {\begin{minipage}[lt]{71.4pt}\setlength\topsep{0pt}
\begin{center}
(10$\displaystyle -n$)d Sugra
\end{center}
\end{minipage}};
\draw    (210.5,45.88) -- (319.5,45.88) -- (319.5,76.88) -- (210.5,76.88) -- cycle  ;
\draw (265,61.38) node   [align=left] {\begin{minipage}[lt]{71.4pt}\setlength\topsep{0pt}
\begin{center}
DFT
\end{center}
\end{minipage}};
\draw    (210.5,191.38) -- (319.5,191.38) -- (319.5,245.38) -- (210.5,245.38) -- cycle  ;
\draw (265,218.38) node   [align=left] {\begin{minipage}[lt]{71.4pt}\setlength\topsep{0pt}
\begin{center}
(10$\displaystyle -n$)d gauged Sugra
\end{center}
\end{minipage}};
\draw (172.5,86.5) node [anchor=north west][inner sep=0.75pt]  [font=\scriptsize] [align=left] {\begin{minipage}[lt]{33.681216pt}\setlength\topsep{0pt}
\begin{center}
Strong\\constraint
\end{center}
\end{minipage}};
\draw (46.5,50) node [anchor=north west][inner sep=0.75pt]  [font=\scriptsize] [align=left] {\begin{minipage}[lt]{24.554392000000004pt}\setlength\topsep{0pt}
\begin{center}
Low\\energy
\end{center}
\end{minipage}};
\draw (154.5,215) node [anchor=north west][inner sep=0.75pt]  [font=\scriptsize] [align=left] {\begin{minipage}[lt]{28.135pt}\setlength\topsep{0pt}
\begin{center}
gauging
\end{center}
\end{minipage}};
\draw (260,114.5) node [anchor=north west][inner sep=0.75pt]  [font=\scriptsize] [align=left] {\begin{minipage}[lt]{74.70810800000001pt}\setlength\topsep{0pt}
\begin{center}
generalised\\Scherk-Schwarz\\reduction
\end{center}
\end{minipage}};
\draw (178,148) node [anchor=north west][inner sep=0.75pt]  [font=\scriptsize] [align=left] {\begin{minipage}[lt]{48.048108pt}\setlength\topsep{0pt}
\begin{center}
with fluxes\\and torsion
\end{center}
\end{minipage}};
\draw (7,151.5) node [anchor=north west][inner sep=0.75pt]  [font=\scriptsize] [align=left] {\begin{minipage}[lt]{55.98500000000001pt}\setlength\topsep{0pt}
\begin{center}
without fluxes\\and torsion
\end{center}
\end{minipage}};
\draw (162,-7) node [anchor=north west][inner sep=0.75pt]  [font=\scriptsize] [align=left] {\begin{minipage}[lt]{55.581216000000005pt}\setlength\topsep{0pt}
\begin{center}
Truncation of\\Closed String\\Field Theory
\end{center}
\end{minipage}};
\end{tikzpicture}
\caption[Relation between String Theory, Double Field Theory and gauged Supergravity]{Relation between String Theory, Double Field Theory, Supergravity and gauged Supergravity, where T-duality is generally promoted to a gauge symmetry.}\end{figure}
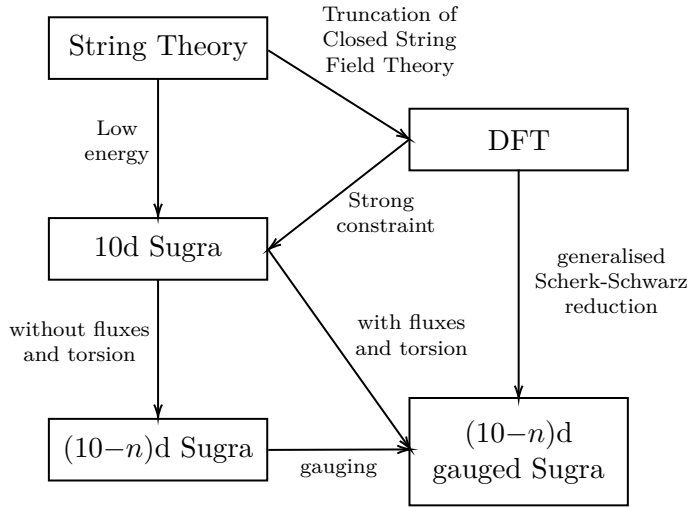

\noindent Notice that, according to chapter \ref{ch:5}, on the left hand side of the reduction we must have a bundle gerbe $\mathscr{G}\twoheadrightarrow M$ on a spacetime $M$ equipped with a generalised metric $\mathscr{G}\xrightarrow{\;\mathcal{G}\;}O(2d)\Struc$. 
Thus, the moduli stack of the global generalised Scherk-Schwarz reductions will be obtained by higher Kaluza-Klein reduction \eqref{eq:KKreduction6} of this higher structure to the base manifold $M_0$ of a $G$-bundle $M\twoheadrightarrow M_0$, which formalises the idea that spacetime is compactified with fibre $G$. \vspace{0.2cm}

\noindent Let us put aside String Theory for a moment and first look at a simpler example of reduction, to gain some intuition about the problem: an electromagnetic field.

\subsection{Toy example: electromagnetic flux compactifications}

Let us consider a spacetime $M$ which is a principal $T^n$-bundle on some $d$-dimensional base manifold $M_0$, i.e.
\begin{equation}
\begin{tikzcd}[row sep=5.5ex, column sep=scriptsize]
    M \arrow[d,two heads, "\pi"'] \arrow[r] & \ast \arrow[d] \\
    M_0 \arrow[r] & \mathbf{B}T^n
\end{tikzcd}
\end{equation}
Let us consider an electromagnetic field on $M$, which is globally given by a principal $U(1)$-bundle $P\twoheadrightarrow M$. Now, we want to study the dimensional reduction of this electromagnetic field, from the total space of the bundle $M$ to the base manifold $M_0$. \vspace{0.2cm}

\noindent In local coordinates of $M$, the operation of dimensional reduction is nothing but the \textit{coordinate split} $\{x^\mu\}_{\mu=1,\dots,d+n} \,\mapsto \{x^\mu, \theta^i\}_{\mu= 1,\dots d}^{i = 1,\dots n}$, where the $\{\theta^i\}^{i = 1,\dots n}$ are the local coordinates of the $T^n$-fibre. Notice that we are not truncating the dependence of the electromagnetic field on the fibre coordinates $\{\theta^i\}$. The stack formalism will allow us to deal with the global geometric picture of such a local coordinates split. \vspace{0.2cm}

\noindent Generally, $P\twoheadrightarrow M_0$ is not a principal $(U(1)\times T^n)$-bundle on the base manifold $M_0$, since the principal $T^n$-action of $M$ cannot generally be lifted to a $T^n$-action on $M$. Therefore, generally, an electromagnetic field $P\twoheadrightarrow M$ is not dimensional reduced to a well-defined electromagnetic field on the base manifold $M_0$. 
In this general case, a dimensional reduction of a principal $U(1)$-bundle $P\twoheadrightarrow M$ with \v{C}ech cocycle  $M\xrightarrow{\;\,g_{(\alpha\beta)}\,\;}\mathbf{B}U(1)$ is given as follows:
\begin{equation}
    \left(\begin{tikzcd}[row sep=7ex, column sep=6ex]
    M \arrow[r, "g_{(\alpha\beta)}"] & \mathbf{B}U(1)
    \end{tikzcd}\right) \;\;\overset{\!\cong}{\longmapsto}\;\; \left(\begin{tikzcd}[row sep=7ex, column sep=5ex]
    & \left[T^n,\mathbf{B}U(1)\right]\!/T^n \arrow[d, two heads]\\
    M_0 \arrow[ru]\arrow[r, ""] & \mathbf{B}T^n
    \end{tikzcd}\right),
\end{equation}
where $\left[T^n,\mathbf{B}U(1)\right]\!/T^n$ is, essentially by definition, the moduli stack of the dimensionally reduced circle bundles. The stack $\left[G,-\right]\!/G$ is a generalization of the \textit{cyclic loop space}: in fact in the particular case where $G=T^n$ and $N$ is a smooth manifold, this reduces to the space $\left[T^n,N\right]\!/T^n \,=\, \Coo(T^n,\,N)/T^n$.\vspace{0.2cm}

\noindent However, there is a particular case where $P$ is a proper principal $(U(1)\times T^n)$-bundle on the base manifold $M_0$. This happens when the $U(1)$-bundle $P\twoheadrightarrow M$ is equivariant under the principal $T^n$-action of $M$, i.e. when the transition functions $g_{(\alpha\beta)}$ of the $U(1)$-bundle are independent from the fibre coordinates $\{\theta^i\}$. In this case, the dimensional reduction has the simple form
\begin{equation}
    \Big(\; T^n\text{-equivariant } U(1)\text{-bundle on }M \;\Big) \;\;\overset{\cong}{\longmapsto}\;\; \Big(\; (U(1)\times{T}^n)\text{-bundle on }M_0 \;\Big).
\end{equation}
In other words, if we consider a $T^n$-equivariant cocycle $f_{(\alpha\beta)}^{\mathrm{eqv}}:M \rightarrow\mathbf{B}U(1)$, this will be dimensionally reduced as follows
\begin{equation}
    \left(\begin{tikzcd}[row sep=7ex, column sep=6ex]
    M \arrow[r, "g_{(\alpha\beta)}^{\mathrm{eqv}}"] & \mathbf{B}U(1)
    \end{tikzcd}\right) \;\;\overset{\!\cong}{\longmapsto}\;\; \left(\begin{tikzcd}[row sep=7ex, column sep=5ex]
    & \mathbf{B}T^{n+1} \arrow[d, two heads]\\
    M_0 \arrow[ru]\arrow[r, ""] & \mathbf{B}T^n
    \end{tikzcd}\right).
\end{equation}
Thus, in this particular case, the electromagnetic field $P\twoheadrightarrow M$ is dimensionally reduced to a globally well-defined  electromagnetic field on the base manifold $M_0$.

\subsection{Moduli stack of Kalb-Ramond flux compactifications}

\noindent Let us now generalise the previous subsection by investigating the Kaluza-Klein reduction the bundle gerbe underlying the Kalb-Ramond flux. \vspace{0.2cm}

\noindent Let us consider a bundle gerbe $\mathscr{G}\xrightarrow{\;\Pi\;} M$ defined by a cocycle $M\xrightarrow{G_{(\alpha\beta\gamma)}}\mathbf{B}^2U(1)$, where spacetime $M$ is itself a principal $T^n$-bundle on a smooth base manifold ${M_0}$, i.e.
\begin{equation}
\begin{tikzcd}[row sep=5.5ex, column sep=scriptsize]
    M \arrow[d,two heads, "\pi"'] \arrow[r] & \ast \arrow[d] \\
    M_0 \arrow[r] & \mathbf{B}T^n
\end{tikzcd}
\end{equation}
\noindent Hence, we need to look at the following Kaluza-Klein reduction:
\begin{equation}
    \left(\begin{tikzcd}[row sep=9ex, column sep=7ex]
    M \arrow[r, "G_{(\alpha\beta\gamma)}"] & \mathbf{B}^2U(1)
    \end{tikzcd}\right) \;\;\overset{\!\cong}{\longmapsto}\;\; \left(\begin{tikzcd}[row sep=9ex, column sep=5ex]
    & {[T^n,\mathbf{B}^2U(1)]/T^n} \arrow[d, two heads, ""]\\
    M_0 \arrow[ru, "{?}"]\arrow[r] & \mathbf{B}T^n
    \end{tikzcd}\right),
\end{equation}
Since $M$ is a principal $T^n$-bundle over $M_0$, we can choose a good cover $\mathcal{V}=\{V_\alpha\}$ for $M$ such that $\mathcal{U}=\{U_\alpha\}$ with $U_\alpha=\pi(V_\alpha)$ is a good cover for the base $M_0$. Since we are not working with a good cover for $M$, we will consider differential forms which are allowed to have integral periods.
We can use definition \ref{def:dimred} to Higher Kaluza-Klein reduce the doubled space to the base $M_0$ by
\begin{equation}
    \begin{aligned}
    \mathbf{H}\big(M,\,\mathbf{B}^2U(1)_{\mathrm{conn}}\big) \;&\cong\; \mathbf{H}\big(M_0,\,\big[T^n,\,\mathbf{B}^2U(1)_{\mathrm{conn}}\big]/T^n\big) \\[0.2cm]
    \Big(\bigsp{B}_{(\alpha)},\,\bigsp{\Lambda}_{(\alpha\beta)},\,\bigsp{G}_{(\alpha\beta\gamma)}\Big) \;&\mapsto\; \Big(B_{(\alpha)}^{(2)},\,B_{(\alpha)}^{(1)},\,B_{(\alpha)}^{(0)},\,\Lambda^{(1)}_{(\alpha\beta)},\,\Lambda^{(0)}_{(\alpha\beta)},\,G_{(\alpha\beta\gamma)}\Big).
    \end{aligned}
\end{equation}
We can split the curvature $\bigsp{H}\in\Omega^3_{\mathrm{cl}}(M)$ of the doubled space in horizontal and vertical components by
\begin{equation}\label{splitH}
    \bigsp{H} \;=\; H^{(3)} + H^{(2)}_i\wedge \xi^i + \frac{1}{2} H^{(1)}_{ij}\wedge\xi^i\wedge\xi^j +\frac{1}{3!} H^{(0)}_{ijk}\xi^i\wedge\xi^j\wedge\xi^k
\end{equation}
 where $H^{(k)}$ are globally defined $\wedge^{3-k}\mathbb{R}^n$-valued differential $k$-forms on the base manifold $M_0$. Now on patches $V_\alpha$ and overlaps $V_\alpha \cap V_\beta$ of $M$ we can use the connection of the torus bundle to split the connections of the gerbe in a horizontal and vertical part
\begin{equation}
\begin{aligned}
    \bigsp{B}_{(\alpha)} \;&=\; B_{(\alpha)}^{(2)} +  B_{(\alpha)i}^{(1)}\wedge \xi^i + \frac{1}{2} B_{(\alpha)ij}^{(0)} \xi^i\wedge\xi^j \\
    \bigsp{\Lambda}_{(\alpha\beta)} \;&=\; \Lambda_{(\alpha\beta)}^{(1)} + \Lambda_{(\alpha\beta)i}^{(0)}\xi^i
\end{aligned}
\end{equation}
where $\bigsp{B}^{(k)}_{(\alpha)}$, and $\bigsp{\Lambda}^{(k)}_{(\alpha\beta)}$ are all local horizontal differential forms on spacetime $M$. For the following calculations we will follow the ones in \cite{BelHulMin07}. The expression of the curvature of the doubled space becomes
\begin{equation}
    \begin{aligned}
    H^{(0)}_{ijk} \;&=\; \partial_{[i}B^{(0)}_{(\alpha)jk]} \\
    H^{(1)}_{ij} \;&=\; {\di}B^{(0)}_{(\alpha)ij} - \partial_{[i}B^{(1)}_{(\alpha)j]} \\
    H^{(2)}_i \;&=\; {\di}B^{(1)}_{(\alpha)i} + \mathcal{L}_{\partial_i}B^{(2)}_{(\alpha)} - B^{(0)}_{(\alpha)ij} F^j \\
    H^{(3)} \;&=\; {\di}B_{(\alpha)}^{(2)} -  B^{(1)}_{(\alpha)i} F^i
    \end{aligned}
\end{equation}
where ${\di}:\Omega^p(M_0)\rightarrow\Omega^{p+1}(M_0)$ is the exterior derivative on the base manifold $M_0$. The patching conditions of the connection $2$-form now become 
\begin{equation}
    \begin{aligned}
    B_{(\beta)ij}^{(0)} - B_{(\alpha)ij}^{(0)} \;&=\; \partial_{[i}\Lambda^{(0)}_{(\alpha\beta)j]} \\
    B_{(\beta)i}^{(1)} - B_{(\alpha)i}^{(1)} \;&=\; {\di}\Lambda^{(0)}_{(\alpha\beta)i} - \mathcal{L}_{\partial_i}\Lambda^{(1)}_{(\alpha\beta)} \\
    B_{(\beta)}^{(2)} - B_{(\alpha)}^{(2)} \;&=\; {\di}\Lambda^{(1)}_{(\alpha\beta)} +  \Lambda^{(0)}_{(\alpha\beta)i} F^i
    \end{aligned}
\end{equation}
where $F=\bigsp{\di}\xi\in\Omega^2(M_0)$ is the curvature of the torus bundle $M\xrightarrow{\pi}M_0$. Finally on three-fold overlaps we get
\begin{equation}
    \begin{aligned}
       \Lambda_{(\alpha\beta)i}^{(0)} + \Lambda_{(\beta\gamma)i}^{(0)} + \Lambda_{(\gamma\alpha) i}^{(0)} \;&=\; \frac{\partial}{\partial\theta^i}G_{(\alpha\beta\gamma)} \\
    \Lambda_{(\alpha\beta)}^{(1)} + \Lambda_{(\beta\gamma)}^{(1)} + \Lambda_{(\gamma\alpha)}^{(1)} \;&=\; {\di}G_{(\alpha\beta\gamma)}
    \end{aligned}
\end{equation}
where $\theta_{(\alpha)}^i$ are the adapted coordinates on the torus fibre and $\partial_i := \partial/\partial\theta^i$.

\noindent Let us now focus on the most interesting part of the field content of Double Field Theory: the tensor hierarchy.

\subsection{Moduli stack of global tensor hierarchies}

In this subsection we specialise our discussion to the tensor hierarchy and we show that our definition by the reduction \eqref{eq:gssred} is more general than a higher gauge theory. This global definition will allow us to identify the field content of a T-fold (which is not a higher gauge theory) with a global tensor hierarchy, as we will see later. \vspace{0.2cm}

\noindent Recall that in literature tensor hierarchies are usually defined as higher gauge theories, as we have reviewed in chapter \ref{ch:4}. 

\paragraph{Motivations for a slightly more general definition.}
\begin{itemize}
    \item In chapter \ref{ch:4}, we reviewed the definition of tensor hierarchy as a higher gauge theory. This definition can be immediately globalised, since higher gauge theories are globally well-defined. Moreover, the fields of a higher gauge theories can be identified with the connection of some (non-abelian) bundle gerbe.
    This would suggest that any global tensor hierarchy is geometrised by a (non-abelian) bundle gerbe structure.
    However, in \cite{HohSam13KK}, tensor hierarchies are introduced as the result of a general dimensional reduction (i.e. split of the coordinates) of a doubled space. 
    As we have seen, the dimensional reduction of a geometric structure does not give, in general, something as nice and regular as a globally well-defined bundle, but the split coordinates will be generally glued by monodromies.
    Therefore, generally, the dimensional reduction of the doubled space will not lead to a globally well-defined higher gauge theory.
    \item The doubled space encoding a T-fold can be obtained by dimensionally reducing a doubled space \cite{BelHulMin07}. Therefore, by following the idea by \cite{HohSam13KK}, the doubled space underlying the T-fold should be considered a global tensor hierarchy.
    However, such doubled spaces cannot be obtained by gauging the local tensor hierarchy algebra of chapter \ref{ch:4}, i.e. the local prestack $\Omega\big(U,\mathscr{D}(\mathbb{R}^{2n})\big)$, because the fields are patched also by cocycles of monodromies. Therefore, they are not globally given by a higher gauge theory.
\end{itemize}
These two points suggest that, even if the picture of tensor hierarchy as higher gauge theory holds locally, it could be not the most general global picture. \vspace{0.25cm}

\noindent Let us recall that tensor hierarchies require the strong constraint to be well-defined. We can thus replace the C-bracket with the anti-symmetrised Roytenberg bracket of Generalised Geometry, which is the direct generalisation of the Courant bracket where all the non-geometric fluxes are allowed. We can then solve the strong constraint and obtain locally the curvature as $\mathscr{D}_{\mathrm{sc}}(\mathbb{R}^{2n})$-valued differential forms (see chapter \ref{ch:4} for details)
\begin{equation}
    \begin{aligned}
    \mathcal{F} \;&=\; \di \mathcal{A} - [ \mathcal{A} \,\overset{\wedge}{,}\, \mathcal{A}]_{\mathrm{Roy}} + \mathfrak{D}\mathcal{B}, \\[0.2em]
    \mathcal{H} \;&=\;  \mathrm{D}\mathcal{B} + \frac{1}{2}\langle \mathcal{A} \,\overset{\wedge}{,}\, \di \mathcal{A}\rangle - \frac{1}{3!} \langle \mathcal{A} \,\overset{\wedge}{,}\,   [ \mathcal{A} \,\overset{\wedge}{,}\, \mathcal{A}]_{\mathrm{Roy}} \rangle ,
    \end{aligned}
\end{equation}
where now the bracket $[ -,-]_{\mathrm{Roy}}$ is the anti-symmetrised Roytenberg bracket. In coordinates this corresponds to setting $\widetilde{\partial}^i(-)=0$ on any field so we will also have $(\mathfrak{D}\mathcal{B})_i = \partial_i\mathcal{B}$, which implies $\mathrm{D}\mathcal{B} \;=\; \di \mathcal{B} + \mathcal{A}^i\wedge \partial_i \mathcal{B}$. Analogously for all the others $\mathscr{D}_{\mathrm{sc}}(\mathbb{R}^{2n})$-valued differential forms.

\paragraph{Global tensor hierarchies.}
The definition of global tensor hierarchy which we can extract from \eqref{eq:KKreduction6} is the following. 
\textit{A global DFT tensor hierarchy on $M_0$ is the result of the dimensional reduction of the connection of a bundle gerbe on a principal $G$-bundle $M\twoheadrightarrow M_0$.} Therefore the $2$-groupoid $\tenhie^{\,G}_{\!\mathrm{sc}}(M_0)$ of global DFT tensor hierarchies on $M_0$ is given as follows:
\begin{equation}\label{eq:globtenhie}
    \tenhie^{\,G}_{\!\mathrm{sc}}(M_0) \;:=\; \left\{\begin{tikzcd}[row sep=7ex, column sep=5ex]
    & \left[G,\mathbf{B}^2U(1)\right]\!/G \arrow[d]\\
    M_0 \arrow[ru]\arrow[r, ""] & \mathbf{B}G
    \end{tikzcd}\right\}.
\end{equation}
where $\left[G,\mathbf{B}^2U(1)\right]\!/G$ is the moduli stack of dimensionally reduced bundle gerbes we already used in this chapter. \vspace{0.2cm}

\noindent From the proposed definition of global tensor hierarchy \eqref{eq:globtenhie} and from the definition of Kaluza-Klein reduction \eqref{eq:KKreduction6} we immediately have the natural isomorphism of $2$-groupoids
\begin{equation}
    \bigsqcup_{\begin{subarray}{c}M\text{ s.t. }
    M\twoheadrightarrow M_0\\\text{is a }G\text{-bundle}\end{subarray}} \!\!\!\!\mathbf{B}U(1)\mathrm{Bund}(M) \;\;\cong\;\; \tenhie^{\,G}_{\!\mathrm{sc}}(M_0). 
\end{equation}
In other words, this means that a global strong constrained tensor hierarchy on $M_0$ is identified with a bundle gerbe on some $G$-bundle $M\twoheadrightarrow M_0$. 

\section{Topology of a bundle gerbe on a $T^n$-bundle}

As seen in chapter \ref{ch:3}, bundle gerbes $\mathscr{G}$ over a base manifold $M$ are topologically classified by their Dixmier-Douady class $\di\di(\mathscr{G})\in H^3(M,\mathbb{Z})$, which physically encodes the $H$-flux. In this section we will follow \cite{Bou04} to introduce the machinery which we will need to describe the topological data of a bundle gerbe on a torus bundle. This will be particularly useful in dealing with abelian T-duality. Let us start from the trivial example.

\begin{example}[Bundle gerbe on a trivial torus bundle]
For trivial torus bundles $M=M_0\times T^n$ one can just use K\"unneth theorem to rewrite the $3$rd cohomology group as
\begin{equation}\label{eq:kunneth}
    H^3(M_0\times T^n,\mathbb{Z}) \;\;\cong\; \bigoplus_{k=0}^3H^{3-k}(M_0,\mathbb{Z})\otimes_\mathbb{Z} H^{k}(T^n,\mathbb{Z})
\end{equation}
The cohomology ring of the torus is $H^\ast(T^n,\mathbb{Z})\cong\mathbb{Z}[u_1,\dots,u_n]/\langle u^2_1,\dots,u^2_n\rangle$ where $u_i$ are generators of $H^1(S^1,\mathbb{Z})$. Hence we have that $u_i:=[\mathrm{d}\theta^i]$ for $i=1,\dots,n$ are the generators of $H^1(T^n,\mathbb{Z})$, while the cup products $[\mathrm{d}\theta^i]\smile[\mathrm{d}\theta^j]=[\mathrm{d}\theta^i\wedge\mathrm{d}\theta^j]$ are the generators of $H^2(T^n,\mathbb{Z})$ and so on. Therefore we can expand the Dixmier-Douady class as
\begin{equation*}\label{eq:topgerbe}
    [H] = [H^{(3)}] + \, \langle [H^{(2)}]\smile[\mathrm{d}\theta] \rangle \, + \, \frac{1}{2}\langle[H^{(1)}]\smile[\mathrm{d}\theta\wedge\mathrm{d}\theta]\rangle \, + \, \frac{1}{3!}\langle[H^{(0)}]\smile[\mathrm{d}\theta\wedge\mathrm{d}\theta\wedge\mathrm{d}\theta]\rangle,
\end{equation*}
where $H^{(k)}\in\Omega^k(M_0,\wedge^{3-k}\mathbb{R}^n)$ are differential forms on the base manifold $M_0$ and $\langle-,-\rangle$ is just the contraction of all the indices of the exterior  algebra $\wedge\,\mathbb{R}^n$. Hence the topology of this doubled space is encoded by the cohomology classes $[H^{(k)}]\in H^k(M_0,\wedge^{3-k}\mathbb{Z}^n)$ with $k=0,1,2,3$.
\end{example}

\noindent However this construction cannot be immediately extended to non-trivial torus bundles.

\begin{definition}[$T^n$-invariant forms]
We define the \textit{$T^n$-invariant forms} $\Omega^{p}(M)^{T^n}$ as the subset of $1$-forms $\alpha\in\Omega^{p}(M)$ such that $\mathcal{L}_{\partial_i}\alpha = 0$, with $\partial_i := \partial/\partial\theta^i$ in adapted coordinates $\{\theta^i\}$. Let us define the differential forms
\begin{equation}
    \Omega^{p,q}(M_0,\wedge\,\mathbb{R}^n) \;:=\; \bigoplus_{k=0}^{p} \Omega^{k}(M_0,\,\wedge^{p+q-k}\mathbb{R}^n)
\end{equation}
Let $\xi\in\Omega^1(M,\mathbb{R}^n)$ be the connection $1$-form of the torus bundle $M\xrightarrow{\pi}M_0$. There is a natural isomorphism $\Omega^{p}(M)^{T^n} \cong \Omega^{p,0}(M_0,\wedge\,\mathbb{R}^n)$ given by
\begin{equation}\label{eq:isomorphism}
    \mathcal{I}:\,\, \alpha = \sum_{k=0}^p\frac{1}{(p-k)!}\langle\alpha^{(k)}\,\overset{\wedge}{,}\, \xi\wedge\cdots\wedge\xi\rangle \;\,\,\mapsto\;\,\, \big(\alpha^{(p)},\,\alpha^{(p-1)},\,\cdots,\,\alpha^{(0)}\big) 
\end{equation}
and we call $D := \mathcal{I}\circ \mathrm{d} \circ \mathcal{I}^{-1}$ the differential under the isomorphism. The sequence $\Omega^{\bullet,q}(M_0,\wedge\,\mathbb{R}^n)$ for a fixed $q$ equipped with differential $D$ defines a cochain complex, whose integer cohomology we call $H^{\bullet,q}_D(M_0,\wedge\,\mathbb{Z}^n)$. Therefore we have the isomorphism  $H^p(M,\mathbb{Z})^{T^n} \cong H^{p,0}_D(M_0,\wedge\,\mathbb{Z}^n)$.
\end{definition}

\noindent It is also possible to prove that there is isomorphism $\Omega^{2}(M,\mathbb{R}^n)^{T^n} \cong \Omega^{2,1}(M_0,\wedge\,\mathbb{R}^n)$ for $T^n$-invariant $\mathbb{R}^n$-valued $2$-forms, which imply $H^{2}(M,\mathbb{Z}^n)^{T^n} \cong H^{2,1}_D(M_0,\wedge\,\mathbb{Z}^n)$.

\begin{theorem}[$T^n$-invariant representatives]
Any element of the cohomology group $H^p(M,\mathbb{Z})$ of a $T^n$-bundle $M\xrightarrow{\pi}M_0$ can be represented by a closed $T^n$-invariant form, i.e. there is an isomorphism $H^p(M,\mathbb{Z})\cong H^p(M,\mathbb{Z})^{T^n}$.
\end{theorem}

\begin{remark}[Dimensionally reduced Gysin sequence]
Any torus bundle $M\xrightarrow{\pi}M_0$ comes with a long exact sequence, which is called \textit{dimensionally reduced Gysin sequence}. This is given by
\begin{equation*}
    \cdots \rightarrow{} H^p(M_0,\mathbb{Z}) \xrightarrow{\pi^\ast} H^{p,0}_D(M_0,\wedge\,\mathbb{Z}^n) \xrightarrow{\pi_\ast} H^{p-1,1}_D(M_0,\wedge\,\mathbb{Z}^n) \xrightarrow{\langle\,-\,\smile\,[F]\,\rangle} H^{p+1}(M_0,\mathbb{Z}) \rightarrow{} \cdots
\end{equation*}
where $[F]$ is the first Chern class of the bundle, while $\pi^\ast$ on a given representative $\alpha$ is just the injection $\alpha \mapsto (\alpha,0,\cdots,0)$, while $\pi_\ast$ on a given representative $\alpha$ is the integration along each circle of the fibre $S^1_i\subset T^n$, which depends only on the homology class $[S^1_i]\in H_1(M,\mathbb{Z})$ of the circle and not on its particular representative. Hence map $\pi_\ast$ will be given on the representative by $(\alpha^{(p)},\alpha^{(p-1)},\cdots,\alpha^{(0)}) \mapsto (\alpha^{(p-1)},\cdots,\alpha^{(0)})$. 
\end{remark}

\noindent The Dixmier-Douady class $[\bigsp{H}]\in H^3(M,\mathbb{Z})$ of a bundle gerbe on $M\xrightarrow{\pi}M_0$ then corresponds to a cohomology class $[(H^{(3)},H^{(2)},H^{(1)},H^{(0)})]\in H^{3,0}_D(M_0,\wedge\,\mathbb{Z}^n)$ given by
\begin{equation}
    \bigsp{H} \;=\; H^{(3)} +  H^{(2)}_{i}\wedge\xi^i + \frac{1}{2} H^{(1)}_{ij}\wedge\xi^i\wedge\xi^j + \frac{1}{6} H^{(0)}_{ijk}\xi^i\wedge\xi^j\wedge\xi^k.
\end{equation}
In the following discussion, for clarity, we will always underline the differential forms on the total space $M$ of the bundle.

\begin{remark}[Closedness]
If $\bigsp{H}\in\Omega^3(M)^{T^n}$ is closed on $M$ we have $\bigsp{\di}\bigsp{H}=0$, which is translated under isomorphism \eqref{eq:isomorphism} to $D(H^{(3)},H^{(2)},H^{(1)},H^{(0)})=0$. Hence we get the equations
\begin{equation}\label{eq:closeness}
    {\di}H^{(p)} +\langle H^{(p-1)}\,\overset{\wedge}{,}\, F \rangle =0,
\end{equation}
where ${\di}$ is the differential on the base $M_0$ and $F\in\Omega^2(M_0,\mathbb{R}^n)$ is the curvature of the $T^n$-bundle $M\xrightarrow{\pi}M_0$. Notice we recover the trivial case \eqref{eq:kunneth} for $F=0$. Also notice that $H^{(0)}$ is always closed on $M_0$, while $H^{(1)}$ is closed on $M_0$ either if the torus bundle is trivial or if $H^{(0)}=0$.
\end{remark}

\begin{remark}[Exactness]
If $\bigsp{H}\in\Omega^3(U)^{T^n}$ is exact on $U\subset M$, there exists a $2$-form $\bigsp{B}\in\Omega^2(U)$ such that $\bigsp{H}=\bigsp{\mathrm{d}}\bigsp{B}$ on $U$, which is translated under isomorphism \eqref{eq:isomorphism} to
\begin{equation}\label{eq:exactness}
    \begin{aligned}
    H^{(0)}_{ijk} \;&=\; \mathcal{L}_{\partial_{[i}}B^{(0)}_{jk]}, \\
    H^{(1)}_{ij} \;&=\; {\di}B^{(0)}_{ij} - \mathcal{L}_{\partial_{[i}}B^{(1)}_{j]}, \\
    H^{(2)}_i \;&=\; {\di}B^{(1)}_i + \mathcal{L}_{\partial_{i}}B^{(2)} - B^{(0)ij} F^j, \\
    H^{(3)} \;&=\; {\di}B^{(2)} - B^{(1)i}\wedge F^i,
    \end{aligned}
\end{equation}
Notice that we did not require $B$ to be $T^n$-invariant here. This will be useful later.
\end{remark}

\section{Geometrisation of T-duality}\label{td2}

It was understood by \cite{BelHulMin07} that a gerbe structure over a principal torus $T^n$-bundle, if it is equivariant under its principal torus action, automatically defines a principal $T^{2n}$-bundle over its base manifold. This bundle is nothing but the \textit{correspondence space} of a T-duality, also known as \textit{doubled torus bundle} in DFT literature. In this section we will explain how the correspondence space can be recovered from our doubled space by higher Kaluza-Klein reduction and how T-duality is naturally encoded.

\subsection{Correspondence space}

\begin{theorem}[Topological T-duality]\label{thm:corrspace}
Let $\mathscr{G}\xtwoheadrightarrow{\Pi}M$ be a bundle gerbe whose base manifold is itself a principal $T^n$-bundle $M\xtwoheadrightarrow{\pi}M_0$. 
Now, if the bundle gerbe $\mathscr{G}\xrightarrow{\bbpi}M$ is equivariant under the principal $T^n$-action of $M$, then it is Kaluza-Klein reduced to a $\mathrm{String}(T^n\times \widetilde{T}^n)$-bundle on $M_0$ by
    \begin{equation}\label{eq:emcorrspace}
    \left(\begin{tikzcd}[row sep=9ex, column sep=7ex]
    M \arrow[r, "G_{(\alpha\beta\gamma)}"] & \mathbf{B}^2U(1)
    \end{tikzcd}\right) \;\;\overset{\!\cong}{\longmapsto}\;\; \left(\begin{tikzcd}[row sep=9ex, column sep=5ex]
    & \mathbf{B}\String(T^n\times\widetilde{T}^n) \arrow[d, two heads, ""]\\
    M_0 \arrow[ru, ""]\arrow[r, "f_{(\alpha\beta)}"] & \mathbf{B}T^n
    \end{tikzcd}\right),
\end{equation}
where the Lie $2$-group $\mathrm{String}(T^n\times \widetilde{T}^n)$ is defined by the following fibration
\begin{equation}
        \begin{tikzcd}[row sep=12ex, column sep=10ex]
        \mathbf{B}\mathrm{String}(T^n\times \widetilde{T}^n) \arrow[r]\arrow[d, "\mathrm{hofib}\left({\langle\,\mathrm{c}_1\,\smile\,{\mathrm{c}}_1\,\rangle}\right)"']&\ast\arrow[d] \\ 
        \mathbf{B}(T^n\times \widetilde{T}^n) \arrow[r, "{\langle\,\mathrm{c}_1\,\smile\,{\mathrm{c}}_1\,\rangle}"] & \mathbf{B}^3U(1),
        \end{tikzcd}
\end{equation}
where ${\langle\,\mathrm{c}_1\smile{\mathrm{c}}_1\,\rangle}:\mathbf{B}(T^n\times \widetilde{T}^n)\rightarrow\mathbf{B}^3U(1)$ is the map which sends a $(T^n\times \widetilde{T}^n)$-bundle with curvature $(F^i,\,\widetilde{F}_i)$ to the bundle $2$-gerbe with curvature $\widetilde{F}_i\wedge F^i$.

\noindent The principal $(T^{n}\times\widetilde{T}^n)$-bundle $K := M\times_{M_0}\widetilde{M}$ on the base manifold $M_0$ underlying the $\mathrm{String}(T^n\times \widetilde{T}^n)$-bundle \eqref{eq:emcorrspace} has first Chern classes $\mathrm{c}_1(M)=[F]$ and ${\mathrm{c}}_1(\widetilde{M})=[\pi_\ast H]$ and is known in the literature as the \textit{correspondence space} of a couple of T-dual spacetimes $M$ and $\widetilde{M}$. 
We have the diagram:
\begin{equation}\label{diag:corrspace}
\begin{tikzcd}[row sep={9ex,between origins}, column sep={10ex,between origins}]
 & & M\times_{M_0}\widetilde{M} \arrow[dl, "1\otimes\widetilde{\pi}"', two heads]\arrow[dr, "\pi\otimes 1", two heads] & & \\
 & M \arrow[dr, "\pi"', two heads] & & \widetilde{M}\arrow[dl, "\widetilde{\pi}", two heads] &   \\
 & & M_0 & &
\end{tikzcd}
\end{equation}
\end{theorem}

\begin{proof}
By using the connection $\xi\in\Omega^1(M,\mathbb{R}^n)$ of the torus bundle $M\xtwoheadrightarrow{\pi}M_0$, we can split the gerbe connection in horizontal and vertical components
\begin{equation}
    \begin{aligned}
        \bigsp{B}_{(\alpha)} \;&=\; \pi^\ast B_{(\alpha)}^{(2)} + \pi^\ast B_{(\alpha)}^{(1)i}\wedge \xi^i  + \frac{1}{2}  
        \pi^\ast B_{(\alpha)}^{(0)ij} \xi^i\wedge\xi^j, 
    \end{aligned}
\end{equation}
where $B^{(k)}_{(\alpha)}$ are local $k$-forms on patches $U_\alpha$. We can do the same for the \v{C}ech cocycle $(\bigsp{\Lambda}_{(\alpha\beta)},\bigsp{G}_{(\alpha\beta\gamma)})$ of the bundle gerbe
\begin{equation}
    \begin{aligned}
        \bigsp{\Lambda}_{(\alpha\beta)} \;&=\; \pi^\ast \lambda_{(\alpha\beta)} + \pi^\ast \widetilde{f}_{(\alpha\beta)i} \xi^i  \\
        \bigsp{G}_{(\alpha\beta\gamma)} \;&=\; \pi^\ast g_{(\alpha\beta\gamma)}
    \end{aligned}
\end{equation}
where $\lambda_{(\alpha\beta)}$ is a local $1$-form on two-fold overlaps of patches $U_\alpha\cap U_\beta$ and $\widetilde{f}_{(\alpha\beta)}$ and $g_{(\alpha\beta\gamma)}$ are local functions respectively on two-fold and three-fold overlaps of patches. The patching condition $\bigsp{B}_{(\beta)}-\bigsp{B}_{(\alpha)} = \bigsp{\mathrm{d}}\bigsp{\Lambda}_{(\alpha\beta)}$ on two-fold overlaps of patches becomes
\begin{equation}\label{eq:nonobpatched}
    \begin{aligned}
    B_{(\beta)ij}^{(0)} - B_{(\alpha)ij}^{(0)} &= \,0 \\
    B_{(\beta)i}^{(1)} \,-  B_{(\alpha)i}^{(1)} \,\, &= \,{\di}\widetilde{f}_{(\alpha\beta)i} \\
    B_{(\beta)}^{(2)} \,- B_{(\alpha)}^{(2)} \,\,&=\, {\di}\lambda_{(\alpha\beta)} + \widetilde{f}_{(\alpha\beta)i} F^i 
    \end{aligned}
\end{equation}
where $F=\mathrm{d}\xi$ is the curvature of $M\rightarrow M_0$, while the patching condition $\bigsp{\Lambda}_{(\alpha\beta)} + \bigsp{\Lambda}_{(\beta\gamma)} + \bigsp{\Lambda}_{(\gamma\alpha)} = \bigsp{\mathrm{d}}\bigsp{G}_{(\alpha\beta\gamma)}$ on three-fold overlaps of patches becomes
\begin{equation}\label{eq:nonobpatched2}
    \begin{aligned}
    \widetilde{f}_{(\alpha\beta)i}+\widetilde{f}_{(\beta\gamma)\,i}+\widetilde{f}_{(\gamma\alpha) i} \;&=\;0 \\
    \lambda_{(\alpha\beta)}+\lambda_{(\beta\gamma)}+\lambda_{(\gamma\alpha)} \;&=\;{\di}g_{(\alpha\beta\gamma)}
    \end{aligned}
\end{equation}
From \eqref{eq:nonobpatched} we get that $B^{(0)}$ are globally defined scalar fields on the base manifold $M_0$ and hence $H^{(1)}=\mathrm{d}B^{(0)}$ globally, which assures $\big[H^{(1)}\big]=0$.
From \eqref{eq:nonobpatched} and \eqref{eq:nonobpatched2} we get that $\big(B^{(1)}_{(\alpha)},\widetilde{f}_{(\alpha\beta)}\big)$ is a cocycle defining a torus $T^n$-bundle $\widetilde{M}$ with connection on the base manifold $M_0$. Together with the torus bundle $M$ given by the cocycle $(A_{(\alpha)},f_{(\alpha\beta)})$ we have the following two torus bundles (or equivalently a single $T^{2n}$-bundle) on $M_0$ with local data
\begin{equation}
\begin{aligned}
    A_{(\beta)} - A_{(\alpha)} \;&=\; {\di}f_{(\alpha\beta)}, & \quad
    B_{(\beta)}^{(1)} - B_{(\alpha)}^{(1)} \;&=\; {\di}\widetilde{f}_{(\alpha\beta)}, \\
    f_{(\alpha\beta)}+f_{(\beta\gamma)} + f_{(\gamma\alpha)} \;&=\; 0, & \quad
    \widetilde{f}_{(\alpha\beta)}+\widetilde{f}_{(\beta\gamma)} + \widetilde{f}_{(\gamma\alpha)} \;&=\; 0,
\end{aligned}
\end{equation}
with first Chern classes given by
\begin{equation}
    \mathrm{c}_1(M)=\big[{\di}A_{(\alpha)}\big]=\big[F\big], \qquad \mathrm{c}_1(\widetilde{M})=\big[{\di}B^{(1)}_{(\alpha)}\big]=\big[ H^{(2)}+\langle B^{(0)},F\rangle\big] = \big[\pi_\ast H\big]
\end{equation}
The coordinates on $T^n\times\widetilde{T}^n$ are given by $\theta_{(\beta)} - \theta_{(\alpha)} = f_{(\alpha\beta)}$ and $\widetilde{\theta}_{(\beta)} - \widetilde{\theta}_{(\alpha)} = \widetilde{f}_{(\alpha\beta)}$. 
Finally, notice that the $3$-form $H^{(3)}\in\Omega^3(M_0)$ is related to the connection by the equation 
\begin{equation}
    H^{(3)} \;=\; {\di}B^{(2)}_{(\alpha)} -  B^{(1)}_{(\alpha)i}\wedge F^i,
\end{equation}
which immediately leads to the Bianchi identity
\begin{equation}
   {\di} H^{(3)} \;=\;  F^i \wedge \widetilde{F}_i.
\end{equation}
By putting all together we obtain the cocycle $$(f_{(\alpha\beta)},\,\widetilde{f}_{(\alpha\beta)},\,g_{(\alpha\beta\gamma)}):M_0\longrightarrow \mathbf{B}\String(T^n\times \widetilde{T}^n),$$
which defines a $\String(T^n\times \widetilde{T}^n)$-bundle and, hence, the conclusion of the lemma.
\end{proof}

\begin{digression}[$H$-flux and $F$-flux]
The Dixmier-Douady class $[H]\in H^3(M,\mathbb{Z})$ and the first Chern class $[F_i]\in H^2(M_0,\mathbb{Z})$ are respectively called $H$-flux and $F$-flux (or geometric flux) in String Theory literature. Moreover the first Chern class $[\pi_\ast H]_i=[H^{(2)}_i+\langle B^{(0)},F\rangle_i]\in H^2(M_0,\mathbb{Z})$, which is given by the integral of $[H]$ on a basis of $1$-cycles $[S_i^1]\in H_1(T^n,\mathbb{Z})$ of the torus fibre, represents a non-trivial flux compactification of the Dixmier-Douady class.
\end{digression}

\begin{remark}[Geometrical interpretation of fluxes]
Notice that in our Higher-Kaluza Klein framework the $H$-flux and the $F$-flux are not something one puts on the doubled space by hand (for instance by defining some $3$-form on a $2d$-dimensional manifold like in the usual approach), but they are natural \textit{topological properties} of the geometry itself of the doubled space $\mathcal{M}$.
\end{remark}

\begin{remark}[Bundle gerbe curvature]\label{rem:globgeomcurv}
The equation for the curvature $\bigsp{H}=\bigsp{\mathrm{d}}\bigsp{B}_{(\alpha)}$ under isomorphism \eqref{eq:isomorphism} becomes $(H^{(3)},H^{(2)},H^{(1)},H^{(0)})=D(B^{(2)}_{(\alpha)},B^{(1)}_{(\alpha)},B^{(0)})$, which is equivalent to
\begin{equation}
    \begin{aligned}
    H^{(0)}_{ijk} \;&=\; 0, \\
    H^{(1)}_{ij} \;&=\; {\di}B^{(0)}_{ij}, \\
    H^{(2)}_{i} \;&=\; {\di}B^{(1)}_{(\alpha)i} - B^{(0)}_{ij} F^j, \\
    H^{(3)} \;&=\; {\di}B_{(\alpha)}^{(2)} - B^{(1)}_{(\alpha)i}\wedge F^i.
    \end{aligned}
\end{equation}
\end{remark}

\noindent We can refine the lemma \ref{thm:corrspace} by explicitly writing the higher Kaluza-Klein reduction of the generalised metric.

\begin{theorem}[T-duality]
In the hypothesis of lemma \ref{thm:corrspace} the generalised metric structure $\mathscr{G}\xrightarrow{\;\mathcal{G}\;} O(2d+2n)\Struc$ is Kaluza-Klein as follows:
\begin{equation*}
    \left(\begin{tikzcd}[row sep=9ex, column sep=3ex]
    \mathscr{G} \arrow[r, "\mathcal{G}"] &  O(2d+2n)\Struc
    \end{tikzcd}\right) \;\overset{\!\cong}{\longmapsto}\; \left(\begin{tikzcd}[row sep=12ex, column sep=-5ex]
    &  O(d)\Struc\times\mathbf{B}\mathrm{String}(T^n\times T^n)_\mathrm{conn}\times O(n,n) \arrow[d, two heads, ""]\\
    M_0 \arrow[ru]\arrow[r, "f_{(\alpha\beta)}"] & \mathbf{B}T^n
    \end{tikzcd}\right)\!,
\end{equation*}
where $M_0\xrightarrow{\;g\;} O(d)\Struc$ is a Riemannian metric, $M_0\rightarrow\mathbf{B}\mathrm{String}(T^n\times T^n)_\mathrm{conn}$ is the bundle of lemma \ref{thm:corrspace} with connection and $M_0\xrightarrow{\;\mathcal{G}^{(0)}\;} O(n,n)$ is a moduli field.
\end{theorem}

\begin{proof}
We can immediately split the generalised metric $\mathcal{G}$ of $\mathscr{G}$ in a Riemannian metric $g$ and a gerbe connection $B_{(\alpha)}$ on $M$. Now, by using the torus connection $\xi\in\Omega^1(M,\mathbb{R}^n)$, we can split these metric and gerbe connection in horizontal and vertical components
\begin{equation}
    \begin{aligned}
        \bigsp{g} \;&=\; \pi^\ast g^{(2)} + \langle \pi^\ast g^{(0)},\xi\odot\xi\rangle \\
        \bigsp{B}_{(\alpha)} \;&=\; \pi^\ast B_{(\alpha)}^{(2)} + \langle \pi^\ast B_{(\alpha)}^{(1)}\,\overset{\wedge}{,}\, \xi \rangle + \frac{1}{2}\langle  
        \pi^\ast B_{(\alpha)}^{(0)}, \xi\wedge\xi \rangle
    \end{aligned}
\end{equation}
where $g^{(2)}$ and $g^{(0)}$ are respectively a metric and a set of moduli fields on $M_0$, while $B^{(k)}_{(\alpha)}$ are local $k$-forms on patches $U_\alpha$. We can do the same for the \v{C}ech cocycle $(\bigsp{\Lambda}_{(\alpha\beta)},\bigsp{G}_{(\alpha\beta\gamma)})$ of the bundle gerbe
\begin{equation}
    \begin{aligned}
        \bigsp{\Lambda}_{(\alpha\beta)} \;&=\; \pi^\ast \lambda_{(\alpha\beta)} + \langle \pi^\ast \widetilde{f}_{(\alpha\beta)}, \xi \rangle \\
        \bigsp{G}_{(\alpha\beta\gamma)} \;&=\; \pi^\ast g_{(\alpha\beta\gamma)}
    \end{aligned}
\end{equation}
where $\lambda_{(\alpha\beta)}$ is a local $1$-form on two-fold overlaps of patches $U_\alpha\cap U_\beta$ and $\widetilde{f}_{(\alpha\beta)}$ and $g_{(\alpha\beta\gamma)}$ are local functions respectively on two-fold and three-fold overlaps of patches. The patching conditions of the cocycle $(\bigsp{B}_{(\alpha)},\bigsp{\Lambda}_{(\alpha\beta)},\bigsp{G}_{(\alpha\beta\gamma)})$ of the bundle gerbe are the ones of lemma \ref{thm:corrspace}.
Thus the generalised metric $\mathcal{G}$ reduces on the base $M_0$ to: a Riemannian metric $g^{(2)}$, a $\String(T^n\times \widetilde{T}^n)$-connection $(A^i_{(\alpha)}, B^{(1)}_{(\alpha)i},B_{(\alpha)}^{(2)})$ and a set of global moduli fields $(g^{(0)}_{ij},B^{(0)}_{ij})$.
Now, we can pack these fields again as follows:
\begin{equation*}
      \Theta_{(\alpha)} := \begin{pmatrix}
 \theta_{(\alpha)} \\[0.2cm]
 \widetilde{\theta}_{(\alpha)}
 \end{pmatrix}, \quad \mathcal{G}^{(0)} := \begin{pmatrix}
 g^{(0)}_{ij} - B_{ik}^{(0)}g^{(0)kl}B_{lj}^{(0)} & B_{ik}^{(0)}g^{(0)kj} \\[0.2cm]
 -g^{(0)ik}B_{kj}^{(0)} & g^{(0)ij}
 \end{pmatrix}, \quad \mathcal{A}_{(\alpha)} := \begin{pmatrix}
 A_{(\alpha)} \\[0.2cm]
 B^{(1)}_{(\alpha)}
 \end{pmatrix},
\end{equation*}
which are respectively the coordinates of the fibres of the $(T^{n}\times \widetilde{T}^n)$-bundle, the globally-defined moduli field of the generalised metric on the base manifold $M_0$ and the principal $(T^{n}\times \widetilde{T}^n)$-connection on $M_0$. Thus we have the conclusion.
\end{proof}

\begin{remark}[Interpretation of T-duality]
Notice that the higher structure which we obtain by Kaluza-Klein reducing the bundle gerbe equipped with generalised metric contains an ambiguity. This structure, indeed, can be obtained both by reducing a bundle gerbe $\mathscr{G}\twoheadrightarrow M$ or its T-dual $\widetilde{\mathscr{G}}\twoheadrightarrow \widetilde{M}$ to the base $M_0$, up to a gauge transformation.
The fact that $H^{(3)}\in\Omega^3(M_0)$ is the same in both pictures implies that we can write both $H^{(3)}= {\di}B^{(2)}_{(\alpha)}-\langle B^{(1)}_{(\alpha)}\,\overset{\wedge}{,}\, F\rangle$ and $H^{(3)}=  {\di}\widetilde{B}^{(2)}_{(\alpha)}-\langle A_{(\alpha)}\,\overset{\wedge}{,}\, \pi_\ast H\rangle$ where $\widetilde{B}^{(2)}_{(\alpha)}$ is the T-dual of ${B}^{(2)}_{(\alpha)}$. But this means that we recover the familiar relation 
\begin{equation}
    \widetilde{B}_{(\alpha)}^{(2)} \;=\; B_{(\alpha)}^{(2)}+A_{(\alpha)}^i\wedge B^{(1)}_{(\alpha)i}.
\end{equation}
\end{remark} 

\noindent Now, there is a natural group action $O(n,n;\mathbb{Z})$, whose elements $\mathcal{O}$ act by
\begin{equation}
      \Theta_{(\alpha)}\mapsto \mathcal{O}^{-1}\Theta_{(\alpha)}, \quad \mathcal{G}^{(0)}\mapsto \mathcal{O}^{\mathrm{T}}\mathcal{G}^{(0)}\mathcal{O}, \quad \mathcal{A}_{(\alpha)}\mapsto \mathcal{O}^{-1}\mathcal{A}_{(\alpha)},
\end{equation}
so that we recover exactly the Buscher rules. Notice that the first Chern class of the correspondence space is rotated by $[{\di}\mathcal{A}_{(\alpha)}]\mapsto \mathcal{O}^{-1}[{\di}\mathcal{A}_{(\alpha)}]$, while the component $H^{(3)}$ is invariant. Hence the \textit{doubled torus bundle} introduced by \cite{Hull06} is nothing but the principal $(T^{n}\times \widetilde{T}^n)$-bundle $K\twoheadrightarrow M_0$ that we obtain by higher Kaluza-Klein reduction of the bundle gerbe $\mathscr{G}$ to the base manifold $M_0$. \vspace{0.2cm}

\noindent Notice that the connection of the $\String(T^n\times \widetilde{T}^n)$-bundle recovers a global \textit{differential T-duality structure} as defined by \cite{KahVal09}.

\subsection{Atlas formulation }

\noindent Recall that the doubled space $\M$, i.e. the atlas of the bundle gerbe is naturally a para-Hermitian manifold.

\begin{theorem}[Topological T-duality on the doubled space]
Let $\mathscr{G}\xtwoheadrightarrow{\;\Pi\;} M$ and $\widetilde{\mathscr{G}}\xtwoheadrightarrow{\;\widetilde{\Pi}\;} \widetilde{M}$ be two topological T-dual bundle gerbes. Then their atlases, respectively $(\M,J,\omega)$ and $(\M,\widetilde{J},\widetilde{\omega})$, are related by a para-Hermitian isometry, i.e. a change of polarisation as defined by \cite{MarSza18}.
\end{theorem}

\begin{proof}
Let us start from the T-duality diagram of two topologically T-dual bundle gerbes. The atlas will sit on top of the diagram as follows:
\begin{equation}
    \begin{tikzcd}[row sep={11ex,between origins}, column sep={12ex,between origins}]
    & & \M \arrow[dr, shorten <= 0.1em, shorten >= 0.1em]\arrow[ld, shorten <= 0.1em, shorten >= 0.1em]\arrow[ddll, bend right=50, "\Phi"', two heads]\arrow[ddrr, bend left=50, "\widetilde{\Phi}", two heads] & & \\
    & \mathscr{G}\times_{M_0}\widetilde{M} \arrow[dr, "\Pi"', shorten <= 0.1em, shorten >= 0.1em]\arrow[dl, "\pi", shorten <= 0.1em, shorten >= 0.1em] & & M\times_{M_0}\widetilde{\mathscr{G}} \arrow[dr, "\widetilde{\pi}"', shorten <= 0.1em, shorten >= 0.1em]\arrow[dl, "\widetilde{\Pi}", shorten <= 0.1em, shorten >= 0.1em] \\
    \mathscr{G} \arrow[dr, "\Pi"', shorten <= 0.1em, shorten >= 0.1em] & & M\times_{M_0}\widetilde{M} \arrow[dr, "\pi"', shorten <= 0.1em, shorten >= 0.1em]\arrow[dl, "\widetilde{\pi}", shorten <= 0.1em, shorten >= 0.1em] & & [-2.5em]\widetilde{\mathscr{G}} \arrow[dl, "\widetilde{\Pi}", shorten <= 0.1em, shorten >= 0.1em] \\
    & M \arrow[dr, "\pi"', shorten <= 0.1em, shorten >= 0.1em] & & \widetilde{M} \arrow[dl, "\widetilde{\pi}", shorten <= 0.1em, shorten >= 0.1em] & \\
    & & M_0 & &
    \end{tikzcd}
\end{equation}
Let us consider the atlas $(\M,J,\omega)$ of the bundle gerbe $\mathscr{G}\twoheadrightarrow M$. Let $e^i\in\Omega^1(M)$ be the connection of the $T^n$-bundle $M\twoheadrightarrow M_0$. As shown in \cite[p.$\,$46]{Alf19}, we can expand the local $2$-form potential of the bundle gerbe in the connection $e^i\in\Omega^1(M)$ by 
\begin{equation}
    B_{(\alpha)} \;=\; B^{(0)}_{ij}e^i \wedge e^j + B^{(1)}_{(\alpha)\mu i}\di x^\mu \wedge e^i + {B}^{(2)}_{(\alpha)\mu\nu}\di x^\mu \wedge \di x^\nu
\end{equation}
where $B^{(0)}_{ij}$ is a globally defined scalar moduli field on $M$ and, therefore, we omitted the $\alpha$-index.
The corresponding fundamental $2$-form on the atlas $\M$ will be
\begin{equation}
    \omega \;=\; \big(\widetilde{e}_i+ B^{(0)}_{ij}e^j\big) \wedge e^i  + \widetilde{e}_\mu\wedge e^\mu,
\end{equation}
where we patch-wise defined the following global $1$-forms on the atlas:
\begin{equation}
    \begin{aligned}
        e^\mu \;&=\; \di x^\mu &\quad e^i \;&=\; \di\theta_{(\alpha)}^i + A^i_{(\alpha)\mu}\di x^\mu \\[0.4em]
        \widetilde{e}_{\mu} \;&=\; \di \widetilde{x}_{(\alpha)\mu}+ B^{(2)}_{(\alpha)\mu\nu}\di x^\nu \quad & \widetilde{e}_{i} \;&=\; \di\widetilde{\theta}_{(\alpha)i} + B^{(1)}_{(\alpha)i\mu}\di x^\mu
    \end{aligned}
\end{equation}
Let us explicitly construct the para-Hermitian metric $\eta$ of the atlas. This will globally be
\begin{equation}
    \eta(-,-) \;:=\; \omega(J-,-) \quad\; \Rightarrow \quad\; \eta \;=\; \widetilde{e}_i \odot e^i + \widetilde{e}_\mu \odot e^\mu
\end{equation}
Since $b:=B^{(0)}_{ij}e^i\wedge e^j\in\Omega^2(M)$ is a global $2$-form, the moduli field $B^{(0)}_{ij}\in\Coo(M,\mathfrak{so}(n))$ can be interpreted as a global B-shift. Thus, there exists an isometry of our para-Hermitian manifold \cite[p.$\,$15]{MarSza18} given by
\begin{equation}
    \omega' \;=\; e^b\,\omega \;=\; \widetilde{e}_i \wedge e^i  + \widetilde{e}_\mu\wedge e^\mu,
\end{equation}
By using this isometry, we forgot the moduli field and we retained only the topologically relevant component of the connection.
Now, let $(\M,\widetilde{J},\widetilde{\omega})$ be the atlas of the bundle gerbe $\widetilde{\mathscr{G}}\twoheadrightarrow \widetilde{M}$. Since we started from a couple of T-dual geometric backgrounds ${\mathscr{G}}$ and $\widetilde{\mathscr{G}}$, we already know that the potential $2$-form of the latter is
\begin{equation}
    \widetilde{B}_{(\alpha)} \;=\; \widetilde{B}^{(0)ij}\widetilde{e}_i \wedge \widetilde{e}_j + A_{(\alpha)\mu}^i\di x^\mu \wedge \widetilde{e}_i + {B}^{(2)}_{(\alpha)\mu\nu}\di x^\mu \wedge \di x^\nu
\end{equation}
where $\widetilde{B}^{(0)ij}$ is a global moduli field (which can be explicitly obtained by using the Buscher rules) and $A_{(\alpha)\mu}^i$ is the $1$-form potential of the $T^n$-bundle $M\twoheadrightarrow M_0$. Therefore, the T-dual corresponding fundamental $2$-form will be
\begin{equation}
    \widetilde{\omega} \;=\; \big(e^i+\widetilde{B}^{(0)ij}\widetilde{e}_j\big)\wedge \widetilde{e}_i + \widetilde{e}_\mu\wedge e^\mu
\end{equation}
Similarly to the first bundle gerbe, $\widetilde{b}:=\widetilde{B}^{(0)ij}\widetilde{e}_i\wedge\widetilde{e}_j$ is a global $2$-form and, thus, the map
\begin{equation}
    \widetilde{\omega}' \;=\; e^{\widetilde{b}}\,\widetilde{\omega} \;=\; e^i\wedge \widetilde{e}_i + \widetilde{e}_\mu\wedge e^\mu 
\end{equation}
is an isometry of the para-Hermitian metric. 
Now, let us call $J'$ and $\widetilde{J}'$ the para-complex structures corresponding to $\omega'$ and $\widetilde{\omega}'$. We need to find a morphism of para-Hermitian manifolds $f:(\M,J',\omega')\longrightarrow(\M,\widetilde{J}',\widetilde{\omega}')$ such that $\widetilde{\omega}' \;=\; f^\ast\omega'$ and check that it is an isometry. This is immediately the map  $f:\big(x_{(\alpha)},\widetilde{x}_{(\alpha)},\theta_{(\alpha)},\widetilde{\theta}_{(\alpha)}\big)\mapsto \big(x_{(\alpha)},\widetilde{x}_{(\alpha)},\widetilde{\theta}_{(\alpha)},\theta_{(\alpha)}\big)$, which is given by the exchange of the torus coordinates $\theta$ and $\widetilde{\theta}$ on each chart and is clearly an isometry. Therefore, by composition, we obtained an isometry $e^{b}\circ f\circ e^{-\widetilde{b}}:(\M,J,\omega)\longrightarrow(\M,\widetilde{J},\widetilde{\omega})$.
\end{proof}

\begin{remark}[Buscher rules]
It is not hard to see that the Buscher transformations $(g^{(0)}_{ij}, B^{(0)}_{ij})\mapsto(\widetilde{g}^{(0)ij}, \widetilde{B}^{(0)ij})$ of the moduli field of the metric and the Kalb-Ramond field follow directly from applying the isometry of the lemma to the generalised metric, i.e. $\widetilde{\mathcal{G}}=f^\ast\mathcal{G}$.
\end{remark}

\subsection{Global tensor hierarchy}

In this subsection we will interpret the case \eqref{eq:emcorrspace} as a simple example of globally-defined tensor hierarchy. \vspace{0.2cm}

\noindent For any Lie group $G$ let us call $G\text{-}\mathrm{equiv}\mathbf{B}U(1)\mathrm{Bund}(M)$ the groupoid of $G$-equivariant gerbes on $M$ and, for any Lie $\infty$-group $H$, let us call $H\mathrm{Bund}(M_0)$ the groupoid of principal $H$-bundles on $M_0$. As we have seen, we have the equivalence
\begin{equation}
    \bigsqcup_{\begin{subarray}{c}M\text{ s.t. }
    M\twoheadrightarrow M_0\\\text{is a }T^n\text{-bundle}\end{subarray}} \!\!\!\!T^n\text{-}\mathrm{equiv}\mathbf{B}U(1)\mathrm{Bund}(M) \;\;\cong\;\; \String(T^n\!\times \widetilde{T}^n)\mathrm{Bund}(M_0).
\end{equation}
This reads as follows: any gerbe on the total space of a $T^n$-bundle $M\twoheadrightarrow M_0$ which is equivariant under the principal $T^n$-action is equivalently a $\String(T^n\!\times \widetilde{T}^n)$-bundle on the base manifold $M_0$. If we forget the higher form fields, we remain with a $T^n\times \widetilde{T}^n$-bundle, which is nothing but the correspondence space $K=M\times_{M_0}\widetilde{M}$ of a topological T-duality.\vspace{0.2cm}

\noindent Let us make the following field redefinitions in lemma \ref{thm:corrspace}:
\begin{equation}
    \begin{aligned}
    \mathcal{F}^I_{(\alpha)} \,&:=\, \begin{pmatrix}\delta^i_j & 0 \\[0.6em] B^{(0)}_{(\alpha)ij} & \delta^j_i \end{pmatrix} \begin{pmatrix} F^j \\[0.6em] H^{(2)}_j \end{pmatrix}, \; &\mathcal{H}\,&:=\, H^{(3)}, \\[0.4em]
    \mathcal{A}^I_{(\alpha)} \,&:=\, \begin{pmatrix} A_{(\alpha)} \\[0.6em] B^{(1)}_{(\alpha)} \end{pmatrix} , \; & \mathcal{B}_{(\alpha)}\,&:=\, B^{(2)}_{(\alpha)}, \\[0.3em]
    \lambda_{(\alpha\beta)}^I \,&:=\, \begin{pmatrix} \lambda_{(\alpha\beta)}^i \\[0.7em] \Lambda^{(0)}_{(\alpha\beta)i} \end{pmatrix} , \; &\Xi_{(\alpha\beta)} \,&:=\, \Lambda^{(1)}_{(\alpha\beta)}.
    \end{aligned}
\end{equation}
where the local $1$-forms $A_{(\alpha)}^i$ and scalars $\lambda_{(\alpha\beta)}^i$ are respectively the local potential and the transition functions of the original torus bundle $M\twoheadrightarrow M_0$. 
The Bianchi equation of the bundle gerbe curvature, together with the Bianchi equation $\di F = 0$ of the curvature of the torus bundle can now be equivalently rewritten as
\begin{equation}
    \begin{aligned}
    \di \mathcal{F}_{(\alpha)} \;&=\; 0, \\
    \di \mathcal{H} - \frac{1}{2}\langle \mathcal{F}_{(\alpha)} \;\overset{\wedge}{,}\; \mathcal{F}_{(\alpha)} \rangle\;&=\;  0,
    \end{aligned}
\end{equation}
where, as usual, we defined the product $\langle X, Y\rangle = \eta_{IJ}X^IY^J$.
These are a particular case of the Bianchi equations of a particularly simple abelian tensor hierarchy. We can now rewrite all the patching conditions in the following equivalent form:
\begin{equation}
    \begin{aligned}
    \mathcal{F} \;&=\; \di \mathcal{A}_{(\alpha)} ,\\
    \mathcal{H} \;&=\;  \di \mathcal{B}_{(\alpha)} + \frac{1}{2}\langle \mathcal{A}_{(\alpha)} \,\overset{\wedge}{,}\, \mathcal{F}\rangle ,\\[0.8em]
    \mathcal{A}_{(\alpha)} - \mathcal{A}_{(\beta)} \;&=\; \di\lambda_{(\alpha\beta)} ,\\
    \mathcal{B}_{(\alpha)} - \mathcal{B}_{(\beta)} \;&=\; \di \Xi_{(\alpha\beta)} - \langle\lambda_{(\alpha\beta)},\mathcal{F}\rangle ,\\[0.8em]
    \lambda_{(\alpha\beta)}+\lambda_{(\beta\gamma)}+\lambda_{(\gamma\alpha)} \;&=\; 0,\\
    \Xi_{(\alpha\beta)} + \Xi_{(\beta\gamma)} + \Xi_{(\gamma\alpha)} \;&=\; \di g_{(\alpha\beta\gamma)} ,\\[0.8em]
    G_{(\alpha\beta\gamma)} - G_{(\beta\gamma\delta)} +  G_{(\gamma\delta\alpha)} -  G_{(\delta\alpha\beta)}\;&\in\;2\pi\mathbb{Z}.
    \end{aligned}
\end{equation}
This way, we explicitly expressed the connection $\{\mathcal{A}^I_\mu,\mathcal{B}_{\mu\nu}\}$ of the $\String(T^n\times\widetilde{T}^n)$-bundle obtained by Kaluza-Klein reduction of the bundle gerbe as a global tensor hierarchy.

\subsection{Relation with pre-NQP-manifolds}

\noindent Let us now give a brief look to the relation between our proposal and the local geometry for DFT developed by \cite{DesSae18} and \cite{DesSae19}, that is called Extended Riemannian Geometry. 

\begin{digression}[Recovering Extended Riemannian Geometry by \cite{DesSae18}]
Given a patch $U\subset M_0$, we can write the local algebroids of the infinitesimal symmetries of the bundle gerbe $\mathscr{G}$ pulled back to the correspondence space $K$, on the base $M$ and Kaluza-Klein reduced to $M_0$ by
\begin{equation*}
\begin{tikzcd}[row sep=9ex,column sep=2ex,nodes={inner sep=4pt}]
   \widetilde{\pi}^\ast\mathscr{G}  \ar[dotted]{d} & \mathscr{G}  \ar["\text{reduction}"']{r} \ar[dotted]{d} \ar[leftarrow, "\widetilde{\pi}_\ast"]{l} & [5ex]\mathscr{G}/T^n  \ar[dotted]{d}  \\
       T^\ast[2]T[1](U\times T^{2n})\ar["\Coo(-)/\langle\mathrm{d}\theta^i\rangle"']{d} \ar{r} & T^\ast[2]T[1](U\times T^n)  \ar{r} & [5ex]T^\ast[2]T[1]U\oplus \mathbb{R}^{2n}[1],  \\
      T^\ast[2]T[1]U\oplus (T[1]\oplus T^\ast[2]) T^{2n} \ar["p"]{ru} && 
 \end{tikzcd}
\end{equation*}
where the dotted maps send a doubled space $\mathcal{N}$ to the differential-graded manifold $N$ which describes its local symmetries, i.e. which satisfies $\Coo(N)=\mathrm{CE}\big(\mathfrak{at}(\mathcal{N}|_V)\big)$ on any patch $V$ of the considered base manifold. Notice that the submanifold $T^\ast[2]T[1]U\oplus (T[1]\oplus T[2]^\ast) T^{2n}$ over a patch $U\times T^{2n}$ of the correspondence space $K$ is the structure considered by \cite{DesSae18} and \cite{DesSae19}. Now notice that the $0$-degree space of functions on this manifold are sections $\Gamma\big(TU\oplus T^\ast U \oplus TT^{2n}\big)\cong\Gamma\big(T(U\times T^{n}) \oplus T^\ast (U\times T^{n})\big)$. Hence it is isomorphic to the one of $T^\ast[2]T[1](U\times T^n)$ and this isomorphism defines the projection $p$ in the diagram. Of course these local patches can be glued to give a differential-graded fibration on $K$.
What is indeed called "doubled space" by the references is exactly the correspondence space $K=M\times_{M_0}\widetilde{M}$, but equipped with this graded bundle structure. Hence Extended Riemannian Geometry can be seen as an infinitesimal description of the doubled space $\M$ pulled back to the correspondence space $K$.
\end{digression}

\section{Geometrisation of non-geometric T-duality}\label{td3}

In this section we will relax the hypothesis of lemma \ref{thm:corrspace}: this time the bundle gerbe $\mathscr{G}\xrightarrow{\bbpi} M$ will not be necessarily equivariant under the principal $T^n$-action of $M$. In the literature the case where the gerbe connection is only required to satisfy $\mathcal{L}_{k_i}\bigsp{B}_{(\alpha)}=0$ on each patch is called \textit{(globally) non-geometric T-duality}. In terms of transition functions, the differential forms $(\bigsp{\Lambda}_{(\alpha\beta)}, \bigsp{G}_{(\alpha\beta\gamma)})$ are allowed to depend on the coordinates of the fibres as long as $\mathcal{L}_{k_i}\bigsp{B}_{(\alpha)}=0$ is satisfied. From \eqref{eq:exactness} we get that the class $[H^{(0)}]$ is still trivial, but $[H^{(1)}]$ generally is not.

\subsection{Generalised correspondence space formulation}

\begin{theorem}[Globally non-geometric T-duality]\label{thm:gencorrspace}
Let $\mathscr{G}$ be a bundle gerbe with generalised metric $\mathcal{G}$ which satisfies the strong constraint and such that the base manifold of $\mathscr{G}\xrightarrow{\Pi}M$ is itself a principal $T^n$-bundle $M\xrightarrow{\pi}M_0$. 
Now, if the automorphisms $\Coo(M_0,T^n)$ of $M\xrightarrow{\pi}M_0$ are lifted to isometries of the generalised metric structure, by applying higher Kaluza-Klein reduction \ref{eq:KKreduction6} we have the following.
\begin{enumerate}
    \item[(a)] If we forget the higher part of the bundle gerbe $\mathscr{G}$ with connective structure, we are left with an ordinary principal $T^n$-bundle $K\twoheadrightarrow M$ on spacetime, with first Chern class $\mathrm{c}_1(K)=[\pi_\ast \bigsp{H}]=[H^{(2)}-\langle H^{(1)}\,\overset{\wedge}{,}\,\xi\rangle]\in H^2(M,\mathbb{Z})^n$ given by the $H$-flux.
    Thus, we have a diagram:
\begin{equation}
\begin{tikzcd}[row sep={9ex,between origins}, column sep={10ex,between origins}]
 & & K \arrow[dl, two heads] & & \\
  & M \arrow[dr, "\pi"', two heads] & & \quad\; & \quad \\
 & & M_0 & &
\end{tikzcd}
\end{equation}
This can be equivalently seen as an affine $T^{2n}$-bundle over the base manifold $M_0$, known in the literature as the \textit{generalised correspondence space} (see digression \ref{rem:tfold}).
\item[(b)] The generalised metric $\mathcal{G}$ reduces on $M_0$ to a metric $g^{(0)}$, a Kalb-Ramond field $B_{(\alpha)}^{(2)}$, a $T^{n}$-connection $A^i_{(\alpha)}$ for $M$, a $T^{n}$-connection $\widetilde{A}_{(\alpha)i}$ for $K$ and a set of global moduli fields $g^{(0)}_{ij}$.
\end{enumerate}
\end{theorem}

\begin{digression}[T-fold]\label{rem:tfold}
The generalised correspondence space $K$ is an affine (non principal) $T^{2n}$-bundle over the base manifold $M_0$. Recall that the affine group of the torus is $\mathrm{Aff}(T^{2n})=GL(2n,\mathbb{Z})\ltimes T^{2n}$ and that an affine torus bundle is defined as the associated bundle $K:=Q\times_{\mathrm{Aff}(T^{2n})}T^{2n}$ to some principal $\mathrm{Aff}(T^{2n})$-bundle $Q$. Generalised correspondence spaces are a special class of affine torus bundles where the structure group is restricted to $\wedge^2\mathbb{Z}^n\ltimes T^{2n}\subset \mathrm{Aff}(T^{2n})$, where $\wedge^2\mathbb{Z}^n$ encodes the monodromy. Since $K$ has monodromy there is no well-defined torus subbundle $\widetilde{M}\rightarrow M_0$ which can be seen as the T-dual to the starting $M\rightarrow M_0$. In fact we could perform T-duality on each patch $U_\alpha$ of $M$, but we would obtain a collection of string-background patches which cannot be glued together. In DFT literature this object has been named \textit{T-fold}. Morally speaking we would have a diagram generalising \eqref{diag:corrspace} of the form
\begin{equation}\label{eq:tfolddiagram}
\begin{tikzcd}[row sep={9ex,between origins}, column sep={10ex,between origins}]
 & & K \arrow[dl, two heads]\arrow[dr, dotted] & \\
 & M \arrow[dr, "\pi"', two heads] & & [-1ex]\text{T-fold}\arrow[dl, dotted] \\
 & & M_0 &
\end{tikzcd}
\end{equation}
where the dotted arrows are not actual maps between spaces, but only indicative ones. Since there is no well defined dual manifold, this T-duality has \textit{no underlying topological T-duality}. We will explain what are the differential data of this kind of T-duality in remark \ref{rem:tdualityoftfold}.
\end{digression}

\begin{proof}[Proof of lemma \ref{thm:gencorrspace}]
The assumption that the automorphisms of $M\xrightarrow{\pi}M_0$ are the isometries of the generalised metric, i.e.  $\mathbf{Iso}(\mathscr{G},\mathcal{G})=\Coo(M_0,T^n)$, implies that $\mathcal{L}_{k_i}B_{(\alpha)}=0$ on each $U_\alpha$ where $\{k^i\}$ are the fundamental vectors. This assures that $\widetilde{F}_i := \iota_{k_i}\bigsp{H}$ is a closed $2$-form on $M$ by
\begin{equation}
\begin{aligned}
    \widetilde{F}_i \;&=\; -\mathrm{d}\iota_{k_i}\bigsp{B}_{(\alpha)} \\ 
    \;&=\; \mathrm{d}(B_{(\alpha)i}^{(1)}-B_{(\alpha)ij}^{(0)}\xi^j)
\end{aligned}
\end{equation}
Now we can define on each patch the local connection $1$-forms for this curvature
\begin{equation}
    \widetilde{A}_{(\alpha)i} \,:=\, -\iota_{k_i}\bigsp{B}_{(\alpha)} \,=\, B_{(\alpha)i}^{(1)}-B_{(\alpha)ij}^{(0)}\xi^j.
\end{equation}
We will closely follow \cite{BelHulMin07} for the next calculations. In the reference it is proven that these $1$-forms are indeed patched like the connection of a $T^n$-bundle as
\begin{equation}
    \begin{aligned}
        \widetilde{A}_{(\beta)i}-\widetilde{A}_{(\alpha)i} \;&=\; {\di}\Lambda^{(0)}_{(\alpha\beta)i} - 2(\partial_{i}\Lambda^{(0)}_{(\alpha\beta)j})\xi^j - \partial_{i}\Lambda^{(1)}_{(\alpha\beta)} \\
        &=\; {\di}\widetilde{f}_{(\alpha\beta)i} - n_{(\alpha\beta)ij}\Big(\mathrm{d}\theta_{(\beta)}^j+\frac{1}{2}{\di}f_{(\beta\alpha)}^j\Big) \\
        &=\; \bigsp{\di}\Big(\widetilde{f}_{(\alpha\beta)i} - n_{(\alpha\beta)ij}\Big(\theta_{(\beta)}^j+\frac{1}{2}f_{(\beta\alpha)}^j\Big)\Big).
    \end{aligned}
\end{equation}
The principal connection $\Xi\in\Omega^1(K,\mathbb{R}^n)$ of the generalised correspondence space $K\rightarrow M$ seen as a $T^n$-bundle over $M$ is then
\begin{equation}
    \Xi_i \,:=\, \mathrm{d}\widetilde{\theta}_{(\alpha)i} + B^{(1)}_{(\alpha)i}-B^{(0)}_{(\alpha)ij}\xi^j.
\end{equation}
If in analogy with geometric T-duality we define the local differential $1$-form $\widetilde{\xi}_{(\alpha)} := \mathrm{d}\widetilde{\theta}_{(\alpha)} + B^{(1)}_{(\alpha)}$, this cannot clearly be globalised. Since we know that $H^{(1)}$ is a closed $1$-form on $M_0$, i.e. ${\di}H^{(1)}=0$, this will define a \v{C}ech cocycle with patching conditions
\begin{equation}
    \begin{aligned}
        H^{(1)} &= {\di}B^{(0)}_{(\alpha)} \\
        B^{(0)}_{(\beta)} - B^{(0)}_{(\alpha)} &= n_{(\alpha\beta)}
    \end{aligned}
\end{equation}
where $n_{(\alpha\beta) ij}\in 2\pi\mathbb{Z}$.
On the other hand the principal connection $\xi=\mathrm{d}\theta_{(\alpha)}+A_{(\alpha)}$ is global on $M$. Therefore from $B_{(\beta)}^{(0)}-B_{(\alpha)}^{(0)} = n_{(\alpha\beta)}$ we get the patching conditions
\begin{equation}\label{eq:hflux}
\begin{pmatrix}
 \xi_{(\beta)} \\[0.2cm]
 \widetilde{\xi}_{(\beta)}
 \end{pmatrix} = \begin{pmatrix}
 1 & 0 \\[0.2cm]
 n_{(\alpha\beta)} & 1
 \end{pmatrix} \begin{pmatrix}
 \xi_{(\alpha)} \\[0.2cm]
 \widetilde{\xi}_{(\alpha)}
 \end{pmatrix}
\end{equation}
Where $n_{(\alpha\beta)}$ is the monodromy matrix of the dual torus fibres. Hence $K$ is equivalently an affine torus $T^{2n}$-bundle on the base manifold $M_0$. The affine transitions functions can be written as
\begin{equation}
\begin{pmatrix}
 \theta_{(\beta)} - \frac{1}{2}f_{(\alpha\beta)} \\[0.2cm]
 \widetilde{\theta}_{(\beta)} - \frac{1}{2}\widetilde{f}_{(\alpha\beta)}
 \end{pmatrix} = \begin{pmatrix}
 1 & 0 \\[0.2cm]
 n_{(\alpha\beta)} & 1
 \end{pmatrix} \begin{pmatrix}
 \theta_{(\alpha)} - \frac{1}{2}f_{(\beta\alpha)} \\[0.2cm]
 \widetilde{\theta}_{(\alpha)} - \frac{1}{2}\widetilde{f}_{(\beta\alpha)}
 \end{pmatrix}
\end{equation}
It has been also proven by \cite{BelHulMin07} that the horizontal components of $B_{(\alpha)}$ are patched by
\begin{equation*}
    B^{(2)}_{(\beta)} - B^{(2)}_{(\alpha)} = {\di}\lambda_{(\alpha\beta)} + \langle \widetilde{f}_{(\alpha\beta)},F \rangle + 
    \frac{1}{2}\left\langle n_{(\alpha\beta)}, \, \left(A_{(\beta)}-\frac{1}{2}{\di}f_{(\alpha\beta)}\right) \wedge \left(A_{(\beta)}-\frac{1}{2}{\di}f_{(\alpha\beta)}\right) \right\rangle
\end{equation*}
on two-fold overlaps, while on three-fold overlaps the $1$-forms $\lambda_{(\alpha\beta)}\in\Omega^1(U_\alpha\cap U_\beta)$ by
\begin{equation*}
\begin{aligned}
    &\lambda_{(\alpha\beta)} + \lambda_{(\beta\gamma)} + \lambda_{(\gamma\alpha)} \;=\; {\di}g_{(\alpha\beta\gamma)} + f_{(\beta\alpha)}(n_{(\alpha\beta)}+n_{(\gamma\beta)})f_{(\beta\gamma)} + \\
    &-\frac{1}{8}(f_{(\beta\alpha)}n_{(\gamma\beta)}{\di}f_{(\beta\alpha)} + f_{(\beta\alpha)}n_{(\gamma\alpha)}{\di}f_{(\beta\gamma)} + f_{(\beta\gamma)}n_{(\beta\alpha)}{\di}f_{(\beta\gamma)} + f_{(\beta\gamma)}n_{(\gamma\alpha)}{\di}f_{(\beta\alpha)})
\end{aligned}
\end{equation*}
and on four-fold overlaps the local functions $g_{(\alpha\beta\gamma)}\in\Coo(U_\alpha\cap U_\beta \cap U_\gamma)$ are patched by
\begin{equation*}
\begin{aligned}
    g_{(\alpha\beta\gamma)}-g_{(\beta\gamma\delta)}+g_{(\gamma\delta\alpha)}-g_{(\delta\alpha\beta)} = -\frac{1}{6}(f_{(\delta\gamma)}n_{(\delta\beta)}f_{(\alpha\delta)} - f_{(\beta\gamma)}n_{(\alpha\delta)}f_{(\delta\gamma)}+ f_{(\beta\delta)}n_{(\gamma\delta)}f_{(\delta\alpha)}).
\end{aligned}
\end{equation*}
It is clear that $(\lambda_{(\alpha\beta)},g_{(\alpha\beta\gamma)})$ is not the local data of a gerbe.
Finally, a global section $\Gamma(M_0,K)$ of the generalised correspondence space will be of the form $(\theta_{(\alpha)},\widetilde{\theta}_{(\alpha)})$, patched as follows:
\begin{equation}
    \begin{aligned}
    \theta_{(\beta)}^{i} \,-\, \theta_{(\alpha)}^{i} \;&=\; f_{(\beta\alpha)}^i, \\
    \widetilde{\theta}_{(\beta)i} - \widetilde{\theta}_{(\alpha)i}  \,&=\; \widetilde{f}_{(\beta\alpha) i} - n_{(\alpha\beta) ij}\left(\theta_{(\beta)}^j-\frac{1}{2}f_{(\beta\alpha)}^j\right).
    \end{aligned}
\end{equation}
Hence we have an affine $T^{2n}$-bundle $K\rightarrow M_0$ and we find the conclusion of the lemma.
\end{proof}

\noindent Notice that in the particular case of a trivial class $[H^{(1)}]=0$ we have $n_{(\alpha\beta)}=0$ and hence we recover exactly the global geometric case discussed in the previous section.

\begin{remark}[$H$-flux and $F$-flux]\label{rem:hfflux}
The $H$-flux and the $F$-flux are still respectively the Dixmier-Douady class $[\bigsp{H}]$ and the first Chern class $[F]$.
However in this case the $H$-flux compactification on $1$-cycles $[S^1_i]\in H_1(T^n,\mathbb{Z})$ cannot be seen as a first Chern class on the base manifold $M_0$ like in the previous section, but it is a cohomology class $[(H^{(2)},H^{(1)},0)]\in H^{2,1}_D(M_0,\,\wedge\,\mathbb{Z}^n)$. Notice the geometric $F$-flux can be also seen as a class $[(F,0,0)]\in H^{2,1}_D(M_0,\,\wedge\,\mathbb{Z}^n)$. Moreover the integration of the Dixmier-Douady class $[H]$ along $2$-cycles $[T^2_{ij}]\in H_2(T^n,\mathbb{Z})$ in the fibre is now non-trivial and it is given by the integral cohomology class $\big[H^{(1)}_{ij}\big]\in H^1(M_0,\mathbb{Z})$.
\end{remark}

\noindent Let us now generalize our discussion to the less simple case (but still abelian) of spacetime being a torus bundle with monodromy.

\begin{remark}[Torus bundle with monodromy]
Let us now generalize our torus bundle spacetime $M\xrightarrow{\pi}M_0$ to a torus bundle with monodromy given by the matrix $n_{(\alpha\beta)}^F\in\wedge^2\mathbb{Z}^n$. This can be seen as a cohomology class $[F^{(1)}]\in H^1(M_0,\mathbb{Z})$ such that we have the \v{C}ech cocycle
\begin{equation}
    \begin{aligned}
        F^{(1)}\, &=\, {\di}A^{(0)}_{(\alpha)}, \\
        A^{(0)}_{(\beta)} - A^{(0)}_{(\alpha)} \,&=\, n^F_{(\alpha\beta)}.
    \end{aligned}
\end{equation}
Hence we can update the $F$-flux in remark \ref{rem:hfflux} by adding the monodromy class to the curvature to obtain the class $[(F^{(2)},F^{(1)},0)]\in H^{2,1}_D(M_0,\,\wedge\,\mathbb{Z}^n)$.
Now, by looking at \eqref{eq:hflux}, we can generalize the patching conditions of the generalised correspondence space $K\twoheadrightarrow M_0$ by 
\begin{equation}\label{eq:monodromieshf}
\begin{pmatrix}
 \xi_{(\alpha)} \\[0.2cm]
 \widetilde{\xi}_{(\alpha)}
 \end{pmatrix} = \left(\begin{array}{@{}cc@{}}
 1+n_{(\alpha\beta)}^{F} &  0 \\[0.2cm] 
 n_{(\alpha\beta)}^{H} & 1-(n_{(\alpha\beta)}^{F})^{\mathrm{T}}
 \end{array}\right) \begin{pmatrix}
 \xi_{(\beta)} \\[0.2cm]
 \widetilde{\xi}_{(\beta)}
 \end{pmatrix},
\end{equation}
where $n_{(\alpha\beta)}^H$ is the monodromy matrix given by the $H$-flux $[H^{(1)}]\in H^1(M_0,\wedge^2\mathbb{Z}^n)$ and $n_{(\alpha\beta)}^F$ is the one given by the $F$-flux $[F^{(1)}]\in H^1(M_0,\wedge^2\mathbb{Z}^n)$. Hence the generalised correspondence space $K\rightarrow M_0$ will be an affine torus bundle patched by transition functions in the subgroup $\big(GL(n;\mathbb{Z})\ltimes\wedge^2\mathbb{Z}\big)\ltimes T^{2n}\subset \mathrm{Aff}(T^{2n})$ on overlaps $U_\alpha\cap U_\beta \subset M_0$, as found by \cite{Hul09}.
\end{remark}

\begin{remark}[T-duality $O(n,n;\mathbb{Z})$-action]\label{rem:tdualityoftfold}
Similarly to the previous section, we can still write our moduli fields of the Kaluza-Klein reduction by using the generalised metric as follows
\begin{equation*}
      \Theta_{(\alpha)} := \begin{pmatrix}
 \theta_{(\alpha)} \\[0.2cm]
 \widetilde{\theta}_{(\alpha)}
 \end{pmatrix}, \quad \mathcal{G}^{(0)}_{(\alpha)} := \begin{pmatrix}
 g^{(0)}_{ij} - B_{(\alpha) ik}^{(0)}g^{(0)kl}B_{(\alpha) lj}^{(0)} & B_{(\alpha) ik}^{(0)}g^{(0)kj} \\[0.2cm]
 -g^{(0)ik}B_{(\alpha) kj}^{(0)} & g^{(0)ij}
 \end{pmatrix}, \quad \mathcal{A}_{(\alpha)} := \begin{pmatrix}
 A_{(\alpha)} \\[0.2cm]
 B^{(1)}_{(\alpha)}
 \end{pmatrix}.
\end{equation*}
We still have a natural $O(n,n;\mathbb{Z})$ group action, whose elements $\mathcal{O}$ act as follows:
\begin{equation*}
      \Theta_{(\alpha)}\mapsto \mathcal{O}^{-1}\Theta_{(\alpha)}, \quad \mathcal{G}^{(0)}_{(\alpha)}\mapsto \mathcal{O}^{\mathrm{T}}\mathcal{G}^{(0)}_{(\alpha)}\mathcal{O}, \quad \mathcal{A}_{(\alpha)}\mapsto \mathcal{O}^{-1}\mathcal{A}_{(\alpha)}.
\end{equation*}
However now the $O(n,n)$-moduli field $\mathcal{G}^{(0)}_{(\alpha)}$ is not globally defined on the base manifold $M_0$. This means that by applying a general $O(n,n;\mathbb{Z})$ transformation we obtain new differential data $\widetilde{A}_{(\alpha)}$, $\widetilde{B}_{(\alpha)}^{(0)}$, $\widetilde{B}_{(\alpha)}^{(1)}$ which in general cannot be interpreted anymore as a $T^n$-bundle with gerbe connection. Only a transformation belonging to the \textit{geometric subgroup} $\mathcal{O}\in GL(d,\mathbb{Z})\ltimes\wedge^2\mathbb{Z}\subset O(n,n;\mathbb{Z})$ will send a background consisting of global $T^n$-bundle $M$ with gerbe connection to another one consisting of a $T^n$-bundle $\widetilde{M}$ with gerbe connection.
\end{remark}

\noindent Moreover notice that the transition functions \eqref{eq:monodromieshf} are not closed under $O(n,n;\mathbb{Z})$-action on the torus fibre, but only under its geometric subgroup. Hence if we want an interpretation for the non-geometric T-duals need to introduce the $Q$-flux which encodes T-folds.

\begin{remark}[$Q$-flux]
We can perform a general $O(n,n;\mathbb{Z})$ transformation of the transition functions \eqref{eq:hflux} to the correspondence space $K\twoheadrightarrow M_0$ to obtain the following new ones
\begin{equation}
\begin{pmatrix}
 \xi_{(\alpha)}' \\[0.2cm]
 \widetilde{\xi}_{(\alpha)}'
 \end{pmatrix} = \left(\begin{array}{@{}cc@{}}
 1+n_{(\alpha\beta)}^{F\,\prime} &  n_{(\alpha\beta)}^{Q\,\prime} \\[0.2cm] 
 n_{(\alpha\beta)}^{H\,\prime} & 1-(n_{(\alpha\beta)}^{F\,\prime})^{\mathrm{T}}
 \end{array}\right) \begin{pmatrix}
 \xi_{(\beta)}' \\[0.2cm]
 \widetilde{\xi}_{(\beta)}'
 \end{pmatrix}.
\end{equation}
Notice we have a new monodromy matrix $n^{Q\, ij}_{(\alpha\beta)}$, which patches the dual $\widetilde{B}^{(0)ij}_{(\beta)} - \widetilde{B}^{(0)ij}_{(\alpha)} = n^{Q\, ij}_{(\alpha\beta)}$
and hence it is a cohomology class.
Therefore the dual background has a new flux, which we call \textit{locally non-geometric flux} or just $Q$-flux, given by the cohomology class $\big[Q^{(1)ij}\big]\in H^{1}(M_0,\mathbb{Z})$. The \v{C}ech cocycle of the $Q$-flux is then by construction
\begin{equation}
    \begin{aligned}
        Q^{(1)}\, &=\, {\di}\widetilde{B}^{(0)}_{(\alpha)}, \\
        \widetilde{B}^{(0)}_{(\beta)} - \widetilde{B}^{(0)}_{(\alpha)} \,&=\, n^Q_{(\alpha\beta)},
    \end{aligned}
\end{equation}
where the moduli field $\widetilde{B}^{(0)}_{(\alpha)}$ is the dual of the original moduli field ${B}^{(0)}_{(\alpha)}$.
\end{remark}

\noindent Let us resume everything in a familiar example. If we start from a background with only a $H$-flux $[H^{(1)}_{ij}]\in H^1(M_0,\mathbb{Z})$ and we perform a T-duality along the $i$-th circle $S^1_i$ of the torus fibre we get a background with $F$-flux $[F^{(1)\;\,j}_{\quad \,i}]\in H^1(M_0,\mathbb{Z})$ on the dual $[\widetilde{S}^1_i]\in H_1(M_0,\mathbb{Z})$ circle. If now we perform another T-duality along the $j$-th circle $S^1_j$ we end up with non-trivial $Q$-flux $[Q^{(1)ij}]\in H^1(M_0,\mathbb{Z})$ on the dual torus $[\widetilde{S}^1_i\times\widetilde{S}^1_j]\in H_2(M_0,\mathbb{Z})$ of the fibre. This argument can be condensed in the following commuting diagram
\vspace{-1.5cm}\begin{equation}
    \begin{tikzcd}[row sep=13ex, column sep=10ex]
    \left[H^{(1)}_{ij}\right] \arrow[out=180,in=90,loop, leftrightarrow, "\mathcal{O}_B"] \arrow[r, leftrightarrow, "\mathcal{T}_i"] \arrow[d, leftrightarrow, "\mathcal{T}_j"] & \left[F^{(1)\;\,j}_{\quad \,i}\right] \arrow[d, leftrightarrow, "\mathcal{T}_j"']\arrow[out=90,in=0,loop, leftrightarrow, "\mathcal{O}_{N_i}"] \\
    \left[F^{(1)i}_{\quad\;\; j}\right] \arrow[out=270,in=180,loop, leftrightarrow, "\mathcal{O}_{N_j}"] \arrow[r, leftrightarrow, "\mathcal{T}_i"] & \left[Q^{(1)ij}\right] \arrow[out=270,in=360,loop, leftrightarrow, "\mathcal{O}_{\widetilde{B}}"']
\end{tikzcd}\vspace{-1.5cm}
\end{equation}
where $\mathcal{O}_{B}:=\left(\begin{smallmatrix*}1&0\\b&1\end{smallmatrix*}\right)\in O(n,n;\mathbb{Z})$ is any  $B$-shift, while $\mathcal{O}_{N_i} := \mathcal{T}_i^\mathrm{T}\mathcal{O}_B\mathcal{T}_i$, $\mathcal{O}_{N_j} := \mathcal{T}_j^\mathrm{T}\mathcal{O}_B\mathcal{T}_j$ and $\mathcal{O}_{\widetilde{B}} := (\mathcal{T}_j\mathcal{T}_i)^\mathrm{T}\mathcal{O}_B(\mathcal{T}_j\mathcal{T}_i)$ are transformations which preserve the respective fluxes.

\begin{remark}[Geometrical interpretation of fluxes]
We remark again that the $H$-flux, the geometric flux and the locally non geometric flux are not put on the doubled space by hand like in the usual approaches, but they are \textit{topological properties} of the bundle gerbe $\mathscr{G}$ itself.
\end{remark}

\subsection{Atlas formulation}

We identified the isometries of our atlas $(\M,J,\omega)$ with changes of polarisation, i.e. with changes of T-duality frame. However, in general, it is not be possible to identify the image $(\M,\widetilde{J},\widetilde{\omega})$ of an isometry with the atlas of another bundle gerbe. In general, we can also obtain an almost para-complex structure $\widetilde{J}$ which is not integrable. In this case, the background described by the transformed atlas $(\M,\widetilde{J},\widetilde{\omega})$ is, then, a non-geometric background.\vspace{0.15cm}

\noindent For example, consider a bundle gerbe $\mathscr{G}\twoheadrightarrow M$ such that (1) its base $M\twoheadrightarrow M_0$ is a $T^n$-bundle with connection $e^i$ and (2) its $2$-form potential satisfies the equation $\mathcal{L}_{k_i}B_{(\alpha)}=0$ for vector fields $k_i$ dual to $e^i$. 
Thus, we have a T-duality of the form
\begin{equation}
    \begin{tikzcd}[row sep={9ex,between origins}, column sep={10ex,between origins}]
    & & \M \arrow[ld, shorten <= 0.1em, shorten >= 0.1em]\arrow[ddll, bend right=50, "\Phi"', two heads]\arrow[ddrr, dotted]  & & \\
    & \mathscr{G}\times_{M}K \arrow[dr, "\Pi"', shorten <= 0.1em, shorten >= 0.1em]\arrow[dl, "\pi", shorten <= 0.1em, shorten >= 0.1em] & &  \\
    \mathscr{G} \arrow[dr, "\Pi"', shorten <= 0.1em, shorten >= 0.1em] & & K \arrow[dl, "\widetilde{\pi}", shorten <= 0.1em, shorten >= 0.1em] & & [-2.5em] \begin{array}{c}\text{non-geometric}\\\text{background}\end{array}\arrow[ddll, dotted] \\
    & M \arrow[dr, "\pi"', shorten <= 0.1em, shorten >= 0.1em] & &  & \\
    & & M_0 & &
    \end{tikzcd}
\end{equation}
where the dotted arrows are not actual maps, but just qualitative relations.

\subsection{T-fold as global tensor hierarchy}

In this subsection we will briefly explain the global geometry of a T-fold, which is obtained by dimensionally reducing a bundle gerbe on a torus bundle spacetime.
Then we will explain how the geometric structure underlying the T-fold can be naturally interpreted as a particular case of the global tensor hierarchies we defined.
Crucially, these T-fold geometries cannot be obtained by gauging the algebra of local tensor hierarchies from chapter \ref{ch:4}. This will give practical motivation to the definition of the previous subsection.

\paragraph{The generalised correspondence space of T-duality.}\label{abeliantfold}
Let us start from the $T^n$-bundle $M\twoheadrightarrow M_0$, whose total space $M$ is equipped with a Riemannian metric $\bigsp{g}$ and gerbe structure with curvature $\bigsp{H}\in\Omega^3_{\mathrm{cl}}(M)$. In the following we will use the underlined notation for the fields living on the total space $M$. 
We can now use the principal connection $\xi\in\Omega^1(M,\mathbb{R}^n)$ of the torus bundle to expand metric $\bigsp{g}$ and curvature $\bigsp{H}$ in horizontal and vertical components respect to the fibration. We will obtain
\begin{align}
    \bigsp{g} \;&=\; g^{(2)} + g^{(0)}_{ij}\xi^i\odot\xi^j,\\
    \bigsp{H} \;&=\; H^{(3)} + H^{(2)}_{i}\wedge\xi^i  + \frac{1}{2} H^{(1)}_{ij}\wedge\xi^i\wedge\xi^j +\frac{1}{3!} H^{(0)}_{ijk}\xi^i\wedge\xi^j\wedge\xi^k,
\end{align}
where we can choose $H^{(3)},H^{(2)}_{i},H^{(1)}_{ij}, H^{(0)}_{ijk}$ as globally defined differential forms which are pullbacks from base manifold $M_0$, so that they do not depend on the torus coordinates. 
Recall that the differential data of a bundle gerbe on $M$ with curvature $\bigsp{H}\in\Omega^3_{\mathrm{cl}}(M)$ is embodied by a \v{C}ech cocycle $\big(\bigsp{B}_{(\alpha)},\bigsp{\Lambda}_{(\alpha\beta)},g_{(\alpha\beta\gamma)}\big)$ satisfying the following patching conditions:
\begin{equation}
    \begin{aligned}
        \bigsp{H} \;&=\; \bigsp{\di}\bigsp{B}_{(\alpha)},\\
        \bigsp{B}_{(\alpha)} - \bigsp{B}_{(\beta)} \;&=\; \bigsp{\di}\bigsp{\Lambda}_{(\alpha\beta)}, \\
        \bigsp{\Lambda}_{(\alpha\beta)} + \bigsp{\Lambda}_{(\beta\gamma)} + \bigsp{\Lambda}_{(\gamma\alpha)} \;&=\; \bigsp{\di} g_{(\alpha\beta\gamma)}, \\
        g_{(\alpha\beta\gamma)}-g_{(\beta\gamma\delta)}-g_{(\gamma\delta\alpha)}+g_{(\delta\alpha\beta)} \;&\in\; 2\pi\mathbb{Z}.
    \end{aligned}
\end{equation}
Now, on patches and two-fold overlaps of patches of a good cover of $M$ we can use the connection of the torus bundle to split the differential local data of the connection of the gerbe in horizontal and vertical part too. We obtain
\begin{equation}\label{splithv}
\begin{aligned}
    \bigsp{B}_{(\alpha)} &= B_{(\alpha)}^{(2)} + B_{(\alpha)i}^{(1)}\wedge\xi^i + \frac{1}{2} B_{(\alpha)ij}^{(0)} \xi^i\wedge\xi^j, \\
    \bigsp{\Lambda}_{(\alpha\beta)} &= \Lambda_{(\alpha\beta)}^{(1)} + \Lambda_{(\alpha\beta)i}^{(0)} \xi^i.
\end{aligned}
\end{equation}
The Bianchi identity of the gerbe on the total space $M$ reduces to the base $M_0$ as follows:
\begin{equation}\label{eq:bianchit}
    \bigsp{\di} \bigsp{H} \,=\, 0 \hspace{0.4cm} \implies \hspace{0.4cm} \left\{ \;
    \begin{aligned}
    \di H^{(0)}_{ijk} \,&=\, 0, \\[0.5em]
    \di H^{(1)}_{ij} + H^{(0)}_{ijk}\wedge F^k \,&=\, 0, \\[0.5em]
    \di H^{(2)}_{i} + H^{(1)}_{ij}\wedge F^j \,&=\, 0, \\[0.5em]
    \di H^{(3)} + H^{(2)}_i\wedge F^i \,&=\, 0,
    \end{aligned} 
    \right.
\end{equation}
where $\bigsp{\di}$ and $\di$ are respectively the exterior derivative on the total space $M$ and on the base manifold $M_0$. Analogously, the expression of the curvature of bundle gerbe on local patches becomes 
\begin{equation}
    \bigsp{H} \,=\, \bigsp{\di} \bigsp{B}_{(\alpha)} \hspace{0.4cm} \implies \hspace{0.4cm} \left\{ \;
    \begin{aligned}
    H^{(0)}_{ijk} \,&=\, \partial_{[i}B^{(0)}_{(\alpha)jk]}, \\[0.5em]
    H^{(1)}_{ij} \,&=\, \di B^{(0)}_{(\alpha)ij} - \partial_{[i}B^{(1)}_{(\alpha)j]}, \\[0.5em]
    H^{(2)}_i \,&=\, \di B^{(1)}_{(\alpha)i} + \partial_iB^{(2)}_{(\alpha)} - B^{(0)}_{(\alpha)ij} F^j ,\\[0.5em]
    H^{(3)} \,&=\, \di B_{(\alpha)}^{(2)} - B^{(1)}_{(\alpha)i}\wedge F^i,
    \end{aligned} 
    \right.
\end{equation}
where $\partial_i = \partial/\partial\theta^i$ is the derivative respect to the $i$-th coordinate of the torus fibre.
The patching conditions of the connection $2$-form on two-fold overlaps of patches are as following:
\begin{equation}\label{eq:bshifts}
    \bigsp{B}_{(\beta)} - \bigsp{B}_{(\alpha)} \,=\, \bigsp{\di} \bigsp{\Lambda}_{(\alpha\beta)} \hspace{0.4cm} \implies \hspace{0.4cm} \left\{ \;
    \begin{aligned}
    B_{(\beta)ij}^{(0)} - B_{(\alpha)ij}^{(0)} \,&=\, \partial_{[i}\Lambda^{(0)}_{(\alpha\beta)j]} ,\\[0.5em]
    B_{(\beta)i}^{(1)} - B_{(\alpha)i}^{(1)} \,&=\, \di \Lambda^{(0)}_{(\alpha\beta)i} - \partial_i\Lambda^{(1)}_{(\alpha\beta)} ,\\[0.5em]
    B_{(\beta)}^{(2)} - B_{(\alpha)}^{(2)} \,&=\, \di \Lambda^{(1)}_{(\alpha\beta)} +  \Lambda^{(0)}_{(\alpha\beta)i}F^i.
    \end{aligned} 
    \right.
\end{equation}
And the patching conditions of the $1$-forms on three-fold overlaps of patches become
\begin{equation*}
    \bigsp{\Lambda}_{(\alpha\beta)} + \bigsp{\Lambda}_{(\beta\gamma)} + \bigsp{\Lambda}_{(\gamma\alpha)} \,=\, \bigsp{\di} g_{(\alpha\beta\gamma)} \hspace{0.4cm} \implies \hspace{0.4cm} \left\{ \;
    \begin{aligned}
       \Lambda_{(\alpha\beta)i}^{(0)} + \Lambda_{(\beta\gamma)i}^{(0)} + \Lambda_{(\gamma\alpha)i}^{(0)} &= \partial_ig_{(\alpha\beta\gamma)} ,\\
    \Lambda_{(\alpha\beta)}^{(1)} + \Lambda_{(\beta\gamma)}^{(1)} + \Lambda_{(\gamma\alpha)}^{(1)} &= \di g_{(\alpha\beta\gamma)}.
    \end{aligned}\right.
\end{equation*}

\paragraph{The tensor hierarchy of a T-fold.}
To show that this geometric structure we obtained from the reduction of the bundle gerbe is a particular case of global tensor hierarchy, let us make the following redefinitions to match with the notation we used in chapter \ref{ch:4}:
\begin{equation}\label{eq:maptolanguage}
    \begin{aligned}
    \mathcal{F}^I_{(\alpha)} \,&:=\, \begin{pmatrix}\delta^i_j & 0 \\[0.6em] B^{(0)}_{(\alpha)ij} & \delta^j_i \end{pmatrix} \begin{pmatrix} F^j \\[0.6em] H^{(2)}_j \end{pmatrix}, \; &\mathcal{H}\,&:=\, H^{(3)}, \\[0.4em]
    \mathcal{A}^I_{(\alpha)} \,&:=\, \begin{pmatrix} A_{(\alpha)} \\[0.6em] B^{(1)}_{(\alpha)} \end{pmatrix} , \; & \mathcal{B}_{(\alpha)}\,&:=\, B^{(2)}_{(\alpha)}, \\[0.3em]
    \lambda_{(\alpha\beta)}^I \,&:=\, \begin{pmatrix} \lambda_{(\alpha\beta)}^i \\[0.7em] \Lambda^{(0)}_{(\alpha\beta)i} \end{pmatrix} , \; &\Xi_{(\alpha\beta)} \,&:=\, \Lambda^{(1)}_{(\alpha\beta)}.
    \end{aligned}
\end{equation}
where the local $1$-forms $A_{(\alpha)}^i$ and scalars $\lambda_{(\alpha\beta)}^i$ are respectively the local potential and the transition functions of the original torus bundle $M\twoheadrightarrow M_0$.
Since our fields are assumed to be strong constrained, we well have simply $\mathfrak{D}^I=(0,\,\partial_i)$ with $\partial_i=\partial/\partial\theta^i$ on horizontal forms. The Bianchi equation \eqref{eq:bianchit} of the gerbe curvature, together with the Bianchi equation $\di F = 0$ of the curvature of the torus bundle can now be equivalently rewritten as
\begin{equation}
    \begin{aligned}
    \di \mathcal{F}_{(\alpha)} \;&=\; 0, \\
    \di \mathcal{H} - \frac{1}{2}\langle \mathcal{F}_{(\alpha)} \;\overset{\wedge}{,}\; \mathcal{F}_{(\alpha)} \rangle\;&=\;  0,
    \end{aligned}
\end{equation}
which are a particular case of the Bianchi equations of a tensor hierarchy. We can now rewrite all the patching conditions in the following equivalent form:
\begin{equation}\label{eq:tfoldpat}
    \begin{aligned}
    \mathcal{F}_{(\alpha)} \;&=\; \di \mathcal{A}_{(\alpha)}  + \mathfrak{D}\mathcal{B}_{(\alpha)} ,\\
    \mathcal{H} \;&=\;  \di \mathcal{B}_{(\alpha)} + \frac{1}{2}\langle \mathcal{A}_{(\alpha)} \,\overset{\wedge}{,}\, \mathcal{F}_{(\alpha)}\rangle ,\\[0.8em]
    \mathcal{A}_{(\alpha)} - \mathcal{A}_{(\beta)} \;&=\; \di\lambda_{(\alpha\beta)} + \mathfrak{D}\Xi_{(\alpha\beta)} ,\\
    \mathcal{B}_{(\alpha)} - \mathcal{B}_{(\beta)} \;&=\; \di \Xi_{(\alpha\beta)} - \langle\lambda_{(\alpha\beta)},\mathcal{F}_{(\alpha)}\rangle ,\\[0.8em]
    \lambda_{(\alpha\beta)}+\lambda_{(\beta\gamma)}+\lambda_{(\gamma\alpha)} \;&=\; \mathfrak{D} g_{(\alpha\beta\gamma)} ,\\
    \Xi_{(\alpha\beta)} + \Xi_{(\beta\gamma)} + \Xi_{(\gamma\alpha)} \;&=\; \di g_{(\alpha\beta\gamma)} ,\\[0.8em]
    g_{(\alpha\beta\gamma)} - g_{(\beta\gamma\delta)} +  g_{(\gamma\delta\alpha)} -  g_{(\delta\alpha\beta)}\;&\in\;2\pi\mathbb{Z},
    \end{aligned}
\end{equation}
which, at first look, appears a particular (and strong constrained) case of the global tensor hierarchy in \eqref{eq:tensorhipatched}. However we will see in the following that it is not completely the case. This will motivate more the identification of a T-fold with an element of $\tenhie^{\,T^n}_{\!\mathrm{sc}}\!(M_0)$. 
\vspace{0.25cm}

\noindent It is well-known that, to be T-dualizable, the string background we started with must satisfy the T-duality condition $\mathcal{L}_{\partial_i}\bigsp{H}=0$ on the curvature of the bundle gerbe. From now on we will assume a simple solution for this equation: the invariance of Kalb-Ramond field under the torus action. In other words we will require $\mathcal{L}_{\partial_i}\bigsp{B}_{(\alpha)}=0$, but the other differential data $\bigsp{\Lambda}_{(\alpha\beta)}$, $g_{(\alpha\beta\gamma)}$ of the gerbe are still allowed to depend on the torus coordinates. Notice that this immediately implies that $\mathcal{F}_{(\alpha)} = \di \mathcal{A}_{(\alpha)}$ in equation \eqref{eq:tfoldpat}.

\paragraph{Topology of the tensor hierarchy of a T-fold.}
Now it is important to show that the curvature $\mathcal{F}_{(\alpha)}^I$\textit{ is not in general the curvature of a }$T^{2n}$\textit{-bundle on the }$(d-n)$\textit{-dimensional base manifold }$M_0$. To see this let us split $\mathcal{F}_{(\alpha)}^I=(\mathcal{F}_{(\alpha)}^i,\,\widetilde{\mathcal{F}}_{(\alpha)i})$ and consider $[\pi_\ast H] \in H^2(M,\mathbb{Z}^n)$. We can see that
\begin{equation}\label{eq:bigneq}
    [(\pi_\ast H)_i] \,=\, \left[H^{(2)}_i-H^{(1)}_{ij}\xi^j\right] \;\neq\; \left[H^{(2)}_i+B^{(0)}_{(\alpha)ij}F^j\right] \,=\, [\widetilde{\mathcal{F}}_{(\alpha)i}].
\end{equation}
Notice that the inequality \eqref{eq:bigneq} becomes an equality if and only if $H^{(1)}$ is an exact form on the base manifold $M_0$, as we have seen. In this case we would have $H^{(0)}=0$ and $H^{(1)}=\di B^{(0)}$, where $B^{(0)}$ would be a global $\wedge^2\mathbb{R}^n$-valued scalar on $M_0$, which indeed implies $[\di B^{(0)}_{ij}\xi^j]=-[B^{(0)}_{ij}F^j]$. As explained in the previous section, this particular case corresponds to \textit{geometric T-duality}, which is the case where the T-dual spacetime is a well-defined manifold and not a non-geometric T-fold. Thus geometric T-duality is exactly the special case where $\mathcal{F}_{(\alpha)}^I$ is the curvature of a $T^{2n}$-bundle on $M_0$. But what is the geometric picture for a T-fold? 
\vspace{0.25cm}

\noindent In the T-fold case, as we have seen, we can think about $[\pi_\ast H] \in H^2(M_0,\mathbb{Z}^n)$ as the curvature of a $T^n$-bundle $K\twoheadrightarrow M$ over the total spacetime $M$. The total space $K$ is called \textit{generalised correspondence space}.
\begin{equation}
    \begin{tikzcd}[row sep=5ex, column sep=5ex]
    & K \arrow[d, two heads] & & \;\;\;\,\mathrm{c}_1(K)=[\pi_\ast H], \\
    & M \arrow[d, two heads, "\pi"] \arrow[r] & \mathbf{B}\widetilde{T}^n & \mathrm{c}_1(M)=[F]. \\
    & M_0 \arrow[r] & \mathbf{B}T^n &
    \end{tikzcd}
\end{equation}

\noindent Now the picture of the doubled torus bundle holds only locally on $U_{\alpha}\times T^{2n}$ for each patch $U_{\alpha}\subset M_0$. Now, if we call collectively $\Theta_{(\alpha)}^I:=\big(\theta^i_{(\alpha)},\,\widetilde{\theta}_{(\alpha)i}\big)$ the $2n$ coordinates of the fibre $T^{2n}$, we can construct the Ehresmann connection of any local doubled torus bundle $U_{\alpha}\times T^{2n}$ by 
\begin{equation}\label{eq:localconn}
    \di\Theta_{(\alpha)}^I +\mathcal{A}^I_{(\alpha)} \,\;\in \;\Omega^{1}\!\left(U_{\alpha}\times T^{2n}\right).
\end{equation}
The geometrical meaning of the curvature $\mathcal{F}_{(\alpha)}\in\Omega^{2}_{\mathrm{cl}}\!\left(U_{\alpha}\times T^{2n}\right)$ is being at every patch the curvature of the local torus bundle $U_{\alpha}\times T^{2n}$, even if these ones are not globally glued to be a $T^n$-bundle on $M_0$. This corresponds indeed to the well-known fact that a \textit{non-geometry is a global property}. In fact we can always perform geometric T-duality if we restrict ourselves on any local patch: the problem is that all these T-dualised patches will in general not glue together. 
\vspace{0.25cm}

\noindent As derived by \cite{BelHulMin07} and more recently by \cite{NikWal18}, T-folds are characterised by a \textit{monodromy matrix} cocycle $\big[n_{(\alpha\beta)}\big]$, which is a collection of an anti-symmetric integer-valued matrix $n_{(\alpha\beta)}$ at each two-fold overlap of patches, satisfying the cocycle condition $n_{(\alpha\beta)}+n_{(\beta\gamma)}+n_{(\gamma\alpha)}=0$ on each three-fold overlap of patches. The monodromy matrix cocycle is nothing but the gluing data for the local $B^{(0)}_{(\alpha)}$ moduli fields, i.e. it encodes integer $B$-shifts $B^{(0)}_{(\alpha)}-B^{(0)}_{(\beta)}=n_{(\alpha\beta)}$ on each two-fold overlap of patches. This arises from equation \eqref{eq:bshifts} combined with the T-dualizability condition $\mathcal{L}_{\partial_i}\bigsp{B}_{(\alpha)}$. The consequence of the presence of the monodromy matrix cocycle is that the local connections \eqref{eq:localconn} are glued on two-fold overlaps of patches $(U_{\alpha}\cap U_{\beta})\times T^{2n}$ by
\begin{equation}
    \big(\di\Theta_{(\alpha)} +\mathcal{A}_{(\alpha)}\big)^I \; = \; \big(e^{n_{(\alpha\beta)}}\big)^I_{\;J} \, \big(\di\Theta_{(\beta)} +\mathcal{A}_{(\beta)} \big)^J.
\end{equation}
This immediately comes from the definition of these connections in equation \eqref{eq:maptolanguage}. Moreover this immediately implies that the curvature is glued by the monodromy matrix cocycle too as 
\begin{equation}
\mathcal{F}_{(\alpha)}^I \,=\, \big(e^{n_{(\alpha\beta)}}\big)^I_{\;J} \,\mathcal{F}_{(\beta)}^J.
\end{equation}
In this sense, a T-fold is patched by a cocycle $e^{n_{(\alpha\beta)}}\in O(n,n;\mathbb{Z})$ valued in the T-duality group.
\vspace{0.2cm}

\noindent If, instead, we want to look at the T-fold as a globally defined $T^n$-bundle $K\twoheadrightarrow M$ with first Chern class $[\pi_\ast H]\in H^2(M,\mathbb{Z}^n)$, we can easily construct its connection by noticing that the following $1$-form is global on the total space $K$ of the bundle:
\begin{equation}
    \Big(e^{-B_{(\alpha)}^{(0)}}\Big)^{\! I}_{\;J} \big(\di\Theta_{(\alpha)} +\mathcal{A}_{(\alpha)}\big)^J \;\, = \,\; \Big(e^{-B_{(\beta)}^{(0)}}\Big)^{\! I}_{\;J} \big(\di\Theta_{(\beta)} +\mathcal{A}_{(\beta)} \big)^J.
\end{equation}
We can thus define the global $1$-form
\begin{equation}\label{eq:defrealconnection}
    \Xi^I \;:=\; \Big(e^{-B_{(\alpha)}^{(0)}}\Big)^{\! I}_{\;J} \big(\di\Theta_{(\alpha)} +\mathcal{A}_{(\alpha)}\big)^J \;\;\in\,\Omega^1(K,\mathbb{R}^{2n}),
\end{equation}
whose first $n$ components are just the pullback of connection $\Xi^i=\xi^i$ of spacetime $M\twoheadrightarrow M_0$ and whose last $n$ components $\Xi_i$ are the wanted connection of the generalised correspondence space $K\twoheadrightarrow M$. As desired, the differential $\di\Xi_i$ on $K$ gives the pullback on $K$ of the globally-defined curvature $\pi_\ast H\in\Omega^2_{\mathrm{cl}}(M,\mathbb{R}^n)$.
\vspace{0.25cm}

\noindent Similarly to $\mathcal{F}_{(\alpha)}$, the moduli field $\mathcal{G}_{(\alpha)IJ}$ of the generalised metric is not a global $O(n,n)$-valued scalar on the base manifold $M_0$, but it is glued on two-fold overlaps of patches $U_\alpha\cap U_\beta \subset M_0$ by the integer $B$-shifts encoded by the monodromy matrix $n_{(\alpha\beta)}$ of the T-fold as
\begin{equation}
    \mathcal{G}_{(\alpha)IJ} \;\,=\,\; \big(e^{n_{(\alpha\beta)}}\big)^K_{\;I}\; \mathcal{G}_{(\beta)KL}\; \big(e^{n_{(\alpha\beta)}}\big)^L_{\;J}.
\end{equation}
Only the $3$-form field $\mathcal{H}$ of the tensor hierarchy, as we have seen, is a globally defined (but not closed) differential form on the $(d-n)$-dimensional base manifold $M_0$.

\begin{figure}[!ht]\centering
\vspace{0.2cm}
\tikzset{every picture/.style={line width=0.75pt}} 
\begin{tikzpicture}[x=0.75pt,y=0.75pt,yscale=-1,xscale=1]
\draw  [fill={rgb, 255:red, 0; green, 0; blue, 0 }  ,fill opacity=0.05 ] (15.22,311.35) .. controls (6.75,279.73) and (46.17,241.7) .. (103.27,226.4) .. controls (160.36,211.1) and (213.52,224.33) .. (221.99,255.95) .. controls (230.46,287.56) and (191.04,325.59) .. (133.94,340.89) .. controls (76.85,356.19) and (23.69,342.96) .. (15.22,311.35) -- cycle ;
\draw  [fill={rgb, 255:red, 0; green, 0; blue, 0 }  ,fill opacity=0.05 ] (147.59,255.24) .. controls (156.06,223.63) and (209.21,210.4) .. (266.31,225.7) .. controls (323.41,241) and (362.83,279.03) .. (354.35,310.65) .. controls (345.88,342.26) and (292.73,355.49) .. (235.63,340.19) .. controls (178.54,324.89) and (139.12,286.86) .. (147.59,255.24) -- cycle ;
\draw  [dash pattern={on 0.84pt off 2.51pt}]  (8.65,153.88) -- (80.09,286.94) ;
\draw  [dash pattern={on 0.84pt off 2.51pt}]  (80.09,286.94) -- (143.12,150.37) ;
\draw  [dash pattern={on 0.84pt off 2.51pt}]  (254.33,169.67) -- (291.29,288.34) ;
\draw  [dash pattern={on 0.84pt off 2.51pt}]  (291.29,288.34) -- (324.33,162.33) ;
\draw   (8.65,146.1) .. controls (8.65,128.01) and (38.58,113.33) .. (75.49,113.33) .. controls (112.41,113.33) and (142.33,128.01) .. (142.33,146.1) .. controls (142.33,164.2) and (112.41,178.88) .. (75.49,178.88) .. controls (38.58,178.88) and (8.65,164.2) .. (8.65,146.1) -- cycle ;
\draw   (250.67,142.5) .. controls (250.67,105.96) and (267.31,76.33) .. (287.83,76.33) .. controls (308.36,76.33) and (325,105.96) .. (325,142.5) .. controls (325,179.04) and (308.36,208.67) .. (287.83,208.67) .. controls (267.31,208.67) and (250.67,179.04) .. (250.67,142.5) -- cycle ;
\draw  [draw opacity=0] (103.66,139.28) .. controls (100.05,147.07) and (88.51,152.77) .. (74.83,152.77) .. controls (61.17,152.77) and (49.65,147.1) .. (46.02,139.33) -- (74.83,134.1) -- cycle ; \draw   (103.66,139.28) .. controls (100.05,147.07) and (88.51,152.77) .. (74.83,152.77) .. controls (61.17,152.77) and (49.65,147.1) .. (46.02,139.33) ;
\draw  [draw opacity=0] (53.07,145.7) .. controls (57.98,140.67) and (65.6,137.44) .. (74.17,137.44) .. controls (83.55,137.44) and (91.81,141.32) .. (96.6,147.19) -- (74.17,159.05) -- cycle ; \draw   (53.07,145.7) .. controls (57.98,140.67) and (65.6,137.44) .. (74.17,137.44) .. controls (83.55,137.44) and (91.81,141.32) .. (96.6,147.19) ;
\draw  [draw opacity=0] (295.98,113.17) .. controls (295.66,122.96) and (292.17,130.68) .. (287.91,130.68) .. controls (283.66,130.68) and (280.17,122.97) .. (279.85,113.19) -- (287.91,111.72) -- cycle ; \draw   (295.98,113.17) .. controls (295.66,122.96) and (292.17,130.68) .. (287.91,130.68) .. controls (283.66,130.68) and (280.17,122.97) .. (279.85,113.19) ;
\draw  [draw opacity=0] (281.67,125.07) .. controls (282.97,119.07) and (285.2,115.11) .. (287.74,115.11) .. controls (290.48,115.11) and (292.87,119.75) .. (294.09,126.58) -- (287.74,137.05) -- cycle ; \draw   (281.67,125.07) .. controls (282.97,119.07) and (285.2,115.11) .. (287.74,115.11) .. controls (290.48,115.11) and (292.87,119.75) .. (294.09,126.58) ;
\draw  [color={rgb, 255:red, 208; green, 2; blue, 27 }  ,draw opacity=1 ][line width=0.75]  (67.33,165.27) .. controls (67.33,158.31) and (70.98,152.67) .. (75.49,152.67) .. controls (80,152.67) and (83.66,158.31) .. (83.66,165.27) .. controls (83.66,172.23) and (80,177.88) .. (75.49,177.88) .. controls (70.98,177.88) and (67.33,172.23) .. (67.33,165.27) -- cycle ;
\draw  [color={rgb, 255:red, 208; green, 2; blue, 27 }  ,draw opacity=1 ][line width=0.75]  (279.67,169.5) .. controls (279.67,147.87) and (283.32,130.33) .. (287.83,130.33) .. controls (292.34,130.33) and (296,147.87) .. (296,169.5) .. controls (296,191.13) and (292.34,208.67) .. (287.83,208.67) .. controls (283.32,208.67) and (279.67,191.13) .. (279.67,169.5) -- cycle ;
\draw  [color={rgb, 255:red, 74; green, 144; blue, 226 }  ,draw opacity=1 ] (26.6,144.76) .. controls (26.6,132.38) and (49.08,122.33) .. (76.8,122.33) .. controls (104.53,122.33) and (127,132.38) .. (127,144.76) .. controls (127,157.15) and (104.53,167.19) .. (76.8,167.19) .. controls (49.08,167.19) and (26.6,157.15) .. (26.6,144.76) -- cycle ;
\draw  [color={rgb, 255:red, 74; green, 144; blue, 226 }  ,draw opacity=1 ] (270,130.17) .. controls (270,111.67) and (278.36,96.67) .. (288.67,96.67) .. controls (298.98,96.67) and (307.33,111.67) .. (307.33,130.17) .. controls (307.33,148.67) and (298.98,163.67) .. (288.67,163.67) .. controls (278.36,163.67) and (270,148.67) .. (270,130.17) -- cycle ;
\draw [line width=1.5]    (158.67,143.67) -- (229,143.67) ;
\draw [shift={(233,143.67)}, rotate = 180] [fill={rgb, 255:red, 0; green, 0; blue, 0 }  ][line width=0.08]  [draw opacity=0] (6.97,-3.35) -- (0,0) -- (6.97,3.35) -- cycle    ;
\draw [shift={(154.67,143.67)}, rotate = 0] [fill={rgb, 255:red, 0; green, 0; blue, 0 }  ][line width=0.08]  [draw opacity=0] (6.97,-3.35) -- (0,0) -- (6.97,3.35) -- cycle    ;
\draw (83.05,314.99) node  [font=\small]  {$U_{\alpha }$};
\draw (292.26,312.72) node  [font=\small]  {$U_{\beta }$};
\draw (184.59,263.37) node  [font=\small]  {$U_{\alpha } \cap U_{\beta }$};
\draw (145.67,123) node [anchor=north west][inner sep=0.75pt]  [font=\footnotesize]  {$e^{n_{( \alpha \beta )}} \!\!\in\! O(n,n;\mathbb{Z})$};
\draw (175,150) node [anchor=north west][inner sep=0.75pt]  [font=\footnotesize]  {gluing};
\draw (-23,135) node [anchor=north west][inner sep=0.75pt]  [font=\small]  {$T^{2n}$};
\draw (334,135) node [anchor=north west][inner sep=0.75pt]  [font=\small]  {$T^{2n}$};
\draw (46,84) node [anchor=north west][inner sep=0.75pt]  [font=\small]  {$\mathcal{F}_{( \alpha )} ,\,\mathcal{G}_{( \alpha )}$};
\draw (258.67,48) node [anchor=north west][inner sep=0.75pt]  [font=\small]  {$\mathcal{F}_{( \beta )},\,\mathcal{G}_{( \beta )}$};
\end{tikzpicture}\caption[Gluing conditions of a T-fold]{The gluing conditions on the $T^{2n}$ fibres and the fields $\mathcal{F}_{( \alpha )} $, $\mathcal{G}_{( \alpha )}$ for a simple T-fold.}\end{figure}
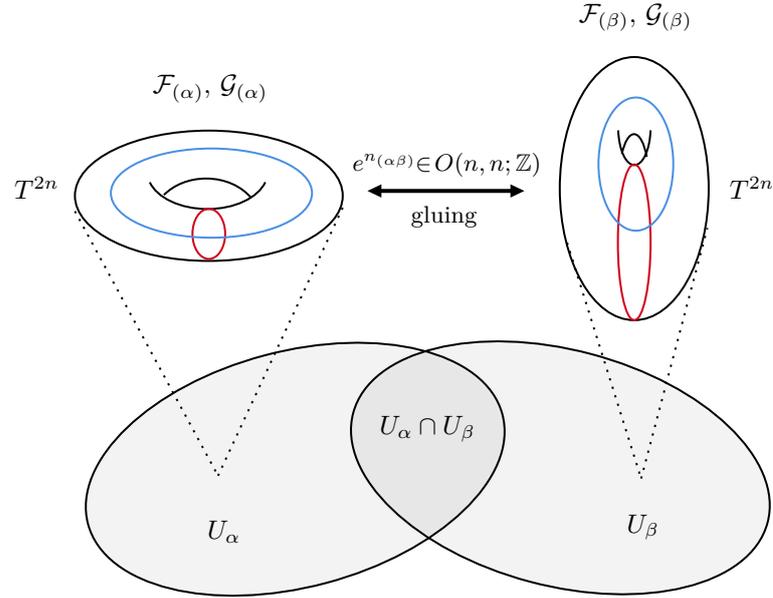

\paragraph{T-duality on the tensor hierarchy of a T-fold.}
Given any element $\mathcal{O}\in O(n,n;\mathbb{Z})$ of the T-duality group, we can see that there is a natural action on the local $2n$ coordinates of the torus by $\Theta_{(\alpha)}^I\mapsto \mathcal{O}^{I}_{\;J}\Theta_{(\alpha)}^J$ and on the fields of the tensor hierarchy by
\begin{equation}
     \mathcal{F}_{(\alpha)}^I \,\mapsto\, \mathcal{O}^{I}_{\;J}\mathcal{F}_{(\alpha)}^J, \qquad \mathcal{H} \,\mapsto\, \mathcal{H}, \qquad \mathcal{G}_{(\alpha)IJ} \,\mapsto\, \mathcal{O}^{K}_{\;\,I}\,\mathcal{G}_{(\alpha)KL}\,\mathcal{O}^{L}_{\;J}.
\end{equation}
If we include also the dimensionally reduced global pseudo-Riemannian metric $g$ on $M_0$, we can sum how the fields transform under T-duality in the table \ref{tab:t}.\vspace{0.2cm}

\begin{table}[h!]\begin{center}
\begin{center}
 \begin{tabular}{|| c |  c | c||} 
 \hline
 \multicolumn{3}{||c||}{Fields under $O(n,n;\mathbb{Z})$-action} \\
 \hline
 \hspace{0.8cm}Singlet rep.\hspace{0.8cm} & Fundamental rep. & \hspace{0.5cm}Adjoint rep.$\quad$ \\ [0.5ex] 
 \hline
 $g, \mathcal{B}_{(\alpha)}$ & $\mathcal{A}^I_{(\alpha)}$ & $\mathcal{G}_{(\alpha)IJ}$   \\[0.8ex]  
 \hline
\end{tabular}
\end{center}
\caption{\label{tab:t}A summary of the fields transforming under T-duality.}\vspace{0.0cm}
\end{center}\end{table}

\noindent Clearly, together with the torus coordinates, this will transform our monodromy matrix cocycle $(e^{n_{(\alpha\beta)}})^I_{\;J}$ to a new $O(d,d;\mathbb{Z})$-cocycle $\mathcal{O}^{I}_{\;K}(e^{n_{(\alpha\beta)}})^K_{\;J}$, but, since the fields depend only on the physical coordinates, the global geometric interpretation in the generalised correspondence space does not change.\vspace{0.25cm}

\noindent On the other hand, if we want an intrinsic and "non-doubled" description of the T-dual string backgrounds, as explained by \cite{Bou08}, we obtain a \textit{non-commutative} $T^n$\textit{-bundle} on $M_0$ which is classified by the cocycle $\big[n_{(\alpha\beta)}\big]\in H^1(M_0,\,\wedge^2\mathbb{Z}^2)$ on its base manifold. \vspace{0.25cm}

\noindent We can finally summarize these two equivalent descriptions within the table \ref{tab:t-fold}. \newpage

\begin{table}[h!]\begin{center}
\vspace{0.25cm}
  \setlength\extrarowheight{1.5ex}
 \begin{tabular}{|c||c|c||} 
 \hline
 & \multicolumn{2}{c||}{\makecell{\textbf{T-fold}}}\\[0.5ex] 
 \hline \hline
 & $\qquad$\makecell{Tensor\\hierarchy\\description}$\qquad$ & \makecell{Generalised\\correspondence space\\description} \\[0.5ex] 
 \hline\hline
 \makecell{Space of\\the T-fold} & \makecell{Local $U_\alpha\times T^{2n}$\\glued by $e^{n_{(\alpha\beta)}}$ } & \makecell{Global $T^n$-bundle\\$K\twoheadrightarrow M$\\ on spacetime $M$} \\[0.5ex]
 \hline
 \makecell{Connection\\data} & \makecell{$\di\Theta_{(\alpha)}^I+\mathcal{A}^I_{(\alpha)}$}  & \makecell{$\Xi^I$} \\[0.5ex]
 \hline
 \makecell{Curvature\\data} & \makecell{$\mathcal{F}^I_{(\alpha)}$}  & \makecell{$F^i,\,(\pi_\ast \bigsp{H})_i$} \\[0.5ex]
 \hline
 \makecell{Main\\ feature} & \makecell{Fields of the \\tensor hierarchy\\are manifest}  & \makecell{Topology (including\\non-geometry)\\is manifest} \\[0.5ex]
 \hline
 \makecell{Relation} & \multicolumn{2}{c||}{Related by a $e^{B_{(\alpha)}^{(0)}}$-twist on each patch}\\[0.5ex]
 \hline
\end{tabular}
\caption{\label{tab:t-fold}A brief summary of the two descriptions for the T-fold.}
\end{center}\end{table}

\paragraph{T-fold as global tensor hierarchy.}
There is an important subtlety in this discussion: \textit{the most general tensor hierarchy of the T-fold does not arise by gauging the tensor hierarchy $2$-algebra}. This means that \textit{the tensor hierarchy of the T-fold is not a higher gauge theory on $M_0$}. This is due to the presence of the monodromy matrix cocycle $n_{(\alpha\beta)}$, which glues the curvatures by $\mathcal{F}_{(\alpha)}^I = \big(e^{n_{(\alpha\beta)}}\big)^{\!I}_{\,J} \mathcal{F}_{(\beta)}^J$ and which is not a gauge transformation of the connection.
\vspace{0.25cm}

\noindent Thus the most general global formalisation of the tensor hierarchy of a T-fold must then be given by the dimensional reduction of a bundle gerbe, i.e. by a diagram of the form
\begin{equation}
    \tenhie^{\,T^n}_{\!\mathrm{sc}}\!(M_0) \;=\; \left\{\begin{tikzcd}[row sep=7ex, column sep=5ex]
    & \left[T^n,\mathbf{B}^2U(1)\right]\!/T^n \arrow[d]\\
    M_0 \arrow[ru]\arrow[r, ""] & \mathbf{B}T^n
    \end{tikzcd}\right\}.
\end{equation}

\section{Geometrisation of general abelian T-duality}\label{td4}

Until now we investigated simple examples of T-dualities. In this section we want to give some insight of the general case by starting from a result by \cite{BelHulMin07}. In the previous sections we assumed the invariance of the gerbe connection under the principal torus action. This condition can be immediately relaxed by requiring just $\mathcal{L}_{k_i}\bigsp{H}=0$ and hence that the connection gauge transforms like $\mathcal{L}_{k_i}\bigsp{B}_{(\alpha)} = \mathrm{d}\eta_{(\alpha)}$ for some local $1$-form $\eta_{(\alpha)}$ under it.

\begin{theorem}[Bundle gerbe with $T^n$-invariant curvature]\label{lemma:flatlie}
It was proven by \cite{BelHulMin07} that the principal $T^n$-action on $M$ can be lifted to a $T^n$-action on a gerbe with $T^n$-invariant curvature. The local data of this action on a good cover $\{U_\alpha\}$ for $M$ are given by a collection of $1$-forms $\eta_{(\alpha)}\in\Omega^1(U_\alpha,\mathbb{R}^n)$ on patches, of functions $\eta_{(\alpha\beta)}\in\Coo(U_\alpha\cap U_\beta,\mathbb{R}^n)$ on two-fold overlaps of patches and of constants $c_{(\alpha\beta\gamma)}\in\mathbb{R}^n$ on three-fold overlaps of patches which satisfy
\begin{equation}\label{eq:hullaction}
    \begin{aligned}
    \mathcal{L}_{k_i}\bigsp{B}_{(\alpha)} \,&=\, \mathrm{d}\eta_{(\alpha) i} \\
    \mathcal{L}_{k_i}\bigsp{\Lambda}_{(\alpha\beta)}  \,&=\, \eta_{(\beta) i} - \eta_{(\alpha) i} -\mathrm{d}\eta_{(\alpha\beta) i}\\
    \mathcal{L}_{k_i}\bigsp{G}_{(\alpha\beta\gamma)} \,&=\, \eta_{(\alpha\beta) i} + \eta_{(\beta\gamma) i} + \eta_{(\gamma\alpha) i} + c_{(\alpha\beta\gamma) i} \\
    & \quad\;\; c_{(\alpha\beta\gamma) i} - c_{(\beta\gamma\delta) i} + c_{(\gamma\delta\alpha) i} - c_{(\delta\alpha\beta) i} \,\in\, 2\pi\mathbb{Z}.
    \end{aligned}
\end{equation}
\end{theorem}
\begin{proof}
We can see $\mathcal{L}_{k_i}\bigsp{H}=0$ as the curvature of a flat bundle gerbe on $M$ for any $i=1,\dots,n$. Therefore, we can obtain the gluing conditions for the differential data $(\mathcal{L}_{k_i}\bigsp{B}_{(\alpha)},\,\mathcal{L}_{k_i}\bigsp{\Lambda}_{(\alpha\beta)},\,\mathcal{L}_{k_i}\bigsp{G}_{(\alpha\beta\gamma)})$ from equation \eqref{eq:flatgerbe}.
\end{proof}

\noindent Notice that the cohomology class $[c_{(\alpha\beta\gamma)}]\in H^2(M,T^n)$ can be interpreted as the flat holonomy class (definition \ref{eq:holonomyclass}) of this flat bundle gerbe. 

\vspace{0.2cm}

\noindent The case in which the bundle gerbe connection simply satisfies $\mathcal{L}_{k_i}\bigsp{B}_{(\alpha)}=0$ on each patch $U_\alpha$ is clearly a particular case of the general case \eqref{eq:hullaction}.

\begin{remark}[Underlying generalised vector]\label{rem:translation}
If the holonomy class $[c_{(\alpha\beta\gamma)}]$ is trivial, then the collection $(k_i,\, \eta_{(\alpha\beta) i},\, \eta_{(\alpha)i})$ is exactly the \v{C}ech data of a section of the stack $T\mathscr{G}$ of the form \eqref{eq:localsymdata} from lemma \ref{thm:dtbild}. If we reparametrize the scalars by $\hat{\eta}_{(\alpha\beta) i} := \eta_{(\alpha\beta) i} - \iota_{k_i}\bigsp{\Lambda}_{(\alpha\beta)}$ according to lemma \ref{thm:dtbild} we get the local data of a global generalised vector $\mathbbvar{k}_i:=(k_i+\eta_{(\alpha)i},\, \hat{\eta}_{(\alpha\beta) i})\in\Gamma\big(M,\mathfrak{at}(\mathscr{G})\big)$.
\end{remark}

\noindent Notice this is an application where our global definition of doubled vector (see lemma \ref{thm:dtbild}), which is equipped with scalars $\hat{\eta}_{(\alpha\beta)}^i$ on two-fold overlaps of patches, is indispensable. Indeed global differential T-duality is formalised in terms of our generalised vectors $\mathbbvar{k}_i:=(k_i+\eta_{(\alpha)i},\, \hat{\eta}_{\alpha\beta i})$ (see definition \ref{def:courantalg}), but not of the usual generalised vectors from Generalised Geometry.

\begin{definition}[Fundamental generalised vector]
The principal torus action on $M$ induces a Lie algebra homomorphism $\mathfrak{u}(1)^n \rightarrow\mathfrak{X}(M)$ which maps an element of the algebra to a fundamental vector $k_i$. Thus this can be lifted to a Lie $2$-algebra homomorphism
\begin{equation}\label{eq:alghom}
    \mathfrak{u}(1)^n \longrightarrow \Gamma\big(M,\mathfrak{at}(\mathscr{G})\big)
\end{equation}
which maps an element of the algebra in a generalised vector $\mathbbvar{k}_i:=(k_i+\eta_{(\alpha)}^i,\, \hat{\eta}_{(\alpha\beta)}^i)$, where $k_i$ is the fundamental vector of the action $\mathfrak{u}(1)^n \rightarrow\mathfrak{X}(M)$ and the local data $(\eta_{(\alpha)}^i,\, \hat{\eta}_{(\alpha\beta)}^i)$ are defined by conditions \eqref{eq:hullaction} with redefinition $\hat{\eta}_{(\alpha\beta) i} := \eta_{(\alpha\beta) i} - \iota_{k_i}\bigsp{\Lambda}_{(\alpha\beta)}$ of remark \ref{rem:translation}. 
\end{definition}

\noindent It is easy to check that $\llbracket\mathbbvar{k}_i,\mathbbvar{k}_j\rrbracket = 0$ for $i,j=1,\dots,n$.

\begin{definition}[Killing generalised vector]
A generalised vector $\mathbbvar{k}$ is Killing if $\llbracket \mathbbvar{k}, \mathbbvar{e}_\mathrm{I} \rrbracket =0$. We will use the symbol $\mathfrak{iso}(\mathscr{G},\mathcal{G})\subset\mathfrak{at}(\mathcal{M})$ for the sub-$2$-algebra of Killing generalised vectors.
\end{definition}

\begin{theorem}[General T-duality on the doubled space]\label{thm:generaldifftduality}
There exists a $T^n$-bundle $K\rightarrow M$ with first Chern class $\mathrm{c}_1(K)=[\pi_\ast \bigsp{H}]=[H^{(2)}-\langle H^{(1)}\wedge\xi\rangle]\in H^2(M,\mathbb{Z})^n$ if and only if the fundamental generalised vectors $\{\mathbbvar{k}_i\}$ of the principal $T^n$-action on $M$ are Killing.
\end{theorem}

\begin{proof}
From lemma \ref{lemma:flatlie} we get the patching conditions for the closed form $\widetilde{F}_i := \iota_{k_i}\bigsp{H}$
\begin{equation}\label{eq:torusinthecore}
    \begin{aligned}
        \widetilde{F}_i \,&=\, \mathrm{d}(\eta_{(\alpha)i} - \iota_{k_i}\bigsp{B}_{(\alpha)}) \\
        (\eta_{(\beta)i} - \iota_{k_i}\bigsp{B}_{(\beta)}) - (\eta_{(\alpha)i} - \iota_{k_i}\bigsp{B}_{(\alpha)}) \,&=\, \mathrm{d}(\iota_{k_i}\bigsp{\Lambda}_{(\alpha\beta)}+\eta_{(\alpha\beta)i}) \\
        (\iota_{k_i}\bigsp{\Lambda}_{(\alpha\beta)}+\eta_{(\alpha\beta)i}) + (\iota_{k_i}\bigsp{\Lambda}_{(\beta\gamma)}+\eta_{(\beta\gamma)i}) + (\iota_{k_i}\bigsp{\Lambda}_{(\gamma\alpha)}+\eta_{(\gamma\alpha) i}) \,&=\, c_{(\alpha\beta\gamma) i}
    \end{aligned}
\end{equation}
These are precisely the \v{C}ech data for a principal $T^n$-bundle on $M$ if and only if $c_{(\alpha\beta\gamma)i}\in 2\pi\mathbb{Z}$.
If the fundamental generalised vectors $\{\mathbbvar{k}_i\}$ are Killing, then ordinary vectors $\{k_i\}$ are Killing respect to the ordinary metric $g$, i.e. $\mathcal{L}_{k_i}\bigsp{g}=0$. Also the flat gerbes defined by $\mathcal{L}_{k_i}\bigsp{H}=0$ in lemma \ref{lemma:flatlie} are trivial, which is equivalent to the fact that the flat holonomy classes $[c_{\alpha\beta\gamma i}]$ are trivial. This means that the local data \eqref{eq:torusinthecore} define a $T^n$-bundle and hence the conclusion.
\end{proof}

\begin{remark}[Generalised correspondence space]\label{rem:gencorrspacegen}
From the proof of lemma \ref{thm:generaldifftduality} we obtain the \v{C}ech data of a principal $T^n$-bundle $K\rightarrow M$, which we call \textit{generalised correspondence space}. This has first Chern class $\big[\widetilde{F}_i\big]=\big[\iota_{k_i}\bigsp{H}\big]$ and global connection $1$-form $\Xi\in\Omega^1(K,\mathbb{R}^n)$ given by
\begin{equation}
\begin{aligned}
    \Xi_i \,:=\, \mathrm{d}\widetilde{\theta}_{(\alpha)i} + \eta_{(\alpha)i}^{(1)} + B^{(1)}_{(\alpha)i} + (\eta_{(\alpha)ij}^{(0)} - B^{(0)}_{(\alpha)ij})\xi^j
\end{aligned}
\end{equation}
where we called its local fibre coordinates $(\widetilde{\theta}_{(\alpha)})$ and we split $\eta_{(\alpha)i}=\eta_{(\alpha)i}^{(1)}+ \eta_{(\alpha)ij}^{(0)}\xi^j$ in horizontal and vertical components. Again the picture \eqref{eq:tfolddiagram} holds:
\begin{equation}
\begin{tikzcd}[row sep={9ex,between origins}, column sep={10ex,between origins}]
 & & K \arrow[dl, two heads]\arrow[dr, dotted] & \\
 T^n \arrow[r, hook] & M \arrow[dr, "\pi"', two heads] & & \text{T-fold}\arrow[dl, dotted] \\
 & & M_0 &
\end{tikzcd}
\end{equation}
\end{remark}

\noindent Recall that generalised vectors are infinitesimal automorphisms of the doubled space. Notice that Killing generalised vectors $\mathfrak{iso}(\mathscr{G},\mathcal{G})$ are infinitesimal isometries of the doubled space and therefore they can be integrated to finite isometries of the generalised metric structure (remark \ref{rem:isometry}), i.e. to elements of $\mathbf{Iso}(\mathscr{G},\mathcal{G})=\mathrm{Iso}(M,g)\ltimes\mathbf{H}(M,\BU)$. By integrating lemma \ref{thm:generaldifftduality} we get the following statement.

\begin{theorem}[General T-duality on the doubled space, finite formulation]
There exists a $T^n$-bundle $K\rightarrow M$ with first Chern class $\mathrm{c}_1(K)=[\pi_\ast \bigsp{H}]=[H^{(2)}-\langle H^{(1)}\wedge\xi\rangle]\in H^2(M,\mathbb{Z})^n$ if and only if any automorphism in $\Coo(M_0,\,T^n)$ of spacetime $M\xrightarrow{\pi}M_0$ is lifted to an isometry in $\mathbf{Iso}(\mathscr{G},\mathcal{G})$ of the doubled space.
\end{theorem}

\noindent More generally this suggests that the \textit{presence of finite isometries of the doubled space implies T-duality}. Therefore again the generalised correspondence space $K$ is inside the total space of the doubled space, and it is well-defined whenever the bundle gerbe with generalised metric has an isometry.

\begin{figure}[ht]\centering
\includegraphics[scale=0.18]{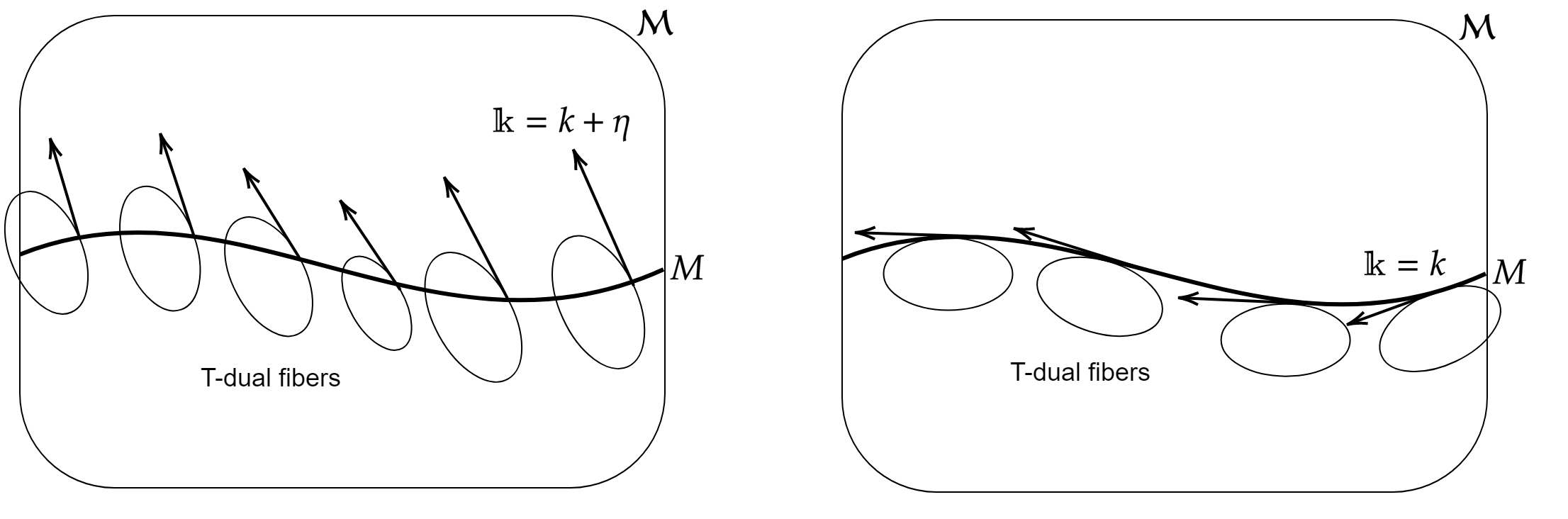}
\caption[Generalised correspondence space]{Generalised correspondence space $K$ is defined by Killing generalised vectors.}\label{FIGtduality}\end{figure}

\begin{remark}[Physical interpretation of general T-duality conditions]
The holonomy of the gerbe is nothing but a global expression for the Wess-Zumino-Witten action of a world-sheet
\begin{equation}
    \exp 2\pi i S_{\mathrm{WZ}}(\Sigma) := \mathrm{Hol}_{(\bigsp{B}_{(\alpha)},\bigsp{\Lambda}_{(\alpha\beta)},\bigsp{G}_{(\alpha\beta\gamma)})}(\Sigma).
\end{equation}
Then the holonomy of the gerbe of lemma \ref{lemma:flatlie} is the global variation of the Wess-Zumino term
\begin{equation}
    \exp 2\pi i \delta_j S_{\mathrm{WZ}}(\Sigma) := \mathrm{Hol}_{(\mathcal{L}_{k_j}\!\bigsp{B}_{(\alpha)},\, \mathcal{L}_{k_j}\!\bigsp{\Lambda}_{(\alpha\beta)},\, \mathcal{L}_{k_j}\!\bigsp{G}_{(\alpha\beta\gamma)})}(\Sigma).
\end{equation}
This needs to vanish for any closed surface $\Sigma\subset M$ to make the background T-dualizable.
\end{remark}

\noindent Again, even if para-Hermitian geometry cannot geometrize the whole gerbe data, it is enough to encode the data of the generalised correspondence space. Similarly to the previous section the para-Hermitian $T^{2n}$-fibre is non-trivially fibrated over the base manifold $M$.

\noindent After digression \ref{dig:preq} we will now highlight another analogy between doubled geometry and higher geometric quantisation, this time regarding the conditions for a general T-duality.

\begin{digression}[A formal similarity with $2$-plectic geometry]
In \cite{Rog11} and \cite{Rog13} Higher Prequantisation of a $2$-plectic manifold $(M,\omega)$ is presented, where the $2$-plectic form $\omega\in\Omega^3_{\mathrm{cl}}(M)$ is non-degenerate and closed. In the references the $2$-algebra of Hamiltonian forms $\Coo(U)\xrightarrow{\mathrm{d}}\Omega^1_{\mathrm{Ham}}(U)$ on a patch $U\subset M$ is defined. Notice this is a sub-$2$-algebra of $\mathbf{H}(U,\BU)\cong\big(\Coo(U)\xrightarrow{\mathrm{d}}\Omega^1(U)\big)$. Hence the stackification on $M$ of this $2$-algebra of Hamiltonian forms will be a sub-$2$-algebra of $\mathbf{H}(M,\BU)$ of circle bundles $L$ satisfying
\begin{equation}\label{eq:higherprequantum}
    \iota_{X_L}\omega = \mathrm{curv}(L)
\end{equation}
for some vector $X_L\in\mathfrak{X}(M)$ that we will call Hamiltonian vector field of $L$. If we interpret our background $(M,H)$ as a $2$-plectic manifold we can see that the T-duality condition
\begin{equation}
    \iota_{k_i}H = \mathrm{curv}(K)_i
\end{equation}
is formally identical to \eqref{eq:higherprequantum}, where $\mathrm{curv}(K)_i=\widetilde{F}_i$ is the curvature of the generalised correspondence space $K\rightarrow M$ from remark \ref{rem:gencorrspacegen}. Therefore we can reformulate the conditions for T-duality in the language of Higher Prequantum Geometry as follows: to have T-duality the fundamental vector fields $\{k_i\}$ of the bundle $M\xrightarrow{\pi}M_0$ must be both Killing and Hamiltonian.
\end{digression}

\section{Geometrisation of non-abelian T-duality}\label{td5}

In this section we will briefly deal with non-abelian T-duality, to show that it is encompassed in our formalism. This means that in the following discussion we can drop at once the assumptions that $[H^{(0)}]=0$ and that spacetime is an abelian principal bundle.

\vspace{0.2cm}
\noindent Let us assume that the spacetime $M$ is a principal $G$-bundle over a smooth base manifold, i.e.
\begin{equation}
\begin{tikzcd}[row sep={11ex,between origins}, column sep={12ex,between origins}]
    M \arrow[d,two heads, "\pi"'] \arrow[r] & \ast \arrow[d] \\
    M_0 \arrow[r] & \mathbf{B}G
\end{tikzcd}
\end{equation}
The bundle gerbe $\mathscr{G}\xrightarrow{\bbpi}M$ must now be reduced by a composition of an Higher Kaluza-Klein reduction from $\mathscr{G}$ to its base $M$ and an ordinary non-abelian Kaluza-Klein reduction from $M$ to its base $M_0$. See the following pullback diagram for the total reduction
\begin{equation}
    \left(\begin{tikzcd}[row sep=7ex, column sep=8ex]
    M \arrow[r, "G_{(\alpha\beta\gamma)}"] & \mathbf{B}^2U(1)
    \end{tikzcd}\right) \;\;\overset{\!\cong}{\longmapsto}\;\; \left(\begin{tikzcd}[row sep=7ex, column sep=5ex]
    & \left[G,\mathbf{B}^2U(1)\right]\!/G \arrow[d, two heads]\\
    M_0 \arrow[ru]\arrow[r, ""] & \mathbf{B}G
    \end{tikzcd}\right),
\end{equation}
The curvature $3$-form of the doubled space can as usual be expanded in the connection $1$-form $\xi\in\Omega^1(M,\mathfrak{g})$ of the $G$-bundle $M$ by
\begin{equation}
    \bigsp{H} = H^{(3)} + H^{(2)}_{i}\wedge\xi^i + \frac{1}{2}H^{(1)}_{ij}\wedge\xi^i\wedge\xi^j + \frac{1}{6}H^{(0)}_{ijk}\xi^i\wedge\xi^j\wedge\xi^k
\end{equation}
Now we can define the usual $2$-form dual curvature by $\widetilde{F}_i := \iota_{k_i}H$. Therefore we obtain the following $2$-form curvature on spacetime $M$
\begin{equation}
\begin{aligned}
    F^i &= \mathrm{d}\xi^i &+&\, \frac{1}{2} \,F^{(0)\, i}_{\;\;jk}\xi^j\wedge\xi^k \\
    \widetilde{F}_i &= \big(H^{(2)}_{i} - H^{(1)}_{ij}\wedge\xi^j\big) \!\!\!\!&+&\, \frac{1}{2} \, H^{(0)}_{ijk}\,\xi^j\wedge\xi^k
\end{aligned}
\end{equation}
We assume that there is a non-abelian group of isometries of the generalised metric space $\mathbf{Iso}(\mathscr{G},\mathcal{G})=\Gamma\big(M_0,\,\mathrm{Ad}(M)\big)$ given by the group of automorphisms of the $G$-bundle $M\xrightarrow{\pi} M_0$. We will see that in this case the generalised correspondence space will be a $T^n$-bundle $K\rightarrow M$ on spacetime, where we defined $n:=\mathrm{dim}\,G$.
Let us expand the differential data of the doubled space
\begin{equation}\label{eq:nonabelianb}
    \begin{aligned}
    \bigsp{B}_{(\alpha)} \;&=\; B^{(2)}_{(\alpha)} + B^{(1)}_{(\alpha)i}\wedge\xi^i + \frac{1}{2} B^{(0)}_{ij}\xi^i\wedge\xi^j, \\
    \bigsp{\mathrm{d}}\bigsp{\Lambda}_{(\alpha\beta)} &= {\di}\Lambda_{(\alpha\beta)}^{(1)} + \Lambda_{(\alpha\beta)}^{(0)i}F^i + {\di}\Lambda_{(\alpha\beta)}^{(0)i}\xi^i -\frac{1}{2} \Lambda_{(\alpha\beta)i}^{(0)} F^{(0)\,i}_{\;\;jk}\xi^j\wedge\xi^k ,
    \end{aligned}
\end{equation}
where a new final vertical term appears respect to the abelian case.
Now, to hugely simplify the discussion, let us assume that $(\bigsp{\Lambda}_{(\alpha\beta)},\,\bigsp{G}_{(\alpha\beta\gamma)}):M\rightarrow\mathbf{B}(\BU)$ is an equivariant structure under the principal $G$-action on $M$. The patching condition $\bigsp{B}_{(\beta)} -\bigsp{B}_{(\alpha)} = \bigsp{\mathrm{d}}\bigsp{\Lambda}_{(\alpha\beta)}$ becomes
\begin{equation}
    \begin{aligned}
        B^{(0)}_{(\beta)ij} - B^{(0)}_{(\alpha)ij}\, &\,=\, -\varepsilon_{ij}^{\;\;\,k}\widetilde{f}_{\alpha\beta k} , \\
        B^{(1)}_{(\beta)i} - B^{(1)}_{(\alpha)i} &\,=\,  {\di}\widetilde{f}_{\alpha\beta i} ,\\
        B^{(2)}_{(\beta)ij} - B^{(2)}_{(\alpha)}\,\, &\,=\, {\di}\lambda_{(\alpha\beta)} +  \widetilde{f}_{(\alpha\beta)i}F^i ,
    \end{aligned}
\end{equation}
where $F$ is the curvature of the principal $G$-bundle and where we called $\Lambda_{(\alpha\beta)}^{(0)}=:\widetilde{f}_{(\alpha\beta)}$ in analogy with the abelian case. On three-fold overlaps we have the simple patching conditions
\begin{equation}
    \begin{aligned}
    \widetilde{f}_{(\alpha\beta)i}+\widetilde{f}_{(\beta\gamma)i}+\widetilde{f}_{(\gamma\alpha)i} &=0 \\
    \lambda_{(\alpha\beta)}+\lambda_{(\beta\gamma)}+\lambda_{(\gamma\alpha)} &={\di}g_{(\alpha\beta\gamma)}
    \end{aligned}
\end{equation}
We can define the $1$-forms $\widetilde{A}_{(\alpha)}:=-\iota_{k_i}\bigsp{B}_{(\alpha)}$ on spacetime, which must be patched by the $1$-forms \eqref{eq:nonabelianb} as
\begin{equation}
    \widetilde{A}_{(\beta)i} - \widetilde{A}_{(\alpha)i} = {\di}\Lambda_{\alpha\beta i}^{(0)} + [\xi,\Lambda_{(\alpha\beta)}^{(0)}]_i - 2\mathcal{L}_{k_i}\Lambda_{\alpha\beta j}^{(0)}\xi^j -\mathcal{L}_{k_i}\Lambda_{(\alpha\beta)}^{(1)}
\end{equation}
In the simple case where the doubled space is equivariant under the principal $G$-action we only have $\widetilde{A}_{(\beta)} - \widetilde{A}_{(\alpha)} = {\di}\Lambda_{(\alpha\beta)}^{(0)} + [\xi,\Lambda_{(\alpha\beta)}^{(0)}]$, which can be written as
\begin{equation}
    \widetilde{A}_{(\beta)} - \widetilde{A}_{(\alpha)} = \mathrm{D}_\xi\Lambda_{(\alpha\beta)}^{(0)}
\end{equation}
where $\mathrm{D}_\xi$ is the covariant derivative defined by the connection $\xi\in\Omega^1(M,\mathfrak{g})$.
The principal connection $\Xi\in\Omega^1(K,\mathbb{R}^3)$ on the non-abelian generalised correspondence space $K\twoheadrightarrow M$, seen as a torus $T^n$-bundle on spacetime $M$, will be the usual $1$-form
\begin{equation}
    \Xi \,=\, \mathrm{d}\widetilde{\theta}_{(\alpha)} + \widetilde{A}_{(\alpha)} \,=\, \widetilde{\xi}_{(\alpha)} - B^{(0)}_{(\alpha)ij}\xi^j
\end{equation}
where we called $\widetilde{\xi}_{(\alpha)} := \mathrm{d}\widetilde{\theta}_{(\alpha)} + B^{(1)}_{(\alpha)}$.
According to \eqref{eq:exactness} the $[H^{(0)}]$ class of the $H$-flux satisfies
\begin{equation}
    H^{(0)}_{ijk} \,=\, \mathcal{L}_{k_{[i}} B^{(0)}_{(\alpha)ij]} + B^{(0)}_{(\alpha)[i|\ell}F^{(0)\,\ell}_{\;\,|jk]}
\end{equation}
which can be solved by imposing
\begin{equation}
    B^{(0)}_{(\alpha)ij} = b_{(\alpha)ij} + H_{ijk}^{(0)}\theta^k_{(\alpha)}
\end{equation}
where $b_{(\alpha)ij}$ are scalars depending only on the base manifold $U_\alpha$, where $\theta_{(\alpha)}$ is a local coordinate of the fibre $G$ and where $\widetilde{\theta}_{(\alpha)}$ must satisfy $\widetilde{\theta}_{(\alpha)}-\widetilde{\theta}_{(\beta)} = \widetilde{f}_{(\alpha\beta)}$. Hence $\widetilde{\theta}_{(\alpha)}$ are the coordinates of the linear $\mathbb{R}^{n}$ fibre of the local principal bundles $U_\alpha \times \mathbb{R}^{n}$ with local connection $\widetilde{\xi}_{(\alpha)} := \mathrm{d}\widetilde{\theta}_{(\alpha)} + B^{(1)}_{(\alpha)}$. 

\begin{remark}[Non-abelian T-fold]
The concept of non-abelian T-fold was recently introduced for $S^3$ in \cite{Bug19}. We will directly generalize this idea to a $G$-bundle spacetime, in analogy with the abelian T-fold. The non-abelian T-dual space will necessarily have a non-trivial locally non-geometric $Q$-flux, which implies it will be a T-fold. We then have a picture
\begin{equation}
\begin{tikzcd}
 & & K \arrow[dl, two heads]\arrow[dr, dotted, end anchor={[yshift=-2ex]}] & \\
  & M \arrow[dr, "\pi"', two heads] & & \begin{tabular}{c}non-abelian \\T-fold\end{tabular} \arrow[dl, dotted, start anchor={[yshift=2ex]}] \\
 & & M_0 &
\end{tikzcd}
\end{equation}
Here again, like for abelian T-fold, the arrows on the right side are not actual maps between actual spaces, but they are only indicative. This is because the non-abelian T-fold, similarly to its abelian version, can only be geometrised inside the non-principal $G\ltimes T^n$-bundle $K\rightarrow M_0$.
\end{remark}

\begin{remark}[Recovering the usual formulation of non-abelian T-duality]\label{rem:natdtraditional}
We use the basis $\xi^I_{(\alpha)}:=(\xi^i,\,\widetilde{\xi}_{(\alpha)i})$ of local connections of respectively $U_\alpha\times G$ and $U_\alpha\times \mathbb{R}^n$. In this basis we must then write the moduli field $\mathcal{G}_{(\alpha)}^{(0)}$ of the generalised metric by using the moduli field of the Kalb-Ramond field $B^{(0)}_{(\alpha)ij}$. Now we can express non-abelian T-duality as the following $O(n,n)$-transformation
\begin{equation}
    \mathcal{T}_{\mathrm{NATD}} \;:=\;
    \begin{pmatrix}
 0 & 1 \\
 1 & F_{\;\;\, ij}^{(0)\,k}\widetilde{\theta}_{(\alpha)k}
 \end{pmatrix}
\end{equation}
which encodes the shift $B^{(0)}_{(\alpha)ij} \mapsto B^{(0)}_{(\alpha)ij}+F_{\;\;\, ij}^{(0)\,k}\widetilde{\theta}_{(\alpha)k}$. Therefore the non-abelian T-dual of the moduli field of the generalised metric in this basis will be $\widetilde{\mathcal{G}}_{(\alpha)}^{(0)} = \mathcal{T}_{\mathrm{NATD}}^{\mathrm{T}}\mathcal{G}_{(\alpha)}^{(0)}\mathcal{T}_{\mathrm{NATD}}$. Its local components $\widetilde{g}^{(0)}_{(\alpha)}$ and $\widetilde{B}^{(0)}_{(\alpha)}$ will be the local data of the non-abelian T-fold.
\end{remark}

\noindent We will now explore a simple example of background which has non-abelian T-duality.

\begin{example}[$3$-sphere bundle]
Let us consider an $S^3$-bundle spacetime given by the sequence of manifolds $S^3\hookrightarrow M \twoheadrightarrow M_0$. Recall that $H^3(S^3,\mathbb{Z})\cong\mathbb{Z}$, so $[H^{(0)}]$ is determined by a single integer. For simplicity let us assume the doubled space is trivial with Dixmier-Douady class $[H]=0$. Even if the gerbe structure is trivial on the $3$-sphere fibre, in contrast with the abelian case, the moduli fields $B_{(\alpha)}^{(0)}$ are not globally defined scalars. $B_{(\beta)ij}^{(0)}-B_{(\alpha)ij}^{(0)} = \varepsilon_{ij}^{\;\;\, k} (\widetilde{\theta}_{(\beta)k}-\widetilde{\theta}_{(\alpha)k})$ 
Thus the connections will be patched on two-fold overlaps by
\begin{equation}
    \begin{pmatrix}
 \xi_{(\alpha)} \\
 \widetilde{\xi}_{(\alpha)}
 \end{pmatrix} =
    \begin{pmatrix}
 1 & 0 \\
 \varepsilon_{ij}^{\;\;\, k}\widetilde{f}_{(\alpha\beta) k} & 1
 \end{pmatrix}    \begin{pmatrix}
 \xi_{(\beta)} \\
 \widetilde{\xi}_{(\beta)}
 \end{pmatrix}
\end{equation}
where $\widetilde{f}_{(\alpha\beta)}=\widetilde{\theta}_{(\alpha)}-\widetilde{\theta}_{(\beta)}$ are the transition functions for the dual torus coordinates. Therefore the monodromy matrix of these coordinates will be given by
\begin{equation}
    n_{(\alpha\beta)}^H = \begin{pmatrix}
 0 & \widetilde{f}_{(\alpha\beta) 3} & -\widetilde{f}_{(\alpha\beta) 2} \\
 -\widetilde{f}_{(\alpha\beta) 3} & 0 & \widetilde{f}_{(\alpha\beta) 1} \\
 \widetilde{f}_{(\alpha\beta) 2} & -\widetilde{f}_{(\alpha\beta) 1} & 0
 \end{pmatrix}.
\end{equation}
Hence actually the $\mathbb{R}^3$ fibres are not non-compact, but they are glued as a non-abelian T-fold. Differently from its abelian counterpart, notice that in the non-abelian T-fold the monodromy matrix is in general not constant.
Now let us write the locally defined moduli field $\mathcal{G}^{(0)}_{(\alpha)}$ of the reduction of the generalised metric to the base manifold $M_0$.
We will suppress few indices, by defining $b := \star_{S^3}B^{(0)}_{(\alpha)}$, so we can write
\begin{equation}
    \mathcal{G}^{(0)}_{(\alpha)} \,=\, 
    \left(\begin{array}{@{}ccc|ccc@{}}
    1+b_2^2+b_3^2&-b_1b_2&-b_1b_3&0&-b_3&b_2\\
    -b_1b_2&1+b_1^2+b_3^2&-b_2b_3&b_3&0&-b_1\\
    -b_1b_3&-b_2b_3&1+b_1^2+b_2^2&-b_2&b_1&0\\\hline
    0&b_3&-b_2&1&0&0\\
    -b_3&0&b_1&0&1&0\\
    b_2&-b_1&0&0&0&1
    \end{array}\right)
\end{equation}
where the metric is just the round metric of the $3$-sphere.
Let us call $B_i := b_i + \widetilde{\theta}_i$ the Hodge dual of the full $B^{(0)}_{(\alpha)}$ moduli field. Now we can perform the T-duality transformation $\mathcal{T}_{S^3}=\left(\begin{smallmatrix*}0&1\\1&0\end{smallmatrix*}\right)$ to obtain the non-abelian T-dual generalised metric moduli field
\begin{equation}
    \widetilde{\mathcal{G}}^{(0)}_{(\alpha)} \,=\, 
   \left(\begin{array}{@{}ccc|ccc@{}}
     1&0&0&0&B_3&-B_2\\
    0&1&0&-B_3&0&B_1\\
    0&0&1&B_2&-B_1&0 \\\hline
    0&-B_3&B_2&1+B_2^2+B_3^2&-B_1B_2&-B_1B_3\\
    B_3&0&-B_1&-B_1B_2&1+B_1^2+B_3^2&-B_2B_3\\
    -B_2&B_1&0&-B_1B_3&-B_2B_3&1+B_1^2+B_2^2
    \end{array}\right)
\end{equation}
Thus the non-abelian T-dual background takes the following familiar expression
\begin{equation}
    \begin{aligned}
    \widetilde{g}^{(0)}_{(\alpha)} &\,=\, \frac{1}{1+B_1^2+B^2_2+B_3^2} \begin{pmatrix}
 1+B_1^2 & B_1B_2 & B_1B_3 \\
 B_1B_2 & 1+B_2^2 & B_2B_3\\
 B_1B_3 & B_2B_3 & 1+B_3^2
 \end{pmatrix}, \\
    \widetilde{B}^{(0)}_{(\alpha)} &\,=\, \frac{1}{1+B_1^2+B^2_2+B_3^2} \begin{pmatrix}
 0 & -B_3 & B_2 \\
 B_3 & 0 & -B_1\\
 -B_2 & B_1 & 0
 \end{pmatrix}.
        \end{aligned}
\end{equation}
The new local metric and Kalb-Ramond field will be respectively $ \widetilde{g}^{(0)}_{(\alpha)} =  \widetilde{g}^{(0)ij}_{(\alpha)}\widetilde{\xi}_{(\alpha)i}\odot\widetilde{\xi}_{(\alpha)j}$ and $ \widetilde{B}^{(0)}_{(\alpha)} =  \widetilde{B}^{(0)ij}_{(\alpha)}\widetilde{\xi}_{(\alpha)i}\wedge\widetilde{\xi}_{(\alpha)j}$. These are the differential data of the fibres of our non-abelian T-fold and indeed they cannot be patched globally on the base manifold.
\end{example}

\begin{example}[Twisted torus bundle]
In the case of the twisted torus bundle $G\hookrightarrow M \twoheadrightarrow M_0$, where $G$ is a twisted torus with $\mathrm{dim}(G)>2$, we recover the general commuting diagram of T-dualities by \cite{Bug19} along any circle of the fibre, i.e.
\begin{equation}
   \begin{tikzcd}[row sep=38,column sep=28]
    & H^{(0)}_{ijk}  \arrow[rr, "\mathcal{T}_i"] \arrow[dd, "\mathcal{T}_j" near start] \arrow[dl, "\mathcal{T}_k"'] & &   F^{(0)i}_{\quad\;\, jk}  \arrow[dd, "\mathcal{T}_j"] \arrow[dl, "\mathcal{T}_k"] \\
    F^{(0)\;\;k}_{\quad ij} \ar[crossing over, "\mathcal{T}_i" near start]{rr} \arrow[dd, "\mathcal{T}_j"'] & & Q^{(0)i\; k}_{\quad\; j} \\
      &  F^{(0)\;j}_{\quad\; i\; k}  \arrow[rr, "\mathcal{T}_i" near start] \arrow[dl, "\mathcal{T}_k"'] & &  Q^{(0)ij}_{\quad\;\; k}  \arrow[dl, "\mathcal{T}_k"] \\
    Q^{(0)\;jk}_{\quad i} \arrow[rr, "\mathcal{T}_i"'] && R^{(0)ijk} \arrow[from=uu,crossing over, "\mathcal{T}_j" near start]
 \end{tikzcd} 
\end{equation}
\end{example}

\noindent Let us now give a quick final look to the general case.

\begin{remark}[General non-abelian case]
In the general case, where we require just the local $\mathcal{L}_{k_i}\bigsp{B}_{(\alpha)} = \mathrm{d}\eta^i_{(\alpha)}$, analogously to its abelian version, we obtain a connection $1$-form $\Xi\in\Omega^1(K,\mathbb{R}^n)$ of a principal $T^n$-bundle over spacetime $M$ by
\begin{equation}
\begin{aligned}
    \Xi_i \,=\, \mathrm{d}\widetilde{\theta}_{(\alpha)i} + \eta_{(\alpha)i}^{(1)} + B^{(1)}_{(\alpha)i} + \big(\eta_{(\alpha)ij}^{(0)} - B^{(0)}_{(\alpha)ij}\big) \xi^j
\end{aligned}
\end{equation}
where $\eta_{(\alpha)i} = \eta_{(\alpha)i}^{(1)} + \eta_{(\alpha)ij}^{(0)}\xi^j$ is split in horizontal and vertical part.
Again the generalised vectors $\mathbbvar{k}_i:=(k_i+\eta_{(\alpha)i},\,\eta_{(\alpha\beta) i})$ are Killing and hence therefore can be integrated to finite isometries of the doubled space in $\Gamma\big(M_0, \, \mathrm{Ad}(M) \big)\ltimes\mathbf{H}(M,\BU)\cong\mathbf{Iso}(\mathscr{G},\mathcal{G})$.
\end{remark}

\subsection{Non-abelian T-fold as global tensor hierarchy}\label{subnatfold}

In this subsection we will consider (1) a spacetime which is a general $S^3$-fibration $M\twoheadrightarrow M_0$ on some base manifold $M_0$ and (2) a gerbe bundle $\mathcal{M}\twoheadrightarrow M$ with satisfies the simple T-dualisation condition $\mathcal{L}_{e_i}\bigsp{B}_{(\alpha)}=0$, where $\bigsp{B}_{(\alpha)}$ is the gerbe connection and $\{e_i\}$ are a basis of $SU(2)$-left invariant vectors on spacetime $M$.\vspace{0.2cm}

\noindent As seen in the previous subsection the dimensional reduction of these gerbe contains a global bundle $K$, the generalised correspondence space, defined by the following diagram:
\begin{equation}
    \begin{tikzcd}[row sep={11ex,between origins}, column sep={12ex,between origins}]
    K \arrow[d, two heads] & \\
    M \arrow[d, two heads] \arrow[r] & \mathbf{B}T^3 \\
    M_0 \arrow[r] & \mathbf{B}SU(2).
    \end{tikzcd}
\end{equation}
This case is often called \textit{semi-abelian}, because spacetime is a principal fibration whose algebra has non-zero structure constants $[e_i,\, e_j]_{\mathfrak{su}(2)}=\epsilon^k_{\;ij}\,e_k$, but the dual ones $\widetilde{C}^{jk}_{\;\; i}=0$ vanish.\vspace{0.25cm}

\noindent From calculations which are analogous to the ones for the abelian T-fold we find that the moduli of the flux are related to the moduli of the Kalb-Ramond field by $H^{(0)}_{ijk}\,=\, \mathfrak{D}_{[i}B^{(0)}_{(\alpha)jk]} + B^{(0)}_{(\alpha)[i|\ell}\,\epsilon^{\;\;\;\ell}_{|ij]}$  where $\mathfrak{D}_i = \mathcal{L}_{e_i}$. Also notice that the moduli of the Kalb-Ramond field is patched on overlaps of patches by 
$B^{(0)}_{(\beta)ij}-B^{(0)}_{(\alpha)ij} \,=\, \mathfrak{D}_{[i}\Lambda^{(0)}_{(\alpha\beta)j]} + \Lambda^{(0)}_{(\alpha\beta)\ell}\,\epsilon^{\;\;\ell}_{ij}$. This will be useful very soon.
We must now apply all the machinery from the previous subsection to this particular example.

\begin{itemize}
\item We can combine the pullback on $K$ of the global connection of the $SU(2)$-bundle $M\twoheadrightarrow M_0$ and the global connection of the $T^3$-bundle $K\twoheadrightarrow M$ in a single object by
\begin{equation}
    \Xi^I\;\, = \,\;\begin{pmatrix} \xi^i \\[0.8em] \di \widetilde{\theta}_{(\alpha)i} + B^{(1)}_{(\alpha)i} - B^{(0)}_{(\alpha)ij}\xi^j \end{pmatrix}\;\;\in \, \Omega^1(K,\,\mathfrak{su}(2)\oplus\mathbb{R}^3) .
\end{equation}

\item Now we can consider a local patch $U_\alpha\subset M_0$ of the base manifold. The total space $K$ restricted on this local patch will be isomorphic to $K|_{U_\alpha} = U_\alpha\times SU(2) \times T^3$. These local bundles can be equipped with local connections
\begin{equation}\label{eq:natdconn2}
    \begin{pmatrix} \xi^i \\[0.8em] \di \widetilde{\theta}_{(\alpha)i} + B^{(1)}_{(\alpha)i}  \end{pmatrix} \;\;\in \, \Omega^1\big(U_\alpha\times SU(2)\times T^3,\,\mathfrak{su}(2)\oplus\mathbb{R}^3\big). 
\end{equation}
As derived in \cite[pag.\,61]{Alf19}, these local connections are glued on two-fold overlaps of patches $(U_\alpha\cap U_\beta)\times SU(2) \times T^3$ by a cocycle of $B$-shifts of the form
\begin{equation}\label{eq:natdpat2}
    \begin{pmatrix} \xi^i \\[0.8em] \di \widetilde{\theta}_{(\alpha)i} + B^{(1)}_{(\alpha)i}  \end{pmatrix} \;=\; \begin{pmatrix} \delta^i_{\;j} & 0 \\[0.8em] n_{(\alpha\beta)ij}+ \epsilon^{\;\;\,k}_{ij}\widetilde{\lambda}_{(\alpha\beta)k} & \delta_i^{\;j} \end{pmatrix}  \begin{pmatrix} \xi^j \\[0.8em] \di \widetilde{\theta}_j + B^{(1)}_{(\beta)j}  \end{pmatrix} ,
\end{equation}
where we defined the matrix $n_{(\alpha\beta)ij} := \mathfrak{D}_{[i}\Lambda^{(0)}_{(\alpha\beta)j]}$. Notice that the T-dualizability condition we imposed on the gerbe implies that $n_{(\alpha\beta)ij}$ is a $\wedge^2\mathbb{Z}^3$-valued \v{C}ech cocycle, similarly to the monodromy matrix cocycle appearing in the abelian T-fold. \vspace{0.1cm}

\noindent As we will explain later, these are the local connection used in most of the non-abelian T-duality literature (before the introduction of Drinfel'd doubles). As noticed by \cite[pag.\,13]{Bug19} they look very similar to an abelian T-fold, but with the monodromy depending on the coordinates via the term $\epsilon^{\;\;\,k}_{ij}\Lambda^{(0)}_{(\alpha\beta)k}$. However we will see that it is better to construct and use proper local $D$-connections to make the tensor hierarchy really manifest.

\item Now we must use the fact that the adjoint action of $T^3$ on $D=SU(2)\ltimes T^3$ is specified by setting $\widetilde{u}_{(\alpha)j}^{\;\;\;i}=\delta^i_{\;j}$ and $\widetilde{b}_{(\alpha)ij}=\epsilon^{\;\;\,k}_{ij}\widetilde{\theta}_{(\alpha)k}$. Thus we can construct the local connection for each local $D$-bundle $U_\alpha\times D$ by $\widetilde{b}_{(\alpha)}$-twisting the local connection \eqref{eq:natdconn2}, obtaining
\begin{equation}\label{eq:natdpat3}
    \left(\Gamma_{(\alpha)}^{-1}\di\Gamma_{(\alpha)} + \mathrm{Ad}_{\Gamma_{(\alpha)}^{-1}}\mathcal{A}_{(\alpha)} \right)^{\! \ind{I}} \;=\; \begin{pmatrix} \xi^i \\[0.8em] \di \widetilde{\theta}_i + B^{(1)}_{(\alpha)i} + \epsilon^{\;\;\,k}_{ij}\widetilde{\theta}_{(\alpha)k}\xi^j  \end{pmatrix} \;\; \in\,\Omega^1(U_\alpha\times D,\, \mathfrak{d}).
\end{equation}
Crucially, this new $1$-form is invariant under gauge transformation $\bigsp{B}_{(\alpha)}\mapsto\bigsp{B}_{(\alpha)}+\bigsp{\di}\bigsp{\eta}_{(\alpha)}$ of the original gerbe bundle on $M$, with gauge parameter $\bigsp{\eta}_{(\alpha)}\in\Omega^1(U_\alpha\times G)^G$. Thus this gives a proper local connection for the internal manifold-fibration rising from the vertical part of the dimensional reduction of the bundle gerbe.
\end{itemize}

\noindent To verify that the $1$-form \eqref{eq:natdpat3} is a proper connection we need to verify that the local potential $\mathcal{A}_{(\alpha)}\in\Omega^1(U_{\alpha},\,\mathfrak{d})$ is actually the pullback of a $1$-form from the base $U_\alpha$. Since for $D=SU(2)\ltimes T^3$ we have $\beta^{ij}=0$, the first component $\mathcal{A}^i=A^i$ is just the local potential of the $SU(2)$-bundle. To check the second component $\widetilde{\mathcal{A}}_{(\alpha)i}$, let us notice that the T-dualizability condition $\mathcal{L}_{e_i}\bigsp{B}_{(\alpha)}=0$ on the gerbe immediately implies $\mathcal{L}_{e_i}(B^{(1)}_{(\alpha)k}\wedge \xi^k)=0$. Now notice that, since the matrix $u^{\mathrm{T}}_{(\alpha)}$ encompasses the adjoint action of the inverse of $\gamma_{(\alpha)}=\exp(\theta^i_{(\alpha)}e_i)$, it must be equal to the exponential of the matrix $\epsilon^{i}_{\;jk}\theta^k_{(\alpha)}$. Therefore we can re-write the $1$-form $B^{(1)}_{(\alpha)k} = \widetilde{\mathcal{A}}_{(\alpha)i}(u^{\mathrm{T}}_{(\alpha)})^i_{\;k}$, where the $\widetilde{\mathcal{A}}_{(\alpha)i}$ depend only on the base $U_\alpha$.
\vspace{0.25cm}

\noindent The field strengths of these principal connections is then given by their covariant derivative
\begin{equation}
    \mathcal{F}_{(\alpha)} \;=\; \di \mathcal{A}_{(\alpha)} + \big[\mathcal{A}_{(\alpha)}\,\overset{\wedge}{,}\,\mathcal{A}_{(\alpha)}\big]_{\mathfrak{d}} \;\;\in\,\Omega^2(U_\alpha,\,\mathfrak{d}).
\end{equation}
In components of the generators of the algebra $\mathfrak{d}$, these assume the following form:
\begin{equation}
    \begin{aligned}
    \mathcal{F}^i_{(\alpha)} \;&=\; \di \mathcal{A}^i_{(\alpha)} + \epsilon^i_{\;jk}\mathcal{A}^j_{(\alpha)}\wedge \mathcal{A}^k_{(\alpha)} ,\\[0.2em]
    \widetilde{\mathcal{F}}_{(\alpha)i} \;&=\; \di \widetilde{\mathcal{A}}_{(\alpha)i} + \epsilon^k_{\;ij}\mathcal{A}^j_{(\alpha)}\wedge \widetilde{\mathcal{A}}_{(\alpha)k} .\\
    \end{aligned}
\end{equation}
What we need to find out now is how these local $D$-bundles are globally glued together.

\noindent We can immediately see that the global connections of the generalised correspondence space encoded in the global $1$-form $\Xi^I$ are related to the local $SU(2)\ltimes T^3$-connections \eqref{eq:natdpat3} by a local patch-wise $B$-shift $U^{\;\;\;\,I}_{(\alpha)\,\ind{J}}$ of the following form:
\begin{equation}\label{eq:natdloctoglob}
    \Xi^I\; = \; \begin{pmatrix} \delta^i_{\;j} & 0 \\[0.8em] -\big(B^{(0)}_{(\alpha)ij}+\epsilon_{\;ij}^{k}\widetilde{\theta}_{(\alpha)k}\big) & \delta_i^{\; j} \end{pmatrix} \left(\Gamma_{(\alpha)}^{-1}\di\Gamma_{(\alpha)} + \mathrm{Ad}_{\Gamma_{(\alpha)}^{-1}}\mathcal{A}_{(\alpha)} \right)^{\! \ind{J}}
\end{equation}
which, crucially, depends both on the physical and on the extra coordinates. The geometric flux is here just given by the structure constants $\epsilon_{ij}^{\;\;\,k}$ of $\mathfrak{su}(2)$. Thus from this expression we immediately get that the wanted patching condition on twofold overlaps of patches are
\begin{equation}\label{eq:natdpat3new}
    \left(\Gamma_{(\alpha)}^{-1}\di\Gamma_{(\alpha)} + \mathrm{Ad}_{\Gamma_{(\alpha)}^{-1}}\mathcal{A}_{(\alpha)} \right)^{\! \ind{I}} \;=\; \big(e^{n_{(\alpha\beta)}}\big)^\ind{I}_{\;\ind{J}} \left(\Gamma_{(\beta)}^{-1}\di\Gamma_{(\beta)} + \mathrm{Ad}_{\Gamma_{(\beta)}^{-1}}\mathcal{A}_{(\beta)} \right)^{\! \ind{J}},
\end{equation}
where, \textit{generalising the abelian case, the monodromy cocycle $e^{n_{(\alpha\beta)}}\in\mathrm{Aut}(D;\,\mathbb{Z})$ is an integer-valued automorphism of the Drinfel'd double}. The interesting point is that semi-abelian T-folds are still glued by integer $B$-shifts, similarly to the abelian ones, if we consider local $D$-bundles. Consequently the curvatures are glued by
\begin{equation}
\mathcal{F}_{(\alpha)}^\ind{I} \,=\, \big(e^{n_{(\alpha\beta)}}\big)^\ind{I}_{\;\ind{J}} \,\Big(\mathrm{Ad}_{\lambda_{(\alpha\beta)}^{-1}}\mathcal{F}_{(\beta)}\Big)^\ind{J},
\end{equation}
where $\lambda_{(\alpha\beta)}^i$ are the expected transition functions of the $SU(2)$-bundle $M\twoheadrightarrow M_0$.
As usual, if we want to include also the higher form field of which characterizes a tensor hierarchy
\[
\mathcal{H} \;=\;  \di \mathcal{B}_{(\alpha)} + \frac{1}{2}\big\langle \mathcal{A}_{(\alpha)}\,\overset{\wedge}{,}\, \di \mathcal{A}_{(\alpha)}\big\rangle - \frac{1}{3!}\big\langle \mathcal{A}_{(\alpha)}\,\overset{\wedge}{,}\,[\mathcal{A}_{(\alpha)} \,\overset{\wedge}{,}\, \mathcal{A}_{(\alpha)}]_{\mathfrak{d}} \big\rangle,
\]
this will be a globally defined, but not closed, $3$-form on the base manifold.

\paragraph{Non-abelian T-duality in the literature.}
We will now explain how the conventional non-abelian T-duality picture we are used in the literature, see for instance \cite{KLMC15} and \cite[pag.\,13]{Bug19}, emerges and is clarified.
Usually, in the literature, (1) we start from the moduli field of the Kalb-Ramond field $B^{(0)}_{(\alpha)ij}\xi^i\wedge\xi^j$ and of the metric $g^{(0)}_{ij}\xi^i\otimes\xi^j$. Then (2) we perform a shift which depends on the dual coordinates $B^{(0)}_{(\alpha)ij}\xi^i\wedge\xi^j \,\mapsto\,  \big(B^{(0)}_{(\alpha)ij}+\epsilon^{\;\;\,k}_{ij}\widetilde{\theta}_{(\alpha)k}\big)\xi^i\wedge\xi^j$. Finally (3) we perform the proper inversion of the matrix of the moduli field by
\begin{equation}\label{literature}
    \widetilde{g}^{(0)ij}+\widetilde{B}^{(0)ij}_{(\alpha)} \;:=\; \Big(g_{ij}^{(0)}+B^{(0)}_{(\alpha)ij}\,+\,\epsilon^{\;\;\,k}_{ij}\widetilde{\theta}_{(\alpha)k}\Big)^{-1},
\end{equation}
to obtain the non-abelian T-dual Kalb-Ramond field
$\widetilde{B}^{(0)ij}_{(\alpha)}\widetilde{\xi}_{(\alpha)i}\wedge\widetilde{\xi}_{(\alpha)j}$ where the role of the connection is now played by the local $1$-forms $\widetilde{\xi}_{(\alpha)i}=\di\widetilde{\theta}_{(\alpha)i}+B^{(1)}_{(\alpha)i}$, which are nothing but the last three components of \eqref{eq:natdconn2}. As we know, these connections are not globally defined and hence, as already observed in \cite[pag.\,13]{Bug19}, we obtain a particular T-fold. If we explicit the transition functions $\widetilde{\lambda}_{(\alpha\beta)}=\widetilde{\theta}_{(\beta)}-\widetilde{\theta}_{(\alpha)}$ of $K$ and we consider the simplest case with $n_{(\alpha\beta)}=0$, we indeed notice that the $B$-shifts \eqref{eq:natdpat2} which glue the $\xi^i$ and $\widetilde{\xi}_{(\alpha)i}$ reduce to the transformations $\epsilon_{ij}^{\;\;\,k}\big(\widetilde{\theta}_{(\beta)}-\widetilde{\theta}_{(\alpha)}\big)_k$ which were firstly illustrated by \cite[pag.\,13]{Bug19} for $S^3$.\vspace{0.25cm}

\noindent However these operations and their underlying geometry can be better understood in terms of local $D$-bundles $U_\alpha\times D$. This is because, as seen in equation \eqref{eq:natdloctoglob}, the "effective" moduli field of the Kalb-Ramond field is the sum $B^{(0)}_{(\alpha)ij}+\epsilon^{\;\;\,k}_{ij}\widetilde{\theta}_{(\alpha)k}$ and T-duality is nothing but an automorphism of the fibre $D$. As we have shown, working by considering local $D$-bundles leads to the simpler patching conditions \eqref{eq:natdpat3}, than the ones \eqref{eq:natdpat2} that are obtained by analogy with abelian T-folds.

\paragraph{The geometric case: the global $\String(SU(2)\ltimes T^3)$-bundle.}\label{geometriccase}
When our bundle gerbe $\mathcal{M}\twoheadrightarrow M$ is equivariant under the principal $SU(2)$-action of $M$, we get that the flux is just $H^{(0)}_{ijk}=B^{(0)}_{(\alpha)[i|\ell}\,\epsilon^{\ell}_{\;|jk]}$ and the monodromy matrix cocycle $n_{(\alpha\beta)}=0$ vanishes. This means that the local $D$-bundles $U_\alpha\times D$ are glued together to form a global $D$-bundle on the base manifold $M_0$. If we include also the higher form field $\mathcal{H}$, we will have a global $\String(D)$-bundle on the base manifold $M_0$. In other words we have the following equivalence of $2$-groupoids
\begin{equation}
    \bigsqcup_{\begin{subarray}{c}M\text{ s.t. }
    M\twoheadrightarrow M_0\\\text{is a }SU(2)\text{-bundle}\end{subarray}} \!\!\!\!SU(2)\text{-}\mathrm{equiv}\mathbf{B}U(1)\mathrm{Bund}(M) \;\;\cong\;\; \String(SU(2)\!\ltimes \widetilde{T}^3)\mathrm{Bund}(M_0),
\end{equation}
where we called $SU(2)\text{-}\mathrm{equiv}\mathbf{B}U(1)\mathrm{Bund}(M)$ the $2$-groupoid of $SU(2)$-equivariant gerbes on $M$ and $ \String(SU(2)\!\ltimes \widetilde{T}^3)\mathrm{Bund}(M_0)$ the $2$-groupoid of $\String(SU(2)\!\ltimes \widetilde{T}^3)$-bundles on the base $M_0$.\vspace{0.25cm}

\noindent For example, we can look at the case where both our spacetime $M=M_0\times S^3$ is a trivial fibration and our gerbe bundle $\mathcal{M}\twoheadrightarrow M$ is topologically trivial with vanishing Dixmier-Douady class $[H]=0\in H^3(M,\mathbb{Z})$. In this case we obtain a trivially-fibred generalised correspondence space $K=M_0\times D$ with doubled fibre $D=SU(2)\ltimes T^3$. Consequently the global tensor hierarchy rising from the dimensional reduction of this trivial gerbe will be just the connection of the trivial principal $\infty$-bundle $M_0\times \String(D)\twoheadrightarrow M_0$.

\section{Geometrisation of Poisson-Lie T-duality}\label{td6}

In this section we will briefly deal with the generalised correspondence space underlying Poisson-Lie T-duality in our framework. See \cite{Has17} for an introduction of Poisson-Lie T-duality in Double Field Theory.
\vspace{0.2cm}

\noindent Let spacetime be a principal $G$-bundle $M\xrightarrow{\pi}M_0$. Let us assume that the fundamental vector fields $\{k_i\}\subset \mathfrak{X}(M)$ are Killing, i.e. $\mathcal{L}_{k_i}\bigsp{g}=0$, but such that the gerbe connection satisfies
\begin{equation}\label{eq:plcond}
    \mathcal{L}_{k_i}\bigsp{B}_{(\alpha)} = \big[\widetilde{A}_{(\alpha)}\,\overset{\wedge}{,}\,\widetilde{A}_{(\alpha)}\big]_i^{\widetilde{\mathfrak{g}}}
\end{equation}
where $\widetilde{A}_{(\alpha)i}:=-\iota_{k_i}\bigsp{B}_{(\alpha)}$ is a local $\mathfrak{g}^\ast$-valued $1$-form on the total spacetime $M$ and $[-,-]^{\widetilde{\mathfrak{g}}}$ is the commutator of some Lie algebra $\widetilde{\mathfrak{g}}$ whose underlying vector space is $\mathfrak{g}^\ast$. Notice that this implies that the automorphisms to the $G$-bundle are not lifted to isometries of the doubled space, i.e. $\Gamma\big(M_0,\,\mathrm{Ad}(M)\big)\not\subset\mathbf{Iso}(\mathscr{G},\mathcal{G})$. However these transformations defy the isometry in a very controlled way. The equation \eqref{eq:plcond} implies that $\widetilde{F}_i:=\iota_{k_i}\bigsp{H}$ can be seen as the curvature of a principal $\widetilde{G}$-bundle with $\mathrm{Lie}(\widetilde{G})=\widetilde{\mathfrak{g}}$, indeed we have
\begin{equation}
    \begin{aligned}
     \widetilde{F}_i&\,:=\,\iota_{k_i}\bigsp{H}\\
     &\;\,=\,\mathcal{L}_{k_i}\bigsp{B}_{(\alpha)} - \mathrm{d}\iota_{k_i}\bigsp{B}_{(\alpha)} \\
     &\;\,=\, \big[\widetilde{A}_{(\alpha)}\,\overset{\wedge}{,}\,\widetilde{A}_{(\alpha)}\big]_i^{\widetilde{\mathfrak{g}}} + \mathrm{d}\widetilde{A}_{(\alpha)i}
    \end{aligned}
\end{equation}
Hence the generalised correspondence space $K\rightarrow M$ will be a principal $\widetilde{G}$-bundle on spacetime $M$ with curvature $\widetilde{F}_i$. Notice we have $\widetilde{A}_{(\alpha)i}: = \iota_{k_i}\bigsp{B}_{(\alpha)} = B^{(1)}_{(\alpha)i} - B^{(0)}_{(\alpha)ij}\xi^j$ where $\xi\in\Omega^1(M,\mathfrak{g})$ is the connection of the $G$-bundle $M$. Hence it can be interpreted as the pullback $\widetilde{A}_{(\alpha)}=\sigma^\ast_{(\alpha)}\Xi$ of a global $1$-form connection $\Xi\in\Omega^1(K,\widetilde{\mathfrak{g}})$ by a choice of local sections $\sigma_{(\alpha)}$.

\begin{remark}[$Q$-flux]
Hence we have both the geometric flux
$F^{(0)\,k}_{\;\;\, ij} := \varepsilon^{\;\;\,k}_{ij}$, given by the bracket $[-,-]_{\mathfrak{g}}$, and locally non-geometric flux
$Q^{(0)ij}_{\quad\;\; k} := \widetilde{\varepsilon}^{ij}_{\;\; k}$, given by the bracket $[-,-]^{\widetilde{\mathfrak{g}}}$. We will see now that this means that we have a Manin triple of Lie algebras $\mathfrak{d}:=\mathfrak{g}\oplus\widetilde{\mathfrak{g}}$ which can be integrated to a Drinfel'd double $D=G\bowtie\widetilde{G}$.
\end{remark}

\begin{remark}[Poisson-Lie T-fold]
Now the generalised correspondence space $K$ will be a principal $\widetilde{G}$-bundle on spacetime $M$ with connection $1$-form $\Xi\in\Omega^1(K,\widetilde{\mathfrak{g}})$ given by
\begin{equation}
    \Xi_i \,=\,  \widetilde{\theta}_{(\alpha)}^{\ast}\widetilde{\tau}_i + \mathrm{Ad}_{\widetilde{\theta}_{(\alpha)}^{-1}}\big( B^{(1)}_{(\alpha)i} - B^{(0)}_{(\alpha)ij}\xi^j \big)
\end{equation}
where $\widetilde{\theta}_{(\alpha)}:U_\alpha\times \widetilde{G}\rightarrow \widetilde{G}$ is a canonical local trivialisation of the $\widetilde{G}$-bundle given by $\sigma_{(\alpha)}$ and where $\widetilde{\tau}$ is the left-invariant Maurer-Cartan form on $\widetilde{G}$. Locally the generalised correspondence space $K$ will be given by patches of the form $U_\alpha\times D$ where $U_\alpha$ is a patch of $M_0$ and the group $D:= G\bowtie\widetilde{G}$ is a Drinfel'd double. In the special case where $\widetilde{A}_{(\alpha)}$ is the pull-back of a local $1$-form on the base manifold $M_0$, then we have a proper principal $D$-bundle on $M_0$, otherwise the patching conditions will be more complicated. We have the following picture
\begin{equation}
\begin{tikzcd}[row sep=4.5ex, column sep=5.3ex]
 & & K \arrow[dl, two heads]\arrow[dr, dotted, end anchor={[yshift=-2ex]}] & \\
 G \arrow[r, hook] & M \arrow[dr, "\pi"', two heads] & & [-2ex]\begin{tabular}{c}Poisson-Lie \\T-fold\end{tabular} \arrow[dl, dotted, start anchor={[yshift=2ex]}] \\
 & & M_0 &
\end{tikzcd}
\end{equation}
where the dotted arrows are again not actual maps, but they are only indicative.
\end{remark}

\begin{remark}[Recovering abelian and non-abelian T-duality]
Notice that in the particular case of a Drinfel'd double $D = T^{n}\times \widetilde{T}^n$ we recover exactly abelian T-duality.
Moreover in the particular case of a Drinfel'd double $D = T^\ast G \cong G\ltimes\mathbb{R}^{\mathrm{dim}\,G}$ with dual fibre $\widetilde{G}:=\mathbb{R}^{\mathrm{dim}\,G}$ we recover exactly the non-abelian T-duality of the previous section. 
\end{remark}

\subsection{Poisson-Lie T-folds as non-abelian global tensor hierarchies}

Non-abelian T-duality is a generalisation of abelian T-duality for string backgrounds whose group of isometries is non-abelian. \textit{Poisson-Lie T-duality} is a further generalisation of this concept where the string background is not even required to have isometries, but which relies on the existence of a more subtle rigid group structure. See \cite{BPV20} for discussion of Poisson-Lie T-duality of a $\sigma$-model in a group manifold and \cite{Has17, DFTWZW19, Demulder:2019bha} for discussion of Poisson-Lie T-duality in DFT. For recent applications concerning the Drinfel'd double $SL(2,\mathbb{C})=SU(2)\bowtie SB(2,\mathbb{C})$ see \cite{Ba19, Vit19, Ba20}. \vspace{0.2cm}

\noindent In this subsection we will introduce the notion of Poisson-Lie T-fold and we will show that it can be formalised by our definition of global tensor hierarchy.

\paragraph{The generalised correspondence space of Poisson-Lie T-duality.}
As we have just seen, a bundle gerbe on a $T^n$-bundle spacetime $\pi:M\twoheadrightarrow M_0$ is abelian T-dualizable if $[\pi_\ast \bigsp{H}]\in H^2(M,\mathbb{Z}^n)$, so that $\pi_\ast \bigsp{H}$ becomes the curvature of a $T^n$-bundle $K\twoheadrightarrow M$, the generalised correspondence space. In the simplest case we examined, we had $(\pi_\ast \bigsp{H})_i = \di(-\iota_{e_i}\bigsp{B}_{(\alpha)})$, which makes $\widetilde{\Xi}_i=\di\widetilde{\theta}_{(\alpha)}-\iota_{e_i}\bigsp{B}_{(\alpha)}$ the global connection $1$-form of the bundle $K$. Now we want to study the generalisation of this from abelian T-duality to Poisson-Lie T-duality.
\vspace{0.25cm}

\noindent We say that \textit{a bundle gerbe $\mathscr{G}\twoheadrightarrow M$ on a $G$-bundle spacetime  $\pi\!:\!M\twoheadrightarrow M_0$ is Poisson-Lie T-dualizable if $\pi_\ast \bigsp{H}$ is the curvature of a $\widetilde{G}$-bundle $K\twoheadrightarrow M$}, where $\widetilde{G}$ is some Lie group with the same dimension $\mathrm{dim}\, G = \mathrm{dim}\, \widetilde{G}$. Here, $(\pi_\ast)_i=\iota_{e_i}$, where $\{e_i\}_{i=1,\dots,\mathrm{dim}\,G}$ is a basis of vertical $G$-left-invariant vectors on $M$. In the simple case we examined, this implies that
\begin{equation}
    (\pi_\ast \bigsp{H})_i \;=\; \di(-\iota_{e_i}\bigsp{B}_{(\alpha)}) + \frac{1}{2}\widetilde{C}_{i}^{\;jk}\, (-\iota_{e_j}\bigsp{B}_{(\alpha)}) \wedge (-\iota_{e_k}\bigsp{B}_{(\alpha)}),
\end{equation}
where $\widetilde{C}_{i}^{\;jk}$ are the structure constants of the Lie algebra $\widetilde{\mathfrak{g}}:=\mathrm{Lie}(\widetilde{G})$. Notice that the local $1$-form $-\iota_{e_i}\bigsp{B}_{(\alpha)}$ is now the local potential of a non-abelian principal bundle. In analogy with abelian T-duality we call the total space $K$ the \textit{generalised correspondence space} of the Poisson-Lie T-duality. Therefore we have a diagram of the following form:
\begin{equation}
    \begin{tikzcd}[row sep={10ex,between origins}, column sep={11ex,between origins}]
    K \arrow[d, two heads] \\
    M \arrow[d, two heads] \arrow[r] & \mathbf{B}\widetilde{G} \\
    M_0 \arrow[r] & \mathbf{B}G.
    \end{tikzcd}
\end{equation}
Crucially the composition $K\twoheadrightarrow M_0$ is a fibre bundle on $M_0$ with fibre $G\times \widetilde{G}$, but it is not a principal bundle. However, for any good cover $\{U_\alpha\}$ of the base manifold $M_0$, the total space $K$ will be still locally of the form $K|_{U_\alpha}\cong U_\alpha \times G \times \widetilde{G}$.

\paragraph{The hidden Drinfel'd double fibre.}
The generalised correspondence space $K$, on any patch $U_\alpha$ of the base manifold $M_0$, can be restricted to a local trivial bundle $K|_{U_\alpha}\cong U_\alpha\times G \times \widetilde{G}$. For the fibre $G\times \widetilde{G}$ we can introduced the parametrisation defined by $\gamma_{(\alpha)}=\exp(\theta_{(\alpha)}^ie_i)$ and $\widetilde{\gamma}_{(\alpha)}=\exp(\widetilde{\theta}_{(\alpha)i}\widetilde{e}^i)$, where $\theta_{(\alpha)}$ and $\widetilde{\theta}_{(\alpha)}$ are local coordinates on the group $G\times \widetilde{G}$ near the identity element.
Now, on each trivial local $G\times\widetilde{G}$-bundle $K|_{U_\alpha} \cong U_\alpha\times G \times \widetilde{G}$ we can construct the following local $\mathfrak{g}\oplus\widetilde{\mathfrak{g}}$-valued differential $1$-form:
\begin{equation}\label{eq:locon}
    \begin{pmatrix}\xi_{(\alpha)}\\[0.4em] \widetilde{\xi}_{(\alpha)} \end{pmatrix}\;:=\; \begin{pmatrix} \gamma_{(\alpha)}^{-1}\di\gamma_{(\alpha)} + \mathrm{Ad}_{\gamma_{(\alpha)}^{-1}}A_{(\alpha)}  \\[1em] \widetilde{\gamma}_{(\alpha)}^{-1}\di\widetilde{\gamma}_{(\alpha)} + \mathrm{Ad}_{\widetilde{\gamma}_{(\alpha)}^{-1}}B^{(1)}_{(\alpha)}  \end{pmatrix} \;\;\in\, \Omega^1\big(U_\alpha\times G \times \widetilde{G},\,\mathfrak{g}\oplus\widetilde{\mathfrak{g}}\big),
\end{equation}
where we called $A_{(\alpha)}=A_{(\alpha)}^i\otimes e_i$ and $B_{(\alpha)}^{(1)}=B_{(\alpha)i}^{(1)}\otimes \widetilde{e}^i$. Here we used the vector notation for elements of $\mathfrak{g}\oplus\widetilde{\mathfrak{g}}$.
Notice that this is a local $G\times\widetilde{G}$-connection $1$-form for our local bundle $U_\alpha\times G\times\widetilde{G}$.
\vspace{0.25cm}

\noindent Now, the global connection data of the generalised correspondence space $K$ is given by the connection of the $G$-bundle $M\twoheadrightarrow M_0$ and the one of the $\widetilde{G}$-bundle $K\twoheadrightarrow M$. We can combine them in a global $\mathfrak{g}\oplus\widetilde{\mathfrak{g}}$-valued $1$-form on the total space $K$ as follows
\begin{equation}
    \Xi\; := \; \begin{pmatrix}\gamma_{(\alpha)}^{-1}\di\gamma_{(\alpha)} + \mathrm{Ad}_{\gamma_{(\alpha)}^{-1}}A_{(\alpha)} \\[1em] \widetilde{\gamma}_{(\alpha)}^{-1}\di\widetilde{\gamma}_{(\alpha)} + \mathrm{Ad}_{\widetilde{\gamma}_{(\alpha)}^{-1}}\big(B^{(1)}_{(\alpha)}-B^{(0)}_{(\alpha)k}\xi^k\big) \end{pmatrix} \;\;\in\, \Omega^1\big(K,\,\mathfrak{g}\oplus\widetilde{\mathfrak{g}}\big).
\end{equation}
The relation between the global $1$-form $\Xi$ encoding the global connection data of the generalised correspondence space and the local $G\times\widetilde{G}$-connections $\xi_{(\alpha)},\widetilde{\xi}_{(\alpha)}$ defined in \eqref{eq:locon} is given by 
\begin{equation}
\begin{aligned}
    \Xi\; = \; \begin{pmatrix}\xi \\[0.3em] \widetilde{\xi}_{(\alpha)} - \mathrm{Ad}_{\widetilde{\gamma}_{(\alpha)}^{-1}}\big(B^{(0)}_{(\alpha)k}\xi^k\big) \end{pmatrix}.
    \end{aligned}
\end{equation}
We can rewrite the relation by making the generators $\{e_i,\,\widetilde{e}^i\}_{i=1,\dots,n}$ of the algebra $\mathfrak{g}\oplus\widetilde{\mathfrak{g}}$ explicit
\begin{equation}\label{eq:d1}
\begin{aligned}
    \Xi\; = \; \begin{pmatrix}\xi^i, & \widetilde{\xi}_{(\alpha)i}\end{pmatrix} \begin{pmatrix}\delta_i^{\;j} & -B^{(0)}_{(\alpha)\ell i}\, \widetilde{u}^{\;\;\;\ell}_{(\alpha)j} \\[0.3em] 0 & \delta^{i}_{\;j} \end{pmatrix}
    \begin{pmatrix}e_j \\[0.3em] \widetilde{e}^{j}\end{pmatrix} ,
    \end{aligned}
\end{equation}
where the matrix $\widetilde{u}^{\;\;\;i}_{(\alpha)j}$ is defined by the adjoint action  $\mathrm{Ad}_{\widetilde{\gamma}_{(\alpha)}^{-1}}(\widetilde{e}^i)= \widetilde{u}^{\;\;\;i}_{(\alpha)j}\,\widetilde{e}^j$ of the group $\widetilde{G}$ on its algebra $\widetilde{\mathfrak{g}}$ and depends only on the local coordinates $\widetilde{\theta}_{(\alpha)}$.
\vspace{0.25cm}

\noindent Let us not introduce the concept of \textit{Drinfel'd double} $D\, :=\,  G \bowtie\widetilde{G}$. A Drinfel'd double $D$ is defined as an even-dimensional Lie group whose Lie algebra $\mathfrak{d}=\mathrm{Lie}(D)$ has underlying vector space $\mathfrak{g}\oplus\widetilde{\mathfrak{g}}$ generated by the generators $\{e_i,\,\widetilde{e}^i\}_{n=1,\dots,i}$ and bracket structure given by
\begin{equation}
    \begin{aligned}
    [e_i,\, e_j]_{\mathfrak{d}} \;&=\; C_{ij}^{\;\;\,k}e_k,\\
    [e_i,\, \widetilde{e}^j]_{\mathfrak{d}} \;&=\; C_{ki}^{\;\;\,j}\widetilde{e}^k - \widetilde{C}^{kj}_{\;\;\;i}e^k,\\
    [\widetilde{e}^i,\, \widetilde{e}^j]_{\mathfrak{d}} \;&=\; \widetilde{C}^{ij}_{\;\;\,k}\widetilde{e}^k.
    \end{aligned}
\end{equation}
Thus we can write its Lie algebra in the split form $\mathfrak{d}=\mathfrak{g}\bowtie\widetilde{\mathfrak{g}}$, where $\mathfrak{g}$ is the subalgebra generated by the generators $\{e_i\}_{n=1,\dots,i}$ and $\widetilde{\mathfrak{g}}$ is the subalgebra generated by the generators $\{\widetilde{e}^i\}_{i=1,\dots,n}$. 
We can also write $D\, =\,  G \bowtie\widetilde{G}$, where $G$ is the Lie group integrating $\mathfrak{g}$ and $\widetilde{G}$ is the one integrating $\widetilde{\mathfrak{g}}$. 
Notice that the manifold underlying the Drinfel'd double group $D\, =\,  G \bowtie\widetilde{G}$ is the same manifold underlying the direct product $G\times \widetilde{G}$, but it crucially comes equipped with a different Lie group structure. 
Now let us choose the parametrisation $\Gamma_{(\alpha)}\,:=\,\gamma_{(\alpha)}\widetilde{\gamma}_{(\alpha)}\in D$ for the Drinfel'd double $D=G\bowtie\widetilde{G}$, where we are still calling $\gamma_{(\alpha)}=\exp(\theta_{(\alpha)}^ie_i)\in G$ and $\widetilde{\gamma}_{(\alpha)}=\exp(\widetilde{\theta}_{(\alpha)i}\widetilde{e}^i)\in \widetilde{G}$ with $\big(\theta_{(\alpha)},\,\widetilde{\theta}_{(\alpha)}\big)$ local coordinates on the product manifold $G\times \widetilde{G}$.
As explained by \cite{Hull07}, the adjoint action of the subgroup $\widetilde{G}$ on the Lie algebra $\mathfrak{d}$ of the full Drinfel'd double $D=G\bowtie\widetilde{G}$ is specified on the generators by the following matrix 
\begin{equation}
    \widetilde{\gamma}_{(\alpha)}^{-1} \begin{pmatrix}e_i \\[0.5em] \widetilde{e}^i \end{pmatrix} \widetilde{\gamma}_{(\alpha)} \;=\; \begin{pmatrix} (\widetilde{u}_{(\alpha)}^{-\mathrm{T}})_i^{\;j} & \widetilde{b}_{(\alpha)ij} \\[0.5em] 0 & \widetilde{u}^{\;\;\;i}_{(\alpha)j} \end{pmatrix} \begin{pmatrix}e_j \\[0.5em] \widetilde{e}^j \end{pmatrix},
\end{equation}
where the submatrices $\widetilde{u}_{(\alpha)}$ and $\widetilde{b}_{(\alpha)}$ depend only on the local coordinates of $\widetilde{G}$ and $\widetilde{b}_{(\alpha)}$ is skew-symmetric.
Similarly, the adjoint action of the subgroup $G$ on the Lie algebra $\mathfrak{d}$ is given on generators by
\begin{equation}
    \gamma_{(\alpha)}^{-1} \begin{pmatrix}e_i \\[0.5em] \widetilde{e}^i \end{pmatrix} \gamma_{(\alpha)} \;=\; \begin{pmatrix} (u_{(\alpha)})_i^{\;j} & 0 \\[0.5em] \beta_{(\alpha)}^{ij} & (u_{(\alpha)}^{-\mathrm{T}})_{\;j}^{i} \end{pmatrix} \begin{pmatrix}e_j \\[0.5em] \widetilde{e}^j \end{pmatrix},
\end{equation}
where this time the matrices ${u}_{(\alpha)}$ and $\beta_{(\alpha)}$ depend only on the local coordinates of $G$ and $\beta_{(\alpha)}$ is skew-symmetric. \vspace{0.25cm}

\noindent Recall that we are parametrising the points of our local bundle by $\big(x_{(\alpha)},\,\Gamma_{(\alpha)}\big)\in U_\alpha\times D$. It was shown by \cite{Hul09} that on each $D$ fibre the Maurer-Cartan $1$-form is given by
\begin{equation}
\begin{aligned}
    \Gamma_{(\alpha)}^{-1}\di\Gamma_{(\alpha)} \;&=\; \begin{pmatrix}\big(\gamma_{(\alpha)}^{-1}\di\gamma_{(\alpha)}\big)^i, & \big(\widetilde{\gamma}_{(\alpha)}^{-1}\di\widetilde{\gamma}_{(\alpha)}\big)_i\end{pmatrix} \begin{pmatrix}\big(\widetilde{u}_{(\alpha)}^{-\mathrm{T}}\big)^{\;j}_{i} & \widetilde{b}_{(\alpha)ij} \\[0.5em] 0 & \delta^{i}_{\;j} \end{pmatrix}
    \begin{pmatrix}e_j \\[0.3em] \widetilde{e}^j\end{pmatrix},
    \end{aligned}
\end{equation}
where $\gamma_{(\alpha)}^{-1}\di\gamma_{(\alpha)}$ is the Maurer-Cartan $1$-form on the subgroup $G$ and $\widetilde{\gamma}_{(\alpha)}^{-1}\di\widetilde{\gamma}_{(\alpha)}$ is the one on $\widetilde{G}$. If we write the Maurer-Cartan $1$-form in terms of the generators of the Drinfel'd double we get
\begin{equation}
\begin{aligned}
    \left(\Gamma_{(\alpha)}^{-1}\di\Gamma_{(\alpha)}\right)^{\!\ind{J}} \;&=\; \begin{pmatrix}\big(\widetilde{u}_{(\alpha)}^{-\mathrm{T}}\big)^{\;j}_{i} \big(\gamma_{(\alpha)}^{-1}\di\gamma_{(\alpha)}\big)^i \\[0.8em] \big(\widetilde{\gamma}_{(\alpha)}^{-1}\di\widetilde{\gamma}_{(\alpha)}\big)_i + \widetilde{b}_{(\alpha)ij} \big(\gamma_{(\alpha)}^{-1}\di\gamma_{(\alpha)}\big)^i \end{pmatrix}.
    \end{aligned}
\end{equation}
Now, on our local bundle $K|_{U_\alpha}\cong U_\alpha\times D$, we can define the following local $\mathfrak{d}$-valued $1$-form
\begin{equation}
    \Gamma_{(\alpha)}^{-1}\di\Gamma_{(\alpha)} + \mathrm{Ad}_{\Gamma_{(\alpha)}^{-1}}\mathcal{A}_{(\alpha)} \;\;\in\,\Omega^1(U_\alpha\times D,\,\mathfrak{d})
\end{equation}
by requiring the identity
\begin{equation}\label{eq:plconn}
\begin{aligned}
    \Gamma_{(\alpha)}^{-1}\di\Gamma_{(\alpha)} + \mathrm{Ad}_{\Gamma_{(\alpha)}^{-1}}\mathcal{A}_{(\alpha)} \;\,&:=\,\; \begin{pmatrix}\xi^i, & \widetilde{\xi}_{(\alpha)i}\end{pmatrix} \begin{pmatrix}\big(\widetilde{u}_{(\alpha)}^{-\mathrm{T}}\big)^{\;j}_{i} & \widetilde{b}_{(\alpha)ij} \\[0.5em] 0 & \delta^{i}_{\;j} \end{pmatrix}
    \begin{pmatrix}e_j \\[0.3em] \widetilde{e}^j\end{pmatrix},
    \end{aligned}
\end{equation}
where $\xi^i$ and $\widetilde{\xi}_{(\alpha)i}$ are the local connections defined in \eqref{eq:locon}. This $1$-form can be seen as a $D$-connection (not necessarily principal) on each local bundle $U_\alpha\times D$. In terms of generators of the Drinfel'd double we have the $1$-forms
\begin{equation}
\begin{aligned}
    \left(\Gamma_{(\alpha)}^{-1}\di\Gamma_{(\alpha)} + \mathrm{Ad}_{\Gamma_{(\alpha)}^{-1}}\mathcal{A}_{(\alpha)}\right)^{\!\ind{J}} \;&=\; \begin{pmatrix}\big(\widetilde{u}_{(\alpha)}^{-\mathrm{T}}\big)^{\;j}_{i} \xi^i \\[0.8em] \widetilde{\xi}_i + \widetilde{b}_{(\alpha)ij} \,\xi^i \end{pmatrix}.
    \end{aligned}
\end{equation}
From equation \eqref{eq:plconn} combined with the identity $\mathrm{Ad}_{\Gamma_{(\alpha)}^{-1}}=\mathrm{Ad}_{\widetilde{\gamma}_{(\alpha)}^{-1}}\circ\mathrm{Ad}_{\gamma_{(\alpha)}^{-1}}$ we can get an explicit expression for the local $1$-form $\mathcal{A}_{(\alpha)}=\mathcal{A}_{(\alpha)}^\ind{I}\otimes E_\ind{I}$ where $\{E_\ind{I}\}$ are collectively the generators of $\mathfrak{d}$. We find that the relation between $\mathcal{A}_{(\alpha)}$ and the potential $A_{(\alpha)}^i$ and the component $B^{(1)}_{(\alpha)i}$ of the Kalb-Ramond field is
\begin{equation}\label{eq:dconn}
\begin{aligned}
    \mathcal{A}^i_{(\alpha)} \,=\, A^i - B^{(1)}_{(\alpha)k}(u_{(\alpha)}^{\mathrm{T}})^k_{\;\ell}\, \beta^{\ell n}_{(\alpha)}\, (u_{(\alpha)}^{-1})^i_{\;n} \quad \text{ and } \quad \widetilde{\mathcal{A}}_{(\alpha)i} \,=\, B^{(1)}_{(\alpha)k}(u_{(\alpha)}^{\mathrm{T}})^k_{\;i}.
\end{aligned}
\end{equation}
When the Drinfel'd double $D$ is an abelian group we immediately recover the usual abelian $1$-form potentials $\mathcal{A}^I_{(\alpha)}=\big(A^i_{(\alpha)},\,B^{(1)}_{(\alpha)i}\big)\in\Omega^1(U_\alpha,\,\mathbb{R}^{2n})$ of the abelian T-fold.
\vspace{0.25cm}

\noindent We can now combine equation \eqref{eq:d1} with equation \eqref{eq:plconn} to find the relation between the global $1$-form $\Xi^I$, encoding the global connection data of the generalised correspondence space, and the local $D$-bundle connection in \eqref{eq:dconn}. The relation is thus given as follows:
\begin{equation}
    \Xi^I\; = \; U^{\;\;\;\,I}_{(\alpha)\,\ind{J}}  \left(\Gamma_{(\alpha)}^{-1}\di\Gamma_{(\alpha)} + \mathrm{Ad}_{\Gamma_{(\alpha)}^{-1}}\mathcal{A}_{(\alpha)} \right)^{\! \ind{J}},
\end{equation}
where we defined the following matrix
\begin{equation}
    U^{\;\;\;\,I}_{(\alpha)\,\ind{J}} \;\,:=\,\; \begin{pmatrix} \widetilde{u}^{\;\;\;i}_{(\alpha)j} & 0 \\[0.6em] \big(\widetilde{b}_{(\alpha)i\ell}-\widetilde{u}^{\;\;k}_{(\alpha)i} B_{(\alpha)k\ell}^{(0)}\big)\widetilde{u}^{\;\;\ell}_{(\alpha)j} & \delta_{i}^{\; j} \end{pmatrix},
\end{equation}
which generally depends on both the local coordinates $\big(\theta_{(\alpha)}^i,\,\widetilde{\theta}_{(\alpha)i}\big)$ of the fibres. Now we must calculate its inverse matrix and find
\begin{equation}
    \left(U_{(\alpha)}^{-1}\right)_{\;J}^{\!\ind{I}} \;\,=\,\; \begin{pmatrix} (\widetilde{u}_{(\alpha)}^{-1})_{\;j}^{i} & 0 \\[0.5em] \widetilde{u}^{\;\;k}_{(\alpha)i} B_{(\alpha)kj}^{(0)}-\widetilde{b}_{(\alpha)ij} & \delta_{i}^{\;j} \end{pmatrix}.
\end{equation}
Finally we can define a \v{C}ech cocycle which is given on two-fold overlaps of patches by
\begin{equation}
    N_{(\alpha\beta)}\;:=\; U_{(\alpha)}^{-1}\,U_{(\beta)}.
\end{equation}
We can calculate this matrix and find
\begin{equation}
   N_{(\alpha\beta)\,\ind{J}}^{\quad\;\ind{I}} \;\,=\,\; \begin{pmatrix} (\widetilde{u}_{(\alpha)}^{-1})_{\;k}^{i} \, \widetilde{u}^{\;\;k}_{(\beta)j} & 0 \\[0.8em] \big[ \big(\widetilde{u}^{\;\;m}_{(\alpha)i} B_{(\alpha)m\ell}^{(0)}-\widetilde{b}_{(\alpha)i\ell}\big) - \big(\widetilde{u}^{\;\;m}_{(\beta)i} B_{(\beta)m\ell}^{(0)}-\widetilde{b}_{(\beta)i\ell}\big)\big]\widetilde{u}^{\;\;\ell}_{(\beta)j} & \delta_{i}^{\; j} \end{pmatrix}.
\end{equation}
Thus we can finally write the patching conditions for our local $D$-bundle connections by
\begin{equation}\label{eq:obstru}
    \left(\Gamma_{(\alpha)}^{-1}\di\Gamma_{(\alpha)} + \mathrm{Ad}_{\Gamma_{(\alpha)}^{-1}}\mathcal{A}_{(\alpha)} \right)^{\! \ind{I}} \;=\; N_{(\alpha\beta)\,\ind{J}}^{\quad\;\ind{I}}\, \left(\Gamma_{(\beta)}^{-1}\di\Gamma_{(\beta)} + \mathrm{Ad}_{\Gamma_{(\beta)}^{-1}}\mathcal{A}_{(\beta)} \right)^{\! \ind{J}}.
\end{equation}
Therefore \textit{the cocycle $N_{(\alpha\beta)}$ represents the obstruction of the generalised correspondence space $K$ from being a global $D$-bundle on the base manifold $M_0$}. In physical terms this means that, whenever the cocycle $N_{(\alpha\beta)}$ is non-trivial,\textit{ the Poisson-Lie T-dual spacetime is not a geometric background, but a T-fold}. This is directly analogous to how the abelian T-fold rises from the generalised correspondence space not being a global $T^{2n}$-bundle (see previous section).
\vspace{0.25cm}

\noindent Moreover, if we include the higher form field, we have that the cocycle $N_{(\alpha\beta)}$ is also the obstruction of the bundle gerbe $\mathscr{G}$ from being equivalent to a global $\String(G\bowtie\widetilde{G})$-bundle on $M_0$. This observation is the key to understand how the tensor hierarchy of the Poisson-Lie T-fold is globalised on the base manifold.
\vspace{0.25cm}

\noindent In the next part of the subsection, like we did for the abelian T-fold, we will compare this structure with the one emerging by gauging the algebra of tensor hierarchies we defined in chapter \ref{ch:4}. We will briefly show that they do not perfectly match, like in the abelian case.

\paragraph{Tensor hierarchy of a Poisson-Lie T-fold.}
Let us define the $2$-algebra of doubled vectors $\mathscr{D}(D)$ on the Drinfel'd double $D=G\bowtie \widetilde{G}$ by directly generalising the $2$-algebra $\mathscr{D}(\mathbb{R}^{2n})$ we saw in \eqref{eq:dofd}. We can then consider the $2$-algebra
\begin{equation}
    \mathscr{D}(D) \;:=\; \Big( \Coo(D) \xrightarrow{\;\;\mathfrak{D}\;\;} \mathfrak{X}(D)\Big).
\end{equation}
The manifestly strong constrained version of the $2$-algebra $\mathscr{D}(D)$ will be given by the following
\begin{equation}
    \mathscr{D}_{\mathrm{sc}}(D) \;=\; \Big( \Coo(G) \xrightarrow{\;\;\mathfrak{D}\;\;} \Gamma(G,\,TG\oplus T^\ast G)\Big),
\end{equation}
since there exists an isomorphism $T^\ast G \cong T\widetilde{G}$, which implies $TG\oplus T^\ast G \,\cong\, TG\oplus T\widetilde{G} \,\cong\, TD$. The algebroid $TG\oplus T^\ast G$ is then generally equipped with anti-symmetrised Roytenberg brackets. However notice that, for a frame $\{E_{\ind{I}}\}$ of $D$-left-invariant generalised vectors, the anti-symmetrised Roytenberg bracket are just
\begin{equation}
    [ E_{\ind{I}}, E_\ind{J} ]_{\mathrm{Roy}} \;=\; C^\ind{K}_{\;\;\,\ind{I}\ind{J}\,}E_\ind{K},
\end{equation}
where the $C^\ind{K}_{\;\;\,\ind{I}\ind{J}}$ are the structure constants of the Drinfel'd double algebra $\mathfrak{d}=\mathfrak{g}\bowtie \widetilde{\mathfrak{g}}$. This means that for such generalised vectors the anti-symmetrised Roytenberg bracket reduces to the Lie bracket $[ -, - ]_{\mathrm{Roy}} = [ -, - ]_{\mathfrak{d}}$ of the Drinfel'd double algebra $\mathfrak{d}$. \vspace{0.25cm}

\begin{remark}[Comparison with higher gauge theory formulation of tensor hierarchies]
Now let us briefly mention how our construction of tensor hierarchies based on higher Kaluza-Klein reduction can be related to a more usual definition of tensor hierarchies as a higher gauge theory.
Let $\exp \mathscr{D}_{\mathrm{sc}}(D)\in\mathbf{H}$ be the Lie integration of the Lie $2$-algebra $\mathscr{D}_{\mathrm{sc}}(D)$. As we have seen in chapters \ref{ch:3} and \ref{ch:4}, we can define the higher gauge theory
\begin{equation}
    \mathscr{T\!\!H}:\, M \, \longmapsto \, \mathbf{H}\Big( \mathscr{P}(M),\, \mathbf{B}\mathrm{Inn}_{\mathrm{adj}}\big(\exp \mathscr{D}_{\mathrm{sc}}(D)\big) \Big),
\end{equation}
where $\mathscr{P}(M)$ is the path $\infty$-groupoid of the smooth manifold $M$. The local data of the connection of such a higher gauge theory will be of the form
\begin{equation}
    \begin{aligned}
        \mathcal{A}_{(\alpha)}=\mathcal{A}_{(\alpha)}^\ind{I}\otimes E_\ind{I}&\in\Omega^1(U_\alpha)\otimes \Gamma(G,TG\oplus T^\ast G), & \mathcal{B}_{(\alpha)}&\in\Omega^2(U_\alpha)\otimes \Coo(G),
    \end{aligned}
\end{equation}
where $\{E_\ind{I}\}$ is a frame of vertical $D$-left-invariant generalised vectors, and the local data of the curvature will be given by the forms
\begin{equation*}
    \begin{aligned}
    \mathcal{F}_{(\alpha)} \;&=\; \di \mathcal{A}_{(\alpha)}  + \big[ \mathcal{A}_{(\alpha)}\,\overset{\wedge}{,}\, \mathcal{A}_{(\alpha)}\big]_{\mathfrak{d}} + \mathfrak{D}\mathcal{B}_{(\alpha)} ,\\
    \mathcal{H} \;&=\;  \di \mathcal{B}_{(\alpha)} + \frac{1}{2}\big\langle \mathcal{A}_{(\alpha)} \,\overset{\wedge}{,}\, \mathcal{F}_{(\alpha)}\big\rangle + \frac{1}{3!}\big\langle \mathcal{A}_{(\alpha)} \,\overset{\wedge}{,}\,\big[ \mathcal{A}_{(\alpha)}\,\overset{\wedge}{,}\, \mathcal{A}_{(\alpha)}\big]_{\mathfrak{d}} \big\rangle .
    \end{aligned}
\end{equation*}
However, by higher Kaluza-Klein reduction we can obtain global tensor hierarchies which are not, strictly speaking, higher gauge theories. This is because the higher gauge theory formulation does not take into account the obstruction $N_{(\alpha\beta)}$ cocycle, appearing in equation \eqref{eq:obstru}, which we get by dimensional reduction of the bundle gerbe.
\end{remark}

\paragraph{Poisson-Lie T-fold as global tensor hierarchy.}
Thus this discussion motivates again the definition of global strong constrained tensor hierarchy by the following dimensional reduction of a bundle gerbe:
\begin{equation}
    \tenhie^{\,G}_{\!\mathrm{sc}}(M_0) \;=\; \left\{\begin{tikzcd}[row sep=7ex, column sep=5ex]
    & \left[G,\mathbf{B}^2U(1)\right]\!/T^n \arrow[d]\\
    M_0 \arrow[ru]\arrow[r, ""] & \mathbf{B}G
    \end{tikzcd}\right\}.
\end{equation}

\section{Physical insights from global DFT compactifications}\label{td7}

Let us conclude this chapter with some remarks about applications to String Theory.

\subsection{The puzzle of the T-dual fibre}
For simplicity, let us consider a $SU(2)$-equivariant bundle gerbe $\mathscr{G}\twoheadrightarrow M$ on a spacetime which is an $SU(2)$-bundle $SU(2)\hookrightarrow M\twoheadrightarrow M_0$. As seen in subsection \ref{geometriccase}, this induces a generalised correspondence space $K$ that is a principal $D$-bundle on $M_0$, with $D=SU(2)\ltimes T^3$.
We can now verify that the extended fibre is compact and, in particular, a $3$-torus $T^3$. \vspace{0.2cm}

\noindent Let us start from the patching equation $\underline{B}_{(\beta)}-\underline{B}_{(\alpha)}=\underline{\di}\underline{\Lambda}_{(\alpha\beta)}$ on $M$. Similarly to subsection \ref{abeliantfold}, we can expand both the differential forms in the connection $\xi\in\Omega^1(M,\mathfrak{su}(2))$ of the $SU(2)$-bundle $M$. As firstly worked out in \cite[p.$\,$59]{Alf19}, this leads to the patching conditions
\begin{equation}
    \begin{aligned}
    B^{(0)}_{(\beta)ij} - B^{(0)}_{(\alpha)ij} \;&=\; -\epsilon^k_{\;\,ij} \Lambda_{(\alpha\beta)k}^{(0)}, \\[0.5ex]
    B^{(1)}_{(\beta)i} - B^{(1)}_{(\alpha)i} \;&=\; \di\Lambda_{(\alpha\beta)i}^{(0)}, \\[0.5ex]
    B^{(2)}_{(\beta)} - B^{(2)}_{(\alpha)} \;&=\; \di\Lambda_{(\alpha\beta)}^{(1)} + \Lambda_{(\alpha\beta)k}^{(0)}F^k.
    \end{aligned}
\end{equation}
Analogously, the patching condition $\underline{\Lambda}_{(\alpha\beta)}+\underline{\Lambda}_{(\beta\gamma)}+ \underline{\Lambda}_{(\gamma\alpha)}=\underline{\di}g_{(\alpha\beta)}$ reduces to
\begin{equation}
    \begin{aligned}
    \Lambda_{(\alpha\beta)i}^{(0)} + \Lambda_{(\beta\gamma)i}^{(0)} + \Lambda_{(\gamma\alpha)i}^{(0)} \;&=\; 0, \\[0.5ex]
    \Lambda_{(\alpha\beta)}^{(1)} + \Lambda_{(\beta\gamma)}^{(1)} + \Lambda_{(\gamma\alpha)}^{(1)} \;&=\; \di g_{(\alpha\beta\gamma)},
    \end{aligned}
\end{equation}
where $g_{(\alpha\beta\gamma)}$ is the \v{C}ech cocycle corresponding to the bundle gerbe $\mathscr{G}\twoheadrightarrow M$. Notice that, with this assumptions, the $1$-form $B^{(1)}_{(\alpha)i}$ is the connection of a principal $T^3$-bundle. A similar, but more complicated, statement will hold for a general T-dualizable bundle gerbe on $M$. 
In the rest of this subsection we will attempt to clarify other geometric aspects of non-abelian T-duality.

\subsection{Application to holographic backgrounds}
Non-abelian T-duality has been used a fundamental tool in studying the structure of AdS/CFT correspondence and in generating new solutions. See seminal work by \cite{Lozano:2012au, Itsios:2013wd}. Moreover, T-duality in AdS/CFT correspondence is also closely related to the fundamental notion of integrability e.g. see \cite{Thompson:2015lzd, Hoare:2016wsk, Demulder:2018lmj}. We redirect to \cite{Thompson:2019ipl} for a broad introduction to these topics. \vspace{0.15cm}

\noindent Let us consider the spacetime $M=\mathrm{AdS}_3\times S^3\times T^4$, which underlies the geometry of a set of NS5-branes wrapped on a $4$-torus $T^4$ and of fundamental strings smeared on the same $T^4$ such that they are all located at the same point in the transverse space.
The $S^3$-bundle $\mathrm{AdS}_3\times S^3\times T^4 \longtwoheadrightarrow \mathrm{AdS}_3\times T^4$ is immediately topologically trivial. Therefore, to investigate its non-abelian T-dual, we need to apply our semi-abelian T-fold construction to a trivial $S^3$-bundle. In particular, we can focus on the $3$-sphere and consider a $S^3$-bundle over the point $S^3\twoheadrightarrow\ast$, i.e. where the base manifold $M_0=\{0\}$ is just a point. In this particular case, the  generalised correspondence space will be just the Lie group $D=SU(2)\ltimes T^3$.
\vspace{0.2cm}

\noindent Now, let us momentarily forget our geometric construction and follow the literature. This will help us to underline some new insights. As discussed in equation \eqref{literature}, in the literature on non-abelian T-duality, one commonly starts from a metric $g=g_{ij}\xi^i\otimes\xi^j$ and the Kalb-Ramond field $B=B_{ij}\xi^i\wedge\xi^j$ on $S^3$, such that $g_{ij}$ and $B_{ij}$ are constant matrices and $\xi^i$ are a basis of left $SU(2)$-invariant $1$-forms. Then, one can T-dualize by 
\begin{equation}\label{eq:explainednatd}
    \widetilde{g}^{ij}+\widetilde{B}^{ij} \;:=\; \Big(g_{ij}+B_{ij}\,+\,\epsilon^{\;\;\,k}_{ij}\widetilde{\theta}_{k}\Big)^{-1},
\end{equation}
So that we obtain the T-dual tensors $\widetilde{g}=\widetilde{g}^{ij}(\widetilde{\theta})\di\widetilde{\theta}_i\otimes\di\widetilde{\theta}_j$ and $\widetilde{B}=\widetilde{B}^{ij}(\widetilde{\theta})\di\widetilde{\theta}_i\wedge\di\widetilde{\theta}_j$. For simplicity, in the following discussion we can choose $B_{ij}=0$.
Commonly, one defines a new set of coordinates
\begin{equation}
    \begin{aligned}
    \widetilde{\theta}_1 \;&=\; r\sin\vartheta, \\
    \widetilde{\theta}_2 \;&=\; r\cos\vartheta\sin\phi, \\
    \widetilde{\theta}_3 \;&=\; r\cos\vartheta\cos\phi, \\
    \end{aligned}
\end{equation}
so that the T-dual metric and Kalb-Ramond field take the following simple form:
\begin{equation}
    \widetilde{g}\,=\, \di r^2 + \frac{r^2}{1+r^2}(\di\vartheta^2 + \sin^2\vartheta \,\di\phi^2), \qquad \widetilde{B}\,=\, -\frac{r^3}{1+r^2}\sin\vartheta\,\di\vartheta \wedge  \di\phi.
\end{equation}
This fact would lead one to think that the new fibre is a non-compact space $\mathbb{R}^3$. However, as noticed by \cite{Bug19}, there is no diffeomorphism relating our tensors at $(r,\vartheta,\phi)$ and at $(r+\Delta r,\vartheta,\phi)$ for any choice of radius $\Delta r$. Let us now na\"{i}vely combine these matrices $\widetilde{g}_{ij}$ and $\widetilde{B}_{ij}$ in a $O(3,3)$-covariant matrix $\widetilde{\mathcal{G}}_{IJ}$. 
We immediately recognize that we can recast T-duality \eqref{eq:explainednatd} as
\begin{equation}
     \widetilde{\mathcal{G}}_{IJ}(\widetilde{\theta}) \;=\; \begin{pmatrix} \delta^\ell_{\;i} & 0 \\[0.8em] \epsilon^{\;\;\,k}_{\ell i}\widetilde{\theta}_{k} & \delta_\ell^{\;i} \end{pmatrix} \mathcal{G}_{LM} \begin{pmatrix} \delta^m_{\;\;j} & 0 \\[0.8em] \epsilon^{\;\;\;\;k}_{m j}\widetilde{\theta}_{k} & \delta_m^{\;\;j} \end{pmatrix}.
\end{equation}
However, given any translation $\Delta\widetilde{\theta}_i$ in the coordinates $(\widetilde{\theta}_1,\widetilde{\theta}_2,\widetilde{\theta}_3)$, one can calculate that
\begin{equation}
     \widetilde{\mathcal{G}}_{IJ}(\widetilde{\theta}+\Delta\widetilde{\theta}) \;=\; \begin{pmatrix} \delta^\ell_{\;i} & 0 \\[0.8em] \epsilon^{\;\;\,k}_{\ell i}\Delta\widetilde{\theta}_{k} & \delta_\ell^{\;i} \end{pmatrix} \widetilde{\mathcal{G}}_{LM}(\widetilde{\theta}) \begin{pmatrix} \delta^m_{\;\;j} & 0 \\[0.8em] \epsilon^{\;\;\;\;k}_{m j}\Delta\widetilde{\theta}_{k} & \delta_m^{\;\;j} \end{pmatrix}.
\end{equation}
At this point, one can have intuition of the fact that we are dealing with a T-fold or, in other words, that the T-dual space is compactified, but in an exotic non-geometric way. Can our geometric construction from subsection \ref{subnatfold} shed some light on this puzzle? \vspace{0.25cm}

\noindent
The generalised correspondence space will be a bundle over the point $D=SU(2)\ltimes T^3\twoheadrightarrow \ast$. Now, the basis of $1$-forms
\begin{equation}
    \begin{pmatrix} \xi^i \\[0.2em] \di \widetilde{\theta}_i  \end{pmatrix}
\end{equation}
that we were trying to use in the previous paragraph do not give a connection for this bundle, since they are not left $D$-invariant. If we try to use these forms as a basis for our doubled geometry, we obtain monodromies of the form
\begin{equation}
    \left.\begin{pmatrix} \xi^i \\[0.2em] \di \widetilde{\theta}_i  \end{pmatrix}\right|_{\widetilde{\theta}+\Delta\widetilde{\theta}} \;=\; \begin{pmatrix} \delta^i_{\;j} & 0 \\[0.8em] \epsilon^{\;\;\,k}_{ij}\Delta\widetilde{\theta}_{k} & \delta_i^{\;j} \end{pmatrix} \left.\begin{pmatrix} \xi^j \\[0.2em] \di \widetilde{\theta}_j  \end{pmatrix}\right|_{\widetilde{\theta}}.
\end{equation}
This comes directly from equation \eqref{eq:natdpat2}, in our particular trivial case.
On the other hand,
\begin{equation}
    (\Gamma^{-1}\di\Gamma)^{ \ind{I}} \;:=\; \begin{pmatrix} \xi^i \\[0.8em] \di \widetilde{\theta}_i + \epsilon^{\;\;\,k}_{ij}\widetilde{\theta}_k\xi^j  \end{pmatrix} \;\; \in\,\Omega^1(D,\, \mathfrak{d})
\end{equation}
is a left $D$-invariant globally-defined $1$-form on $D=SU(2)\ltimes T^3$. Consequently, on the generalised correspondence space the monodromy disappears, i.e.
\begin{equation}
    \left.\begin{pmatrix} \xi^i \\[0.2em] \di \widetilde{\theta}_i + \epsilon^{\;\;\,k}_{ij}\widetilde{\theta}_k\xi^j \end{pmatrix}\right|_{\widetilde{\theta}+\Delta\widetilde{\theta}} \;=\; \left.\begin{pmatrix} \xi^i \\[0.2em] \di \widetilde{\theta}_i + \epsilon^{\;\;\,k}_{ij}\widetilde{\theta}_k\xi^j \end{pmatrix}\right|_{\widetilde{\theta}}.
\end{equation}
The T-dual of $S^3$ is a T-fold, hence there is no obvious undoubled description for it. In other words, since the T-dual background is non-geometric, we can only describe it in the generalised correspondence space $D=SU(2)\ltimes T^3$. The total generalised correspondence space will thus be the manifold $K = \mathrm{AdS}_{3}\times (SU(2)\ltimes T^3) \times T^4$. 
A completely analogous statement will hold for the manifold formalising the near horizon geometry of a $\frac{1}{2}$BPS NS5-brane, i.e. $M=\mathrm{AdS}_{7}\times S^3$. In this case, the generalised correspondence space will be, immediately, $K = \mathrm{AdS}_{7}\times (SU(2)\ltimes T^3)$. 

\subsection{Application to general brane configurations}
The spacetime underlying the $\frac{1}{2}$BPS NS5-brane is, near horizon, $M=\mathrm{AdS}_7\times S^3$, where the $7$-dimensional Anti-de Sitter space $\mathrm{AdS}_7\cong\mathbb{R}^{1,5}\times\mathbb{R}^+$ is diffeomorphic to the product of a world-volume $\mathbb{R}^{1,5}$ and a radial direction $\mathbb{R}^+:=(0,+\infty)\subset\mathbb{R}$.
However, a general spacetime $M$ underlying a fivebrane charged under the Kalb-Ramond field is only required to satisfy the condition $M/W_{\mathrm{NS5}}\cong \mathbb{R}^4-\{0\}$, where $W_{\mathrm{NS5}}$ is a general world-volume of an NS5-brane. Since $\mathbb{R}^4-\{0\}\cong S^3\times \mathbb{R}^+$, our spacetime will be, in general, a fibration of the following form:
\begin{equation}\label{eq:brane}\begin{tikzcd}[row sep=6ex, column sep=8ex]
    M \arrow[d, two heads] \arrow[r] & \ast \arrow[d] \\
    W_{\mathrm{NS5}}\times\mathbb{R}^+ \arrow[r] & \mathbf{B}S^3.
    \end{tikzcd}\end{equation}
This allows us to formalize the spacetime underlying more general fivebrane configurations by generalising the highly symmetric $\frac{1}{2}$BPS configuration $\mathrm{AdS}_7\times S^3$.
For example, we can consider the near horizon geometry of an NS5-brane wrapping a Riemannian surface $\Sigma_g$ with genus $g$. Such a spacetime will be $M=\mathrm{AdS}_{5}\times_\mathrm{w} N$, where the transverse space $N$ is given by a fibration
\begin{equation}\begin{tikzcd}[row sep=6ex, column sep=5ex]
    N \arrow[d, two heads] \arrow[r] & \ast \arrow[d] \\
    \Sigma_g \arrow[r] & \mathbf{B}S^3,
    \end{tikzcd}\end{equation}
which is, generally, a topologically non-trivial $S^3$-bundle on the Riemannian manifold $\Sigma_g$. Now, if we consider a bundle gerbe $\mathscr{G}\twoheadrightarrow M$ on the manifold \eqref{eq:brane} satisfying the T-dualisation condition, this will define a generalised correspondence space $K$ of the form
\begin{equation}
    \begin{tikzcd}[row sep=7ex, column sep=8ex]
    K \arrow[d, two heads] \\
    M \arrow[d, two heads] \arrow[r] & \mathbf{B}T^3 \\
    W_{\mathrm{NS5}}\times\mathbb{R}^+ \arrow[r] & \mathbf{B}S^3 ,
    \end{tikzcd}
\end{equation}
Therefore, the generalised correspondence space $K$ is not generally a principal $D$-bundle on $W_{\mathrm{NS5}}\times\mathbb{R}^+$. As we have seen, it will be generally patched by a cocycle of monodromy matrices $n_{(\alpha\beta)}$ on the base manifold $W_{\mathrm{NS5}}\times\mathbb{R}^+$, on the world-volume. In general, then, the field content on the base manifold $W_{\mathrm{NS5}}\times\mathbb{R}^+$ will be given by $\big\{g, \mathcal{B}_{(\alpha)}, \mathcal{A}_{(\alpha)}^\ind{I},\mathcal{G}_{(\alpha)\ind{IJ}}\big\}$, where $g$ is the reduced metric, $\mathcal{G}_{(\alpha)\ind{IJ}}$ is the moduli field of the generalised metric and $\big\{\mathcal{B}_{(\alpha)}, \mathcal{A}_{(\alpha)}^\ind{I}\big\}$ is the tensor hierarchy, whose fields correspond respectively to the singlet and fundamental representations of the Drinfel'd double $D$. See table \ref{tab:pt3} for a sum.
Now, the fields which carry $D$-indices, i.e. the moduli of the generalised metric $\mathcal{G}_{(\alpha)\ind{IJ}}$ and the field $\mathcal{A}_{(\alpha)}^{\ind{I}}$, will be globally patched by a cocycle $n_{(\alpha\beta)}\in H^1(W_{\mathrm{NS5}}\times\mathbb{R}^+\!,\,\wedge^2\mathbb{Z}^n)$ of monodromy matrices
\begin{equation}\label{eq:thonns5}
\begin{aligned}
    \left(\Gamma_{(\alpha)}^{-1}\di\Gamma_{(\alpha)} + \mathrm{Ad}_{\Gamma_{(\alpha)}^{-1}}\mathcal{A}_{(\alpha)} \right)^{\! \ind{I}} \;&=\; \big(e^{n_{(\alpha\beta)}}\big)^\ind{I}_{\;\ind{J}} \left(\Gamma_{(\beta)}^{-1}\di\Gamma_{(\beta)} + \mathrm{Ad}_{\Gamma_{(\beta)}^{-1}}\mathcal{A}_{(\beta)} \right)^{\! \ind{J}}, \\[0.5ex]
    \mathcal{G}_{(\alpha)\ind{IJ}} \;&=\; \big(e^{n_{(\alpha\beta)}}\big)^\ind{K}_{\;\ind{I}} \big(e^{n_{(\alpha\beta)}}\big)^\ind{L}_{\;\ind{J}} \, \mathcal{G}_{(\beta)\ind{KL}},
\end{aligned}
\end{equation}
on the overlaps of patches of the base manifold $W_{\mathrm{NS5}}\times \mathbb{R}^+$.
Notice that, if the $S^3$-fibration is trivial, e.g. the example $M=\mathrm{AdS}_3\times S^3\times T^4$ we previously discussed, the cocycle $n_{(\alpha\beta)}$ is immediately trivial. \vspace{0.15cm}

\noindent In the context of holography, the global structure of the duality-covariant fields on the general base manifold $W_{\mathrm{NS5}}\times \mathbb{R}^+$, which is given in equation \ref{eq:thonns5}, will be relevant for the understanding beyond the current known examples.
\begin{table}[h!]\begin{center}
\begin{center}
 \begin{tabular}{|| c |  c | c||} 
 \hline
 \hspace{0.8cm}Singlet rep.\hspace{0.8cm} & Fundamental rep. & \hspace{0.5cm}Adjoint rep.$\quad$ \\ [0.5ex] 
 \hline
 $g, \mathcal{B}_{(\alpha)}$ & $\mathcal{A}^I_{(\alpha)}$ & $\mathcal{G}_{(\alpha)IJ}$   \\[0.8ex]  
 \hline
\end{tabular}
\end{center}
\caption{\label{tab:pt3}A summary of the fields transforming under non-abelian T-duality.}\vspace{0.0cm}
\end{center}\end{table}

\noindent Finally, since the global geometry of the moduli space of the string compactifications is supposed to be related to non-perturbative effects in String Theory, the investigation of the global properties of such geometric and non-geometric compactifications is likely to have some relevance in the study of the String Landscape, or in the understanding the Swampland.
\begin{savequote}[8cm]
{\textgreekfont Τὰ μέλλοντα μὴ ταρασσέτω· ἥξεις γὰρ ἐπ' αὐτά, ἐὰν δεήσῃ, φέρων τὸν αὐτὸν λόγον ᾧ νῦν πρὸς τὰ παρόντα χρᾷ.}

Never let the future disturb you. You will meet it, if you have to, with the same weapons of reason which today arm you against the present.
  \qauthor{--- Marcus Aurelius, \textit{Meditations}}
\end{savequote}

\chapter{\label{ch:7}Towards global super-Exceptional Field Theory}

\minitoc

\noindent Higher Kaluza-Klein does not only provide a formal underlying explanation for the existing geometric features of DFT (e.g. para-Hermitian geometry), but it can also overcome their difficulty in being directly generalised to M-theory. Indeed the higher Kaluza-Klein theory we presented in this dissertation has a large number of immediate natural generalisations:
\begin{itemize}
    \item The principal $\infty$-bundle has a structure $n$-group with $n>2$: this could allow us to formulate global ExFT.
    \item The principal $\infty$-bundle has a non-abelian structure $n$-group: this can not only allow us to formulate global heterotic DFT, but also to go slightly beyond Exceptional Field Theory to embody the global spin-twisted structures by \cite{Sat09} and thus to geometrise the complicated interplay of gravity and bundle gerbe. 
    \item The base manifold is a super-manifold: this can immediately allow us to generalise everything we mentioned to their global super-space formulation.
\end{itemize}
We are intrigued by the possibility that a super (non-abelian) higher Kaluza-Klein theory on the total space of the M2/M5-brane twisted $\infty$-bundle on the $11d$ super-spacetime of \cite{FSS18} can be something closer to a geometrised M-theory than what previously allowed. \vspace{0.2cm}

\noindent In this section we will take the first important steps in each of these directions of generalisation. Most of the content of this section was not previously published.

\section{Towards global heterotic Double Field Theory}
Generalising our discussion of Double Field Theory on bundle gerbes to a principal bundle with non-abelian structure $2$-group can lead to a global formulation of heterotic Double Field Theory as introduced by \cite{Hohm11}. 

\subsection{An atlas for heterotic Double Field Theory}
The heterotic doubled space will be identified with the total space of a principal string-bundle $\mathscr{G}_{\mathrm{het}}\twoheadrightarrow M$ on spacetime $M$.

\begin{definition}[Heterotic $\mathrm{String}_{\mathrm{het}\!}^{\mathbf{c}_2}(d)$-bundle]
The string $2$-group $\mathrm{String}_{\mathrm{het}\!}^{\mathbf{c}_2}(d)$ which encodes the higher gauge theory of heterotic Supergravity is defined by the following commuting diagram:
\begin{equation}\begin{tikzcd}[row sep=12ex, column sep=10ex]
\mathbf{B}\mathrm{String}_{\mathrm{het}\!}^{\mathbf{c}_2}(d) \arrow[d, "\mathrm{hofib}\!\,\left(\frac{1}{2}\mathbf{p}_1-\mathbf{c}_2\right)"']\arrow[r] &\ast \arrow[d]  \\
\mathbf{B}\big(\mathrm{Spin}(d)\times G\big)\arrow[r, "\frac{1}{2}\mathbf{p}_1-\mathbf{c}_2"] &\mathbf{B}^3U(1),
\end{tikzcd}\end{equation}
where 
\begin{itemize}
    \item the map $\frac{1}{2}\mathbf{p}_1: \mathbf{B}\mathrm{Spin}(1,9) \rightarrow \mathbf{B}^3U(1)$ is the smooth refinement of the $1$st fractional Pontryagin class of the frame bundle $FM\twoheadrightarrow M$ given by a non-abelian cocycle $M\rightarrow \mathbf{B}\mathrm{Spin}(1,9)$,
    \item the map $\mathbf{c}_2: \mathbf{B}G \rightarrow \mathbf{B}^3U(1)$ is the smooth refinement of the second Chern class of a principal $G$-bundle $P\twoheadrightarrow M$ given by a cocycle $M\rightarrow \mathbf{B}G$.
\end{itemize}
and where we called $\mathbf{B}\mathrm{String}_{\mathrm{het}\!}^{\mathbf{c}_2}(d)$ the moduli-stack of $\mathbf{B}\mathrm{String}_{\mathrm{het}\!}^{\mathbf{c}_2}(d)$-bundles. 
\end{definition}

\begin{remark}[Heterotic Supergravity]
The choices of gauge group in heterotic String Theory are the following:
\begin{equation}
    G \;:=\; SO(32) \quad \text{or} \quad E_8\times E_8.
\end{equation}
Thus, the $\mathrm{String}_{\mathrm{het}\!}^{\mathbf{c}_2}(d)$-bundle $\mathscr{G}_{\mathrm{het}}\twoheadrightarrow M$ will be a twisted $\infty$-bundle of the following form:
\begin{equation}\begin{tikzcd}[row sep=11ex, column sep=4ex]
\mathscr{G}_{\mathrm{het}}\arrow[d]\arrow[r] &\ast \arrow[d] \\ 
FM\!\times_{M}\!P  \arrow[d]\arrow[r] &\mathbf{B}^2U(1) \arrow[d]\arrow[r] & \ast \arrow[d]  \\
M\arrow[r, "f"] & \mathbf{B}\mathrm{String}_{\mathrm{het}\!}^{\mathbf{c}_2}(d) \arrow[r] &\mathbf{B}(\mathrm{Spin}(d)\times G).
\end{tikzcd}\end{equation}
Locally on a patch $U_\alpha\subset M$, the connection of the $\mathrm{String}_{\mathrm{het}\!}^{\mathbf{c}_2}(d)$-bundle is given by a triple $\big(\omega_{(\alpha)},\,A_{(\alpha)},\,B_{(\alpha)}\big)$, where
\begin{equation}
\begin{aligned}
     \omega_{(\alpha)} \,&\in\, \Omega^1(U_\alpha,\mathfrak{spin}(d)),\\
     A_{(\alpha)}\,&\in\, \Omega^1(U_\alpha,\mathfrak{g}), \\
     B_{(\alpha)}\,&\in\, \Omega^2(U_\alpha).
\end{aligned}
\end{equation}
The curvature of a $\mathrm{String}_{\mathrm{het}\!}^{\mathbf{c}_2}(d)$-bundle will be given by
\begin{equation}
    \begin{aligned}
    R_{(\alpha)b}^{\;\;\,a} \;&=\; \di  \omega_{(\alpha)b}^{\;\;\,a} + \omega_{(\alpha)c}^{\;\;\,a} \wedge  \omega_{(\alpha)b}^{\;\;\,c}, \\[0.15cm]
    F_{(\alpha)} \;&=\; \di  A_{(\alpha)} + \big[ A_{(\alpha)} \wedge  A_{(\alpha)} \big]_{\mathfrak{g}}, \\
    H_{(\alpha)} \;&=\; \di B_{(\alpha)} - \mathrm{cs}_3\big(A_{(\alpha)}\big) +  \mathrm{cs}_3\big(\omega_{(\alpha)}\big),
    \end{aligned}
\end{equation}
where $\mathrm{cs}_3\big(A_{(\alpha)}\big)$ and $\mathrm{cs}_3\big(A_{(\alpha)}\big)$ are the Chern-Simons super $3$-forms of the two connections, which are defined by the expressions
\begin{equation}
    \begin{aligned}
    \mathrm{cs}_3\big(A_{(\alpha)}\big) \;&:=\; \tr_{\mathfrak{g}}\Big(A_{(\alpha)}\wedge F_{(\alpha)} +\frac{2}{3}A_{(\alpha)}\wedge \big[A_{(\alpha)}\wedge A_{(\alpha)}\big]_{\mathfrak{g}}\Big), \\
        \mathrm{cs}_3\big(\omega_{(\alpha)}\big) \;&:=\; \tr_{\mathfrak{spin}(d)}\Big(\omega_{(\alpha)}\wedge R_{(\alpha)} +\frac{2}{3}\omega_{(\alpha)}\wedge \big[\omega_{(\alpha)}\wedge \omega_{(\alpha)}\big]_{\mathfrak{spin}(1,9)}\Big).
    \end{aligned}
\end{equation}
\end{remark}

\begin{remark}[Heterotic $L_\infty$-algebra]
Let us consider the $L_\infty$-algebra $\mathfrak{string}_\mathrm{het}^{\mathbf{c}_2}(d)$ of the Lie $\infty$-group $\String_\mathrm{het}^{\mathbf{c}_2}(d)$. Let us define the $L_\infty$-algebra $\mathfrak{string}_\mathrm{het} := \mathbb{R}^d\times \mathfrak{string}_\mathrm{het}^{\mathbf{c}_2}(d)$, whose Chevalley-Eilenberg dg-algebra is the following:
\begin{equation*}
    \mathrm{CE}\big(\mathfrak{string}_\mathrm{het}\big) = \mathbb{R}\!\left[e^\mu,\tau^a,t^i,B\right]/ \!\left(\begin{array}{l}\! \di e^\mu = 0, \\[0.5ex] \di \tau^a = C^a_{\;\,bc}\tau^b\wedge \tau^c, \\[0.5ex] \di t^i \,=\, C^i_{\;\,jk}t^j\wedge t^k, \\[0.5ex] \di B \,=  \kappa_{aa'}C^{a'}_{\;\,bc}\tau^a\wedge \tau^b\wedge \tau^c - \kappa_{ii'}C^{i'}_{\;jk}t^i\wedge t^j\wedge t^k \!\end{array}\right),
\end{equation*}
where $C^a_{\;\,bc}$ and $C^i_{\;\,jk}$ are respectively the structure constants of $\mathfrak{spin}(d)$ and $\mathfrak{g}$, while the degree of the generators are $\deg(e^\mu)=1$, $\deg(\tau^a)=1$, $\deg(t^i)=1$ and $\deg(B)=2$.
\end{remark}

\noindent The $L_\infty$-algebra $\mathfrak{string}_\mathrm{het}$ can be thought as a linearised version of a $\mathrm{String}_\mathrm{het}^{\mathbf{c}_2}(d)$-bundle.

\begin{remark}[Heterotic doubled space]
In perfect analogy with the abelian case, let us now find an atlas for the $L_\infty$-algebra $\mathfrak{string}_\mathrm{het}$. The underlying space of such atlas will be our local chart of heterotic doubled space.
Thus, we have an atlas of the form:
\begin{equation}
    \phi:\,\mathfrak{double}_\mathrm{het}\, \longrightarrow\, \mathfrak{string}_\mathrm{het}.
\end{equation}
It is not difficult to derive that $\mathfrak{double}_\mathrm{het}$ must have a Chevalley-Eilenberg dg-algebra of the following form:
\begin{equation*}
    \mathrm{CE}\big(\mathfrak{double}_\mathrm{het}\big) = \mathbb{R}\!\left[e^\mu, \widetilde{e}_\mu,\tau^a,t^i\right]/ \!\left(\begin{array}{l} \di e^\mu = 0, \\[0.5ex] \di \widetilde{e}_\mu = 0, \\[0.5ex] \di \tau^a = C^a_{\;\,bc}\tau^b\wedge \tau^c, \\[0.5ex] \di t^i \,=\, C^i_{\;\,jk}t^j\wedge t^k \end{array}\right),
\end{equation*}
where all the generators are in degree $1$. As in the abelian case, we have a $2$-degree element $\omega\in\mathrm{CE}(\mathfrak{double}_\mathrm{het})$ given by
\begin{equation}
    \omega \;:=\; \widetilde{e}_\mu\wedge e^\mu + \kappa_{ab} \tau^a\wedge \tau^b - \kappa_{ij} t^i\wedge t^j,
\end{equation}
whose differential satisfies the same equation of the generator $B\in\mathrm{CE}(\mathfrak{string}_\mathrm{het})$, i.e.
\begin{equation}
    \di\omega \;=\; \kappa_{aa'}C^{a'}_{\;\,bc}\tau^a\wedge \tau^b\wedge \tau^c - \kappa_{ii'}C^{i'}_{\;jk}t^i\wedge t^j\wedge t^k.
\end{equation}
Therefore, the element $\omega\in\mathrm{CE}(\mathfrak{double}_\mathrm{het})$ is the transgression of the generator $B\in\mathrm{CE}(\mathfrak{string}_\mathrm{het})$ to the atlas $\mathfrak{double}_\mathrm{het}$.
\end{remark}

\begin{remark}[Heterotic para-Hermitian geometry]
As in the abelian case, the atlas $\mathfrak{double}_\mathrm{het}$ is equipped with the following coordinates
\begin{equation}
    \Coo(\mathfrak{double}_\mathrm{het}) = \langle x^\mu,\widetilde{x}_\mu,\vartheta^a,\theta^i\rangle
\end{equation}
Recall that there exists an isomorphism $\mathrm{CE}(\mathfrak{double}_\mathrm{het})\cong\Omega^\bullet_{\mathrm{li}}(\mathrm{Double}_\mathrm{het})$ where $\mathrm{Double}_\mathrm{het}$ is the Lie group which exponentiates the Lie algebra $\mathfrak{double}_\mathrm{het}$. This isomorphism is given, in local coordinates, by
\begin{equation}
    \begin{aligned}
    \di x^\mu \;&=\; e^\mu, \\
    \di \widetilde{x}_\mu \;&=\; \widetilde{e}_\mu, \\
    \di\vartheta^a \;&=\; \tau^a, \\
    \di\theta^i \;&=\; t^i.
    \end{aligned}
\end{equation}
Thus, the element $\omega\in\mathrm{CE}(\mathfrak{double}_\mathrm{het})$ can be interpreted as a left-invariant differential form $\omega\in\Omega^2_{\mathrm{li}}(\mathrm{Double}_\mathrm{het})$ by
\begin{equation}
    \omega \;=\; \di \widetilde{x}_\mu\wedge \di x^\mu + \kappa_{ab} \di\vartheta^a\wedge \di\vartheta^b - \kappa_{ij} \di\theta^i\wedge \di\theta^j.
\end{equation}
This is nothing but a heterotic generalisation of the fundamental form of para-Hermitian geometry.
\end{remark}

\noindent This means that the string-bundle $\mathscr{G}_{\mathrm{het}}$ can be covered by a collection of manifolds $\mathrm{Double}_\mathrm{het}=\mathbb{R}^{d,d} \times \mathrm{Spin}(d)\times G$. These are glued such that the submanifolds $\mathbb{R}^d\times \mathrm{Spin}(d)\times G$ make up an atlas for the $(\mathrm{Spin}(d)\times G)$-bundle $FM\times_M P\twoheadrightarrow M$, while the $\mathbb{R}^{d,d}$ are, intuitively, patched with bundle gerbe $2$-gauge transformations twisted by non-abelian $(\mathrm{Spin}(d)\times G)$-gauge transformations. 

\subsection{Atiyah $L_\infty$-algebroid for heterotic Double Field Theory}

\begin{remark}[Heterotic Courant algebroid]
The $2$-group of automorphisms of this heterotic string-bundle will extend not only the group of diffeomorphisms of the base manifold $M$, but also the group of automorphisms of the principal $(\mathrm{Spin}(d)\times G)$-bundle $P\rightarrow M$ (see example \ref{ex:nonabelianaut}). Explicitly, for a string-bundle given by $f:M\rightarrow \mathbf{B}\mathrm{String}_\mathrm{het}^{\mathbf{c}_2}(d)$, the $2$-group of its automorphisms will be
\begin{equation}
    \Aut_{\mathbf{H}_/}(f) \,=\, \Big(\Diff(M)\ltimes\Gamma\big(M,\,\mathrm{Ad}(P)\big)\Big)\ltimes\mathbf{H}(M,\,\BU).
\end{equation}
Let us consider the Atiyah $L_\infty$-algebroid $\mathfrak{at}(\mathscr{G}_{\mathrm{het}})=\mathrm{Lie}\big(\Aut_{\mathbf{H}_/}(f)\big)$ of our string-bundle. 
Notice that, on local patches $U\subset M$, this will have the following underlying complex:
\begin{equation}
    \mathfrak{at}(\mathscr{G}_{\mathrm{het}})|_U \,=\, \Big( \Coo(U) \xrightarrow{\;\;\;\di\;\;\;} \Gamma(TU\oplus\mathrm{ad}(P)\oplus T^\ast U) \Big).
\end{equation}
This can be interpreted as a heterotic Courant $2$-algebroid.
\end{remark}

\begin{remark}[Heterotic strong constraint]
Notice that we will have a non-abelian global strong constraint
\begin{equation}
    \mathscr{G}_{\mathrm{het}} /\!/\!_\rho \,\mathrm{String}_{\mathrm{het}\!}^{\mathbf{c}_2}(d) \;\cong\; M,
\end{equation}
which remarkably combines the abelian strong constraint with the cylindricity condition on the $\mathrm{Spin}(d)$-bundle.
Thus doubled metric will depend only on physical coordinates of spacetime $M$.
\end{remark}

\section{Towards global super-Double Field Theory}

See appendix \ref{app:2} for an introduction to supergeometry and Type II Supergravity.

\subsection{An atlas for super-Double Field Theory}

\begin{definition}[Type IIA string super Lie $2$-algebra]
Following \cite{Fiorenza:2013nha}, let us define $\mathfrak{string}_\IIA$ as the super $L_\infty$-algebra whose Chevalley-Eilenberg dg-algebra is the following:
\begin{equation}
    \mathrm{CE}(\mathfrak{string}_\IIA) \;=\; \mathbb{R}\!\left[e^\mu,\psi^\upalpha,B\right]/ \!\left(\begin{array}{l} \di e^\mu = \bar{\psi}\Gamma^\mu_\IIA\psi, \\[0.5ex] \di \psi^\upalpha = 0, \\[0.5ex] \di B = i\bar{\psi}\Gamma_\mu^\IIA\Gamma_{10}\psi\wedge e^\mu \end{array}\right),
\end{equation}
where the degree of the generators are $\deg(e^\mu)=(1,\mathrm{even})$, $\deg(\psi^\upalpha)=(1,\mathrm{odd})$ and $\deg(B)=(2,\mathrm{even})$.
\end{definition}

\begin{definition}[Doubled superspace]
Let us define $\mathfrak{double}_\II$ as the ordinary Lie super-algebra whose Chevalley-Eilenberg dg-algebra is the following:
\begin{equation}
    \mathrm{CE}(\mathfrak{double}_\II) \;=\; \mathbb{R}\!\left[e^\mu,\widetilde{e}_\mu,\psi^\upalpha,\widetilde{\psi}_\upalpha\right]/ \!\left(\begin{array}{l} \di e^\mu = \bar{\psi}\Gamma^\mu_\IIA\psi, \\[0.5ex] \di \widetilde{e}_\mu = i\bar{\psi}\Gamma_\mu^\IIA\Gamma_{10}\psi, \\[0.5ex] \di \psi^\upalpha = 0, \\[0.5ex] \di \widetilde{\psi}_\upalpha = (\bar{\psi}\Gamma_\IIA^\mu)_\upalpha\wedge\widetilde{e}_\mu \end{array}\right),
\end{equation}
where $\deg(e^\mu)=\deg(\widetilde{e}_\mu)=(1,\mathrm{even})$ and $\deg(\psi^\upalpha)=\deg(\widetilde{\psi}_\upalpha)=(1,\mathrm{odd})$.
\end{definition}

\begin{theorem}[$\mathfrak{double}_\II/\mathfrak{string}_\IIA$ correspondence]
The atlas of the super Lie $2$-algebra $\mathfrak{string}_\IIA$ is the super Lie-algebra $\mathfrak{double}_\II$, equipped with a fundamental $2$-form $\omega_\IIA$ which is given by the transgression of the higher generator of $\mathfrak{string}_\IIA$, i.e. such that
\begin{equation}
    \di\omega_\IIA \;=\;  i\bar{\psi}\Gamma_\mu^\IIA\Gamma_{10}\psi\wedge e^\mu.
\end{equation}
\end{theorem}
\begin{proof}
Let us consider a homomorphism of $L_\infty$-algebras of the form
\begin{equation}
    \phi_\IIA:\mathfrak{double}_\IIA \longtwoheadrightarrow \mathfrak{string}_\IIA.
\end{equation}
Let us consider its dual homomorphism of Chevalley-Eilenberg dg-algebras, i.e.
\begin{equation}
    \phi_\IIA^\ast: \mathrm{CE}(\mathfrak{string}_\IIA) \longhookrightarrow \mathrm{CE}(\mathfrak{double}_\IIA),
\end{equation}
which maps the low-degree generators by $e^\mu\mapsto e^\mu$ and $\psi^\upalpha\mapsto\psi^\upalpha$.
Moreover, we must find an element $\omega_\IIA$ with degree $\deg(\omega_\IIA)=(2,\mathrm{even})$ such that
\begin{itemize}
    \item $\omega_\IIA = \phi^\ast_\IIA(B) \in \mathrm{CE}(\mathfrak{double}_\IIA)$,
    \item $\di\omega_\IIA = \phi^\ast_\IIA(\di B) \in \mathrm{CE}(\mathfrak{double}_\IIA)$.
\end{itemize}
Let us posit
\begin{equation}
    \omega_\IIA \;:=\; \widetilde{e}_\mu\wedge e^\mu + {\widetilde{\psi}}_\upalpha\wedge \psi^\upalpha .
\end{equation}
The generators $\widetilde{\psi}$ satisfy the differential equation $\di \widetilde{\psi} = \bar{\psi}\Gamma_\IIA^\mu\wedge\tilde{e}_\mu$. Since $\di\psi=0$, we have $\di(\widetilde{\psi}_\upalpha\wedge \psi^\upalpha) = \di(\widetilde{\psi}_\upalpha)\wedge \psi^\upalpha = (\bar{\psi}\Gamma_\IIA^\mu)_\upalpha\wedge\tilde{e}_\mu\wedge \psi^\upalpha = - (\bar{\psi}\Gamma_\IIA^\mu)_\upalpha\wedge\psi^\upalpha\wedge\tilde{e}_\mu = - (\bar{\psi}\Gamma_\IIA^\mu\psi)\wedge\tilde{e}_\mu$.
Therefore, we have that it satisfies
\begin{equation}
\begin{aligned}
    \di\omega_\IIA \;&=\;  \di\widetilde{e}_\mu\wedge e^\mu - \widetilde{e}_\mu\wedge \di e^\mu + \di(\overline{\widetilde{\psi}}_\upalpha\wedge \psi^\upalpha) \\
    \;&=\;  i\bar{\psi}\Gamma_\mu^\IIA\Gamma_{10}\psi\wedge e^\mu + \di e^\mu \wedge \widetilde{e}_\mu - \bar{\psi}\Gamma_\IIA^\mu\psi\wedge\tilde{e}_\mu \\
    \;&=\;  i\bar{\psi}\Gamma_\mu^\IIA\Gamma_{10}\psi\wedge e^\mu.
\end{aligned}
\end{equation}
Moreover, we can check that the generators $\widetilde{\psi}$ satisfy the property $\di^2=0$ of a differential-graded algebra, i.e.
\begin{equation}
    \di^2\widetilde{\psi}_\upalpha \,=\, i(\Gamma^\mu_\IIA \psi)_\upalpha \wedge \bar{\psi}\Gamma_\mu^\IIA\Gamma_{10}  \psi \,=\, (\di H)_{\upalpha\upbeta\upgamma\updelta}\,\psi^\upbeta\wedge \psi^\upgamma \wedge \psi^\updelta \,=\, 0.
\end{equation}
Therefore we have that $\phi_\IIA:\mathfrak{double}_\IIA \longtwoheadrightarrow \mathfrak{string}_\IIA$ is a well-defined atlas.
\end{proof}

\begin{remark}[Coordinates for the doubled superspace]
The underlying super vector space of the super Lie algebra $\mathfrak{double}_\II$ will be $(20|64)$-dimensional and naturally equipped with the following coordinates:
\begin{equation}
    \Coo( \mathfrak{double}_\II ) \;=\; \left\langle x^\mu, \, \widetilde{x}_\mu, \, \vartheta^\upalpha, \, \widetilde{\vartheta}_\upalpha \right\rangle .
\end{equation}
Now, let $\mathrm{Double}_\II$ be the super Lie group given by Lie integrating the super Lie algebra $\mathfrak{double}_\II$. Then, the Chevalley-Eilenberg dg-algebra of $\mathfrak{double}_\II$ can be interpreted as the dg-algebra of the left-invariant differential forms on the super Lie group $\mathrm{Double}_\II$, i.e.
\begin{equation}
    \mathrm{CE}(\mathfrak{double}_\II) \;\cong\; \big(\,\Omega^\bullet_{\mathrm{li}}(\mathrm{Double}_\II),\,\di\,\big).
\end{equation}
This algebra is generated by the following $1$-forms, expressed in the local coordinates as
\begin{equation}
    \begin{aligned}
    e^\mu &= \di x^\mu + \bar{\vartheta}\Gamma^\mu_\IIA \di\vartheta , \\
    \widetilde{e}_\mu &=  \di \tilde{x}_\mu + i\bar{\vartheta}\Gamma_\mu^\IIA\Gamma_{10}\di\vartheta ,\\
    \psi^\upalpha &= \di \vartheta^{\upalpha}, \\
    \widetilde{\psi}_\upalpha &= \di\widetilde{\vartheta}_{\upalpha} + (\bar{\vartheta}\Gamma_\IIA^\mu)_\upalpha\widetilde{e}_\mu.
    \end{aligned}
\end{equation}
These can be interpreted as super-vielbein for the flat doubled superspace.
Notice that, consistently with the literature, not only the bosonic coordinates are doubled, but also the fermionic ones.
In coordinates, we can rewrite the transgression element by
\begin{equation}
    \omega_\IIA \,=\, (\di \tilde{x}_\mu + i\bar{\vartheta}\Gamma_\mu^\IIA\Gamma_{10}\di\vartheta)\wedge (\di x^\mu + \bar{\vartheta}\Gamma^\mu_\IIA \di\vartheta) + \widetilde{\psi}_\upalpha \wedge \psi^\upalpha .
\end{equation}
\end{remark}

\begin{remark}[Super para-Hermitian geometry]
Notice that the underlying super vector space of $\mathfrak{double}_\II$ is isomorphic to $\mathbb{R}^{10,10|64}$. Moreover, let $J$ be the automorphism induced by the natural splitting $\mathbb{R}^{10,10|64}=\mathbb{R}^{10|32}\oplus (\mathbb{R}^{10|32})^\ast$. Now, notice that the triple $(\mathbb{R}^{10,10|64},J,\omega_\IIA)$ can be interpreted as a super-geometric generalisation of an almost para-Hermitian vector space.
\end{remark}

\noindent Notice that, analogously to \cite{Cederwall:2016ukd}, we can introduce a generalised metric as an $OSp(10,10|64)$-structure, which restricts the natural $GL(10|32)\times GL(10|32)$ frame.

\begin{definition}[Type IIB string super Lie $2$-algebra]
Let us define $\mathfrak{string}_\IIB$ as the super $L_\infty$-algebra whose Chevalley-Eilenberg dg-algebra is the following:
\begin{equation}
    \mathrm{CE}(\mathfrak{string}_\IIB) \;=\; \mathbb{R}\!\left[e^\mu,\psi^\upalpha,B\right]/ \!\left(\begin{array}{l} \di e^\mu = \bar{\psi}\Gamma^\mu_\IIB\psi, \\[0.5ex] \di \psi^\upalpha = 0, \\[0.5ex] \di B = i\bar{\psi}\Gamma_\mu^\IIB\Gamma_{10}\psi\wedge e^\mu \end{array}\right),
\end{equation}
where the degree of the generators are $\deg(e^\mu)=(1,\mathrm{even})$, $\deg(\psi^\upalpha)=(1,\mathrm{odd})$ and $\deg(B)=(2,\mathrm{even})$.
\end{definition}

\subsection{T-duality on the doubled superspace}

\begin{theorem}[T-duality on the doubled superspace]\label{eq:theosuper}
The super Lie algebra $\mathfrak{double}_\II$ is an atlas for both the Type IIA and IIB string $2$-algebras, i.e. there exists a span 
\begin{equation}
    \begin{tikzcd}[row sep=5ex, column sep=2ex] 
    & \mathfrak{double}_\II \arrow[dr, two heads, "\phi_\IIB"]\arrow[dl, two heads, "\phi_\IIA"']& \\
    \mathfrak{string}_\IIA & & \mathfrak{string}_\IIB .  
\end{tikzcd}
\end{equation}
\end{theorem}

\begin{proof}
Let us start from our definition $\mathfrak{double}_\II$ and let us split the bosonic part of the basis of left-invariant forms as $\{e^\mu,\widetilde{e}_\mu\}_{\mu=0,\dots,9}=\{e^{\hat{\mu}},e^9,\widetilde{e}_{\hat{\mu}},\widetilde{e}_9\}_{\mu=0,\dots,8}$. These will immediately satisfy the following differential equations:
\begin{equation}
\begin{aligned}
    \di e^{\hat{\mu}} \;&=\; \bar{\psi}\Gamma^{\hat{\mu}}\psi && (\hat{\mu}=0,\dots,8), \\
    \di \widetilde{e}_{\hat{\mu}} \;&=\; i\bar{\psi}\Gamma_{\hat{\mu}}\Gamma_{10}\psi && (\hat{\mu}=0,\dots,8), \\
    \di e^9 \;&=\; \bar{\psi}\Gamma_\IIA^9\psi && (\mu=9), \\
    \di \widetilde{e}_9 \;&=\; \bar{\psi}\Gamma^\IIB_9\psi&& (\mu=9), \\
    \di \psi^\upalpha \;&=\; 0, \\
    \di \widetilde{\psi}_\upalpha \;&=\; (\bar{\psi}\Gamma^{\hat{\mu}})_\upalpha\wedge\widetilde{e}_{\hat{\mu}} + (\bar{\psi}\Gamma_\IIA^9)_\upalpha\wedge\widetilde{e}_9,
\end{aligned}
\end{equation}
where we used the identity
\begin{equation}
    i\Gamma_9^\IIA\Gamma_{10} \;=\; \Gamma_9^\IIB.
\end{equation}
Now, the algebra $\mathfrak{double}_\II$ can be an atlas for $\mathfrak{string}_\IIA$ if there exists a $(2,\mathrm{even})$-degree element $\omega_\IIB\in\mathrm{CE}(\mathfrak{double}_\II)$ such that
\begin{equation}\label{eq:superIIBpara}
    \di\omega_\IIB \;=\;  i\bar{\psi}\Gamma_\mu^\IIB\Gamma_{10}\psi\wedge e^\mu.
\end{equation}
Notice that, if we define the element $\omega_\IIB\in\mathrm{CE}(\mathfrak{double}_\II)$ by
\begin{equation}
    \omega_\IIA - \omega_\IIB  \;=\; \widetilde{e}_9\wedge e^9,
\end{equation}
then it satisfies the required condition \eqref{eq:superIIBpara}.
Since $e^9$ and $\widetilde{e}_9$ are both $(1,\mathrm{even})$-degree elements of the dg-algebra $\mathrm{CE}(\mathfrak{double}_\II)$ and $\omega_\IIA$ is a $(2,\mathrm{even})$-degree element of the same dg-algebra, then we have that $\omega_\IIB\in\mathrm{CE}(\mathfrak{double}_\II)$.
\end{proof}
\noindent Notice that we can express the super Lie algebra $\mathfrak{double}_\II$ by choosing generators adapted to the super Lie $2$-algebra $\mathfrak{string}_\IIB$ as follows:
\begin{equation}
    \mathrm{CE}(\mathfrak{double}_\II) \;=\; \mathbb{R}\!\left[e^\mu,\widetilde{e}_\mu,\psi^\upalpha,\widetilde{\psi}_\upalpha\right]/ \!\left(\begin{array}{l} \di e^\mu = \bar{\psi}\Gamma^\mu_\IIB\psi, \\[0.5ex] \di \widetilde{e}_\mu = i\bar{\psi}\Gamma_\mu^\IIB\Gamma_{10}\psi, \\[0.5ex] \di \psi^\upalpha = 0, \\[0.5ex] \di \widetilde{\psi}_\upalpha = (\bar{\psi}\Gamma_\IIB^\mu)_\upalpha\wedge\widetilde{e}_\mu \end{array}\right).
\end{equation}

\begin{remark}[T-duality of Type IIA/B string super Lie $2$-algebras]
Thus, the doubled superspace $\mathfrak{double}_\II$ we defined fits in following diagram:
\begin{equation}\label{diag:supertduality}
    \begin{tikzcd}[row sep={15ex,between origins}, column sep={16ex,between origins}]
    & & \mathfrak{double}_\II \arrow[dr, shorten <= 0.8em, shorten >= 0.8em]\arrow[ld, shorten <= 0.8em, shorten >= 0.8em] \arrow[ddll, bend right=50, "\phi_\IIA"', shorten <= 0.8em, shorten >= 0.8em] \arrow[ddrr, bend left=50, "\phi_\IIB", shorten <= 0.8em, shorten >= 0.8em] & & \\
    & \widetilde{\pi}^\ast\mathfrak{string}_\IIA \arrow[dr, "\Pi"', shorten <= 0.8em, shorten >= 0.8em]\arrow[dl, "\widetilde{\pi}", shorten <= 0.8em, shorten >= 0.8em] & & \pi^\ast\mathfrak{string}_\IIB \arrow[dr, "\pi"', shorten <= 0.8em, shorten >= 0.8em]\arrow[dl, "\widetilde{\Pi}", shorten <= 0.8em, shorten >= 0.8em] \\
    \mathfrak{string}_\IIA \arrow[dr, "\Pi"', shorten <= 0.8em, shorten >= 0.8em] & & \mathbb{R}^{1,9|\mathbf{16}\oplus\overline{\mathbf{16}}}\!\!\!\!\!\!\!\!\!\underset{\mathbb{R}^{1,8|\mathbf{16}\oplus\mathbf{16}}}{\times}\!\!\!\!\!\!\!\!\!\mathbb{R}^{1,9|\mathbf{16}\oplus\mathbf{16}} \arrow[dr, "\pi"', shorten <= 0.8em, shorten >= 0.8em]\arrow[dl, "\widetilde{\pi}", shorten <= 0.8em, shorten >= 0.8em] & & [-2.5em]\mathfrak{string}_\IIB \arrow[dl, "\widetilde{\Pi}", shorten <= 0.8em, shorten >= 0.8em] \\
    & \mathbb{R}^{1,9|\mathbf{16}\oplus\overline{\mathbf{16}}} \arrow[dr, "\pi"', shorten <= 0.8em, shorten >= 0.8em] & & \mathbb{R}^{1,9|\mathbf{16}\oplus\mathbf{16}} \arrow[dl, "\widetilde{\pi}", shorten <= 0.8em, shorten >= 0.8em] & \\
    & & \mathbb{R}^{1,8|\mathbf{16}\oplus\mathbf{16}} & &
    \end{tikzcd}
\end{equation}
where, for visual clarity, we used the symbol $\mathfrak{g}\underset{\mathfrak{h}}{\times}\widetilde{\mathfrak{g}}$ for the fiber product of two super Lie algebras $\mathfrak{g}$ and $\widetilde{\mathfrak{g}}$ over a base $\mathfrak{h}$.
\end{remark}

\noindent Similarly to the bosonic case, in principle, we can integrate this diagram to obtain the diagram of topological T-duality of super bundle gerbes.

\begin{definition}[Type IIA/B string super Lie $2$-groups]
Let $\String_\IIA$ and $\String_\IIB$ be the super Lie $2$-groups which integrate respectively the super Lie $2$-algebras $\mathfrak{string}_\IIA$ and $\mathfrak{string}_\IIB$.
\end{definition}

\begin{remark}[T-duality of Type IIA/B string super Lie $2$-groups]
By Lie integrating the super $L_\infty$-algebras and the homomorphisms in lemma \ref{eq:theosuper}, we obtain the atlas
\begin{equation}
    \begin{tikzcd}[row sep=5ex, column sep=2ex] 
    & \mathrm{Double}_\II \arrow[dr, two heads, "\Phi_\IIB"]\arrow[dl, two heads, "\Phi_\IIA"']& \\
    \String_\IIA & & \String_\IIB.
\end{tikzcd}
\end{equation}
Notice that $\Pi:\String_{\mathrm{IIA}}\longtwoheadrightarrow \mathbb{R}^{1,9|\mathbf{16}\oplus\overline{\mathbf{16}}}$ can be interpreted as the bundle gerbe encoding the Type IIA string background at the vacuum and analogously for $\String_{\mathrm{IIB}}$. Thus, the doubled superspace $\mathrm{Double}_\II$ is the atlas of both these bundle gerbes.
\end{remark}

\begin{remark}[T-duality of Type II of bundle super-gerbes]
By integrating the diagram \eqref{diag:supertduality} of super $L_\infty$-algebras to a diagram of super bundle gerbes, where we denote the atlas by $\mathcal{M}$, we find the following:
\begin{equation}\label{eq:supercorrspace}
    \begin{tikzcd}[row sep={12ex,between origins}, column sep={14ex,between origins}]
    & & \M \arrow[dr ]\arrow[ld ] \arrow[ddll, bend right=50, "\Phi"' ] \arrow[ddrr, bend left=50, "\widetilde{\Phi}" ] & & \\
    & \widetilde{\pi}^\ast\mathscr{G}_\IIA \arrow[dr, "\Pi"' ]\arrow[dl, "\widetilde{\pi}" ] & & \pi^\ast\widetilde{\mathscr{G}}_\IIB \arrow[dr, "\pi"' ]\arrow[dl, "\widetilde{\Pi}" ] \\
    \mathscr{G}_\IIA \arrow[dr, "\Pi"' ] & & M\times_{M_0}\widetilde{M} \arrow[dr, "\pi"' ]\arrow[dl, "\widetilde{\pi}" ] & & [-2.5em]\widetilde{\mathscr{G}}_\IIB \arrow[dl, "\widetilde{\Pi}" ] \\
    & M \arrow[dr, "\pi"' ] & & \widetilde{M} \arrow[dl, "\widetilde{\pi}" ] & \\
    & & M_0 & &
    \end{tikzcd}
\end{equation}
where $\mathscr{G}_\IIA\twoheadrightarrow M$ and $\mathscr{G}_\IIB\twoheadrightarrow \widetilde{M}$ are two topologically T-dual super bundle gerbes.
\end{remark}

\begin{remark}[Correspondence superspace]
The principal $T^{2n}$-bundle $M\times_{M_0}\widetilde{M} \twoheadrightarrow M_0$ appearing in the centre of the diagram \eqref{eq:supercorrspace} is, thus, the \textit{correspondence superspace} of the topological T-duality between the Type IIA and IIB spacetimes $M$ and $\widetilde{M}$.
\end{remark}

\section{Towards global Exceptional Field Theory}

The correspondence between doubled spaces and bundle gerbes we explored in this dissertation sheds new light on the global geometry underlying Double Field Theory. Moreover, it provides a higher geometric explanation for the appearance of the extra coordinates and for para-Hermitian geometry. These results are particularly important for the investigation of the other extended geometries, i.e. the exceptional geometries underlying Exceptional Field Theories, whose globalisation is significantly more obscure. In particular, the higher geometric perspective will allow to find a generalisation of para-Hermitian geometry for Exceptional Field Theory. Even if exceptional generalised geometry \cite{Wald08E, Wald11, Wald12} is well-understood, such a generalisation is still quite unclear. A generalised para-Hermitian formalism would be extremely fruitful, for example, in the current research in exceptional Drinfel'd geometries\cite{Sakatani:2019zrs, Malek:2019xrf, Blair:2020ndg, Musaev:2020nrt, Sakatani:2020wah, Malek:2020hpo}.

\subsection{An atlas for Exceptional Field Theory}

\begin{remark}[Twisted $\infty$-bundle structure of $11$d Supergravity]
The higher structure which encompasses the global geometry of the $C$-field of $11$-dimensional supergravity can be seen as a bundle $5$-gerbe twisted by a bundle $2$-gerbe \cite{FSS15x}, which gives rise to the following diagram:
\begin{equation}
\begin{tikzcd}[row sep={18ex,between origins}, column sep={20ex,between origins}]
\mathscr{G}_{\mathrm{M5}}\arrow[r]\arrow[d, "\Pi_{\mathrm{M5}}"'] & \ast \arrow[d] & \\
\mathscr{G}_{\mathrm{M2}}\arrow[r, "f_{\mathrm{M5}}"]\arrow[d, "\Pi_{\mathrm{M2}}"'] & \mathbf{B}^6U(1) \arrow[r]\arrow[d] & \ast \arrow[d]\\
M \arrow[r, "f_{\mathrm{M2/5}}"]\arrow[rr, "f_{\mathrm{M2}}", bend right=30]& \mathbf{B}^6U(1)/\!/\mathbf{B}^2U(1), \arrow[r] & \mathbf{B}^3U(1),
\end{tikzcd}
\end{equation}
where the twisted cocycle $f_{\mathrm{M2/5}}$ can be also generalised to a $4$-cohomotopy cocycle $M\xrightarrow{\;\;f_{\mathrm{M2/5}}\;\;}S^4$, where the $4$-sphere can be given in terms of its minimal Sullivan dg-algebra by
\begin{equation}\label{sphere}
    \mathrm{CE}(S^4)\;=\; \mathbb{R}[g_4,g_7]/\langle\di g_4=0,\;\di g_7 + g_4\wedge g_4=0\rangle,
\end{equation}
where $g_4$ and $g_7$ are respectively $4$- and $7$-degree generators.
\end{remark} 

\begin{remark}[Atlas for ExFT]
The total space of the twisted $\infty$-bundle $\mathscr{G}_\mathrm{M5}\longtwoheadrightarrow M$ can be locally trivialised by the Lie $\infty$-group $\mathbb{R}^{1,10} \times \mathbf{B}^2U(1) \times \mathbf{B}^5U(1)$, whose $L_\infty$-algebra is immediately $\mathbb{R}^{1,10} \oplus \mathbf{b}^2\mathfrak{u}(1)\oplus \mathbf{b}^5\mathfrak{u}(1)$.
It is possible to define an atlas for this $L_\infty$-algebra by
\begin{equation}
    \phi:\,\mathbb{R}^{1,10}_{\mathrm{ex}}\;\longtwoheadrightarrow\;\mathbb{R}^{1,10} \oplus \mathbf{b}^2\mathfrak{u}(1)\oplus \mathbf{b}^5\mathfrak{u}(1),
\end{equation}
whose underlying vector space is given by
\begin{equation}\label{eq:atlasexft}
    \mathbb{R}^{1,10}_{\mathrm{ex}} \;=\;  \mathbb{R}^{1,10}\oplus \wedge^2(\mathbb{R}^{1,10})^\ast \oplus\wedge^5(\mathbb{R}^{1,10})^\ast.
\end{equation}
This can also be interpreted as the atlas of the linearised version of the bundle gerbe $\mathscr{G}_\mathrm{M5}\twoheadrightarrow M$.
In the context of super $L_\infty$-algebras, a notion of super exceptional space $\mathbb{R}^{1,10|\mathbf{32}}_{\mathrm{ex}}$ has been defined by \cite{FSS18, FSS19x, FSS20}. Interestingly, the bosonic form of $\mathbb{R}^{1,10|\mathbf{32}}_{\mathrm{ex}}$ is exactly $\mathbb{R}^{1,10}_{\mathrm{ex}}$ from equation \eqref{eq:atlasexft}.
If we split the base space in time and space by $\mathbb{R}^{1,10}=\mathbb{R}^{1}_{\mathrm{t}}\oplus \mathbb{R}^{10}$, we obtain the decomposition
\begin{equation}\label{eq:excalg}
    \mathbb{R}^{1,10}_{\mathrm{ex}} \;=\; \underbrace{ \mathbb{R}^{10}}_{\text{pp-wave}}\!\oplus\, \underbrace{\wedge^2(\mathbb{R}^{10})^\ast}_{\text{M2-brane}} \oplus \underbrace{ \wedge^2\mathbb{R}^{10} }_{\text{M9-brane}} \oplus \underbrace{\wedge^5(\mathbb{R}^{10})^\ast}_{\text{M5-brane}} \,\oplus\!\!\!  \underbrace{\wedge^6\mathbb{R}^{10}}_{\text{KK-monopole}} ,
\end{equation}
which agrees with the description of brane charges in M-theory \cite{Hull:1997kt}.
Notice that, if we split the base space in an internal and external space by $\mathbb{R}^{1,10}=\mathbb{R}^{1,3}\oplus \mathbb{R}^{7}$, we obtain
\begin{equation}
    \mathbb{R}^{1,10}_{\mathrm{ex}} \;=\; \mathbb{R}^{1,3} \oplus \Big(\mathbb{R}^{7}\oplus \wedge^2(\mathbb{R}^{7})^\ast \oplus\wedge^5(\mathbb{R}^{7})^\ast \oplus \wedge^6\mathbb{R}^{7} \Big) \oplus \,\cdots ,
\end{equation}
where the terms we explicitly wrote correspond to the $(4+56)$-dimensional extended space underlying $E_{7(7)}$ Exceptional Field Theory \cite{Hohm:2013uia}. Moreover, the terms we omitted are mixed terms involving wedge products between $\mathbb{R}^{1,3}$ and $\mathbb{R}^{7}$ which correspond to tensor hierarchies \cite{Cagnacci:2018buk, Hohm19DFT} at $0$-degree. 
\end{remark}

\subsection{Atiyah $L_\infty$-algebroid for Exceptional Field Theory}

\begin{remark}[Exceptional generalised tangent bundle]
The $5$-group $\mathrm{Aut}_{\mathbf{H}_/}(f_\mathrm{M5})$ of automorphisms of the twisted bundle $\mathscr{G}_{\mathrm{M5}}\twoheadrightarrow M$ will be defined defined by the following short exact sequences 
\begin{equation}
\begin{tikzcd}[row sep=2ex, column sep=6ex]
1 \arrow[r, hook] & \mathbf{H}\big(M,\mathbf{B}^2U(1)_{\mathrm{conn}}\big) \arrow[r, hook] & \mathrm{Aut}_{\mathbf{H}_/}(f_\mathrm{M2}) \arrow[r, two heads] & \Diff(M)  \arrow[r, two heads] & 1,\\
1 \arrow[r, hook] & \mathbf{H}\big(M,\mathbf{B}^5U(1)_{\mathrm{conn}}\big) \arrow[r, hook] & \mathrm{Aut}_{\mathbf{H}_/}(f_\mathrm{M5}) \arrow[r, two heads] & \mathrm{Aut}_{\mathbf{H}_/}(f_\mathrm{M2})  \arrow[r, two heads] & 1.
\end{tikzcd}
\end{equation}
These can be immediately recognised as the finite version of the short exact sequences defining the Exceptional algebroid $E_{\text{M5}}\rightarrow M$ which appears in exceptional generalised geometry:
\begin{equation}
\begin{tikzcd}[row sep=2ex, column sep=6ex]
0 \arrow[r, hook] & \wedge^2T^\ast M \arrow[r, hook] & E_{\text{M2}} \arrow[r, two heads] & TM  \arrow[r, two heads] & 0,\\
0 \arrow[r, hook] & \wedge^5T^\ast M \arrow[r, hook] & E_{\text{M5}} \arrow[r, two heads] & E_{\text{M2}}  \arrow[r, two heads] & 0.
\end{tikzcd}
\end{equation}
See \cite{Wald08E} for a comparison.
\end{remark}

\subsection{U-duality from higher geometry}

\begin{remark}[U-duality and tensor hierarchy]
The twisted $\infty$-bundle $\mathscr{G}_{\mathrm{M5}}\twoheadrightarrow M$, in analogy with doubled space in chapter \ref{ch:5}, can be interpreted as the extended space of Exceptional Field Theory. The differential data of such a higher structure will be encoded by a particular \v{C}ech cocycle, which includes a collection of $2$-forms $\Lambda_{(\alpha\beta)}^{\mathrm{M2}}$ and $5$-forms $\Lambda_{(\alpha\beta)}^{\mathrm{M5}}$ on two-fold overlaps of patches $U_\alpha\cap U_\beta$. 
By generalising section \ref{ch:6}, we can obtain the field content of a general tensor hierarchy and its underlying manifold by dimensionally reducing the twisted $\infty$-bundle $\mathscr{G}_{\mathrm{M5}}\twoheadrightarrow M$ on a spacetime which is a torus bundle on some base manifold $M_0$.
\end{remark} 

\begin{example}[ExFT in $5$d]
Now assume that spacetime $M$ is itself a torus $T^6$-bundle over a $5d$ base manifold $M_0$ with transition functions $f_{(\alpha\beta)}^i$ and that the gerbe is equivariant under its torus action. Analogously to the DFT case the $2$-forms will dimensionally reduce to a collection of $2,1,0$-forms $\Lambda_{(\alpha\beta)}^{\mathrm{M2}(2)}$, $\Lambda_{(\alpha\beta) i}^{\mathrm{M2}(1)}$ and $\Lambda_{(\alpha\beta) ij}^{\mathrm{M2}(0)}$ on the base $M_0$ describing the winding modes of the M2-brane on the base. Similarly the $5$-forms on $M$ will split in a collection of $5,4,3,2,1,0$-forms on $M_0$ describing the winded M5-brane on the base. Let us simplify the assumptions by requiring that the $5$-gerbe is not twisted by the $2$-gerbe. Hence transition functions $f_{(\alpha\beta)}^i$ will describe a $T^{6}$-bundle, while $\Lambda_{(\alpha\beta) ij}^{(0)\mathrm{M2}}$ a $T^{15}$-bundle and $\Lambda_{(\alpha\beta) ijkln}^{(0)\mathrm{M5}}$ a $T^6$-bundle on the base $M_0$. In total this is a $T^{27}$-bundle over the $5$d manifold $M_0$ and can be interpreted as the \textit{extended manifold} of ExFT in $5$d (see \cite{HohSam14}) where the $T^{27}$ fiber is the so-called \textit{internal space}. We can then equip it with a $E_{6(6)}(\mathbb{Z})$-action which mixes, topologically, the first Chern classes of the $T^{27}$-bundle and, differentially, the components of the moduli fields $g_{ij}^{(0)}$, $C^{(0)}_{ijk}$ and $\star C^{(0)}_{ijklnm}$. Hence the extended manifold (and generally a U-fold) will be possible to be understood through Higher Kaluza-Klein Theory. Moreover the $1,2,3,4,5,6$-form components of the dimensional reduction of the connection of the gerbe and its $n$-gauge transformations can describe ExFT tensor hierarchy. See \cite{Hohm19DFT} for remarkable hints in this direction, in infinitesimal fashion. If we drop the simplifying unphysical assumption that the $5$-gerbe is not twisted, the extended manifold will be a more complicated globalisation of $U_\alpha\times T^{27}$.
\end{example}

\begin{example}[ExFT in $7$d]
Let us make another simpler example. If we choose $M$ to be a $T^4$-bundle over a $7$d base manifold $M_0$ with transition functions $f_{(\alpha\beta)}^i$, the local data $\Lambda_{(\alpha\beta)}^{\mathrm{M2}}$ will include the transition functions $\Lambda_{(\alpha\beta) ij}^{\mathrm{M2}(0)}$ of a $T^6$-bundle over $M_0$. On the other hand the local data $\Lambda_{(\alpha\beta)}^{\mathrm{M5}}$ will not. Hence, coherently with ExFT in $7$d (see \cite{BlaMal15}), the transition functions $f_{(\alpha\beta)}^i$ and $\Lambda_{(\alpha\beta) ij}^{\mathrm{M2}(0)}$ will define a just a $T^{10}$-bundle over $M_0$, which can be equipped with a $SL(5,\mathbb{Z})$-action mixing the components of the moduli fields $\{g_{ij}^{(0)},C^{(0)}_{ijk}\}$.
\end{example}

\section{Towards global super-Exceptional Field Theory}

\subsection{An atlas for super-Exceptional Field Theory}

Recall, from chapter \ref{ch:3}, that the higher structure which encodes the geometry of the fluxes of $11$d Supergravity can be linearised by the $L_\infty$-brane $\mathfrak{m5brane}$ \eqref{eq:m5branelinftyalgebra}, which can be dually defined by its Chevalley-Eilenberg dg-algebra as follows:
\begin{equation*}
    \mathrm{CE}(\mathfrak{m5brane}) \;=\; \frac{\mathbb{R}\!\left[e^a,\psi^\upalpha, c_3,c_6\right]}{ \!\left(\begin{array}{l} \di e^a = \overline{\psi}\Gamma^a\psi, \\[0.5ex] \di \psi^\upalpha = 0, \\[0.5ex] \di c_3 = \overline{\psi}\Gamma_{ab}\psi\wedge e^a\wedge e^b , \\[0.5ex] \di c_6 = c_3\wedge \!\left(\overline{\psi}\Gamma_{ab}\psi\wedge e^a\wedge e^b\right) + \frac{1}{2}\overline{\psi}\Gamma_{a_1\cdots a_5}\psi\wedge e^{a_1}\wedge \cdots \wedge e^{a_5}\end{array}\right)}.
\end{equation*}
Just like the previous cases, we want to find an atlas for this super $L_\infty$-algebra of the form
\begin{equation}\label{eq:atlasm5}
    \phi:\, \mathbb{R}^{1,10|\mathbf{32}}_{\mathrm{exc}}\;\longrightarrow\; \mathfrak{m5brane},
\end{equation}
where $\mathbb{R}^{1,10|\mathbf{32}}_{\mathrm{exc}}$ is some super Lie algebra.
In \cite{FSS19x,FSS20} the following super Lie algebra was proposed:
\begin{equation*}
    \mathrm{CE}\big(\mathbb{R}^{1,10|\mathbf{32}}_{\mathrm{exc}}\big) \;=\;\frac{\mathbb{R}\!\left[e^a,\widetilde{e}_{a_1\cdots a_5},\psi^\upalpha,\widetilde{\psi}^\upalpha\right]}{ \!\left(\begin{array}{l} \di e^a\,= \overline{\psi}\Gamma^a\psi, \\[0.8ex] \di\widetilde{e}_{ab}=\frac{i}{2}\overline{\psi}\Gamma_{a_b}\psi, \\[0.8ex]\di\widetilde{e}_{a_1\cdots a_5}= \frac{1}{5!}\overline{\psi}\Gamma_{a_1\cdots a_5}\psi, \\[0.8ex] \di\psi\, = 0, \\[0.8ex] \di\widetilde{\psi}=(s+1)ie^a\Gamma_a\psi + \widetilde{e}_{ab}\Gamma^{ab}\psi +(1+\frac{s}{6})\widetilde{e}_{a_1\cdots a_5}\Gamma^{a_1\cdots a_5}\psi\end{array}\right)}.
\end{equation*}
Notice that the Lie algebra $\mathbb{R}^{1,10}_{\mathrm{exc}}$ previously defined in \eqref{eq:excalg} is exactly the bosonic component of the super Lie algebra $\mathbb{R}^{1,10|\mathbf{32}}_{\mathrm{exc}}$.
In \cite{FSS19x} it was also shown that there exists a $(3,\mathrm{even})$-degree element $\omega_{\mathrm{M2}}\in\mathrm{CE}\big(\mathbb{R}^{1,10|\mathbf{32}}_{\mathrm{exc}}\big)$ which transgresses the M2-brane cocycle to the atlas, i.e. such that
\begin{equation}
    \di \omega_\mathrm{M2} \;=\;  \overline{\psi}\Gamma_{ab}\psi\wedge e^a\wedge e^b.
\end{equation}
In \cite{FSS19x} it was shown that such an element will be of the following form:
\begin{equation}
\begin{aligned}
    \omega_{\mathrm{M2}} \;&:=\; k_0(s)\,\widetilde{e}_{ab}\wedge e^a \wedge e^b \; +\\
    &\;+\; k_1(s)\, \widetilde{e}^{a_1}_{\;\;a_2}\wedge \widetilde{e}^{a_2}_{\;\;a_3}\wedge \widetilde{e}^{a_3}_{\;\;a_1} \; + \\
    &\;+\; k_2(s)\, \epsilon_{a_1\cdots a_5b_1\cdots b_5c}\, \widetilde{e}^{a_1\cdots a_5}\wedge \widetilde{e}^{b_1\cdots b_5} \wedge e^c \; +\\
    &\;+\; k_3(s)\, \epsilon_{a_1\cdots a_6b_1\cdots b_5}\, \widetilde{e}^{a_1a_2a_3}_{\qquad\;\, c_1c_2} \wedge \widetilde{e}^{a_4a_5a_6c_1c_2} \wedge \widetilde{e}^{b_1\cdots b_5}\; +\\
    &\;+\; -\frac{1}{2}\overline{\eta}_\upbeta\wedge \psi^\upalpha \wedge \!\left(k_4(s)\,\Gamma_{a\;\upbeta}^{\;\upalpha} e^a + k_5(s)\,\Gamma_{\;\;\;\;\;\;\,\upbeta}^{ab\,\upalpha} \widetilde{e}_{ab} + k_6(s)\,\Gamma_{\;\;\;\;\;\;\;\;\;\;\;\;\;\,\upbeta}^{a_1\cdots a_5\,\upalpha} \widetilde{e}_{a_1\cdots a_5} \right),
\end{aligned}
\end{equation}
where the $k_i(s)$ are analytic functions of the parameter $s\in\mathbb{R}-\{0\}$.

\subsection{Relation with D'Auria-Fr\'{e} algebra, M-algebra and $\mathfrak{osp}(1|\mathbf{32})$}

\noindent Now, we will briefly explain how the super algebra $\mathbb{R}^{1,10|\mathbf{32}}_{\mathrm{exc}}$, which underlies Exceptional Field Theory, is related to the D'Auria-Fr\'{e} super algebra, which was independently discovered by studying $11$d Supergravity, without U-duality.

\begin{remark}[Relation with D'Auria-Fr\'{e} super algebra]
The $L_\infty$-algebra $\mathfrak{m5brane}$ can be seen as a linearised version of twisted $\infty$-bundle of $11$d supergravity. However, the full field content of $11$d supergravity includes a spin-connection $\omega^a_{\;b}$, whose curvature $R^a_{\;b}=\di\omega^a_{\;b}+\omega^a_{\;c}\wedge \omega^c_{\;b}$ can be identified with the notion of curvature of general relativity. Thus, the full field content of $11$d supergravity will be given by a $\mathfrak{sugra}_{11}$-valued connection where the $L_\infty$-algebra $\mathfrak{sugra}_{11}$ is dually defined as follows:
\begin{equation*}
    \mathrm{CE}\big(\mathfrak{sugra}_{11}\big) \;=\; \frac{\mathbb{R}\!\left[\omega^a_{\;b},e^\mu,\psi^\upalpha, c_3,c_6\right]}{ \!\left(\begin{array}{l} D\omega^a_{\;b}=0, \\[0.5ex] D e^\mu = \overline{\psi}\Gamma^\mu\psi, \\[0.5ex] D \psi^\upalpha = 0, \\[0.5ex] \di c_3 = \overline{\psi}\Gamma_{ab}\psi\wedge e^a\wedge e^b , \\[0.5ex] \di c_6 = c_3\wedge \!\left(\overline{\psi}\Gamma_{ab}\psi\wedge e^a\wedge e^b\right) + \frac{1}{2}\overline{\psi}\Gamma_{a_1\cdots a_5}\psi\wedge e^{a_1}\wedge \cdots \wedge e^{a_5}\end{array}\right)}.
\end{equation*}
where $D$ is the spin-covariant derivative. This super $L_\infty$-algebra underlying $11$d Supergravity was introduced by \cite{Fiorenza:2013nha}.
Notice that there is a natural projection $\mathfrak{sugra}_{11}\twoheadrightarrow\mathfrak{m5brane}$ of super $L_\infty$-algebras given by the quotient map
\begin{equation}
    \mathfrak{m5brane}\;\cong \;\mathfrak{sugra}_{11}/\mathfrak{so}(1,10).
\end{equation}
The D'Auria-Fr\'{e} super algebra $\mathfrak{df}\text{-}\mathfrak{algebra}$, introduced in \cite{DAuria:1982uck}, can be dually defined by its Chevalley-Eilenberg dg-algebra:
\begin{equation*}
    \mathrm{CE}\big(\mathfrak{df}\text{-}\mathfrak{algebra}\big) \;=\;\frac{\mathbb{R}\!\left[\omega^a_{\;b},e^a,\widetilde{e}_{ab},\widetilde{e}_{a_1\cdots a_5},\psi^\upalpha,\widetilde{\psi}^\upalpha\right]}{ \!\left(\begin{array}{l} D\omega^a_{\;b} =0, \\[0.8ex] De^a\,= \overline{\psi}\Gamma^a\psi, \\[0.8ex] D\widetilde{e}_{ab}=\frac{i}{2}\overline{\psi}\Gamma_{a_b}\psi, \\[0.8ex]D\widetilde{e}_{a_1\cdots a_5}= \frac{1}{5!}\overline{\psi}\Gamma_{a_1\cdots a_5}\psi, \\[0.8ex] D\psi\, = 0, \\[0.8ex] D\widetilde{\psi}=(s+1)ie^a\Gamma_a\psi + \widetilde{e}_{ab}\Gamma^{ab}\psi +(1+\frac{s}{6})\widetilde{e}_{a_1\cdots a_5}\Gamma^{a_1\cdots a_5}\psi\end{array}\right)}
\end{equation*}
where $D$ is the spin-covariant derivative and $s\in\mathbb{R}-\{0\}$ is a constant.
Notice that there is a natural projection $\mathfrak{df}\text{-}\mathfrak{algebra}\twoheadrightarrow\mathbb{R}^{1,10|\mathbf{32}}_{\mathrm{exc}}$ of super Lie algebras given by the quotient map
\begin{equation}
    \mathbb{R}^{1,10|\mathbf{32}}_{\mathrm{exc}}\;\cong \;\mathfrak{df}\text{-}\mathfrak{algebra}/\mathfrak{so}(1,10),
\end{equation}
where $\mathbb{R}^{1,10|\mathbf{32}}_{\mathrm{exc}}$ is the super-exceptional space proposed appearing in \cite{FSS19x}.
The results of \cite{DAuria:1982uck} imply that there exists an atlas
\begin{equation}
    \phi':\,\mathfrak{df}\text{-}\mathfrak{algebra} \;\longtwoheadrightarrow\; \mathfrak{sugra}_{11}.
\end{equation}
This atlas naturally extends the atlas \eqref{eq:atlasm5} we constructed in the previous paragraph.
\end{remark}

\noindent Now, we will illustrate the interesting relation between these algebras and other independently studied super algebras in M-theory: the M-algebra and $\mathfrak{osp}(1|\mathbf{32})$.

\begin{remark}[Relation with M-algebra and $\mathfrak{osp}(1|\mathbf{32})$]
The M-algebra was defined in \cite{Sezgin:1996cj} and it can be dually given by its Chevalley-Eilenberg dg-algebra
\begin{equation*}
    \mathrm{CE}\big(\mathfrak{m}\text{-}\mathfrak{algebra}\big) \;=\;\frac{\mathbb{R}\!\left[\omega^a_{\;b},e^a,\widetilde{e}_{ab},\widetilde{e}_{a_1\cdots a_5},\psi^\upalpha\right]}{ \!\left(\begin{array}{l} D\omega^a_{\;b} =0, \\[0.8ex] De^a\,= \overline{\psi}\Gamma^a\psi, \\[0.8ex] D\widetilde{e}_{ab}=\frac{i}{2}\overline{\psi}\Gamma_{a_b}\psi, \\[0.8ex]D\widetilde{e}_{a_1\cdots a_5}= \frac{1}{5!}\overline{\psi}\Gamma_{a_1\cdots a_5}\psi, \\[0.8ex] D\psi\, = 0\end{array}\right)},
\end{equation*}
where $D$ is the spin-covariant derivative.
In terms of generators of $\mathfrak{m}\text{-}\mathfrak{algebra}$, where we call $\{\mathsf{q}_\upalpha,\mathsf{p}_a,\mathsf{z}_{ab},\mathsf{z}_{a_1\cdots a_5}\}$ the dual generators respectively of $\{\psi^\upalpha,e^a,\widetilde{e}^{ab},\widetilde{e}^{a_1\cdots a_5}\}$, we obtain the usual commutation relation
\begin{equation}
    \{\mathsf{q},\mathsf{q}\} \;=\; i(C\Gamma^a)\mathsf{p}_a +  \frac{1}{2}(C\Gamma^{ab})\mathsf{z}_{ab}+  \frac{i}{5!}(C\Gamma^{a_1\cdots a_5})\mathsf{z}_{a_1\cdots a_5}.
\end{equation}
Notice that the $\mathfrak{m}\text{-}\mathfrak{algebra}$ can be immediately embedded in the $\mathfrak{df}\text{-}\mathfrak{algebra}$.
In fact, the $\mathfrak{df}\text{-}\mathfrak{algebra}$ can be obtained by extension of the $\mathfrak{m}\text{-}\mathfrak{algebra}$ with the extra odd generators $\widetilde{\psi}^\upalpha$.
Now, the super Lie algebra $\mathfrak{osp}(1|\mathbf{32})$, decomposed in terms of its subalgebra $\mathfrak{so}(1,10)$, is dually given by its Chevalley-Eilenberg dg-algebra
\begin{equation*}
    \mathrm{CE}\big(\mathfrak{osp}(1|\mathbf{32})\big) \;=\; \frac{\mathbb{R}\!\left[\omega^a_{\;b},e^a,\widetilde{e}_{a_1\cdots a_5},\psi^\upalpha\right]}{ \!\left(\begin{array}{l} D\omega^a_{\;b} =- s^2e^a\wedge e_b - \frac{s^2}{4!}\widetilde{e}^{ab_1\cdots b_4}\wedge\widetilde{e}^{b}_{\;b_1\cdots b_4} - \frac{s}{2}\overline{\psi}\Gamma^a_{\;b}\psi, \\[0.9ex] De^a\,= \frac{i}{2}\overline{\psi}\Gamma^a\psi + \frac{s}{2(5!)^2}\epsilon^{ab_1\cdots b_5c_1\cdots c_5}\widetilde{e}_{b_1\cdots b_5}\wedge \widetilde{e}_{c_1\cdots c_5} , \\[0.9ex] D\widetilde{e}^{a_1\cdots a_5}= \frac{s}{5!}\epsilon^{a_1\cdots a_5b_1\cdots b_6}\widetilde{e}_{b_1\cdots b_5}\wedge e_{b_6} +\\[0.9ex] \qquad\qquad\! - \frac{5s}{6!}\epsilon^{a_1\cdots a_5b_1\cdots b_6}\widetilde{e}^{c_1c_2}_{\quad\;\; b_1b_2b_3}\wedge \widetilde{e}_{c_1c_2b_4\cdots b_6} + \frac{i}{2}\overline{\psi}\Gamma^{a_1\cdots a_5}\psi, \\[0.9ex] D\psi\, = \frac{i}{2}\Gamma_a\psi\wedge e^a + \frac{is}{2\cdot 5!}\Gamma_{a_1\cdots a_5}\psi\wedge\widetilde{e}^{a_1\cdots a_5} \end{array}\right)},
\end{equation*}
where $D$ is the spin-covariant derivative and $s$ is a constant.
The super Lie algebra $\mathfrak{osp}(1|\mathbf{32})$ was related to the M-algebra by a procedure known as S-expansion by \cite{Izaurieta:2006zz}.
\end{remark}

\begin{figure}[!ht]\hspace{-5ex}\centering
\tikzset{every picture/.style={line width=0.75pt}} 
\begin{tikzpicture}[x=0.75pt,y=0.75pt,yscale=-1,xscale=1]

\draw    (87,72.5) -- (87,115.75) ;
\draw [shift={(87,117.75)}, rotate = 270] [color={rgb, 255:red, 0; green, 0; blue, 0 }  ][line width=0.75]    (8.74,-2.63) .. controls (5.56,-1.12) and (2.65,-0.24) .. (0,0) .. controls (2.65,0.24) and (5.56,1.12) .. (8.74,2.63)   ;
\draw    (87,186.5) -- (87,231.75) ;
\draw [shift={(87,186.5)}, rotate = 90] [color={rgb, 255:red, 0; green, 0; blue, 0 }  ][line width=0.75]    (14.11,-2.63) .. controls (10.93,-1.12) and (8.02,-0.24) .. (5.37,0) .. controls (8.02,0.24) and (10.93,1.12) .. (14.11,2.63)(8.74,-2.63) .. controls (5.56,-1.12) and (2.65,-0.24) .. (0,0) .. controls (2.65,0.24) and (5.56,1.12) .. (8.74,2.63)   ;
\draw    (87,298) -- (87,343.25) ;
\draw [shift={(87,343.25)}, rotate = 270] [color={rgb, 255:red, 0; green, 0; blue, 0 }  ][line width=0.75]    (14.11,-2.63) .. controls (10.93,-1.12) and (8.02,-0.24) .. (5.37,0) .. controls (8.02,0.24) and (10.93,1.12) .. (14.11,2.63)(8.74,-2.63) .. controls (5.56,-1.12) and (2.65,-0.24) .. (0,0) .. controls (2.65,0.24) and (5.56,1.12) .. (8.74,2.63)   ;
\draw    (305.5,297.5) -- (305.5,342.75) ;
\draw [shift={(305.5,342.75)}, rotate = 270] [color={rgb, 255:red, 0; green, 0; blue, 0 }  ][line width=0.75]    (14.11,-2.63) .. controls (10.93,-1.12) and (8.02,-0.24) .. (5.37,0) .. controls (8.02,0.24) and (10.93,1.12) .. (14.11,2.63)(8.74,-2.63) .. controls (5.56,-1.12) and (2.65,-0.24) .. (0,0) .. controls (2.65,0.24) and (5.56,1.12) .. (8.74,2.63)   ;
\draw    (231,245.75) -- (171,245.75) ;
\draw [shift={(231,245.75)}, rotate = 180] [color={rgb, 255:red, 0; green, 0; blue, 0 }  ][line width=0.75]    (14.11,-2.63) .. controls (10.93,-1.12) and (8.02,-0.24) .. (5.37,0) .. controls (8.02,0.24) and (10.93,1.12) .. (14.11,2.63)(8.74,-2.63) .. controls (5.56,-1.12) and (2.65,-0.24) .. (0,0) .. controls (2.65,0.24) and (5.56,1.12) .. (8.74,2.63)   ;
\draw    (232.5,371.25) -- (172.5,371.25) ;
\draw [shift={(232.5,371.25)}, rotate = 180] [color={rgb, 255:red, 0; green, 0; blue, 0 }  ][line width=0.75]    (14.11,-2.63) .. controls (10.93,-1.12) and (8.02,-0.24) .. (5.37,0) .. controls (8.02,0.24) and (10.93,1.12) .. (14.11,2.63)(8.74,-2.63) .. controls (5.56,-1.12) and (2.65,-0.24) .. (0,0) .. controls (2.65,0.24) and (5.56,1.12) .. (8.74,2.63)   ;
\draw    (528,297.5) -- (528,342.75) ;
\draw [shift={(528,342.75)}, rotate = 270] [color={rgb, 255:red, 0; green, 0; blue, 0 }  ][line width=0.75]    (14.11,-2.63) .. controls (10.93,-1.12) and (8.02,-0.24) .. (5.37,0) .. controls (8.02,0.24) and (10.93,1.12) .. (14.11,2.63)(8.74,-2.63) .. controls (5.56,-1.12) and (2.65,-0.24) .. (0,0) .. controls (2.65,0.24) and (5.56,1.12) .. (8.74,2.63)   ;
\draw    (438.5,245.75) -- (378.5,245.75) ;
\draw [shift={(438.5,245.75)}, rotate = 180] [color={rgb, 255:red, 0; green, 0; blue, 0 }  ][line width=0.75]    (14.11,-2.63) .. controls (10.93,-1.12) and (8.02,-0.24) .. (5.37,0) .. controls (8.02,0.24) and (10.93,1.12) .. (14.11,2.63)(8.74,-2.63) .. controls (5.56,-1.12) and (2.65,-0.24) .. (0,0) .. controls (2.65,0.24) and (5.56,1.12) .. (8.74,2.63)   ;
\draw    (440,371.25) -- (380,371.25) ;
\draw [shift={(440,371.25)}, rotate = 180] [color={rgb, 255:red, 0; green, 0; blue, 0 }  ][line width=0.75]    (14.11,-2.63) .. controls (10.93,-1.12) and (8.02,-0.24) .. (5.37,0) .. controls (8.02,0.24) and (10.93,1.12) .. (14.11,2.63)(8.74,-2.63) .. controls (5.56,-1.12) and (2.65,-0.24) .. (0,0) .. controls (2.65,0.24) and (5.56,1.12) .. (8.74,2.63)   ;

\draw (24,4.5) node [anchor=north west][inner sep=0.75pt]  [font=\small] [align=left] {\begin{minipage}[lt]{88.52784pt}\setlength\topsep{0pt}

\begin{equation*}
\mathfrak{osp}( 1|\mathbf{32})
\end{equation*}
\begin{center}

$\displaystyle \left\{\omega ^{a}_{\; b} ,e^{a} ,\tilde{e}_{{a_{1} \cdots a_{5}}} ,\psi \right\}$
\end{center}

\end{minipage}};
\draw (10.5,116.5) node [anchor=north west][inner sep=0.75pt]  [font=\small] [align=left] {\begin{minipage}[lt]{104.75196000000001pt}\setlength\topsep{0pt}

\begin{equation*}
\mathfrak{m\text{\mbox{-}} algebra}
\end{equation*}
\begin{center}

$\displaystyle \left\{\omega ^{a}_{\; b} ,e^{a} ,\tilde{e}_{ab} ,\tilde{e}_{{a_{1} \cdots a_{5}}} ,\psi \right\}$
\end{center}

\end{minipage}};
\draw (7,232) node [anchor=north west][inner sep=0.75pt]  [font=\small] [align=left] {\begin{minipage}[lt]{116.65196pt}\setlength\topsep{0pt}

\begin{equation*}
\mathfrak{df}\text{-}\mathfrak{algebra}
\end{equation*}
\begin{center}

$\displaystyle \left\{\omega ^{a}_{\; b} ,e^{a} ,\tilde{e}_{ab} ,\tilde{e}_{{a_{1} \cdots a_{5}}} ,\psi ,\tilde{\psi }\right\}$
\end{center}

\end{minipage}};
\draw (30.5,353.5) node [anchor=north west][inner sep=0.75pt]  [font=\small] [align=left] {\begin{minipage}[lt]{83.66108000000001pt}\setlength\topsep{0pt}

\begin{equation*}
\mathfrak{sugra}_{11}
\end{equation*}
\begin{center}\vspace{0.15cm}

$\displaystyle \left\{\omega ^{a}_{\;b} ,e^{a} ,c_{3} ,c_{6} ,\psi \right\}$
\end{center}

\end{minipage}};
\draw (236,232) node [anchor=north west][inner sep=0.75pt]  [font=\small] [align=left] {\begin{minipage}[lt]{96.61304000000001pt}\setlength\topsep{0pt}

\begin{equation*}
\mathbb{R}^{1,10|\mathbf{32}}_{\mathrm{exc}}
\end{equation*}
\begin{center}

$\displaystyle \left\{e^{a} ,\tilde{e}_{ab} ,\tilde{e}_{{a_{1} \cdots a_{5}}} ,\psi ,\tilde{\psi }\right\}$
\end{center}

\end{minipage}};
\draw (259.5,340.5) node [anchor=north west][inner sep=0.75pt]  [font=\small] [align=left] {\begin{minipage}[lt]{63.6225pt}\setlength\topsep{0pt}

\begin{equation*}
\mathfrak{m5brane}
\end{equation*}
\begin{center}

$\displaystyle \left\{e^{a} ,c_{3} ,c_{6} ,\psi \right\}$
\end{center}

\end{minipage}};
\draw (476.5,232) node [anchor=north west][inner sep=0.75pt]  [font=\small] [align=left] {\begin{minipage}[lt]{72.81304pt}\setlength\topsep{0pt}

\begin{equation*}
\mathbb{R}^{1,10}_{\mathrm{exc}}
\end{equation*}
\begin{center}

$\displaystyle \left\{e^{a} ,\tilde{e}_{ab} ,\tilde{e}_{{a_{1} \cdots a_{5}}}\right\}$
\end{center}

\end{minipage}};
\draw (455,340.5) node [anchor=north west][inner sep=0.75pt]  [font=\small] [align=left] {\begin{minipage}[lt]{103.49804pt}\setlength\topsep{0pt}

\begin{equation*}
\mathbb{R}^{1,10}\! \oplus\! \mathbf{b}^{2}\mathfrak{u}( 1)\! \oplus\! \mathbf{b}^{5}\mathfrak{u}( 1)
\end{equation*}
\begin{center}

$\displaystyle \left\{e^{a} ,c_{3} ,c_{6}\right\}$
\end{center}

\end{minipage}};
\draw (93,86.67) node [anchor=north west][inner sep=0.75pt]  [font=\scriptsize] [align=left] {S-expansion};
\draw (94.33,197) node [anchor=north west][inner sep=0.75pt]  [font=\scriptsize] [align=left] {Forget $\displaystyle \tilde{\psi }$};
\draw (96.67,315) node [anchor=north west][inner sep=0.75pt]  [font=\scriptsize] [align=left] {atlas};
\draw (313,315) node [anchor=north west][inner sep=0.75pt]  [font=\scriptsize] [align=left] {atlas};
\draw (535.67,315) node [anchor=north west][inner sep=0.75pt]  [font=\scriptsize] [align=left] {atlas};
\draw (172.33,220.33) node [anchor=north west][inner sep=0.75pt]  [font=\scriptsize] [align=left] {Forget $\displaystyle \omega $};
\draw (175,346) node [anchor=north west][inner sep=0.75pt]  [font=\scriptsize] [align=left] {Forget $\displaystyle \omega $};
\draw (374,215.67) node [anchor=north west][inner sep=0.75pt]  [font=\scriptsize] [align=left] {Forget $\displaystyle \psi ,\tilde{\psi }$};
\draw (382,346) node [anchor=north west][inner sep=0.75pt]  [font=\scriptsize] [align=left] {Forget $\displaystyle \psi $};
\end{tikzpicture}
\caption[Relations between $\mathbb{R}^{1,10|\mathbf{32}}_{\mathrm{exc}}$, $\mathfrak{sugra}_{11}$, $\mathfrak{m}\text{-}\mathfrak{algebra}$, $\mathfrak{osp}(1|\mathbf{32})$]{Web of relations between the main super Lie algebras and super $L_\infty$-algebras underlying M-theory. The $L_\infty$-algebra $\mathfrak{sugra}_{11}$ underlies the super-Cartan geometry of $11$d Supergravity. Its atlas is the super algebra $\mathfrak{df}\text{-}\mathfrak{algebra}$ \cite{DAuria:1982uck}, which can be also obtained as a $\widetilde{\psi}$-extension of the $\mathfrak{m}\text{-}\mathfrak{algebra}$ \cite{Sezgin:1996cj}. The super algebra $\mathfrak{osp}(1|\mathbf{32})$ is related to the $\mathfrak{m}\text{-}\mathfrak{algebra}$ by S-expansion \cite{Izaurieta:2006zz}. By considering the $\mathfrak{df}\text{-}\mathfrak{algebra}$ and forgetting the Lorentz generators $\omega^a_{\;b}$ we obtain the super algebra $\mathbb{R}^{1,10|\mathbf{32}}_{\mathrm{exc}}$, which is the atlas of $\mathfrak{m5brane}$. By forgetting also the odd generators, we obtain the bosonic algebra $\mathbb{R}^{1,10}_{\mathrm{exc}}$, which can be seen as a linearised extended space.
}\end{figure}
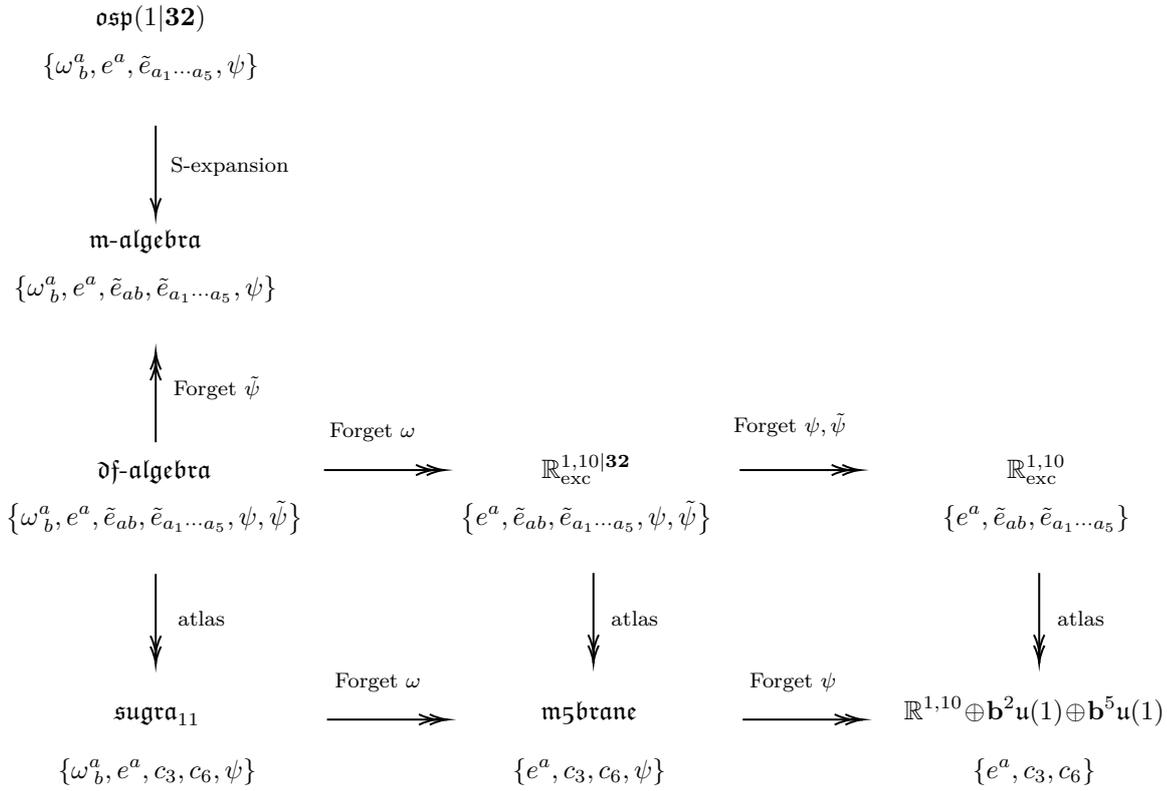

\begin{savequote}[8cm]
{\textgreekfont Ὁδὸς ἄνω κάτω μία καὶ ὡυτή.}

The road up and the road down is one and the same.
  \qauthor{--- Heraclitus, \textit{Fragments}}
\end{savequote}

\chapter{\label{ch:8}Geometric quantisation of Double Field Theory}

\minitoc


\noindent In this chapter we want to discuss the geometric quantisation of a string, its relation with the doubled space and with the bundle gerbe which underlies it.
Recall from chapter \ref{ch:3} our discussion of higher geometric quantisation.
Let a prequantum bundle gerbe $\mathscr{G}\twoheadrightarrow M$ be given by a cocycle $M\xrightarrow{\;f\;} \mathbf{B}^2U(1)$ on a phase space $M$.
Let us focus on a closed string, whose worldsheet would be of the form $\Sigma=\mathbb{R}\times S^1$.
We can transgress this cocycle to a $U(1)$-bundle $P$ on the loop space $\mathcal{L}M$ of the phase space by 
\begin{equation}
    \mathcal{L}M \,=\, [S^1,M] \;\xrightarrow{\;[S^1,f]\;}\; [S^1,\mathbf{B}^2U(1)] \,\cong\, \mathbf{B}U(1).
\end{equation}
Consequently, the $\mathbf{B}U$-associated bundle $\mathscr{G}\times_{\mathbf{B}U(1)}\mathbf{B}U \twoheadrightarrow\ M$ given by a cocycle
\begin{equation}
    M\;\longrightarrow \; \mathbf{B}U/\!/\mathbf{B}U(1)
\end{equation}
is transgressed exactly to an ordinary $\mathbb{C}$-associated bundle $P\times_{U(1)}\mathbb{C}\twoheadrightarrow\mathcal{L}M$ given by the transgressed cocycle
\begin{equation}
     [S^1,M]\;\longrightarrow \; \mathbb{C}/\!/U(1)
\end{equation}
on the loop space $\mathcal{L}M=[S^1,M]$ of the smooth manifold $M$. 
Thus, we can use the machinery of ordinary geometric quantisation on the phase loop space $\mathcal{L}M$ of the closed string.
\vspace{1cm}

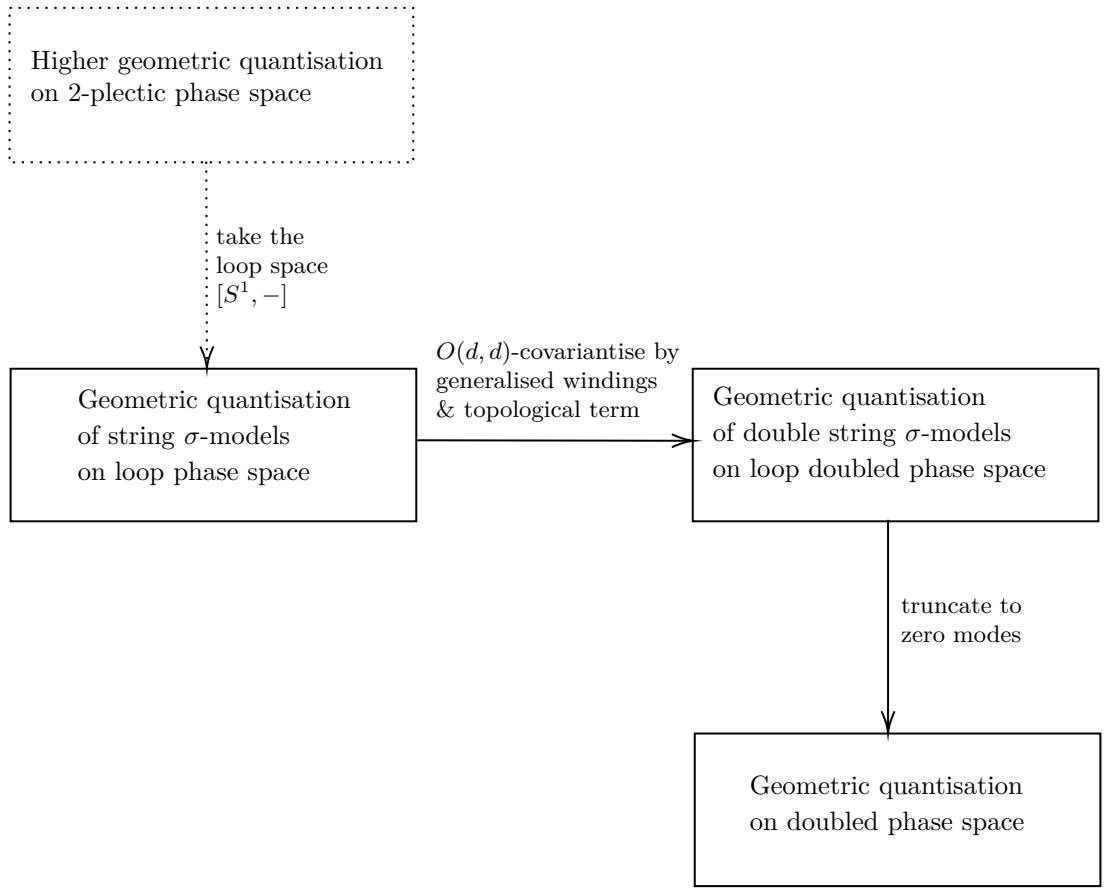
\begin{figure}[h]\begin{center}
\tikzset{every picture/.style={line width=0.75pt}} 
\begin{tikzpicture}[x=0.75pt,y=0.75pt,yscale=-1,xscale=1]
\draw  [dash pattern={on 0.84pt off 2.51pt}] (1.5,1.5) -- (203,1.5) -- (203,79.17) -- (1.5,79.17) -- cycle ;
\draw   (2,183) -- (204.5,183) -- (204.5,259.75) -- (2,259.75) -- cycle ;
\draw  [dash pattern={on 0.84pt off 2.51pt}]  (99.67,78.83) -- (99.99,180.75) ;
\draw [shift={(100,182.75)}, rotate = 269.82] [color={rgb, 255:red, 0; green, 0; blue, 0 }  ][line width=0.75]    (10.93,-3.29) .. controls (6.95,-1.4) and (3.31,-0.3) .. (0,0) .. controls (3.31,0.3) and (6.95,1.4) .. (10.93,3.29)   ;
\draw   (342.5,183) -- (545,183) -- (545,259.75) -- (342.5,259.75) -- cycle ;
\draw    (205,219) -- (341,219.25) ;
\draw [shift={(343,219.25)}, rotate = 180.1] [color={rgb, 255:red, 0; green, 0; blue, 0 }  ][line width=0.75]    (10.93,-3.29) .. controls (6.95,-1.4) and (3.31,-0.3) .. (0,0) .. controls (3.31,0.3) and (6.95,1.4) .. (10.93,3.29)   ;
\draw   (343.5,366.5) -- (546,366.5) -- (546,443.25) -- (343.5,443.25) -- cycle ;
\draw    (440,259) -- (440,364.25) ;
\draw [shift={(440,366.25)}, rotate = 270] [color={rgb, 255:red, 0; green, 0; blue, 0 }  ][line width=0.75]    (10.93,-3.29) .. controls (6.95,-1.4) and (3.31,-0.3) .. (0,0) .. controls (3.31,0.3) and (6.95,1.4) .. (10.93,3.29)   ;
\draw (10.67,21.5) node [anchor=north west][inner sep=0.75pt]  [font=\small] [align=left] {Higher geometric quantisation\\on $\displaystyle 2$-plectic phase space};
\draw (34,192) node [anchor=north west][inner sep=0.75pt]   [align=left] {{\small Geometric quantisation}\\{\small of string $\displaystyle \sigma $-models}\\{\small on loop phase space }};
\draw (103.17,110.5) node [anchor=north west][inner sep=0.75pt]  [font=\footnotesize] [align=left] {{\footnotesize take the}\\{\footnotesize loop space}\\{\footnotesize $[S^1,-]$}};
\draw (351,190.5) node [anchor=north west][inner sep=0.75pt]   [align=left] {{\small Geometric quantisation}\\{\small of double string $\displaystyle \sigma $-models}\\{\small on loop doubled phase space }};
\draw (213,167.5) node [anchor=north west][inner sep=0.75pt]  [font=\footnotesize] [align=left] {{\footnotesize $\displaystyle O( d,d)$-covariantise by}\\{\footnotesize generalised windings}\\{\footnotesize \& topological term}};
\draw (369.5,386) node [anchor=north west][inner sep=0.75pt]   [align=left] {{\small Geometric quantisation}\\{\small on doubled phase space }};
\draw (445,297) node [anchor=north west][inner sep=0.75pt]  [font=\footnotesize] [align=left] {{\footnotesize truncate to}\\{\footnotesize zero modes}};
\end{tikzpicture}
\caption[T-duality covariant geometric quantisation]{Concept map of the T-duality covariant geometric quantisation of a closed string.}
\end{center}\end{figure}

\noindent The zero modes of the phase loop space will give us the doubled phase space of DFT and the choice of polarisation in the quantisation will provide the T-duality frame. We will construct transformations between T-dual descriptions based on the transformations induced by different choices of polarisation. This will lead to the idea of a coherent state in the doubled space that saturates the uncertainty bound on distance in the doubled space. The geometric quantisation will result in a noncommutative algebra associated to the doubled phase space. As usual, the noncommutative nature of position and momentum is controlled by the parameter $\hbar$. In addition, we will also have a deformation parameter $\ell_s^2=\hbar \alpha'$ which controls the noncommutativity of coordinates $x^\mu$ and their T-duals $\tilde{x}_\mu$.

\section{Quantum geometry of the closed string}

\subsection{The phase space of the closed string}

\paragraph{The configuration space of $\sigma$-models.}
The fields $X^\mu(\sigma,\tau)$ are embeddings from a surface $\Sigma$ into a target space $M$, i.e. smooth maps $\Coo(\Sigma, M)$, denoted by 
\begin{equation}
    \begin{aligned}
    X^\mu:\, \Sigma \;&\lhook\joinrel\longrightarrow\; M \\
    (\sigma,\tau) \;&\longmapsto\; X^\mu(\sigma,\tau)
    \end{aligned}
\end{equation}

\paragraph{The configuration space of the closed string.}
Consider a surface of the form $\Sigma\simeq\mathbb{R}\times S^1$ with coordinates $\sigma\in[0,2\pi)$ and $\tau\in\mathbb{R}$. The fields $X^\mu(\sigma,\tau)$ of the $\sigma$-model can now be seen as curves $\Coo(\mathbb{R},\mathcal{L}M)$ on the \textit{free loop space} $\mathcal{L}M := \Coo(S^1,M)$ of the original manifold $M$. This will be denoted as follows:
\begin{equation}
    \begin{aligned}
    X^\mu(\sigma):\, \mathbb{R} \;&\lhook\joinrel\longrightarrow\; \mathcal{L}M \\
    \tau \;&\longmapsto\; X^\mu(\sigma,\tau)
    \end{aligned}
\end{equation}
where $X^\mu(\sigma,\tau)$ is a loop for any fixed $\tau\in\mathbb{R}$. In other words we have
\begin{equation}
    \Coo(\mathbb{R},\mathcal{L}M) \;\cong\; \Coo(\Sigma,M)
\end{equation}
This is why the configuration space for the closed string can be identified with the free loop space $\mathcal{L}M$ of the spacetime manifold $M$.

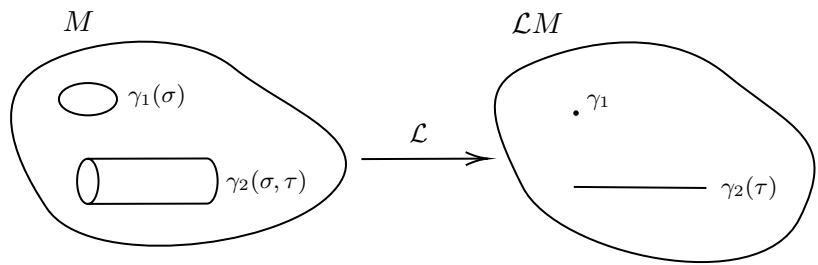
\begin{figure}[h]\begin{center}
\tikzset{every picture/.style={line width=0.75pt}} 
\begin{tikzpicture}[x=0.75pt,y=0.75pt,yscale=-1,xscale=1]
\draw   (29.33,40.33) .. controls (50.17,26.42) and (101.17,18.92) .. (127.5,40.25) .. controls (153.83,61.58) and (203.5,77.42) .. (174.5,104.75) .. controls (145.5,132.08) and (54,140.75) .. (34,110.75) .. controls (14,80.75) and (8.5,54.25) .. (29.33,40.33) -- cycle ;
\draw   (271.5,40) .. controls (291.5,30) and (339.5,15.75) .. (373,43.25) .. controls (406.5,70.75) and (430,75.75) .. (409.5,116.25) .. controls (389,156.75) and (288.5,133.75) .. (271.5,100) .. controls (254.5,66.25) and (251.5,50) .. (271.5,40) -- cycle ;
\draw   (40,55.88) .. controls (40,51.39) and (46.49,47.75) .. (54.5,47.75) .. controls (62.51,47.75) and (69,51.39) .. (69,55.88) .. controls (69,60.36) and (62.51,64) .. (54.5,64) .. controls (46.49,64) and (40,60.36) .. (40,55.88) -- cycle ;
\draw   (54.4,85.83) -- (113.84,85.69) .. controls (116.8,85.68) and (119.2,90.76) .. (119.22,97.03) .. controls (119.23,103.3) and (116.85,108.38) .. (113.9,108.39) -- (54.45,108.54) .. controls (51.5,108.54) and (49.1,103.47) .. (49.08,97.2) .. controls (49.06,90.93) and (51.44,85.84) .. (54.4,85.83) .. controls (57.35,85.83) and (59.75,90.9) .. (59.77,97.17) .. controls (59.78,103.44) and (57.4,108.53) .. (54.45,108.54) ;
\draw    (191,86) -- (252,85.76) ;
\draw [shift={(254,85.75)}, rotate = 539.77] [color={rgb, 255:red, 0; green, 0; blue, 0 }  ][line width=0.75]    (10.93,-3.29) .. controls (6.95,-1.4) and (3.31,-0.3) .. (0,0) .. controls (3.31,0.3) and (6.95,1.4) .. (10.93,3.29)   ;
\draw  [fill={rgb, 255:red, 0; green, 0; blue, 0 }  ,fill opacity=1 ] (297.25,62.88) .. controls (297.25,62.39) and (297.64,62) .. (298.13,62) .. controls (298.61,62) and (299,62.39) .. (299,62.88) .. controls (299,63.36) and (298.61,63.75) .. (298.13,63.75) .. controls (297.64,63.75) and (297.25,63.36) .. (297.25,62.88) -- cycle ;
\draw    (297,100) -- (363.22,100.53) ;
\draw (40.5,9.9) node [anchor=north west][inner sep=0.75pt]    {$M$};
\draw (264.5,10.9) node [anchor=north west][inner sep=0.75pt]    {$\mathcal{L} M$};
\draw (73,46.9) node [anchor=north west][inner sep=0.75pt]  [font=\footnotesize]  {$\gamma _{1}( \sigma )$};
\draw (122,89.9) node [anchor=north west][inner sep=0.75pt]  [font=\footnotesize]  {$\gamma _{2}( \sigma ,\tau )$};
\draw (213.5,66.9) node [anchor=north west][inner sep=0.75pt]  [font=\small]  {$\mathcal{L}$};
\draw (302,51.9) node [anchor=north west][inner sep=0.75pt]  [font=\footnotesize]  {$\gamma _{1}$};
\draw (369,92.9) node [anchor=north west][inner sep=0.75pt]  [font=\footnotesize]  {$\gamma _{2}( \tau )$};
\end{tikzpicture}
\caption[Curves on a loop space $\mathcal{L}M$]{A loop $\gamma_1:S^1\rightarrow M$ on the smooth manifold $M$ corresponds to a point on the loop space $\mathcal{L}M$. Similarly, a cylinder $\gamma_2:S^1\times[0,1]\rightarrow M$ on $M$ corresponds to a curve on $\mathcal{L}M$.}
\end{center}\end{figure}

\noindent Fortunately, for any given smooth manifold $M$, the free loop space $\mathcal{L}M$ is a Fr\'{e}chet manifold and this assures that there will be a well-defined notion of differential geometry on it. \vspace{0.2cm}


\noindent For any loop $X(\sigma):S^1\rightarrow M$ of $\mathcal{L}M$ we can consider the space of sections $\Gamma(S^1,X^\ast TM)$. This is homeomorphic to the loop space $\mathcal{L}\mathbb{R}^n$ where $n=\mathrm{dim}(M)$, which, thus, plays the role analogous to a local patch.\vspace{0.2cm}

\noindent Since the points of the loop space are loops $X(\sigma)$ in $M$, a smooth function $F\in\Coo(\mathcal{L}M)$ can be identified with a functional $F[X(\sigma)]$. Similarly, a vector field $V\in T(\mathcal{L}M)$ will be given by a functional operator of the form
\begin{equation}
    V[X(\sigma)] \;=\; \oint\di\sigma \, V^\mu[X(\sigma)](\sigma)\, \frac{\delta}{\delta X^\mu(\sigma)}.
\end{equation}
For a wider and deeper exploration of use of loop spaces to formalise some kinds of path integrals in physics, see \cite{Sza96}.

\paragraph{The transgression functor.}
For any $n$-form $\xi=\frac{1}{n!}\xi_{\mu_1\dots\mu_n}\di x^{\mu_1}\wedge\cdots\wedge\di x^{\mu_n}$, given in local coordinates $\{x^\mu\}$ of $M$, there exist a map, named \textit{transgression functor}, from the complex of differential forms on $M$ to the one of the differential forms on the loop space $\mathcal{L}M$:
\begin{equation}
    \begin{aligned}
    \mathfrak{T}:\; \Omega^n(M) \;&\longrightarrow\; \Omega^{n-1}(\mathcal{L}M)\\
    \xi \;&\longmapsto\; \oint\di\sigma \frac{1}{(n-1)!}\xi_{\mu_1\dots\mu_{n}}\!(X(\sigma))\,\frac{\partial X^{\mu_1}(\sigma)}{\partial \sigma}\,\delta X^{\mu_2}(\sigma)\wedge\cdots\wedge\delta X^{\mu_{n}}(\sigma)
    \end{aligned}
\end{equation}
Crucially, it satisfies the following functorial property:
\begin{equation}
    \delta\, \mathfrak{T} \,=\, \mathfrak{T}\,\di
\end{equation}

\paragraph{The phase space of the closed string.}
Thus, the choice for phase space of a string on spacetime $M$ will be the cotangent bundle $T^\ast (\mathcal{L} M)$. By definition, the tangent bundle of $\mathcal{L}M$ can be identified with
\begin{equation}
    T\mathcal{L}M \,=\, \mathcal{L}(T M).
\end{equation}
By considering the dual bundle of the tangent bundle $T\mathcal{L}M$, we can define the cotangent bundle $T^\ast\mathcal{L}M$. This space comes equipped with a canonical symplectic form:
\begin{equation}\label{eq:omega}
    \Omega \,:=\, \oint\di\sigma \, \delta P_\mu(\sigma) \wedge \delta X^\mu(\sigma) \;\;\in\,\Omega^2(T^\ast\mathcal{L}M).
\end{equation}
In the next paragraph we will illustrate that the phase space of the closed string is exactly the infinite-dimensional symplectic manifold $(T^\ast\mathcal{L}M,\,\Omega)$ we just described.\vspace{0.2cm}

\noindent We can now define a Liouville potential $\Theta$ such that its derivative is the canonical symplectic form $\Omega\in\Omega^2(T^\ast\mathcal{L}M)$. Thus we have
\begin{equation}
    \Theta \;:=\, \oint\di\sigma\, P_\mu(\sigma)\, \delta X^\mu(\sigma)\;\;\in\,\Omega^1(T^\ast\mathcal{L}U).
\end{equation}
We can verify that $\Omega = \delta\Theta$ by calculating:
\begin{equation}
    \begin{aligned}
    \delta\Theta \,&=\, \oint\di\sigma\left(\delta X^\mu(\sigma)\wedge\frac{\delta \Theta}{\delta X^\mu(\sigma)} + \delta P_\mu(\sigma)\wedge\frac{\delta \Theta}{\delta P_\mu(\sigma)}\right) \\
    &= \oint\di\sigma\oint\di\sigma'\, \delta(\sigma-\sigma')\,\delta P_\mu(\sigma) \wedge \delta X^\mu(\sigma') \\
    &= \oint\di\sigma \, \delta P_\mu(\sigma) \wedge \delta X^\mu(\sigma) \;=\; \Omega.
    \end{aligned}
\end{equation}

\paragraph{The closed string as classical system.}
Let us start from the action of the closed string
\begin{gather}\label{eq:closedaction}
\begin{aligned}
    S[X(\sigma,\tau),P(\sigma,\tau)] \;&=\; \frac{1}{2}\int\di\tau \oint\di\sigma \Big( P_\mu \dot{X}^\mu \,+\, \\
    & -\; g^{\mu\nu}(X)\big(P_\mu-B_{\mu\lambda}(X)X^{\prime\lambda}\big)\big(P_\nu-B_{\nu\lambda}(X)X^{\prime\lambda}\big) + g_{\mu\nu}(X)X^{\prime\mu}X^{\prime\nu} \Big).
\end{aligned}
\raisetag{1.6cm}
\end{gather}
Recall that the $\sigma$-model of a closed string $\big(X(\sigma,\tau),P(\sigma,\tau)\big):\Sigma\simeq \mathbb{R}\times S^1 \rightarrow T^\ast M$ can be equivalently expressed as a path $\mathbb{R}\rightarrow T^\ast\mathcal{L}M$ on the cotangent bundle of the loop space. Recall also that the Lagrangian density $\mathfrak{L}\in\Omega^1(\mathbb{R})$ of a classical system, consisting of a phase space $(T^\ast\mathcal{L}M,\Omega)$ and a Hamiltonian function $H\in\Coo(T^\ast\mathcal{L}M)$, is given by $\mathfrak{L}=(\iota_{V_H}\Theta - H)\di\tau$, where $V_H\in\mathfrak{ham}(T^\ast\mathcal{L}M,\Omega)$ is the Hamiltonian vector of $H$. This means that the action will be
\begin{equation}\label{eq:aaction}
    S[X(\sigma,\tau),P(\sigma,\tau)] \;=\; \int_{\tau_0}^{\tau_1}\di\tau \big(\iota_{V_H}\Theta - H \big), 
\end{equation}
where $\Theta$ is the Liouville potential of the symplectic form $\Omega$.
Now we want to check that the symplectic structure $\Omega\in\Omega^2(T^\ast\mathcal{L}M)$ of the phase space of the closed string is exactly the canonical symplectic structure \eqref{eq:omega} on $T^\ast\mathcal{L}M$. To do that, we can assume that the Hamiltonian vector field is just the translation along proper time. In other words we impose
\begin{equation}\label{eq:ttrans}
\begin{aligned}
    V_H \;&:=\; \frac{\di}{\di\tau} \\
    \;&=\; \oint\di\sigma\left( \dot{X}^\mu(\sigma)\frac{\delta}{\delta X^\mu(\sigma)} + \dot{P}_\mu(\sigma)\frac{\delta}{\delta P_\mu(\sigma)} \right).
\end{aligned}
\end{equation}
By putting together definition \eqref{eq:ttrans} and equation \eqref{eq:aaction} we immediately get the equation
\begin{equation}
    \iota_{V_H}\Theta \;=\; \oint\di\sigma\,\dot{X}^\mu(\sigma) P_\mu(\sigma),
\end{equation}
which is solved by the Liouville potential
\begin{equation}
    \Theta \;=\; \oint\di\sigma\, P_\mu(\sigma)\, \delta X^\mu(\sigma)\;\;\in\,\Omega^1(T^\ast\mathcal{L}U).
\end{equation}
Its differential is, indeed, exactly the canonical symplectic form \eqref{eq:omega}, i.e.
\begin{equation}
    \begin{aligned}
    \Omega \;&=\; \delta\Theta \\
    &=\; \oint\di\sigma\,\delta P_\mu(\sigma)\wedge \delta X^\mu(\sigma).
    \end{aligned}
\end{equation}
Moreover, by combining the equation \eqref{eq:aaction} with the action \eqref{eq:closedaction}, we can immediately find the Hamiltonian of a closed string:
\begin{equation*}
\begin{aligned}
    H[X(\sigma),P(\sigma)] &=\oint\di\sigma\,  \frac{1}{2}\Big(g^{\mu\nu}(X)\big(P_\mu-B_{\mu\lambda}(X)X^{\prime\lambda}\big)\big(P_\nu-B_{\nu\lambda}(X)X^{\prime\lambda}\big) + g_{\mu\nu}(X)X^{\prime\mu}X^{\prime\nu}\Big).
\end{aligned}
\end{equation*}
We can formally pack together the momentum $P(\sigma)$ and the derivative $X'(\sigma)$ in the following doubled vector:
\begin{equation}
    \mathbb{P}^M(\sigma) := \begin{pmatrix}X^{\prime\mu}(\sigma)\\P_\mu(\sigma)\end{pmatrix}
\end{equation}
with $M=1,\dots,2n$. Notice that $\mathbb{P}^M(\sigma)$ is uniquely defined at any given loop $(X(\sigma),P(\sigma))$ in the phase space.
Thus, we can rewrite the Hamiltonian of the string as
\begin{equation}\label{eq:h}
    H[X(\sigma),P(\sigma)] = \oint\di\sigma\,\frac{1}{2} \mathbb{P}^M(\sigma)\,\mathcal{H}_{MN}(X(\sigma))\,\mathbb{P}^N(\sigma),
\end{equation}
where the matrix $\mathcal{H}_{MN}$ is defined by
\begin{equation}
    \mathcal{H}_{MN} \;:=\; \begin{pmatrix}g_{\mu\nu}- B_{\mu\lambda}g^{\lambda\rho}B_{\rho\nu} & B_{\mu\lambda}g^{\mu\nu} \\-g^{\mu\lambda}B_{\lambda\mu} & g^{\mu\nu} \end{pmatrix} . \label{genmet1}
\end{equation}
In conclusion, by putting everything together, we can see that a closed string is a classical system $(T^\ast\mathcal{L}M,\Omega,H)$, where $\Omega$ is the canonical symplectic form on $T^\ast\mathcal{L}M$ and the Hamiltonian $H$ is given by definition \eqref{eq:h}. We now see the appearance of the generalised metric, described by matrix \eqref{genmet1}. This metric is a representative of an $O(d,d)/O(d)\times O(d)$ coset, and defines the generalised metric of generalised geometry. As such the Hamiltonian \eqref{eq:h} has a manifest $O(d,d)$ symmetry. This is of course the T-duality symmetry of the string. As discussed in the introduction, the Hamiltonian will often exhibit the symmetries not present in the Lagrangian and T-duality is one of these symmetries. 

\subsection{Generalised coordinates and the Kalb-Ramond field}

\begin{example}[Generalised coordinates for a charged particle]
In the geometric quantisation of an ordinary particle we have, in local Darboux coordinates, a local Liouville potential given by $\theta = p_\mu\di x^\mu$, where $p_\mu$ is the canonical momentum. In presence of an electromagnetic field with a minimally coupled $1$-form potential $A$, the canonical momentum $p_\mu$ which is defined from the Lagrangian perspective by $p_\mu= \frac{\partial \mathcal{L}}{\partial \dot{q^\mu}}$ is given by: $p_\mu=k_\mu + eA_\mu$. (We have used $k_\mu$ to denote the naive non-canonical momentum, also sometimes called the kinetic momentum).  \vspace{0.15cm}

\noindent Then the Liouville potential can be rewritten as $\theta=k_\mu\di x^\mu + eA$, with $A$ is the pullback of the electromagnetic potential to the phase space. Consequently the symplectic form takes the form $\omega=\di k_\mu \wedge \di x^\mu + eF$. Let us call the $2$-form $\omega_{A=0} := \di k_\mu\wedge\di x^\mu$. Thus, the geometric prequantisation condition $[\omega]=[\omega_{A=0}]+e[F]\in H^2(T^\ast M,\mathbb{Z})$ of the symplectic form on the phase space implies the Dirac quantisation condition $e[F]\in H^2(M,\mathbb{Z})$ of the electromagnetic field on spacetime. 

\noindent Notice that, in canonical coordinates, we have a Hamiltonian $H = g^{\mu\nu}(p_\mu-eA_\mu)(p_\nu-eA_\nu)$ and the commutation relations
\begin{equation}
    [\hat{x}^\mu,\,\hat{x}^\nu] = 0, \quad [\hat{p}_\mu,\,\hat{x}^\nu] = i\hbar\delta_\mu^\nu, \quad [\hat{p}_\mu,\,\hat{p}_\nu] = 0.
\end{equation}
On the other hand, in terms of the kinetic non-canonical coordinates, we have the Hamiltonian $H = g^{\mu\nu}k_\mu k_\nu$ and commutation relations
\begin{equation}
    [\hat{x}^\mu,\,\hat{x}^\nu] = 0, \quad [\hat{k}_\mu,\,\hat{x}^\nu] = i\hbar\delta_\mu^\nu, \quad [\hat{k}_\mu,\,\hat{k}_\nu] = i\hbar eF_{\mu\nu}.
\end{equation}
It is worth observing that the space coordinates do not commute anymore and the non-commutativity term is proportional to the field strength of the electromagnetic field. A similar picture will hold for strings.
\end{example}

\paragraph{Kinetic coordinates for a charged string.}
Similarly to the charged particle, for a string we require our $\Omega$ defined by equation \eqref{eq:omega} to be quantised as $[\Omega]\in H^2(T^\ast\mathcal{L}M,\mathbb{Z})$. We will now see that this implies, similarly to the electromagnetic field, the quantisation of the Kalb-Ramond field flux $[H]\in H^3(M,\mathbb{Z})$. \vspace{0.2cm}

\noindent Let us recall that an abelian gerbe with Dixmier-Douady class $[H]\in H^3(M,\mathbb{Z})$ on the base manifold $M$ is encoded by the following patching conditions:
\begin{equation}
\begin{aligned}
    H \,&=\, \di B_{(\alpha)} &\;\in\Omega^3_{\mathrm{cl}}(M),\\
    B_{(\alpha)} - B_{(\beta)} \,&=\, \di \Lambda_{(\alpha\beta)} &\;\in\Omega^2(U_\alpha\cap U_\beta),\\
    \Lambda_{(\alpha\beta)} + \Lambda_{(\beta\gamma)} + \Lambda_{(\gamma\alpha)} \,&=\, \di G_{(\alpha\beta\gamma)} &\;\in\Omega^1(U_\alpha\cap U_\beta \cap U_\gamma).
\end{aligned}
\end{equation}
By using the properties of the transgression functor from $M$ to its loop space $\mathcal{L}M$, we immediately obtain the new patching conditions
\begin{equation}
\begin{aligned}
    \mathfrak{T}H \,&=\, \delta (\mathfrak{T}B_{(\alpha)}) &\;\in\Omega^2_{\mathrm{cl}}(\mathcal{L}M)\\
    \mathfrak{T}B_{(\alpha)} - \mathfrak{T}B_{(\beta)} \,&=\, \delta (\mathfrak{T}\Lambda_{(\alpha\beta)}) &\;\in\Omega^1(\mathcal{L}U_\alpha\cap \mathcal{L}U_\beta)
\end{aligned}
\end{equation}
on $\mathcal{L}M$.
Therefore, the transgression functor sends a gerbe on a manifold $M$ to a circle bundle on its loop space $\mathcal{L}M$, i.e. in other words we have an equivalence
\begin{equation}
    \mathfrak{T}:\, \mathbf{B}U(1)\mathrm{Bund}(M) \,\xrightarrow{\;\;\;\cong\;\;\;}\, U(1)\mathrm{Bund}(\mathcal{L}M),
\end{equation}
where the first Chern class of the circle bundle is $\mathfrak{T}H\in H^2(\mathcal{L}M,\mathbb{Z})$. \vspace{0.2cm}

\noindent Now we can decompose the canonical symplectic form of the phase space of the closed string $(T^\ast\mathcal{L}M,\Omega)$ by
\begin{equation}
    \Omega \;=\; \Omega_{B=0} + \mathfrak{T}H.
\end{equation}
We can write $\Omega_{B=0}:=  \oint\di\sigma\, \delta K_\mu(\sigma) \wedge\delta X^\mu(\sigma) $, so we find that the symplectic form can be expressed by
\begin{equation}
    \Omega \;=\; \oint\di\sigma\, \delta\Big( K_\nu(\sigma) + B_{\mu\nu}\big(X(\sigma)\big)X^{\prime \mu}(\sigma)\Big)\wedge\delta X^\nu(\sigma) ,
\end{equation}
where $K_{\nu}(\sigma) := P_\nu(\sigma) - B_{\mu\nu}\big(X(\sigma)\big)X^{\prime \mu}(\sigma)$ is the \textit{non-canonical momentum} of the string and $P_\mu(\sigma)$ is its canonical momentum. 
The relevance of the transgression of gerbes to the loop space in dealing with T-duality and, more generally, with Double Field Theory was underlined by \cite{BelHulMin07}. \vspace{0.2cm}

\noindent For a clarification on the relation between the phase space of a closed string, seen as a loop space, and the Courant algebroids of supergravity, see \cite{Ost19, Osten:2021fil}.

\subsection{The algebra of operators of a closed string}

The machinery of geometric quantisation can now be applied on the phase space $(T^\ast\mathcal{L}M,\Omega)$ of the closed string. See also \cite{SaSza13} for a different quantisation approach on a loop space.
Given our choice of gauge for the Liouville potential $\Theta=\oint\di\sigma P_\mu(\sigma)\delta X^\mu(\sigma)$, we can now determine the algebra $\mathfrak{heis}(T^\ast\mathcal{L}M,\Omega)$ of quantum observables defined by
\begin{equation}
    \hat{f} \;=\; -i\hbar\nabla_{V_f} + f
\end{equation}
for any classical observable $f\in\Coo(T^\ast\mathcal{L}M)$. 
For the classical observables corresponding to the canonical coordinates $X^\mu(\sigma),P_\mu(\sigma)$ of the phase space of the closed string, we have the following quantum observables:
\begin{equation}
    \hat{P}_\mu(\sigma) \,=\, -i\hbar\frac{\delta}{\delta X^\mu(\sigma)}, \qquad \hat{X}^\mu(\sigma) \,=\, i\hbar\frac{\delta}{\delta P_\mu(\sigma)} + X^\mu(\sigma).
\end{equation}
If we choose the polarisation determined by the Lagrangian subbundle $L=T\mathcal{L}M$ with the corresponding basis $\big\{\Ket{X(\sigma)}\big\}_{X(\sigma)\in\mathcal{L}M}$, we have the following operators acting on wave-functional $\Psi[X(\sigma)]=\Braket{X(\sigma)|\Psi}$
\begin{equation}
    \begin{aligned}
    \Braket{X(\sigma)|\hat{P}_\mu(\sigma)|\Psi} \,&=\, -i\hbar\frac{\delta}{\delta X^\mu(\sigma)} \Psi[X^\mu(\sigma)], \\
     \Braket{X(\sigma)|\hat{X}^\mu(\sigma)|\Psi} \,&=\, X^\mu(\sigma)\Psi[X^\mu(\sigma)].
    \end{aligned}
\end{equation}
The commutation relations of these operators will then as follows:
\begin{equation}
    \begin{aligned}
    \big[\hat{P}_\mu(\sigma),\, \hat{X}^\nu(\sigma')\big] \,&=\, 2\pi i\hbar\delta_{\mu}^{\;\nu}\,\delta(\sigma-\sigma'), \\
    \big[\hat{X}^\mu(\sigma),\, \hat{X}^\nu(\sigma')\big] \,&=\, 0, \\
    \big[\hat{P}_\mu(\sigma),\, \hat{P}_\nu(\sigma')\big] \,&=\, 0.
    \end{aligned}
\end{equation}
These define the Heisenberg algebra $\mathfrak{heis}(T^\ast\mathcal{L}M,\Omega)$ of quantum observables on the phase space of the closed string. \vspace{0.2cm}

\noindent On the other hand, if we use the non-canonical, kinetic coordinates $(K(\sigma), X(\sigma))$, we obtain commutation relations of the form
\begin{equation}
    \begin{aligned}
    \big[\hat{K}_\mu(\sigma),\, \hat{X}^\nu(\sigma')\big] \,&=\, 2\pi i\hbar\delta_{\mu}^{\;\nu}\,\delta(\sigma-\sigma'), \\
    \big[\hat{X}^\mu(\sigma),\, \hat{X}^\nu(\sigma')\big] \,&=\, 0, \\
    \big[\hat{K}_\mu(\sigma),\, \hat{K}_\nu(\sigma')\big] \,&=\, H_{\mu\nu\lambda}\big(X(\sigma)\big)X^{\prime\lambda}(\sigma)\,\delta(\sigma-\sigma').
    \end{aligned}
\end{equation}

\subsection{The phase space of the closed string on a torus}

\paragraph{Closed string on the torus.}
Let us consider a closed string propagating in the background $M=T^n$ with constant metric $g_{\mu\nu}$ and constant Kalb-Ramond field $B_{\mu\nu}$. As explained by \cite{KugZwi92}, the compactification condition $x^\mu = x^\mu + 2\pi$ of a torus target space is background-independent, i.e. it does not depend on the bosonic background fields $g_{\mu\nu}$ and $B_{\mu\nu}$. In fact, the coordinates $x^\mu$ are periodic with radius $2\pi$ and the physical radii of the compactification are given by $R_\mu := \int_{S^1_\mu}\!\sqrt{g_{\mu\mu}}\,\di x^\mu = \sqrt{g_{\mu\mu}}$. \vspace{0.2cm}

\noindent Consider the phase space $T^\ast\mathcal{L}M\cong \mathcal{L}(T^n\times \mathbb{R}^n)$ of a closed string on a torus background. Let us call the matrix $E:=g+B$, so that its transpose is $E^\mathrm{T}= g-B$. Let us also define the generalised metric by the constant matrix
\begin{equation}
    \mathcal{H}_{MN} \;=\; \begin{pmatrix}g_{\mu\nu} - B_{\mu\lambda}g^{\lambda\rho}B_{\rho\nu} & -g^{\mu\lambda}B_{\lambda\nu}\\ B_{\mu\lambda}g^{\lambda\nu} & g^{\mu\nu}\end{pmatrix}.
\end{equation}
We can, now, explicitly expand position and momentum in $\sigma$ as follows:
\begin{equation}
    \begin{aligned}
    X^\mu(\sigma) \; &=\; x^\mu + \alpha' w^\mu\sigma + \!\!\sum_{n\in\mathbb{N}\backslash\{0\}}\!\! \frac{1}{n}\Big( \alpha^\mu_n(\mathcal{H})e^{in\sigma} + \bar{\alpha}^\mu_n(\mathcal{H})e^{-in\sigma} \Big), \\[0.5em]
    P_\mu(\sigma) \; &=\; p_\mu + \!\!\sum_{n\in\mathbb{N}\backslash\{0\}}\!\! \Big( E_{\mu\nu}^{\mathrm{T}}\alpha^\nu_n(\mathcal{H})e^{in\sigma} + E_{\mu\nu}\bar{\alpha}^\nu_n(\mathcal{H})e^{-in\sigma} \Big),
    \end{aligned}
\end{equation}
where we used the notation $\alpha^\mu_n=\alpha^\mu_n(\mathcal{H})$  and $\bar{\alpha}^\mu_n=\bar{\alpha}^\mu_n(\mathcal{H})$ to indicate that the higher-modes depend on the background, encoded by the generalised metric. The operators $\hat{\alpha}^\mu_n$ and $\hat{\bar{\alpha}}^\mu_n$ are the creation and annihilation operators for the excited states of the string, which depend on the background. However, ss pointed out by \cite{KugZwi92}, the operators $\hat{X}^\mu(\sigma)$ and $\hat{P}_\mu(\sigma)$ must be thought as background-independent objects. From the Fourier expansion, we also immediately obtain that the zero-modes coordinate $p_\mu$ and $w^\mu$ are integers because of the periodicity of $x^i$. \vspace{0.2cm}

\noindent The T-dual coordinates $\widetilde{X}_\mu(\sigma)$ are defined by,
 \begin{equation}
\widetilde{X}^{\prime }_\mu(\sigma)=P_\mu(\sigma)\, ,
\end{equation} 
and so we must have $\widetilde{X}_i(\sigma)=\tilde{x}_i+\int_0^\sigma \di\sigma'P_i(\sigma')$. Therefore, we obtain the expressions
\begin{equation}
    \begin{aligned}
    \widetilde{X}_\mu(\sigma) \; &=\; \tilde{x}_\mu + \alpha' p_\mu\sigma + \!\!\sum_{n\in\mathbb{N}\backslash\{0\}}\!\! \frac{1}{n}\Big( -E_{\mu\nu}^{\mathrm{T}}\alpha^\nu_n(\mathcal{H})e^{in\sigma} + E_{\mu\nu}\bar{\alpha}^\nu_n(\mathcal{H})e^{-in\sigma} \Big), \\
    \widetilde{P}^\mu(\sigma) \; &=\; w^\mu + \!\!\sum_{n\in\mathbb{N}\backslash\{0\}}\!\! \frac{1}{n}\Big( \alpha^\mu_n(\mathcal{H})e^{in\sigma} + \bar{\alpha}^\mu_n(\mathcal{H})e^{-in\sigma} \Big).
    \end{aligned}
\end{equation}
Notice that the T-dual coordinate $\widetilde{X}_\mu(\sigma)$ is also periodic with period $2\pi$. Moreover the new coordinates $\widetilde{X}(\sigma)$ and $\widetilde{P}(\sigma)$ are independent from the background.
The commutation relations 
\begin{equation}
[\hat{X}^\mu(\sigma),\hat{P}_\nu(\sigma')]= i\delta^\mu_\nu\delta(\sigma-\sigma')
\end{equation} become, on zero and higher modes,
\begin{equation}
    \begin{aligned}
    [\hat{x}^\mu,\hat{p}_\nu] \,&=\, i\delta^\mu_{\;\nu}, \\
    [\hat{\alpha}^\mu_n(\mathcal{H}),\hat{\alpha}^\nu_m(\mathcal{H})] \,&=\, n\delta_{n+m,0}\,g^{\mu\nu}.
    \end{aligned}
\end{equation}

\noindent As we have seen in the first section, the action is given on the phase space by the equation $S:= \oint\di\sigma\,\dot{X}^{\mu}(\sigma,\tau)\,P_{\mu}(\sigma,\tau) - H[X(\sigma,\tau),P(\sigma,\tau)]$. We can, thus, rewrite the action of the closed string on a torus as
\begin{equation}
    S[X(\sigma,\tau),P(\sigma,\tau)] \;=\; \int\di\tau \oint\di\sigma\left( \dot{X}^\mu P_\mu -  \frac{1}{2}\mathbb{P}^M\mathcal{H}_{MN}\,\mathbb{P}^N \right).
\end{equation}

\subsection{T-duality and background independence}

\paragraph{T-duality as a symplectomorphism.}
We will now show that it is possible to interpret T-duality as a symplectomorphism of phase spaces of two closed strings of the form
\begin{equation}\begin{aligned} 
    f:\;\quad(T^\ast\mathcal{L}M,\,\Omega)\,&\longrightarrow\,(T^\ast\mathcal{L}\widetilde{M},\,\widetilde{\Omega}) \\
     \big(X^\mu(\sigma),P_\mu(\sigma)\big)\,&\longmapsto\, \big(\widetilde{X}_\mu(\sigma),\widetilde{P}^\mu(\sigma)\big).
\end{aligned}\end{equation}
In fact, in \cite{AAL94}, it was firstly argued that T-duality can be seen a canonical transformation. However, a canonical transformation with generating functional $F[X(\sigma),\widetilde{X}(\sigma)]$ is nothing but the symplectomorphism $f$ associated to the following Lagrangian correspondence of the form \eqref{diag:lagrangiancorr}
\begin{equation} \label{lagcor2}
    \begin{tikzcd}[row sep={13ex,between origins}, column sep={13ex,between origins}]
    &  \big(T^\ast\mathcal{L}(M\times\widetilde{M}),\, \pi^\ast\Omega-\widetilde{\pi}^{\ast}\widetilde{\Omega}\big) \arrow[rd, "\widetilde{\pi}", two heads]\arrow[ld, "\pi"', two heads] & \\
   (T^\ast\mathcal{L}M,\,\Omega) \arrow[rr, "f"] && (T^\ast\mathcal{L}\widetilde{M},\,\widetilde{\Omega}),
    \end{tikzcd}
\end{equation}
which satisfies the following trivialisation condition for Liouville potential:
\begin{equation}\label{eq:trivcond}
    \begin{aligned}
        \pi^\ast\Theta \,-\, \widetilde{\pi}^\ast\widetilde{\Theta} \;=\; \delta F.
    \end{aligned}
\end{equation}
We can immediately check that, if we substitute the expression for the Liouville potentials, we get the expression
\begin{equation*}
    \begin{aligned}
        \oint\di\sigma\Big(P_\mu(\sigma)\delta X^\mu(\sigma)-\widetilde{P}^\mu(\sigma)\delta \widetilde{X}_\mu(\sigma) \Big) \;=\; \oint\di\sigma\left(\frac{\delta F}{\delta X^\mu(\sigma)}\delta X^\mu(\sigma) + \frac{\delta F}{\delta \widetilde{X}_\mu(\sigma)}\delta \widetilde{X}_\mu(\sigma)\right)
    \end{aligned}
\end{equation*}
and hence we recover exactly the equations of the canonical transformation
\begin{equation}
    \begin{aligned}
        P_\mu(\sigma) \;=\; \frac{\delta F}{\delta X^\mu(\sigma)}  , \qquad
        \widetilde{P}^\mu(\sigma) \;=\; -\frac{\delta F}{\delta \widetilde{X}_\mu(\sigma)}  .
        \end{aligned}
\end{equation}
By considering the generating functional
\begin{equation}\label{eq:genfunF}
    \begin{aligned}
    F[X(\sigma),\widetilde{X}(\sigma)] \;:&=\;\int_{D,\,\partial D=S^1}  \!\di \widetilde{X}_\mu \wedge \di X^\mu \\[0.1ex]
    &=\; \frac{1}{2} \oint\di\sigma\big( X^{\prime\mu}(\sigma)\widetilde{X}_{\mu}(\sigma) - X^{\mu}(\sigma)\widetilde{X}'_{\mu}(\sigma)\big)
    \end{aligned}
\end{equation}
which was originally proposed by \cite{AAL94, LAL94}, we obtain exactly T-duality on the phase space:
\begin{equation}
    \begin{aligned}
        P_\mu(\sigma) \;=\; \widetilde{X}_\mu'(\sigma)  , \qquad
        \widetilde{P}^\mu(\sigma) \;=\; X^{\prime\mu}(\sigma)  .
        \end{aligned}
\end{equation}
The Lagrangian correspondence space  in \eqref{lagcor2} is then the loop space of the doubled space of DFT. 
We can notice that, in this simple case, the doubled space can be identified with the correspondence space of a topological T-duality \cite{Bou03, Bou03x, Bou03xx, Bou04, Bou08} over a base point.
A similar observation was made in \cite{Pap14}.

\paragraph{Relation with the symplectic form on the doubled space.}
Let us consider the symplectic $2$-form $\varpi:=\di x^\mu\wedge\di \widetilde{x}_\mu\in\Omega^2(M\times\widetilde{M})$ on the product space, where $\{x^\mu,\tilde{x}_\mu\}$ are local coordinates on $M\times\widetilde{M}$. Notice that, for such a symplectic form, we can choose a Liouville potential of the form $\frac{1}{2}(\tilde{x}_\mu\di x^\mu - x^\mu\di\tilde{x}_\mu)$. (This is like the choice in Weyl quantisation or when we construct a Fock space). Now, we can immediately recognise that the generating functional \eqref{eq:genfunF} is nothing but the transgression of this Liouville potential to the loop space $\mathcal{L}(M\times\widetilde{M})$, i.e. we have
\begin{equation}
    F[X(\sigma),\widetilde{X}(\sigma)] \; = \; \frac{1}{2}\,\mathfrak{T}(\tilde{x}_\mu\di x^\mu - x^\mu\di\tilde{x}_\mu).
\end{equation}
By using the functorial property $\delta \mathfrak{T}=\mathfrak{T}\di$ of the transgression functor, we can rewrite the trivialisation condition \eqref{eq:trivcond} of T-duality by
\begin{equation}
    \pi^\ast\Theta \,-\, \widetilde{\pi}^\ast\widetilde{\Theta} \;=\; \mathfrak{T}(\varpi).
\end{equation}
Therefore, T-duality on the phase space is associated to the symplectic $2$-form $\varpi$ on the \textit{doubled space} $M\times\widetilde{M}$. Notice that this $2$-form is a particular and simple case of the fundamental $2$-form considered by \cite{Svo17, Svo18, Svo19}.

\paragraph{Background independence.}
Since the two loop phase spaces $T^\ast\mathcal{L}M$ and $T^\ast\mathcal{L}\widetilde{M}$ are symplectomorphic, they can be effectively considered the same symplectic $\infty$-dimensional Fr\'{e}chet manifold. In this "passive" symplectomorphism perspective, the T-duality from $(X(\sigma),P(\sigma))$ to $(\widetilde{X}(\sigma),\widetilde{P}(\sigma))$ can be interpreted as a change of coordinates on the phase space of the closed string. The Hamiltonian formulation of the closed string on the phase space is thus T-duality invariant.

\paragraph{T-duality as isomorphism of classical systems.}
T-duality, seen as a symplectomorphism $f:(T^\ast\mathcal{L}M,\,\Omega)\rightarrow(T^\ast\mathcal{L}\widetilde{M},\,\widetilde{\Omega})$ of the phase space of the closed string, does also preserve the Hamiltonian of the closed string, i.e. we have
\begin{equation}
    f^\ast H \,=\, H.
\end{equation}
In other words a T-duality is not just a symplectomorphism of our phase space $(T^\ast\mathcal{L}M,\Omega)$, but also an isomorphism of the classical system $(T^\ast\mathcal{L}M,\Omega,H)$ of the closed string. \vspace{0.2cm}

\noindent In general we can T-dualise the Hamiltonian of the closed string by applying a transformation $\mathcal{O}\in O(n,n;\mathbb{Z})$ to the doubled metric $\mathcal{H}$. Notice that the Hamiltonian functional does not change under such transformations. We have, in fact,
\begin{equation*}
    {H}[\widetilde{X},\widetilde{P}] = \oint\di\sigma\,\frac{1}{2} (\mathcal{O}\mathbb{P})^{M}(\mathcal{O}^T\mathcal{H}\mathcal{O})_{MN}(\mathcal{O}\mathbb{P})^N = \oint\di\sigma\,\frac{1}{2} \mathbb{P}^M\mathcal{H}_{MN}\mathbb{P}^N = H[X,P].
\end{equation*}

\paragraph{T-duality as change of basis on the Hilbert space.}
The Lagrangian correspondence \eqref{diag:lagcorstring} induces a diagram of quantum Hilbert spaces
\begin{equation}
    \begin{tikzcd}[row sep={10ex,between origins}, column sep={10ex,between origins}]
    &  \mathfrak{H}_{T\Gamma_f} \arrow[rd, "\pi^{\prime\ast}", hookleftarrow]\arrow[ld, "\pi^\ast"', hookleftarrow] & \\
    \mathfrak{H}_L  && \mathfrak{H}_{\widetilde{L}} \arrow[ll, "f^\ast"]
    \end{tikzcd}
\end{equation}
where $\mathfrak{H}_L$ and $\mathfrak{H}_{\widetilde{L}}$ are respectively polarised along the Lagrangian subbundles $L=T(\mathcal{L}M)$ and $\widetilde{L}=T(\mathcal{L}\widetilde{M})$. Now, as we have seen in \eqref{diag:hilbertcorr}, the map $(f^\ast)^{-1}$ $\mathfrak{H}_L\cong\mathfrak{H}_{\widetilde{L}}$ is an isomorphism $\mathfrak{H}_L \cong \mathfrak{H}_{\widetilde{L}}$ of Hilbert spaces. Therefore we can use just the notation $\mathfrak{H}$ for the abstract quantum Hilbert space.  \vspace{0.2cm}

\noindent Any quantum state $\Ket{\Psi}\in\mathfrak{H}$ can be expressed in the two basses defined by the two different polarisations:
\begin{equation}
    \ket{\Psi} = \int\mathcal{D}X(\sigma)\,\Psi[X(\sigma)]\,\ket{X(\sigma)}, \qquad \ket{\Psi} = \int\mathcal{D}\widetilde{X}(\sigma)\,\widetilde{\Psi}[\widetilde{X}(\sigma)]\,\ket{\widetilde{X}(\sigma)},
\end{equation}
where we called
\begin{equation}
    \braket{X(\sigma)|\Psi} \,=:\, \Psi[X(\sigma)], \qquad \braket{\widetilde{X}(\sigma)|\Psi} \,=:\, \widetilde{\Psi}[\widetilde{X}(\sigma)].
\end{equation}
The expansions in different basses will be then related by the Fourier-like transformation $(f^\ast)^{-1}$ of string wave-functionals, given by
\begin{equation}\label{eq:tdualityasunitarytransformation}
    \widetilde{\Psi}[\widetilde{X}(\sigma)] \,=\, \int_{\mathcal{L}M} \mathcal{D}X(\sigma)\, e^{\frac{i}{\hbar}F[X(\sigma),\widetilde{X}(\sigma)]}\, \Psi[X(\sigma)],
\end{equation}
in accord with \cite{AAL94}. We can also explicitly write the matrix of the change of basis on the Hilbert space $\mathfrak{H}$ by 
\begin{equation}
    \braket{X(\sigma)|\widetilde{X}(\sigma)} =  e^{\frac{i}{\hbar}F[X(\sigma),\widetilde{X}(\sigma)]}.
\end{equation}
Interestingly, this isomorphism is naturally defined by lifting the polarised wave functionals $\Psi[X(\sigma)]\in\mathfrak{H}_L$ and $\widetilde{\Psi}[\widetilde{X}(\sigma)]\in\mathfrak{H}_{\widetilde{L}}$ to wave-functionals ${\Psi}[\mathbb{X}(\sigma)]$ on the doubled space and by considering their Hermitian product in the Hilbert space of the doubled space. In double field theory solving the so called strong constraint provides the choice of polarisation. Here the quantisation procedure itself demands  a polarisation choice and the strong constraint is solved automatically. It is interesting to consider the weak constraint from this perspective but this is beyond the goals of this dissertation.

\paragraph{T-duality invariant dynamics.}
The dynamics of the quantised closed string is encoded by the background independent equation
\begin{equation}
    i\hbar\frac{\partial}{\partial \tau}\ket{\Psi} + \hat{H}\ket{\Psi} = 0.
\end{equation}
Let us consider, for simplicity, that we are starting from a Minkowski flat background with $g_{\mu\nu}=\eta_{\mu\nu}$ and $B_{\mu\nu}=0$. Then we will have a trivial doubled metric $\mathcal{H}_{MN}=\delta_{MN}$. Therefore, the equation of motion can be expressed in the basis $\big\{\ket{X(\sigma)}\big\}_{X(\sigma)\in\mathcal{L}M}$ by 
\begin{equation}
     i\hbar\frac{\partial}{\partial\tau}\Psi[X(\sigma)] + \oint\di\sigma\frac{1}{2}\left( -\hbar^2\frac{\delta^2\;\;}{\delta X(\sigma)^2} + X'(\sigma)^2 \right)\Psi[X(\sigma)] \,=\, 0,
\end{equation}
but immediately also in the T-dual basis $\big\{\ket{\widetilde{X}(\sigma)}\big\}_{X(\sigma)\in\mathcal{L}\widetilde{M}}$ by
\begin{equation}
     i\hbar\frac{\partial}{\partial\tau}\widetilde{\Psi}[\widetilde{X}(\sigma)] + \oint\di\sigma\frac{1}{2}\left( \widetilde{X}'(\sigma)^2 -\hbar^2\frac{\delta^2\;\;}{\delta \widetilde{X}(\sigma)^2} \right)\widetilde{\Psi}[\widetilde{X}(\sigma)] \,=\, 0.
\end{equation}

\paragraph{T-duality as a symplectomorphism for torus bundles.}
Let us conclude this section by considering a slightly more general class of examples: (geometric) T-duality of torus bundles.  \vspace{0.2cm}

\noindent Let $M\twoheadrightarrow N$ and $\widetilde{M}\twoheadrightarrow N$ be two principal $T^n$-bundles on a common base manifold $N$. T-duality can be still seen as a symplectomorphism between loop phase spaces $T^\ast\mathcal{L}M\rightarrow T^\ast\mathcal{L}\widetilde{M}$ and we can still employ the machinery of Lagrangian correspondence \eqref{diag:lagrangiancorr}. Now, the Lagrangian correspondence of the T-duality on the phase space of the closed string is
\begin{equation}\label{diag:lagcorstring}
    \begin{tikzcd}[row sep={13ex,between origins}, column sep={13ex,between origins}]
    &  \big(T^\ast \mathcal{L}(M\times_{N}\widetilde{M}), \, \pi^\ast\Omega-\widetilde{\pi}^{\ast}\widetilde{\Omega}\big) \arrow[rd, "\pi'", two heads]\arrow[ld, "\pi"', two heads] & \\
    (T^\ast \mathcal{L}M,\,\Omega) \arrow[rr, "f"] && (T^\ast \mathcal{L}\widetilde{M},\,\widetilde{\Omega})
    \end{tikzcd}
\end{equation}
where the fiber product $M\times_{N}\widetilde{M}$ can be naturally seen as the doubled torus bundle of the duality. 
For the torus fibration, we have the following equation for the Liouville potential:
\begin{equation}
    \pi^\ast\Theta \,-\, \widetilde{\pi}^\ast\widetilde{\Theta} \,=\, \mathfrak{T}(\varpi),
\end{equation}
where $\mathfrak{T}(\varpi)$ is the transgression to the loop space of the \textit{fundamental }$2$\textit{-form} $\varpi\in\Omega^2(M\times_{N}\widetilde{M})$, which lives on the doubled torus bundle $M\times_{N}\widetilde{M}$ and it is given by
\begin{equation}
    \varpi \;=\; (\di x^i + A^i) \wedge (\di \widetilde{x}_i + \widetilde{A}_i) \;\;\in\,\Omega^2(M\times_{N}\widetilde{M})
\end{equation}
If the bundle $M\times_{N}\widetilde{M}\twoheadrightarrow N$ is trivial, then we recover the symplectic form $\varpi = \di x^i\wedge\di \widetilde{x}_i$ from the previous paragraphs. Notice that, by moving to the generalised coordinates, we can easily recover the equation characterising topological T-duality by \cite{Bou04}, i.e. we have
\begin{equation}
    H - \widetilde{H} \;=\; \di (A^i\wedge \widetilde{A}_i)
\end{equation}
on the doubled torus bundle $M\times_N\widetilde{M}$.
In this class of cases, we notice that the doubled space can be identified with the correspondence space $M\times_N\widetilde{M}$ of a topological T-duality \cite{Bou03, Bou03x, Bou03xx, Bou04, Bou08} over a base manifold $M$.

\section{Quantum geometry of the doubled string}

\subsection{The generalised boundary conditions of the doubled string}

\paragraph{The phase space and the doubled space.}
To describe doubled strings, we introduce new coordinates $\widetilde{X}_\mu(\sigma)$ which satisfy the equation $P_\mu(\sigma)=\widetilde{X}^{\prime}_\mu(\sigma)$. 
Let us define the following doubled loop-space vectors:
\begin{equation}
    \mathbb{X}^M(\sigma) \,:=\, \begin{pmatrix}X^\mu(\sigma)\\[0.5ex] \widetilde{X}_\mu(\sigma)\end{pmatrix}, \qquad \mathbb{P}^M(\sigma) \,:=\, \begin{pmatrix}X^{\prime\mu}(\sigma)\\[0.5ex] {P}_\mu(\sigma)\end{pmatrix} \,=\, \mathbb{X}^{\prime M}(\sigma).
\end{equation}
Therefore, for a doubled string the doubled momentum $\mathbb{P}^M(\sigma)$ coincides with the derivative along the circle of the doubled position vector $\mathbb{X}^M(\sigma)$. 
Thus, instead of encoding the $\sigma$-model of the closed string by an embedding $\left(X^\mu(\sigma),\, P_\mu(\sigma)\right)$ into the phase space, we can encode it by an embedding $\mathbb{X}^M(\sigma)=(X^\mu(\sigma),\, \widetilde{X}_\mu(\sigma))$ into a doubled position space. Our objective is, then, be able to reformulate a string wave-functional $\Psi\!\left[X^\mu(\sigma),P_\mu(\sigma)\right]$ in terms of doubled fields as a wave-functional of the form $\Psi\big[\mathbb{X}^M(\sigma)\big]$. \vspace{0.15cm}

\noindent However, notice that, since the new coordinates $\widetilde{X}_\mu(\sigma)$ are the integral of the momenta of the string, specifying $\widetilde{X}_\mu(\sigma)$ is a stronger statement than specifying $P_\mu(\sigma)=\widetilde{X}_\mu'(\sigma)$. This observation is crucial when considering the possible boundary conditions of the doubled string $\sigma$-model. \vspace{0.15cm}

\noindent Let us define the following zero-modes of the doubled loop-space vectors:
\begin{equation}
    \mathbbvar{x}^M \,:=\, \frac{1}{2\pi}\oint\di\sigma\, \mathbb{X}^M(\sigma), \qquad \mathbbvar{p}^M \,:=\,\frac{1}{2\pi\alpha'} \oint\di\sigma\, \mathbb{X}^{\prime M}(\sigma),
\end{equation}
which, in components, read
\begin{equation}
    \mathbbvar{x}^M \,=\, \begin{pmatrix}x^\mu\\[0.1ex] \tilde{x}_\mu\end{pmatrix}, \qquad \mathbbvar{p}^M \,=\, \begin{pmatrix}\tilde{p}^{\mu}\\[0.1ex] p_\mu\end{pmatrix} \,\equiv\, \begin{pmatrix}w^{\mu}\\[0.1ex] \tilde{w}_\mu\end{pmatrix}.
\end{equation}
By using the new coordinate $\widetilde{X}_\mu$, we can rewrite the action of a closed string by
\begin{equation}
    S_{\mathrm{string}}[X(\sigma,\tau),P(\sigma,\tau)] \;=\; \frac{1}{2\pi\alpha'}\int\di\tau \oint\di\sigma\left( \dot{X}^\mu \widetilde{X}'_\mu -  \frac{1}{2}\mathbb{X}^{\prime M}\mathcal{H}_{MN}\,\mathbb{X}^{\prime N} \right) . \label{stringdoubledaction}
\end{equation}
Let us use the following notation for the derivatives
\begin{equation}
    \dot{\mathbb{X}}(\sigma,\tau) \,:=\, \frac{\partial \mathbb{X}(\sigma,\tau)}{\partial \tau}, \qquad {\mathbb{X}}'(\sigma,\tau) \,:=\, \frac{\partial \mathbb{X}(\sigma,\tau)}{\partial \sigma}.
\end{equation}

\paragraph{The generalised boundary conditions.}
Since in the action of the closed string the field $\mathbb{X}^{M}(\sigma)$ never appears, but only its derivatives $\mathbb{X}^{\prime M}(\sigma)$, we only need to require that the latter are periodic, i.e.
\begin{equation}
    \mathbb{X}^{\prime M}(\sigma+2\pi) \;=\; \mathbb{X}^{\prime M}(\sigma).
\end{equation}
This implies that the generalised boundary conditions are
\begin{equation}
    \mathbb{X}^M(\sigma+2\pi,\tau) \;=\; \mathbb{X}^M(\sigma,\tau) + 2\pi\alpha' \mathbbvar{p}^M(\tau),
\end{equation}
where the quasi-period $\mathbbvar{p}^M(\tau)$ can, in general, be dynamical and depend on proper time.\vspace{0.15cm}

\noindent Let us define the \textit{quasi-loop space} $\mathcal{L}_{\mathrm{Q}}\mathcal{M}$ of a manifold $\mathcal{M}$ as follows:
\begin{equation}
    \mathcal{L}_{\mathrm{Q}}\mathcal{M} \;:=\; \big\{\, \mathbb{X}:[0,2\pi)\rightarrow \mathcal{M} \,\;\big|\;\, \di\mathbb{X}(2\pi)=\di\mathbb{X}(0)\,\big\},
\end{equation}
where we used the simple identity $\mathbb{X}^{\prime M}\!(\sigma) \,\di\sigma = \di\mathbb{X}^M(\sigma)$. \vspace{0.15cm}

\noindent The phase space of the doubled string will be a symplectic manifold $(\mathcal{L}_{\mathrm{Q}}\mathcal{M}, \mathbbvar{\Omega})$ where the symplectic form $\mathbbvar{\Omega}\in\Omega^2(\mathcal{L}_{\mathrm{Q}}\mathcal{M})$ will be determined in the following subsection.

\paragraph{A remark on the global geometry of the doubled space.}
Given local coordinates $\mathbbvar{x}^M$ on the doubled space we can express the vector $\mathbb{X}'(\sigma)$ by
\begin{equation}
    \mathbb{X}^{\prime M}\!(\sigma) \,\di\sigma \;=\; \di\mathbb{X}^M(\sigma)  \;=\; \mathbb{X}^\ast\big(\di \mathbbvar{x}^M\big) \, ,
\end{equation}
where $\mathbb{X}^\ast\big(\di \mathbbvar{x}^M\big)$ denotes the pullback of $\di \mathbbvar{x}^M$ to the quasi-loop space. Therefore, the requirement that $\mathbb{X}'(\sigma)$ is periodic can be immediately recasted as the requirement that the pullback of $\di\mathbbvar{x}^M$ is periodic.
Chapter \ref{ch:5} explores the idea that the doubled space $\mathcal{M}$ is globally not a smooth manifold, but a the atlas of a bundle gerbe. In particular, in chapter \ref{ch:5}, it is derived that the patching conditions for local coordinate patches $\mathcal{U}_{(\alpha)}$ and $\mathcal{U}_{(\beta)}$ of the doubled space $\mathcal{M}$ should be of the form $\mathbbvar{x}^M_{(\beta)}=\mathbbvar{x}^M_{(\alpha)} + \Lambda_{(\alpha\beta)}^M + \partial^M\!\phi_{(\alpha\beta)}$, where we have an additional gauge-like transformation $\phi_{(\alpha\beta)}$ on the overlap of patches. Notice that this implies the patching conditions $\di\mathbbvar{x}^M_{(\beta)}=\di\mathbbvar{x}^M_{(\alpha)} + \di\Lambda_{(\alpha\beta)}^M$ and therefore the gauge transformations $\phi_{(\alpha\beta)}$ do not appear for the differential. Since the \v{C}ech cocycle condition $\di\Lambda_{(\alpha\beta)}^M + \di\Lambda_{(\beta\gamma)}^M + \di\Lambda_{(\gamma\alpha)}^M = 0$ is satisfied, we do not encounter problems for $\mathbb{X}^\ast\!\big(\di \mathbbvar{x}^M\big)=\di\mathbb{X}(\sigma)$ being periodic.
In other words, a doubled string can naturally live on a doubled space $\mathcal{M}$ that is patched in a more general way than a manifold (like the proposal in chapters \ref{ch:5} and \ref{ch:6}) exactly because $\mathbb{X}(\sigma)$ does not appear in the action, but only $\mathbb{X}'(\sigma)$ does.


\subsection{The symplectic structure of the doubled string.}
Recall that the Lagrangian density is related to the Liouville potential $\mathbbvar{\Theta}$ by
\begin{equation}
\begin{aligned}
    \mathfrak{L}_H \;=\; ( \iota_{V_H}\mathbbvar{\Theta} - H)\di\tau.
\end{aligned}
\end{equation}
Therefore, we can find the symplectic structure $\mathbbvar{\Omega}=\delta\mathbbvar{\Theta}$ on the phase space of the doubled string from its full Lagrangian.
One should also note here that the different choices of Liouville potential will give different results corresponding to either a particular choice of duality frame or a duality symmetric frame.

\paragraph{The Tseytlin action.}
The doubled string $\sigma$-model  as first constructed by Tseytlin is by now well known \cite{Tse90, Tse90b} and there are many routes one might take to its construction. Here, since we already have the doubled perspective in place for the Hamiltonian, the immediate method is to take the action as given by \eqref{stringdoubledaction} and then allow the dual variables to also be dynamical by augmenting the term $\dot{X}^\mu P_\mu$ term in the action with its dual equivalent: $\dot{\widetilde{X}}_\mu \widetilde{P}^\mu $. This is like picking a duality symmetric choice for the Liouville potential. Once this term is included then one can simply substitute the expressions for the doubled vectors into the action. It is these $\dot{X} P$ terms that produce the Legendre transformation between the Hamiltonian and the Lagrangian. They are sometimes called the {\it{abbreviated action}} and from now on we will adopt this nomenclature. The duality augmented abbreviated action is then:
\begin{equation}
\begin{aligned}
     S_{abb}&=\;  \frac{1}{4\pi\alpha'}\int\di\tau \oint\di\sigma\left(\dot{{X}}^\mu {P}_\mu + \dot{\widetilde{X}}_\mu \widetilde{P}^\mu \right),
    \end{aligned}
\end{equation}
which, using the expressions for the doubled vectors, becomes (up to total derivative  terms of which we will discuss more later) the $O(d,d)$ manifestly symmetric term:
\begin{equation}\label{eq:kin3}
\begin{aligned}
   S_{abb}=  \int\di\tau \oint\di\sigma\,\frac{1}{4\pi\alpha'}\!\left( \dot{\mathbb{X}}^{M}\eta_{MN}\mathbb{X}^{\prime N} \right) \; .
    \end{aligned}
\end{equation}
Combining this with the Hamiltonian to produce the total action $S=S_{abb}-H$ gives the Tseytlin action \cite{Tse90, Tse90b}: 
\begin{equation}
    S_{\mathrm{Tsey}}[\mathbb{X}(\sigma,\tau)] \;=\; \frac{1}{4\pi\alpha'}\int\di\tau \oint\di\sigma\left( \dot{\mathbb{X}}^{M}\eta_{MN}\mathbb{X}^{\prime N} -  \mathbb{X}^{\prime M}\mathcal{H}_{MN}\mathbb{X}^{\prime N}\right) \, .
\end{equation}
This action has been the subject of much study and we will return to the quantum equivalence to the usual string action later. Let us first examine this action taking care with the important property that the fields are quasi-periodic in $\sigma$ with quasi-period $\mathbbvar{p}^M$ which implies $\mathbb{X}^M(\sigma+2\pi,\tau)=\mathbb{X}^M(\sigma,\tau)+2\pi\alpha'\mathbbvar{p}^M(\tau)$.
This means that although the world sheet is periodic and has no boundary the total derivative terms of such quasi periodic fields can contribute to the action. These contributions to the Hamiltonian produce the important zero mode contributions to the Hamiltonian from the winding and momenta.
What follows is an analysis of the doubled abbreviated action with quasi-periodic fields.

\paragraph{The total derivative contributions to the Tseytlin action.}
When integrating with respect to $\sigma$ we may write $\oint\di\sigma$ as $\int_{\sigma_0}^{\sigma_0+2\pi} \di\sigma$. Then the integral of a total derivative is $ \int \di\sigma \frac{\di}{\di \sigma} f(\sigma)= f(2\pi+\sigma_0)- f(\sigma_0)$. For any periodic function $f(\sigma)$ this then vanishes as it should. However for a quasi-periodic function this integral will be non zero  e.g. $\mathbbvar{p}^M=\frac{1}{2\pi\alpha'}\oint\di\sigma\,\mathbb{X}^{\prime M}(\sigma)$. Note, that this is still independent of $\sigma_0$ as it should be since $\sigma_0$ is an entirely arbitrary choice of coordinate origin in the loop. \vspace{0.15cm}

\noindent Let us write the doubled abbreviated action explicitly with the manifest dependence on $\sigma_0$ as follows:
\begin{equation}
    S_{\mathrm{Tsey}}(\sigma_0) \;=\; \frac{1}{4\pi\alpha'}\int\di\tau\int_{\sigma_0}^{\sigma_0+2\pi}\!\!\!\!\!\di\sigma \,\dot{\mathbb{X}}^M(\sigma,\tau)\eta_{MN}\mathbb{X}^{\prime N}(\sigma,\tau) \, .
\end{equation}
Then taking the derivative with respect to $\sigma_0$ produces:
\begin{equation}
\begin{aligned}
    \frac{\di S_{\mathrm{Tsey}}}{\di \sigma_0} \;&=\; \frac{1}{4\pi\alpha'}\int\di\tau \left(\dot{\mathbb{X}}^M(\sigma_0+2\pi,\tau)- \dot{\mathbb{X}}^M(\sigma_0,\tau)\right)\eta_{MN}\mathbb{X}^{\prime N}(\sigma_0,\tau) \\
    &=\; \frac{1}{2}\int\di\tau\, \dot{\mathbbvar{p}}^M(\tau)\eta_{MN}\mathbb{X}^{\prime N}(\sigma_0,\tau) \, .
\end{aligned}
\end{equation}
To get to the second line we have used the periodicity of $\mathbb{X}^{\prime N}(\sigma+2\pi,\tau) = \mathbb{X}^{\prime N}(\sigma,\tau)$ and the quasi-periodicity of $\mathbb{X}^{N}(\sigma)$.
We thus have an anomaly. The action now depends on on the arbitrary choice of $\sigma_0$ when we allow quasi-periodic fields to encode the zero modes in the doubled space. The proposal in \cite{Bla14} is then to add an explicit "boundary term" to the Tseytlin action to cancel this piece as follows:
\begin{equation}\label{TseyBoundaryTerm}
    S_{\partial\mathrm{Tsey}}[\mathbbvar{p}(\tau),\,\mathbb{X}(\sigma_0,\tau)] \;:=\; -\frac{1}{2}\int\di\tau\, \dot{\mathbbvar{p}}^M(\tau)\eta_{MN}\mathbb{X}^{N}(\sigma_0,\tau).
\end{equation}
The full action, therefore, does not depend on the choice of $\sigma_0$, i.e.
\begin{equation}
    \frac{\di}{\di \sigma_0} (S_{\mathrm{Tsey}} + S_{\partial\mathrm{Tsey}}) \;=\; 0,
\end{equation}
and thus diffeomorphism-invariance of the doubled string is restored. \vspace{0.15cm}

\noindent Putting together all these terms we obtain the action
\begin{gather}
    \begin{aligned}  \label{tcorrect}
    {S}[\mathbb{X}(\sigma,\tau)] \,&=\, \frac{1}{4\pi\alpha'}\int\di\tau \!\oint\di\sigma \left( \dot{\mathbb{X}}^{M}(\sigma,\tau)\eta_{MN}\mathbb{X}^{\prime N}(\sigma,\tau) -  \mathbb{X}^{\prime M}(\sigma,\tau)\mathcal{H}_{MN}\mathbb{X}^{\prime N}(\sigma,\tau)\right) \\ 
    \;&-\, \frac{1}{2}\int\di\tau\,\mathbbvar{p}^{M}(\tau)\eta_{MN}\dot{\mathbb{X}}^N(0,\tau) \, .
    \end{aligned}
\raisetag{0.65cm}
\end{gather}
This doubled action is now world sheet diffeomorphism invariant for quasi-periodic fields but how does it relate to the original string action? Recall that the in writing down the Tseytlin action total derivative terms were neglected. We will now examine the relationship between the Tseytlin string and ordinary string with quasi-periodic fields. \vspace{0.35cm}

\begin{figure}[h]\begin{center}
\tikzset{every picture/.style={line width=0.75pt}} 
\begin{tikzpicture}[x=0.75pt,y=0.75pt,yscale=-1,xscale=1]
\draw   (21.57,6) -- (205.43,6) .. controls (212.92,6) and (219,26.26) .. (219,51.25) .. controls (219,76.24) and (212.92,96.5) .. (205.43,96.5) -- (21.57,96.5) .. controls (14.08,96.5) and (8,76.24) .. (8,51.25) .. controls (8,26.26) and (14.08,6) .. (21.57,6) .. controls (29.07,6) and (35.15,26.26) .. (35.15,51.25) .. controls (35.15,76.24) and (29.07,96.5) .. (21.57,96.5) ;
\draw    (35,63.5) .. controls (36.67,61.84) and (38.34,61.84) .. (40,63.51) .. controls (41.66,65.18) and (43.33,65.19) .. (45,63.53) .. controls (46.67,61.87) and (48.33,61.87) .. (50,63.54) .. controls (51.67,65.21) and (53.33,65.21) .. (55,63.55) .. controls (56.67,61.89) and (58.34,61.9) .. (60,63.57) .. controls (61.67,65.24) and (63.33,65.24) .. (65,63.58) .. controls (66.67,61.92) and (68.34,61.93) .. (70,63.6) .. controls (71.67,65.27) and (73.33,65.27) .. (75,63.61) .. controls (76.67,61.95) and (78.33,61.95) .. (80,63.62) .. controls (81.66,65.29) and (83.33,65.3) .. (85,63.64) .. controls (86.67,61.98) and (88.33,61.98) .. (90,63.65) .. controls (91.67,65.32) and (93.33,65.32) .. (95,63.66) .. controls (96.67,62) and (98.34,62.01) .. (100,63.68) .. controls (101.67,65.35) and (103.33,65.35) .. (105,63.69) .. controls (106.67,62.03) and (108.33,62.03) .. (110,63.7) .. controls (111.66,65.37) and (113.33,65.38) .. (115,63.72) .. controls (116.67,62.06) and (118.33,62.06) .. (120,63.73) .. controls (121.66,65.4) and (123.33,65.41) .. (125,63.75) .. controls (126.67,62.09) and (128.33,62.09) .. (130,63.76) .. controls (131.67,65.43) and (133.33,65.43) .. (135,63.77) .. controls (136.67,62.11) and (138.34,62.12) .. (140,63.79) .. controls (141.67,65.46) and (143.33,65.46) .. (145,63.8) .. controls (146.67,62.14) and (148.33,62.14) .. (150,63.81) .. controls (151.66,65.48) and (153.33,65.49) .. (155,63.83) .. controls (156.67,62.17) and (158.33,62.17) .. (160,63.84) .. controls (161.66,65.51) and (163.33,65.52) .. (165,63.86) .. controls (166.67,62.2) and (168.33,62.2) .. (170,63.87) .. controls (171.67,65.54) and (173.33,65.54) .. (175,63.88) .. controls (176.67,62.22) and (178.34,62.23) .. (180,63.9) .. controls (181.67,65.57) and (183.33,65.57) .. (185,63.91) .. controls (186.67,62.25) and (188.33,62.25) .. (190,63.92) .. controls (191.66,65.59) and (193.33,65.6) .. (195,63.94) .. controls (196.67,62.28) and (198.33,62.28) .. (200,63.95) .. controls (201.67,65.62) and (203.33,65.62) .. (205,63.96) .. controls (206.67,62.3) and (208.34,62.31) .. (210,63.98) .. controls (211.67,65.65) and (213.33,65.65) .. (215,63.99) -- (218,64) -- (218,64)(35,66.5) .. controls (36.67,64.84) and (38.33,64.84) .. (40,66.51) .. controls (41.66,68.18) and (43.33,68.19) .. (45,66.53) .. controls (46.67,64.87) and (48.33,64.87) .. (50,66.54) .. controls (51.67,68.21) and (53.33,68.21) .. (55,66.55) .. controls (56.67,64.89) and (58.34,64.9) .. (60,66.57) .. controls (61.67,68.24) and (63.33,68.24) .. (65,66.58) .. controls (66.67,64.92) and (68.34,64.93) .. (70,66.6) .. controls (71.67,68.27) and (73.33,68.27) .. (75,66.61) .. controls (76.67,64.95) and (78.33,64.95) .. (80,66.62) .. controls (81.66,68.29) and (83.33,68.3) .. (85,66.64) .. controls (86.67,64.98) and (88.33,64.98) .. (90,66.65) .. controls (91.67,68.32) and (93.33,68.32) .. (95,66.66) .. controls (96.67,65) and (98.34,65.01) .. (100,66.68) .. controls (101.67,68.35) and (103.33,68.35) .. (105,66.69) .. controls (106.67,65.03) and (108.33,65.03) .. (110,66.7) .. controls (111.66,68.37) and (113.33,68.38) .. (115,66.72) .. controls (116.67,65.06) and (118.33,65.06) .. (120,66.73) .. controls (121.66,68.4) and (123.33,68.41) .. (125,66.75) .. controls (126.67,65.09) and (128.33,65.09) .. (130,66.76) .. controls (131.67,68.43) and (133.33,68.43) .. (135,66.77) .. controls (136.67,65.11) and (138.34,65.12) .. (140,66.79) .. controls (141.67,68.46) and (143.33,68.46) .. (145,66.8) .. controls (146.67,65.14) and (148.33,65.14) .. (150,66.81) .. controls (151.66,68.48) and (153.33,68.49) .. (155,66.83) .. controls (156.67,65.17) and (158.33,65.17) .. (160,66.84) .. controls (161.66,68.51) and (163.33,68.52) .. (165,66.86) .. controls (166.67,65.2) and (168.33,65.2) .. (170,66.87) .. controls (171.67,68.54) and (173.33,68.54) .. (175,66.88) .. controls (176.67,65.22) and (178.34,65.23) .. (180,66.9) .. controls (181.67,68.57) and (183.33,68.57) .. (185,66.91) .. controls (186.67,65.25) and (188.33,65.25) .. (190,66.92) .. controls (191.66,68.59) and (193.33,68.6) .. (195,66.94) .. controls (196.67,65.28) and (198.33,65.28) .. (200,66.95) .. controls (201.67,68.62) and (203.33,68.62) .. (205,66.96) .. controls (206.67,65.3) and (208.34,65.31) .. (210,66.98) .. controls (211.67,68.65) and (213.33,68.65) .. (215,66.99) -- (218,67) -- (218,67) ;
\draw (221,62) node [anchor=north west][inner sep=0.75pt]    {$\sigma _{0}$};
\end{tikzpicture}
\caption{The "cut" on the worldsheet at $\sigma=\sigma_0$.}
\end{center}\end{figure}
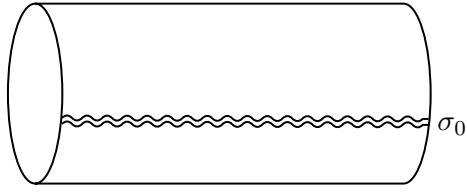\vspace{-0.2cm}

\paragraph{The Relation between Tseytlin string and ordinary string.}
Let us begin with the usual abbreviated action for the ordinary string, and then integrate by parts keeping the $\sigma$ total derivatives, we may neglect the total derivatives: in $\tau$ since there is no quasi periodicity in this variable:
\begin{gather}
\begin{aligned}
    \frac{1}{2\pi\alpha'}\int\di\tau\oint\di\sigma\, \dot{X}^\mu P_\mu &= \frac{1}{2\pi\alpha'}\int\di\tau\oint\di\sigma\, \dot{X}^\mu \widetilde{X}'_\mu \\
    &= \frac{1}{4\pi\alpha'}\int\di\tau\oint\di\sigma\, ( \dot{X}^\mu \widetilde{X}'_\mu + \dot{X}^\mu\widetilde{X}'_\mu )\\
    &= \frac{1}{4\pi\alpha'}\int\di\tau\oint\di\sigma\, (\dot{X}^\mu \widetilde{X}'_\mu + X'^\mu \dot{\widetilde{X}}_\mu + \frac{\di}{\di \sigma} (\dot{X}^\mu \widetilde{X}_\mu) ) \\
    &= \frac{1}{4\pi\alpha'}\int\di\tau\oint\di\sigma\, \dot{\mathbb{X}}^M\eta_{MN}\mathbb{X}^{\prime N} \\
    &+ \frac{1}{4\pi\alpha'}\int\di\tau\!\left( \dot{X}^\mu(2\pi+\sigma_0)\widetilde{X}_\mu(2\pi+\sigma_0)- \dot{X}^\mu(\sigma_0)\widetilde{X}_\mu(\sigma_0) \right)
    \end{aligned}.\raisetag{4.5cm}  \label{tdaction}
\end{gather}
Here we remark that we have also here made a choice in boundary total derivative term for $\sigma$. We are free to exchange it with:
\begin{equation}
-\frac{\di}{\di \sigma} ({X}^\mu\dot{\widetilde{X}}_\mu) \, .
\end{equation}
This choice is simply related by a neglected total derivative in $\tau$. A natural possibility is also the duality symmetric combination:
\begin{equation}
\frac{1}{2}\frac{\di}{\di \sigma} \left(\dot{X}^\mu{\widetilde{X}}_\mu- {X}^\mu\dot{\widetilde{X}}_\mu \right) \, .
\end{equation}
Now, the quasi-periodicity of the fields is:
\begin{equation}
    \mathbb{X}^M(2\pi+\sigma_0) = \mathbb{X}^M(\sigma_0) + 2\pi\alpha'\mathbbvar{p}^M
\end{equation}
so that we may evaluate the final term in \eqref{tdaction} as follows:
\begin{equation}
    \dot{X}(2\pi+\sigma_0)\widetilde{X}(2\pi+\sigma_0)- \dot{X}(\sigma_0)\widetilde{X}(\sigma_0) \;=\; 2\pi\alpha'\left(\dot{\tilde{p}}{\widetilde{X}}(\sigma_0) + {p}\dot{X}(\sigma_0) + 2\pi\alpha'\dot{\tilde{p}}p\right).
\end{equation}
Substituting this into \eqref{tdaction} produces:
\begin{equation}\label{eq:totderivative}
\begin{aligned}
    \frac{1}{2\pi\alpha'}\int\di\tau\oint\di\sigma\, \dot{X}^\mu P_\mu 
    &= \frac{1}{4\pi\alpha'}\int\di\tau\oint\di\sigma\, \dot{\mathbb{X}}^M\eta_{MN}\mathbb{X}^{\prime N} \\ &+ \int\di\tau\!\left(\frac{1}{2}\dot{\tilde{p}}{\widetilde{X}}(\sigma_0) + \frac{1}{2}{p}\dot{X}(\sigma_0) + \pi\alpha'\dot{\tilde{p}}p\right).
    \end{aligned}
\end{equation}

\noindent Now, $\dot{X}(2\pi) = \dot{X}(2\pi) + 2\pi\alpha'\dot{\tilde{p}}$ is quasi-periodic, so the initial term $\int\di\tau\oint\di\sigma\, \dot{X}P $ suffers from the same anomaly as we described in the previous section. We need to also add a "boundary term" for $\int\di\tau\oint\di\sigma\, \dot{X}P $ so that there is no dependence on $\sigma_0$. This implies that the abbreviated action on the LHS of \eqref{tdaction} must be changed to
\begin{equation}\label{Sabb}
    S_\mathrm{abb} \;:=\; \frac{1}{2\pi\alpha'}\int\di\tau\oint\di\sigma\, \dot{X}^\mu P_\mu - \int\di\tau\, \dot{\tilde{p}}^{\mu}\widetilde{X}_\mu(\sigma_0),
\end{equation}
so there is no overall $\sigma_0$ dependence.
Then including the same term to the RHS of \eqref{tdaction} will produce:
\begin{equation}\label{SabbEquation}
\begin{aligned}
    S_\mathrm{abb} & \;=\; \frac{1}{4\pi\alpha'}\int\di\tau\oint\di\sigma\, \dot{\mathbb{X}}^M\eta_{MN}\mathbb{X}^{\prime N} \\ & \;+ \; \int\di\tau\!\left(-\frac{1}{2}\dot{\tilde{p}}^\mu{\widetilde{X}}_\mu(\sigma_0) \; + \;\frac{1}{2}{p}_\mu\dot{X}^\mu(\sigma_0) + \pi\alpha'\dot{\tilde{p}}^\mu p_\mu \right).
    \end{aligned}
\end{equation}
We may now write this, using integration by parts and deleting total derivatives in $\tau$, or by using the duality symmetric choice of total derivative terms in $\sigma$, to give a duality symmetric action
\begin{equation*}
\begin{aligned}
    S_\mathrm{abb} \;=\; \frac{1}{4\pi\alpha'}\int\di\tau\oint\di\sigma\, \dot{\mathbb{X}}^M\eta_{MN}\mathbb{X}^{\prime N} + \int\di\tau\!\left(+\frac{1}{2}{\tilde{p}}\dot{\widetilde{X}}(\sigma_0) + \frac{1}{2}{p}\dot{X}(\sigma_0) + \frac{\pi\alpha'}{2}(\dot{\tilde{p}}p - \tilde{p}\dot{p})\right),
    \end{aligned}
\end{equation*}
that we may write in a using doubled vectors in an doubled fashion, to give 
\begin{equation*}
\begin{aligned}
    S_\mathrm{abb} \;=\; \frac{1}{4\pi\alpha'}\int\di\tau\oint\di\sigma\, \dot{\mathbb{X}}^M\eta_{MN}\mathbb{X}^{\prime N} + \int\di\tau\!\left(\frac{1}{2}\mathbbvar{p}^M\eta_{MN}\dot{\mathbb{X}}^N(\sigma_0) + \frac{\pi\alpha'}{2}\dot{\mathbbvar{p}}^M\omega_{MN}\mathbbvar{p}^N\right).  \label{tcorrect1}
    \end{aligned}
\end{equation*}

\noindent We may now identify the $\int\di\tau\frac{1}{2}\mathbbvar{p}^M\eta_{MN}\dot{\mathbb{X}}^N(0)$ term as the "boundary term" \eqref{TseyBoundaryTerm} we introduced earlier to make the Tseytlin action independent of $\sigma_0$. Thus when the dust settles we see that the abbreviated action of the string, including the "boundary" piece, produces the Tseytlin string (including the doubled "boundary" term) and a correction term from the final term in \eqref{tcorrect1}.
Thus, the relation between $S_\mathrm{abb}$ of the usual string and the Tseytlin abbreviated action $S_{\mathrm{Tsey,abb}}$ (including "boundary" pieces) is
\begin{equation}
    S_\mathrm{abb} \;=\; S_{\mathrm{Tsey,abb}} +\int\di\tau\,\frac{\pi\alpha'}{2}\dot{\mathbbvar{p}}^M\omega_{MN}\mathbbvar{p}^N \, .
\end{equation}

\paragraph{The dual picture.}
 Let us define the abbreviated action of the T-dual string, including the "boundary" piece, as follows: 
\begin{equation}
    \widetilde{S}_\mathrm{abb} \;:=\; \frac{1}{2\pi\alpha'}\int\di\tau\oint\di\sigma\, \dot{\widetilde{X}}_\mu \widetilde{P}^\mu - \int\di\tau\, \dot{{p}}_\mu{X}^\mu(\sigma_0)  \, .
\end{equation}
Now, we can repeat all this procedure with this "dual" abbreviated action. Then the difference between the two duality related frames for the action will be the difference of these abbreviated actions (the Hamiltonian being invariant), thus: 
\begin{equation}
\begin{aligned}
    S_\mathrm{abb} - \widetilde{S}_\mathrm{abb}\;=\; \pi\alpha'\int\di\tau\,\dot{\mathbbvar{p}}^M\omega_{MN}\mathbbvar{p}^N  \, .
    \end{aligned}
\end{equation}
The implication is that ordinary string and its dual are related by a phase shift:
\begin{equation}
\begin{aligned}
    \exp\!\left({\frac{i}{\hbar}S_\mathrm{abb}}\right)\;=\; \exp\!\left({\frac{i\pi\alpha'}{\hbar}\int\di\tau\,\dot{\mathbbvar{p}}^M\omega_{MN}\mathbbvar{p}^N }\right) \exp\!\left({\frac{i}{\hbar}\widetilde{S}_\mathrm{abb}} \right) \, .
    \end{aligned}
\end{equation}

\noindent Let us make a sanity check. For an ordinary toroidal space $\mathbbvar{p}$ is constant in $\tau$ so it is zero for ordinary strings on a torus. The reader at this point may feel frustrated that after some considerable care with total derivatives and quasi-periodic fields we have generated a term that vanishes. However, crucially this term will not vanish for strings in backgrounds where the winding number is not conserved. Such a situation is exactly where double field theory is most useful captures the dynamic nature of winding. For example this occurs with a string in a Kaluza-Klein monopole background; a set up that was first considered in \cite{Gregory:1997te} and studied using double field theory in \cite{Jen11}. This term will also make a contribution if there is no globally defined duality frame and so one needs to form a good cover over the space and choose a duality frame in each patch. Such spaces with no globally defined T-duality frame are called T-folds. The above phase shift will then be part of the transition function between different patches acting on the string wavefunction. From the geometric quantisation perspective this is reminiscent of the Maslov correction.


\paragraph{The Hull "topological" term and its role.}
In addition to the discussion above, Hull proposed the addition of a "topological" to Tseytlin action based on global requirements for a gauging procedure \cite{Hull06}. The importance of this term for the partition function was emphasised in \cite{Berman:2007vi,Tan:2014mba}. This term is given by:
\begin{equation}\label{eq:topterm}
    S_{\mathrm{top}}[\mathbb{X}(\sigma,\tau)] \;=\; \frac{1}{4\pi\alpha'}\int\di\tau \oint\di\sigma\, \dot{\mathbb{X}}^M\omega_{MN} \mathbb{X}^{\prime N}.
\end{equation}
Now, we can use Stokes' theorem as follows:
\begin{gather}\label{eq:topExpanded}
    \begin{aligned}
    S_{\mathrm{top}}[\mathbb{X}(\sigma,\tau)] \;&=\; \frac{1}{4\pi\alpha'}\int\di\tau \oint\di\sigma\, \dot{\mathbb{X}}^{M}\omega_{MN}\mathbb{X}^{\prime N} \\
    \;&=\; \frac{1}{2\pi\alpha'}\int\!\!\!\!\oint_\Sigma \di{X}^{\mu}\wedge\di\widetilde{X}_\mu   \\
    \;&=\; \frac{1}{2\pi\alpha'} \int\di\tau\left({X}^{\mu}(\sigma_0+2\pi,\tau)\dot{\widetilde{X}}_\mu(\sigma_0+2\pi,\tau) - {X}^{\mu}(\sigma_0,\tau)\dot{\widetilde{X}}_\mu(\sigma_0,\tau)  \right)  \\
    &= \frac{1}{2} \int\di\tau\left(\dot{\tilde{p}}{\widetilde{X}}(\sigma_0) + {p}\dot{X}(\sigma_0) + 2\pi\alpha'\dot{\tilde{p}}p\right)
    \end{aligned}
    \raisetag{0.7cm}
\end{gather}
where, without any loss of generality, we have chosen the gauge $X^{\mu}\di\widetilde{X}_\mu$ for the potential of the $2$-form $\di{X}^{\mu}\wedge\di\widetilde{X}_\mu$.
Just like for the Tseytlin action, we can define a boundary term for the topological term
\begin{equation}
     S_{\partial\mathrm{top}}[\mathbb{X}(0,\tau),\mathbbvar{p}(\tau)] \;=\; -\frac{1}{2}\int\di\tau \,\dot{\mathbbvar{p}}^M(\tau)\omega_{MN}{\mathbb{X}}^N(\sigma_0,\tau)
\end{equation}
to remove the dependence on the cut $\sigma_0$.
Recall the equation \eqref{eq:totderivative} relating the Tseytlin abbreviated action of a doubled string with the term $\int\di\tau\oint\di\sigma\,\dot{X}^\mu P_\mu$. By combining equation \eqref{eq:totderivative} with equation \eqref{eq:topExpanded}, we immediately obtain the relation
\begin{equation}
     \frac{1}{2\pi\alpha'}\int\di\tau\oint\di\sigma\,\dot{X}^\mu P_\mu \;=\; \frac{1}{4\pi\alpha'}\int\di\tau \oint\di\sigma\, \dot{\mathbb{X}}^M(\eta_{MN} + \omega_{MN}) \mathbb{X}^{\prime N} .
\end{equation}
At this point, we can provide both sides of the equation with the boundary term, so that we have
\begin{equation}\label{SabbTop}
     S_{\mathrm{abb}} \;=\; \frac{1}{4\pi\alpha'}\int\di\tau \oint\di\sigma\, \dot{\mathbb{X}}^M(\eta_{MN} + \omega_{MN}) \mathbb{X}^{\prime N} - \frac{1}{2}\int\di\tau\,\mathbbvar{p}^M(\eta_{MN}+\omega_{MN})\dot{\mathbb{X}}^N(\sigma_0)
\end{equation}
where we recall the definition \eqref{Sabb} for the abbreviated action (including the boundary term) of the ordinary string:
\begin{equation}
    S_\mathrm{abb} \;:=\; \frac{1}{2\pi\alpha'}\int\di\tau\oint\di\sigma\, \dot{X}^\mu P_\mu - \int\di\tau\, \dot{\tilde{p}}^{\mu}\widetilde{X}_\mu(\sigma_0)
\end{equation}
This equation can be interpreted as the fact that we need to add the boundary term $-\int\di\tau\, \dot{\tilde{p}}^{\mu}\widetilde{X}_\mu(\sigma_0)$ to the usual abbreviated action $\frac{1}{2\pi\alpha'}\int\di\tau\oint\di\sigma\, \dot{X}^\mu P_\mu$ of an ordinary string to obtain the manifestly $O(n,n)$-covariant abbreviated action \eqref{SabbTop}.

\paragraph{The total action.}
Finally, putting all this together, the total action of the doubled string will be given as follows
\begin{gather}
    \begin{aligned}
    \mathbb{S}[\mathbb{X}(\sigma,\tau)] \,&=\, \frac{1}{4\pi\alpha'}\int\di\tau \!\oint\di\sigma\left( \dot{\mathbb{X}}^{M}(\sigma,\tau)(\eta_{MN}+\omega_{MN} )\mathbb{X}^{\prime N}(\sigma,\tau) -  \mathbb{X}^{\prime M}(\sigma,\tau)\mathcal{H}_{MN}\mathbb{X}^{\prime N}(\sigma,\tau)\right) \\
    \,&-\, \frac{1}{2}\int\di\tau\,\mathbbvar{p}^M(\eta_{MN}+\omega_{MN})\dot{\mathbb{X}}^N(\sigma_0).
    \end{aligned}
    \raisetag{0.7cm}
\end{gather}

\paragraph{Fourier expansion of the kinetic term.}
Now, recall that the Hamiltonian of the doubled string is the functional
\begin{equation}
    H[\mathbb{X}(\sigma)] \;=\; \frac{1}{4\pi\alpha'}\oint\di\sigma\,\mathbb{X}^{\prime M}(\sigma)\mathcal{H}_{MN}\big(\mathbb{X}(\sigma)\big)\mathbb{X}^{\prime N}(\sigma).
\end{equation}
We can now expand $\mathbb{X}^M(\sigma)$ in $\sigma$ by
\begin{equation}
    \mathbb{X}^M(\sigma) \;=\; \mathbbvar{x}^M + \alpha'\sigma\mathbbvar{p}^M + \!\!\sum_{n\in\mathbb{Z}\backslash\{0\}}\!\frac{1}{n}\bbalpha^M_n e^{in\sigma},
\end{equation}
where $\bbalpha^M_n$ satisfies the identity $\bbalpha^M_{-n}= \bar{\bbalpha}^M_n$ and it can be decomposed as $\bbalpha^M_n = (\alpha_n^\mu,\,\widetilde{\alpha}_{n\mu})$ with
\begin{equation}
    \widetilde{\alpha}_{n\mu} \;=\; \begin{cases}-E_{\mu\nu}^{\mathrm{T}}\,\alpha_n^\nu, & n>0\\+E_{\mu\nu}\,\alpha_n^\nu, & n<0\end{cases}.
\end{equation}
The coordinates of the phase space can be taken as $\{\mathbbvar{x}^M,\mathbbvar{p}^M,\alpha_k^\mu,\bar{\alpha}_k^\mu\}_{k\in\mathbb{N}\backslash\{0\}}$.
Now we can explicitly express the kinetic part of the action of the doubled string in these coordinates.
Firstly, we calculate the mode expansion of the abbreviated Tseytlin action together with the topological term:
\begin{equation}
\begin{aligned}
    &\frac{1}{4\pi\alpha'}\int\di\tau \oint\di\sigma\, \dot{\mathbb{X}}^{M}(\eta_{MN}+\omega_{MN})\mathbb{X}^{\prime N} \, \, \\ 
    &=\, \frac{1}{2} \int\di\tau\!\left(p_\mu\dot{x}^\mu + \pi\alpha'\dot{\tilde{p}}^\mu p_\mu + \dot{\tilde{p}}^\mu\!\!\!\sum_{n\in\mathbb{Z}\backslash\{0\}}\!\frac{1}{n}\dot{\tilde{\alpha}}_{n\mu} +i \!\!\!\sum_{n\in\mathbb{Z}\backslash\{0\}}\!\frac{1}{n}\,\omega _{MN}\,\dot{\bbalpha}_{-n}^M\bbalpha_n^N \right).
\end{aligned}
\end{equation}
Then, we expand the boundary terms:
\begin{equation}
    \begin{aligned}
    (S_{\partial\mathrm{Tsey}}+S_{\partial\mathrm{top}})[\mathbb{X}(0,\tau),\mathbbvar{p}(\tau)] \;&=\; -\int\di\tau \,\frac{1}{2}\mathbbvar{p}^M(\tau)(\eta_{MN}+\omega_{MN} )\dot{\mathbb{X}}^N(\sigma_0,\tau) \\
    \;&=\; -\int\di\tau\,\frac{1}{2}\left( \tilde{p}^\mu\dot{\tilde{x}}_\mu + \tilde{p}^\mu\!\!\!\sum_{n\in\mathbb{Z}\backslash\{0\}}\!\frac{1}{n}\dot{\tilde{\alpha}}_{n\mu} \right)
    \end{aligned}
\end{equation}
By adding these terms together, we obtain the the mode expansion of the abbreviated action of the doubled string, which is the following:
\begin{equation*}
\begin{aligned}
    \mathbb{S}[\mathbb{X}(\sigma,\tau)]+  \int\di\tau H[\mathbb{X}(\sigma,\tau)] &= \int\di\tau \left( \mathbbvar{p}_M\dot{\mathbbvar{x}}^M - \frac{\pi\alpha'}{2}\omega^{MN}\mathbbvar{p}_M\dot{\mathbbvar{p}}_N +i \!\!\!\sum_{n\in\mathbb{Z}\backslash\{0\}}\!\frac{1}{n}\,\omega _{MN}\,\dot{\bbalpha}_{-n}^M\bbalpha_n^N \right)
\end{aligned}
\end{equation*}

\paragraph{The symplectic structure of the doubled string.}
Now we can use the equation
\begin{equation}\label{eq:Liouvillepot}
\begin{aligned}
    \mathbb{S}[\mathbb{X}(\sigma,\tau)]+ \int\di\tau H[\mathbb{X}(\sigma,\tau)] \;&=\; \int\di\tau\,\iota_{V_H}\mathbbvar{\Theta}
\end{aligned}
\end{equation}
to determine the Liouville potential $\mathbbvar{\Theta}$ on the phase space of the doubled string and, hence, the its symplectic structure. 
To solve the equation, we choose again the Hamiltonian vector $V_H$ associated to the time flow, which, in the new coordinates $\{\mathbbvar{x}^M,\mathbbvar{p}_M,\alpha_k^\mu,\bar{\alpha}_k^\mu\}_{n\in\mathbb{Z}\backslash\{0\}}$ of the phase space, takes the form
\begin{equation}
\begin{aligned}
    V_H \;&=\; \frac{\di}{\di\tau} \\
    &=\; \dot{\mathbbvar{x}}^M\frac{\partial}{\partial\mathbbvar{x}^M} + \dot{\mathbbvar{p}}^M\frac{\partial}{\partial\mathbbvar{p}^M} + \sum_{n>0} \left(\dot{\alpha}_n^\mu \frac{\partial}{\partial\alpha_n^\mu} + \dot{\bar{\alpha}}_n^\mu \frac{\partial}{\partial\bar{\alpha}_n^\mu} \right).
    \end{aligned}
\end{equation}
Now, by solving the equation \eqref{eq:Liouvillepot} we obtain the following Liouville potential:
\begin{equation}
\begin{aligned}
    \mathbbvar{\Theta} \;&=\; \mathbbvar{p}_M \di\mathbbvar{x}^M -\frac{\pi\alpha'}{2} \omega^{MN}\mathbbvar{p}_M\di{\mathbbvar{p}}_N +i \!\!\!\sum_{n\in\mathbb{Z}\backslash\{0\}}\!\frac{1}{n}\,\omega _{MN}\,\di\bbalpha_{-n}^M\bbalpha_n^N.
\end{aligned}
\end{equation}
By calculating the differential $\mathbbvar{\Omega}=\delta\mathbbvar{\Theta}$, we finally obtain the symplectic form
\begin{equation}
    \mathbbvar{\Omega} \;=\; \di\mathbbvar{p}_M\wedge\di\mathbbvar{x}^M - \frac{\pi\alpha'}{2}\omega^{MN}  \di\mathbbvar{p}_M\wedge\di\mathbbvar{p}_N +i \!\!\!\sum_{n\in\mathbb{N}\backslash\{0\}}\!\frac{1}{n}\,\omega _{MN}\,\di\bbalpha_{-n}^M\wedge\di\bbalpha_n^N.
\end{equation}
This is, therefore, the symplectic form of the phase space $(\mathcal{L}_\mathrm{Q}\mathcal{M},\,\mathbbvar{\Omega})$ of the doubled string.
Notice that we can also rewrite this symplectic form as
\begin{equation}
    \mathbbvar{\Omega} \;=\; \oint\di\sigma\,\frac{1}{2} \omega_{MN}\,\delta\mathbb{X}^M(\sigma)\wedge\delta\mathbb{X}^{\prime N}(\sigma) + \di\mathbbvar{p}_M\wedge\di\mathbb{X}^M(\sigma_0).
\end{equation}
Notice that, if we ignore the second term encoding the boundary, the first term can be immediately given by a potential $\oint\di\sigma\,\frac{1}{2} \omega_{MN}\,\delta\mathbb{X}^M(\sigma)\mathbb{X}^{\prime N}(\sigma)$, which is nothing but the transgression of a symplectic form
$\varpi := \frac{1}{2}\omega_{MN}\,\di\mathbbvar{x}^M\wedge\di\mathbbvar{x}^{N} = \di x^\mu\wedge \di \tilde{x}_\mu$ defined on the doubled space. Notice that this is still a particular example of the fundamental $2$-form which appears in Born geometry in \cite{Svo17, Svo18, Svo19} for a background without fluxes.

\subsection{Algebra of observables}
We want to determine the algebra $\mathfrak{heis}(\mathcal{L}_\mathrm{Q}\mathcal{M},\mathbbvar{\Omega})$ of quantum observables of the phase space of the doubled string.
\begin{equation}
    \hat{f} \;=\; -i\hbar\nabla_{V_f} + f
\end{equation}
We, thus, obtain the following commutation relations:
\begin{equation}\label{eq:commsigma}
    \big[\hat{\mathbb{X}}^M(\sigma)\,,\hat{\mathbb{X}}^N(\sigma')\big] \;=\; i\pi\hbar\alpha'\omega^{MN} -i\hbar\eta^{MN}\varepsilon(\sigma-\sigma') .
\end{equation}
where the function $\varepsilon(\sigma)$ is the quasi-periodic function defined by
\begin{equation}
    \varepsilon(\sigma) \;:=\; \sigma - i\!\!\!\sum_{n\in\mathbb{Z}\backslash\{0\}}\!\!\!\frac{e^{in\sigma}}{n}
\end{equation}
and it satisfies the following properties: firstly, its derivative $\varepsilon'(\sigma)=\delta(\sigma)$ is the Dirac comb; secondly, it satisfies the boundary condition $\varepsilon(\sigma+2\pi n)=\varepsilon(\sigma)+2\pi n$ and, finally, it is an odd function, i.e. $\varepsilon(-\sigma)=-\varepsilon(\sigma)$. \vspace{0.15cm}

\noindent The fact that the operators associated with $X^\mu(\sigma)$ and $\widetilde{X}_\mu(\sigma')$ do not commute by a term $\propto\epsilon(\sigma-\sigma')$ was already observed as far as in \cite{DGMP13}. However, the commutation relations \eqref{eq:commsigma} contain a new term: the skew-symmetric constant matrix $\propto\pi\hbar\alpha'\omega^{MN}$, originating from the topological term \eqref{eq:topterm} in the total action of the doubled string and totally analogous of the one recently observed by \cite{Fre17a, Fre17b}.
We can also easily derive the commutation relations for the higher modes
\begin{equation}
    \left[\hat{\alpha}_n^\mu, \hat{\alpha}_m^\nu\right] \;=\; ng^{\mu\nu}\delta_{m+n,0}  \;=\; \left[\hat{\bar{\alpha}}_n^\mu, \hat{\bar{\alpha}}_m^\nu\right].
\end{equation}

\paragraph{Limits of the algebra of observables} It is worth remarking the role of the dimensionful constants $\hbar$ and $\alpha'$ in providing the deformation to the classical algebra. We are used to seeing $\hbar$ as a quantum deformation parameter but here we also see $\hbar \alpha'$ as another quantum deformation parameter. This suggests interesting limits. The classical limit is the obvious limit given by $\hbar \rightarrow 0$. The particle limit is $\hbar$ fixed but $\alpha' \rightarrow 0$. This algebra suggests a new limit:
\begin{equation}
\hbar \rightarrow 0 \, ; \qquad \alpha' \rightarrow \infty\, ; \qquad \alpha' \hbar \,\,\, \text{fixed}.
\end{equation}
This would be a classical stringy limit where we keep the stringy deformation but remove the quantum deformation. It would be interesting to study the system further in this limit to identify the pure string deformation based effects.

\section{Geometric quantisation on the doubled space}

\subsection{The phase space of the zero-mode string}
Recall that we can expand the fields $\mathbb{X}^M(\sigma)$ of our doubled string $\sigma$-model by
\begin{equation}
    \mathbb{X}^M(\sigma) \;=\; \mathbbvar{x}^M + \alpha'\sigma\mathbbvar{p}^M + \!\!\sum_{n\in\mathbb{Z}\backslash\{0\}}\!\frac{1}{n}\bbalpha^M_n e^{in\sigma},
\end{equation}
where the coordinates of the phase space of a doubled string are $\{\mathbbvar{x}^M,\mathbbvar{p}^M,\alpha_k^\mu,\bar{\alpha}_k^\mu\}$, with the former $\{\mathbbvar{x}^M,\mathbbvar{p}^M\}$ co-ordinatising the zero-modes of the string and the latter $\{\alpha_k^\mu,\bar{\alpha}_k^\mu\}$ co-ordinatising its higher-modes.
A zero-mode truncated doubled string is a doubled string where we are neglecting the higher-modes and it can be seen as a simple embedding of the form
\begin{equation}
    \mathbb{X}^M(\sigma) \;=\; \mathbbvar{x}^M + \alpha'\sigma\mathbbvar{p}^M.
\end{equation}
The zero-modes of a doubled string $\mathbb{X}^M(\sigma,\tau)$ can be thought as a particle in a doubled phase space $(\mathbbvar{x}^M(\tau), \mathbbvar{p}_M(\tau))$. Similarly we expect that the wave-functional $\Psi[\mathbb{X}(\sigma)]$ at zero modes is just a wave-function $ \psi(\mathbbvar{x},\mathbbvar{p})$ on the doubled phase space of zero modes:
\begin{equation}
    \Psi[\mathbb{X}(\sigma)] \;\; \xrightarrow{\;{0\text{ modes}}\;} \;\; \psi(\mathbbvar{x},\mathbbvar{p}), \qquad \mathbbvar{\Omega} \;\; \xrightarrow{\;{0\text{ modes}}\;} \;\; \bbomega.
\end{equation}
The phase space of the zero-modes of a doubled string is, therefore, a $4n$-dimensional symplectic manifold $(\mathcal{P},\bbomega)$ with symplectic form
\begin{equation}
    \bbomega \;=\; \eta_{MN}\, \di\mathbbvar{p}^M\wedge\di\mathbbvar{x}^N - \frac{\pi\alpha'}{2}\omega_{MN} \, \di\mathbbvar{p}^M\wedge\di\mathbbvar{p}^N
\end{equation}
and underlying smooth manifold $\mathcal{P}=\mathbb{R}^{4n}$. Notice that this $2$-form, obtained by a Hamiltonian treatment of the total action of a doubled string $\sigma$-model, exactly agrees with the symplectic form found by \cite{Fre17a, Fre17b} by starting from vertex algebra arguments. \vspace{0.15cm}

\noindent Now, we can apply the machinery of geometric quantisation to this symplectic manifold $(\mathcal{P},\bbomega)$ to quantise the zero-modes of a doubled string.

\paragraph{Kinetic coordinates for the doubled phase space.}
Let us change the canonical momentum coordinates with the untwisted non-canonical momentum coordinates $\mathbbvar{k}^M=(e^{-B})^M_{\;\;\,N}\mathbbvar{p}^N$. Given a doubled string $\sigma$-model $\mathbb{X}(\sigma,\tau)$, these will be related by
\begin{equation}
    k_\mu(\sigma,\tau) = \oint \di\sigma\, g_{\mu\nu}\dot{X}^\nu(\sigma,\tau), \qquad \tilde{k}^\mu(\sigma,\tau) = \oint \di\sigma\, g^{\mu\nu}\dot{\widetilde{X}}_\nu(\sigma,\tau).
\end{equation}
We can rotate the doubled coordinates, accordingly $\mathbbvar{x}^M\mapsto(e^{-B})^M_{\;\;\,N}\mathbbvar{x}^N$ to the untwisted frame.
We can now rewrite the symplectic form in the kinetic coordinates $\{\mathbbvar{x}^M,\mathbbvar{k}^M\}$ and have
\begin{equation}
    \bbomega \;=\; \eta_{MN}\, \di\mathbbvar{k}^M\wedge\di\mathbbvar{x}^N - \frac{\pi\alpha'}{2}\omega_{MN}^{(B)}\, \di\mathbbvar{k}^M\wedge\di\mathbbvar{k}^N
\end{equation}
where we called the matrix
\begin{equation}
    \omega^{(B)}_{MN} \;=\;  \begin{pmatrix}B_{\mu\nu} & \delta_\mu^{\;\,\nu} \\-\delta_{\;\,\nu}^{\mu} & 0 \end{pmatrix}.
\end{equation}
Then, we can choose the following gauge for the Liouville potential:
\begin{equation}
    \bbtheta \;=\; \eta_{MN}\, \mathbbvar{k}^M\di\mathbbvar{x}^N - \frac{\pi\alpha'}{2}\omega_{MN}^{(B)}\, \mathbbvar{k}^M\di\mathbbvar{k}^N.
\end{equation}

\paragraph{The action of the zero-mode string.}
As we remarked, in geometric quantisation the Lagrangian density $\mathfrak{L}\in\Omega^1(\gamma)$ of a particle is related to the Liouville potential $\bbtheta\in\Omega^1(\mathcal{P})$ by the equation 
\begin{equation}
    \mathfrak{L}_H \,=\, (\iota_{V_H}\bbtheta-H)\di\tau.
\end{equation}
We can then immediately use it, in the form
\begin{equation}
    S[\mathbbvar{x}(\tau),\mathbbvar{k}(\tau)] \,=\, \int_\gamma\di\tau\big(\iota_{V_H}\bbtheta-H\big) \qquad \text{with} \qquad V_H=\dot{\mathbbvar{x}}^M\frac{\partial}{\partial \mathbbvar{x}^M} + \dot{\mathbbvar{k}}^M\frac{\partial}{\partial \mathbbvar{k}^M},
\end{equation}
to find the action of the zero-mode doubled string: 
\begin{equation}
    S[\mathbbvar{x}(\tau),\mathbbvar{k}(\tau)] \,=\, \int_\gamma \di\tau \left(\eta_{MN}\mathbbvar{k}^M\dot{\mathbbvar{x}}^N- \frac{\pi\alpha'}{2}\omega_{MN}^{(B)}\, \mathbbvar{k}^M\dot{\mathbbvar{k}}^N -\mathcal{H}_{MN}^{(0)} \mathbbvar{k}^M\mathbbvar{k}^N\right)
\end{equation}
where we called the matrix
\begin{equation}
    \mathcal{H}^{(0)}_{MN} \;=\;  \begin{pmatrix}g_{\mu\nu} & 0 \\0 & g^{\mu\nu} \end{pmatrix}.
\end{equation}

\subsection{Algebra of the observables}

Recall that in geometric quantisation a quantum observable $\hat{f}\in\mathrm{Aut}(\mathfrak{H})$ is a linear automorphism of the Hilbert space, obtained from the corresponding classic observable $f\in\mathcal{C}^\infty(\mathcal{P})$ by the following identification:
\begin{equation}\label{eq:op}
    \hat{f} \;:=\; -i\hbar \nabla_{V_f} + f,
\end{equation}
where the vector $V_f\in\mathfrak{X}(\mathcal{P})$ is the Hamiltonian vector with Hamiltonian function $f$, i.e. the vector which solves the Hamilton equation
\begin{equation}\label{eq:hamilton}
    \iota_{V_f}\bbomega\,=\,\di f.
\end{equation}
In this subsection we want to determine the Lie algebra of quantum observables  $\mathfrak{heis}(\mathcal{P},\bbomega)$ on the doubled phase space.

\paragraph{Hamiltonian vector fields.}
Let us first solve the Hamilton equation \eqref{eq:hamilton} for a generic Hamiltonian function $f\in\mathcal{C}^\infty(\mathcal{P})$. We expand the vector $V_f\in\mathfrak{X}(\mathcal{P})$ in the kinetic coordinates
\begin{equation}
    V_f \;=\; V_{f,\mathbbvar{x}}^M\,\frac{\partial}{\partial\mathbbvar{x}^M} \,+\, V_{f,\mathbbvar{k}}^M\,\frac{\partial}{\partial\mathbbvar{k}^M}.
\end{equation}
Hence, the Hamilton equation \eqref{eq:hamilton}, in coordinates, becomes
\begin{equation}\begin{aligned}
   \eta_{MN}\,( V_{f,\mathbbvar{k}}^M \,\di\mathbbvar{x}^N -V_{f,\mathbbvar{x}}^M \,\di\mathbbvar{k}^N ) - \pi\alpha'\omega_{MN}^{(B)}\, V_{f,\mathbbvar{k}}^M\,\di\mathbbvar{k}^N \;&=\; \frac{\partial f}{\partial\mathbbvar{x}^M}\di\mathbbvar{x}^M + \frac{\partial f}{\partial\mathbbvar{k}^M}\di\mathbbvar{k}^M.
\end{aligned}\end{equation}
Therefore, the Hamiltonian vector field $V_f$ with Hamiltonian $f$ is given by
\begin{equation}
    \begin{aligned}
    V_{f} \;=\; \Big( \eta^{MN}\frac{\partial f}{\partial\mathbbvar{x}^N} \Big) \frac{\partial}{\partial\mathbbvar{k}^M} + \Big( -\eta^{MN}\frac{\partial f}{\partial\mathbbvar{k}^N} - \pi\alpha'\omega^{MN}_{(B)}\, \frac{\partial f}{\partial\mathbbvar{x}^N} \Big) \frac{\partial}{\partial\mathbbvar{x}^M}
    \end{aligned}
\end{equation}
where we called $\omega^{MN}_{(B)}:=\eta^{ML}\, \omega_{LP}^{(B)}\, \eta^{PN}$. 
In particular the Hamiltonian vector fields corresponding to the classical observables of the kinetic coordinates $\mathbbvar{x}^M$ and $\mathbbvar{k}^M$ are
\begin{equation}
    \begin{aligned}
    V_{f=\mathbbvar{x}^N} \;&=\; \eta^{MN}\frac{\partial }{\partial\mathbbvar{k}^M} - \pi\alpha'\omega^{MN}_{(B)}\, \frac{\partial }{\partial\mathbbvar{x}^M}, \\
    V_{f=\mathbbvar{k}^N}  \;&=\; -\eta^{MN}\frac{\partial }{\partial\mathbbvar{x}^M}.
    \end{aligned}
\end{equation}

\paragraph{Non-commutative Heisenberg algebra.}
By applying the definition \eqref{eq:op} of quantum observable, we find that the operators associated to the kinetic coordinates are the following:
\begin{equation}
\begin{aligned}
        \hat{\mathbbvar{x}}^M \,&=\,  i\hbar\eta^{NM}\frac{\partial}{\partial\mathbbvar{k}^N}  -i\hbar\frac{\pi\alpha'}{2}\omega_{(B)}^{NM}\frac{\partial}{\partial\mathbbvar{x}^N} + \mathbbvar{x}^M, \\
        \hat{\mathbbvar{k}}^M \,&=\,  -i\hbar\eta^{NM}\frac{\partial}{\partial\mathbbvar{x}^N}.
\end{aligned}
\end{equation}
Therefore the commutation relations between the coordinates operators are the following:
\begin{equation}
    [\hat{\mathbbvar{x}}^M,\hat{\mathbbvar{x}}^N] \,=\, \pi i\hbar\alpha' \omega^{MN}_{(B)}, \quad [\hat{\mathbbvar{x}}^M,\hat{\mathbbvar{k}}^N] \,=\, i\hbar\eta^{MN}, \quad [\hat{\mathbbvar{k}}^M,\hat{\mathbbvar{k}}^N] \,=\,0.
\end{equation}
Thus, the $4n$-dimensional Lie algebra $\mathfrak{heis}(\mathcal{P},\bbomega)$ can be regarded as a non-commutative version of the usual Heisenberg algebra, where the position operators do not generally commute. \vspace{0.15cm}

\noindent Explicitly, in undoubled notation, we have the following commutation relations:
\begin{equation}\label{eq:commrelund}
    \begin{aligned}
    [\hat{x}^\mu,\hat{x}^\nu]&=0,&\quad [\hat{x}^\mu,\hat{\tilde{x}}_\nu]&=\pi i\hbar \alpha'\delta^\mu_{\;\nu},& \quad [\hat{\tilde{x}}_\mu,\hat{\tilde{x}}_\nu]&=-2\pi i\hbar\alpha'B_{\mu\nu}, \\[0.5ex]
    [\hat{k}_\mu,\hat{k}_\nu]&=0,&\quad [\hat{k}_\mu,\hat{\tilde{k}}^\nu]&=0,& \quad [\hat{\tilde{k}}^\mu,\hat{\tilde{k}}^\nu]&=0, \\[0.5ex]
    [\hat{x}^\mu,\hat{\tilde{k}}^\nu]=[\hat{\tilde{x}}_\mu,\hat{k}_\nu]&=0,&\quad [\hat{x}^\mu,\hat{k}_\nu] &= i\hbar\delta^\mu_{\;\nu},& \quad [\hat{\tilde{x}}_\mu,\hat{\tilde{k}}^\nu] &= i\hbar\delta_\mu^{\;\nu}.
    \end{aligned}
\end{equation}
Examining this algebra from the perspective of the limits we discussed earlier we see that $\hbar$ controls the noncommutativity of the position with the momentum and $\hbar \alpha'$ the noncommutativity of the coordinates and their duals. Finally, $\alpha' B$ the noncommutativity of the spacetime coordinates. Thus when the B-field is included we have three noncommutativity parameters.

\paragraph{Uncertainty principle on the doubled space.}
Following standard text book techniques applied to the commutation relations \eqref{eq:commrelund}, we can immediately show that any position coordinate $x^\mu$ and its dual $\tilde{x}_\mu$ satisfy the following uncertainty relation:
\begin{equation}
    \Delta x \, \Delta\tilde{x} \;\geq\; \frac{\pi\hbar}{2}\alpha'.
\end{equation}
This means that $x^\mu$ and $\tilde{x}_\mu$ cannot be measured with absolute precision at the same time, but there will be always a minimum uncertainty proportional to the area $\hbar\alpha'$. This provides support to the intuition of a minimal distance scale in string theory. The standard lore is that for small distances one goes to the T-dual frame and the distances will always be larger than the string scale.\vspace{0.15cm}

\noindent In addition, both the couples $(x,p)$ and $(\tilde{x},\tilde{p})$ satisfy the usual uncertainty relation between position and momentum:
\begin{equation}
    \Delta x \, \Delta p \;\geq\; \frac{\hbar}{2}, \qquad \Delta \tilde{x} \, \Delta\tilde{p} \;\geq\; \frac{\hbar}{2}.
\end{equation}
However, it is worth noticing that the momentum and its dual can be measured at the same time:
\begin{equation}
    \Delta p \, \Delta\tilde{p} \;\geq\; 0.
\end{equation}

\paragraph{Hamiltonian.}
Notice that the Hamiltonian operator of the zero-mode doubled string will be given by
\begin{equation}
    \hat{H} \,=\, \mathcal{H}_{MN}^{(0)} \hat{\mathbbvar{k}}^M\hat{\mathbbvar{k}}^N \,=\, -\hbar^2\, \mathcal{H}^{MN}_{(0)} \frac{\partial}{\partial\mathbbvar{x}^M}\frac{\partial}{\partial\mathbbvar{x}^N},
\end{equation}
where we called the matrix $\mathcal{H}^{MN}_{(0)}:=\eta^{ML}\eta^{NP}\mathcal{H}_{LP}^{(0)}$.

\paragraph{Non-commutative Heisenberg algebra in canonical coordinates.}
In the zero-mode string canonical coordinates $\{\mathbbvar{x}^M,\mathbbvar{p}_M\}$ we obtain the following operators:
\begin{equation}
\begin{aligned}
        \hat{\mathbbvar{x}}^M \,&=\,  i\hbar\frac{\partial}{\partial\mathbbvar{p}_M}  -i\hbar\frac{\pi\alpha'}{2}\omega^{NM}\frac{\partial}{\partial\mathbbvar{x}^N} + \mathbbvar{x}^M, \\
        \hat{\mathbbvar{p}}_M \,&=\,  -i\hbar\frac{\partial}{\partial\mathbbvar{x}^M}.
\end{aligned}
\end{equation}
Therefore the commutation relations between the canonical coordinates observables are 
\begin{equation}
    [\hat{\mathbbvar{x}}^M,\hat{\mathbbvar{x}}^N] \,=\, \pi i\hbar\alpha' \omega^{MN}, \quad [\hat{\mathbbvar{x}}^M,\hat{\mathbbvar{p}}_N] \,=\, i\hbar\delta^{M}_{\;\;N}, \quad [\hat{\mathbbvar{p}}_M,\hat{\mathbbvar{p}}_N] \,=\,0.
\end{equation}

\paragraph{Relation with the symplectic structure of the doubled space.}
Let us now focus on the subalgebra generated by the operators $\hat{x}^\mu$ and $\hat{\tilde{x}}_\mu$. This will be given by the following commutation relations:
\begin{equation}
    \begin{aligned}
    [\hat{x}^\mu,\hat{x}^\nu]=0,\quad [\hat{x}^\mu,\hat{\tilde{x}}_\nu]=\pi i\hbar \alpha'\delta^\mu_{\;\nu}, \quad [\hat{\tilde{x}}_\mu,\hat{\tilde{x}}_\nu]=0.
    \end{aligned}
\end{equation}
Notice that this can be seen as an ordinary $2n$-dimensional Heisenberg algebra $\mathfrak{h}(2n)$. This means that such an algebra is immediately given by a symplectic manifold $(\mathcal{M},\varpi)$ with $\mathcal{M}\cong\mathbb{R}^{2n}$ and symplectic form $\varpi := \pi\hbar\alpha'\di x^\mu\wedge \di \tilde{x}_\mu$. This symplectic structure on the doubled space is exactly the one introduced by \cite{Vai12}.

\subsection{T-duality and the string deformed Fourier transform}

In ordinary quantum mechanics we choose to represent the wavefunctions in either the position or the momentum basis and it is the Fourier transform that maps the wavefunction in one basis to the other basis. From the persepective of geometric quantisation this is the transformation between elements of the Hilbert spaces constructed with different choices of Lagrangian submanifold i.e. different polarisations. T-duality is a change in our choice of polarisation. We can then follow the pairing construction used in \cite{WeiGQ, Kostant, Souriau, Woodhouse:1980pa} to construct the transformation for the string wavefunction moving between different duality frames. This will produce a string deformed Fourier transform (that reduces to the usual Fourier transform in the $\alpha' \rightarrow 0$ limit). From the geometric quantisation perspective these transformations are known as Blatter-Kostant-Sternberg \cite{WeiGQ} kernel's. 

\paragraph{Polarisations.}
In the geometric quantisation of a symplectic space $(\mathcal{P},\bbomega)$, a polarisation corresponds to a choice of an integrable Lagrangian subspace $L\subset \mathcal{P}$. 
Since $\mathcal{P}$ is a vector space, the first Chern class of the prequantum $U(1)$-bundle whose curvature is the symplectic form $\bbomega\in\Omega^2(\mathcal{P})$, is necessarily trivial.
In geometric quantisation, this implies that the Hilbert space of the quantised system is defined by the space of the complex $\mathrm{L}^2$-functions on the Lagrangian submanifold $L\subset\mathcal{P}$, i.e. by
\begin{equation}
    \mathfrak{H} \;:=\; \mathrm{L}^2(L,\mathbb{C}).
\end{equation}
Remarkably, this does not depend on the choice of polarisation $L\subset\mathcal{P}$ and it is possible to prove that, for any other Lagrangian subspace $L'\subset\mathcal{P}$, we would have an isomorphism of Hilbert spaces $\mathfrak{H} \cong \mathrm{L}^2(L',\mathbb{C})$.

\paragraph{T-duality as a change of polarisation.}
Let us rewrite the symplectic form $\bbomega\in\Omega^2(\mathcal{P})$ in canonical coordinates $(\mathbbvar{x}^M,\mathbbvar{p}^M)$, i.e.
\begin{equation}
    \bbomega \;=\; \eta_{MN}\,\di\mathbbvar{p}^M\wedge\di\mathbbvar{x}^N - \frac{\pi\alpha'}{2}\omega_{MN}\, \di\mathbbvar{p}^M\wedge\di\mathbbvar{p}^N,
\end{equation}
and let us recall that the momenta doubled vector can be interpreted as the doubled vector of winding numbers $\mathbbvar{p}^M=(w^\mu,\,\tilde{w}_\mu)$. 
It is now immediate that the vector spaces
\begin{equation}
    L \;:=\; \mathrm{Span}(x^\mu,w^\mu), \qquad \widetilde{L} \;:=\; \mathrm{Span}(\tilde{x}_\mu,\tilde{w}_\mu)
\end{equation}
are Lagrangian subspaces of the symplectic space $(\mathcal{P},\bbomega)$. This means that we will have two polarisations corresponding to the two T-duality frames $(x^\mu,w^\mu)$ and $(\tilde{x}_\mu,\tilde{w}_\mu)$.
We can thus define two basis $\{\Ket{{x},{w}}\}_{({x},{w})\in{L}}$ and $\{\Ket{\tilde{x},\tilde{w}}\}_{(\tilde{x},\tilde{w})\in\widetilde{L}}$ for our Hilbert space $\mathfrak{H}$.
If we consider a generic state $\Ket{\psi}\in\mathfrak{H}$ of our Hilbert space, we can now express it in the basis associated to both the T-duality frames by
\begin{equation}
    \psi_w(x) \;:=\; \Braket{x,w|\psi}, \qquad \widetilde{\psi}_{\tilde{w}}(\tilde{x}) \;:=\; \Braket{\tilde{x},\tilde{w}|\psi}.
\end{equation}
Now, we want to explicitly find the isomorphism $\mathrm{L}^{\!2}(L;\mathbb{C})\cong\mathrm{L}^2(\widetilde{L};\mathbb{C})$ between wave-functions in the two T-duality frames. Let us expand our zero-mode truncated string by $X^\mu(\sigma)=x^\mu + \alpha'\sigma \tilde{p}^\mu$ and $\widetilde{X}_\mu(\sigma)=\tilde{x}_\mu +\alpha'\sigma p_\mu$ and call $\mathbb{X}^M(\sigma)=\big(X^\mu(\sigma),\,\widetilde{X}_\mu(\sigma)\big)$. A zero-mode truncated string $\mathbb{X}^M(\sigma)=\mathbbvar{x}^M+\alpha'\sigma\mathbbvar{p}^M$ is represented by a point $(\mathbbvar{x}^M,\mathbbvar{p}_M)\in\mathcal{P}$ of the phase space of the zero-mode doubled string. \vspace{0.2cm}

\noindent Now, notice that the symplectic form immediately satisfies the following identity:
\begin{equation}
\begin{aligned}
    \bbomega \;=\;  \di_L\di_{\widetilde{L}}\Big(\tilde{p}^\mu\tilde{x}_\mu - p_\mu x^\mu + \pi\alpha'p_\mu\tilde{p}^\mu\Big),
    \end{aligned}
\end{equation}
where $\di_L$ and $\di_{\widetilde{L}}$ are respectively the differentials on the Lagrangian subspaces $L = \mathrm{Span}(x^\mu,\tilde{p}^\mu)$ and $\widetilde{L} = \mathrm{Span}(\tilde{x}_\mu,p_\mu)$.
Thus, we can express the symplectomorphism $f:\mathcal{P}\rightarrow\mathcal{P}$ encoding T-duality on the phase space of the zero-mode doubled string by the generating function
\begin{equation}\label{eq:genfundou}
\begin{aligned}
    F(\mathbbvar{x},\mathbbvar{p})\;&=\; \tilde{p}^\mu\tilde{x}_\mu - p_\mu x^\mu + \pi\alpha'p_\mu\tilde{p}^\mu \\
    \;&=\; \omega_{MN} \mathbbvar{p}^M\mathbbvar{x}^N + \frac{\pi\alpha'}{2}\eta_{MN}\mathbbvar{p}^M\mathbbvar{p}^N,
    \end{aligned}
\end{equation}
which is nothing but the zero-mode truncation of the lift to the phase space of the doubled string of the action functional \eqref{eq:genfunF}.
Such a symplectomorphism is simply the $O(n,n)$ transformation of the doubled coordinates and momenta by $(\mathbbvar{x}^M,\mathbbvar{p}^M) \mapsto (\eta_{MN}\mathbbvar{x}^N,\,\eta_{MN}\mathbbvar{p}^N)$.
Now, by applying the machinery of geometric quantisation, the matrix of the change of basis on the Hilbert space $\mathfrak{H}$ will be given by the generating function \eqref{eq:genfundou} as follows:
\begin{equation}
\begin{aligned}
    \Braket{x,w|\tilde{x},\tilde{w}} \;&=\; \exp\frac{i}{\hbar}\!\left(p_\mu x^\mu - \tilde{p}^\mu\tilde{x}_\mu + \pi\alpha'p_\mu\tilde{p}^\mu\right) \\[0.5ex]
    \;&=\; \exp\frac{i}{\hbar}\!\left(\omega_{MN} \mathbbvar{p}^M\mathbbvar{x}^N + \frac{\pi\alpha'}{2}\eta_{MN}\mathbbvar{p}^M\mathbbvar{p}^N\right),
    \end{aligned}
\end{equation}
where $w^\mu \equiv \tilde{p}^\mu$ and $\tilde{w}_\mu \equiv p_\mu$. 
Therefore we can equivalently rewrite the transformation
\begin{equation}
    \Braket{\tilde{x},\tilde{w}|\psi} \;=\; \int_L\di^n x\, \di^n w \Braket{\tilde{x},\tilde{w}|x,w}\Braket{x,w|\psi}
\end{equation}
as the following, {\it{{{stringy}} Fourier transformation}}:
\begin{equation}
    \widetilde{\psi}_{\tilde{w}}(\tilde{x}) \;=\; \int_L \di^n x\, \di^n w \exp\frac{i}{\hbar}\!\left(\omega_{MN} \mathbbvar{p}^M\mathbbvar{x}^N  + \frac{\pi\alpha'}{2}\eta_{MN}\mathbbvar{p}^M\mathbbvar{p}^N\right)\psi_w(x)  \, .
\end{equation}
This is the transformation between the wavefunctions in different duality frames. Mathematically it is the isomorphism $\mathrm{L}^{\!2}(L,\mathbb{C})\cong\mathrm{L}^2(\widetilde{L},\mathbb{C})$. In undoubled coordinates we can explicitly rewrite such a stringy Fourier transformation as follows: 
\begin{equation}
    \widetilde{\psi}_{\tilde{w}}(\tilde{x}) \;=\; \int_L \di^n x\, \di^n w \exp\frac{i}{\hbar}\!\left(\tilde{w}_\mu x^\mu - w^\mu\tilde{x}_\mu + \pi\alpha'\tilde{w}_\mu w^\mu\right)\psi_w(x),
\end{equation}
where we used the identities $w^\mu\equiv \tilde{p}^\mu$ and $\tilde{w}_\mu\equiv p_\mu$. Notice that, even if the form of the symplectomorphism $f$ is particularly simple, the transformation for wave-function is more complicated than just a Fourier transform. The difference from the usual Fourier transformation is given by the additional $\frac{\pi\alpha'}{2}\eta_{MN}\mathbbvar{p}^M\mathbbvar{p}^N$ term. This then will reduces to a standard Fourier transform in the limit $\alpha'\rightarrow 0$. \vspace{0.15cm}

\noindent In terms of basis, this transformation can be also be expressed by
\begin{equation}
    \Ket{\tilde{x},\tilde{w}} \;=\; \int_L \di^n x\, \di^n w \exp\frac{i}{\hbar}\!\left(\omega_{MN} \mathbbvar{p}^M\mathbbvar{x}^N  + \frac{\pi\alpha'}{2}\eta_{MN}\mathbbvar{p}^M\mathbbvar{p}^N\right) \Ket{x,w}.
\end{equation}

\paragraph{A phase term in the change of polarisation.} 
Finally, notice that, if we restrict our generalised winding to ordinary integer winding $w,\tilde{w}\in\mathbb{Z}^{n}$, we will obtain a change of polarisation of the form
\begin{equation}
    \widetilde{\psi}_{\tilde{w}}(\tilde{x}) \,=\, \sum_{w\in\mathbb{Z}^n}e^{\frac{i}{\hbar}\pi\alpha'\tilde{w}_\mu w^\mu} \! \int_M \di^n x\, e^{\frac{i}{\hbar}\!\left(\tilde{w}_\mu x^\mu - w^\mu\tilde{x}_\mu\right)}\psi_w(x).
\end{equation}
In this context, as firstly noticed with different arguments by \cite{Fre17b}, T-duality does not simply act as a "double" Fourier transformation of the wave-function of a string, because there will be an extra phase contribution given by $\exp\!\big(i\pi{\frac{\alpha'}{\hbar}\tilde{w}_\mu w^\mu}\big)$ for any term with $w,\tilde{w}\neq 0$.
Since we are restricting now to the case where $w,\tilde{w}$ are integers and $\sqrt{\hbar/\alpha'}$ is just the unit of momentum, we immediately conclude that the only possible phase contributions are $\exp\!\big(i\pi{\frac{\alpha'}{\hbar}\tilde{w}_\mu w^\mu}\big)\in \{+1,-1\}$, depending on the product $\tilde{w}_\mu w^\mu\equiv p_\mu w^\mu$ being even or odd.
Notice that the presence of the topological term in the action induces a very similar phase term in the partition function of a string with an analogous role, as seen by \cite{Berman:2007vi}.

\paragraph{Darboux coordinates for the zero-mode string.}
Let us find the Darboux coordinates on the manifold $\mathcal{P}$ for the symplectic form $\bbomega\in\Omega^2(\mathcal{P})$. If we define the new coordinates
\begin{equation}
    \begin{aligned}
        q^\mu \;&:=\; x^\mu \\
        \tilde{q}_\mu \;&:=\; \tilde{x}_\mu - \pi\alpha'p_\mu
    \end{aligned}
\end{equation}
and we pack them together as $\mathbbvar{q}^M:=(q^\mu,\tilde{q}_\mu)$, we can rewrite the symplectic form simply as
\begin{equation}
    \bbomega \;=\; \di\mathbbvar{p}_M\wedge\di\mathbbvar{q}^M
\end{equation}
with $\mathbbvar{p}_M=\eta_{MN}\mathbbvar{p}^N=(p_\mu,\tilde{p}^\mu)$. Therefore the conjugate variable on the phase to the canonical momenta $\mathbbvar{p}_M$ of the zero-mode string is the new coordinate $\mathbbvar{q}^M$. Notice that this variable is not the proper position $\mathbbvar{x}^M$ on the doubled space, but a mix of position and momentum. This change of coordinates is intimately related to what is known as Bopp's shift in non-commutative quantum mechanics.

\subsection{Relation with non-commutativity induced by fluxes}

The non-commutativity we are exploring follows that in \cite{Bla14}, but is different from (though close to) the one introduced by \cite{Lus10} and further explored by \cite{ALLP13}, where the non-commutativity of the doubled space is induced by the presence of fluxes. For a more recent account see \cite{Sza18} (who also discusses the appearance of non-associativity) and \cite{Ost19}. The notion of non-commutativity we are considering is completely independent by the presence of fluxes and characterises even flat and topologically trivial doubled spaces. As we will see in the next section, the non-commutativity between a physical coordinate and its T-dual is intrinsic and linked to the existence of a minimal length $\ell_s =\sqrt{\hbar\alpha'}$ on the doubled space.\vspace{0.15cm}

\noindent The link between the two notions of non-commutativity is provided by \cite{Bla14}.
The presence of flux implies monodromies for the generalised metric of the form
\begin{equation}
    \mathcal{H}(x + 2\pi) \;=\; \mathcal{O}\mathcal{H}(x)\mathcal{O}^\mathrm{T}
\end{equation}
with monodromy matrix $\mathcal{O}\in O(n,n)$. Thus, as explained by \cite{Bla14}, we need to consider the further generalised boundary conditions
\begin{equation}
    \mathbb{X}^M(\sigma+2\pi) \;=\; \mathcal{O}^M_{\;\;\;N}  \mathbb{X}^N(\sigma) + 2\pi\mathbbvar{p}^M
\end{equation}
for our doubled string $\sigma$-model. When we write the Tseytlin action, we then have to generalise its boundary term accordingly. As seen by \cite{Bla14}, the new action produces the non-commutativity given by the fluxes on the doubled phase space.

\section{Non-commutative QM of the zero-mode string}

Non-commutative quantum mechanics was introduced as far as in 1947 by \cite{Sny47}. The fundamental idea, at the time, was to quantize flat spacetime by introducing a minimal length and generalising the uncertainty principle to make it fuzzy. \vspace{0.15cm}

\noindent As we saw in the previous section, the zero-mode truncation of the pre-quantised wave-functional $\Psi[\mathbb{X}(\sigma)]$ of a doubled string can be seen as a conventional pre-quantised wave-function $\psi(\mathbbvar{x},\mathbbvar{p})$ of a particle in a doubled space. In other words, the zero-modes of strings behave like particles in a double space.
However, such a doubled space is intrinsically non-commutative. As we derived, indeed, the commutation relations of the position operators are of the form
\begin{equation}
    [\hat{\mathbbvar{x}}^M,\hat{\mathbbvar{x}}^N] \,=\, i\vartheta^{MN}  \quad\text{ with }\quad \vartheta^{MN} := \pi\hbar\alpha'\omega^{MN} .
\end{equation}
This means that the Quantum Mechanics of the string zero-modes will be non-commutative (NCQM). Let us choose units where we only require $c=1$. This way the two physical dimensions of length and energy are explicitly parametrised by the two universal constants as follows:
\begin{equation}
    \left[\hbar\alpha'\right] \,=\, \mathrm{length}^2, \qquad \left[\frac{\hbar}{\alpha'}\right] \,=\, \mathrm{energy}^2
\end{equation}
and thus the string scale must be expressed as $\ell_s=\sqrt{\hbar\alpha'}$. We notice that any couple of T-dual coordinates fails in commuting by an area which is proportional to the string scale, i.e. we have
\begin{equation}
    \big[\hat{x}^\mu,\,\hat{\tilde{x}}_\mu\big] \,=\, i\pi\ell_s^2
\end{equation}
for any fixed $\mu=1,\dots,n$.
In this context $\pi\ell^2_s$ can be interpreted as a minimal area of the doubled space.


\subsection{Non-commutative coherent states of zero-mode strings}

Let us start from the non-commutative Heisenberg algebra $\mathfrak{heis}(\bbomega,\mathcal{P})$ of the phase space $(\bbomega,\mathcal{P})$ of the zero-modes truncated doubled string:
\begin{equation}
    [\hat{\mathbbvar{x}}^M,\hat{\mathbbvar{x}}^N] \,=\, \pi i\hbar\alpha' \omega^{MN} , \quad [\hat{\mathbbvar{x}}^M,\hat{\mathbbvar{p}}_N] \,=\, i\hbar\delta^{M}_{\;\, N}, \quad [\hat{\mathbbvar{p}}_M,\hat{\mathbbvar{p}}_N] \,=\,0.
\end{equation}
Notice that the subspace of $\mathcal{P}=\mathbb{R}^{4n}$ spanned by $(x^\mu,\tilde{x}_\mu)$ is not a Lagrangian subspace, therefore there is no well-defined notion of wave-function of the form $\psi(x^\mu,\tilde{x}_\mu)$. In quantum mechanical terms, since T-dual coordinates $[\hat{x}^\mu,\hat{\tilde{x}}_\nu]\neq 0$ do not generally commute, there exists no basis $\big\{\Ket{x^\mu,\tilde{x}_\mu}\big\}\not\subset \mathfrak{H}$ of eigenstates of the doubled position operators.

\paragraph{Coherent states.}
However, we can define the following annihilation and creation operators
\begin{equation}
    \begin{aligned}
    \hat{\mathbbvar{z}}^\mu \;&=\; \frac{1}{\sqrt{2\pi\hbar}}\left(\hat{x}^\mu + i \hat{\tilde{x}}_\mu\right) \\
    \hat{\mathbbvar{z}}^{\dagger\mu} \;&=\; \frac{1}{\sqrt{2\pi\hbar}}\left(\hat{x}^\mu - i \hat{\tilde{x}}_\mu\right) 
    \end{aligned}
\end{equation}
whose commutator
\begin{equation}
    \left[\hat{\mathbbvar{z}}^\mu,\,\hat{\mathbbvar{z}}^{\dagger\nu}\right] \;=\; \alpha'\delta^{\mu\nu}
\end{equation}
satisfies the commutation relations of the Fock algebra.
Thus the non-commutative quantum configuration space is a bosonic Fock space
\begin{equation}
    \mathbf{F}_{\mathrm{cs}} \;:=\; \bigodot_{k\in\mathbb{N}} \mathbb{C}^n \;=\;  \mathbb{C} \oplus \mathbb{C}^n \oplus (\mathbb{C}^n\odot\mathbb{C}^n) \oplus (\mathbb{C}^n\odot\mathbb{C}^n\odot\mathbb{C}^n) \oplus \dots
\end{equation}
generated by vectors of the form
\begin{equation}
    \Ket{0}, \quad\hat{\mathbbvar{z}}^{\dagger\mu}\Ket{0}, \quad\frac{1}{\sqrt{2}} \hat{\mathbbvar{z}}^{\dagger\mu_1}\hat{\mathbbvar{z}}^{\dagger\mu_2}\Ket{0}, \quad \frac{1}{\sqrt{3!}} \hat{\mathbbvar{z}}^{\dagger\mu_1}\hat{\mathbbvar{z}}^{\dagger\mu_2}\hat{\mathbbvar{z}}^{\dagger\mu_3}\Ket{0}, \quad \dots
\end{equation}
where the vacuum state $\Ket{0}$ is defined by the equation $\hat{\mathbbvar{z}}^\mu\Ket{0}=0$ for all $\mu=1,\dots,n$.\vspace{0.15cm}

\noindent The important aspect of working with the creation and annihilation operators is that there exist eigenstates $\Ket{\mathbbvar{z}^1,\cdots,\mathbbvar{z}^n}$ for all the operators $\hat{\mathbbvar{z}}^\mu$ with $\mu=1,\dots,n$. These satisfy the following defining properties:
\begin{equation}
    \begin{aligned}
    \hat{\mathbbvar{z}}^\mu\Ket{\mathbbvar{z}^1,\cdots,\mathbbvar{z}^n} \;&=\; \mathbbvar{z}^\mu\Ket{\mathbbvar{z}^1,\cdots,\mathbbvar{z}^n} \\[0.7ex]
   \Bra{\mathbbvar{z}^1,\cdots,\mathbbvar{z}^d} \hat{\mathbbvar{z}}^{\dagger\mu} \;&=\; \Bra{\mathbbvar{z}^1,\cdots,\mathbbvar{z}^d} \bar{\mathbbvar{z}}^{\mu}
    \end{aligned}
\end{equation}
for eigenvalues $(\mathbbvar{z}^1, \cdots, \mathbbvar{z}^n)\in\mathbb{C}^n$. These states are called \textit{coherent states}. \vspace{0.15cm}

\noindent Let us now use the compact notation $\Ket{\mathbbvar{z}} := \Ket{\mathbbvar{z}^1,\cdots,\mathbbvar{z}^n}$ for coherent states. A normalised coherent state can be expressed by
\begin{equation}
    \Ket{\mathbbvar{z}} \;:=\; \exp\!\left(-\frac{\delta_{\mu\nu}}{2\alpha'}\,\mathbbvar{z}^\mu\,\bar{\mathbbvar{z}}^\nu\right) \exp\!\left(\frac{\delta_{\mu\nu}}{\alpha'}\,\mathbbvar{z}^\mu\,\hat{\mathbbvar{z}}^{\dagger\nu}\right) \Ket{0},
\end{equation}
where the vacuum state $\Ket{0}\in\mathbf{F}_{\mathrm{cs}}$ is defined as previously.
These states constitute an overcomplete basis on the Fock space space $\mathbf{F}_{\mathrm{cs}}$, since they satisfy the property
\begin{equation}
    \frac{1}{(2\pi\alpha')^n}\int_{\mathbb{C}^n}\di^n\mathbbvar{z}\,\di^n\bar{\mathbbvar{z}} \,\Ket{\mathbbvar{z}}\Bra{\mathbbvar{z}} \;=\; 1.
\end{equation}
There is an isomorphism between the non-commutative quantum configuration space $\mathbf{F}_{\mathrm{cs}}$ and the quantum Hilbert space $\mathfrak{H}$, i.e.
\begin{equation}
    \mathbf{F}_{\mathrm{cs}} \,\cong\, \mathfrak{H}.
\end{equation}
Such an isomorphism will be explicitly presented in equation \eqref{eq:NCFourier}.

\paragraph{Mean position of a coherent state.}
The expectation value of the non-commutative position operators on a coherent state $\Ket{\mathbbvar{z}}$ can be found by
\begin{equation}
    \braket{\mathbbvar{z}|\hat{x}^\mu|\mathbbvar{z}} \,=\, \sqrt{2\pi\hbar}\;\mathfrak{Re}(\mathbbvar{z}^\mu) \,=:\, x^\mu, \qquad \braket{\mathbbvar{z}|\hat{\tilde{x}}_\mu|\mathbbvar{z}} \,=\, \sqrt{2\pi\hbar}\;\mathfrak{Im}(\mathbbvar{z}^\mu) \,=:\, \tilde{x}_\mu.
\end{equation}
The doubled vector $\mathbbvar{x}^M=(x^\mu,\,\tilde{x}_\mu)$ is the \textit{mean position} of the coherent state $\ket{\mathbbvar{z}}$ on the doubled space $\mathbb{R}^{2d}$, also known as quasi-coordinate vector. It is important to remark that $\mathbbvar{x}^M$ are not coordinates, i.e. they are not eigenvalues of the operators $\hat{\mathbbvar{x}}^M$. Thus, by working with coherent states $\ket{\mathbbvar{z}}$ with mean position $\mathbbvar{x}^M$, we can bypass the problem of not being able to work with eigenstates of the position operators. In general, any operator $\hat{f}(\hat{\mathbbvar{x}}^M)$ can be expressed as a function of the mean positions of a coherent state by $F({\mathbbvar{x}}^M) := \braket{\mathbbvar{z}|\hat{f}(\hat{\mathbbvar{x}}^M)|\mathbbvar{z}}$.

\paragraph{Minimal uncertainty.} Coherent states minimize the uncertainty between a coordinate operator of the doubled space and its T-dual, i.e.
\begin{equation}
    \Delta x^\mu\, \Delta\tilde{x}_\nu=\frac{\pi\ell^2_s}{2}\delta^\mu_{\;\nu}.
\end{equation}
The coherent states of the quantum configuration space can then be interpreted as states which are approximately localised at a point $\mathbbvar{x}^M$ of the doubled space, the mean position. 

\subsection{Free particles on the doubled space}

\paragraph{Plane waves.}
The mean value of a plane wave operator on a coherent state is given by
\begin{equation}
    \Braket{\mathbbvar{z}|\exp\!\left(\frac{i}{\hbar}\mathbbvar{p}_M\hat{\mathbbvar{x}}^M\right)|\mathbbvar{z}} \;=\; \exp\!\left(\frac{i}{\hbar}\mathbbvar{p}_M\mathbbvar{x}^M-\frac{\pi\alpha'}{4\hbar}\delta^{MN}\mathbbvar{p}_M\mathbbvar{p}_N \right).
\end{equation}
This can be immediately proved by defining the complex momentum operators
\begin{equation}
    \begin{aligned}
    \hat{\mathbbvar{p}}_\mu^+ \;&=\; \sqrt{\frac{\pi}{{2\hbar}}}\left(\hat{p}_\mu + i \hat{\tilde{p}}^\mu\right), \\
    \hat{\mathbbvar{p}}^{-}_\mu \;&=\; \sqrt{\frac{\pi}{{2\hbar}}}\left(\hat{p}_\mu - i \hat{\tilde{p}}^\mu\right) ,
    \end{aligned}
\end{equation}
and by applying the Baker–Campbell–Hausdorff formula as follows:
\begin{equation}
    \begin{aligned}
    \exp\!\left(\frac{i}{\hbar}\mathbbvar{p}_M\hat{\mathbbvar{x}}^M\right) \;&=\; \exp\!\left(i\mathbbvar{p}_\mu^+\hat{\mathbbvar{z}}^{\dagger\mu} + i\mathbbvar{p}_\mu^-\hat{\mathbbvar{z}}^{\mu} \right) \\
    \;&=\; \exp\!\left(i\mathbbvar{p}_\mu^+\hat{\mathbbvar{z}}^{\dagger\mu} \right) \exp\!\left(i\mathbbvar{p}_\mu^-\hat{\mathbbvar{z}}^{\mu} \right) \exp\!\left(-\frac{\pi\alpha'}{4\hbar}\delta^{MN}\mathbbvar{p}_M\mathbbvar{p}_N \right).
    \end{aligned}
\end{equation}

\paragraph{The Hilbert space of a free particle.}
The subspace $L_\mathbbvar{p}=\mathrm{Span}(\mathbbvar{p}^M)\subset\mathcal{P}$ is Lagrangian. This can be immediately understood by writing the symplectic form in Darboux coordinates as $\bbomega = \di\mathbbvar{p}_M\wedge \di \mathbbvar{q}^M$. Therefore we can express our Hilbert space $\mathfrak{H}$ as the space of complex $\mathrm{L}^2$-functions on $L_\mathbbvar{p}$, i.e. as
\begin{equation}
    \mathfrak{H} \;\cong\; \mathrm{L}^{\!2}(L_{\mathbbvar{p}};\mathbb{C})
\end{equation}
Since the doubled momenta commute, i.e. $[\hat{\mathbbvar{p}}_M,\hat{\mathbbvar{p}}_N] \,=\,0$, we can define a basis of eigenstates $\{\Ket{\mathbbvar{p}}\}_{\mathbbvar{p}\in L_\mathbbvar{p}}\subset \mathfrak{H}$ of the doubled momentum operators $\hat{\mathbbvar{p}}_M$ without the problems encountered with the doubled position operator. These satisfy $\hat{\mathbbvar{p}}_M\Ket{\mathbbvar{p}} = \mathbbvar{p}_M\Ket{\mathbbvar{p}}$ for any $M=1,\dots,2d$. If we choose the basis of coherent states $\left\{\Ket{\mathbbvar{z}}\right\}_{\mathbbvar{z}\in\mathbb{C}^n}$, a doubled momentum eigenstate can then be expressed by
\begin{equation}\label{eq:NCfreeparticle}
    \Braket{\mathbbvar{z}|\mathbbvar{p}} \;=\; \exp\!\left(\frac{i}{\hbar}\mathbbvar{p}_M\mathbbvar{x}^M-\frac{\pi\alpha'}{4\hbar}\delta^{MN}\mathbbvar{p}_M\mathbbvar{p}_N \right)
\end{equation}
where $\mathbbvar{x}^M$ is the mean position of the coherent state $\Ket{\mathbbvar{z}}$. This can be interpreted as the expression of a free particle state on the doubled space, where we are using the mean position as a variable.

\paragraph{Strings are waves.}
Interestingly, if we choose the basis $\Ket{x,w}$ with $w^\mu=\tilde{p}^\mu$, which diagonalizes the commuting operators of the physical position $\hat{x}^\mu$ and winding $\hat{\tilde{p}}^\mu$, a doubled momentum eigenstate can be expressed just as a free particle in the wave-function on the physical space
\begin{equation}\label{eq:planewave}
    \Braket{x,w|\mathbbvar{p}} \;=\; \frac{1}{(2\pi\hbar)^{\frac{n}{2}}}\exp\!\left(\frac{i}{\hbar} p_\mu x^\mu \right)
\end{equation}
where now $x^\mu$ are proper eigenvalues of the position operator.
Analogously, in the T-dual frame $\Ket{\tilde{x},\tilde{w}}$ with $\tilde{w}_\mu=p_\mu$, we recover a free particle on the T-dual space
\begin{equation}
    \Braket{\tilde{x},\tilde{w}|\mathbbvar{p}} \;=\; \frac{1}{(2\pi\hbar)^{\frac{n}{2}}}\exp\!\left(\frac{i}{\hbar} \tilde{p}^\mu \tilde{x}_\mu \right)
\end{equation}
The interpretation of this fact is that a free particle, i.e. a plane wave, on the doubled space with fixed doubled momentum $\mathbbvar{p}_M = (p_\mu,\,\tilde{p}^\mu)$ can be quantum-mechanically interpreted as 
\begin{itemize}
    \item a free string on the physical space with fixed momentum $p_\mu$ and winding $w^\mu = \tilde{p}^\mu$,
    \item a free string on the T-dual space with fixed momentum $\tilde{p}^\mu$ and winding $\tilde{w}_\mu = p_\mu$.
\end{itemize}
As classically derived in \cite{BBR14}, this implies that a plane wave with doubled momentum $\mathbbvar{p}_M = (p_\mu,\,0)$ on the doubled space reduces to a plane wave with momentum $p_\mu$ on the physical space and one with $\mathbbvar{p}_M = (0,\,\tilde{p}^\mu)$ reduces to a standing string with winding $w^\mu=\tilde{p}^\mu$.

\paragraph{Probability distribution.}
Let us calculate the probability distribution of the wave-function \eqref{eq:NCfreeparticle} of a free particle on the doubled space:
\begin{equation}
    \left|\Braket{\mathbbvar{z}|\mathbbvar{p}}\right|^2 \;=\; \left(\frac{\alpha'}{2\hbar}\right)^{\!\!n}\exp\!\left(-\frac{\pi\alpha'}{2\hbar}\delta^{MN}\mathbbvar{p}_M\mathbbvar{p}_N\right).
\end{equation}
Hence, the probability of measuring the doubled momentum $\mathbbvar{p}_M$ (or equivalently a string with momentum $p_\mu$ and winding $\tilde{p}^\mu$) is not uniform, but it exponentially decays far from zero.

\paragraph{Coherent state as superposition of strings.}
Now, the eigenstate $\Ket{\mathbbvar{p}}$ can be interpreted as a free string for which we know with certainty the momentum $p_\mu$ and the winding number $w^\mu=\tilde{p}^\mu$. The equation \eqref{eq:NCfreeparticle} can be immediately interpreted as the expansion of a coherent state $\Ket{\mathbbvar{z}}$ in the basis $\Ket{\mathbbvar{p}}$, i.e.
\begin{equation}\label{eq:NCFourier}
    \Ket{\mathbbvar{z}} \;=\; \int\frac{\di^{2n}\mathbbvar{p}}{(2\pi\hbar)^{2n}}\exp\!\left(\frac{i}{\hbar}\mathbbvar{p}_M\mathbbvar{x}^M-\frac{\pi\alpha'}{4\hbar}\delta^{MN}\mathbbvar{p}_M\mathbbvar{p}_N \right) \Ket{\mathbbvar{p}}
\end{equation}
where $\mathbbvar{x}^M$ is the mean doubled position of the coherent state $\Ket{\mathbbvar{z}}$.

\paragraph{Hamiltonian as number operator.}
Observe that the Hamiltonian operator is given by
\begin{equation}
    \hat{H} \;=\; \mathcal{H}^{MN}\hat{\mathbbvar{p}}_M\hat{\mathbbvar{p}}_N
\end{equation}
Let us consider the simple case where the generalised metric is trivial, i.e. $\mathcal{H}^{MN}=\delta^{MN}$.
We can then define a number operator $\hat{N}_\mu \,:=\, \hat{\mathbbvar{p}}_\mu^-\hat{\mathbbvar{p}}_\mu^+$ for any fixed $\mu=1,\dots,n$ and the total number operator as a sum $\hat{N}= \sum_{\mu=1}^{n}\hat{N}_\mu$. What we obtain is that the Hamiltonian is proportional to the number operator by $\hat{H}=\frac{2}{\pi}\hbar\hat{N}$.

\subsection{Minimal scale of the doubled space}

\paragraph{Non-commutative Fourier transform.}
Let us consider a general string state $\Ket{\psi}\in\mathfrak{H}$. We can express this state as a wave function $\psi(\mathbbvar{p})=\Braket{\mathbbvar{p}|\psi}$ on the momentum space. Thus, if we want to express it in the coherent states basis $\Ket{\mathbbvar{z}}$, we need to use equation \eqref{eq:NCFourier} as follows:
\begin{equation}
    \Braket{\mathbbvar{z}|\psi} \;=\; \int\frac{\di^{2n}\mathbbvar{p}}{(2\pi\hbar)^{2n}}\exp\!\left(\frac{i}{\hbar}\mathbbvar{p}_M\mathbbvar{x}^M-\frac{\pi\alpha'}{4\hbar}\delta^{MN}\mathbbvar{p}_M\mathbbvar{p}_N \right) \Braket{\mathbbvar{p}|\psi}.
\end{equation}
Now we can transform wave-functions $\psi(\mathbbvar{p}):=\Braket{\mathbbvar{p}|\psi}$ on the doubled momentum space to wavefunctions $\psi(\mathbbvar{x}):=\Braket{\mathbbvar{z}|\psi}$ expressed in the  basis of the coherent states. (Here $\mathbbvar{x}^M$ denotes the mean position of $\Ket{\mathbbvar{z}}$ and is not a coordinate.) This is effectively a non-commutative version of the Fourier transform. \vspace{0.2cm}

\noindent Let us now mention an example of this non-commutative Fourier transform which is useful to develop some intuition about the non-commutative nature of the doubled space. Let us choose a wave-function $\psi(\mathbbvar{p})=1$ on the doubled momentum space, which, in some sense, means that the doubled momentum is maximally spread. The transformation \eqref{eq:NCFourier}, applied to $\psi(\mathbbvar{p})=1$, gives
\begin{equation}
\begin{aligned}
    \psi(\mathbbvar{x}) \;&=\; \int\frac{\di^{2n}\mathbbvar{p}}{(2\pi\hbar)^{2n}}\exp\!\left(\frac{i}{\hbar}\mathbbvar{p}_M\mathbbvar{x}^M-\frac{\pi\alpha'}{4\hbar}\delta^{MN}\mathbbvar{p}_M\mathbbvar{p}_N \right) \\[0.1cm]
    \;&=\; \frac{1}{(\pi^2\hbar\alpha')^n}\exp\Bigg(-\frac{\big|\mathbbvar{x}^M\big|^2}{\pi\hbar\alpha'}\Bigg),
    \end{aligned}
\end{equation}
which is a Gaussian distribution on the doubled space and not a delta function. This, on an intuitive level, means that, even if the doubled momentum is maximally spread, the uncertainty on the doubled coordinates cannot be zero. This is because each couple of T-dual coordinates can shrink only to a minimal area proportional to $\ell^2_s=\hbar\alpha'$. This is an interesting manifestation of the fuzziness of doubled space between physical and T-dual coordinates, which is parametrised by $\alpha'$.

\paragraph{Amplitude between coherent states.}
Let us consider two coherent states $\Ket{\mathbbvar{z}_1}$ and $\Ket{\mathbbvar{z}_2}$, respectively with mean position $\mathbbvar{x}_1^M$ and $\mathbbvar{x}_2^M$. We want now to calculate the scattering amplitude $\Braket{\mathbbvar{z}_2|\mathbbvar{z}_1}$ between such states\footnote{We thank Kevin T. Grosvenor for extremely helpful discussion, which led to the improvement of this section in the manuscript \cite{Alfonsi:2021uwh}. In particular, the calculation of the amplitude between coherent states will follow the calculation proposed in \cite{Kev21}.}.
First, notice that the following identity holds:
\begin{equation}
    \exp\Bigg(-\frac{\delta_{\mu\nu}}{\alpha'}\mathbbvar{z}^\mu\hat{\mathbbvar{z}}^{\dagger\nu}\Bigg) \hat{\mathbbvar{z}}^\lambda \exp\Bigg(+\frac{\delta_{\mu\nu}}{\alpha'}\mathbbvar{z}^\mu\hat{\mathbbvar{z}}^{\dagger\nu}\Bigg) \;=\; \hat{\mathbbvar{z}}^\lambda + \mathbbvar{z}^\lambda.
\end{equation}
By combining this identity with the definition of coherent state, we get the equation
\begin{equation}
    \Braket{\mathbbvar{z}_2|\mathbbvar{z}_1} \;=\; \exp\Bigg(-\frac{\big|\mathbbvar{z}_2|^2+|\mathbbvar{z}_1\big|^2}{2\alpha'}\Bigg) \Braket{0| \exp\Bigg(\frac{\delta_{\mu\nu}}{\alpha'}\bar{\mathbbvar{z}}^\mu_2\hat{\mathbbvar{z}}^{\nu}_2\Bigg) \exp\Bigg(\frac{\delta_{\mu\nu}}{\alpha'}\mathbbvar{z}^\mu_1\hat{\mathbbvar{z}}^{\dagger\nu}_1\Bigg) |0}.
\end{equation}
By using the Baker–Campbell–Hausdorff formula, we have
\begin{equation*}
    \exp\Bigg(\frac{\delta_{\mu\nu}}{\alpha'}\bar{\mathbbvar{z}}^\mu_2\hat{\mathbbvar{z}}^{\nu}_2\Bigg) \exp\Bigg(\frac{\delta_{\mu\nu}}{\alpha'}\mathbbvar{z}^\mu_1\hat{\mathbbvar{z}}^{\dagger\nu}_1\Bigg) \;=\; \exp\Bigg(\frac{\delta_{\mu\nu}}{\alpha'}\mathbbvar{z}^\mu_1\hat{\mathbbvar{z}}^{\dagger\nu}_1\Bigg) \exp\Bigg(\frac{\delta_{\mu\nu}}{\alpha'}\bar{\mathbbvar{z}}^\mu_2\hat{\mathbbvar{z}}^{\nu}_2\Bigg) \exp\Bigg(\frac{\delta_{\mu\nu}}{\alpha'}\bar{\mathbbvar{z}}^\mu_2\mathbbvar{z}^{\nu}_1\Bigg).
\end{equation*}
From this, we finally find the amplitude:
\begin{equation}
    \Braket{\mathbbvar{z}_2|\mathbbvar{z}_1} \;=\; \exp\Bigg(-\frac{\big|\mathbbvar{z}_2-\mathbbvar{z}_1\big|^2 + \delta_{\mu\nu}(\mathbbvar{z}_2^\mu\bar{\mathbbvar{z}}_1^\nu - \bar{\mathbbvar{z}}_2^\mu\mathbbvar{z}_1^\nu)}{2\alpha'}\Bigg).
\end{equation}
Thus, we immediately get the result that the amplitude between two coherent states with different mean positions $\mathbbvar{x}_1^M$ and $\mathbbvar{x}_2^M$ is given by
\begin{equation}\label{eq:amplitudez}
    \Braket{\mathbbvar{z}_2|\mathbbvar{z}_1} \;=\; \exp\Bigg(-\frac{\big|\mathbbvar{x}_2-\mathbbvar{x}_1\big|^2 - i\omega_{MN}\mathbbvar{x}_2^M\mathbbvar{x}_1^N}{\pi\hbar\alpha'}\Bigg).
\end{equation}
Thus, the ordinary Dirac delta function $\braket{x_2|x_1}\sim\delta(x_2-x_1)$ between eigenstates of the coordinates operator of commutative Quantum Mechanics is replaced by the function \eqref{eq:amplitudez}, whose width is proportional to string length scale $\ell_s=\sqrt{\hbar\alpha'}$.
The squared amplitude between two coherent states is, then, the Gaussian distribution
\begin{equation}
    |\Braket{\mathbbvar{z}_2|\mathbbvar{z}_1}|^2 \;=\; \exp\bigg(-\frac{2}{\pi\hbar\alpha'}\big|\mathbbvar{x}_2-\mathbbvar{x}_1\big|^2\bigg).
\end{equation}
Physically, this means that the probability is high if the distance between the mean positions $\mathbbvar{x}_1$ and $\mathbbvar{x}_2$ of the respective coherent states $\Ket{\mathbbvar{z}_1}$ and $\Ket{\mathbbvar{z}_2}$ is smaller than $\sqrt{\pi}\ell_s$.

\paragraph{The $\alpha'\rightarrow 0$ limit.}
If we take the limit $\alpha'\rightarrow 0$ the fuzziness of the doubled space disappears. 
The non-commutative Heisenberg algebra of quantum observables reduces to an ordinary commutative $4n$-dimensional Heisenberg algebra, whose commutation relations are given by 
\begin{equation}
    \lim_{\alpha'\rightarrow\,0}\,[\hat{\mathbbvar{x}}^M,\hat{\mathbbvar{x}}^N] \,=\, 0 , \;\quad \lim_{\alpha'\rightarrow\,0}\,[\hat{\mathbbvar{x}}^M,\hat{\mathbbvar{p}}_N] \,=\, i\hbar\delta^{M}_{\;\, N}, \;\quad \lim_{\alpha'\rightarrow\,0}\,[\hat{\mathbbvar{p}}_M,\hat{\mathbbvar{p}}_N] \,=\,0.
\end{equation}
Consequently the minimal uncertainty in measuring a coordinate and its dual vanishes.
The basis of coherent states $\Ket{\mathbbvar{z}}$ reduces to a basis of eigenstates $\Ket{x,\tilde{x}}$ of the position operator $\hat{\mathbbvar{x}}$, which are now well-defined.
Moreover, the scattering amplitudes shrink to
\begin{equation}
    \lim_{\alpha'\rightarrow\,0}\Braket{\mathbbvar{z}_2|\mathbbvar{z}_1} \;=\; \delta(\mathbbvar{x}_2-\mathbbvar{x}_1).
\end{equation}
In the limit $\alpha'\rightarrow 0$, the quantum mechanics on the doubled space becomes ordinary commutative quantum mechanics on a $2n$-dimensional spacetime.

\subsection{Polarisation of coherent states in a T-duality frame}

Let us consider on our Hilbert space the basis $\{\Ket{x,w}\}_{(x,w)\in L}\subset\mathfrak{H}$, which corresponds to the T-duality frame given by the Lagrangian subspace $L\subset\mathcal{P}$ with coordinates $(x^\mu,w^\mu)$. Recall that, given a string state $\Ket{\psi}\in\mathfrak{H}$, we can express it as a wave-function on the Lagrangian subspace $L\subset\mathcal{P}$ by $\psi_w(x) \,:=\, \Braket{x,w|\psi}$, where $w^\mu:=\tilde{p}^\mu$ is the generalised winding number. Thus $|\psi_w(x)|^2$ can be interpreted as the probability of measuring a string at the point $x^\mu$ on physical spacetime with winding number $w^\mu$. Now, we want express a coherent state $\Ket{\mathbbvar{z}}$ in this basis. In other words we want to calculate
\begin{equation}
    \psi^{\mathrm{coh}}_w(x) \;:=\; \Braket{x,w|\mathbbvar{z}}.
\end{equation}
To do that, we can use the fact that $\Ket{\mathbbvar{p}}$ are a complete basis for the Hilbert space $\mathfrak{H}$ and write
\begin{equation}\label{eq:cohchangeofbasis}
    \Braket{x,w|\mathbbvar{z}} \;=\; \int\frac{\di^{2n}\mathbbvar{p}'}{(2\pi)^{2n}} \Braket{x,w|\mathbbvar{p}'} \Braket{\mathbbvar{p}'|\mathbbvar{z}}.
\end{equation}
Let us now use the notation $\braket{\mathbbvar{x}^M}:=\braket{\mathbbvar{z}|\hat{\mathbbvar{x}}^M|\mathbbvar{z}}$ for the mean position of a coherent sate $\Ket{\mathbbvar{z}}$ and again $(x^\mu,w^\mu)$ for the coordinates of the Lagrangian subspace $L\subset\mathcal{P}$ where the polarised wave-function lives. Let us use the expressions \eqref{eq:NCfreeparticle} and \eqref{eq:planewave} of a free particle in the doubled space to calculate the intermediate terms
\begin{equation}
    \begin{aligned}
    \Braket{x,w|\mathbbvar{p}'} \;&=\; \exp\!\left(\frac{i}{\hbar}p_\mu^{\prime}x^\mu\right)\delta(w - \tilde{p}^{\prime}) \\
    \Braket{\mathbbvar{p}'|\mathbbvar{z}} \;&=\; \exp\!\left(-\frac{i}{\hbar}\mathbbvar{p}'_M\!\Braket{\mathbbvar{x}^M}-\frac{\pi\alpha'}{4\hbar}\delta^{MN}\mathbbvar{p}'_M\mathbbvar{p}'_N \right).
    \end{aligned}
\end{equation}
Hence the integral \eqref{eq:cohchangeofbasis} becomes a Fourier transform in the physical momentum $p_\mu'$ only
\begin{equation*}
    \Braket{x,w|\mathbbvar{z}} \,=\, \left(\int\frac{\di^{n}p'}{(2\pi)^{n}} \exp\!\left(-\frac{i}{\hbar}p'_\mu\!(x^\mu - \Braket{x^\mu}) - \frac{\pi\alpha'}{4\hbar}\left|p_\mu'\right|^2 \right)\right) \exp\!\left(\frac{i}{\hbar}w^\mu\!\Braket{\tilde{x}_\mu} - \frac{\pi\alpha'}{4\hbar}\left|w^\mu\right|^2 \right).
\end{equation*}
Thus we obtain the following wave-function:
\begin{equation}\label{eq:cohwavefunction}
    \psi^{\mathrm{coh}}_w(x) \;=\; \exp\!\left(-\frac{\left|x^\mu-\braket{x^\mu}\right|^2}{\pi\hbar\alpha'} \right) \exp\!\left(\frac{i}{\hbar}w^\mu\!\Braket{\tilde{x}_\mu} - \frac{\pi\alpha'}{4\hbar}\left|w^\mu\right|^2 \right).
\end{equation}
We notice that the first term of this wave-function is a Gaussian on the physical position space and that the second term contains an exponential cut-off for large winding numbers.

\paragraph{Probability distribution.}
The probability of measuring a string with position $x^\mu$ and winding number $w^\mu$, for a given coherent state $\Ket{\mathbbvar{z}}$ with mean doubled position $\Braket{\mathbbvar{x}^M}$, will be immediately given by
\begin{equation}
    \left|\psi^{\mathrm{coh}}_w(x)\right|^2 \;=\; \exp\!\left(-\frac{2}{\pi\hbar\alpha'}\left|x^\mu-\braket{x^\mu}\right|^2  - \frac{\pi\alpha'}{2\hbar}\left|w^\mu\right|^2 \right).
\end{equation}
This probability distribution exponentially decays by going away from the mean position $\Braket{x^\mu}$ on the physical position space and from zero on the winding number space. 

\paragraph{Limit $\alpha'\rightarrow 0$.}
It is immediate to notice that, in the limit $\alpha'\rightarrow 0$, the probability distribution spreads in the winding space and localizes in the physical position space. In other words, our probability distribution shrinks to a Dirac delta $\left|\psi^{\mathrm{coh}}_w(x)\right|^2=\delta(x-\Braket{x})$.

\paragraph{Change of T-duality frame.}
Now we want to find the T-dual wave-function of \eqref{eq:cohwavefunction}. To do so, we only need to express the same coherent state $\Ket{\mathbbvar{z}}\in\mathfrak{H}$ in another basis of our Hilbert space, the basis $\Ket{\tilde{x},\tilde{w}}$ corresponding to the complementary Lagrangian subspace $\widetilde{L}\subset\mathcal{P}$. In other words we must calculate $\widetilde{\psi}^{\mathrm{coh}}_{\tilde{w}}(\tilde{x}) \;:=\; \Braket{\tilde{x},\tilde{w}|\mathbbvar{z}}$. 
We immediately obtain the following wave-function on the dual Lagrangian subspace $\widetilde{L}\subset\mathcal{P}$
\begin{equation}
    \widetilde{\psi}^{\mathrm{coh}}_{\tilde{w}}(\tilde{x}) \;=\; \exp\!\left(-\frac{\left|\tilde{x}_\mu-\!\braket{\tilde{x}_\mu}\right|^2}{\pi\hbar\alpha'} \right) \exp\!\left(\frac{i}{\hbar}\tilde{w}_\mu\Braket{x^\mu} - \frac{\pi\alpha'}{4\hbar}\left|\tilde{w}_\mu\right|^2 \right),
\end{equation}
where the role of the physical and T-dual coordinates is exchanged.

\section{Metaplectic structure}

Let us focus on the doubled space $\mathcal{M}$. We observed that it comes, at least locally, equipped with a canonical symplectic form $\varpi$.
Let now assume that $(\mathcal{M},\varpi)$ is simply a $2n$-dimensional symplectic manifold and $L\subset T\mathcal{M}$ be a Lagrangian subbundle.

\paragraph{The metaplectic structure.}
The \textit{metaplectic group} $Mp(2n,\mathbb{R})$ is the universal double cover of the symplectic group $Sp(2n,\mathbb{R})$. It is, thus, given by a group extension of the form
\begin{equation}
    \begin{tikzcd}[row sep=7ex, column sep=5ex]
    0\arrow[r, hook] & \mathbb{Z}_2 \arrow[r, hook] & Mp(2n,\mathbb{R})\arrow[r, two heads] & Sp(2n,\mathbb{R}). \arrow[r, two heads] & 0
    \end{tikzcd}.
\end{equation}
A \textit{metaplectic structure} on a symplectic manifold $(\mathcal{M},\varpi)$ is defined as the lift of the structure group $Sp(2n,\mathbb{R})$ of the bundle $T\mathcal{M}$ along the group extension $Mp(2n,\mathbb{R})\twoheadrightarrow Sp(2n,\mathbb{R})$. \vspace{0.2cm}

\noindent There is a lemma (see \cite{WeiGQ}) which states that $T\mathcal{M}$ admits a metaplectic structure if and only if $L$ admits a metalinear structure.
Another result states \cite{WeiGQ} that the existence of a metalinear structure on a bundle $E$ is equivalent to the existence of the square root bundle $\sqrt{\mathrm{det}(E)}$.
By putting these two lemmas together we obtain that $T\mathcal{M}$ admits a metaplectic structure if and only if $\sqrt{\mathrm{det}({L})}$ exists. \vspace{0.15cm}

\noindent The existence of a metaplectic structure is intimately linked to the definition of the canonical $\mathrm{Spin}(d,d)$ spinor bundle
\begin{equation}
    S_\mathcal{M} \;=\; \wedge^\bullet \widetilde{L} \otimes \sqrt{\mathrm{det}({L})}.
\end{equation}
If the Lagrangian subbundle $L$ is integrable, there exists a submanifold $M\subset\mathcal{M}$ such that $L=TM$, i.e. the physical spacetime. In this case, the  canonical $\mathrm{Spin}(d,d)$ spinor bundle is isomorphic to the spinor bundle of generalised geometry on $M$, which is defined in \cite{Gualtieri:2007ng}. Thus, the isomorphism $L\oplus L^\ast$ by a B-shift $L\oplus\widetilde{L} \xrightarrow{ e^{-B} } L\oplus L^\ast$ can be immediately extended to an isomorphism
\begin{equation}
    S_\mathcal{M} \;\cong\; \wedge^\bullet T^\ast M \otimes \sqrt{\mathrm{det}({TM})},
\end{equation}
given by the untwist $\Phi \mapsto e^{-B}\wedge\Phi$ on polyforms $\Phi\in \wedge^\bullet \widetilde{L} \otimes \sqrt{\mathrm{det}({L})}$. Notice that this recovers a construction which is analogous to \cite{Wald08}.

\paragraph{The quantum Hilbert space.}
The physical necessity for the existence of  $\sqrt{\mathrm{det}({L})}$ is that it is this measure that is used to construct the quantum  Hilbert space. In half-form quantisation, one thinks of a state as the combination of the wavefunction with the half-form used to construct its norm. \vspace{0.15cm}

\noindent Thus, the quantum Hilbert space of this symplectic manifold, which will be:
\begin{equation}
    \mathfrak{H} \;=\; \bigg\{\psi\in\Gamma\big(\mathcal{M},\,\mathcal{E}\otimes\sqrt{\mathrm{det}({L})}\big) \;\bigg|\; \nabla_V\psi=0 \;\; \forall V\in L \bigg\}.
\end{equation}
Let us now consider sections of the form $e^{-\phi}\sqrt{\mathrm{vol}_M} \in \Gamma\big(\mathcal{M},\sqrt{\mathrm{det}({L})}\big)$, where the top form is the Riemannian volume form $\mathrm{vol}_M := \sqrt{\mathrm{det}(g)}\, \di x^1 \wedge \dots \wedge \di x^n$ and $\phi\in\Coo(M)$ is just a function. Any section $\Ket{\psi}\in\mathfrak{H}$ can be uniquely written as: 
\begin{equation}
\Ket{\psi} = \psi e^{-\phi} \otimes \sqrt{\mathrm{vol}_M} \,.
\end{equation}
With $\psi$ and $\phi$ obeying the polarisation condition. \vspace{0.15cm}

\noindent For $A\in GL(n;\mathbb{R})$ acting on the bundle $L$, we have that sections of the square root bundle transform accordingly by
\begin{equation}
    e^{-\phi}\sqrt{\mathrm{vol}_M} \;\longmapsto\; \sqrt{\mathrm{det}(A)}\,(e^{-\phi}\sqrt{\mathrm{vol}_M})
\end{equation}
Consider a state $\ket{\psi}\in\mathfrak{H}$. Let us call simply $\psi$ the corresponding wave-function. We, thus, have a Hilbert product given by
\begin{equation}
    \Braket{\psi|\psi} \;=\; \int_M \psi^\dagger\psi \,\sqrt{\mathrm{det}(g)}\, e^{-2\phi} \di x^1 \wedge \dots \wedge \di x^n
\end{equation}
where $\sqrt{\mathrm{det}(g)}\, e^{-2\phi}$ is nothing but the string frame measure and it is T-duality invariant.
By following the literature we can, define a T-duality invariant dilaton by
\begin{equation}
    d \;:=\; \phi - \frac{1}{2}\ln{\mathrm{det}(g)},
\end{equation}
so that we can rewrite the measure as $\sqrt{\mathrm{det}(g)}\, e^{-2\phi} = e^{-2d}$. 
Now, notice that a Hilbert product $\Braket{\psi|\psi}$ does not depend on the choice of polarisation and, therefore, it must be invariant under change of T-duality frame. Under the symplectomorphism encoding T-duality we have the volume half form transforming by
\begin{equation}
    e^{-d}\sqrt{\di {x}^1 \wedge \dots \wedge \di {x}^n} \;\mapsto\; e^{-d}\sqrt{\di \tilde{x}^1 \wedge \dots \wedge \di \tilde{x}^n}
\end{equation}
Thus, we can express the same state $\ket{\psi}\in\mathfrak{H}$ as an $\widetilde{L}$-polarised section $\tilde{\psi}e^{-\tilde{\phi}}\otimes \sqrt{\mathrm{vol}_{\widetilde{M}}}$, where the dual measure is $\mathrm{vol}_{\widetilde{M}} := \sqrt{\mathrm{det}(\tilde{g})}\,  \di \tilde{x}^1 \wedge \dots \wedge \di \tilde{x}^n$. In this T-duality frame the Hilbert product will immediately have the following form:
\begin{equation}
    \Braket{\psi|\psi} \;=\; \int_{\widetilde{M}} \tilde{\psi}^\dagger\tilde{\psi} \,\sqrt{\mathrm{det}(\tilde{g})}\, e^{-2\tilde{\phi}} \di \tilde{x}^1 \wedge \dots \wedge \di \tilde{x}^n
\end{equation}
Thus the dilaton transformation arises from the transformation of the measure in the half-form quantisation of the string.

\paragraph{The Metaplectic correction to observables}
There is one further effect associated to the Metaplectic structure of quantisation. When we move to the representation of observables the operators now act on states in $\mathfrak{H}$ i.e. $ \psi e^{-\phi} \otimes \sqrt{\mathrm{vol}_M}$ not just on the wavefunctions $\psi$. Practically that means there may be in additional contribution to an operator given by the Lie derivative generated by the vector field associated to the observable acting on the half form. Contributions of this type occur with holomorphic polarisations in which case the Hamiltonian operator is shifted by $1/2$. For the simple harmonic oscillator in quantum mechanics this is just the usual "zero-point" energy shift. In this context, the Hamiltonian constructed in section 5.24 would receive a zero-point shift. This would be relevant for T-fold type configurations where the space time moves between $x$ and $\tilde{x}$ spaces. Of course, we have only dealt with the bosonic string, it is a open question as to whether fermionic contributions might cancel this shift for the full superstring.

\subsection{The Maslov correction}

A related effect is the Maslov quantisation condition \cite{Arn67} (also known as Einstein–Brillouin–Keller quantisation) is
\begin{equation}
    \frac{1}{2\pi}\oint_{\gamma}\bbtheta \;=\; \hbar\left( n + \frac{\mu(\gamma)}{4} \right),
\end{equation}
where $n\in\mathbb{Z}$ and $\mu(\gamma)$ is the Maslov index of the loop $\gamma$.
Notice that the prequantisation condition $[\bbomega]\in H^2(\mathcal{P},\mathbb{Z})$ alone implies only that $\frac{1}{2\pi}\oint_{\gamma}\bbtheta = \hbar n$ for some integer $n\in\mathbb{Z}$. The Maslov quantisation condition adds an explicit correction to the quantisation procedure depending on the Maslov index of the loop. \vspace{0.2cm}

\noindent These metaplectic/Maslov type corrections really only appear when the polarisation is non-trivial by which we mean in the double field theory context a spacetime that moves between $x$ and $\tilde{x}$ spaces. One expects such a description is needed for a T-fold where no global T-duality frame exists. These subtle "quantum" effects will then change the string spectrum in the T-fold background. We leave the detailed study of the metaplectic/Maslov corrections for T-folds for future work.

\begin{savequote}[7.0cm]
No book can ever be finished.\\While working on it we learn just enough to find it immature the moment we turn away from it.
  \qauthor{--- Karl Popper}
\end{savequote}

\chapter{\label{ch:9-conclusion}Conclusion}

\minitoc

\section{Conclusion}

From a small set of assumptions we developed a global formulation of the geometry underlying Double Field Theory by directly generalising Kaluza-Klein principle to higher gauge fields. 
We posit that the total space of the bundle gerbe should be understood as the extended spacetime where Double Field Theory lives. In chapter \ref{ch:5}, the higher Kaluza-Klein hypothesis found an extremely interesting confirmation in the fact that a bundle gerbe can be covered by an atlas whose charts can be naturally identified with the coordinate charts of Double Field Theory. Crucially, this construction simultaneously geometrises both the Kalb-Ramond flux and T-duality from a single principle.\vspace{0.2cm} 

\noindent From our formalism, we recovered many previous relevant proposals of geometry of Double Field Theory and we clarified how they fit in a global geometric picture. Both in chapter \ref{ch:5} and \ref{ch:6} we illustrated that this proposal of doubled space naturally recovers para-Hermitian geometry \cite{Vai12,Vai13}, including the para-Hermitian structure of Drinfel’d doubles \cite{MarSza18,MarSza19}. We also showed that the infinitesimal symmetries of such doubled space are given by a symplectic Lie $2$-algebroid, which is equivalently a NQP-manifold, and it can be related to the construction by \cite{DesSae18, DesSae18x, DesSae19}.\vspace{0.2cm}

\noindent These geometric results are also able to shed new light on the nature of physical objects in String Theory. In ordinary Kaluza-Klein theory, a magnetic monopole of the gauge field can be embedded in a gravitational monopole on the extended spacetime: the Gross-Perry monopole. In chapter \ref{ch:5} we showed that the NS5-brane of String Theory is nothing but the direct generalisation of the Gross-Perry monopole to the bundle gerbe. This provides a new and clear geometric meaning to the NS5-brane and completes the idea initially proposed by \cite{BR14}. \vspace{0.2cm}

\noindent In chapter \ref{ch:6}, we showed that the dimensional reduction of the bundle gerbe recovers exactly Hull’s doubled torus bundles \cite{Hull07}, with their wanted topologies, monodromy cocycles and T-dualities \cite{BelHulMin07}. 
Moreover, in chapter \ref{ch:6} we derived the notion of global non-abelian T-fold by dimensionally reducing the bundle gerbe and we applied it to the underlying geometry of tensor hierarchies \cite{HohSam13KK}, i.e. the gauge field content of gauged supergravity, where duality generally appears promoted to a local symmetry. From this global version of the generalised Scherk-Schwarz reductions, for the very first time, it was possible to study the topology of some simple examples of tensor hierarchies.\vspace{0.2cm} 

\noindent In chapter \ref{ch:7} we moved some first steps in generalising our proposal to other Extended Field Theories. In particular, we defined a chart of the heterotic doubled space, the Type II super-double space, the exceptional space and, finally, of the super-exceptional space.
\vspace{0.2cm}

\noindent Finally, in chapter \ref{ch:8} we apply geometric quantisation to a closed string and we link it to the doubled space of Double Field Theory. A key result is the identification of the stringy effects linked to the noncommutativity of the doubled space controlled by the string length. The choice of polarisation in quantisation then becomes the choice of duality frame. Transformations between frames is then given geometrically by changing polarisations and constructing the non-local transforms acting on wavefunctions. 
The construction of a double coherent state gives a minimal distance state which we can examine from the point of view of traditional polarisations. Finally, the subtle metaplectic effects may have important consequences for quantising strings on T-folds.

\section{Discussion}

We showed that the current understanding of Extended Field Theory and, in particular for what it concerns its global properties, is far from complete. Moreover, several mathematically interesting results deserve to be investigated further.

\paragraph{Global Exceptional Geometry.}
Recall the definition of the atlas $\mathbb{R}^{1,10}_{\mathrm{exc}}\longrightarrow \mathfrak{m5brane}$ in chapter \ref{ch:7}, which we identified with a local chart of exceptional space.
As already argued by \cite[sec.$\,$9.2]{Arvanitakis:2018hfn}, the naturally expected structure generalising the fundamental $2$-form $\omega$ of para-Hermitian geometry to the exceptional case would generally be an almost $n$-plectic structure. Recently, \cite{Sakatani:2020umt} proposed a local generalisation of the Born $\sigma$-model of the string to the M-branes. These are equipped with $3$- and $6$-forms which appear to be closely related to the transgression of the higher field whose curvature comes from the dg-algebra \eqref{sphere}.
All these are strong hints that the correspondence between extended geometry and higher geometry via atlases can be well-defined for the exceptional cases too.

\paragraph{Relation with representation theory.}
In \cite{Cederwall:2017fjm, Cederwall:2018aab, Cederwall:2018kqk}, extended geometry has been studied in algebraic terms, in the light of representation theory. The extended/higher correspondence will then provide a complementary global geometric perspective to extended geometry, as well as new connections between higher geometry and representation theory.

\paragraph{Exceptional Drinfel'd Algebras.}
It would be interesting to clarify the global picture of the exceptional Drinfel’d algebras \cite{Sakatani:2019zrs, Malek:2019xrf, Blair:2020ndg, Musaev:2020nrt, Sakatani:2020wah, Malek:2020hpo}, which are Leibniz algebras generalising the Drinfel'd algebras of Double Field Theory to Exceptional Field Theory. Since the fluxes of M-theory appear to be formalised in the context of cohomotopy, it would be interesting to investigate the relation between exceptional Drinfel'd algebras and cohomotopy.

\paragraph{Global U-duality and moduli stack of global U-folds.}
A future natural direction will be extending the global formalisation of T-duality we proposed in chapter \ref{ch:6} to global U-duality, including Poisson-Lie U-duality, its underlying globally-defined tensor hierarchies and generalised correspondence spaces.
This can be achieved by by globalising the local chart of exceptional space to a complete atlas of the higher structure of M-theory and, then, by developing a global notion of generalised Scherk-Schwarz reductions.
The generalisation of this construction could lead to the definition of a cohomology theory for global Poisson-Lie U-folds, or, in other words a global classification of such spaces and their fluxes, geometric and non-geometric.
Finally, once the characteristic classes which classify the fluxes of M-theory are established, this procedure could lead to the definition of a notion of topological U-duality, which prescribes how such topological classes are mixed under U-duality.

\paragraph{Non-perturbative quantisation of M-theory.}
Since the non-perturbative quantisation of strings and branes can be achieved by higher geometric quantisation \cite{SaSza11,BSS16,BS16, FSS16} on prequantum bundle $n$-gerbes, the close relation we established between extended and higher geometries will have a profound impact on the problem of quantisation. Moreover, the global properties of the geometric involved structures play a fundamental role in higher geometric quantisation, just like in ordinary geometric quantisation. In this perspective, our proposal of global geometry of Extended Field Theory and duality will likely have an interesting role to play in this kind of non-perturbative quantisation. \vspace{0.2cm}

\noindent The results of chapter \ref{ch:8} lead to some further questions far outside the scope of this thesis. Usually the properties of double field theory are shared with exceptional field theory. Here though seems a mystery. If double field theory is just phase space and its subsequent quantisation then what is exceptional field theory. Is there some sense in which it can be thought of as a more general "quantisation" with the generalised "phase space" being related to the extended space. Spacetime would no longer be a Lagrangian submanifold of the extended space. Perhaps some clue is available in the construction of the basic states of theory as given in \cite{Berman:2014hna} where the branes were again momentum states in the extended space but now also combined with a type of  generalised monopole to give a self-dual configuration. Other mysterious properties of M-theory phase space have been noticed in \cite{Lust:2017bwq}. Other exotica that would be curious to explain from the phase space perspective would be the recently discovered non-Riemannian phase to Double and Exceptional Field Theory as discussed in \cite{Par17xx,Par18xx,Berman:2019izh,Park:2020ixf,Blair:2020gng,Gallegos:2020egk}; this is also somewhat of a mystery from the quantisation perspective. Any insight into such exotic backgrounds from the geometric quantisation approach proposed here would be very interesting and we leave it for future work.

\pdfbookmark[part]{Appendices}{toc}
\startappendices

\begin{savequote}[7.5cm]
{\textgreekfont  Πᾶν τὸ μάθημα γνώσεως ἕνεκα ἐπιτηδευόμενον.}

The knowledge at which geometry aims is knowledge of the eternal.
  \qauthor{--- Plato, \textit{Republic}}
\end{savequote}

\chapter{\label{app:1}Fundamentals of generalised geometry}

\minitoc


\noindent In this appendix we introduce fundamental notions in generalised complex geometry and theory of Courant algebroids.
For information about exceptional generalised geometry, which is not included in this brief discussion, we redirect to the seminal works \cite{Hull07, Wald08, Wald08E, Wald11, Wald12} and \cite{Bugden:2021wxg} for a recent formalisation.

\section{Generalised tangent bundle}

\begin{definition}[Generalised tangent bundle]
A \textit{generalised tangent bundle} $E\twoheadrightarrow M$ is a vector bundle which sits at the center of the following short exact sequence:
\begin{equation}
    \begin{tikzcd}[row sep=5ex, column sep=6ex] 
    T^\ast M \arrow[r, hook, "\rho^\ast"] & E \arrow[r, two heads, "\rho"] & TM,
\end{tikzcd}
\end{equation}
where $\rho:E\twoheadrightarrow TM$ is known as \textit{anchor} map.
\end{definition}

\begin{definition}[Splitting map]
A \textit{splitting map} is a bundle map
\begin{equation}
    \begin{tikzcd}[row sep=5ex, column sep=6ex] 
    T^\ast M \arrow[r, hook, "\rho^\ast"] & E \arrow[r, two heads, "\rho"] & TM\ar[l, "\omega", bend left=65, hook],
\end{tikzcd}
\end{equation}
so that we have the isomorphism of vector bundles
\begin{equation}\label{eq:isocourantlagebroids}
    \omega\oplus\rho^\ast :\; TM\oplus T^\ast M \;\xrightarrow{\;\;\;\cong\;\;\;}\; E.
\end{equation}
\end{definition}

\begin{definition}[Generalised vector]
A generalised vector is a section $V\in\Gamma(M,E)$.
\end{definition}

\begin{remark}[Components of a generalised vector]
Given a local patch $U_{(\alpha)}\subset M$, we can locally express a generalised vector $V\in \Gamma(M,E)$ by
\begin{equation}
    V \;=\; \begin{pmatrix}v_{(\alpha)} \\[0.6ex] \widetilde{v}_{(\alpha)} \end{pmatrix}
\end{equation}
where $v_{(\alpha)}\in\mathfrak{X}(U_\alpha)$ and $\widetilde{v}_{(\alpha)}\in\Omega^1(U_\alpha)$ and their patching conditions are given by
\begin{equation}
    \begin{aligned}
    v_{(\alpha)} \;&=\; v_{(\beta)}, \\
    \widetilde{v}_{(\alpha)} \;&=\; \widetilde{v}_{(\beta)} - \iota_{v_{(\alpha)}}\di\Lambda_{(\alpha\beta)},
    \end{aligned}
\end{equation}
on any overlap of patches $U_{(\alpha)}\cap U_{(\beta)} \subset M$. From the first patching condition, notice that $v\in\mathfrak{X}(M)$ is a globally defined vector field on $M$.
By using the isomorphism $\omega\oplus\rho^\ast: TM\oplus T^\ast M\xrightarrow{\;\cong\;} E$, we can express a generalised vector as $V=\omega\oplus\rho^\ast(v+\mu)$, where $v+\mu\in\Gamma(M,TM\oplus T^\ast M)$, as follows:
\begin{equation}
    V \;=\; \begin{pmatrix}v \\[0.6ex]  \mu + \iota_{v}B_{(\alpha)} \end{pmatrix},
\end{equation}
where $v\in\mathfrak{X}(M)$ and $\mu\in\Omega^1(M)$ are respectively a globally defined vector and $1$-form, while $B_{(\alpha)}\in\Omega^2(U_{(\alpha)})$ is a local $2$-form corresponding to $\omega$ and patched by the condition $B_{(\beta)} - B_{(\alpha)} = \di\Lambda_{(\alpha\beta)}$.
\end{remark}

\begin{definition}[Metric $\eta$ and generalised vielbein]
A metric $\eta$ is a symmetric bilinear form
\begin{equation}
    \eta : \;\Gamma(M,E)\times\Gamma(M,E)\; \longrightarrow \; \Coo(M)
\end{equation}
defined as follows. Let us first define the \textit{generalised vielbein} $E_A\in\Gamma(M,E)$ by 
\begin{equation}
    E_A \;:=\; \begin{cases} E_a=\begin{pmatrix} e_a \\ -\iota_{e_a}B_{(\alpha)} \end{pmatrix} & A=a, \\  \widetilde{E}^a=\begin{pmatrix} 0 \\ e^a \end{pmatrix} & A=d+a.\end{cases}
\end{equation}
Finally, the $\eta$-metric can be defined by the generalised tensor
\begin{equation}
    \eta \;=\; \eta_{AB}E^A \otimes E^B, \qquad \eta\;:=\; \begin{pmatrix} 0&1 \\ 1&0 \end{pmatrix}.
\end{equation}
\end{definition}

\begin{definition}[Generalised frame bundle]
The generalised frame bundle $\mathrm{Fr}(E)$ of a generalised tangent bundle $E$ is the principal $O(d,d)$-bundle defined by
\begin{equation}
    \mathrm{Fr}(E) \;:=\; \bigsqcup_{x\in M}\Big\{ \big(x,\{E_A\}_{A=1,\dots,2d}\big)\;\big|\; x\in M, \;\; \eta(E_A,E_B)=\eta_{AB} \Big\}
\end{equation}
on the base manifold $M$.
\end{definition}

\begin{remark}[Structure group]
The metric $\eta$ together with a choice of orientation reduces the structure group $O(d,d)$ of the generalised tangent bundle $E\twoheadrightarrow M$ to $SO(d,d)$.
\end{remark}

\begin{remark}[Automorphisms of the generalised tangent bundle]
The structure group preserving the $\eta$-metric on $E\twoheadrightarrow M$ and the orientation is $SO(d,d)\cong SO(E)$. Its Lie algebra will be
\begin{equation}
    \mathfrak{so}(E) \;=\; \big\{A\in\Gamma(M,E\otimes E^\ast)\;\big|\; \eta(A-,-)+\eta(-,A-)=0\big\}.
\end{equation}
This Lie algebra is isomorphic to
\begin{equation}
    \mathfrak{so}(E) \;\cong\; \wedge^2 TM \oplus \wedge^2 T^\ast M \oplus \mathrm{End}(TM).
\end{equation}
Thus, an element can be expressed by the block matrix
\begin{equation}
    A \;=\; \begin{pmatrix} N & \beta \\[0.6ex] b & -N^\mathrm{T} \end{pmatrix}
\end{equation}
where $N\in\mathrm{End}(TM)$, $b\in\wedge^2 T^\ast M$ (known as \textit{B-shift}) and $\beta\in \wedge^2 TM$.
\end{remark}

\begin{definition}[Isotropic subbundle]
Let $E\twoheadrightarrow M$ be a generalised tangent bundle. A subbundle $L\subset E$ is called \textit{isotropic subbundle} if it is isotropic under the symmetric bilinear form $\eta$.
\end{definition}

\section{Courant algebroid structure}

\noindent Let us start this section by recalling the definitions and some useful facts about Lie algebroids and Lie bialgebroids.

\subsection{Review of Lie algebroids and bialgebroids}
\begin{definition}
A \textit{Lie algebroid} $\mathfrak{a}$ is a triple $\big(E,[-,-]_\mathfrak{a},\rho\big)$ where
\begin{enumerate}[label=(\alph*)$\;$]
    \item $E\longtwoheadrightarrow M$ is a vector bundle on a smooth manifold $M$,
    \item $[-,-]_\mathfrak{a}:\Gamma(M,E)\times\Gamma(M,E)\rightarrow \Gamma(M,E)$ is a skew-symmetric pairing,
    \item $\rho:E\longtwoheadrightarrow TM$ is a vector bundle morphism, known as \textit{anchor map},
\end{enumerate}
such that the following properties hold:
\begin{enumerate}[label=(\textit{\roman*})$\;$]
    \item $\rho\big([X,Y]_\mathfrak{a}\big) \,=\, \big[\rho(X),\rho(Y)\big]_\mathrm{Lie}$,
    \item $[X,fY]_\mathfrak{a} \,=\, \rho(X)(f)Y + f[X,Y]_\mathfrak{a}\;$ (Leibniz rule),
\end{enumerate}
where $[-,-]_\mathrm{Lie}$ is the usual Lie bracket on $TM$, for any $X,Y\in\Gamma(M,E)$ and $f\in\Coo(M)$.
\end{definition}

\begin{example}[Lie algebra]
Notice that a Lie algebroid $\mathfrak{g}=(E,[-,-]_\mathfrak{g},0)$ over a point $M:=\{0\}$ is exactly a Lie algebra $\mathfrak{g}$, whose underlying vector space is $E$ and Lie bracket is $[-,-]_\mathfrak{g}$. 
\end{example} 

\begin{example}[Tangent bundle]
The tangent bundle $TM\twoheadrightarrow M$ has a natural algebroid structure $(TM,[-,-]_\mathrm{Lie},\mathrm{id}_{TM})$.
\end{example}

\begin{definition}
The \textit{exterior derivative} on a Lie algebroid $\mathfrak{a}=\big(E,[-,-]_\mathfrak{a},\rho\big)$ is the map $\di_\mathfrak{a}:\wedge^\bullet(E^\ast)\rightarrow\wedge^{\bullet+1}(E^\ast)$ such that
\begin{enumerate}[label=(\alph*)$\;$]
    \item $(\di_\mathfrak{a}f)(X) \,=\, \rho(X)(f)$ for any $f\in\mathcal{C}^\infty(M)$,
    \item $(\di_\mathfrak{a}\psi)(X_1,\cdots,X_k) \,=\, \sum_{i=1}^{k} (-1)^i \rho(X_i)\big(\psi(X_1,\cdots, X_{i-1},X_{i+1},\cdots,X_k)\big) \,+$
    \newline $+\, \sum_{1\leq i<j \leq k}(-1)^{i+j}\psi\big([ X_i,X_j]_\mathfrak{a}, X_1,\cdots,X_k\big)$ for any $\psi\in\wedge^k(E^\ast)$.
\end{enumerate}
\end{definition}

\begin{definition}
The \textit{Lie derivative} on the Lie algebroid $\big(E,[-,-]_\mathfrak{a},\rho\big)$ along any section $X\in\Gamma(M,E)$ is defined as the operator $\mathcal{L}^\mathfrak{a}_X:\wedge^\bullet(E^\ast)\rightarrow\wedge^\bullet(E^\ast)$ given by
\begin{equation}
    \mathcal{L}^\mathfrak{a}_X \;:=\; \di_\mathfrak{a}\iota_X+\iota_X\di_\mathfrak{a},
\end{equation}
where $\iota_X$ is the inner product.
\end{definition}

\begin{definition}[Lie bialgebroid]
A \textit{Lie bialgebroid} is the datum of two Lie algebroids $\mathfrak{a}=(E,[-,-]_\mathfrak{a},\rho)$ and $\mathfrak{a}^\ast=(E^\ast,[-,-]_{\mathfrak{a}^\ast},\rho^\ast)$, whose underlying vector bundles $E\twoheadrightarrow M$ and $E^\ast\twoheadrightarrow M$ are dual and such that the following compatibility condition holds
\begin{equation}
    \di_{\mathfrak{a}^\ast}[-,-]_{\mathfrak{a}} \;\,=\,\; [\di_{\mathfrak{a}^\ast}-,-]_{\mathfrak{a}} \,+\, [-,\di_{\mathfrak{a}^\ast}-]_{\mathfrak{a}},
\end{equation}
where $\di_\mathfrak{a}$ and $\di_{\mathfrak{a}^\ast}$ are respectively the differentials of the algebroids $\mathfrak{a}$ and $\mathfrak{a}^\ast$.
\end{definition}

\begin{example}[Lie bialgebra]
Notice that a Lie bialgebroid over a point $M:=\{0\}$ is exactly a Lie bialgebra. 
\end{example}

\subsection{Courant algebroids}

\noindent Let us start this section from the general definition of a Courant algebroid.

\begin{definition}
A \textit{Courant algebroid} $\mathfrak{c}$ is a quadruple $\big(E,\eta, \circ, \rho\big)$ where
\begin{enumerate}[label=(\alph*)$\;$]
    \item $E\twoheadrightarrow M$ is a vector bundle over a manifold $M$,
    \item $\eta:\Gamma(M,E)\times\Gamma(M,E)\rightarrow \Coo(M)$ is a non-degenerate symmetric pairing,
    \item $\circ:\Gamma(M,E)\times\Gamma(M,E)\rightarrow \Gamma(M,E)$ is a pairing (\textit{Dorfman bracket}),
    \item $\rho:E\twoheadrightarrow TM$ is a vector bundle morphism (\textit{anchor map}),
\end{enumerate}
such that the following properties hold:
\begin{enumerate}[label=(\textit{\roman*})$\;$]
    \item $X \circ (Y\circ Z) + Y \circ (Z\circ X) +  Z\circ (X\circ Y)=0$ (Leibniz identity),
    \item $\rho(X)\big(\eta(Y,Z)\big) = \eta( X\circ Y,Z) + \eta( Y,  X\circ Z)$,
    \item $\eta( X\circ X ,Y ) = \frac{1}{2}\rho(Y)\big(\eta( X,X)\big)$,
\end{enumerate}
for any section $X,Y,Z\in\Gamma(M,E)$.
\end{definition}

\begin{example}[Standard Courant algebroid]
The simplest example of Courant algebroid is the \textit{standard Courant algebroid}, which is given by $\big(TM\oplus T^\ast M, \eta, \circ, \mathrm{id}_{TM}\oplus 0\big)$.
\begin{equation}
    \begin{split}
        \eta( v+\lambda,w+\mu) \;&:=\; \iota_v\mu + \iota_w\lambda,\\
        (v+\lambda) \circ (w+\mu) \;&:=\; [v,w]_\mathrm{Lie} +\mathcal{L}_v\mu  - \iota_w\di\lambda,
    \end{split}
\end{equation}
where the bracket are called \textit{standard Dorfman bracket}.
\end{example}

\noindent Let us now generalise this simple example.

\begin{example}[Courant algebroid from a given Lie algebroid]
Given a Lie algebroid $\mathfrak{a}=\big(E,[-,-]_\mathfrak{a},\rho\big)$, we can directly define a Courant algebroid by $\big(E\oplus E^\ast, \eta, \circ, \rho\oplus 0\big)$ where the pairings are canonically given by
\begin{equation}
    \begin{split}
        \eta( v+\lambda,w+\mu) \;&:=\; \iota_v\mu + \iota_w\lambda,\\
        (v+\lambda)\circ(w+\mu) \;&:=\; [v,w]_\mathfrak{a} +\mathcal{L}^\mathfrak{a}_v\mu - \iota_w\di_\mathfrak{a}\lambda.
    \end{split}
\end{equation}
Notice that the standard Courant algebroid is exactly an example of this kind, where the chosen Lie algebroid is the tangent bundle $(TM,[-,-]_\mathrm{Lie},\mathrm{id}_{TM})$.
\end{example}

\begin{definition}[Courant bracket]
Given a Courant algebroid $\mathfrak{c}=\big(E,\eta, \circ, \rho\big)$, we can define the Courant bracket as the anti-symmetrisation of the Dorfman bracket $\circ$, i.e.
\begin{equation}
    [X,Y]_{\mathrm{Cou}} \;:=\; \frac{1}{2}(X\circ Y - Y\circ X),
\end{equation}
for any couple of sections $X,Y\in\Gamma(M,E)$.
\end{definition}

\begin{example}[Standard Courant bracket]
For the standard Courant algebroid $\mathfrak{c}=\big(TM\oplus T^\ast M, \eta, \circ, \mathrm{id}_{TM}\oplus 0\big)$, we obtain the standard Courant bracket
\begin{equation}
    [v+\mu,w+\lambda]_{\mathrm{Cou}} \;=\; [v,w]_{\mathrm{Lie}} + \mathcal{L}_v\lambda - \mathcal{L}_w\mu - \frac{1}{2}\di(\iota_v\lambda-\iota_w\mu),
\end{equation}
for any couple of sections $v+\mu,w+\lambda\in\Gamma(M,TM\oplus T^\ast M)$.
\end{example}

\begin{definition}[Integrable subbundle]
Given a Courant algebroid $\mathfrak{c}=\big(E,\eta, \circ, \rho\big)$, a subbundle $L\subset E$ is an \textit{integrable subbundle} if $\Gamma(M,L)$ is closed under the Courant bracket $[-,-]_{\mathrm{Cou}}$.
\end{definition}

\begin{definition}[Dirac structure]
Given a Courant algebroid $\mathfrak{c}=\big(E,\eta, \circ, \rho\big)$, a \textit{Dirac structure} is a subbundle $L\subset E$ which is both maximally isotropic and integrable.
\end{definition}

\begin{theorem}[Polarisation of a Courant algebroid]
Let $L$ be a Dirac structure of a Courant algebroid $\mathfrak{c}=\big(E,\eta, \circ, \rho\big)$. Then $\mathfrak{c}|_L:=\big(L, [-,-]_{\mathrm{Cou}}, \rho|_L\big)$ is a Lie algebroid.
\end{theorem}

\begin{remark}[Generalised tangent bundle as Courant algebroid]
Given a generalised tangent bundle $E\xtwoheadrightarrow{\;\rho\;} TM$ and a splitting map $\omega:TM\hookrightarrow E$, the isomorphism \eqref{eq:isocourantlagebroids} of vector bundles
\begin{equation}
    \omega\oplus\rho^\ast : TM\oplus T^\ast M \xrightarrow{\;\;\cong\;\;} E
\end{equation}
can be refined to an isomorphism of Courant algebroids from some $\mathfrak{c}=\big(E, \eta_\mathfrak{c}, \circ_\mathfrak{c}, \rho_\mathfrak{c}\big)$ to the standard Courant algebroid $\big(TM\oplus T^\ast M, \eta, \circ, \mathrm{id}_{TM}\oplus 0\big)$. Let us now rename the isomorphism $\mathcal{I}:=(\omega\oplus\rho^\ast)^{-1}$.
On generalised vectors $V\in\Gamma(M,E)$, the isomorphism of vector bundles can be given as follows:
\begin{equation}
    \mathcal{I}: \;V = \begin{pmatrix}v \\[0.6ex] \widetilde{v}_{(\alpha)} \end{pmatrix} \,\longmapsto\, v + \mu \quad\text{with}\quad \mu :=\widetilde{v}_{(\alpha)}-\iota_vB_{(\alpha)}
\end{equation}
where $v\in\mathfrak{X}(M)$ and $\mu\in\Omega^1(M)$. Consequently, the Courant algebroid structure can be defined as pullback of the standard Courant algebroid structure, i.e.
\begin{equation}
    \begin{split}
        \eta_\mathfrak{c} \;&:=\; \mathcal{I}^\ast\eta,\\
        \circ_\mathfrak{c} \;&:=\;\mathcal{I}^\ast\circ, \\
        \rho_\mathfrak{c} \;&:=\; \mathcal{I}^\ast \mathrm{id}_{TM}.
    \end{split}
\end{equation}
Finally, we obtain that the Courant bracket on the generalised tangent bundle is
\begin{equation}
    [V,W]_{\mathrm{Cou}} \;=\; [v,w]_{\mathrm{Lie}} + \mathcal{L}_v\widetilde{w}_{(\alpha)} - \mathcal{L}_w\widetilde{v}_{(\alpha)} - \frac{1}{2}\di\!\left(\iota_v\widetilde{w}_{(\alpha)}-\iota_w\widetilde{v}_{(\alpha)}\right) + \iota_v\iota_w H
\end{equation}
where $V,W\in\Gamma(M,E)$ is any couple of generalised vectors and $H=\di B_{(\alpha)}$ is the curvature of the corresponding bundle gerbe.
\end{remark}

\noindent Notice that the Dorfman bracket on a generalised tangent bundle can be rewritten as
\begin{equation}
    (V \circ W)^M \;=\; V^N\partial_NW^M - (\partial \times_{\mathrm{ad}}V)^M_{\;\;\,N}W^N,
\end{equation}
where we called $\partial_M:=(\partial_\mu,0)$ and we defined the product
\begin{equation}
    \times_{\mathrm{ad}}:\, E^\ast\otimes E \;\longrightarrow\; \mathrm{ad}(\mathrm{Fr}(E)),
\end{equation}
where $\mathrm{ad}(\mathrm{Fr}(E)):=\mathrm{Fr}(E)\times_{\mathrm{ad}}\mathfrak{o}(d,d)$ is the adjoint bundle of the generalised frame bundle $\mathrm{Fr}(E)\longtwoheadrightarrow M$, regarded as an $O(d,d)$-bundle.

\subsection{The double of a Lie bialgebroid as a Courant algebroid}

\noindent This subsection will be devolved to the intimate relation between Courant algebroids and Lie bialgebroids. For more details about this topic, see \cite{liu1997manin}.

\begin{definition}[Double of Lie bialgebroid]
The \textit{double of a Lie bialgebroid}, where the Lie bialgebroid is given by two Lie algebroids $\mathfrak{a}=\big(E,[-,-]_\mathfrak{a},\rho\big)$ and $\mathfrak{a}^\ast=\big(E^\ast,[-,-]_{\mathfrak{a}^\ast},\rho^\ast\big)$, is a Courant algebroid whose underlying vector bundle is $E\oplus E^\ast\twoheadrightarrow M$, whose anchor map is $\rho\oplus\rho^\ast:E\oplus E^\ast\twoheadrightarrow TM$ and which is equipped with
\begin{equation}
\begin{split}
            \eta( v+\lambda,w+\mu ) \;&:=\; \iota_v\mu + \iota_w\lambda, \\
           (v+\lambda)\circ(w+\mu) \;&:=\; [v,w]_\mathfrak{a} + \mathcal{L}^{\mathfrak{a}^\ast}_\lambda w  - \iota_\mu\di_{\mathfrak{a}^\ast}v \,+\\
            &+\; [\lambda,\mu]_{\mathfrak{a}^\ast} +\mathcal{L}^\mathfrak{a}_v\mu - \iota_w\di_\mathfrak{a}\lambda,
\end{split}
\end{equation}
respectively as symmetric and skew-symmetric pairing. Here, the operators $\di_{\mathfrak{a}^\ast}$ and $\mathcal{L}^{\mathfrak{a}^\ast}$ are the algebroid exterior derivative and Lie derivative of the dual Lie algebroid.
\end{definition}

\noindent This last example of Courant algebroid puts on the same footing both Lie algebroids which make up a Lie algebroid and it will be important in Double Field Theory.

\begin{definition}[Manin triple of a Lie bialgebroid]
If the Courant algebroid $\mathfrak{c}$ is the double of the Lie algebroid given by two Lie algebroid structures $\mathfrak{a}$ and $\mathfrak{a}^\ast$, then we can call $(\mathfrak{c},\mathfrak{a},\mathfrak{a}^\ast)$ \textit{Manin triple of the Lie bialgebroid}.
\end{definition}

\begin{theorem}[Polarisation of a Courant algebroid II]
Let $\mathfrak{c}=\big(E,\eta, \circ, \rho\big)$ be a Courant algebroid. Let $L$ and $\widetilde{L}$ be Dirac structures which are transversal to each other, i.e. such that $E = L \oplus \widetilde{L}$. Then, the couple $\mathfrak{c}|_L$ and $\mathfrak{c}|_{\widetilde{L}}$ is the datum of a Lie bialgebroid, where $\widetilde{L}$ is interpreted as the dual bundle of $L$ under the pairing $\eta$.
\end{theorem}

\section{Generalised complex structure}

\begin{definition}[Generalised complex structure]
A \textit{generalised complex structure} on a generalised tangent bundle $E\xtwoheadrightarrow{\,\rho\,}TM$ is a bundle automorphism $I:E\rightarrow E$ such that 
\begin{itemize}
    \item it is an isometry, i.e. $\eta(I-,I-)=\eta(-,-)$,
    \item it satisfies the equation $I^2=-\mathrm{id}_E$.
\end{itemize}
\end{definition}

\begin{remark}[Structure group]
The existence of a generalised complex structure $I$ on $E\xtwoheadrightarrow{\;\rho\;}TM$ implies that the dimension of the base manifold is even. Moreover, it reduces the structure group $O(d,d)$ of the generalised tangent bundle to $U(\nicefrac{d}{2},\nicefrac{d}{2})$. This is analogous to how an ordinary complex structure on $TM$ reduces the structure group $GL(n)$ to $U(\nicefrac{d}{2})$.
\end{remark}

\noindent We can use the isomorphism $\rho\oplus\omega:E\xrightarrow{\;\cong\;}TM\oplus T^\ast M$ to express a generalised complex structure $I\in\mathrm{Aut}(E)$ in blocks by
\begin{equation}
    I \;=\; \begin{pmatrix}-j & \beta\\ b & j^\mathrm{T}\end{pmatrix},
\end{equation}
where $j\in\mathrm{Aut}(TM)$, $b\in\wedge^2T^\ast M$ and $\beta\in\wedge^2TM$. In this terms, the defining equation of the generalised complex structure becomes
\begin{equation}
    I^2=-\mathrm{id}_E \qquad \Longrightarrow \qquad \begin{aligned}j^2 + \beta b &= -\mathrm{id}_{TM}, \\
    -j\beta+\beta j^\mathrm{T} &= 0, \\
    -b j + j^\mathrm{T}b &= 0.\end{aligned}
\end{equation}

\begin{example}[Complex structure]
The generalised complex structure defined by
\begin{equation}
    I_j \;=\; \begin{pmatrix} -j & 0 \\ 0 & j^\mathrm{T} \end{pmatrix}
\end{equation}
is, equivalently, a complex structure $j\in\mathrm{Aut}(TM)$ on the base manifold.
\end{example}

\begin{example}[Symplectic structure]
The generalised complex structure defined by
\begin{equation}
    I_\omega \;=\; \begin{pmatrix} 0 & \omega^{-1} \\ -\omega & 0 \end{pmatrix}
\end{equation}
is, equivalently, a symplectic structure $\omega\in\Omega^2(M)$ on the base manifold.
\end{example}

\begin{remark}[$\pm i$-eigenbundles]
Let us define the complexification of the generalised tangent bundle by $E_\mathbb{C}:=E\otimes\mathbb{C}$. Then, the generalised complex structure has two eigenbundles with eigenvalues $+i$ and $-i$, i.e.
\begin{equation}
    E_\mathbb{C} \;\cong\; L_{+i} \oplus L_{-i}
\end{equation}
where $L_{-i}=\overline{L_{+i}}$.
\end{remark}

\begin{theorem}[$\pm i$-eigenbundles are maximally isotropic]
The eigenbundles $L_{+i}$ and $L_{-i}$ are \textit{maximally isotropic} subbundles of $E$, i.e. $\eta|_{L_{\pm i}}=0$ and $\mathrm{rank}(L_{\pm i})=d$.
\end{theorem}
\begin{proof}
For any couple of generalised vectors $V,W\in\Gamma(M,L_{\pm i})$, we have $\eta(V,W) = \eta(IV,IW) = \eta(\pm iV, \pm iW) = (\pm i)^2\eta(V,W) = -\eta(V,W)$.
\end{proof}

\noindent The integrability of an ordinary complex structure $j$ is equivalent to the involutivity of the $i$-eigenbundle of $TM_\mathbb{C}$ respect to the Lie bracket of the algebroid $(TM,[-,-]_\mathrm{Lie})$. If this condition is satisfied, we can call the couple $(M,j)$ complex manifold. In generalised complex geometry, this statement is generalised as follows.

\begin{definition}[Integrable generalised complex structure]
A generalised complex structure $I\in\mathrm{Aut}(E)$ is integrable if and if the $i$-eigenbundle $L_{+i}$ of $E_\mathbb{C}=E\otimes\mathbb{C}$ is closed under the Courant bracket, i.e.
\begin{equation}
    \big[ \Gamma(M,L_{+i}), \Gamma(M,L_{+i}) \big]_\mathrm{Cou} \;\subset\; \Gamma(M,L_{+i}).
\end{equation}
A smooth manifold $M$ equipped with an integrable generalised complex structure $I\in\mathrm{Aut}(E)$ can be called \textit{generalised complex manifold} $(M,I)$.
\end{definition}

\begin{remark}[Dirac structures and integrability]
Notice that, if $I\in\mathrm{Aut}(E)$ is an integrable generalised complex structure, then $L_{\pm i}$ are transversal Dirac structures.
\end{remark}

\begin{savequote}[8.7cm]
At ubi materia, ibi Geometria.

Where there is matter, there is geometry.
  \qauthor{--- Johannes Kepler, \textit{De Fundamentis Astrologi\ae{} Certioribus}}
\end{savequote}

\chapter{\label{app:2}Fundamentals of supergeometry}

\minitoc

\noindent In this appendix we provide an introduction to supergeometry, which underlies Supergravity. 
The focus will be given on the global higher structures which characterise the Supergravity limit of the $10$-dimensional string theories.
The main species of fields we find in Supergravity are collected in the following table.

\begin{table}[h!]\begin{center}
 \begin{tabular}{||c | c || c | c ||} 
 \hline
 \multicolumn{4}{||c||}{Type II and Heterotic Supergravity fields} \\
 \hline
  \textbf{Field} & \textbf{Meaning} & \textbf{Curvature }& \textbf{Meaning}  \\ [0.5ex] 
 \hline\hline
  $e^a$ & Graviton & $T^a$ & Torsion \\ 
  \hline
 $\psi^\upalpha$ & Gravitino & $\rho^\upalpha$ & Gravitino field strength \\ 
 \hline
 $\omega^a_{\;\;b}$ & Spin connection & $R^a_{\;\;b}$ & Ricci curvature \\ 
 \hline\hline
 $\varphi$ & Dilaton & $F_{\mathrm{dil}}$ & Dilaton field strength \\ 
 \hline
 $\chi^\upalpha$ & Dilatino & $\rho_{\mathrm{dil}}^\upalpha$ & Dilatino field strength \\ 
 \hline\hline
 $A^i$ & Yang-Mills field & $F^i$ & F-flux \\ 
 \hline
 $\lambda^{\upalpha i}$ & Gaugino & $\rho^{\upalpha i}_{\mathrm{gau}}$ & Gaugino field strength \\ 
 \hline\hline
 $B$ & Kalb-Ramond field & $H$ & H-flux \\ 
 \hline
\end{tabular}
\caption[Fields of Type II and Heterotic Supergravity]{A sum of the fields of Type II and Heterotic Supergravity, with their curvatures.}
\end{center}\end{table}

\section{Supermanifolds}

\begin{definition}[Type II Dirac representations]
There are two Dirac representations of $\mathrm{Spin}(1,9)$ on the vector space $\mathbb{C}^{16}\oplus\mathbb{C}^{16}$ which we will call $\mathbf{16}\oplus\overline{\mathbf{16}}$ for Type IIA supergravity and $\mathbf{16}\oplus\mathbf{16}$ for Type IIB supergravity. These are given by the following matrices $\{\Gamma_a\}_{a=0,\dots,10}$ acting on $\mathbb{C}^{16}\oplus\mathbb{C}^{16}$
\begin{equation}
    \begin{aligned}
    \Gamma_{a} &=\begin{pmatrix}0 & \gamma^a \\ \gamma^a & 0 \end{pmatrix} \;\;\, \text{ for }a\leq 8, \\
    \Gamma_{9}^{\IIA} &=\begin{pmatrix}0 & 1 \\ -1 & 0 \end{pmatrix}, \quad \Gamma_{9}^{\IIB}=\begin{pmatrix}0 & 1 \\ 1 & 0 \end{pmatrix}, \\
    \Gamma_{10}&=\begin{pmatrix}i & 0 \\ 0 & -i \end{pmatrix},
    \end{aligned}
\end{equation}
where $\{\gamma^a\}$ with $a=0,\dots,8$ are the gamma-matrices of a Dirac representation $\mathbf{16}$ of $\mathrm{Spin}(1,8)$ on the vector space $\mathbb{C}^{16}$.
Let us also define
\begin{equation}
    \Gamma_{a_1a_2\cdots a_k} \;:=\; \frac{1}{k!}\sum_{\sigma\in\mathrm{perm}} (-1)^{|\sigma|}\, \Gamma_{a_{\sigma(1)}} \cdots \Gamma_{a_{\sigma(k)}}
\end{equation}
\end{definition}

\begin{remark}[Identities]
Notice that we have the following useful identities:
\begin{equation}
    \begin{aligned}
    \Gamma_{9}^{\mathrm{IIA}} &= i\Gamma_{9}^{\mathrm{IIB}}\Gamma_{10}, \\
    \Gamma_{9}^{\mathrm{IIB}} &= i\Gamma_{9}^{\mathrm{IIA}}\Gamma_{10}.
    \end{aligned}
\end{equation}
\end{remark}

\begin{definition}[Super-Minkowski space]
A $(1+d|N)$-dimensional super Minkowski space $\mathbb{R}^{1,d|\mathbf{N}}$, with a real spinor representation $\mathbf{N}$ of $\mathrm{Spin}(1,d)$, is 
\begin{equation}
    \Coo\big(\mathbb{R}^{1,d|\mathbf{N}}\big) \,:=\, \Coo\big(\mathbb{R}^{1,d}\big)\otimes_{\mathbb{R}}\wedge\,\mathbb{R}^N
\end{equation}
where $\wedge\,\mathbb{R}^N$ is the Grassmann (or exterior) algebra of $\mathbb{R}^N$. This means that any function $f\in\Coo\big(\mathbb{R}^{1,d|\mathbf{N}}\big)$ on the super Minkowski space can be expressed by
\begin{equation}
    f(x,\vartheta)= f(x) + \sum_{k=1}^{N}\;\sum_{\upalpha_1<\dots<\upalpha_k} \!\!\! f_{\upalpha_1\dots\upalpha_k}(x)\vartheta^{\upalpha_1}\cdots\vartheta^{\upalpha_k}
\end{equation}
where $f,f_{\upalpha_1\dots\upalpha_k}\in\Coo(\mathbb{R}^{1,d})$ are functions on $\mathbb{R}^{1,d}$.
\end{definition}

\begin{definition}[Supertranslation supergroup]
The super-Minkowski space is equipped with a supergroup structure
\begin{equation}
    \begin{aligned}
    +:\;\; \mathbb{R}^{1,d|\mathbf{N}}\,&\times\,\mathbb{R}^{1,d|\mathbf{N}}\;\xrightarrow{\quad\;\;\,}\;\mathbb{R}^{1,d|\mathbf{N}} \\
    (x^\mu,\, \vartheta^\upalpha)\,&+\,(x^{\prime\mu},\, \vartheta^{\prime\upalpha})\, =\,(x^\mu+x^{\prime\mu}+\bar{\vartheta}\Gamma^\mu\vartheta^\prime,\, \vartheta^{\upalpha}+\vartheta^{\prime\upalpha})
    \end{aligned}
\end{equation}
\end{definition}

\begin{definition}[Supersymmetry]
In theoretical physics we call \textit{supersymmetry} a supertranslation in the odd coordinates of the form 
\begin{equation}
    (x^\mu,\, \vartheta^\upalpha)+(0,\, \epsilon^{\upalpha}) \;=\; (x^\mu+\bar{\vartheta}\Gamma^\mu\epsilon,\, \vartheta^{\upalpha}+\epsilon^{\upalpha}).
\end{equation}
\end{definition}

\begin{definition}[Supermanifold]
A $(d+1|N)$-dimensional supermanifold $M$ is defined as a locally ringed space $M:=\left(|M|,\Coo\right)$ where $|M|$ is a topological space and $\Coo(-)$ is a sheaf on $|M|$ which is given on open subsets $|U|\subset |M|$ by 
\begin{equation}
    \Coo(U) \;:=\; \Coo\big(\mathbb{R}^{1,d}\big)\otimes_{\mathbb{R}}\wedge\,\mathbb{R}^N,
\end{equation}
where the space $\mathbb{R}^N$ is spanned by the odd coordinates $\{\vartheta^\upalpha\}_{\upalpha=1,\dots,N}$.
\end{definition}

\begin{definition}[Superdiffeomorphism]
A \textit{superdifferentiable map} $M\rightarrow N$ of supermanifolds is a homomorphisms of locally ringed spaces.
The set of all superdifferentiable maps $M\rightarrow N$ will be denoted by 
\begin{equation}
    \SDiff(M,N) \; := \; \mathrm{Hom}\big((|M|,\Coo),\,(|N|,\Coo)\big).
\end{equation}
In particular, a \textit{superdiffeomorphism} of supermanifolds is an isomorphism of locally ringed space.
The supergroup of all superdiffeomorphisms of a supermanifold $M$ will be simply denoted by $\SDiff(M)$.
\end{definition}

\begin{example}[Superdifferential maps of super-Minkowski spaces]
Superdifferential maps $\mathbb{R}^{1,d|\mathbf{N}}\xrightarrow{\;f\;}\mathbb{R}^{1,d'|\mathbf{N}'}$ between super-Minkowski spaces are given by the following transformations of coordinates:
\begin{equation}\label{eq:sdiffcoords}
    \begin{aligned}
    x^{\prime \mu} &= f^\mu(x)+ \sum_{k=1}^{N}\;\sum_{\upalpha_1<\dots<\upalpha_k} \!\!\! f^\mu_{\upalpha_1\dots\upalpha_k}(x)\vartheta^{\upalpha_1}\cdots\vartheta^{\upalpha_k}, \\
    \vartheta^{\prime \upalpha} &= f^\upalpha(x)+ \sum_{k=1}^{N}\;\sum_{\upalpha_1<\dots<\upalpha_k} \!\!\! f^\upalpha_{\upalpha_1\dots\upalpha_k}(x)\vartheta^{\upalpha_1}\cdots\vartheta^{\upalpha_k}.
    \end{aligned}
\end{equation}
\end{example}

\begin{remark}[Gluing a supermanifold]
The local coordinate patches $(U_\alpha,z_{(\alpha)})$ of a supermanifold $M$ are glued by 
\begin{equation}
    z_{(\alpha)} \,=\, f_{(\alpha\beta)}\big(z_{(\beta)}\big),
\end{equation}
where $\SDiff\big(U_{\alpha}\cap U_{\beta}\big)\ni f_{(\alpha\beta)}:U_\alpha\cap U_\beta\longrightarrow U_\beta\cap U_\alpha$ are superdiffeomorphisms.
In coordinates, such maps can be expanded as follows:
\begin{equation}\label{eq:supergluing}
    \begin{aligned}
    x^{\mu}_{(\alpha)} \;&=\; f^\mu_{(\alpha\beta)}\big(x_{(\beta)}\big)\,+\, \sum_{k=1}^{N}\;\sum_{\upalpha_1<\dots<\upalpha_k} \!\!\! f^\mu_{(\alpha\beta)\upalpha_1\dots\upalpha_k}\big(x_{(\beta)}\big)\vartheta^{\upalpha_1}_{(\beta)}\cdots\vartheta^{\upalpha_k}_{(\beta)}, \\
    \vartheta^{\upalpha}_{(\alpha)} \;&=\; f^\upalpha_{(\alpha\beta)}\big(x_{(\beta)}\big)\,+\, \sum_{k=1}^{N}\;\sum_{\upalpha_1<\dots<\upalpha_k} \!\!\! f^\upalpha_{(\alpha\beta)\upalpha_1\dots\upalpha_k}\big(x_{(\beta)}\big)\vartheta^{\upalpha_1}_{(\beta)}\cdots\vartheta^{\upalpha_k}_{(\beta)}.
    \end{aligned}
\end{equation}
\end{remark}

\begin{definition}[Reduced manifold]
Given any supermanifold $M$, its \textit{reduced manifold} $\bosonic{M}$ is an ordinary manifold defined by $\bosonic{M} = \left( |M|,\, \Coo/\mathcal{I} \right)$, where the presheaf $\mathcal{I}$ is the nilpotent ideal of $\Coo$ which is given on local patches by functions of the odd coordinates 
\begin{equation}
    \mathcal{I}(U) \;:=\; \frac{\mathbb{R}[\vartheta^1,\vartheta^2,\dots,\vartheta^N]}{\langle\vartheta^\upalpha\vartheta^\upbeta+\vartheta^\upbeta\vartheta^\upalpha\rangle} \;\;\subset\;\Coo(U).
\end{equation}
\end{definition}

\begin{theorem}[Canonical embedding of reduced manifold]
Given a supermanifold $M$, its reduced manifold $\bosonic{M}$ is a canonically embedded in the former by a superdifferential map
\begin{equation}
    \begin{tikzcd}[row sep=scriptsize, column sep=8ex] 
        \bosonic{M} \arrow[r, "\iota", hook] & M.
    \end{tikzcd}
\end{equation}
\end{theorem}

\begin{proof}
There is a canonical morphism of ringed spaces $\iota:\left(|M|,\Coo/\mathcal{I}\right)\hookrightarrow\left(|M|,\Coo\right)$ given by a pair $\iota=\big(\mathrm{id}_{|M|},\iota^\sharp\big)$ where $\mathrm{id}_{|M|}:|M|\xrightarrow{\cong}|M|$ is the identity map on the topological space $|M|$ and $\iota^\sharp:\Coo\twoheadrightarrow\Coo/\mathcal{I}$ is a quotient map of sheaves on $|M|$.
\end{proof}

\begin{example}[Reduced Minkowski space]
The reduced manifold of a super-Minkowski space is a Minkowski space, i.e. $\bosonic{\mathbb{R}^{1,d|\mathbf{N}}}=\mathbb{R}^{1,d}$.
\end{example}

\begin{definition}[Reduced differentiable map]
Given a superdifferentiable map $f:M\rightarrow N$, we define its \textit{reduced differentiable map} $\bosonic{f}:\bosonic{M}\rightarrow \bosonic{N}$ as the map which makes the following diagram commute
\begin{equation}
    \begin{tikzcd}[row sep=8ex, column sep=8ex] 
        \bosonic{M} \arrow[d, "\iota", hook]\arrow[r, "\bosonic{f}"] & \bosonic{N}\arrow[d, "\iota", hook]  \\
        M \arrow[r, "f"] & N.
    \end{tikzcd}
\end{equation}
\end{definition}

\begin{remark}[Reduced differential maps of Minkowski spaces]
Given a superdifferentiable map $f:\mathbb{R}^{1,d|\mathbf{N}}\rightarrow\mathbb{R}^{1,d'|\mathbf{N}'}$, its reduced differentiable map is $\bosonic{f}:\mathbb{R}^{1,d}\rightarrow\mathbb{R}^{1,d'}$ and it is given by the first term $x^{\prime\mu}=f^\mu(x)$ of equation \eqref{eq:sdiffcoords}
\end{remark}

\begin{theorem}[Gluing the reduced manifold]
If $\bigsqcup_\alpha\!U_{\alpha}$ is a cover for a supermanifold $M$, then $\bigsqcup_\alpha\!\bosonic{U}_{(\alpha)}$ is a cover for its reduced manifold $\bosonic{M}$.
By looking at the patching conditions \eqref{eq:supergluing} of $M$, the reduced manifold $\bosonic{M}$ will be glued by the component
\begin{equation}
    x^{\mu}_{(\alpha)} \;=\; f^\mu_{(\alpha\beta)}\big(x_{(\beta)}\big).
\end{equation}
\end{theorem}

\noindent Notice that the kernel of the quotient map  $\iota^\sharp:\Coo(\mathbb{R}^{1,d|\mathbf{N}})\twoheadrightarrow\Coo(\mathbb{R}^{1,d})$ is a nilpotent ideal $\Coo(\mathbb{R}^{0|\mathbf{N}})$ generated by the odd coordinates. This implies that the ring $\Coo(\mathbb{R}^{1,d|\mathbf{N}})$ is an \textit{infinitesimal extension} of $\Coo(\mathbb{R}^{1,d})$.

\begin{remark}[Superpoint is infinitesimally thickened point]
Notice that we have
\begin{equation}
        \Coo(\mathbb{R}^{0|\mathbf{1}}) \;=\; \frac{\mathbb{R}[\vartheta]}{\langle\vartheta^2\rangle}.
\end{equation}
Hence the superpoint is the ringed space $\mathbb{R}^{0|\mathbf{1}} = \mathrm{Spec}\big(\mathbb{R}[\vartheta]/\langle\vartheta^2\rangle\big)$, whose underlying topological space is a single point $\big|\mathbb{R}^{0|\mathbf{1}}\big|=\{0\}$, but equipped with a smooth algebra of function given by the ring of dual numbers.

\vspace{0.2cm}
\noindent As pointed out by \cite{Huerta18}, a scalar field on the superpoint (i.e. a differentiable map $\mathbb{R}^{0|\mathbf{1}}\rightarrow \mathbb{R}$) is equivalently a map of smooth algebras of functions $\Coo(\mathbb{R})\rightarrow\Coo(\mathbb{R}^{0|\mathbf{1}})=\mathbb{R}\oplus\mathbb{R}\vartheta$ which can be interpreted as sending $f\mapsto f(0)+f'(0)\vartheta$ for any function $f\in\Coo(\mathbb{R})$. Crucially this map well-behaves respect to the product of scalars, indeed we have $\big(f(0)+f'(0)\vartheta\big)\big(g(0)+g'(0)\vartheta\big)=f(0)g(0) + (f'(0)g(0)+f(0)g'(0))\vartheta=(fg)(0) + (fg)'(0)\vartheta$ for any couple of functions $f,g\in\Coo(\mathbb{R})$.

\vspace{0.2cm}
\noindent More generally, a superdiffeomorphism $\mathbb{R}^{0|\mathbf{1}}\rightarrow M$ to any ordinary manifold $M$ is equivalently a morphism of ringed spaces $(x,X):(\{0\},\Coo)\rightarrow(M,\Coo)$ where $x:\{0\}\hookrightarrow M$ is the embedding of the underlying point $|\mathbb{R}^{0|\mathbf{1}}|=\{0\}$ in the manifold $M$ and where $X:\Coo(M)\rightarrow\mathbb{R}[\vartheta]/\langle\vartheta^2\rangle$ is a morphisms of smooth algebras sending $f\mapsto f(x)+X^\mu\partial_\mu f(x)\vartheta$ for any function $f\in\Coo(M)$. In other words $(x,X)$ is the datum of a point $x\in M$ and a vector $X\in T_xM$. Therefore the space of all superdiffeomorphisms $\mathbb{R}^{0|\mathbf{1}}\xrightarrow{\,(x,X)\,} M$ is exactly the tangent bundle of $M$, i.e.
\begin{equation}
    [\mathbb{R}^{0|\mathbf{1}},\,M]  \;\cong\;  TM
\end{equation}
\end{remark}

\noindent This gives an important insight in the nature of the geometry of supermanifolds.

\begin{remark}[Supermanifolds as infinitesimally thickened manifolds]
Consider
\begin{equation}
        \Coo(\mathbb{R}^{0|\mathbf{N}}) \;=\; \frac{\mathbb{R}[\vartheta^1,\dots,\vartheta^N]}{\langle\vartheta^\upalpha\vartheta^\upbeta+\vartheta^\upbeta\vartheta^\upalpha\rangle}.
\end{equation}
Notice that its underlying topological space is just a point $\big|\mathbb{R}^{0|\mathbf{N}}\big|=\{0\}$.
Thus, morally speaking, a supermanifold can be interpreted as an ordinary manifold with infinitesimal neighborhoods. See \cite{Huerta18} for more details.
\end{remark}

\section{Super-Cartan geometry}

\begin{definition}[Super-Poincar\'{e} group]
We can define the \textit{super Poincar\'{e} algebra} $\mathfrak{iso}(1,d|\mathbf{N})$, where $\mathbf{N}$ is a spin representation, by extending the Poincar\'{e} algebra $\mathfrak{iso}(1,d)$ with new odd generators $\{\mathsf{q}_{\upalpha}\}$ such that they appear in the following commutation relations:
\begin{equation}
    \begin{aligned}
        [\mathsf{p}_{a},\mathsf{q}_\upbeta] \,&=\, 0, \\
        [\mathsf{m}_{ab},\mathsf{q}_\upbeta] \,&=\, \frac{1}{4} \Gamma_{ab\;\upbeta}^{\;\;\,\upalpha} \,\mathsf{q}_\upalpha,\\
        \{\mathsf{q}_\upalpha,\mathsf{q}_\upbeta\} \,&=\, i\Gamma^a_{\upalpha\upbeta} \,\mathsf{p}_a,
    \end{aligned}
\end{equation}
where $\mathsf{p}_{a}$ are the generators of $\mathbb{R}^{1,d}$, $\mathsf{m}_{ab}$ are the generators of $\mathfrak{so}(1,d)$ and $\Gamma^a_{\upalpha\upbeta}$ are the gamma-matrices of the spin representation $\mathbf{N}$.
Finally, the \textit{super Poincar\'{e} group} $\ISO(1,d|\mathbf{N}):=\mathbb{R}^{1,d|\mathbf{N}}\rtimes \SO(1,d)$ is given by the Lie-integration of $\mathfrak{iso}(1,d|\mathbf{N})$.
\end{definition}

\noindent Notice that a $(1+d|\mathbf{N})$-dimensional super-Minkowski space is thus recovered by
\begin{equation}
    \frac{\ISO(1,d|\mathbf{N})}{SO(1,d)} \;\cong\; \mathbb{R}^{1,d|\mathbf{N}}.
\end{equation}

\begin{definition}[Super-Cartan geometry]
A super-Cartan geometry on a $(1+d|\mathbf{N})$-dimensional supermanifold $M$ is the datum of a principal $SO(1,d)$-bundle $FM\rightarrow M$, called \textit{frame bundle}, equipped with a $1$-form 
\begin{equation}
    (\underline{E},\underline{\omega})\, \in\,  \Omega^1\big(FM,\, \mathfrak{iso}(1,d|\mathbf{N})\big)
\end{equation}
called \textit{super-Cartan connection}, which satisfies the following properties:
\begin{itemize}
    \item $(\underline{E},\underline{\omega})|_E: T_EFM\xrightarrow{\;\cong\;}\mathfrak{g}$ is an isomorphism of vector spaces for any $E\in FM$,
    \item $(\underline{E},\underline{\omega})$ maps vertical vectors in $\mathfrak{so}(1,d)\subset \mathfrak{iso}(1,d|\mathbf{N})$ and, in particular, $\underline{\omega}$ restricts to the Maurer-Cartan form $T(F_xM)\cong T_eSO(1,d)\xrightarrow{\;\cong\;}\mathfrak{so}(1,d)$,
    \item $(R_h)_\ast (\underline{E},\underline{\omega}) = \mathrm{Ad}_{h^{-1}} (\underline{E},\underline{\omega})$, where $R_h$ is the right multiplication $\forall h\in SO(1,d)$.
\end{itemize}
\end{definition}

\begin{remark}[Spin connection]
Notice that $\underline{\omega}\in \Omega^1(FM,\,\mathfrak{so}(1,d))$ is exactly an Ehresmann connection for the principal $SO(1,d)$-bundle $FM\rightarrow M$. We will call it \textit{spin connection}.
\end{remark}

\begin{remark}[Coframe field]
Notice that $\underline{E}\in \Omega^1(FM,\,\mathbb{R}^{1,d|\mathbf{N}})$ is exactly a \textit{coframe field} or, in other words, the vielbein.
\end{remark}

\begin{remark}[Local trivialisation of a super-Cartan geometry]
Let $\mathcal{U}:=\{U_{\alpha}\}$ be a good open cover of a supermanifold $M$. A Cartan connection is given in local data by
\begin{equation}
    \begin{aligned}
     \omega_{(\alpha)} \,&\in\, \Omega^1\big( U_{\alpha},\, \mathfrak{so}(1,d) \big), \\
    E_{(\alpha)} \,&\in\, \Omega^1\big( U_{\alpha},\, \mathbb{R}^{1,d|\mathbf{N}} \big), \\
    h_{(\alpha\beta)} \,&\in\, \Coo\big(U_{\alpha}\cap U_{\beta},\, SO(1,d) \big),
    \end{aligned}
\end{equation}
such that they satisfy the following patching conditions:
\begin{equation}\label{eq:patching0}
    \begin{aligned}
    \omega_{(\beta)} \,&=\,  \mathrm{Ad}_{h_{(\alpha\beta)}^{-1}}\omega_{(\alpha)} + h_{(\alpha\beta)}^{-1}\di h_{(\alpha\beta)}, \\
    E_{(\beta)} \,&=\,  h_{(\alpha\beta)}^{-1} \cdot E_{(\alpha)}, \\
    h_{(\alpha\gamma)}\,&=\, h_{(\alpha\beta)}\cdot h_{(\beta\gamma)}.
    \end{aligned}
\end{equation}
A gauge transformation between two Cartan connections is defined as a coboundary $\eta_{(\alpha)}\in\Coo\big(U_\alpha, SO(1,d)\big)$ of the \v{C}ech $SO(1,d)$-cocycle, i.e.
\begin{equation}
    \begin{aligned}
    \omega_{(\alpha)}' \,&=\,  \mathrm{Ad}_{\eta_{(\alpha)}^{-1}}\omega_{(\alpha)} + \eta_{(\alpha)}^{-1}\di \eta_{(\alpha)}, \\
    E_{(\alpha)}' \,&=\,  \eta_{(\alpha)}^{-1} \cdot E_{(\alpha)}, \\
    h_{(\alpha\beta)}' \,&=\, \eta_{(\alpha)}^{-1}\cdot h_{(\alpha\beta)}\cdot \eta_{(\beta)}.
    \end{aligned}
\end{equation}
Notice that we can define a groupoid $\Cartan(M)$ whose objects are Cartan connections on $M$ and whose morphisms are gauge transformations between Cartan connections. In other words, we have
\begin{equation}
    \begin{tikzcd}[row sep=7ex, column sep=14ex]
         M\simeq \check{C}(\mathcal{U}) \arrow[r, bend left=50, ""{name=U, below}, "{(\omega_{(\alpha)},\,E_{(\alpha)},\,h_{(\alpha\beta)})}"]
        \arrow[r, bend right=50, "{(\omega_{(\alpha)}',\,E_{(\alpha)}',\,h_{(\alpha\beta)}')}"', ""{name=D}]
        & \Cartan.
        \arrow[Rightarrow, from=U, to=D, "(\eta_{(\alpha)})"]
    \end{tikzcd}
\end{equation}
\end{remark}

\noindent Notice that the vielbein and the spin connection are \textit{not} globally defined $1$-forms on the base manifold $M$.

\begin{remark}[Supervielbein in local coordinates]
Given a local patch $U\subset M$ with coordinates $\{z^M\}=\{x^\mu,\vartheta^\upalpha\}$ with index $M=(\mu,\upalpha)$, we can express the supervielbein in local coordinates by $E^A=E^A_M\di z^M$ with index $A=(a,\upalpha)$. We can then split the supervielbein in bosonic-valued component $E^a =: e^a = e^a_M\di z^M$ and fermionic-valued component $E^\upalpha =: \psi^\upalpha = \psi^\upalpha_M\di z^M$. We have then the further split in local coordinates:
\begin{equation}\label{eq:supervielbeincomponents}
    \begin{aligned}
    e^a(x,\vartheta) &= \,e^a_\mu(x,\vartheta)\di x^\mu + \,e^a_{{\upbeta}}(x,\vartheta)\di \vartheta^{{\upbeta}} \\
    \psi^\upalpha(x,\vartheta) &= \psi^\upalpha_\mu(x,\vartheta)\di x^\mu + \psi^\upalpha_{{\upbeta}}(x,\vartheta)\di \vartheta^{{\upbeta}}
    \end{aligned}
\end{equation}
In more physical terms, we can identify $e^a$ with the \textit{graviton} and $\psi^\upalpha$ with the \textit{gravitino} field.
\end{remark}

\begin{remark}[Supervielbein commutation/anti-commutation rule]
The commutation/anti-commutation rule of the supervielbein are equivalently the following:
\begin{equation}\label{eq:supervielbeincommutation}
    \begin{aligned}
    e^a\wedge e^b \,&=\, - e^b\wedge e^a, \\
    e^a\wedge \psi^\upbeta \,&=\, -\psi^\upbeta\wedge e^a,\\
    \psi^\upalpha\wedge \psi^\upbeta \,&=\, +\psi^\upbeta\wedge \psi^\upalpha.
    \end{aligned}
\end{equation}
\end{remark}

\begin{remark}[Patching conditions in terms of graviton and gravitino]
The patching conditions \eqref{eq:patching0} become
\begin{equation}\label{eq:patching1}
    \begin{aligned}
    \omega_{(\beta)} \,&=\,  \mathrm{Ad}_{h_{(\alpha\beta)}^{-1}}\omega_{(\alpha)} + h_{(\alpha\beta)}^{-1}\di h_{(\alpha\beta)}, \\
    e_{(\beta)} \,&=\,  h_{(\alpha\beta)}^{-1} \cdot e_{(\alpha)}, \\
    \psi_{(\beta)} \,&=\,  h_{(\alpha\beta)}^{-1} \cdot \psi_{(\alpha)}, \\
    h_{(\alpha\gamma)}\,&=\, h_{(\alpha\beta)}\cdot h_{(\beta\gamma)}.
    \end{aligned}
\end{equation}
\end{remark}

\begin{definition}[Curvature of a super-Cartan geometry]
The curvature of a super-Cartan connection is defined by
\begin{equation}\label{eq:defcurv}
    (\underline{T},\underline{R}) \;:=\; \di (\underline{E},\underline{\omega}) + \left[(\underline{E},\underline{\omega}) \wedge (\underline{E},\underline{\omega})\right]\; \in\,  \Omega^2\big(FM,\, \mathfrak{iso}(1,d|\mathbf{N})\big)
\end{equation}
where $\di$ here is the differential on the manifold $FM$.
\end{definition}

\noindent Notice that the curvature satisfies the property $(R_h)_\ast(\underline{T},\underline{R})=\mathrm{Ad}_{h^{-1}}(\underline{T},\underline{R})$ for any element $h\in SO(1,d)$.

\begin{remark}[Local trivialisation of curvature of a super-Cartan geometry]
By rewriting \eqref{eq:defcurv} in components (and by suppressing the patch index $\alpha$ of $U_{\alpha}$) and by splitting $T^A = (T^a,\rho^\upalpha)$ we get 
\begin{equation}
    \begin{aligned}
    R^a_{\;\,b} \,&=\, \di\omega^a_{\;\,b} + \omega^a_{\;\,c}\wedge \omega^c_{\;\,b} && \in\Omega^2(U,\mathfrak{so}(1,d)) \\[0.15cm]
    T^a \,&=\, \di e^a + \omega^a_{\;\,b}\wedge e^b - \bar{\psi}\Gamma^a \psi && \in\Omega^2(U,\mathbb{R}^{1,d})\\
    \rho^\upalpha  \,&=\, \di\psi^\upalpha +  \frac{1}{4}\omega^{ab}\Gamma_{ab\;\,\upbeta}^{\;\;\;\upalpha}\psi^{\upbeta}  && \in\Omega^2(U,\mathbb{R}^{0|\mathbf{N}})
    \end{aligned}
\end{equation}
By following the physics convention we can identify $R^a_{\;\,b}$ with the \textit{Riemannian curvature}, $T^a$ with the \textit{(super-)torsion} and $\rho^\upalpha$ with the \textit{field strength of the gravitino}. The patching conditions of the curvature on overlaps of patches $U_{\alpha}\cap U_{\beta}\subset M$ are then given by
\begin{equation}
    \begin{aligned}
    R_{(\beta)} \,&=\,  \mathrm{Ad}_{h_{(\alpha\beta)}^{-1}}R_{(\alpha)} \\
    T_{(\beta)} \,&=\,  h_{(\alpha\beta)}^{-1} \cdot T_{(\alpha)} \\[0.05cm]
    \rho_{(\beta)} \,&=\,  h_{(\alpha\beta)}^{-1} \cdot \rho_{(\alpha)}
    \end{aligned}
\end{equation}
where $h_{(\alpha\beta)}\in\Coo\big(U_{\alpha}\cap U_{\beta},\,SO(1,d)\big)$ is the \v{C}ech cocycle of the frame bundle $FM$.
\end{remark}

\begin{digression}[Spin connection convention in literature]
Let us consider a spin connection of the following form, on the supermanifold:
\begin{equation}
    \omega^{A}_{\;\;B} \;:=\; \begin{pmatrix}\omega^{a}_{\;\;b} & 0 \\0& \frac{1}{4}\omega^{ab}\Gamma_{ab\;\,\upbeta}^{\;\;\;\upalpha} \end{pmatrix},
\end{equation}
where $\omega^{a}_{\;\;b}$ is the usual (bosonic) spin connection.
\end{digression}

\begin{definition}[Spin-covariant derivative]
Let us define the spin-covariant derivative:
\begin{equation}
    D\zeta \;:=\; \di\zeta + [\omega\,\overset{\wedge}{,}\,\zeta],
\end{equation}
where $\zeta$ is a $\mathfrak{iso}(1,d|\mathbf{N})$-valued differential form.
\end{definition}

\noindent By using the spin-covariant derivative we can rewrite the curvature in a simple fashion:
\begin{equation}
    \begin{aligned}
    R^a_{\;\,b} \,&=\, D\omega^a_{\;\,b} && \in\Omega^2(U,\mathfrak{so}(1,d)), \\
    T^a \,&=\, De^a - \bar{\psi}\Gamma^a \psi && \in\Omega^2(U,\mathbb{R}^{1,d}),\\
    \rho^\upalpha  \,&=\, D\psi^\upalpha && \in\Omega^2(U,\mathbb{R}^{0|\mathbf{N}}).
    \end{aligned}
\end{equation}

\begin{example}[Supervielbein of a flat superspace]
On a flat super-Minkowski space $\mathbb{R}^{1,d|\mathbf{N}}$ with coordinates $(x^\mu,\vartheta^\upalpha)$ we can solve the zero curvature equations $R=T=\rho=0$ and find
\begin{equation}
    \begin{aligned}
    e^\mu &= \di x^\mu + \bar{\vartheta}\Gamma^\mu \di \vartheta \\
    \psi^\upalpha &= \di \vartheta^{\upalpha}
    \end{aligned}
\end{equation}
We notice that even in a flat superspace, e.g. a super-Minkowski space, the differential of the vielbein $\di e^\mu = \bar{\psi}\Gamma^\mu\psi$ is not zero.
\end{example}

\begin{definition}[Bianchi identities of a super-Cartan geometry]
The curvature of a super-Cartan geometry satisfies the Bianchi identities, i.e.
\begin{equation}
    \di (\underline{T},\underline{R}) + \left[(\underline{T},\underline{R}) \,\overset{\wedge}{,}\, (\underline{T},\underline{R})\right] \,=\,0\; \in\,  \Omega^3\big(FM,\, \mathfrak{iso}(1,d|\mathbf{N})\big)
\end{equation}
where, here, $\di$ is the differential on the manifold $FM$.
\end{definition}

\begin{remark}[Local trivialisation of Bianchi identities of a super-Cartan geometry]
\begin{equation}
    \begin{aligned}
    DR^a_{\;\,b} \,&=\, 0   && \in\Omega^3(U,\mathfrak{so}(1,d))\\[0.15cm]
    DT^a \,&=\, R^a_{\;\,b} e^b - 2\bar{\psi}\Gamma^a \rho && \in\Omega^3(U,\mathbb{R}^{1,d})\\
    D\rho^\upalpha  \,&=\, \frac{1}{4}R^{ab}\Gamma_{ab\;\,\upbeta}^{\;\;\;\upalpha}\psi^{\upbeta} && \in\Omega^3(U,\mathbb{R}^{0|\mathbf{N}})
    \end{aligned}
\end{equation}
\end{remark}

\begin{remark}[Super-Cartan geometry as a $\SO(1,d)$-structure]
By equations \eqref{eq:supervielbeincomponents} and \eqref{eq:supervielbeincommutation} the supervielbein, locally given by $1$-forms $E^A=E^A_{M}\di z^M$, can be seen locally as a $\GL(1+d|\mathbf{N})$-valued functions
\begin{equation}
    E^A_{\;M} \,=\, \begin{pmatrix}e^a_\mu & e^a_\upbeta\\\psi^\upalpha_\mu & \psi^\upalpha_\upbeta \end{pmatrix}
\end{equation}
Thus we have the usual picture of non-coordinate basis for differential forms on $M$:
\begin{equation}
    \begin{pmatrix}e^a\\\psi^\upalpha\end{pmatrix} \,=\, \begin{pmatrix}e^a_\mu & e^a_\upbeta\\\psi^\upalpha_\mu & \psi^\upalpha_\upbeta \end{pmatrix} \begin{pmatrix}\di x^\mu\\ \di\vartheta^\upbeta\end{pmatrix}.
\end{equation}
By applying the theory of $G$-structures from chapter \ref{ch:3}, we can easily see that super-Cartan geometry can be obtained by reduction of the structure group $SO(1,d)\hookrightarrow GL(1+d|\mathbf{N})$, where $GL(1+d|\mathbf{N})$ is the structure group of the frame bundle $FM$ prior of the reduction. 
In other words, we have a structure group reduction of the following form:
\begin{equation}
    \begin{tikzcd}[row sep=7ex, column sep=10ex]
        & \B\SO(1,d) \arrow[d, hook] \\
        M \arrow[r, "f"]\arrow[ur, "\hat{f}"] &  \B\GL(1+d|\mathbf{N}).
    \end{tikzcd}
\end{equation}
\end{remark}

\begin{remark}[Field equations from supergeometry]
Interestingly, the equations of motion of supergravity are typically implied by the Bianchi identities together with a \textit{superconstraint}, i.e. an extra constraint on the field strenghts. In other words, we typically have:
\[\text{Bianchi identities \& superconstraint }\implies\text{ field equations}.\]
The superconstraint often takes the form of a Supergravity Torsion Constraints \cite{Howe:1997he}.
\end{remark}

\subsection{Super-Cartan geometry with dilaton}

Now, we want to add a super-dilaton to our formalisation of Cartan geometry.
We will consider the following extension of the moduli stack of the general linear supergroup:
\begin{equation}
    \mathbf{B}GL(1,d|\mathbf{N})^{\mathrm{dil}} \;:=\; \mathbf{B}GL(1,d|\mathbf{N})\ltimes \mathbb{R}^{1|\mathbf{N}}.
\end{equation}
The super-dilaton $(\varphi,\chi)$ will thus be scalar field encoded by a section of the associated $\mathbb{R}^{1|\mathbf{N}}$-bundle $FM\times_{SO(1,d)}\mathbb{R}^{1|\mathbf{N}}$. The bosonic component, i.e. the \textit{dilaton}, will be then a global scalar field $\varphi\in\Coo(M)$, while the fermionic component, i.e. the \textit{dilatino} (or \textit{gravitello}), will not.

\begin{remark}[Local trivialisation of a super-Cartan geometry with dilaton]
Given a good open cover $\mathcal{U}:=\{U_{\alpha}\}$ of the supermanifold $M$, a Cartan connection with super-dilaton on $M$ will be given by the following differential data
\begin{equation}
    \begin{aligned}
    \omega_{(\alpha)} \,&\in\, \Omega^1\big( U_{\alpha},\, \mathfrak{so}(1,d) \big), \\
    e_{(\alpha)} \,&\in\, \Omega^1\big( U_{\alpha},\, \mathbb{R}^{1,d} \big), \\
    \psi_{(\alpha)} \,&\in\, \Omega^1\big( U_{\alpha},\, \mathbb{R}^{0|\mathbf{N}} \big), \\
    \varphi\;\; \;&\in\, \Coo(M), \\
    \chi_{(\alpha)} \,&\in\, \Coo\big( U_{\alpha},\, \mathbb{R}^{0|\mathbf{N}} \big), \\
    h_{(\alpha\beta)} \,&\in\, \Coo\big(U_{\alpha}\cap U_{\beta},\, SO(1,d) \big).
    \end{aligned}
\end{equation}
The patching conditions are the same as \eqref{eq:patching1} with the addition of the following ones:
\begin{equation}
    \begin{aligned}
    \varphi _{(\beta)} \,&=\,\varphi _{(\alpha)}, \\
    \chi_{(\beta)} \,&=\,  h_{(\alpha\beta)}^{-1} \cdot \chi_{(\alpha)}, 
    \end{aligned}
\end{equation}
for the dilaton and dilatino. 
Similarly, dilaton and dilatino transform under gauge transformations of the Cartan connection as follows:
\begin{equation}
    \begin{aligned}
    \varphi_{(\alpha)}' \,&=\,  \varphi_{(\alpha)}, \\
    \chi_{(\alpha)}' \,&=\,  \eta_{(\alpha)}^{-1} \cdot \chi_{(\alpha)}.
    \end{aligned}
\end{equation}
\end{remark}

\begin{remark}[Curvature of a super-Cartan geometry with dilaton]
On any patch $U\subset M$, the curvature of our Cartan geometry will be given by the forms
\begin{equation}
    \begin{aligned}
    R^a_{\;\,b} \,&=\, D\omega^a_{\;\,b} && \in\Omega^2(U,\mathfrak{so}(1,d)), \\
    T^a \,&=\, De^a - \bar{\psi}\Gamma^a \psi && \in\Omega^2(U,\mathbb{R}^{1,d}),\\
    \rho^\upalpha  \,&=\, D\psi^\upalpha && \in\Omega^2(U,\mathbb{R}^{0|\mathbf{N}}),\\
    F_{\mathrm{dil}} \,&=\, \di\varphi && \in\Omega^1(M),\\
    \rho^\upalpha_{\mathrm{dil}} \,&=\, D\chi^\upalpha && \in\Omega^1(U,\mathbb{R}^{0|\mathbf{N}}),
    \end{aligned}
\end{equation}
where $F_{\mathrm{dil}}$ and $\rho^\upalpha_{\mathrm{dil}}$ are respectively the curvature of the dilaton and the dilatino.
\end{remark}

\begin{remark}[Bianchi identities of a super-Cartan geometry with dilaton]
On any patch $U\subset M$, the Bianchi identities of our Cartan geometry will be given by the equations
\begin{equation}
    \begin{aligned}
    DR^a_{\;\,b} \,&=\, 0   && \in\Omega^3(U,\mathfrak{so}(1,d)),\\
    DT^a \,&=\, R^a_{\;\,b} e^b - 2\bar{\psi}\Gamma^a \rho && \in\Omega^3(U,\mathbb{R}^{1,d}),\\
    D\rho^\upalpha  \,&=\, \frac{1}{4}R^{ab}\Gamma_{ab\;\,\upbeta}^{\;\;\;\upalpha}\psi^{\upbeta} && \in\Omega^3(U,\mathbb{R}^{0|\mathbf{N}}),\\
    \di F_{\mathrm{dil}} \,&=\, 0 && \in\Omega^2(M),\\
    D\rho^\upalpha_{\mathrm{dil}} \,&=\, \frac{1}{4}R^{ab}\Gamma_{ab\;\,\upbeta}^{\;\;\;\upalpha}\chi^{\upbeta} && \in\Omega^2(U,\mathbb{R}^{0|\mathbf{N}}).
    \end{aligned}
\end{equation}
\end{remark}

\begin{remark}[Geometrical and physical meaning of dilaton and dilatino]
Components $\phi = e^{10}_{\;\;\,10}$ and $\chi^\upalpha = \psi^\upalpha_{\;\;10}$ of the supervielbein of a $(1+10|\mathbf{32})$-dimensional supermanifold.
\end{remark}

\begin{remark}[Dilatino or gravitello]
This dissertation has been following the physical convention that the spinor $\chi^\upalpha$ is named \textit{dilatino}, even if this nomenclature is misleading. Indeed, $\chi^\upalpha$ is not the superpartner of the dilaton field (which is instead $\partial_\upalpha\varphi$), but a totally different field coming from the dimensional reduction of the gravitino of $11$-dimensional supergravity.
\end{remark}

\section{Type II super-Cartan geometry}

Type II super-Cartan geometry was developed by \cite{AFGT08}.

\begin{remark}[Fierz identities]
The following identities are called \textit{Fierz identities}:
\begin{equation}
\begin{aligned}
       \Gamma_a^\IIA\psi\wedge\bar{\psi}\Gamma^a_\IIA\psi=0 \quad\; \implies \;\quad i\bar{\psi}\Gamma_a^{\IIA}\psi\wedge e^a \,\in\, \Omega^3_{\mathrm{cl}}\!\left(\mathbb{R}^{1,9|\mathbf{16}\oplus\overline{\mathbf{16}}}\right), \\
       \Gamma_a^\IIB\psi\wedge\bar{\psi}\Gamma^a_\IIB\psi=0 \quad\; \implies \;\quad i\bar{\psi}\Gamma_a^{\IIB}\psi\wedge e^a \,\in\, \Omega^3_{\mathrm{cl}}\!\left(\mathbb{R}^{1,9|\mathbf{16}\oplus{\mathbf{16}}}\right).
\end{aligned}
\end{equation}
\end{remark}
\noindent These Fierz identities identify super-algebra $3$-cocycles on the Type IIA and IIB super-Minkowski spaces. Recall that a $3$-cocycle on a Lie algebra $\mathfrak{g}$ can be dually interpreted as a map to the Chevalley-Eilenberg dg-algebra $\mathrm{CE}(\mathfrak{g})$ from the dg-algebra $\mathrm{CE}\!\left(\mathbf{b}^3\mathfrak{u}(1)\right) = \!\left(\mathbb{R}[3],\,\di=0\right)$.
By Lie integration, we can use such cocycles to define super Lie group extensions of the supertranslation group (i.e. of the super-Minkowski space).

\begin{definition}[Type IIA supergravity $2$-group]
The \textit{Type IIA string $2$-group} $\String_\IIA$ is defined by the following group extension of the Type IIA super-Minkowski space:
\begin{equation}
\begin{tikzcd}[row sep=16ex, column sep=16ex]
\mathbf{B}\String_\IIA \arrow[r]\arrow[d, "\mathrm{hofib}(i\bar{\psi}\Gamma^a_{\mathrm{IIA}}\Gamma_{10}\psi\wedge e^a)"']&\ast\arrow[d] \\ 
\mathbf{B}\mathbb{R}^{1,9|\mathbf{16}\oplus\overline{\mathbf{16}}} \arrow[r, "{i\bar{\psi}\Gamma^a_{\mathrm{IIA}}\Gamma_{10}\psi\wedge e^a}"] & \mathbf{B}^3U(1).
\end{tikzcd}
\end{equation}
See \cite{Fiorenza:2013nha} for details.
The \textit{Type IIA supergravity $2$-group} $\SugraIIA$ is now defined by:
\begin{equation}
    \SugraIIA \;:=\; \String_\IIA \rtimes SO(1,d),
\end{equation}
where $SO(1,d)$ acts on the subgroup $\mathbb{R}^{1,9|\mathbf{16}\oplus\overline{\mathbf{16}}}$.
Notice that we have the coset spaces
\begin{equation}
\begin{aligned}
    \String_\IIA \;&\cong\; \frac{\SugraIIA}{SO(1,9)}, \\[0.1cm]
    \mathbb{R}^{1,9|\mathbf{16}\oplus\overline{\mathbf{16}}} \;&\cong\; \frac{\SugraIIA}{SO(1,9)\times\mathbf{B}U(1)}.
\end{aligned}
\end{equation}
\end{definition}

\begin{definition}[Type IIB supergravity $2$-group]
The \textit{Type IIB string $2$-group} $\SugraIIB$ is defined by the following group extension of the Type IIB super-Minkowski space:
\begin{equation}
\begin{tikzcd}[row sep=16ex, column sep=16ex]
\mathbf{B}\String_\IIB \arrow[r]\arrow[d, "\mathrm{hofib}(i\bar{\psi}\Gamma^a_{\mathrm{IIB}}\Gamma_{10}\psi\wedge e^a)"']&\ast\arrow[d] \\ 
\mathbf{B}\mathbb{R}^{1,9|\mathbf{16}\oplus{\mathbf{16}}} \arrow[r, "i\bar{\psi}\Gamma^a_{\mathrm{IIB}}\Gamma_{10}\psi\wedge e^a"] & \mathbf{B}^3U(1)
\end{tikzcd}
\end{equation}
The \textit{Type IIB supergravity $2$-group} $\SugraIIB$ is now defined by:
\begin{equation}
    \SugraIIA \;:=\; \String_\IIA \rtimes SO(1,d),
\end{equation}
where $SO(1,d)$ acts on the subgroup $\mathbb{R}^{1,9|\mathbf{16}\oplus\mathbf{16}}$.
Notice that we have the coset spaces
\begin{equation}
\begin{aligned}
    \String_\IIB \;&\cong\; \frac{\SugraIIA}{SO(1,9)}, \\[0.1cm]
    \mathbb{R}^{1,9|\mathbf{16}\oplus{\mathbf{16}}} \;&\cong\; \frac{\SugraIIB}{SO(1,9)\times\mathbf{B}U(1)}.
\end{aligned}
\end{equation}
\end{definition}

\begin{remark}[Local trivialisation of a Type IIA Supergravity]
On any local patch $U_\alpha\subset M$ and overlaps of patches, the connection of Type IIA Supergravity is given by the following differential data:
\begin{equation}
    \begin{aligned}
    \omega_{(\alpha)} \,&\in\, \Omega^1\big( U_{\alpha},\, \mathfrak{so}(1,9) \big), \\
    e_{(\alpha)} \,&\in\, \Omega^1\big( U_{\alpha},\, \mathbb{R}^{1,9} \big), \\
    \psi_{(\alpha)} \,&\in\, \Omega^1\big( U_{\alpha},\, \mathbb{R}^{0|\mathbf{16}\oplus\overline{\mathbf{16}}} \big) \\
    B_{(\alpha)}\,&\in\, \Omega^2\big( U_{\alpha}\big), \\
    \varphi\;\; \;&\in\, \Coo(M), \\
    \chi_{(\alpha)} \,&\in\, \Coo\big( U_{\alpha},\, \mathbb{R}^{0|\mathbf{16}\oplus\overline{\mathbf{16}}} \big), \\
    h_{(\alpha\beta)} \,&\in\, \Coo\big(U_{\alpha}\cap U_{\beta},\, SO(1,9) \big), \\
    \Lambda_{(\alpha\beta)} \,&\in\, \Omega^1\big(U_{\alpha}\cap U_{\beta}\big), \\
    G_{(\alpha\beta\gamma)} \,&\in\, \Coo\big(U_{\alpha}\cap U_{\beta}\cap U_{\gamma}\big). 
    \end{aligned}
\end{equation}
\end{remark}

Patching conditions on $n$-fold overlaps of patches are the following:
\begin{equation}\label{eq:patchingIIA}
    \begin{aligned}
    \omega_{(\beta)} \,&=\,  \mathrm{Ad}_{h_{(\alpha\beta)}^{-1}}\omega_{(\alpha)} + h_{(\alpha\beta)}^{-1}\di h_{(\alpha\beta)}, \\
    e_{(\beta)} \,&=\,  h_{(\alpha\beta)}^{-1} \cdot e_{(\alpha)}, \\
    \psi_{(\beta)} \,&=\,  h_{(\alpha\beta)}^{-1} \cdot \psi_{(\alpha)}, \\
    B_{(\beta)} \,&=\,  B_{(\alpha)} + \di \Lambda_{(\alpha\beta)}, \\
    \varphi _{(\beta)} \,&=\,\varphi _{(\alpha)}, \\
    \chi_{(\beta)} \,&=\,  h_{(\alpha\beta)}^{-1} \cdot \chi_{(\alpha)}, \\
    h_{(\alpha\beta)}\cdot h_{(\beta\gamma)}\cdot h_{(\gamma\alpha)} \,&=\, 1, \\
    \Lambda_{(\alpha\beta)} + \Lambda_{(\beta\gamma)} + \Lambda_{(\gamma\alpha)} \,&=\, \di G_{(\alpha\beta\gamma)},  \\
    G_{(\alpha\beta\gamma)}-G_{(\beta\gamma\delta)}+G_{(\gamma\delta\alpha)}-G_{(\delta\alpha\beta)} \,&\in\, 2\pi\mathbb{Z}.
    \end{aligned}
\end{equation}

\begin{remark}[Curvature of Type IIA Supergravity]
On any patch $U\subset M$
\begin{equation}
    \begin{aligned}
    R^a_{\;\,b} \,&=\, D\omega^a_{\;\,b} && \in\Omega^2(U,\mathfrak{so}(1,9)) \\
    T^a \,&=\, De^a - \bar{\psi}\Gamma^a_{\mathrm{IIA}} \psi && \in\Omega^2(U,\mathbb{R}^{1,9})\\
    \rho^\upalpha  \,&=\, D\psi^\upalpha && \in\Omega^2(U,\mathbb{R}^{0|\mathbf{16}\oplus\overline{\mathbf{16}}})\\
    H\,&=\, \di B + i\bar{\psi}\Gamma_a^{\mathrm{IIA}}\Gamma_{10}\psi\wedge e^a  && \in\Omega^3(M) \\
    F_{\mathrm{dil}} \,&=\, \di\varphi && \in\Omega^1(M)\\
    \rho^\upalpha_{\mathrm{dil}} \,&=\, D\chi^\upalpha && \in\Omega^1(U,\mathbb{R}^{0|\mathbf{16}\oplus\overline{\mathbf{16}}})
    \end{aligned}
\end{equation}
\end{remark}

\begin{remark}[Bianchi identities of Type IIA Supergravity]
On any patch $U\subset M$
\begin{equation}
    \begin{aligned}
    DR^a_{\;\,b} \,&=\, 0   && \in\Omega^3(U,\mathfrak{so}(1,9))\\
    DT^a \,&=\, R^a_{\;\,b} e^b - 2\bar{\psi}\Gamma^a_{\mathrm{IIA}} \rho && \in\Omega^3(U,\mathbb{R}^{1,9})\\
    D\rho^\upalpha  \,&=\, \frac{1}{4}R^{ab}\Gamma_{ab\;\;\;\upbeta}^{\IIA\upalpha}\psi^{\upbeta} && \in\Omega^3(U,\mathbb{R}^{0|\mathbf{16}\oplus\overline{\mathbf{16}}})\\
    \di H\,&=\, 2i\bar{\psi}\Gamma_a^{\mathrm{IIA}}\Gamma_{10}\rho\wedge e^a - i \bar{\psi}\Gamma_a^{\mathrm{IIA}}\Gamma_{10}\psi\wedge T^a && \in\Omega^4(M)\\
    \di F_{\mathrm{dil}} \,&=\, 0 && \in\Omega^2(M)\\
    D\rho^\upalpha_{\mathrm{dil}} \,&=\, \frac{1}{4}R^{ab}\Gamma_{ab\;\;\;\upbeta}^{\mathrm{IIA}\upalpha}\chi^{\upbeta} && \in\Omega^2(U,\mathbb{R}^{0|\mathbf{16}\oplus\overline{\mathbf{16}}})
    \end{aligned}
\end{equation}
\end{remark}

\noindent The \textit{string frame} is defined by the following equations
\begin{equation}
    \begin{aligned}
    T^a \,&\, =0, \\
    H \,&\, =\mathscr{H}_{abc}e^a\wedge e^b \wedge e^c,
    \end{aligned}
\end{equation}
for some function $\mathscr{H}_{abc}$.

\begin{remark}[Supervectors]
Consider the push-forward of the canonical bosonic inclusion $\iota:\bosonic{M}\hookrightarrow M$. This will induce a short exact sequence
\begin{equation}
    \begin{tikzcd}[row sep=scriptsize, column sep=8ex] 
    0 \arrow[r, two heads] & \mathrm{Ker}(\iota_\ast) \arrow[r, two heads] & T\bosonic{M} \arrow[r, "\iota_\ast", hook] & TM \arrow[r, hook] & 0.
\end{tikzcd}
\end{equation}
Let us call $S:=\mathrm{Ker}(\iota_\ast)$. We can construct the isomorphism
\begin{equation}
    TM|_{\bosonic{U}} \;\cong\; T\bosonic{U}\oplus S\big|_{\bosonic{U}}.
\end{equation}
Thus, we can write a supervector in components by $X+\epsilon\in\Gamma\big(\bosonic{M},TM\big)$, where $X=X^a\partial_a$ and $\epsilon = \epsilon^\upalpha \partial_\upalpha$.
If we consider a pure spinorial vector $\epsilon = \epsilon^\upalpha \partial_\upalpha$ and we consider infinitesimal translations on a super-Cartan geometry, we find the local supersymmetry:
\end{remark}
\begin{equation}
    \begin{aligned}
    \mathcal{L}_\epsilon e^a &= \bar{\psi}\Gamma^{a}\epsilon, \\
    \mathcal{L}_\epsilon \psi &= D \epsilon, \\
    \mathcal{L}_\epsilon B &= i(\bar{\psi}\Gamma_{a}\Gamma_{10}\epsilon)\wedge e^a, \\
    \mathcal{L}_\epsilon \varphi &= 0, \\
    \mathcal{L}_\epsilon \chi &= 0.
    \end{aligned}
\end{equation}

\section{Super Yang-Mills theory}

Let $\gamma_a$ be the gamma-matrices of the spin representation $\mathbf{16}$ and let $M$ be a $(1+9|\mathbf{16})$-dimensional supermanifold. 
A Super Yang-Mills theory on $M$ is encoded by a $G$-bundle of the following form:
\begin{equation}\begin{tikzcd}[row sep=7ex, column sep=6ex]
P_{\mathrm{SYM}} \arrow[d, "\pi"]\arrow[r] &\ast \arrow[d]  \\
M\arrow[r, "h"] &\mathbf{B}G
\end{tikzcd}\end{equation}
The map $h\in\mathbf{H}(M,\mathbf{B}G)$ is given in local data by a $G$-valued function $h_{(\alpha\beta)}\in\Coo\big(U_{\alpha}\cap U_{\beta}, G\big)$ on each two-fold overlap of patches of $M$ such than they satisfy the \v{C}ech cocycle condition
\begin{equation}
    h_{(\alpha\beta)}\cdot h_{(\beta\gamma)}\cdot h_{(\gamma\alpha)} \;=\; 1
\end{equation}
on each three-fold overlap of patches.

\begin{definition}[Connection and curvature]
A connection for $M\xrightarrow{h}\mathbf{B}G$ is defined by a collection of local $\mathfrak{g}$-valued $1$-forms $A_{(\alpha)}\in\Omega^1(U_{\alpha},\mathfrak{g})$ patched together by
\begin{equation}
    A_{(\beta)}  \;=\; \mathrm{Ad}_{h_{(\alpha\beta)}^{-1}}A_{(\alpha)} +h_{(\alpha\beta)}^{-1}\di h_{(\alpha\beta)}
\end{equation}
on two-fold overlaps of patches. The curvature $F\!\in\!\Omega^2(M,\mathrm{ad}(P_{\mathrm{SYM}}))$ of the bundle is defined as
\begin{equation}\label{eq:SYMcurvature}
    \begin{aligned}
    F_{(\alpha)} \;=\; \di  A_{(\alpha)} + \big[ A_{(\alpha)} \,\overset{\wedge}{,}\,  A_{(\alpha)} \big]_{\mathfrak{g}}
    \end{aligned}
\end{equation}
and it is patched in the adjoint representation of the transition functions by
\begin{equation}
    F_{(\beta)}  \;=\; \mathrm{Ad}_{h_{(\alpha\beta)}^{-1}}F_{(\alpha)}.
\end{equation}
\end{definition}

\noindent The Bianchi identity of the curvature is given by
\begin{equation}
    \nabla F_{(\alpha)} = 0.
\end{equation}
An infinitesimal gauge transformation of the connection is given by $\delta_{\zeta_{(\alpha)}} A_{(\alpha)} = \nabla\zeta_{(\alpha)} = \di\zeta_{(\alpha)} + [A_{(\alpha)}, \zeta_{(\alpha)}]$, where $\zeta_{(\alpha)}\in\Coo(M,\mathfrak{g})$ is an infinitesimal gauge parameter.

\begin{definition}[Gauge transformation]
A gauge transformation is a coboundary of the cocycle $h\in\mathbf{H}(M,\mathbf{B}G)$, i.e. a morphism $(A_{(\alpha)},h_{(\alpha\beta)})\xmapsto{(\zeta_{(\alpha)})} (A_{(\alpha)}',h_{(\alpha\beta)}')$ given in local data by local functions $\zeta_{(\alpha)}\in\Coo(U_{\alpha},G)$ on each patch such that
\begin{equation}
    \begin{aligned}
    h_{(\alpha\beta)}' \;&=\; \zeta_{(\alpha)}^{-1}h_{(\alpha\beta)}\zeta_{(\beta)} \\
    A_{(\alpha)}^\prime  \;&=\; \mathrm{Ad}_{\zeta_{(\alpha)}^{-1}}A_{(\alpha)} +\zeta_{(\alpha)}^{-1}\di\zeta_{(\alpha)}
    \end{aligned}
\end{equation}
\end{definition}

\begin{definition}[Covariant derivative]
The covariant derivative of the connection $A_{(\alpha)}$ of the principal $G$-bundle $P_{\mathrm{SYM}}$ is defined on any $\mathfrak{g}$-valued differential form $\zeta$ by
\begin{equation}
    \nabla\zeta \;:=\; \di\zeta + [A_{(\alpha)}\,\overset{\wedge}{,}\,\zeta]_{\mathfrak{g}}
\end{equation}
This is given in components as follows:
\begin{equation}
    \begin{aligned}
    \nabla_a\zeta &= \;\partial_a\zeta \,+ \,[A_{(\alpha)a},\,\zeta]_{\mathfrak{g}} \\
    \nabla_\upalpha\zeta &= \partial_\upalpha\zeta + \{A_{(\alpha)\upalpha},\zeta\}_{\mathfrak{g}}
    \end{aligned}
\end{equation}
where we decomposed as $\di = e^a\partial_a + \psi^\upalpha \partial_\upalpha$ the differential on the supermanifold $M$.
\end{definition}

\begin{remark}[Connection and curvature in bosonic and fermionic components]
Notice that the following expansion holds:
\begin{equation}
    \begin{aligned}
    A_{(\alpha)} \;&=\; A_{(\alpha)a}\,e^a + A_{(\alpha)\upalpha}\, \psi^\upalpha, \\
    F_{(\alpha)} \;&=\; F_{(\alpha)ab}\,e^a\wedge e^b + F_{(\alpha)a\upbeta}\,e^a\wedge \psi^\upbeta + F_{(\alpha)\upalpha\upbeta}\,\psi^\upalpha \wedge \psi^\upbeta,
    \end{aligned}
\end{equation}
If we also rewrite \eqref{eq:SYMcurvature} in bosonic and fermionic components we obtain the equations
\begin{equation}
    \begin{aligned}\label{eq:forcecomponents}
    F_{(\alpha)ab} \;&=\; \partial_{a}A_{(\alpha)b}-\partial_{b}A_{(\alpha)a} + \big[ A_{(\alpha)a},\,  A_{(\alpha)b} \big]_{\mathfrak{g}}, \\
    F_{(\alpha)a\upbeta} \;&=\; \partial_{a}A_{(\alpha)\upbeta}-\partial_{\upbeta}A_{(\alpha)a} + \big[ A_{(\alpha)a},\,  A_{(\alpha)\upbeta} \big]_{\mathfrak{g}}, \\
    F_{(\alpha)\upalpha\upbeta} \;&=\; \partial_{\upalpha}A_{(\alpha)\upbeta}+\partial_{\upbeta}A_{(\alpha)a}+ \gamma^a_{\upalpha\upbeta}A_{(\alpha)a} + \big\{ A_{(\alpha)\upalpha},\,  A_{(\alpha)\upbeta} \big\}_{\mathfrak{g}},
    \end{aligned}
\end{equation}
where the superderivative is given by $\partial_\upalpha = \partial/\partial\vartheta^\upalpha - (\gamma^a\vartheta)_\upalpha\partial_a$.
\end{remark}

\begin{remark}[Superconstraint]
The superconstraint of Super Yang-Mills theory is the equation $F_{(\alpha)\upalpha\upbeta}=0$. See \cite{Harnad:1985bc} for more details.
\end{remark}

\noindent Let us explicitly show that the differential data of a principal $G$-bundle which satisfies the superconstraint are exactly given by a super Yang-Mills theory.

\begin{theorem}[Super Yang-Mills theory $=$ $G$-bundle with superconstraint]
The connection of a $G$-bundle on a $(1+9|\mathbf{16})$-dimensional supermanifold $M$ satisfying the superconstraint is equivalently given by a gauge field $\mathscr{A}$ and a gaugino field $\mathscr{X}$ on the $(1+9)$-dimensional manifold $\bosonic{M}$ minimizing the super Yang-Mills action
\begin{equation}\label{eq:symaction}
    S[\mathscr{A},\mathscr{X}] \;=\; \int_{\bosonic{M}}\!\di^{10}x\,\, \tr_{\mathfrak{g}}\bigg( -\frac{1}{4}\mathscr{F}^{ab}\mathscr{F}_{ab} + \mathscr{X}\slashed{\nabla}\mathscr{X} \bigg)
\end{equation}
where $\mathscr{F}_{ab}:=\partial_a\mathscr{A}_b-\partial_b\mathscr{A}_a +[\mathscr{A}_a,\mathscr{A}_b]_{\mathfrak{g}}$ is the strength of the gauge field.
\end{theorem}

\begin{proof}[Sketch of the proof]The superconstraint $F_{\upalpha\upbeta}=0$ can be rewritten on any patch $U_{\alpha}\subset M$ by using equation \eqref{eq:forcecomponents} as
\begin{equation}
    D_{\upalpha}A_{(\alpha)\upbeta}+D_{\upbeta}A_{(\alpha)a}+ \gamma^a_{\upalpha\upbeta}A_{(\alpha)a} + \big\{ A_{(\alpha)\upalpha},\,  A_{(\alpha)\upbeta} \big\}_{\mathfrak{g}} =0
\end{equation}
and, more compactly, as $\gamma^a_{\upalpha\upbeta}A_{(\alpha)a} +\nabla_{(\upalpha}A_{(\alpha)\upbeta)}=0$. This immediately implies the relation $A_{(\alpha)a} = -\frac{1}{8}\gamma^{\upalpha\upbeta}_a\nabla_\upalpha A_{(\alpha)\upbeta}$ between bosonic and fermionic components of the connection. Now we can define a $\mathfrak{g}$-valued scalar by $\chi^\upalpha_{(\alpha)} := \nabla^\upbeta \nabla^\upalpha A_{(\alpha)\upbeta}$. If we plug this definition in \eqref{eq:forcecomponents} we can explicit the components of the curvature of the bundle by
\begin{equation}
    \begin{aligned}
    F_{(\alpha)ab} \;&=\; \gamma_{ab\upalpha}^{\quad\;\upbeta}\nabla_\upbeta \chi^\upalpha_{(\alpha)} \\
    F_{(\alpha)a\upbeta} \;&=\; \gamma_{a\upalpha\upbeta} \chi^\upalpha_{(\alpha)} \\
    F_{(\alpha)\upalpha\upbeta} \;&=\; 0.
    \end{aligned}
\end{equation}
Thus the superconstraint implies that the curvature of the bundle is of the form 
\begin{equation}
    F_{(\alpha)} \; =\; \big(\gamma_{ab\upalpha}^{\quad\;\upbeta}\nabla_\upbeta \chi^\upalpha_{(\alpha)}\big) \,e^a\wedge e^b  + \bar{\psi}\gamma_{a}\chi_{(\alpha)}.
\end{equation}
Sketchily, we can expand $A$ and $\chi$ on each patch $U\subset M$ of the base supermanifold with local coordinates $(x,\vartheta)$ in the fermionic coordinates as follows. First, the bosonic components can be expanded as
\begin{equation}
    \begin{aligned}
    A_a(x,\vartheta) \,=\, \mathscr{A}_a(x) + \bar{\vartheta}\gamma_a\mathscr{X}(x) + \mathcal{O}(\vartheta^2).
    \end{aligned}
\end{equation}
From this, by applying equation $A_{a} = -\frac{1}{8}\gamma^{\upalpha\upbeta}_a\nabla_\upalpha A_{\upbeta}$, we can obtain
\begin{equation}
    \begin{aligned}
    A_{\upalpha}(x,\vartheta) \,=\, 0 + (\bar{\vartheta}\gamma^a)_\upalpha\mathscr{A}_a(x) + \frac{1}{2}\vartheta^2\mathscr{X}_\upalpha(x).
    \end{aligned}
\end{equation}
Finally, by applying equation $\chi^\upalpha = \nabla^\upbeta \nabla^\upalpha A_{\upbeta}$, we can obtain
\begin{equation}
    \begin{aligned}
    \chi^\upalpha(x,\vartheta) \,=\, \mathscr{X}^\alpha(x) + (\bar{\vartheta}\gamma^{ab})^\upalpha\mathscr{F}_{ab}(x)+\mathcal{O}(\vartheta^2).
    \end{aligned}
\end{equation}
Now notice that we can recover the gauge and gaugino field of super Yang-Mills theory respectively by
\begin{equation}
    \begin{aligned}
    A(x,\vartheta)\big|_{\!\!{\substack{\vartheta=0\\\;\;\di\vartheta=0}}} =\, \mathscr{A}(x)\in\, \Omega^1\big(\bosonic{U}, \, \mathfrak{g}\big), \qquad
    \chi(x,\vartheta)\big|_{\!\!{\substack{\vartheta=0\\\;\;\di\vartheta=0}}} =\, \mathscr{X}(x) \in\, \gamma\big(\bosonic{U}, \, S\otimes\mathfrak{g}\big),
    \end{aligned}
\end{equation}
in clear compatibility with rheonomy principle. Here $\mathscr{A}$ can be interpreted as the connection of the bosonic principal $G$-bundle $\bosonic{P}_{\mathrm{SYM}}\rightarrow\bosonic{M}$. We can now rewrite the Bianchi identity $\nabla F=0$ in terms of the fields $\mathscr{A}$ and $\mathscr{X}$ by
\begin{equation}
    \nabla^a\mathscr{F}_{ab} \,=\, [\mathscr{X},\, \gamma_b\mathscr{X}]_{\mathfrak{g}}, \quad\; \slashed{\nabla}\mathscr{X} \,=\, 0,
\end{equation}
where $\slashed{\nabla}:=\gamma^a\nabla_a$. These are the super Yang-Mills equations coming from \eqref{eq:symaction}. For a more detailed discussion see \cite{Harnad:1985bc}.
\end{proof}

\section{Heterotic Supergravity}

Let $M$ be a $(1+9|\mathbf{16})$-dimensional supermanifold.
As firstly delineated by \cite{Sat09}, heterotic Supergravity is encoded by a principal $\String^{\mathbf{c}_2}$-bundle of the following form:
\begin{equation}\begin{tikzcd}[row sep=14ex, column sep=5ex]
\mathscr{G}_{\mathrm{het}}\arrow[d, "\Pi"]\arrow[r] &\ast \arrow[d] \\ 
FM\times_{M} \!P  \arrow[d, "\pi"]\arrow[r] &\mathbf{B}^2U(1) \arrow[d]\arrow[r] & \ast \arrow[d]  \\
M\arrow[r] & \mathbf{B}\String^{\mathbf{c}_2}(1,9) \arrow[r] &\mathbf{B}(\mathrm{Spin}(1,9)\times G).
\end{tikzcd}\end{equation}
Let $\gamma_a$ be the gamma-matrices of the spin representation $\mathbf{16}$. 
The curvature of such a bundle must be, on any patch $U_\alpha\subset M$, of the form
\begin{equation}
    \begin{aligned}
    R_{(\alpha)b}^{\;\;\,a} \;&=\; \di  \omega_{(\alpha)b}^{\;\;\,a} + \omega_{(\alpha)c}^{\;\;\,a} \wedge  \omega_{(\alpha)b}^{\;\;\,c}, \\
    T^a_{(\alpha)} \,&=\, \di e^a_{(\alpha)} + \omega^{\;\;\,a}_{(\alpha)b}\wedge e^b_{(\alpha)} - \bar{\psi}_{(\alpha)}\gamma^a \psi_{(\alpha)} \\
    F_{(\alpha)} \;&=\; \di  A_{(\alpha)} + \big[ A_{(\alpha)} \wedge  A_{(\alpha)} \big]_{\mathfrak{g}}, \\
    H_{(\alpha)} \;&=\; \di B_{(\alpha)} + \bar{\psi}_{(\alpha)}\gamma_a\psi_{(\alpha)} \wedge e^a_{(\alpha)} + \mathrm{cs}_3\big(A_{(\alpha)}\big) -  \mathrm{cs}_3\big(\omega_{(\alpha)}\big), \\
    \rho_{(\alpha)} \;&=\; D\psi_{(\alpha)},
    \end{aligned}
\end{equation}
where $\mathrm{cs}_3\big(A_{(\alpha)}\big)$ and $\mathrm{cs}_3\big(A_{(\alpha)}\big)$ are the Chern-Simons super $3$-forms of the two connections, defined by the expressions
\begin{equation}
    \begin{aligned}
    \mathrm{cs}_3\big(A_{(\alpha)}\big) &:= \tr_{\mathfrak{g}}\Big(A_{(\alpha)}\wedge F_{(\alpha)} +\frac{2}{3}A_{(\alpha)}\wedge \big[A_{(\alpha)}\wedge A_{(\alpha)}\big]_{\mathfrak{g}}\Big) \\
        \mathrm{cs}_3\big(\omega_{(\alpha)}\big) &:= \tr_{\mathfrak{spin}(1,9)}\Big(\omega_{(\alpha)}\wedge R_{(\alpha)} +\frac{2}{3}\omega_{(\alpha)}\wedge \big[\omega_{(\alpha)}\wedge \omega_{(\alpha)}\big]_{\mathfrak{spin}(1,9)}\Big). \\
    \end{aligned}
\end{equation}
\noindent The following can be seen as a global closed $3$-form on the total space of the $G$-bundle $P$:
\begin{equation}
    \pi^\ast H+\mathrm{cs}_3(\underline{\omega})-\mathrm{cs}_3(\underline{A}) \;\in\, \Omega^3_{\mathrm{cl}}(FM\times_{M}\! P)
\end{equation}
where $\underline{A}\in\Omega^1(P,\mathfrak{g})$ and $\underline{\omega}\in\Omega^1(FM,\mathfrak{so}(1,9))$ are the global Ehresmann connections respectively of the bundles $P\twoheadrightarrow M$ and $FM\twoheadrightarrow M$.

\begin{remark}[Superconstraint]
The superconstraint of Heterotic Supergravity \cite{Bonora:1990mt} is given by the equations $T^a_{(\alpha)}=0$, $F_{(\alpha)\upalpha\upbeta}=0$ and $H_{(\alpha)\upalpha\upbeta\upgamma}=0$.
\end{remark}

\begin{remark}[Topology of Heterotic doubled space]
Topologically, principal bundles $P\twoheadrightarrow M$ are classified by the $2$nd Chern class $c_2(P)\in H^4(M,\mathbb{Z})$, while the frame bundle $FM\twoheadrightarrow M$ by and $1$st fractional Pontryagin class $\frac{1}{2}p_1(FM)\in H^4(M,\mathbb{Z})$. However, we need to impose the Green-Schwarz anomaly cancellation on the topology of the bundle:
\begin{equation}
    \frac{1}{2}p_1(FM)-c_2(P)=0 \;\,\in\, H^4(M,\mathbb{Z}) \qquad \text{(\textit{Green-Schwarz anomaly}).}
\end{equation}
Thus, the Heterotic doubled spaces will be topologically classified by the following data:
\begin{equation}
    \begin{aligned}
        \frac{1}{2}p_1(FM)=c_2(P) \;&\in\, H^4(M,\mathbb{Z}), \\
        \big[ \pi^\ast H+\mathrm{cs}_3(\underline{\omega})-\mathrm{cs}_3(\underline{A}) \big] \;&\in\, H^3(FM\times_M\! P,\mathbb{Z}).
    \end{aligned}
\end{equation}
\end{remark}

\section{Rheonomy principle}

\begin{remark}[Pullback of the canonical embedding]
The pullback of the canonical embedding of a reduced manifold $\iota:\bosonic{M}\longhookrightarrow M$ induces a pullback
\begin{equation}
    \begin{tikzcd}[row sep=scriptsize, column sep=8ex] 
    \Omega^\bullet(M,\mathfrak{g}) \arrow[r, "\iota^\ast", two heads] & \Omega^\bullet(\bosonic{M},\mathfrak{g})
\end{tikzcd}
\end{equation}
for superdifferential forms valued in any super Lie algebra $\mathfrak{g}$. 
Locally, let $\xi\in\Omega^k(U_{\alpha},\mathfrak{g})$ where the open set $U_{\alpha}\subset M$ is equipped with local coordinates $(x,\vartheta)$. Thus, in local coordinates the pullback is given by
\begin{equation}
\iota^\ast\xi \,=\, \xi\big|_{\!\!{\substack{\vartheta=0\\\;\;\di\vartheta=0}}}.
\end{equation}
More in general, we can patch-wise apply this pullback of superdifferential forms to define a map of Cartan geometries
\begin{equation}
    \begin{tikzcd}[row sep=scriptsize, column sep=8ex] 
    \Cartan^{\mathfrak{g}}(M) \arrow[r, "\iota^\ast", two heads] & \Cartan^{\mathfrak{g}}(\bosonic{M}),
\end{tikzcd}
\end{equation}
where $\Cartan^{\mathfrak{g}}(M)$ is a super-Cartan geometry locally modelled on the super $L_\infty$-algebra $\mathfrak{g}$ (e.g. $\mathfrak{g}=\mathfrak{sugra}_\IIA$ for Type IIA Supergravity) and $\Cartan^{\mathfrak{g}}(\bosonic{M})$ is its bosonic Cartan geometry on the reduced manifold $\bosonic{M}$.
\end{remark}

\begin{example}[Pullback of supergravity fields]
Explicitly, on a patch $U_{\alpha}$ with local coordinates $\{x^\mu,\vartheta^\upalpha\}$, we have that the supergravity fields are mapped by the pullback $\iota^\ast$ as follows:
\begin{equation}
    \begin{aligned}
    \omega_{(\alpha)}&=\omega_{(\alpha)\mu}(x,\vartheta)\di x^{\mu} + \omega_{(\alpha)\upalpha}(x,\vartheta)\di\vartheta^{\upalpha} &&\longmapsto& \omega_{(\alpha)\mu}(x,0)\di x^{\mu} ,\\
    e_{(\alpha)} &= e_{(\alpha)\mu}(x,\vartheta)\di x^{\mu} + e_{(\alpha)\upalpha}(x,\vartheta)\di\vartheta^{\upalpha}   &&\longmapsto& e_{(\alpha)\mu}(x,0)\di x^{\mu} ,\\
    \psi_{(\alpha)} &= \psi_{(\alpha)\mu}(x,\vartheta)\di x^{\mu} + \psi_{(\alpha)\upalpha}(x,\vartheta)\di\vartheta^{\upalpha}  &&\longmapsto&\psi_{(\alpha)\mu}(x,0)\di x^{\mu} ,\\
    B_{(\alpha)} &= B_{(\alpha)\mu\nu}(x,\vartheta)\di x^{\mu}\wedge \di x^{\mu} + && \,& \\
    &+2B_{(\alpha)\mu\upbeta}(x,\vartheta)\di x^\mu \wedge\di\vartheta^{\upbeta} +  && \,& \\
    &+B_{(\alpha)\upalpha\upbeta}(x,\vartheta)\di \vartheta^{\upalpha} \wedge\di\vartheta^{\upbeta}  &&\longmapsto& B_{(\alpha)\mu\nu}(x,0)\di x^{\mu}\wedge \di x^{\nu} ,\\
    \varphi\; &= \varphi(x,\vartheta) \;\;  &&\longmapsto& \varphi(x,0) ,\\
    \chi_{(\alpha)}\, &= \chi_{(\alpha)}(x,\vartheta)  &&\longmapsto& \chi_{(\alpha)}(x,0) ,\\
    h_{(\alpha\beta)} &= h_{(\alpha\beta)}(x,\vartheta)  &&\longmapsto& h_{(\alpha\beta)}(x,0).
    \end{aligned}
\end{equation}
\end{example}

\begin{definition}[Rheonomy]
Let us define the groupoid of rheonomic super-Cartan geometries as a subgroupoid $\Cartan_\mathfrak{Rh}(M)\subset\Cartan(M)$ which satisfies
\begin{equation}
    \iota^\ast\Cartan^{\mathfrak{g}}_\mathfrak{Rh}(M) \;\cong\; \Cartan^{\mathfrak{g}}(\bosonic{M}).
\end{equation}
We can now define the \textit{rheonomy extension mapping} by the inverse of $\iota^\ast$
\begin{equation}
    \mathfrak{Rh}: \,\Cartan^{\mathfrak{g}}(\bosonic{M}) \,\longrightarrow\; \Cartan^{\mathfrak{g}}_\mathfrak{Rh}(M) .
\end{equation}
\end{definition}

\noindent A sufficient condition for the subgroupoid $\Cartan_\mathfrak{Rh}(M)$ to be rheonomic is that the components of the curvature with at least one odd-graded index are linear combinations of the components of the curvature with all even-graded indices.

\begin{figure}[h]
\begin{center}\centering
\tikzset {_umfc898og/.code = {\pgfsetadditionalshadetransform{ \pgftransformshift{\pgfpoint{89.1 bp } { -108.9 bp }  }  \pgftransformscale{1.32 }  }}}
\pgfdeclareradialshading{_rge29gyb9}{\pgfpoint{-72bp}{88bp}}{rgb(0bp)=(1,1,1);
rgb(0bp)=(1,1,1);
rgb(25bp)=(0.62,0.62,0.62);
rgb(400bp)=(0.62,0.62,0.62)}
\tikzset {_r7eek9wa7/.code = {\pgfsetadditionalshadetransform{ \pgftransformshift{\pgfpoint{89.1 bp } { -108.9 bp }  }  \pgftransformscale{1.32 }  }}}
\pgfdeclareradialshading{_7586ca9p7}{\pgfpoint{-72bp}{88bp}}{rgb(0bp)=(1,1,1);
rgb(0bp)=(1,1,1);
rgb(25bp)=(0.62,0.62,0.62);
rgb(400bp)=(0.62,0.62,0.62)}
\tikzset {_hhutq5v7k/.code = {\pgfsetadditionalshadetransform{ \pgftransformshift{\pgfpoint{0 bp } { 0 bp }  }  \pgftransformrotate{0 }  \pgftransformscale{2 }  }}}
\pgfdeclarehorizontalshading{_cfg3rndzo}{150bp}{rgb(0bp)=(1,1,1);
rgb(37.5bp)=(1,1,1);
rgb(62.5bp)=(0.45,0.45,0.45);
rgb(100bp)=(0.45,0.45,0.45)}
\tikzset{_fxqy5hzl1/.code = {\pgfsetadditionalshadetransform{\pgftransformshift{\pgfpoint{0 bp } { 0 bp }  }  \pgftransformrotate{0 }  \pgftransformscale{2 } }}}
\pgfdeclarehorizontalshading{_xjlrcivuk} {150bp} {color(0bp)=(transparent!100);
color(37.5bp)=(transparent!100);
color(62.5bp)=(transparent!88);
color(100bp)=(transparent!88) } 
\pgfdeclarefading{_xn7948ndh}{\tikz \fill[shading=_xjlrcivuk,_fxqy5hzl1] (0,0) rectangle (50bp,50bp); } 
\tikzset{every picture/.style={line width=0.75pt}} 
\begin{tikzpicture}[x=0.75pt,y=0.75pt,yscale=-0.7,xscale=0.7]
\path  [shading=_rge29gyb9,_umfc898og] (516,146) .. controls (536,136) and (606,123) .. (641,137) .. controls (676,151) and (772,227) .. (674,218) .. controls (576,209) and (499,263) .. (479,233) .. controls (459,203) and (496,156) .. (516,146) -- cycle ; 
 \draw   (516,146) .. controls (536,136) and (606,123) .. (641,137) .. controls (676,151) and (772,227) .. (674,218) .. controls (576,209) and (499,263) .. (479,233) .. controls (459,203) and (496,156) .. (516,146) -- cycle ; 
\draw  [fill={rgb, 255:red, 193; green, 127; blue, 252 }  ,fill opacity=1 ] (589.78,122.97) -- (608.07,110.12) -- (661.46,119.48) -- (655.1,155.81) -- (636.8,168.65) -- (583.41,159.3) -- cycle ; \draw   (661.46,119.48) -- (643.16,132.32) -- (589.78,122.97) ; \draw   (643.16,132.32) -- (636.8,168.65) ;
\path  [shading=_7586ca9p7,_r7eek9wa7] (52,126) .. controls (72,116) and (142,103) .. (177,117) .. controls (212,131) and (308,207) .. (210,198) .. controls (112,189) and (35,243) .. (15,213) .. controls (-5,183) and (32,136) .. (52,126) -- cycle ; 
 \draw   (52,126) .. controls (72,116) and (142,103) .. (177,117) .. controls (212,131) and (308,207) .. (210,198) .. controls (112,189) and (35,243) .. (15,213) .. controls (-5,183) and (32,136) .. (52,126) -- cycle ; 
\draw  [color={rgb, 255:red, 0; green, 0; blue, 0 }  ,draw opacity=1 ][fill={rgb, 255:red, 74; green, 144; blue, 226 }  ,fill opacity=1 ] (126.84,109.27) -- (190.39,116.26) -- (180.23,153.06) -- (116.68,146.06) -- cycle ;
\path  [shading=_cfg3rndzo,_hhutq5v7k,path fading= _xn7948ndh ,fading transform={xshift=2}] (717,67.85) -- (717,238.15) .. controls (717,263.47) and (662.38,284) .. (595,284) .. controls (527.62,284) and (473,263.47) .. (473,238.15) -- (473,67.85)(717,67.85) .. controls (717,93.17) and (662.38,113.7) .. (595,113.7) .. controls (527.62,113.7) and (473,93.17) .. (473,67.85) .. controls (473,42.53) and (527.62,22) .. (595,22) .. controls (662.38,22) and (717,42.53) .. (717,67.85) -- cycle ; 
 \draw   (717,67.85) -- (717,238.15) .. controls (717,263.47) and (662.38,284) .. (595,284) .. controls (527.62,284) and (473,263.47) .. (473,238.15) -- (473,67.85)(717,67.85) .. controls (717,93.17) and (662.38,113.7) .. (595,113.7) .. controls (527.62,113.7) and (473,93.17) .. (473,67.85) .. controls (473,42.53) and (527.62,22) .. (595,22) .. controls (662.38,22) and (717,42.53) .. (717,67.85) -- cycle ; 
\draw  [fill={rgb, 255:red, 0; green, 0; blue, 0 }  ,fill opacity=1 ] (319,112.95) -- (369.5,112.95) -- (369.5,107) -- (389,115.5) -- (369.5,124) -- (369.5,118.05) -- (319,118.05) -- cycle ;
\draw  [fill={rgb, 255:red, 0; green, 0; blue, 0 }  ,fill opacity=1 ] (151.23,131.16) .. controls (151.23,129.89) and (152.26,128.86) .. (153.54,128.86) .. controls (154.81,128.86) and (155.84,129.89) .. (155.84,131.16) .. controls (155.84,132.43) and (154.81,133.46) .. (153.54,133.46) .. controls (152.26,133.46) and (151.23,132.43) .. (151.23,131.16) -- cycle ;
\draw  [fill={rgb, 255:red, 0; green, 0; blue, 0 }  ,fill opacity=1 ] (615.23,151.16) .. controls (615.23,149.89) and (616.26,148.86) .. (617.54,148.86) .. controls (618.81,148.86) and (619.84,149.89) .. (619.84,151.16) .. controls (619.84,152.43) and (618.81,153.46) .. (617.54,153.46) .. controls (616.26,153.46) and (615.23,152.43) .. (615.23,151.16) -- cycle ;
\draw  [fill={rgb, 255:red, 0; green, 0; blue, 0 }  ,fill opacity=1 ] (382,221.05) -- (331.5,221.05) -- (331.5,227) -- (312,218.5) -- (331.5,210) -- (331.5,215.95) -- (382,215.95) -- cycle ;
\draw [color={rgb, 255:red, 208; green, 2; blue, 27 }  ,draw opacity=1 ][line width=1.5]    (157.08,94.98) -- (153.54,128.86) ;
\draw [shift={(157.5,91)}, rotate = 95.98] [fill={rgb, 255:red, 208; green, 2; blue, 27 }  ,fill opacity=1 ][line width=0.08]  [draw opacity=0] (11.61,-5.58) -- (0,0) -- (11.61,5.58) -- cycle    ;
\draw (484,30) node    {$M^{d|\mathbf{N}}$};
\draw (26,125) node    {$M^{d}$};
\draw (347,95) node    {$\mathfrak{Rh}$};
\draw (166,127) node  [font=\footnotesize]  {$x$};
\draw (617,139) node  [font=\footnotesize]  {$( x ,\vartheta )$};
\draw (133,258) node   [align=left] {Manifold with spinor bundle};
\draw (600,306) node   [align=left] {Supermanifold};
\draw (215,113) node  [color={rgb, 255:red, 0; green, 100; blue, 225 }  ,opacity=1 ]  {$\;\cong \mathbb{R}^{d}$};
\draw (620,94) node  [color={rgb, 255:red, 144; green, 19; blue, 254 }  ,opacity=1 ]  {$\;\cong \mathbb{R}^{d|\mathbf{N}}$};
\draw (351,143) node  [font=\small] [align=left] {Rheonomy};
\draw (354,198) node    {$\leftsquigarrow $};
\draw (353,253) node  [font=\small] [align=left] {Bosonic};
\draw (182,86) node  [color={rgb, 255:red, 208; green, 2; blue, 27 }  ,opacity=1 ]  {$\cong \mathbf{N}$};
\end{tikzpicture}
\caption{A picture of rheonomy.}
\end{center}\end{figure}

\begin{definition}[Metric]
Given a super-Cartan geometry which satisfies rheonomy, we can define a metric by
\begin{equation}
    g \,=\, e_{(\alpha)}^\ast \eta_{(\alpha)},
\end{equation}
where $\eta_{(\alpha)}\in \bigodot^2T^\ast U_{\alpha}$ is just the local Lorentzian metric $\eta_{(\alpha)}:=\eta_{\mu\nu}\di x^\mu_{(\alpha)}\odot\di x^\nu_{(\alpha)}$.
\end{definition}

\begin{savequote}[7.5cm]
We must be clear that when it comes to atoms, language can be used only as in poetry.
  \qauthor{--- Niels Bohr}
\end{savequote}

\chapter{\label{app:3}Fundamentals of geometric quantisation}

\minitoc


\noindent In this appendix we provide an introduction to geometric quantisation, an approach to quantisation that is underpinned by the symplectic geometry of phase space. Its emergence in the 1970s from the work of Kostant and Souriau has produced a geometric approach to quantisation that provides numerous insights into the quantisation procedure. In particular, it showed how the  symplectomorphism-invariance of phase space is broken in naive quantisation methods, even though the physics is left invariant, and how the underlying symplectomorphism-invariance may be restored. \vspace{0.2cm}

\noindent In more mundane language, classical Hamiltonian physics is invariant under canonical transformations and yet the wavefunctions of quantum mechanics are functions of just half the coordinates of phase space and, thus, they break such invariance. A key part of quantum mechanics is that physics cannot depend on the choice of basis of wavefunctions. We can transform between the coordinate and momentum basis and the physics is invariant. In fact, the coordinate and momentum representations are mutually non-local and to move between different bases requires a non-local transformation (this is the Fourier transform, in a free theory).\vspace{0.2cm}

\noindent Much of this can be found in any one of the books by \cite{WeiGQ, Kostant, Souriau, Woodhouse:1980pa} or the recent review by \cite{Nair:2016ufy}.

\section{Classical physics as symplectic geometry}

\subsection{Hamiltonian mechanics}
Let us recall that a symplectic manifold $(\mathcal{P},\omega)$ is defined as a smooth manifold $\mathcal{P}$ equipped with a closed non-degenerate $2$-form $\omega\in\Omega^2(M)$, called the symplectic form.
In Hamiltonian mechanics, a \textit{classical system} $(\mathcal{P},\omega,H)$ is defined by a symplectic manifold $(\mathcal{P},\omega)$, describing the  \textit{phase space} of the system, and a smooth function $H\in\mathcal{C}^\infty(\mathcal{P})$, called the \textit{Hamiltonian}. In Newtonian terms, the phase space and the Hamiltonian encode respectively the kinematics and the dynamics of a classical system. The equations of motion are described Hamilton's equation as follows:
\begin{equation}
     \iota_{X_H}\omega \,=\, \di H.
\end{equation}
A vector field $X_H\in\mathfrak{X}(\mathcal{P})$ which solves the Hamilton equation is called \textit{Hamiltonian vector} for the Hamiltonian $H$. The flow of a Hamiltonian vector fields describes the motion of the classical system on the phase space. This means if we choose a starting point $\gamma_0\in M$ in phase space, the motion of the classical system will be given by the path
\begin{equation}
\begin{aligned}
    \gamma: \mathbb{R} \;&\longrightarrow\; \mathcal{P} \\
     \tau \;&\longmapsto\; \gamma(\tau) = e^{\tau X_H}\gamma_0  \, 
\end{aligned}
\end{equation}
where $\tau\in\mathbb{R}$ is a $1$-dimensional parameter. \vspace{0.15cm}

\noindent Locally, on an simply connected open subset $U\subset \mathcal{P}$, we can apply Poincar\'{e} lemma to the symplectic form and find $\omega = \di \theta$, where the local $1$-form $\theta\in\Omega^1(U)$ is called Liouville potential. The definition of the Liouville potential is gauge dependent, meaning that any other choice of potential $\theta' = \theta + \di \lambda$ with $\lambda\in\Coo(U)$ equally satisfies $\omega= \di \theta'$. 
Now, given a path $\gamma:\mathbb{R}\rightarrow \mathcal{P}$ on the phase space, we define the Lagrangian $L_H\in\Omega^1(\mathbb{R})$ by
\begin{equation}
    L_H \;=\; \gamma^\ast\theta - H\di\tau \, .
\end{equation}
Where we denote the pull-back of the Liouville one-form $\theta$ to the curve $\gamma$ by $\gamma^\ast\theta$.
The action $S_H[\gamma(\tau)]$ associated to such a Lagrangian will be given by
\begin{equation}
    S_H[\gamma(\tau)] \;=\; \int_\mathbb{R}(\gamma^\ast\theta - H\di\tau)  \, .
\end{equation}
For future use, let us notice that we can rewrite $\gamma^\ast\theta=\iota_{X_H}\theta\,\di\tau$, when restricted on the path $\gamma$. Thus note that the choice of Liouville potential effects the Lagrangian description.

\subsection{Classical algebra of observables}
An observable is defined as a smooth function $f\in\mathcal{C}^\infty(\mathcal{P})$ of the phase space.
Crucially, a symplectic manifold $(\mathcal{P},\omega)$ is canonically also a Poisson manifold $(\mathcal{P}, \{-,-\})$, where the Poisson bracket is given as follows:
\begin{equation}\label{eq:gq1}
    \{f,g\} \;:=\; \omega(X_f,X_g)
\end{equation}
for any pair of observables $f,g\in\mathcal{C}^\infty(\mathcal{P})$. In other words, this means that the observables of a symplectic manifold $(M,\omega)$ constitute a Poisson algebra:
\begin{equation}\label{eq:gq2}
    [X_f,X_g] \;=\; X_{\{f,g\}}  \, .
\end{equation}
Thus, the Hamiltonian vector field on a symplectic manifold $(\mathcal{P},\omega)$ constitute a Lie algebra, which we will denote as $\mathfrak{ham}(\mathcal{P},\omega)$.

\section{Geometric quantisation}

\subsection{Prequantum geometry}
Let us consider the Lie group $U(1)_\hbar:=\mathbb{R}/2\pi\hbar\mathbb{Z}$. The \textit{prequantum bundle} $\mathcal{Q}\twoheadrightarrow \mathcal{P}$ is defined as the principal $U(1)_\hbar$-bundle, whose first Chern class $\mathrm{c}_1(\mathcal{Q})\in H^2(M,\mathbb{Z})$ is the image of the element $[\omega]\in H^2(M,\mathbb{R})$ of the de Rham cohomology group. We can now define the associated bundle $\mathcal{E}\twoheadrightarrow \mathcal{P}$ to the prequantum bundle with fibre $\mathbb{C}$, i.e.
\begin{equation}
    \mathcal{E} \,:=\, \mathcal{Q}\times_{U(1)_\hbar}\!\mathbb{C},
\end{equation}
where the natural action $U(1)_\hbar\times \mathbb{C}\rightarrow\mathbb{C}$ is given by the map $(\phi,z)\mapsto e^{\frac{i}{\hbar}\phi}z$. Now, the \textit{prequantum Hilbert space} of the system is defined by
\begin{equation}
    \mathfrak{H}_{\mathrm{pre}}\;:=\; \mathrm{L}^{\!2}(\mathcal{P},\mathcal{E}),
\end{equation}
i.e. the Hilbert space of $\mathrm{L}^2$-integrable sections of the bundle $\mathcal{E}$ on the base manifold $\mathcal{P}$. Whenever the first Chern class of $\mathcal{Q}$ is trivial, then the bundle $\mathcal{E}=\mathcal{P}\times\mathbb{C}$ is trivial and the prequantum Hilbert space reduces to $\mathfrak{H}_{\mathrm{pre}} = \mathrm{L}^{\!2}(\mathcal{P};\mathbb{C})$, i.e. the Hilbert space of $\mathrm{L}^2$-integrable complex functions.

\subsection{Quantum algebra of observables}
In geometric quantisation, given a classical observable $f\in\Coo(\mathcal{P})$, we define a \textit{quantum observable} $\hat{f}\in\mathrm{Aut}(\mathfrak{H}_{\mathrm{pre}})$ by the expression
\begin{equation}
    \hat{f} \;:=\; -i\hbar \nabla_{V_f} + f,
\end{equation}
where $V_f\in\mathfrak{ham}(\mathcal{P},\omega)$ is the Hamiltonian vector of the Hamiltonian function $f\in\Coo(\mathcal{P})$ and $\nabla$ is the connection on $T\mathcal{P}$ given by the Liouville potential $\theta$.
By using equations \eqref{eq:gq1} and \eqref{eq:gq2}, we find that the commutator of two quantum observables closes. In particular, given $\hat{f}$ and $\hat{g}$, we obtain the commutator
\begin{equation}
    [\hat{f},\hat{g}] \;=\; i\hbar\, \widehat{\{f,g\}}
\end{equation}
where the classical observable of $\widehat{\{f,g\}}$ is the Poisson bracket $\{f,g\}$ of the classical observables $f$ and $g$. Thus, we can use this fact to define the Heisenberg Lie algebra $\mathfrak{heis}(\mathcal{P},\omega)$ of quantum observables on our phase space $(\mathcal{P},\omega)$.

\subsection{Quantum geometry}
Denote the tangent bundle of phase space by  $T\mathcal{P}$.
A \textit{polarisation} of the phase space $(\mathcal{P},\omega)$ is an involutive Lagrangian subbundle $L\subset T\mathcal{P}$, i.e. an $n$-dimensional subbundle of $T\mathcal{P}$ such that $\omega|_{L}=0$ and $[V,W]\subset L $ for any pair of vectors $V,W\in L$. \vspace{0.15cm}

\noindent The \textit{square root bundle} of a line bundle $\mathcal{B}\twoheadrightarrow\mathcal{M}$ is defined as a complex line bundle, which we will denote as $\sqrt{\mathcal{B}}\twoheadrightarrow\mathcal{M}$, equipped with a bundle isomorphism $\sqrt{\mathcal{B}}\otimes \sqrt{\mathcal{B}} \;\xrightarrow{\;\;\simeq\;\;}\; \mathcal{B}$ which sends sections $\sqrt{s}\in\Gamma(\mathcal{M},\sqrt{\mathcal{B}})$ to $\sqrt{s}\otimes\sqrt{s}\mapsto s\in\Gamma(\mathcal{M},\mathcal{B})$. \vspace{0.15cm}

\noindent Let us consider the determinant bundle $\mathrm{det}(L):=\wedge^nL^\ast_{\mathbb{C}}$ of a Lagrangian subbundle $L\subset T\mathcal{P}$ of our phase space, where $n=\mathrm{rank}(L)$. We need now to consider the square root bundle $\sqrt{\mathrm{det}(L)}$ of the determinant bundle, which comes equipped with the isomorphism
\begin{equation}
    \sqrt{\mathrm{det}(L)} \otimes \sqrt{\mathrm{det}(L)} \;\xrightarrow{\;\;\simeq\;\;}\;   \mathrm{det}(L)
\end{equation}
The choice of square root bundle $\sqrt{\mathrm{det}(L)}$ is also related to the \textit{metaplectic correction} which leads to the quantum theory forming a representation of the metaplectic group rather than the symplectic group.
The \textit{quantum Hilbert space} is defined by the following space of sections:
\begin{equation}\label{eq:quantumhilbert}
    \mathfrak{H} \;:=\; \Big\{\; \psi\in\mathrm{L}^{\!2}\big(\mathcal{P},\,\mathcal{E}\otimes\sqrt{\mathrm{det}(L)}\big) \;\Big|\; \nabla_V\psi=0\;\;\,\forall\,V\in L \; \Big\}.
\end{equation}
If the Lagrangian subbundle $L$ is integrable, we can write $L=T\mathcal{M}$ for some $n$-dimensional submanifold $\mathcal{M}\subset \mathcal{P}$ of the phase space. Then, quantum states $\ket{\psi}\in\mathfrak{H}$ can be uniquely chosen of the form
\begin{equation}
    \ket{\psi} = \psi\otimes \sqrt{\mathrm{vol}_\mathcal{M}},
\end{equation}
where $\psi\in\mathrm{L}^{\!2}(\mathcal{M},\mathcal{E})$ is a polarised section and $\sqrt{\mathrm{vol}_\mathcal{M}}\in\Gamma(\mathcal{P},\sqrt{\mathrm{det}(L)})$ is the half-form whose square is a fixed volume form $\mathrm{vol}_\mathcal{M}\in\Omega^n(\mathcal{M})$.
The inner product of the Hilbert space is given by the integral
\begin{equation}
    \Braket{\psi_1|\psi_2} \;=\; \int_\mathcal{M}\psi_1^\ast\psi_2\,\mathrm{vol}_\mathcal{M}
\end{equation}
for any couple of quantum states $\ket{\psi_1} = \psi_1\otimes \sqrt{\mathrm{vol}_\mathcal{M}}$ and $\ket{\psi_2} = \psi_2\otimes \sqrt{\mathrm{vol}_\mathcal{M}}\in\mathfrak{H}$. \vspace{0.15cm}

\noindent Crucially, the Hilbert space defined in \eqref{eq:quantumhilbert} does not depend on the choice of Lagrangian subbundle $L$. If we call $\mathfrak{H}_L$ the quantum Hilbert space polarised along the Lagrangian subbundle $L$ and $\mathfrak{H}_{L'}$ the one along another Lagrangian subbundle $L'$, we have a canonical isomorphism $\mathfrak{H}_L\cong\mathfrak{H}_{L'}$. At the end of this section we will explain why this is the case.

\begin{example}[Wave-functions of QM]
To illustrate the ideas in this section lets look at a simple example with $(M,\omega)=(\mathbb{R}^{2n},\,\di p_\mu \wedge \di x^\mu)$, where $\{p_\mu,x^\mu\}$ are Darboux coordinates on $\mathbb{R}^{2n}$. We can now choose the gauge $\theta = p_\mu\di x^\mu$ for the Liouville potential. We have two perpendicular polarisations defined by the Lagrangian fibrations $L_p:=\mathrm{Span}\!\left(\frac{\partial}{\partial p_\mu}\right)$ and $L_x:=\mathrm{Span}\!\left(\frac{\partial}{\partial x^\mu}\right)$. Recall that the covariant derivative is related to the Liouville potential by $\nabla_V = V - \frac{i}{\hbar}\iota_V\theta$. In our case, this implies
\begin{equation}
    \begin{aligned}
    \nabla_\frac{\partial}{\partial x^\mu} \;&=\; \frac{\partial}{\partial x^\mu} - \frac{i}{\hbar} p_\mu \\[0.4ex]
    \nabla_\frac{\partial}{\partial p_\mu} \;&=\; \frac{\partial}{\partial p_\mu}  .
    \end{aligned}
\end{equation}
Therefore, for the polarisation $L_p$ and $L_x$, we obtain respectively the sections 
\begin{equation}
    \begin{aligned}
    \Ket{\psi} \;&=\; \psi(p)e^{-ip_\mu x^\mu} \otimes \sqrt{\di^n p} \\
    \Ket{\psi} \;&=\; \psi(x) \otimes \sqrt{\di^n x},
    \end{aligned}
\end{equation}
where $\sqrt{\di^n p}$ is the half form such that $\sqrt{\di^n p}\otimes \sqrt{\di^n p}=\di^n p$ and analogously for $\sqrt{\di^n x}$.
\end{example}

\section{Polarisations and canonical transformations}

\subsection{Canonical transformations as symplectomorphisms}
Let us recall that a \textit{symplectomorphism} between two manifolds $(\mathcal{P},\omega)\xrightarrow{\;f\;}(\mathcal{P}',\omega')$ is a diffeomorphism $f:\mathcal{P}\rightarrow \mathcal{P}'$ which maps the symplectic form of the first manifold into the symplectic form of the second one, i.e. such that it satisfies $\omega = f^\ast\omega'$.
According to \cite{Arn89}, what in Hamiltonian physics is known under the name of \textit{canonical transformation} with generating function $F$ is equivalently a symplectomorphism $f:(\mathcal{P},\omega)\rightarrow(\mathcal{P}',\omega')$ such that the Liouville potential is gauge-transformed by $\theta-f^\ast\theta' = \di F$. However, this first formalisation can be significantly refined.

\subsection{Lagrangian correspondence}
Following \cite{WeinLag}, there exists a powerful way to formalise a canonical transformation by using the notion of \textit{Lagrangian correspondence}. To define a Lagrangian correspondence we first need to introduce the \textit{graph} of a symplectomorphism $f:(\mathcal{P},\omega)\rightarrow(\mathcal{P}',\omega')$, which is the submanifold of the product space $\mathcal{P}\times \mathcal{P}'$ given by
\begin{equation}
    \Gamma_f \;:=\; \big\{ (a,b)\in \mathcal{P}\times\mathcal{P}' \;\big|\; b=f(a)\big\}.
\end{equation} 
Let us call $\iota:\Gamma_f\hookrightarrow\mathcal{P}\times \mathcal{P}'$ the inclusion in the product space. Now, a Lagrangian correspondence is defined a correspondence diagram of the form
\begin{equation}\label{diag:lagrangiancorr}
    \begin{tikzcd}[row sep={13ex,between origins}, column sep={13ex,between origins}]
    &  (\mathcal{P}\times \mathcal{P}',\, \pi^\ast\omega-\pi^{\prime\ast}\omega') \arrow[rd, "\pi'", two heads]\arrow[ld, "\pi"', two heads] & \\
    (\mathcal{P},\,\omega) \arrow[rr, "f"] && (\mathcal{P}',\,\omega')
    \end{tikzcd}
\end{equation}
where $f$ is a symplectomorphism and $\pi,\pi'$ are the canonical projections of $\mathcal{P}\times \mathcal{P}'$ onto $\mathcal{P}, \mathcal{P}'$ respectively. The submanifold $\Gamma_f\subset\mathcal{P}\times\mathcal{P}'$ can be immediately recognised as a Lagrangian submanifold of $(\mathcal{P}\times \mathcal{P}',\, \pi^\ast\omega-\pi^{\prime\ast}\omega')$, i.e. the total symplectic form vanishes when restricted on $\Gamma_f$. In other words, we have
\begin{equation}
    \iota^\ast\!\left(\pi^\ast\omega-\pi^{\prime\ast}\omega'\right) \;=\; 0.
\end{equation}
To formalise a canonical transformation, we need to add another condition: the correspondence space $(\mathcal{P}\times \mathcal{P}',\, \pi^\ast\omega-\pi^{\prime\ast}\omega')$ must be symplectomorphic to a symplectic manifold $(T^\ast\mathcal{M},\, \omega_\mathrm{can})$ for some manifold $\mathcal{M}$, where $\omega_\mathrm{can}\in\Omega^2(T^\ast\mathcal{M})$ is just the canonical symplectic form of the cotangent bundle.  \vspace{0.15cm}

\noindent This implies that we can write the combination of Liouville potentials $\pi^\ast\theta - \pi^{\prime\ast}\theta^{\prime}$ as the Liouville 1-form on $\mathcal{P}\times \mathcal{P}'\cong T\mathcal{M}$. Since $\Gamma_f$ is Lagrangian, the Liouville potential can be trivialised on $T\Gamma_f$. In other words, we have the equation
\begin{equation}
    \pi^\ast\theta - \pi^{\prime\ast}\theta^{\prime} \;=\; \di (F\circ \Pi) \qquad \text{on }\;T\Gamma_f,
\end{equation}
where $\Pi:T^\ast\mathcal{M}\twoheadrightarrow\mathcal{M}$ is the canonical projection and where the function $F\in\Coo(\mathcal{M})$ can be interpreted as the \textit{generating function} of the canonical transformation associated to the symplectomorphism $f$.

\begin{example}[Canonical transformations]
For clarity, let us consider a simple example. Let us start from symplectic manifolds which are cotangent bundles of configuration spaces, i.e. $\mathcal{P}=T^\ast M$ and $\mathcal{P}'=T^\ast M'$. Thus we can write the Liouville potential as 
\begin{equation}\label{eq:extype1}
    p_\mu \di x^\mu -  p'_\mu\di x^{\prime\mu} \,=\, \di F
\end{equation}
in local coordinates on the correspondence space $\mathcal{P}\times\mathcal{P}' = T^\ast(M\times M')$. We immediately notice that, in the notation of the previous paragraph, we have $\mathcal{M}:=M\times M'$. Now the generating function $F=F(x,x')$ of a canonical transformation can be properly seen as the pullback of a function of the product manifold $M\times M'$. Equation \eqref{eq:extype1} can be equivalently written as
\begin{equation}
    p_\mu \,=\, \frac{\partial F}{\partial x^\mu}, \qquad p'_\mu \,=\, -\frac{\partial F}{\partial x^{\prime\mu}}.
\end{equation}
In particular, If we choose $M,M'=\mathbb{R}^d$ and $F(x,x')= \delta_{\mu\nu} x^\mu x^{\prime\nu}$, we recover the symplectic linear transformation $(x,p)\mapsto f(x,p) = (p,-x)$.
\end{example}

\subsection{Canonical transformation on the Hilbert space}
So far we formalised canonical transformations as symplectomorphisms. Now, we need to show how these symplectomorphisms give rise to isomorphisms of the corresponding quantum Hilbert spaces. \vspace{0.15cm}

\noindent First of all, we must fix a symplectomorphism $f:\mathcal{P}\rightarrow\mathcal{P}'$, then we must choose two polarisations $L\subset T\mathcal{P}$ and $L'\subset T\mathcal{P}'$ which satisfy $L=f^\ast(L')$. Let us call $\mathfrak{H}_{L}$ and $\mathfrak{H}_{L'}$ the quantum Hilbert spaces corresponding respectively to the $L$ and $L'$ polarisations of the phase space. \vspace{0.15cm}

\noindent Now, notice that $T\Gamma_f\subset T(\mathcal{P}\times\mathcal{P}')$ is a Lagrangian submanifold of the Lagrangian correspondence space $(\mathcal{P}\times\mathcal{P}',\,\pi^\ast\omega- \pi^{\prime\ast}\omega')$.
As observed by \cite{WeiGQ}, The $T\Gamma_f$-polarised Hilbert space $\mathfrak{H}_{T\Gamma_{\!f}}$ of the correspondence space $(\mathcal{P}\times\mathcal{P}',\,\pi^\ast\omega- \pi^{\prime\ast}\omega')$ is isomorphic to the topological tensor product $\mathfrak{H}_{T\Gamma_{\!f}}\cong\mathfrak{H}_{L}\,\widehat{\otimes}\, \mathfrak{H}_{L'}^\ast$, which is nothing but the space of Hilbert–Schmidt operators $\mathfrak{H}_{L'}\longrightarrow \mathfrak{H}_{L}$.
Then, we will obtain the following diagram:
\begin{equation}\label{diag:hilbertcorr}
    \begin{tikzcd}[row sep={10ex,between origins}, column sep={10ex,between origins}]
    &  \mathfrak{H}_{T\Gamma_{\!f}}  \arrow[rd, "\pi^{\prime\ast}", hookleftarrow]\arrow[ld, "\pi^\ast"', hookleftarrow] & \\
    \mathfrak{H}_{L}  && \mathfrak{H}_{L'} \arrow[ll, "f^\ast"]
    \end{tikzcd}
\end{equation}
Now we can lift sections $\psi\in\mathfrak{H}_{L}$ and $\psi'\in\mathfrak{H}_{L'}$ to the Hilbert space $\mathfrak{H}_{T\Gamma_f}$ and consider their products $\Braket{\pi^\ast\psi|\pi^{\prime\ast}\psi'}$ in this space. This, then, naturally defines a pairing $(\!(\;\cdot\;,\;\cdot\;)\!):\mathfrak{H}_{L}\times \mathfrak{H}_{L'}\rightarrow\mathbb{C}$ between the two polarised Hilbert spaces given by
\begin{equation}
    (\!(\;\cdot\;,\;\cdot\;)\!)\; :=\; \Braket{\pi^\ast\;\cdot\;|\pi^{\prime\ast}\;\cdot\;}
\end{equation}
But any such pairing is equivalently a linear isomorphism  $f^\ast:\mathfrak{H}_{L'} \xrightarrow{\;\cong\;} \mathfrak{H}_{L}$ such that
\begin{equation}
    (\!(\;\cdot\;,\;\cdot\;)\!)\; =\; \Braket{\;\cdot\;|f^\ast\;\cdot\;}
\end{equation}
where this time the product on the right hand side is the hermitian product of the first Hilbert space $\mathfrak{H}_{L}$. \vspace{0.15cm}

\noindent Let us workout what this means in coordinates.
Recall that on $T\Gamma_f\subset T(\mathcal{P}\times\mathcal{P}')$ we have the gauge transformation $\pi^{\prime\ast}\theta^\prime = \pi^\ast\theta - \di F $, where $F(x,x')$ is the generating function of the symplectomorphism. Therefore, wave-functions $\psi(x)$ of $\mathfrak{H}_L$ will be lifted by $\psi(x)\mapsto \psi(x,x')=\psi(x)$ and wave-functions $\psi'(x')$ of $\mathfrak{H}_{L'}$ will be lifted by $\psi'(x')\mapsto \psi'(x,x')=\psi(x)e^{-\frac{i}{\hbar}F(x,x')}$ to wave-functions of $\mathfrak{H}_{T\Gamma_f}$.
Thus, the pairing will be given by
\begin{equation}
    (\!(\psi,\psi')\!) \;=\; \int_\mathcal{M}\di^n x\,\di^n x^{\prime }\, \psi^\dagger(x)\psi'(x')e^{-\frac{i}{\hbar}F(x,x')}
\end{equation}
where we called $\mathcal{M}$ the manifold such that $T^\ast\mathcal{M}\cong \mathcal{P}\times \mathcal{P}$. Finally the isomorphism $f^\ast:\mathfrak{H}_{L}\rightarrow \mathfrak{H}_{L'}$ induced by the diffeomorphism $f$ will be given in coordinates by
\begin{equation}
    (f^\ast\psi')(x) \;=\; \int_{M'}\di^nx\, \psi'(x')e^{-\frac{i}{\hbar} F(x,x')}
\end{equation}
where we called $M'$ the manifold such that $T^\ast M' \cong \mathcal{P}'$. Therefore we have a natural isomorphism $\mathfrak{H}_L\cong\mathfrak{H}_{L'}$ any time there is a canonical transformation mapping the Lagrangian subbundle $L$ into $L'$ and thus we are allowed to write just $\mathfrak{H}$ for the Hilbert space of a quantum system, without specifying the polarisation. We will write just
\begin{equation}
    \Ket{\psi} \,\in\,\mathfrak{H}
\end{equation}
for an abstract element of the Hilbert space, independent from the polarisation.

\begin{example}[Quantum canonical transformations]
To give some intuition for this idea, let us consider a simple example. Choose $M,M'=\mathbb{R}^n$ and let the symplectomorphism $f:(\mathbb{R}^{2n},\,\di p_\mu\wedge \di x^\mu)\rightarrow(\mathbb{R}^{2n},\,\di p'_\mu\wedge \di x^{\prime\mu})$ be the linear transformation $f(x,p)=(p,-x)$. This is generated by generating function $F(x,x')=\delta_{\mu\nu}x^\mu x^{\prime\nu}$. Thus, if we substitute $(x',p')=f(x,p)= (p,-x)$, we recover that $(f^\ast)^{-1}$ is exactly the Fourier transformation of wave-functions:
\begin{equation}
    \begin{aligned}
    (f^\ast\psi')(x) \;&=\; \int_{M'}\di^np\, \psi'(p)e^{-\frac{i}{\hbar} p_\mu x^\mu} \\
    \left((f^\ast)^{-1}\psi\right)\!(p) \;&=\; \int_M\di^nx\, \psi(x)e^{\frac{i}{\hbar} p_\mu x^\mu}
    \end{aligned}
\end{equation}
Thus the same quantum state $\Ket{\psi} \,\in\,\mathfrak{H}$ can be represented as a wave-function $\Braket{x|\psi}=\psi(x)$ or as its Fourier transform $\Braket{p|\psi}=\psi(p)$ in the two basis $\big\{\Bra{x}\big\}_{x\in M}$ and $\big\{\Bra{p}\big\}_{p\in M'}$ given by the Lagrangian correspondence.
\end{example}


\setlength{\baselineskip}{0pt} 


{\renewcommand*\MakeUppercase[1]{#1}%
\printbibliography[heading=bibintoc,title={\bibtitle}]}

\end{document}